\documentclass[11pt, a4paper, twoside]{report}
\usepackage[utf8]{inputenc}
\usepackage[english]{babel}
\usepackage{graphicx}
\usepackage{amsmath}
\usepackage{amssymb}
\usepackage{amsfonts}
\usepackage{color}
\usepackage{MnSymbol}
\usepackage{eurosym}
\usepackage{cancel}
\usepackage{fancyhdr}
\usepackage{hyperref}\hypersetup{
colorlinks=true,
linkcolor=blue,
urlcolor=blue,
anchorcolor=blue,
citecolor=blue}
\usepackage{physics}
\usepackage{comment}
\usepackage{float}
\usepackage{subfig}
\usepackage[font={small},labelfont=bf]{caption}
\usepackage{chngpage}
\usepackage{dsfont}
\usepackage{bm}
\usepackage{booktabs}
\usepackage{empheq}
\usepackage{textcomp}
\usepackage{multirow}
\usepackage{array, makecell}
\usepackage{ragged2e}
\usepackage{lipsum} 
\usepackage[textwidth=20mm,textsize=small]{todonotes}
\usepackage{relsize}
\usepackage{pifont}
\usepackage{shuffle}

\usepackage{tcolorbox}
\usepackage{nicematrix}
\NiceMatrixOptions{margin=5pt} %This is to set the space between the parenthesis and the entries in the matrix

\usepackage{tikz,pgfplots}
\usepackage{tikz-feynman}
\usetikzlibrary{shapes,arrows,automata,positioning,backgrounds,tqft}

\definecolor{DarkRed}{RGB}{185,22,22}

\definecolor{darkgreen}{rgb}{0.0, 0.5, 0.0}

\usepackage{etoolbox}
\usepackage{cite}
\apptocmd{\thebibliography}{\justifying}{}{} 

\numberwithin{equation}{chapter}
\counterwithin{figure}{chapter}
\counterwithin{table}{chapter}

\usepackage[top=3cm,bottom=3cm,left=2.5cm,right=2.5cm,asymmetric]{geometry}

\usepackage{fancyhdr}
\usepackage[framemethod=tikz]{mdframed} 
\mdfdefinestyle{info}{%
  topline=false, bottomline=false,
  leftline=false, rightline=false,
  nobreak,
  singleextra={%
    \fill[black](P-|O)circle[radius=0.4em];
    \node at(P-|O){\color{white}\scriptsize\bf i};
    \draw[very thick](P-|O)++(0,-0.8em)--(O);%--(O-|P);
  }
}

\newenvironment{info}[1][Info:]{ % Set the default title to "Info:"
  \medskip
  \begin{mdframed}[style=info]
    \noindent{\textbf{#1}}
  }{
  \end{mdframed}
}

\newenvironment{dedication}
  {\thispagestyle{empty}% no header and footer
   \vspace*{\stretch{1}}% some space at the top
   \itshape             % the text is in italics
   \raggedleft          % flush to the right margin
  }
  {\par % end the paragraph
   \vspace{\stretch{3}} % space at bottom is three times that at the top
   \clearpage           % finish off the page
  }

\usetikzlibrary{arrows}
\newcommand{\midarrow}{\tikz \draw [-{Stealth[length=2.4mm]}](0,0) -- (.1,0);}

\newcommand{\nint}{{n_{\text{int}}}}
\newcommand{\nex}{{n_{\text{ext}}}}
\newcommand{\nISP}{{n_{\text{ISP}}}}
\newcommand{\LS}{{\text{LS}}}

\newcommand{\integralFinal}{\mathcal{J}}
\newcommand{\integralprec}{\mathcal{I}}
\newcommand{\integralseed}{\mathcal{I}_{\text{seed}}}
\newcommand{\sectora}{(\text{a})}
\newcommand{\sectorb}{(\text{b})}
\newcommand{\sectorc}{(\text{c})}
\newcommand{\sectord}{(\text{d})}

\newcommand{\RationalizeRoots}{\texttt{\textup{RationalizeRoots}}}
\newcommand{\DlogBasis}{\texttt{\textup{DlogBasis}}}
\newcommand{\Maple}{\texttt{\textup{Maple}}}

\newcommand{\HyperInt}{\texttt{\textup{HyperInt}}}
\newcommand{\Mathematica}{\texttt{\textup{Mathematica}}}

\newcommand{\Ef}[3]{{\textrm{E}_4}(\begin{smallmatrix}#1\\#2%
\end{smallmatrix};#3)} 
\newcommand{\cEf}[3]{{\mathcal{E}_4}(\begin{smallmatrix}#1\\#2%
\end{smallmatrix};#3)} 
\newcommand{\gamt}[3]{{\widetilde{\Gamma}}(\begin{smallmatrix}#1\\#2%
\end{smallmatrix};#3)}

\newcommand{\diagramnumberingsign}{}
\usepackage{array}
\newcommand{\PreserveBackslash}[1]{\let\temp=\\#1\let\\=\temp}
\newcolumntype{C}[1]{>{\PreserveBackslash\centering}p{#1}}
\newcolumntype{R}[1]{>{\PreserveBackslash\raggedleft}p{#1}}
\newcolumntype{L}[1]{>{\PreserveBackslash\raggedright}p{#1}}

\newcommand{\drawbox}{%
\begin{tikzpicture}[scale=0.25, label distance=-1mm,baseline={([yshift=-0.1cm]current bounding box.center)}]
\clip (-1,-3) rectangle (3,1);
\node (p1) at (2,1) {};
\node (p2) at (3,0) {};
\node (p3) at (3,-2) {};
\node (p4) at (2,-3) {};
\node (p5) at (0,-3) {};
\node (p6) at (-1,-2) {};
\node (p7) at (-1,0) {};
\node (p8) at (0,1) {};
\draw[line width=0.225mm]  (p1.center) edge (p4.center);
\draw[line width=0.225mm]  (p5.center) edge (p8.center);
\draw[line width=0.225mm]  (p2.center) edge (p7.center);
\draw[line width=0.225mm]  (p3.center) edge (p6.center);
\end{tikzpicture}
}

\newcommand{\drawdbten}{%
\begin{tikzpicture}[scale=0.4,label distance=-1mm,baseline={([yshift=-.5ex]current bounding box.center)}]
\clip (-0.25,13.75) rectangle (3.25,16.25);
		\node (0) at (0.5, 15.5) {};
		\node (1) at (1.5, 16.25) {};
		\node (4) at (2.5, 15.5) {};
		\node (5) at (0.5, 14.5) {};
		\node (7) at (1.5, 13.75) {};
		\node (8) at (2.5, 14.5) {};
        \node (11) at (5, 15) {};
		\node (12) at (6, 15) {};
		\draw[line width=0.225mm] (0.center) to (4.center);
		\draw[line width=0.225mm] (5.center) to (8.center);
		\draw[line width=0.225mm] (0.center) to (5.center);
		\draw[line width=0.225mm] (1.center) to (7.center);
		\draw[line width=0.225mm] (4.center) to (8.center);
		\draw[line width=0.225mm] (0.5,15.5) to (-0.25, 15.8);
		\draw[line width=0.225mm] (0.5,15.5) to (0.2, 16.25);
		\draw[line width=0.225mm] (0.5,14.5) to (-0.25, 14.2);
		\draw[line width=0.225mm] (0.5,14.5) to (0.2, 13.75);
		\draw[line width=0.225mm] (2.5,15.5) to (3.25, 15.8);
		\draw[line width=0.225mm] (2.5,15.5) to (2.95, 16.25);
		\draw[line width=0.225mm] (2.5,14.5) to (3.25, 14.2);
		\draw[line width=0.225mm] (2.5,14.5) to (2.95, 13.75);
\end{tikzpicture}%
}

\newcommand{\drawdb}{%
\begin{tikzpicture}[scale=0.4,label distance=-1mm,baseline={([yshift=-.5ex]current bounding box.center)}]
\clip (-0.25,13.75) rectangle (3.25,16.25);
		\node (0) at (0.5, 15.5) {};
		\node (1) at (1.5, 16.25) {};
		\node (4) at (2.5, 15.5) {};
		\node (5) at (0.5, 14.5) {};
		\node (7) at (1.5, 13.75) {};
		\node (8) at (2.5, 14.5) {};
        \node (11) at (5, 15) {};
		\node (12) at (6, 15) {};
		\draw[line width=0.225mm] (0.center) to (4.center);
		\draw[line width=0.225mm] (5.center) to (8.center);
		\draw[line width=0.225mm] (0.center) to (5.center);
		\draw[line width=0.225mm] (4.center) to (8.center);
		\draw[line width=0.225mm] (0.5,15.5) to (-0.25, 15.8);
		\draw[line width=0.225mm] (0.5,15.5) to (0.2, 16.25);
		\draw[line width=0.225mm] (0.5,14.5) to (-0.25, 14.2);
		\draw[line width=0.225mm] (0.5,14.5) to (0.2, 13.75);
        \draw[line width=0.225mm] (1.5,14.5) to (1.5, 15.5);
        \draw[line width=0.225mm] (1.5,14.5) to (1.2, 13.75);
        \draw[line width=0.225mm] (1.5,14.5) to (1.8, 13.75);
        \draw[line width=0.225mm] (1.5, 15.5) to (1.2, 16.25);
        \draw[line width=0.225mm] (1.5, 15.5) to (1.8, 16.25);
		\draw[line width=0.225mm] (2.5,15.5) to (3.25, 15.8);
		\draw[line width=0.225mm] (2.5,15.5) to (2.95, 16.25);
		\draw[line width=0.225mm] (2.5,14.5) to (3.25, 14.2);
		\draw[line width=0.225mm] (2.5,14.5) to (2.95, 13.75);
\end{tikzpicture}%
}

\newcommand{\drawhexten}{%
\begin{tikzpicture}[scale=0.37,label distance=-1mm,baseline={([yshift=-.5ex]current bounding box.center)}]
\clip (4.42,13.77) rectangle (7.02,16.49);
       \draw [line width=0.225mm] (5.027,15.531) -- (5.027,14.731) -- (5.720,14.331) -- (6.413,14.731) -- (6.413,15.531) -- (5.720,15.931) -- (5.027,15.531);
       \draw [line width=0.225mm] (5.027,15.531) -- (4.72,16.06);
       \draw [line width=0.225mm] (5.027,15.531) -- (4.42,15.53);
       \draw [line width=0.225mm] (5.027,14.731) -- (4.42,14.73);
       \draw [line width=0.225mm] (5.027,14.731) -- (4.72,14.21);
       \draw [line width=0.225mm] (5.720,14.331) -- (5.72,13.77);
       \draw [line width=0.225mm] (6.413,14.731) -- (6.72,14.21);
       \draw [line width=0.225mm] (6.413,14.731) -- (7.02,14.73);
       \draw [line width=0.225mm] (6.413,15.531) -- (7.02,15.53);
       \draw [line width=0.225mm] (6.413,15.531) -- (6.72,16.06);
       \draw [line width=0.225mm] (5.720,15.931) -- (5.72,16.49);
		\end{tikzpicture}
 }

 \newcommand{\drawhex}{%
\begin{tikzpicture}[scale=0.37,label distance=-1mm,baseline={([yshift=-.5ex]current bounding box.center)}]
\clip (4.42,13.77) rectangle (7.02,16.49);
       \draw [line width=0.225mm] (5.027,15.531) -- (5.027,14.731) -- (5.720,14.331) -- (6.413,14.731) -- (6.413,15.531) -- (5.720,15.931) -- (5.027,15.531);
       \draw [line width=0.225mm] (5.027,15.531) -- (4.72,16.06);
       \draw [line width=0.225mm] (5.027,15.531) -- (4.42,15.53);
       \draw [line width=0.225mm] (5.027,14.731) -- (4.42,14.73);
       \draw [line width=0.225mm] (5.027,14.731) -- (4.72,14.21);
       \draw [line width=0.225mm] (5.720,14.331) -- (5.42,13.81);
       \draw [line width=0.225mm] (5.720,14.331) -- (6.02,13.81);
       \draw [line width=0.225mm] (6.413,14.731) -- (6.72,14.21);
       \draw [line width=0.225mm] (6.413,14.731) -- (7.02,14.73);
       \draw [line width=0.225mm] (6.413,15.531) -- (7.02,15.53);
       \draw [line width=0.225mm] (6.413,15.531) -- (6.72,16.06);
       \draw [line width=0.225mm] (5.720,15.931) -- (6.02,16.46);
       \draw [line width=0.225mm] (5.720,15.931) -- (5.42,16.46);
		\end{tikzpicture}
}

\begin{document}

\setcounter{tocdepth}{1}

\thispagestyle{empty}
\begin{picture}(0,0)\put(-87.1,-754){\includegraphics{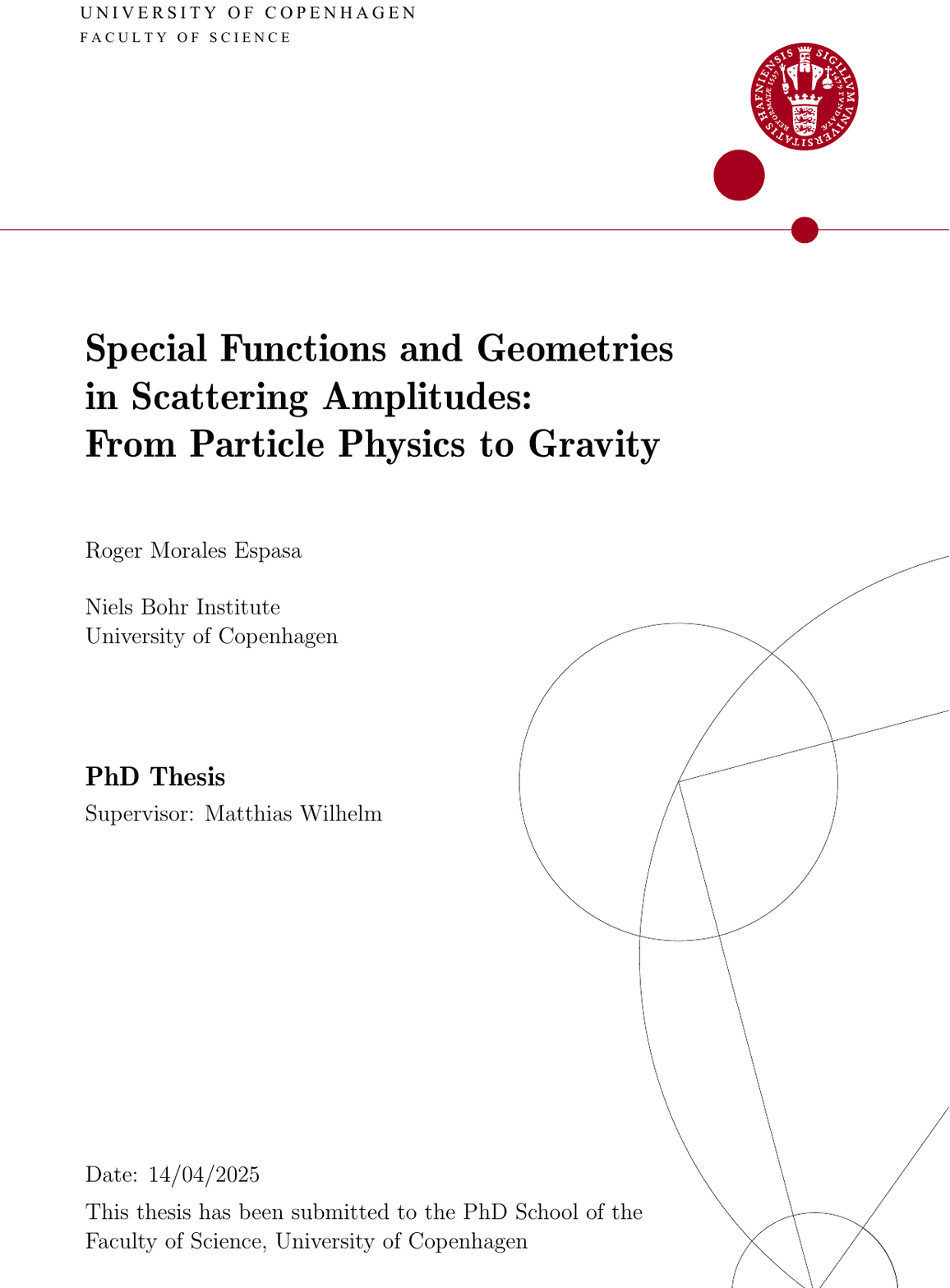}}\end{picture}%
   \clearpage

\newpage\null\thispagestyle{empty}
\noindent Special Functions and Geometries in Scattering Amplitudes: From Particle Physics to Gravity

\noindent Roger Morales Espasa

\vspace{0.5cm}

\noindent PhD Thesis

\vspace{0.5cm}

\noindent Theoretical Particle Physics and Cosmology

\noindent Niels Bohr International Academy

\noindent Niels Bohr Institute

\noindent University of Copenhagen

\vspace{0.5cm}

\noindent Supervisor:

\noindent Assoc. Prof. Matthias Wilhelm

\vspace{0.5cm}

\noindent Co-supervisor:

\noindent Prof. Poul Damgaard

\vspace{0.5cm}

\noindent Assessment committee:

\noindent Assoc. Prof. Emil Bjerrum-Bohr

\noindent Prof. Radu Roiban

\noindent Prof. Lorenzo Tancredi

\newpage
\thispagestyle{empty}

\section*{Abstract}
This thesis focuses on the fields of scattering amplitudes and Feynman integrals, with an emphasis on the geometries and special functions that they involve, and is devoted to two distinct research directions. In the first half of the thesis, we explore elliptic Feynman integrals in maximally supersymmetric Yang--Mills ($\mathcal{N}=4$ SYM) theory. In particular, we study elliptic generalizations of ladder diagrams, identifying the first two families of Feynman integrals involving the same elliptic curve to all loop orders, which provide a great testing ground for developing mathematical tools that can facilitate the calculation of elliptic integrals. In this direction, we initiate the symbol bootstrap for elliptic Feynman integrals, for which we generalize the so-called Schubert analysis to predict elliptic symbol letters. As a proof of principle, we obtain for the first time the symbol of the two-loop twelve-point elliptic double-box integral, resulting in a compact, one-line formula. 

In the second half of the thesis, we pioneer a systematic investigation of the Feynman integral geometries that are relevant to the study of gravitational waves emitted during the inspiral phase of black-hole mergers within the post-Minkowskian expansion of classical gravity. Specifically, we classify the geometries and special functions appearing in the expansion up to four loops. Among other findings, we identify the first Calabi--Yau three-dimensional geometry relevant to gravitational-wave physics. Subsequently, we study the Feynman integral giving rise to this Calabi--Yau geometry in more detail, and solve it by bringing its differential equation into canonical form -- a step that required developing a new method that accommodates for apparent singularities.

\newpage
\thispagestyle{empty}
\section*{Resumé}
Denne afhandling fokuserer på spredningsamplituder og Feynman-integraler med vægt på de geometrier og specielle funktioner, de involverer, og den dækker to forskellige videnskabelige spor. I den første halvdel af afhandlingen udforsker vi elliptiske Feynman-integraler i maksimalt supersymmetrisk Yang--Mills-teori ($\mathcal{N}=4$ SYM). Især studerer vi elliptiske generaliseringer af stigediagrammer og identificerer de første to familier af Feynman-integraler, der involverer den samme elliptiske kurve til alle loop-ordener. Disse udgør en værdifuld testarena for udviklingen af matematiske værktøjer, der kan lette beregningen af elliptiske integraler. I den sammenhæng initierer vi symbol-bootstrap for elliptiske Feynman-integraler, og vi generaliserer den såkaldte Schubert-analyse til at forudsige elliptiske symbolbogstaver. Som det vigtigste eksempel bestemmer vi for første gang symbolet for det elliptiske tolvpunkts to-loop dobbeltboksintegral, hvilket resulterer i en kompakt énlinjesformel.

I anden halvdel af afhandlingen foretager vi, for første gang, en systematisk undersøgelse af Feynman-integralgeometrier, der er relevante for studiet af gravitationsbølger udsendt under inspiralfasen af sorte hullers sammensmeltning i den post-Minkowskiske tilgang til klassisk tyngdekraft. Mere præcist klassificerer vi de geometrier og specielle funktioner, der optræder op til og med fire loops. Et af vores resultater er fundet af den første tredimensionelle Calabi--Yau-geometri med relevans for gravitationsbølgefysik. Derefter undersøger vi det Feynman-integral, der indeholder denne Calabi--Yau-geometri, i større detalje, og vi løser det ved at bringe dets differentialligning på kanonisk form -- et skridt, der nødvendiggjorde udviklingen af en ny metode til håndtering af tilsyneladende singulariteter.

\newpage
\thispagestyle{empty}
\section*{Acknowledgments}
First and foremost, I would like to thank my supervisor Matthias Wilhelm for helping me grow both as a researcher and as a person. Thank you for your guidance and motivation, our relaxed conversations, and for the thorough proofreading of this thesis. You have been a fantastic PhD supervisor, and I wish the best to you and your lovely family in Odense.

I am also indebted to the NBI amplitudes group, particularly to Hjalte Frellesvig, Robin Marzucca, Andrew McLeod, Anne Spiering, Cristian Vergu, Matt von Hippel and Chi Zhang. Thank you for creating such a friendly and welcoming group environment, and for making these years in Denmark particularly enjoyable. Most of the results in this thesis stem from our enlightening office discussions. Similarly, I am grateful to Daniel Brammer, Sebastian Pögel, Florian Seefeld, Stefan Weinzierl and Qinglin Yang for the discussions and fruitful collaborations.

I would also like to express my gratitude towards the NBIA, and especially to Emil Bjerrum-Bohr, Jacob Bourjaily, Adam Chalabi, Gang Chen, Poul Damgaard, Troels Harmark, Charlotte Kristjansen, Zhengwen Liu, Andres Luna, Niels Obers and Marcos Skowroneck for the support, the discussions and, of course, for the Friday sushi together.

I sincerely thank Zvi Bern and the Mani L. Bhaumik Institute at UCLA for their hospitality during my three-month visit, as well as Juan Pablo Gatica, Enrico Herrmann, Callum Jones, Seolhwa Kim, Richard Myers, Michael Ruf and Anna Wolz for making my visit a wonderful experience. I am particularly thankful to Zvi, Enrico and Michael for the insightful discussions. I am also grateful to Lance Dixon and Andrew McLeod for the hospitality received at SLAC and CERN, respectively. 

In addition, I am deeply grateful to Henriette Elvang for our enlightening discussions, and for the opportunity to continue my academic career as a postdoc at the University of Michigan. I look forward to continue working together in Ann Arbor, as well as collaborating with Justin Berman, Aditi Chandra, Nicholas Geiser and Loki Lin.  

I would also like to thank Piotr Bargieła, Ekta Chaubey, Mathieu Giroux, Gustav Jakobsen, Sara Maggio, Franziska Porkert, Benjamin Sauer, Yoann Sohnle, Sven Stawinski and Philip Velie for the numerous discussions, hilarious lunch conversations and for making research and conferences much more enjoyable.

I am also fortunate to have a loving and supportive family, and I am especially grateful to my parents and grandparents for their encouragement and motivation. Similarly, I am very grateful to Àurea Carrizo, Marta Carrizo, Raúl Morral, Carla Muñoz, Carlos Ruiz, Alba Torras and Santiago Vallés for the emotional support and priceless memories over these years. 

Finally, I am profoundly grateful to my wife Mercè Roig -- my PhD partner, cycling partner, and life partner -- for the unconditional support, motivation, and a meticulous proofreading of the thesis. Sharing the adventure of doing the PhD together has been an absolute joy.

\newpage
\begin{dedication}
    Dedicada a mi abuela y a mi abuelo
\end{dedication}

\newpage
\thispagestyle{empty}
\section*{List of publications}
This thesis is based on the following published papers and preprints, as indicated in each chapter:
\begin{itemize}
    \item[\cite{Morales:2022csr}] \textit{Bootstrapping Elliptic Feynman Integrals Using Schubert Analysis},\\
    R. Morales, A. Spiering, M. Wilhelm, Q. Yang, C. Zhang, \\
    \href{https://doi.org/10.1103/PhysRevLett.131.041601}{\textit{Phys. Rev. Lett.} \textbf{131} (2023) 041601} [\href{https://arxiv.org/abs/2212.09762}{2212.09762}].
    
    \item[\cite{McLeod:2023qdf}] \textit{An infinite family of elliptic ladder integrals},\\
    A. McLeod, R. Morales, M. von Hippel, M. Wilhelm and C. Zhang, \\
    \href{https://doi.org/10.1007/JHEP05(2023)236}{\textit{JHEP} \textbf{05} (2023) 236} [\href{https://arxiv.org/abs/2301.07965}{2301.07965}].
    
    \item[\cite{Frellesvig:2023bbf}] \textit{Calabi--Yau Meets Gravity: A Calabi--Yau Three-fold at Fifth Post-Minkowskian Order},\\
    H. Frellesvig, R. Morales and M. Wilhelm, \\
    \href{https://doi.org/10.1103/PhysRevLett.132.201602}{\textit{Phys. Rev. Lett.} \textbf{132} (2024) 201602} [\href{https://arxiv.org/abs/2312.11371}{2312.11371}].
    
    \item[\cite{Frellesvig:2024zph}] \textit{Classifying post-Minkowskian geometries for gravitational waves via loop-by-loop Baikov},\\
    H. Frellesvig, R. Morales and M. Wilhelm, \\
    \href{https://doi.org/10.1007/JHEP08(2024)243}{\textit{JHEP} \textbf{08} (2024) 243} [\href{https://arxiv.org/abs/2405.17255}{2405.17255}].

    \item[\cite{Frellesvig:2024rea}] \textit{Calabi--Yau Feynman integrals in gravity: $\varepsilon$-factorized form for apparent singularities},\\
    H. Frellesvig, R. Morales, S. Pögel, S. Weinzierl and M. Wilhelm, \\
    \href{https://doi.org/10.1007/JHEP02(2025)209}{\textit{JHEP} \textbf{02} (2025) 209} [\href{https://arxiv.org/abs/2412.12057}{2412.12057}].

    \item[\cite{Brammer:2025rqo}] \textit{Classification of Feynman integral geometries for black-hole scattering at 5PM order},\\
    D. Brammer, H. Frellesvig, R. Morales and M. Wilhelm, \\
    \href{https://arxiv.org/abs/2505.10274}{arXiv:2505.10274}.
\end{itemize}

\newpage

{
  \hypersetup{linkcolor=blue}
  \hypersetup{linktoc=page}
  \addtocontents{toc}{\protect\thispagestyle{empty}}
  \tableofcontents
  \thispagestyle{empty}
}

\newpage\thispagestyle{plain}\pagenumbering{arabic}

\addtocontents{toc}{%
 \protect\vspace{1em}%
 \protect\noindent
\textcolor{DarkRed}{\textbf{I
\hspace{.4em} Preliminaries}}
\protect\par
 \protect\vspace{0em}%
}
\part*{Part I \\[1cm] Preliminaries}

\chapter*{Introduction}
\label{intro_thesis}
\addcontentsline{toc}{chapter}{\protect\numberline{}Introduction}
\markboth{Introduction}{Introduction} % Manually set the left and right marks

Scattering amplitudes are the foundation for calculating theoretical predictions using quantum field theory (QFT), as they encode the probability of a scattering process in terms of the masses and momenta of the particles involved. Historically, they have been mostly used to calculate precision predictions for cross sections and decay rates, which are observables that can be measured at particle collider experiments such as the Large Hadron Collider (LHC). Among other recent applications~\cite{Bern:2022jnl}, scattering amplitudes are nowadays also used for gravitational-wave physics, where they can be employed to study the gravitational waves emitted during the inspiral and coalescence of astrophysical compact objects, such as neutron stars or black holes. Specifically, they are used to construct theoretical waveform templates, which can be contrasted with measurements at gravitational-wave detectors such as the Laser Interferometer Gravitational-Wave Observatory (LIGO). 

While the respective discoveries of the Higgs boson~\cite{ATLAS:2012yve,CMS:2012qbp} and gravitational waves~\cite{LIGOScientific:2016aoc} marked two groundbreaking achievements in the history of physics, in order to study in more detail the properties of these exceptionally rare events and attempt to discover more exotic ones, it is necessary to push the precision frontier over the next decade. In particular, the upcoming arrival of a high-luminosity phase at the LHC~\cite{Bruning:2019}, as well as the next generation of gravitational-wave detectors~\cite{Ballmer:2022uxx}, makes it imperative to provide more precise theoretical predictions to keep up with the increasing experimental precision~\cite{Purrer:2019jcp,FebresCordero:2022psq,Buonanno:2022pgc}.

Scattering amplitudes are calculated in QFT as a perturbative loop expansion in terms of the coupling constant, with each successive loop order providing a more precise contribution to the theoretical prediction.~As~a~result, the pursuit of higher-precision predictions drives the need for performing calculations at increasingly higher loop orders. Traditionally, scattering amplitudes are calculated as a sum over Feynman diagrams, which visually encompass all different configurations of particles resulting in the same scattering process at a fixed loop order. While at first glance the calculation for even the simplest scattering processes may seem daunting, the application of appropriate tools~\cite{Dixon:1996wi,Weinzierl:2016bus} reveals an underlying elegance in the structure of scattering amplitudes~\cite{Dixon:2011xs,Arkani-Hamed:2012zlh}, which often leads to surprisingly simple and compact results. One such example is the Parke-Taylor formula~\cite{Parke:1986gb}, which captures the tree-level gluon scattering in quantum chromodynamics (QCD) with remarkable simplicity. Despite the resounding success of QFT and Feynman diagrams,
which have led to some of the most precise predictions in science~\cite{Laporta:2017okg}, calculating scattering amplitudes at high-loop orders becomes a tremendous endeavor. This difficulty is partly due to an exponential growth in the number of Feynman diagrams arising from the combinatorics of the process, but principally because evaluating the associated Feynman integrals becomes increasingly challenging. 

Naturally, this raises the question of whether obtaining analytical expressions for scattering amplitudes is useful (and necessary) given the availability of numerical integration methods~\cite{Smirnov:2021rhf,Heinrich:2023til}. However, an analytical result expressed in terms of a well-defined class of functions offers many advantages, such as an efficient and reliable numerical evaluation~\cite{Vollinga:2004sn,Walden:2020odh}, especially in cases involving branch cuts that require analytic continuation. Furthermore, there exist scattering processes where numerical integration still lacks efficiency. For instance, two-loop amplitudes for phenomenology for $2 \to 3$ processes involving many scales can already require hundreds of thousands and even millions of CPU core hours due to the vast phase space~\cite{FebresCordero:2022psq}. On a similar note, numerical relativity waveforms for black-hole binary systems are computationally very expensive, especially if many gravitational-wave cycles are simulated, and generating them for every possible initial configuration in the parameter space is not feasible~\cite{Boyle:2019kee}. Therefore, analytic calculations for scattering amplitudes and for waveform templates remain both useful and necessary for current and future applications.

One of the objectives of this thesis is precisely to study the classes of analytical functions that arise when calculating Feynman integrals, with the aim of developing tools that can facilitate the calculation of high-loop scattering amplitudes. At tree level there are no integrals over loop momenta, and scattering amplitudes therefore evaluate to rational functions of the kinematics. At one loop, all Feynman integrals can be written in terms of so-called multiple polylogarithms (MPLs)~\cite{Chen:1977oja,Goncharov:1995ifj}, which are iterated integrals over the Riemann sphere that generalize classical polylogarithms. By contrast, at higher loops and principally in cases involving many physical scales, Feynman integrals give rise to a much richer class of transcendental functions, which arise from integrals over non-trivial geometries; see ref.~\cite{Bourjaily:2022bwx} for a recent review. Concretely, there are integrals over elliptic curves~\cite{Sabry:1962rge,Broadhurst:1993mw,Laporta:2004rb,Caron-Huot:2012awx,Adams:2013nia,Bloch:2013tra,Adams:2014vja,Remiddi:2016gno,Adams:2016xah,Broedel:2017siw,Kristensson:2021ani,Giroux:2022wav,Morales:2022csr,McLeod:2023qdf,Stawinski:2023qtw,Giroux:2024yxu} and higher-genus curves~\cite{Huang:2013kh,Marzucca:2023gto,Duhr:2024uid}, along with K3 surfaces and higher-dimensional Calabi--Yau varieties~\cite{Primo:2017ipr,Bourjaily:2018ycu,Bourjaily:2018yfy,Bonisch:2021yfw,Broedel:2021zij,Duhr:2022pch,Lairez:2022zkj,Pogel:2022vat,Duhr:2022dxb,Cao:2023tpx,McLeod:2023doa,Duhr:2023eld,Duhr:2024hjf}. Importantly, such non-trivial geometries have also been found in the perturbative expansion of classical gravity~\cite{Bern:2021dqo,Dlapa:2021npj,Bern:2022jvn,Dlapa:2022wdu,Jakobsen:2023ndj,Frellesvig:2023bbf,Klemm:2024wtd,Driesse:2024xad,Bern:2024adl,Frellesvig:2024zph,Driesse:2024feo}. These complicated geometries are thus an essential feature of Feynman integrals, and appear even in diagrams without any kinematic dependence~\cite{Brown:2010bw,Schnetz:2019cab}.

While there exist elliptic generalizations of MPLs~\cite{LevinRacinet2007,brown2011multiple,Broedel:2014vla,Broedel:2017kkb,Broedel:2017siw,Broedel:2018iwv,Broedel:2018qkq}, the polylogarithmic analogues for higher-genus Riemann surfaces are under construction~\cite{DHoker:2023vax,Baune:2024biq,DHoker:2024ozn,Baune:2024ber}, and the corresponding class of functions necessary for Calabi--Yau geometries is far less understood. Consequently, these non-trivial geometries pose a fundamental challenge to the traditional methods used to evaluate scattering amplitudes, which rapidly become both computationally and conceptually obsolete. Yet, the calculation of these integrals, which regularly appear in the forefront of state-of-the-art calculations, is essential for obtaining precision predictions. As an example, elliptic integrals appear in diagrams relevant to $t\bar{t}$ production via a massive top quark loop at two loops~\cite{Adams:2018bsn,Adams:2018kez,Broedel:2019hyg}, as well as in the $\rho$ parameter of the Standard Model at three loops~\cite{Abreu:2019fgk}. Even more, the three-loop electron and photon self-energies in quantum electrodynamics (QED) were only recently computed analytically~\cite{Duhr:2024bzt,Forner:2024ojj}, due to the presence of integrals over elliptic curves and K3 surfaces therein. Lastly, cutting-edge results in classical gravity at four loops involve integrals over K3 surfaces and Calabi--Yau threefolds~\cite{Driesse:2024feo}, which corresponds to the first instance of special functions related to Calabi--Yau manifolds appearing in observables in nature. Overall, in order to keep pushing the precision frontier, we need to develop methods to calculate these integrals over non-trivial geometries. While Feynman integrals can be calculated using a plethora of methods, see ref.~\cite{Weinzierl:2022eaz} for a recent pedagogical introduction, in this thesis we will focus on a few approaches and applications.

In the first half of the thesis, we focus on Feynman integrals in maximally supersymmetric Yang--Mills ($\mathcal{N}=4$ SYM) theory~\cite{Brink:1976bc}. This is a supersymmetric extension of QCD, the theory that describes strong interactions and is tested with experiments at the LHC. In particular, amplitudes in $\mathcal{N}=4$ SYM theory enjoy both superconformal and dual superconformal symmetry~\cite{Drummond:2007au,Drummond:2008vq}, offering a unique framework that facilitates the development of techniques that can subsequently be applied to optimize the calculations of QCD. Specifically, we will study elliptic generalizations of ladder diagrams in this theory, which have traditionally provided fruitful insight into the all-loop structure of polylogarithmic Feynman integrals~\cite{Usyukina:1993ch,Broadhurst:2010ds,Caron-Huot:2018dsv,He:2020uxy}.

First, in chapter~\ref{ch:chapter2}, we approach these integrals using one of the most naive methods -- via direct integration with Feynman parameters. As we will see, direct integration can be very useful for evaluating certain integrals involving a single square root, even for elliptic curves. In particular, it will allow us to identify the first two families of Feynman integrals that depend on the same elliptic curve to all loop orders, and thus evaluate to the same class of elliptic multiple polylogarithms. Even though this integral family provides a useful testing ground for developing elliptic technology, one rapidly runs into examples with multiple square roots, which prevent direct integration. The first of such examples is the two-loop twelve-point elliptic double-box integral, which is the focus of chapter~\ref{ch:chapter3}. To circumvent the direct integration obstacle, we will focus on the symbol~\cite{Goncharov:2010jf,Duhr:2011zq}, which provides a unique decomposition of the result while encoding its singularity structure. Concretely, we will bootstrap the result for the two-loop twelve-point elliptic double-box symbol by creating an ansatz that can be uniquely fixed with physical constraints using much simpler linear algebra. In the process, we will initiate the symbol bootstrap for elliptic Feynman integrals, which builds upon the success of bootstrapping developments for polylogarithmic integrals~\cite{Goncharov:2010jf,Dixon:2011pw,Dixon:2011nj,Brandhuber:2012vm,Caron-Huot:2016owq,Almelid:2017qju,Henn:2018cdp,Caron-Huot:2019vjl,Dixon:2020bbt,Guo:2021bym,Dixon:2022rse,Dixon:2022xqh,Hannesdottir:2024hke,Basso:2024hlx}, see also refs.~\cite{Caron-Huot:2020bkp,Arkani-Hamed:2022rwr} for a review in $\mathcal{N}=4$ SYM theory.

In the second half of the thesis, we switch topic and study the two-body problem in general relativity. In particular, we focus on the classical dynamics and gravitational waves emitted during the inspiral phase of a black-hole merger. There exist many complementary approaches to model this dynamical system, ranging from numerical relativity~\cite{Pretorius:2005gq,Campanelli:2005dd,Baker:2005vv} to the post-Newtonian~\cite{Goldberger:2004jt,Blanchet:2013haa,Levi:2018nxp},
post-Minkowskian~\cite{Damour:2016gwp,Buonanno:2022pgc} and self-force~\cite{Mino:1996nk,Quinn:1996am,Poisson:2011nh,Barack:2018yvs} expansions. In this thesis, we will focus on the post-Minkowskian (PM) expansion, which treats the two-body problem perturbatively in Newton's constant $G$ while maintaining all-order corrections in the velocity, thus accounting for relativistic effects and enabling the use of QFT techniques, such as scattering amplitudes. In particular, analytic continuation~\cite{Kalin:2019rwq,Kalin:2019inp,Cho:2021arx,Dlapa:2024cje} and the effective one-body formalism~\cite{Damour:2016gwp} allow us to compute corrections to the classical dynamics by calculating the $2 \to 2$ scattering of black holes~\cite{Damour:2016gwp,Bern:2019nnu,Bern:2019crd,Kalin:2020mvi,Mogull:2020sak}; see also refs.~\cite{Bjerrum-Bohr:2022blt,Buonanno:2022pgc} for a recent review. 

Concretely, in chapter~\ref{ch:chapter4}, we will initiate a systematic classification of the geometries that arise in PM Feynman integrals, offering the first broad understanding of the class of functions that can appear up to four-loop order for this scattering process. For that, we will use two different tools -- differential equations~\cite{Kotikov:1990kg} and leading singularities~\cite{Cachazo:2008vp,Arkani-Hamed:2010pyv} -- that will allow us to detect non-trivial geometries before carrying out any calculation. Remarkably, we will identify the first Calabi--Yau threefold appearing in gravitational-wave physics, among other non-trivial geometries. Subsequently, in chapter~\ref{ch:chapter5}, we will characterize the Feynman integral giving rise to this Calabi--Yau geometry in more detail. We will also solve this integral by bringing its differential equation into canonical form~\cite{Henn:2013pwa}, a procedure that required developing a new method adapting to apparent singularities.

Before delving into these very interesting topics, in chapter~\ref{ch:intro} let us introduce in detail the basic concepts that will be used throughout the thesis, as well as give a brief review of Feynman integrals and their properties. Unless stated otherwise, all figures and diagrams in this thesis have been drawn using TikZ~\cite{Tan12} and GeoGebra\textsuperscript{\textregistered}.

\chapter{Introduction to Feynman integrals}
\label{ch:intro}

In this chapter, we provide a brief introduction to Feynman integrals; see also ref.~\cite{Weinzierl:2022eaz} for a recent book on the subject. First, in sec.~\ref{sec:ch1_Feynman_integrals}, we present the notation used for Feynman integrals throughout the thesis, and highlight the connection to special functions and non-trivial geometries. Then, in sec.~\ref{sec:ch1_MPLs}, we introduce the simplest class of such special functions -- multiple polylogarithms~\cite{Chen:1977oja,Goncharov:1995ifj} -- along with some of their properties and, especially relevant for us, their symbol~\cite{Goncharov:2010jf,Duhr:2011zq}. Finally, in sec.~\ref{sec:ch1_DE_LS}, we review the methods of differential equations~\cite{Kotikov:1990kg} and leading singularities~\cite{Cachazo:2008vp,Arkani-Hamed:2010pyv}, and how they can be used to detect, characterize, and calculate Feynman integrals depending on non-trivial geometries.

\section{Feynman integrals and special functions}
\label{sec:ch1_Feynman_integrals}

\begin{figure}[t]
\begin{center}
\includegraphics[height=3.25cm]{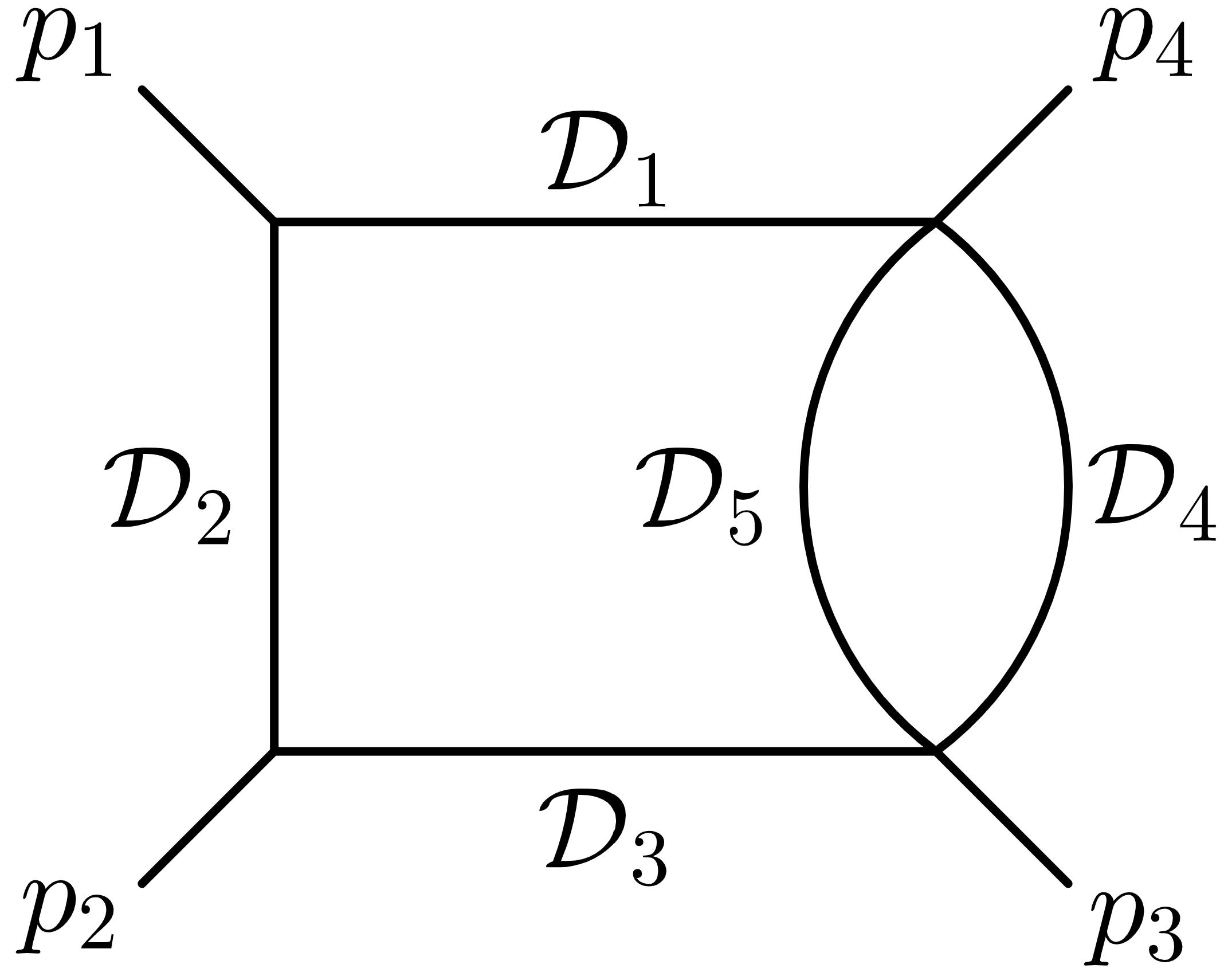}
\caption{Diagram for the two-loop four-point massless box-bubble, which is used in the main text to exemplify the concepts of integral family, sectors and subsectors.}
\label{fig: bubble-box_example}
\end{center}
\end{figure}

The momentum representation of an $L$-loop scalar Feynman integral in $D$ dimensions is
\begin{equation}
\label{eq: loop momentum Feynman integral}
\mathcal{I}_{\nu_1\dots\,\nu_\nint}= \int \prod_{j=1}^L d^D k_j \frac{1}{\mathcal{D}_1^{\nu_1} \, \cdots \, \mathcal{D}_\nint^{\nu_\nint}}= \int \prod_{j=1}^L d^D k_j \frac{1}{\prod_{i=1}^\nint {(Q_i^2-m_i^2)}^{\nu_i}}\,,
\end{equation}
where we have $\nint$ internal propagators $\mathcal{D}_i$ with non-negative integer exponents $\nu_i$. In this expression, we drop some conventional prefactors since they will not play a role in the following. In addition, $m_i$ denotes the mass of the propagator $\mathcal{D}_i$, and $Q_i$ its momentum, which can depend on loop momenta $k_j$ as well as on external momenta $p_j$. The specific form taken by the momentum of each propagator depends on the diagram at hand. Unless stated otherwise, we will work in dimensional regularization $D=4 - 2\varepsilon$, and consider generic kinematics for the $\nex$ external particles, such that $E=\text{dim} \langle p_1, \dots, p_{n_{\text{ext}}} \rangle = \nex - 1$ is the number of independent external momenta. As usual, for $\nu_i>1$, we refer to $\nu_i-1$ as the number of dots in the corresponding internal edge in the diagram. We moreover refer to the collection of Feynman integrals with the same set of internal propagators $\{ \mathcal{D}_1, \dots, \mathcal{D}_\nint \}$ and different values (including zero) for the exponents $\nu_i$ as an integral family. For example, we have the following integral family for the two-loop four-point massless box-bubble of fig.~\ref{fig: bubble-box_example},
\begin{equation}
\mathcal{I}_{\nu_1\dots\,\nu_5}= \int \frac{d^D k_1 \, d^D k_2}{{[k_1^2]}^{\nu_1} {[(k_1-p_1)^2]}^{\nu_2} {[(k_1-p_{1,2})^2]}^{\nu_3} {[(k_1+k_2-p_{1,2,3})^2]}^{\nu_4} {[k_2^2]}^{\nu_5}}\,,
\end{equation}
where the external momenta $p_i$ are all outgoing, and $p_{i,\dots,j} \equiv p_i + \dots + p_j$.

In general, Feynman integrals can also include a numerator factor $\mathcal{N}(k_i \cdot k_j, k_i \cdot p_j)$ that may depend on the $L(L+1)/2$ scalar products $k_i \cdot k_j$ between loop momenta, and on the $E L$ products $k_i \cdot p_j$ with the independent external momenta. In total, there are
\begin{equation}
\label{eq: number of total scalar products}
N_\text{V} = \frac{L(L+1)}{2}+EL
\end{equation}
scalar products involving loop momenta, which can be classified into two types: those already present in the $\nint$ propagators $\mathcal{D}_i$ of the diagram, and additional scalar products not associated with any propagator. The latter are known as irreducible scalar products (ISPs), and at $L$ loops there are a total of $\nISP = N_\text{V} - \nint$. To embed cases involving ISPs within the integral family framework, we will artificially introduce $\nISP$ additional propagators, absent from the original diagram, which appear as factors in the numerator:
\begin{equation}
\label{eq: loop momentum Feynman integral with ISPs}
\mathcal{I}_{\nu_1\dots\,\nu_{N_\text{V}}}= \int \prod_{j=1}^L d^D k_j \frac{\prod_{l=\nint+1}^{N_\text{V}} {(Q_l^2-m_l^2)}^{-\nu_l}}{\prod_{i=1}^\nint {(Q_i^2-m_i^2)}^{\nu_i}}\,,
\end{equation}
where $\nu_{\nint+1},\dots,\nu_{N_\text{V}}\leq0$. For instance, for the example above we have
\begin{equation}
\mathcal{I}_{\nu_1\dots\,\nu_9}= \int \frac{d^D k_1 \, d^D k_2 \, {[(k_1+k_2)^2]}^{-\nu_6} {[(k_2-p_1)^2]}^{-\nu_7} {[(k_2-p_{1,2})^2]}^{-\nu_8} {[(k_2-p_{1,2,3})^2]}^{-\nu_9}}{{[k_1^2]}^{\nu_1} {[(k_1-p_1)^2]}^{\nu_2} {[(k_1-p_{1,2})^2]}^{\nu_3} {[(k_1+k_2-p_{1,2,3})^2]}^{\nu_4} {[k_2^2]}^{\nu_5}}\,.
\end{equation}
In combination with dimension-shift identities~\cite{Tarasov:1996br,Tarasov:1997kx} and tensor reduction~\cite{Passarino:1978jh}, we can use ISPs to express any Feynman integral in terms of a linear combination of scalar integrals within a family. Thus, in the following we will mostly focus on scalar Feynman integrals.

For later convenience, let us also introduce the notion of integral sectors. For a given integral family, we call the top sector the diagram that contains the most amount of propagators, thus all vertices are cubic and $\nint$ is maximal. Then, we call subsectors all Feynman integrals within the family where some of the propagators are absent, that is, some $\nu_i\leq0$ for $1 \leq i \leq \nint$. Graphically, subsectors are obtained from the top sector by contracting (or pinching) the edges with $\nu_i\leq0$ in the diagram. For example,
\begin{equation}
    \parbox[c]{0.2\textwidth}{
            \centering
            \raisebox{-0.5\height}{\includegraphics[height=1.75cm]{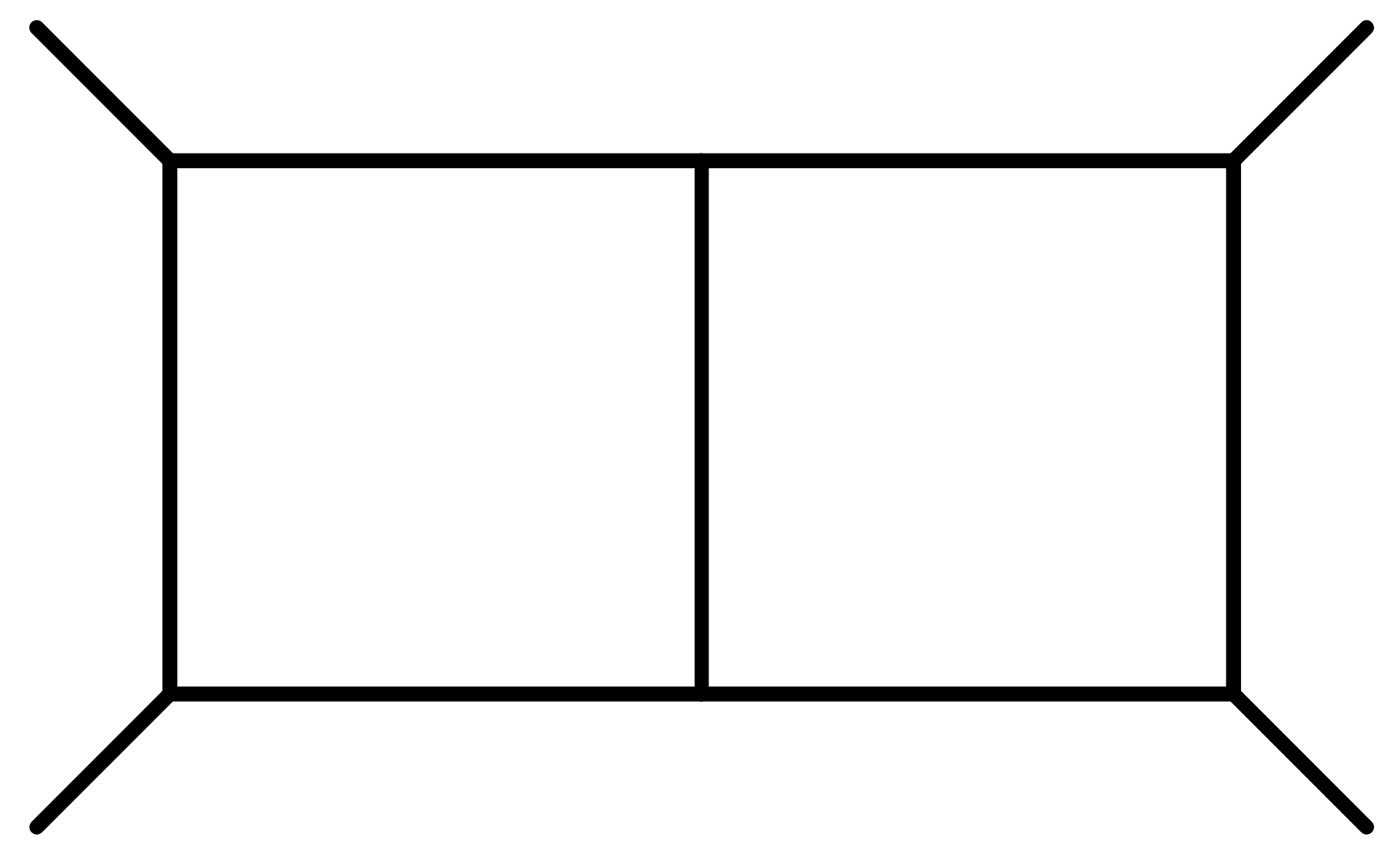}} \\[0.2cm]
            \text{$\mathcal{I}_{\nu_1\nu_2\nu_3\nu_4\nu_5\nu_60\nu_80}$}
        } \ \ \ \parbox[c]{0.1\textwidth}{\raisebox{0.8cm}{$\xleftarrow{\displaystyle \text{Top sector}}$}\\[-0.7cm]
        \raisebox{0.9cm}{$\xrightarrow[\displaystyle \text{Subsector }]{}$}} \ \ \parbox[c]{0.2\textwidth}{
            \centering
            \raisebox{-0.5\height}{\includegraphics[height=1.75cm]{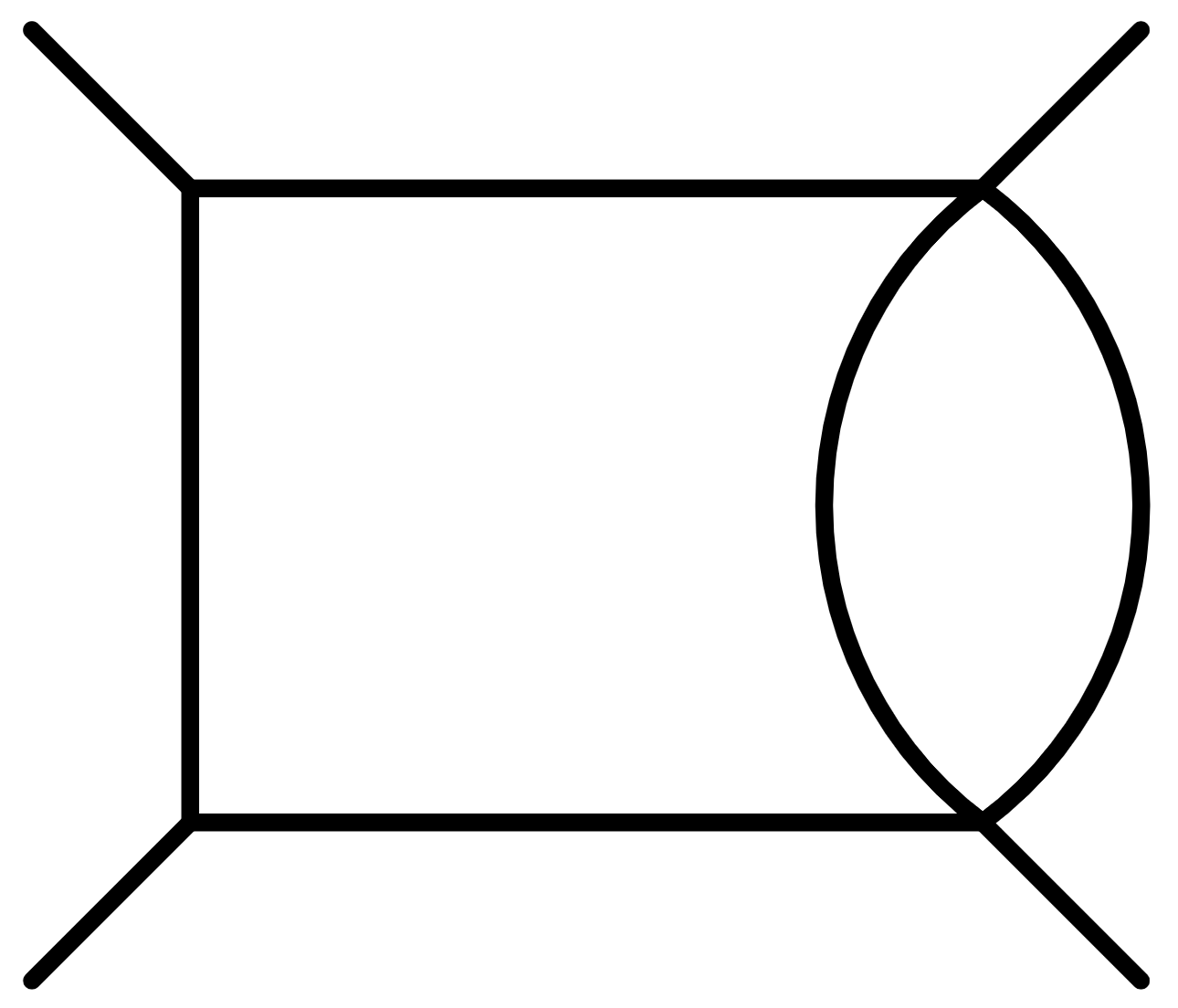}} \\[0.2cm]
            \text{$\mathcal{I}_{\nu_1\nu_2\nu_3\nu_4\nu_50000}$}
        } \ \raisebox{0.3cm}{$\xrightarrow{\displaystyle \text{Subsector}}$} \ \parbox[c]{0.2\textwidth}{
            \centering
            \raisebox{-0.5\height}{\includegraphics[height=1.75cm]{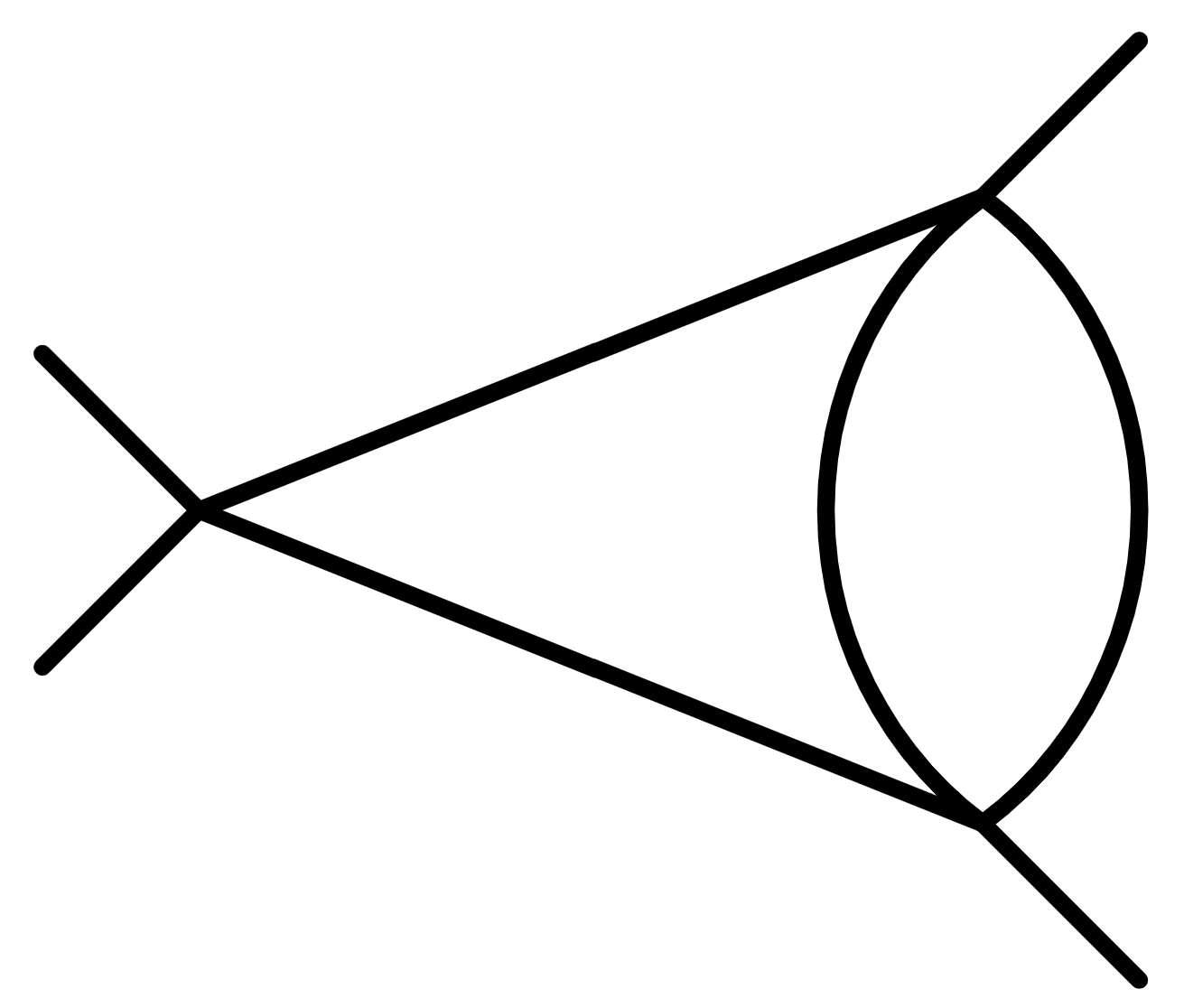}} \\[0.2cm]
            \text{$\mathcal{I}_{\nu_10\nu_3\nu_4\nu_50000}$}
        }\,.
\end{equation}

Throughout the thesis, an integral sector will refer to the set of Feynman integrals where a fixed subset of propagators is absent $(\nu_i\leq0)$, while the remaining exponents can take different non-zero values. Graphically, this corresponds to a specific Feynman diagram with different numbers of dots in the propagators. Therefore, whenever we focus on one specific Feynman integral within a sector, we will generally denote it by drawing the respective diagram, including dots. For instance,
\begin{equation}
    \vcenter{\hbox{\includegraphics[height=1.75cm]{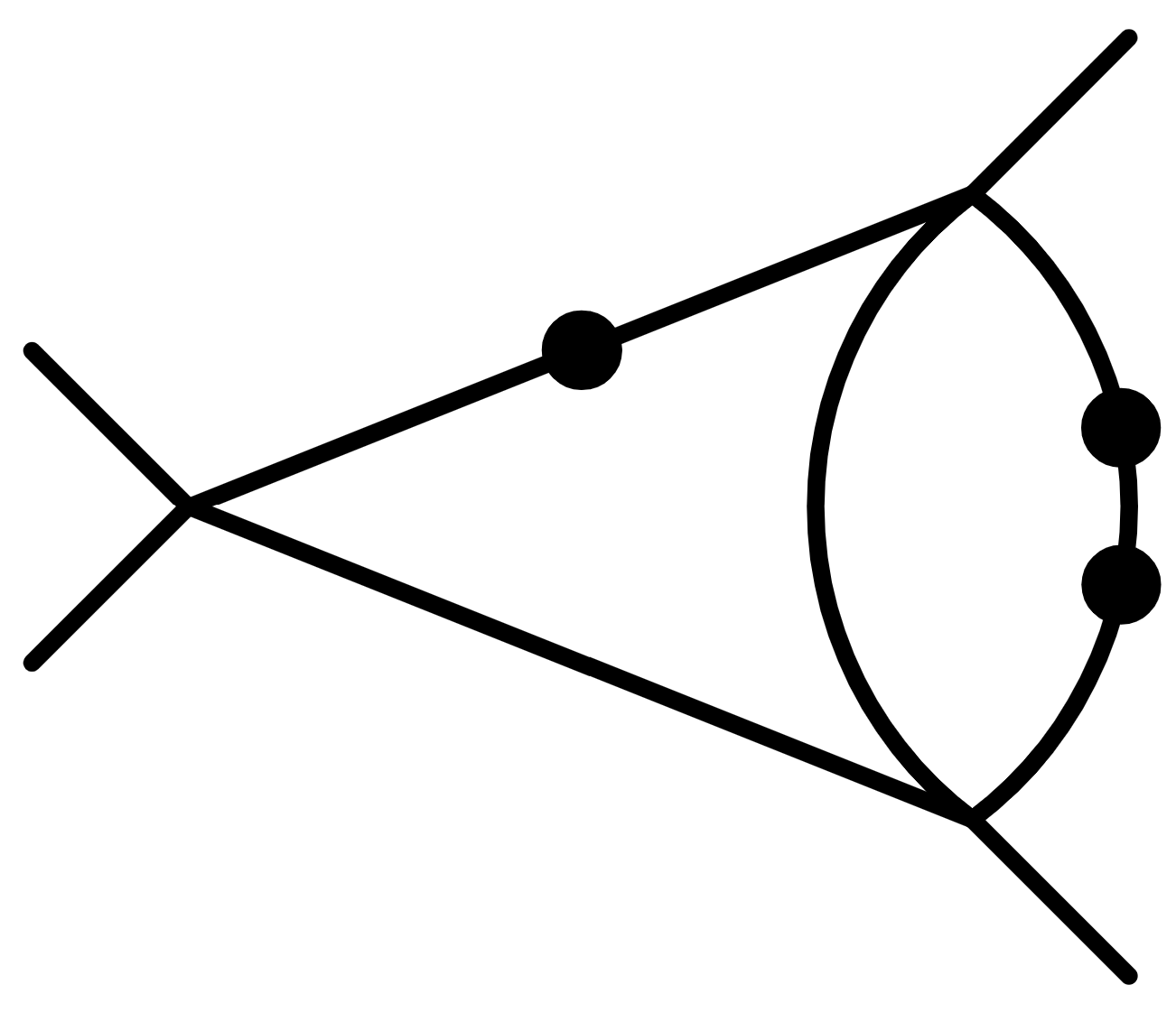}}} \ = \, \mathcal{I}_{201310000} \, .
\end{equation}

While in this thesis we will only touch upon a few approaches, there exist, however, numerous methods that can be used to calculate the result of Feynman integrals; see ref.~\cite{Weinzierl:2022eaz} for an introduction. One of the most natural strategies is to carry out the loop-momentum integrations, which is best achieved using a parametric representation. A well-known example is the Feynman representation, where one uses the identity
\begin{equation}
    \frac{1}{\mathcal{D}_1 \, \cdots \,  \mathcal{D}_n} = \int_{0}^{1} d\alpha_1 \, \cdots \int_{0}^{1} d\alpha_n \, \, \delta \bigg(1 - \sum_{i = 1}^{n} \alpha_i \bigg) \,  \frac{(n - 1)!}{\left(\alpha_1 \mathcal{D}_1 + \cdots + \alpha_n \mathcal{D}_n \right)^n}
\end{equation}
to replace the propagators by Feynman parameters $\alpha_i$, which in low-loop cases often leads to a known integral; see e.g.~chapter~\ref{ch:chapter2} for an application to ladder integrals in $\mathcal{N}=4$ SYM theory. Other parametrizations include the Symanzik, Mellin-Barnes and Lee-Pomeransky representations. Of particular relevance for our purposes will be the Baikov representation~\cite{Baikov:1996iu,Frellesvig:2017aai}, where the integration variables are given by the propagators themselves, see sec.~\ref{sec:ch4_Baikov} for details. Although this representation is not especially suited for the evaluation of the integral, it can nonetheless be used to efficiently identify the underlying geometry. In particular, in chapter~\ref{ch:chapter4}, we will rely on the Baikov representation to classify the Feynman integral geometries appearing in the PM expansion of classical gravity up to four loops.

Alternatively, one can also evaluate Feynman integrals by solving the differential equations that they satisfy~\cite{Kotikov:1990kg}. In particular, if we take the derivative of a Feynman integral with a fixed set of exponents $\nu_i$ with respect to a kinematic variable $x$, we obtain a linear combination of integrals within the same integral family of the form
\begin{equation}
\label{eq: pre_diff_equation}
\frac{d}{dx} \, \mathcal{I}_{\nu_1\dots\,\nu_{N_\text{V}}}= \sum_{i_1,\dots,i_{N_\text{V}}} f_{i_1\dots\, i_{N_\text{V}}}\!(x,\varepsilon) \ \mathcal{I}_{\nu_{i_1}\dots\,\nu_{i_{N_\text{V}}}}.
\end{equation}
Here, $f_{i_1 \dots \, i_{N_\text{V}}}\!(x,\varepsilon)$ denote rational functions which depend on the kinematics and the dimensional regulator $\varepsilon$. In general, the right-hand side may involve integrals with different exponents than the original integral. However, we can facilitate solving this differential equation by reducing the space of integrals in the family to a basis. Specifically, there exist integration-by-parts identities (IBPs)~\cite{Chetyrkin:1981qh} of the type
\begin{equation}
\label{eq: IBP_definition}
0= \int \prod_{j=1}^L d^D k_j \frac{\partial}{\partial k_l^\mu} \left( \frac{v^\mu}{\mathcal{D}_1^{\nu_1} \, \cdots \, \mathcal{D}_{N_\text{V}}^{\nu_{N_\text{V}}}} \right) \,,
\end{equation}
where $v^\mu$ can be any combination of the four-momenta present in the integral and $k_l$ is one of the loop momenta. Solving these IBP relations, all Feynman integrals in a family can be expressed in terms of a minimal set of Feynman integrals, which are commonly known as the master integrals. Note, however, that we can always choose a different basis of master integrals within a family. Moreover, in refs.~\cite{Smirnov:2010hn,Bitoun:2017nre} it was shown that the number of master integrals in a family is always finite. Therefore, we can rewrite the right-hand side of eq.~\eqref{eq: pre_diff_equation} as a linear combination of (fewer) master integrals, such that if we obtain the solution for these master integrals, all Feynman integrals in the family are then determined. This approach will be further developed in sec.~\ref{sec:ch1_DEs}, where we will also present the associated system of coupled first-order differential equations and discuss its relation with non-trivial geometries; see also chapter~\ref{ch:chapter5} for an application in classical gravity.

\subsection{Geometries and special functions}
\label{sec:ch1_geometries}

Having set up the basics for Feynman integrals, let us now review their relation to special functions and algebraic geometry; see ref.~\cite{Bourjaily:2022bwx} for a recent review. To illustrate this connection, we will consider a simple example~\cite{Bourjaily:2018aeq}. Suppose that we have an arbitrary Feynman integral, which has been parametrized using $n+1$ Feynman parameters $\{ \alpha_0, \dots, \alpha_n \}$, leading to an expression of the form
\begin{equation}
\mathcal{I} = \int_0^\infty \frac{d \alpha_0 \, \cdots \, d \alpha_n}{\alpha_0^2 + 2 \alpha_0 f(\vec{\alpha}) +g(\vec{\alpha})}\,,
\end{equation}
where we introduce the short-hand notation $\vec{\alpha} \equiv ( \alpha_1, \dots, \alpha_n )$. Using partial fraction and dropping the dependence on $\vec\alpha$ for ease of notation, the integral over $\alpha_0$ yields
\begin{align}
\label{eq: example special functions}
\mathcal{I} = & \, \int_0^\infty \frac{d \alpha_0 \; \cdots \; d \alpha_n}{2\sqrt{f^2-g}} \left( \frac{1}{\alpha_0 + f - \sqrt{f^2-g}} - \frac{1}{\alpha_0 + f + \sqrt{f^2-g}} \right) \nonumber \\[0.1cm]
= & \, \int_0^\infty \frac{d \alpha_1 \; \cdots \; d \alpha_n}{2 \sqrt{f^2-g}} \log(\frac{f - \sqrt{f^2-g}}{f + \sqrt{f^2-g}}) \nonumber \\
\equiv & \, \int_0^\infty \frac{d \alpha_1 \; \cdots \; d \alpha_n}{2 \sqrt{P(\vec{\alpha})}} \log(\frac{f(\vec{\alpha}) - \sqrt{P(\vec{\alpha})}}{f(\vec{\alpha}) + \sqrt{P(\vec{\alpha})}})\,.
\end{align}
As can be seen, integrating over $\alpha_0$ introduces a square root $\sqrt{P(\vec{\alpha})}$ in the result, where $P(\vec{\alpha}) \equiv f(\vec{\alpha})^2-g(\vec{\alpha})$ is a polynomial in the remaining $n$ Feynman parameters. Notably, the square root in the denominator and the degree of the polynomial $P(\vec{\alpha})$ determine the class of functions that eq.~\eqref{eq: example special functions} evaluates to, since they specify the corresponding geometry that we are integrating over. As we will further investigate in sec.~\ref{sec:ch1_DE_LS}, for the purpose of identifying the non-trivial geometry, we can drop the dependence on the logarithm and focus solely on the denominator,
\begin{equation}
\label{eq: LS_geometry_intro}
    \LS(\mathcal{I}) \propto \int\frac{d \alpha_1 \; \cdots \; d \alpha_n}{\sqrt{P(\vec{\alpha})}}\,.
\end{equation}
Up to constant factors, this operation corresponds to taking the discontinuity across the branch cut of the logarithm, which is related to the so-called leading singularity (LS)~\cite{Cachazo:2008vp,Britto:2024mna}; see sec.~\ref{sec:ch1_LS} for further details. For the purpose of characterizing the non-trivial geometry that the leading singularity depends on, let us define
\begin{equation}
\label{eq: polynomial_algebaric_curve}
    y^2 = P(\vec{\alpha})\,.
\end{equation}
In the trivial case, $n=0$, the leading singularity is algebraic, i.e.~a function of the kinematics without transcendental integrals left. Thus, there is no complicated geometry involved, and the Feynman integral can be written in terms of multiple polylogarithms. For a univariate problem, $n=1$, eq.~\eqref{eq: polynomial_algebaric_curve} defines an algebraic curve. If the degree of the polynomial is $\deg P(\alpha_1) \leq 2$, with an appropriate change of variables to $\tilde{\alpha}_1$, the square root can always be rationalized $\sqrt{P(\alpha_1)} = \widetilde{P}(\tilde{\alpha}_1)$; see the next subsection for examples. In this simple case we can consider the full integral, use partial fraction again and write the result as a sum of integrals over the Riemann sphere, with an integrand consisting of a logarithm and a denominator that is linear in $\tilde{\alpha}_1$,
\begin{equation}
\label{eq: intro_integral_preMPL}
\mathcal{I} = \int_0^\infty \frac{d \tilde{\alpha}_1}{2 \widetilde{P}(\tilde{\alpha}_1)} \log(\frac{f(\tilde{\alpha}_1) - \widetilde{P}(\tilde{\alpha}_1)}{f(\tilde{\alpha}_1) + \widetilde{P}(\tilde{\alpha}_1)}) = \sum_i C_i \int_0^\infty \frac{d \tilde{\alpha}_1}{\tilde{\alpha}_1 - r_i} \log(\frac{f(\tilde{\alpha}_1) - \widetilde{P}(\tilde{\alpha}_1)}{f(\tilde{\alpha}_1) + \widetilde{P}(\tilde{\alpha}_1)})\,,
\end{equation}
where $r_i$ denote the roots, and $C_i$ are some coefficients. As we will introduce in sec.~\ref{sec:ch1_MPLs}, the result can be written in terms of multiple polylogarithms.

Otherwise, if the degree of the polynomial is $\deg P(\alpha_1) > 2$ and unless there are coinciding roots, in general the square root cannot be rationalized. This leads to an integral over a cycle of a non-trivial geometry, one of the so-called periods of the geometry, which evaluates to new transcendental functions; see tab.~\ref{tab:geometries_intro} for a summary. The simplest of such cases is for $\deg P(\alpha_1) = 3$ or $4$, when eq.~\eqref{eq: polynomial_algebaric_curve} defines an elliptic curve. In this case, eq.~\eqref{eq: LS_geometry_intro} gives rise to an elliptic integral, which can also be understood as an integral over a torus (a two-dimensional surface of genus one, see sec.~\ref{sec:ch2_eMPLs_torus} for an explicit connection). Examples of Feynman integrals involving an elliptic curve are the well-known two-loop sunrise in $D=2$~\cite{Broadhurst:1993mw,Adams:2013nia,Bloch:2013tra}, which was identified more than sixty years ago~\cite{Sabry:1962rge}, and the two-loop double-box in $D=4$~\cite{Caron-Huot:2012awx,Kristensson:2021ani,Morales:2022csr}, which we will study in detail in chapters~\ref{ch:chapter2} and~\ref{ch:chapter3}. For $\deg P(\alpha_1) = 5$ or $6$, eq.~\eqref{eq: polynomial_algebaric_curve} defines a hyperelliptic curve, which leads to an integral over a 2-torus, i.e.~a torus of genus 2. One such example is the two-loop massive non-planar double box~\cite{Huang:2013kh,Marzucca:2023gto,Duhr:2024uid}. In general, for $\deg P(\alpha_1) = 2k-1$ or $2k$, we obtain an integral over a surface of genus $k-1$. However, no examples of Feynman integrals involving a surface of arbitrarily high genus are known yet.

More often than not, however, after the integral over $\alpha_0$ has been carried out, we have $n>1$ remaining Feynman parameters. In this case, eq.~\eqref{eq: polynomial_algebaric_curve} no longer defines a curve but an $n$-dimensional manifold. In particular, if $\deg P(\vec\alpha) = 2n+1$ or $2n+2$, this manifold turns out to be quite special. Homogenizing the polynomial with $\alpha_{n+1}$ as $\widetilde{P}(\alpha_1,\dots,\alpha_n,\alpha_{n+1})=P(\alpha_1/\alpha_{n+1},\dots,\alpha_n/\alpha_{n+1})(\alpha_{n+1})^{2n+2}$, we obtain the equation
\begin{equation}
y^2 - \widetilde{P}(\alpha_1,\dots,\alpha_n,\alpha_{n+1}) = 0\,,
\end{equation}
which can be defined in weighted projective space as
\begin{equation}
\label{eq: condition_degree_weight_CY}
[\alpha_1,\dots,\alpha_n,\alpha_{n+1},y]\sim[\lambda^1\alpha_1,\dots,\lambda^1\alpha_n,\lambda^1\alpha_{n+1},\lambda^{n+1}y] \in \mathbb{WP}^{1,\dots,1,1,n+1}\,.
\end{equation}
Since the sum of the projective weights is equal to the degree $2n+2$ of the polynomial, this equation defines a Calabi--Yau (CY) manifold of dimension $n$~\cite{Hubsch:1992nu,Bourjaily:2019hmc}, or Calabi--Yau $n$-fold.\footnote{In fact, it defines a Calabi--Yau variety, as the resulting geometry is normally not smooth, but contains singularities. However, we will not make such a distinction, since it is not necessary to desingularize the CY variety and obtain a manifold in order to solve the Feynman integral via differential equations; see chapter~\ref{ch:chapter5}.} In this case, eq.~\eqref{eq: LS_geometry_intro} defines one of the periods of a CY geometry; see tab.~\ref{tab:geometries_intro} for examples. 

\begin{table}[H]
\begin{center}
\caption{Prototypes of leading singularities depending on different geometries,\protect\footnotemark~where $P_k(\vec{\alpha})$ denotes a polynomial of degree $k$ in $\vec{\alpha}$, as well as some Feynman integral examples involving these geometries.}
\label{tab:geometries_intro}
\begin{tabular}{|c|c|c|}
    \Xhline{0.2ex}
		Leading singularity & Geometry & Feynman integral examples \\ \hline
        & & \\[-0.5cm]
        Algebraic & \multirow{2}{*}{\parbox[c]{0.2\textwidth}{
            \centering
            \raisebox{-0.5\height}{\includegraphics[height=1.5cm]{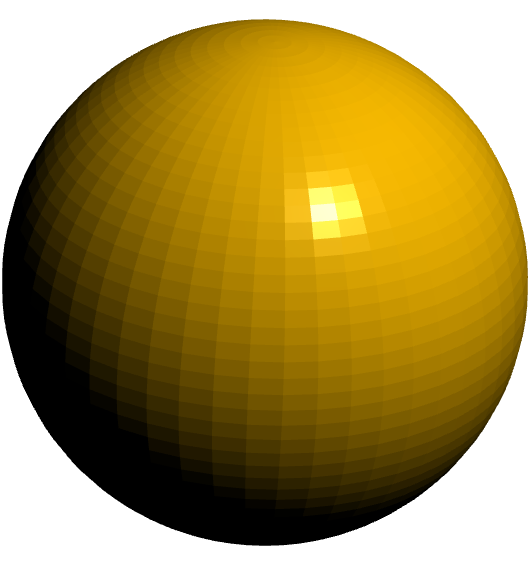}} \\[0.1cm]
            \small Riemann sphere
        }} & \multirow{2}{*}{\parbox[c]{0.2\textwidth}{
            \centering
            \raisebox{-0.5\height}{\includegraphics[height=1.2cm]{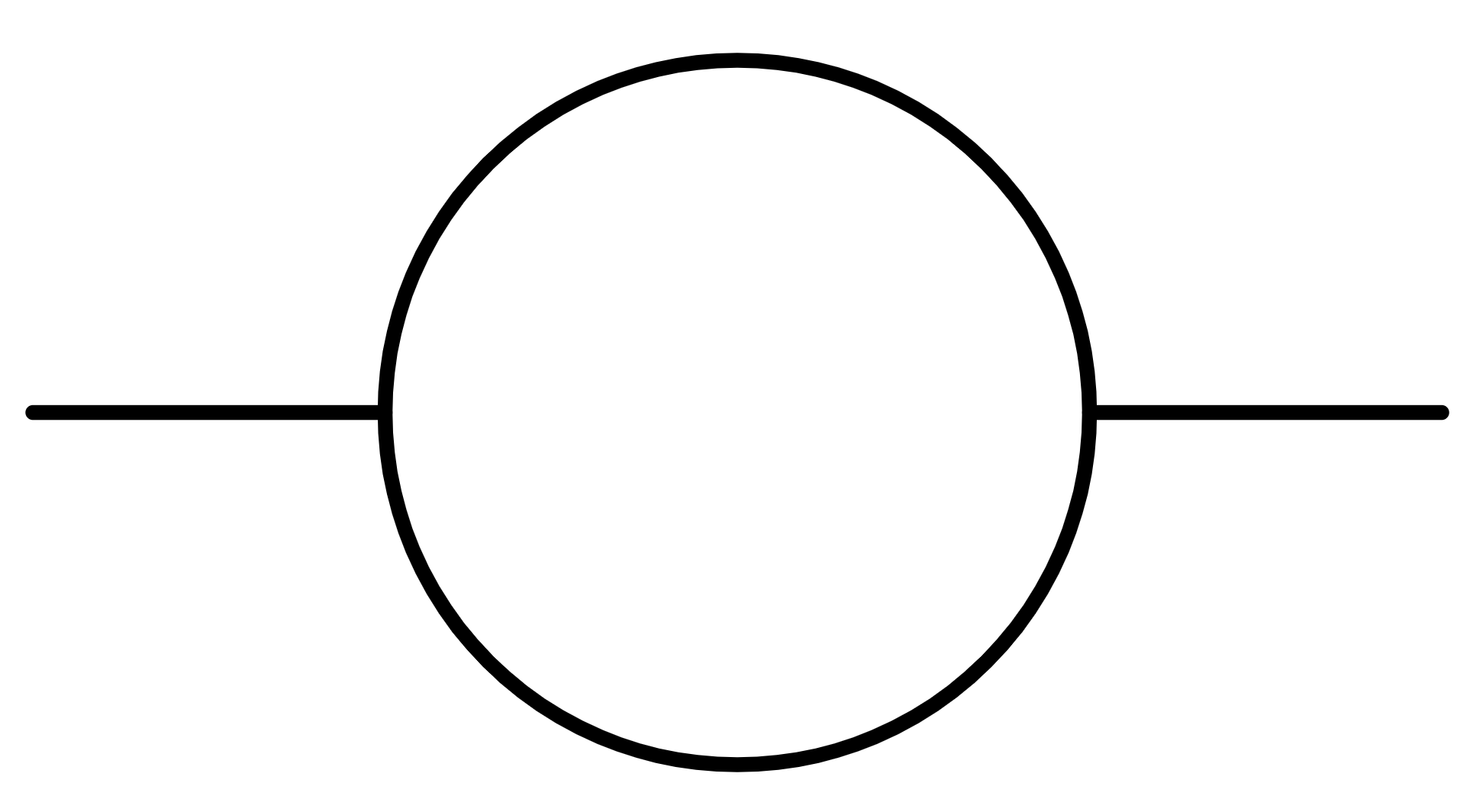}} \\[0.1cm]
            \small Bubble
        } \ \parbox[c]{0.25\textwidth}{
            \centering
            \raisebox{-0.5\height}{\includegraphics[height=1.4cm]{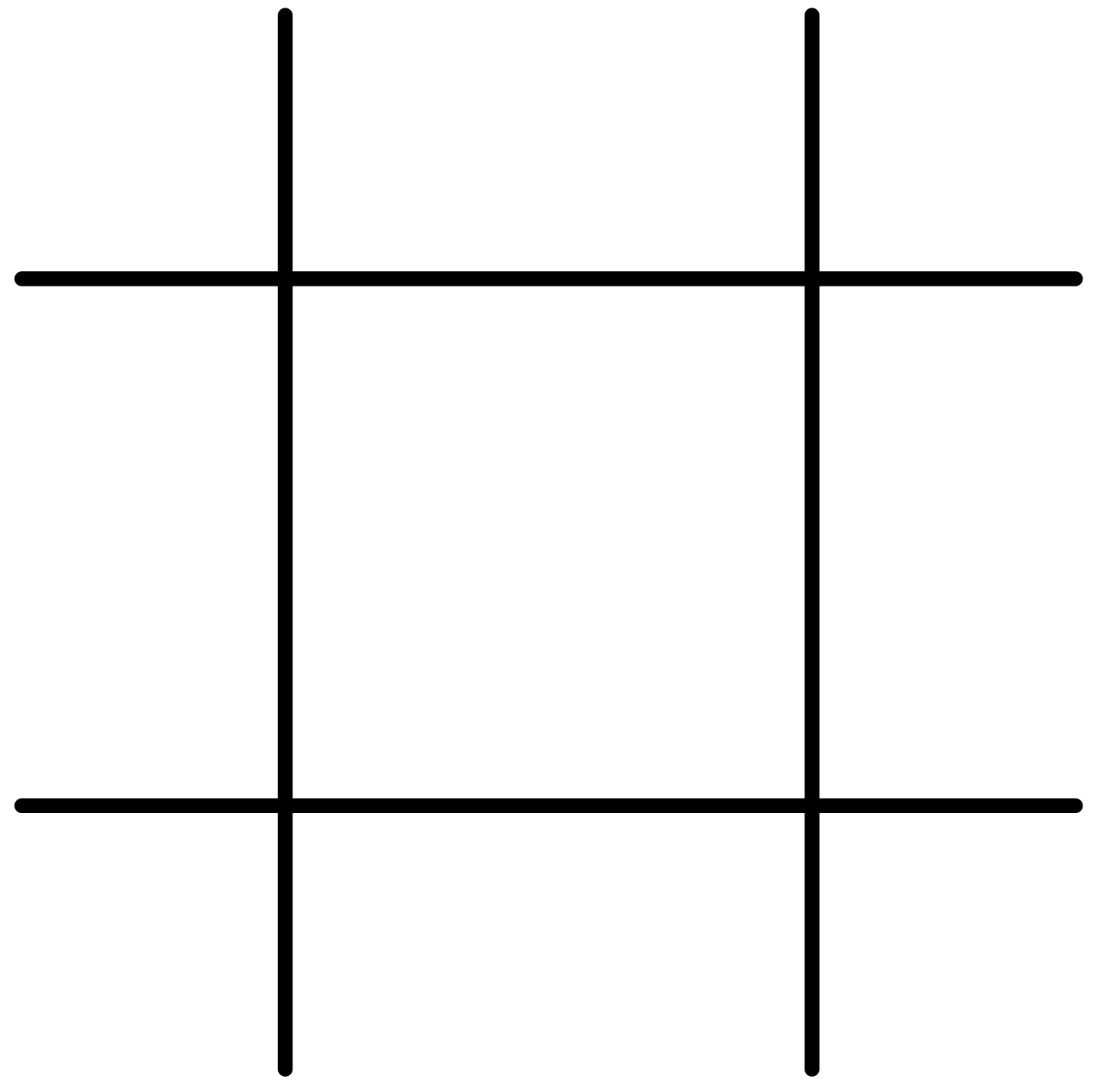}} \\[0.1cm]
            \small Box
        }} \\ \cline{1-1} 
        & & \\[-0.3cm]
		$\displaystyle \int \frac{d \alpha_1}{\sqrt{P_2(\alpha_1)}} \,$ & & \\[0.5cm] \hline
        & & \\[-0.5cm]
        $\displaystyle \int \frac{d \alpha_1}{\sqrt{P_4(\alpha_1)}}$ & \parbox[c]{0.2\textwidth}{
            \centering
            \raisebox{-0.5\height}{\includegraphics[height=1.1cm]{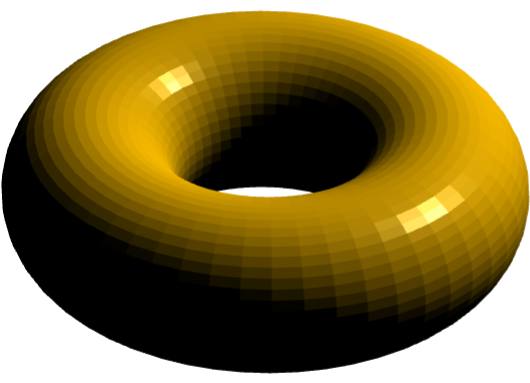}} \\[0.1cm]
            \small Torus
        } & \parbox[c]{0.2\textwidth}{
            \centering
            \raisebox{-0.5\height}{\includegraphics[height=1.2cm]{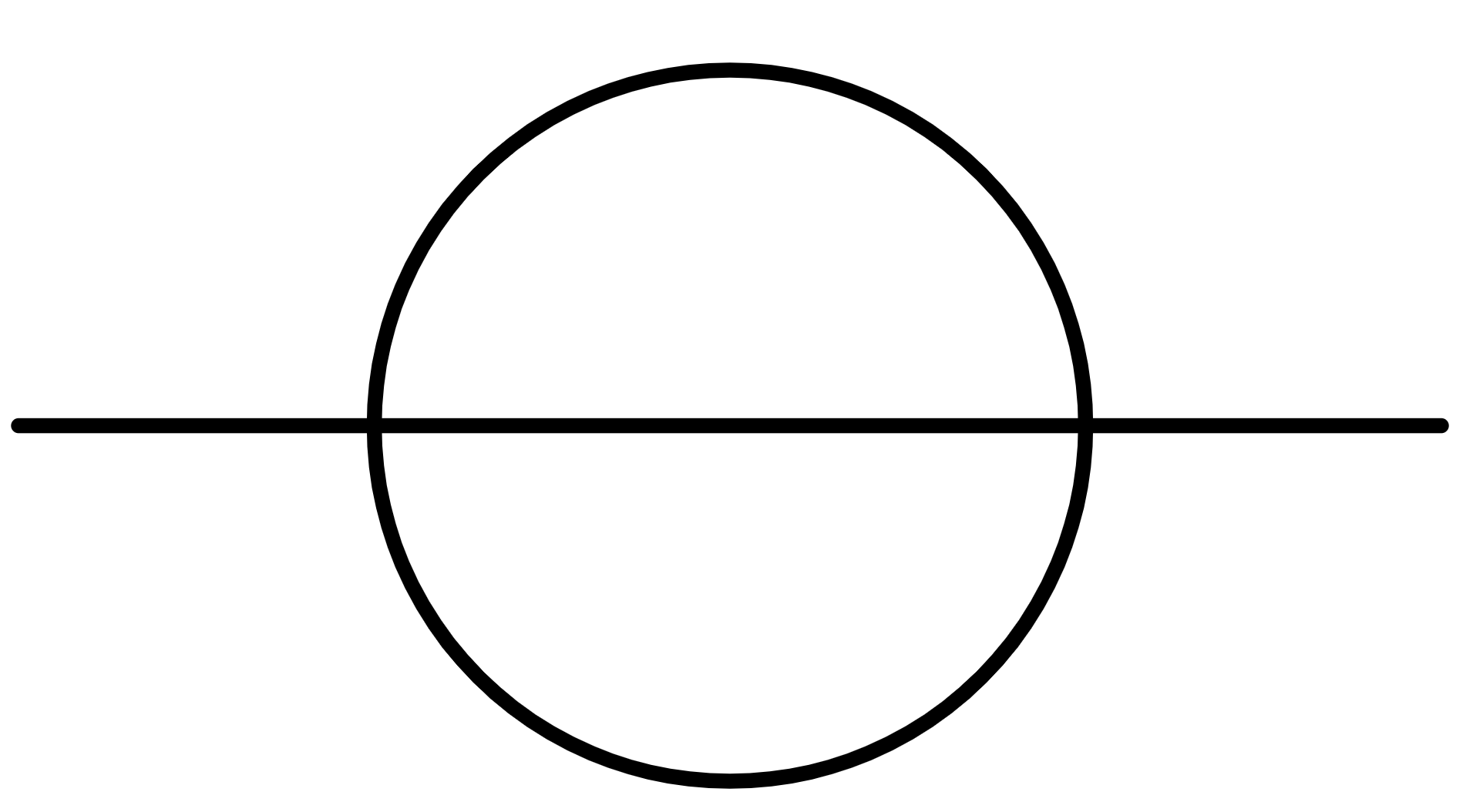}} \\[0.1cm]
            \small Sunrise
        } \ \parbox[c]{0.25\textwidth}{
            \centering
            \raisebox{-0.5\height}{\includegraphics[height=1.3cm]{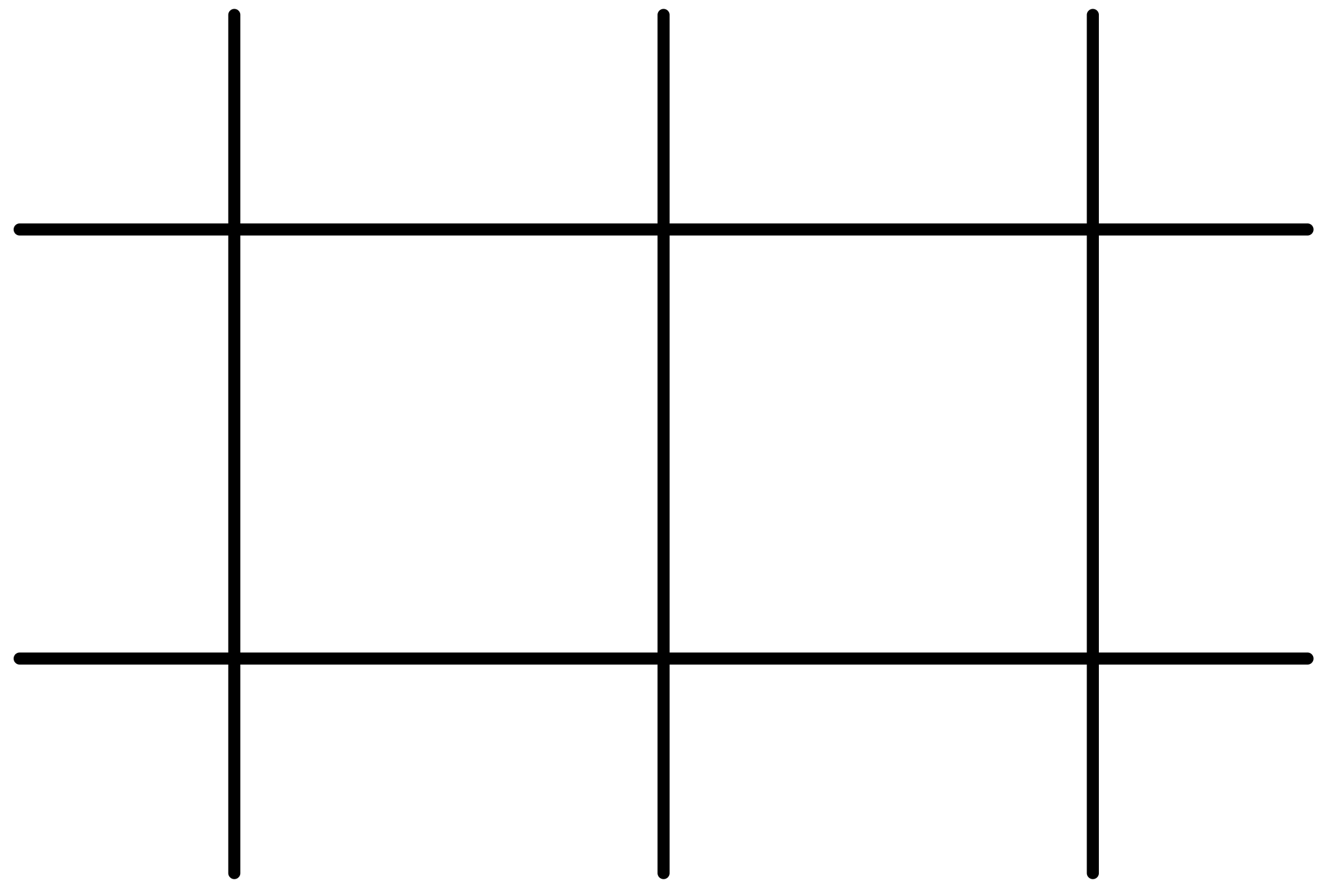}} \\[0.1cm]
            \small Double box
        } \\[0.7cm] \hline
        & & \\
\end{tabular}
\end{center}
\end{table}
While\footnotetext{The geometries have been drawn using \texttt{Asymptote}~\cite{Asymptote}, see \href{http://asymptote.ualberta.ca/}{http://asymptote.ualberta.ca/} for a web application.} the precise definition of Calabi--Yau manifolds will not be relevant for us, they are complex manifolds with certain properties, such as Ricci flatness, which make them among the simplest non-trivial geometries in a given dimension, yet with rich topological and geometric structure. The simplest cases are a CY onefold (an elliptic curve) and a CY twofold, also known as a K3 surface. Examples of Feynman integrals involving CY $L-1$-folds are the $L$-loop banana integrals in $D=2$~\cite{Bonisch:2021yfw,Pogel:2022vat}, which generalize the two-loop sunrise, as well as the $L$-loop traintrack integrals in $D=4$~\cite{Bourjaily:2018ycu,Cao:2023tpx} generalizing the double box; see tab.~\ref{tab:geometries_intro} for reference. Notice, however, that there are also integrals such as the $L$-loop tardigrade in $D=4$, which involve a CY manifold of dimension $2L-2$~\cite{Bourjaily:2018yfy,Lairez:2022zkj}, which grows much faster with the loop order than the banana and traintrack integrals.

To date, all Feynman integrals depending on non-trivial geometries have only been found to involve either higher-genus curves or Calabi--Yau manifolds~\cite{Bourjaily:2022bwx}. Thus, the question of whether other, more general geometries can also arise in the perturbative expansion of scattering amplitudes remains unresolved. See, however, ref.~\cite{Schimmrigk:2024xid} for a first investigation in this direction.

\subsection{Rationalization and changes of variables}
\label{sec:ch1_rationalization}

Before introducing multiple polylogarithms, let us review a key technique -- the rationalization of square roots -- that will play an important role in our toolkit, particularly in chapter~\ref{ch:chapter4} (see e.g.~ref.~\cite{Papathanasiou:2025stn} and the review~\cite{Bourjaily:2022bwx} for some applications in scattering amplitudes). The idea behind rationalization is to make appropriate changes of variables in order to simplify square roots into rational expressions, thus facilitating the calculations and integrations~\cite{Bourjaily:2018aeq}.

\begin{figure}[t]
\begin{center}
\includegraphics[height=5cm]{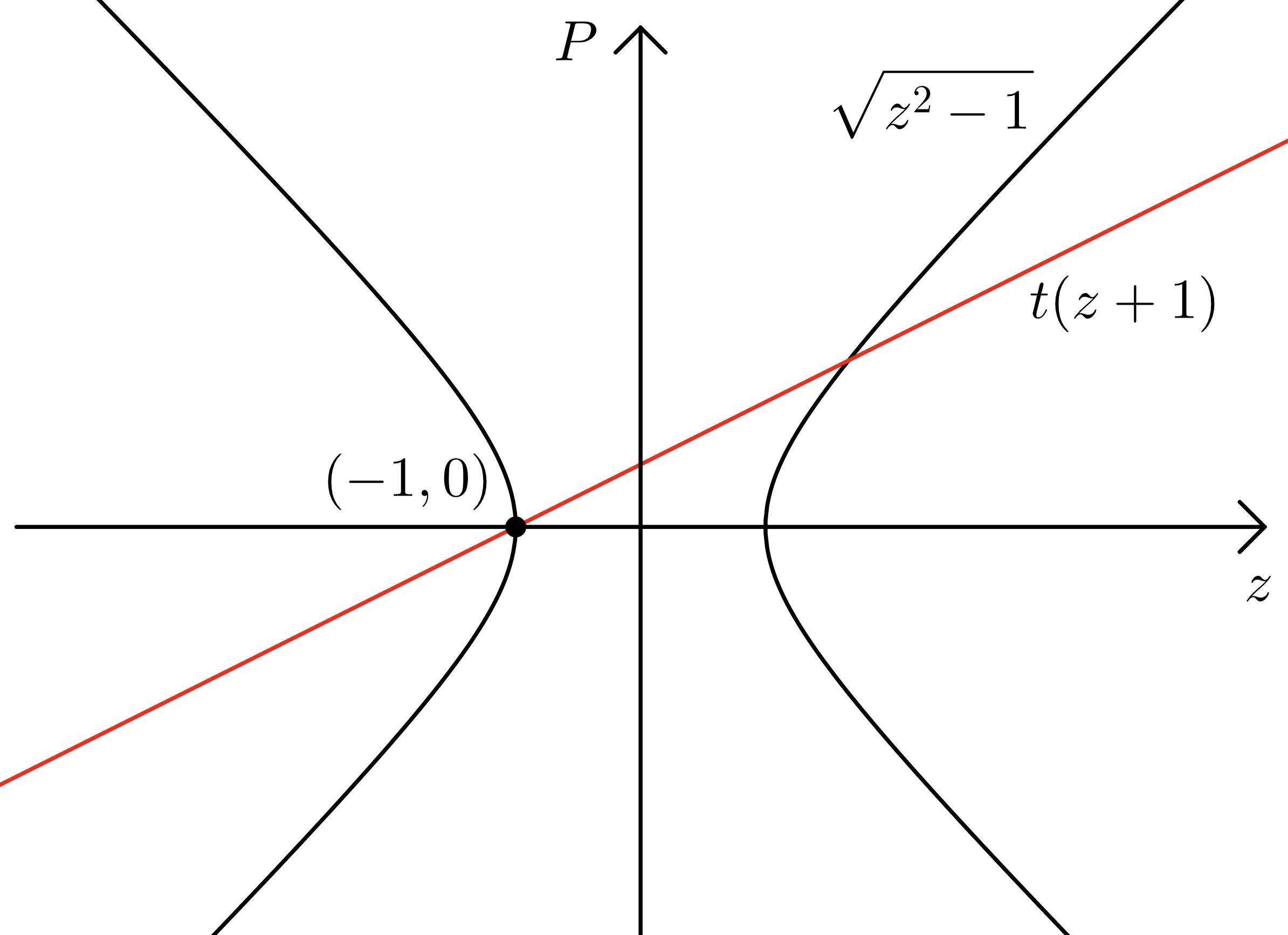}
\caption{Hyperbola $P(z)=\sqrt{z^2-1}$ (shown in black) and line $L(z)=t(z+1)$ (shown in red), used in the main text to exemplify the principle of rationalization of square roots.}
\label{fig: hyperbola_rationalization}
\end{center}
\end{figure}
To exemplify the rationalization of square roots, let us first consider
\begin{equation}
\label{eq: eq_hyperbola_rationalization}
    P(z) = \sqrt{z^2-1}\,,
\end{equation}
which defines the hyperbola shown in fig.~\ref{fig: hyperbola_rationalization}. As we will show below, the square root appears because we did not choose appropriate variables to parametrize the curve. Concretely, we are using Cartesian coordinates $(z,P)$, and the origin $(0,0)$ does not belong to the hyperbola. Let us instead consider the reference point $(-1,0)$, which belongs to the curve, and consider a line $L(z)=t(z+1)$ with slope $t$ going through this point. We may now investigate at which points does this line intersect the hyperbola. Equating $L(z)=P(z)$ and solving for $z$, we obtain two solutions
\begin{equation}
z=-1\,, \qquad \text{and} \qquad z=\frac{1+t^2}{1-t^2}\,.
\end{equation}
The first solution is the reference point $(-1,0)$, while the second one corresponds to another point on the curve, which depends on the slope of the line. Substituting the second solution in eq.~\eqref{eq: eq_hyperbola_rationalization}, we see that the curve
\begin{equation}
P(t)=\sqrt{\frac{4t^2}{(t^2-1)^2}} = \frac{2t}{t^2-1}
\end{equation}
develops a perfect square and becomes rational in this point. Thus, we have found a rational parametrization of the hyperbola. We can further introduce a Möbius transformation $t \to \frac{t-1}{t+1}$ for the slope, such that
\begin{equation}
\label{eq: rationalization_sqrt_root_PM}
z= \frac{1+t^2}{2t}\,, \qquad \qquad P(t) = \frac{1-t^2}{2t}\,.
\end{equation}
This transformation corresponds to the change of variables commonly employed in the post-Minkowskian expansion to rationalize $\sqrt{y^2-1}$ with $y=\frac{1+x^2}{2x}$~\cite{Parra-Martinez:2020dzs}; see also chapters~\ref{ch:chapter4} and~\ref{ch:chapter5}.

The principle behind the previous manipulation is that the reference point is already a rational point of the curve, since $P(-1)=0$. Therefore, as $P(z)$ is a quadratic polynomial, a line going through this rational point will only intersect the curve once again, through another rational point. Hence, finding a rational parametrization for a univariate quadratic polynomial reduces to finding one rational point, which is simple and can be done by inspection.

As a matter of fact, let us show how to rationalize the square root of a generic quadratic polynomial $P(z)=\sqrt{(z-r_1)(z-r_2)}$, where $r_i$ are the roots. First, we can redefine $z \to z + r_2$ and look instead at $\widetilde{P}(z)=\sqrt{z(z-r)}$, with $r \equiv r_2 - r_1$. Among other rational points, it is easy to identify that the origin $(0,0)$ is already a rational point of the curve. Thus, we can consider the line $L(z)=tz$ going through the origin, which leads to the solutions
\begin{equation}
    z=0\,, \qquad \text{and} \qquad z=\frac{r}{1-t^2}\,.
\end{equation}
Introducing a Möbius transformation $t \to \frac{t+1}{t-1}$, we obtain the well-known change of variables~\cite{Bonciani:2010ms,Adams:2018yfj}
\begin{equation}
z= -\frac{r(1-t)^2}{4t}\,, \qquad \qquad \widetilde{P}(t) = \frac{r(t^2-1)}{4t}\,.
\end{equation}
Recovering the original square root, we get a rational parametrization
\begin{equation}
\label{eq: change_of_variables_rationalization_r1r2}
z= r_1-\frac{(r_2-r_1)(1-t)^2}{4t}\,, \qquad \qquad P(t) = \frac{(r_2-r_1)(t^2-1)}{4t}\,.
\end{equation}
This change of variables, which notice is linear in the roots $r_i$, will be recurrently used in chapter~\ref{ch:chapter4} to rationalize square roots of quadratic polynomials when other square roots are simultaneously present. In addition, this change of variables shows that any quadratic polynomial can be rationalized, which was already used in sec.~\ref{sec:ch1_geometries}.

Lastly, let us consider another example which will also be used in chapter~\ref{ch:chapter4}, the square root $P(z)=\sqrt{z-C^2}$. We can take the rational point $(0,i C)$, and a line $L(z)=tz+iC$. Then, we obtain a change of variables
\begin{equation}
\label{eq: change_of_variables_rationalization_C}
z= \frac{1-2itC}{t^2}\,, \qquad \qquad P(t) = \frac{itC-1}{t}\,.
\end{equation}
As opposed to the solution that we would obtain by starting with the naive rational point $(C^2,0)$, with this parametrization we exploit the fact that we have a $C^2$ inside of the square root, such that the change of variables becomes linear in $C$ at the end.

While for lines and quadratic curves it suffices to find one rational point to rationalize them, for higher-degree curves, surfaces and hypersurfaces there are more restrictions, and no rational parametrization exists in general. For a hypersurface of degree $d$, one needs to identify (if there exists) a point of multiplicity $d-1$, i.e.~a point where all $i$-th partial derivatives vanish for $i=1,\dots,d-2$, while at least one $(d-1)$-th partial derivative is non-zero. Then, there exist similar algorithms employing intersections with lines that allow to rationalize such hypersurfaces, see ref.~\cite{Besier:2018jen,Besier:2019kco} as well as the \Mathematica~implementation \RationalizeRoots~\cite{Besier:2019kco}.

\section{Multiple polylogarithms}
\label{sec:ch1_MPLs}

As it has been alluded to in the previous sections, all one-loop Feynman integrals can be written as iterated integrals on the Riemann sphere; see for instance sec.~\ref{sec:ch4_Baikov} for a quick proof using the Baikov representation. To this end, we thus need to define a class of functions that spans the space of all iterated integrals over rational functions on the Riemann sphere; see ref.~\cite{Duhr:2014woa} for a detailed introduction.

Let us follow ref.~\cite{Broedel:2017kkb} and first investigate the type of functions that can arise in such iterated integrals. For that, we start with a rational function $R(x)$ with $n$ poles of multiplicity $m_i$ at some complex values $x=p_i$, $i=1,\dots,n$, which may also depend on the external kinematics. Using partial fraction, we can rewrite the rational function as
\begin{equation}
    R(x) = \sum_{i=1}^n \left[ \frac{c^{(i)}_{1}}{x-p_i} + \frac{c^{(i)}_{2}}{(x-p_i)^{2}} + \dots + \frac{c^{(i)}_{m_i}}{(x-p_i)^{m_i}} \right],
\end{equation}
where the $c^{(i)}_{j}$ are some coefficients depending on the kinematics. Therefore, we can limit our analysis to iterated integrals with integrands of the form $\frac{1}{(x-p_i)^k}$. Integrating once, we obtain
\begin{equation}
    \int \frac{dx}{(x-p_i)^k} = \frac{1}{1-k} \frac{1}{(x-p_i)^{k-1}}\,,
\end{equation}
which can be written again in terms of rational functions. This is, however, not the case for $k=1$, which yields instead
\begin{equation}
    \int \frac{dx}{x-p_i} = \log(x-p_i)\,.
\end{equation}
Thus, we observe that to integrate once over rational functions, we need to enlarge the space of primitives to also incorporate logarithms.

As a consequence, in order to define iterated integrals of rational functions we also need to accommodate for iterated integrals of logarithms. One such class of functions are classical polylogarithms, defined as
\begin{equation}
\label{eq: classical_polylogs}
    \text{Li}_n(x) = \int_0^x \frac{dt}{t} \, \text{Li}_{n-1}(t)\,, \qquad \text{with} \qquad \text{Li}_1(x) = \int_0^x \frac{dt}{1-t} = - \log(1-x)\,.
\end{equation}
In particular, classical polylogarithms also have a series representation
\begin{equation}
    \text{Li}_n(x) = \sum_{k=1}^\infty \frac{x^k}{k^n}\,, \qquad \text{where} \qquad \text{Li}_n(1) = \sum_{k=1}^\infty \frac{1}{k^n} = \zeta(n)\,,
\end{equation}
relating them to the Riemann zeta function $\zeta(n)$.

However, classical polylogarithms are not sufficient to capture the whole space of functions that can arise, since they only allow for integration kernels with simple poles at 0 and 1. Instead, the simplest construction spanning the whole space of iterated integrals over the Riemann sphere are the so-called multiple polylogarithms~\cite{Chen:1977oja,Goncharov:1995ifj}, also known in the literature as hyperlogarithms or Goncharov polylogarithms, which we turn to next.

\subsection{Definition}
\label{sec:ch1_MPLs_definition}

Multiple polylogarithms (MPLs)~\cite{Chen:1977oja,Goncharov:1995ifj}, which we will refer to as multiple polylogs, can be recursively defined via
\begin{equation}
\label{eq: definition_MPLs}
    G(a_1,\dots,a_n;x)= \int_0^x \frac{dt}{t-a_1} \, G(a_2,\dots,a_n;t)\,, \qquad \text{with} \qquad G(;x) = 1,
\end{equation}
where $n$ is known as the transcendental weight of the MPL, and $\vec{a} \equiv (a_1, \dots, a_n)$ is the vector of singularities. Notice that we now allow for integration kernels with simple poles at arbitrary points $a_i$ that can be algebraic, thus extending the construction of classical polylogarithms. Concretely, these ones appear as a special instance of MPLs,
\begin{equation}
    G(\underbrace{0,\dots,0}_{n-1},1;x)= -\text{Li}_n(x)\,.
\end{equation}
For $a_1=\dots=a_n=a$, eq.~\eqref{eq: definition_MPLs} yields
\begin{equation}
    G(\underbrace{a,\dots,a}_{n};x)= \frac{1}{n!} \log^n\Big(1-\frac{x}{a}\Big)\,,
\end{equation}
while for $a_1=\dots=a_n=0$ it becomes divergent and is regulated as
\begin{equation}
    G(\underbrace{0,\dots,0}_{n};x)= \frac{1}{n!} \log^n(x)\,.
\end{equation}
In particular, we see that MPLs can be used to express the result of the Feynman integral in eq.~\eqref{eq: intro_integral_preMPL}.

One of the most significant properties of MPLs is their shuffle algebra, which allows us to rewrite the product of MPLs as a linear combination of other MPLs via
\begin{equation}
    G(\vec{c_1};x) G(\vec{c_2};x) = \sum_{\vec{c}=\vec{c}_1 \shuffle \vec{c}_2} G(\vec{c};x)\,.
\end{equation}
The symbol $\shuffle$ is the shuffle product, which denotes all permutations between the elements of $\vec{c}_1$ and $\vec{c}_2$ while preserving their relative internal order; see the package \texttt{PolyLogTools}~\cite{Duhr:2019tlz} for a \Mathematica~implementation. 

Finally, it has been conjectured that relations among MPLs exist only between terms of the same transcendental weight~\cite{Duhr:2014woa}, such as those derived from their shuffle algebra. For example, at weight two, we encounter the following functional relation:
\begin{equation}
\label{eq: functional_relation_dilog}
\text{Li}_2(1-x) + \text{Li}_2(x) + \log(1-x) \log(x) - \frac{\pi^2}{6} = 0\,,
\end{equation}
where $\text{Li}_2(1)=\zeta(2)=\frac{\pi^2}{6}$ also has transcendental weight two. Although this conjecture greatly simplifies the landscape of possible relations that MPLs satisfy, these functional relations become increasingly intricate and bulky at higher transcendental weight. This complexity has significant implications for Feynman integrals, where results expressed in terms of MPLs often become bloated with terms, making their simplification highly challenging.

To facilitate the task of simplifying equations containing MPLs, we now introduce the concept of the symbol, a powerful mathematical operation that makes functional relations among MPLs manifest. Beyond its utility for simplifying expressions, the symbol will prove crucial for many computations, particularly in chapter~\ref{ch:chapter3}, where it will play a central role in the bootstrap of the two-loop twelve-point double box.

\subsection{The symbol}
\label{sec:ch1_MPLs_symbol}

The space of MPLs respects a Hopf algebra structure~\cite{Goncharov:2005sla}, and in particular there exists a coproduct that allows us to decompose any MPL of weight $n$ into a tensor product of lower-weight components. Iterating the coproduct, we can associate to each MPL a unique decomposition into an $n$-fold tensor product of components of weight one, that is, in terms of ordinary logarithms. This unique decomposition is known as the symbol~\cite{Goncharov:2010jf,Duhr:2011zq}; see ref.~\cite{Duhr:2014woa} for further details.

In particular, for a function $F^{(n)}$ with transcendental weight $n$, whose total differential is
\begin{equation}
\label{eq: differential_pre_symbol}
    d F^{(n)}=\sum_i F^{(n-1)}_i \, d\log(s_i)\,,
\end{equation}
the symbol is recursively defined via
\begin{equation}
\label{eq: symbol}
\mathcal{S}\Big(F^{(n)}\Big)=\sum_i \mathcal{S}\Big(F^{(n-1)}_i\Big)\otimes \log(s_i)\,, \qquad \text{with} \qquad \mathcal{S}(1) = 1\,.
\end{equation} 
The symbol is read from left to right, so that the left-most element in the tensor product is called the first entry, whereas the right-most element is referred to as the last entry. The number of entries in the tensor product is known as the length, and in polylogarithmic cases is equal to the transcendental weight of the function. The entries $\log(s_i)$ that appear in the tensor product are called the symbol letters,\footnote{In the polylog literature sometimes the logarithm is dropped, and the $s_i$ themselves are called the symbol letters. However, we explicitly keep the logarithm to avoid ambiguities with the elliptic symbol in chapter~\ref{ch:chapter3}.} and the set of all symbol letters for a given function is referred to as its symbol alphabet.

For instance, from eq.~\eqref{eq: classical_polylogs} we have that
\begin{equation}
\text{Li}_2(x) = - \int_0^x \frac{dt}{t} \log(1-t) = - \int_0^x d \log(t) \log(1-t)\,.
\end{equation}
Hence,
\begin{equation}
\label{eq: symbol_dilog}
d \, \text{Li}_2(x) = - \log(1-x) \, d \log(x)\,, \quad \Longrightarrow \quad \mathcal{S}(\text{Li}_2(x)) =- \log(1-x) \otimes \log(x)\,,
\end{equation}
and the symbol alphabet for the function $\text{Li}_2(x)$ is given by $\{ \log(1-x), \log(x) \}$.

One of the key properties of the symbol is its ability to make functional relations manifest by reducing them to the familiar logarithmic identity $\log(ab) = \log(a) + \log(b)$. To illustrate this, let us apply the symbol map to the combination $\text{Li}_2(1-x) + \text{Li}_2(x)$. Using eq.~\eqref{eq: symbol_dilog}, we find:
\begin{equation} 
\mathcal{S}\Big(\text{Li}_2(1-x) + \text{Li}_2(x)\Big) = - \log(x) \otimes \log(1-x) - \log(1-x) \otimes \log(x)\,. 
\end{equation}
This result can also be derived from $-\log(1-x) \log(x)$, as the total differential acts first on one logarithm and then on the other (but reversing their order in the tensor product). Consequently, we have
\begin{equation} 
\mathcal{S}\Big(\text{Li}_2(1-x) + \text{Li}_2(x) + \log(1-x) \log(x) \Big) = 0\,, 
\end{equation}
which represents the symbol-level\footnote{To ensure compatibility of the symbol map, we need to work modulo factors of $i \pi$; see ref.~\cite{Duhr:2014woa} for further details. These factors can be recovered when uplifting the symbol to the function level by forming an appropriate ansatz and numerically sampling values.} expression of the functional identity encountered in eq.~\eqref{eq: functional_relation_dilog}. In particular, eq.~\eqref{eq: functional_relation_dilog} manifestly vanishes under the symbol map, implying that any equation involving this combination of terms is automatically simplified. This demonstrates how the symbol can be used to reduce complex MPL expressions and rewrite them in a much simpler form; see the \Mathematica~package \texttt{PolyLogTools}~\cite{Duhr:2019tlz} for a computer implementation. Notably, this idea was first introduced in ref.~\cite{Goncharov:2010jf}, where the original 17-page result for the two-loop six-point remainder function in $\mathcal{N}=4$ SYM~\cite{DelDuca:2010zg} was reduced to just a few concise lines.

Besides the utility of the symbol in simplifying equations involving MPLs, it also plays a central role in different aspects of Feynman integrals. First, the symbol encodes the singularity structure of the integral~\cite{Duhr:2014woa}: the first entry is related to its discontinuities, while the last entry is linked to its derivative. For a generic symbol
\begin{equation}
\label{eq: generic_symbol}
    \mathcal{S}(F^{(n)})= \sum_{i_1, \dots, i_n} c_{i_1 \cdots\, i_n} \, S_{i_1} \otimes \cdots \otimes S_{i_n}
\end{equation}
of length $n$, where the $S_i$ are the symbol letters (containing the logarithms), we have
\begin{align}
\label{eq: disc_symbol}
     \mathcal{S}(\text{Disc}_x \, F^{(n)}) =&\, \sum_{i_1, \dots, i_n} c_{i_1 \cdots\, i_n} \, S_{i_2} \otimes \cdots \otimes S_{i_n} \, \text{Disc}_x S_{i_1}\,, \\
     \mathcal{S} (\partial_x F^{(n)}) =&\, \sum_{i_1, \dots, i_n} c_{i_1 \cdots\, i_n} \, S_{i_1} \otimes \cdots \otimes S_{i_{n-1}} \, \partial_x S_{i_n}\,.
\label{eq: deriv_symbol}
\end{align}
Secondly, the symbol is closely connected to the differential equations obeyed by Feynman integrals, where symbol letters appear in the differential equation matrix; see the next section for details. 

Lastly, one can pose the inverse question: whether a given symbol, expressed as a combination of tensor products, can always be associated to a well-defined function. In general, this is not always possible, since the function must satisfy that partial derivatives commute. This imposes constraints on the associated symbol, known as the integrability conditions~\cite{Chen:1977oja}, which for a generic symbol such as eq.~\eqref{eq: generic_symbol} take the form
\begin{equation}
\label{eq: integrability_condition}
0 = \sum_{i_1,\dots,i_n} c_{i_1 \cdots\, i_n} S_{i_1} \otimes \dots \otimes S_{i_{j-1}} \otimes S_{i_{j+2}} \otimes \dots \otimes S_{i_n} \, \Bigg( \frac{\partial S_{i_j}}{\partial x_k} \, \frac{\partial S_{i_{j+1}}}{\partial x_m}-\frac{\partial S_{i_j}}{\partial x_m} \, \frac{\partial S_{i_{j+1}}}{\partial x_k} \Bigg)\,,
\end{equation}
for all $1\leq j \leq n-1$. Here, the conditions must be satisfied for each of the \(\frac{N(N-1)}{2}\) pairs \(\{ x_k, x_m \}\) of the \(N\) independent kinematic variables, resulting in \(n-1\) constraints per pair. Note that for single-scale Feynman integrals, i.e.~for $N=1$, integrability becomes empty and imposes no constraints.

Together with an educated guess of the symbol alphabet and some physical intuition, the integrability conditions can sometimes be exploited to deduce the symbol of a Feynman integral without actually performing the full calculation -- a strategy known as the bootstrap method. This technique, which circumvents the complicated problem of integration by recasting it as a much simpler linear algebra exercise, has been very successful in the literature~\cite{Goncharov:2010jf,Dixon:2011pw,Dixon:2011nj,Brandhuber:2012vm,Caron-Huot:2016owq,Almelid:2017qju,Henn:2018cdp,Caron-Huot:2019vjl,Dixon:2020bbt,Guo:2021bym,Dixon:2022rse,Dixon:2022xqh,Hannesdottir:2024hke,Basso:2024hlx}, even allowing the computation of some of the highest-loop analytic results in perturbative QFT~\cite{Dixon:2022rse}. In chapter~\ref{ch:chapter3}, we will use the bootstrap approach to obtain for the first time the symbol of the two-loop twelve-point double-box integral, for which we will initiate the symbol bootstrap for elliptic Feynman integrals.

\section{Differential equations and leading singularities}
\label{sec:ch1_DE_LS}

In sec.~\ref{sec:ch1_geometries} we have seen that Feynman integrals are related to geometries and special functions. However, in general, it can be very hard to determine beforehand whether a given high-loop Feynman integral will evaluate to MPLs. One such strategy is proving that the integral is linearly reducible~\cite{Brown:2008um,Brown:2009ta,Panzer:2014gra}, a technique that mimics the integration and, if satisfied, guarantees that the integral can be expressed in terms of MPLs. However, this method only works in special cases, and is very sensitive to changes of variables; see chapter~\ref{ch:chapter2} for an application and concretely sec.~\ref{sec:ch2_polynomial_reduction} for details. Therefore, we would like to seek for other, more general methods to detect the presence of non-trivial geometries; see ref.~\cite{Bourjaily:2022bwx} for an overview.

In this section, we will first introduce a very powerful method -- differential equations for master integrals~\cite{Kotikov:1990kg} -- which can be used to detect non-trivial geometries as well as to calculate the result of Feynman integrals; see chapter~\ref{ch:chapter5} for an application in gravity. In addition, we will review a second technique, based on an analysis of leading singularities~\cite{Cachazo:2008vp,Arkani-Hamed:2010pyv}, which can be used to detect the presence of non-trivial geometries and to facilitate the evaluation of Feynman integrals through differential equations. Although the latter method is more sensitive to changes of variables, it can be much easier and less resource-consuming than differential equations, and it will prove very useful in chapter~\ref{ch:chapter4} to classify the geometries appearing in the post-Minkowskian expansion of classical gravity.

\subsection{Differential equations and Picard--Fuchs operator}
\label{sec:ch1_DEs}

In sec.~\ref{sec:ch1_Feynman_integrals}, we introduced differential equations as a method to calculate the result of individual Feynman integrals, see eq.~\eqref{eq: pre_diff_equation}. Notably, using IBP relations such as eq.~\eqref{eq: IBP_definition}, one can express all Feynman integrals in a family in terms of a basis of master integrals. As a consequence, if we obtain a solution for the master integrals, then we effectively solve any Feynman integral pertaining to the integral family. With that aim, let us first organize the $n$ master integrals into a vector $\vec{\mathcal{I}}$ and take a derivative with respect to $x$, leading to a vector of equations such as eq.~\eqref{eq: pre_diff_equation}. Then, we can use IBPs to rewrite the right-hand side again in terms of the master integrals, and obtain a linear system of coupled first-order differential equations
\begin{equation}
\label{eq: differential_equation_precanonical_intro}
\frac{d}{dx} \, \vec{\mathcal{I}}=\tilde{A}_x(x,\varepsilon) \, \vec{\mathcal{I}}\,.
\end{equation}
Here, $\tilde{A}_x(x,\varepsilon)$ is an $n \times n$ matrix with entries being rational functions of $x$ and $\varepsilon$. 

When taking a derivative, we cannot generate new propagators not already present in the original integral. By contrast, it can result in some propagators being absent due to $\nu_i=0$. Therefore, an integral in a given sector can only couple to other integrals in the same sector and its subsectors (recall the definitions from sec.~\ref{sec:ch1_Feynman_integrals}). Hence, if we organize the master integrals in a family from subsectors to the top sector, i.e.~$\vec{\mathcal{I}}=\left(\mathcal{I}^{(\text{sub})},\dots,\mathcal{I}^{(\text{top})}\right)^T$, the differential equation matrix naturally has a lower-triangular form of the type
\begin{equation}
\label{eq: matrix_shape_diff_eq}
\tilde{A}_x(x,\varepsilon)=
\begin{pNiceMatrix}[columns-width = 0.75em]
\Block[draw=blue,rounded-corners]{1-1}{}
\bullet  &  &  &  &  &  & & &  \\
\bullet & \Block[draw=blue,rounded-corners]{3-3}{} \bullet & \bullet  & \bullet  &   &   &  & &   \\
\bullet & \bullet  & \bullet  & \bullet  &   &   &  & &   \\
\bullet & \bullet  & \bullet  & \bullet  &  &   &  & &   \\
\bullet  & \bullet  & \bullet & \bullet  & \Block[draw=blue,rounded-corners]{2-2}{}\bullet  & \bullet &  & &  \\
\bullet  & \bullet  & \bullet & \bullet  & \bullet  & \bullet &  & &  \\
\bullet  & \bullet  & \bullet & \bullet  & \bullet  & \bullet & \ddots & & \\
 \bullet  & \bullet  & \bullet  & \bullet & \bullet & \bullet  & \bullet & \Block[draw=blue,rounded-corners]{2-2}{} \bullet &\bullet \\
\bullet  & \bullet  & \bullet  & \bullet & \bullet & \bullet  & \bullet &\bullet &\bullet
\end{pNiceMatrix}\,\,,
\end{equation}
where there are zeroes above the diagonal blocks. Each of these blocks in the diagonal corresponds to a different sector, that is, a diagram with different propagators. They are ordered from fewer to more propagators, with the top sector being the last block, which can couple to all of the subsectors. In some cases, however, some of the blocks can totally decouple from the rest, leading to independent sectors. For example, this happens in post-Minkowskian Feynman integrals due to the presence of linearized propagators, which split the integrals into two blocks of different parity~\cite{Parra-Martinez:2020dzs,DiVecchia:2021bdo,Herrmann:2021tct}; see chapter~\ref{ch:chapter4} and concretely sec.~\ref{sec:ch4_soft_expansion} for details. 

Although a priori this system seems more complicated to solve than a single inhomogeneous differential equation, we can change basis of master integrals to a new basis $\vec{\mathcal{J}}$, where the solution is easy to find. In particular, there is the so-called canonical or $\varepsilon$-factorized form~\cite{Henn:2013pwa},
\begin{equation}
\label{eq: differential_equation_canonical_intro}
\frac{d}{dx} \, \vec{\mathcal{J}}=\varepsilon A_x(x) \, \vec{\mathcal{J}}\,,
\end{equation}
where $A_x(x)$ is no longer dependent on $\varepsilon$, which is fully factorized. In such a case, the solution for $\vec{\mathcal{J}}$ can be systematically found in terms of a path-ordered exponential~\cite{Henn:2013pwa}, or order-by-order in $\varepsilon$. Therefore, the problem of evaluating any Feynman integral in the family is reduced to finding an $n \times n$ transformation matrix $U(x,\varepsilon)$, such that the master integrals $\vec{\mathcal{J}}=U \vec{\mathcal{I}}$ obey an $\varepsilon$-factorized differential equation, where the matrices are related via
\begin{equation}
\label{eq: relation_linear_system_intro}
\varepsilon A_x(x) = \frac{d U}{dx} \, U^{-1} + U \tilde{A}_x(x,\varepsilon) \, U^{-1}\,.
\end{equation}

Anticipating changes of variables also in the derivative, which we will use in chapter~\ref{ch:chapter5}, we can also introduce a new variable $\tau$, such that
\begin{equation}
\label{eq: differential_equation_canonical_tau_intro}
J \frac{d}{dx} \, \vec{\mathcal{J}}=\varepsilon A_\tau(\tau) \, \vec{\mathcal{J}}\,, \qquad \text{with} \qquad A_\tau = J A_x\,,
\end{equation}
where we use the Jacobian $J=\frac{dx}{d\tau}$.

The canonical form was first introduced in ref.~\cite{Henn:2013pwa} for polylogarithmic Feynman integrals, and nowadays there exist systematic algorithms that can find the appropriate change of basis to canonical form~\cite{Lee:2014ioa}; see refs.~\cite{Prausa:2017ltv,Gituliar:2017vzm,Meyer:2017joq,Dlapa:2020cwj,Lee:2020zfb} for different computer implementations. In fact, for polylogarithmic integrals one can demand that the entries of the differential equation matrix have at most single poles, such that $A_x(x) dx$ have $d \log$ entries. Comparing with eq.~\eqref{eq: differential_pre_symbol}, we see that in such case the entries correspond to the symbol letters.

Recently, the notion of an $\varepsilon$-factorized form has been also extended to elliptic~\cite{Adams:2018yfj,Dlapa:2020cwj,Dlapa:2022wdu,Jiang:2023jmk,Gorges:2023zgv}, hyperelliptic~\cite{Duhr:2024uid} and some simple single-scale Calabi--Yau cases~\cite{Pogel:2022ken,Pogel:2022vat,Klemm:2024wtd,Driesse:2024feo,Duhr:2025lbz}, although the connection with the symbol is not so well understood yet. As we will investigate in chapter~\ref{ch:chapter5}, the CY integrals studied in the literature so far require a good choice of initial master integrals to be able to find a canonical form, and the transformation matrix $U$ used is limited in scope. In particular, we will show that the methods in refs.~\cite{Pogel:2022ken,Pogel:2022vat} do not accommodate for $\varepsilon$-dependent apparent singularities, which appear for instance in the higher-order differential equation that a four-loop CY Feynman integral in classical gravity obeys. In sec.~\ref{sec:ch5_canonical_form}, we thus generalize the method to allow for such cases, and apply it to the CY integral in gravity in sec.~\ref{sec:ch5_canonical_form_gravity}, where we bring its differential equation into $\varepsilon$-factorized form.

Thus far, we have discussed how to use differential equations to compute Feynman integrals. Nevertheless, they can also be employed to detect non-trivial geometries before evaluating the integrals. Concretely, we can change basis to a derivative basis
\begin{equation}
   \integralseed, \frac{d}{dx}\, \integralseed, \dots, \left( \frac{d}{dx} \right)^{n-1} \integralseed\,,
\end{equation}
where we choose one of the $n$ master integrals in $\vec{\mathcal{I}}$ as the so-called seed integral $\integralseed$. Taking one further derivative, we may transform the system of first-order differential equations in eq.~\eqref{eq: differential_equation_precanonical_intro} into a single higher-order differential equation for the seed integral~\cite{Muller-Stach:2012tgj,Adams:2017tga},
\begin{equation}
\label{eq: PF_definition}
\mathcal{L}_n \, \integralseed = \left( \sum_{j=0}^{n} C_j(x,\varepsilon) \, \frac{d^j}{d x^j} \right) \, \integralseed = \text{inhomogeneity}\,.
\end{equation}
The operator $\mathcal{L}_n$ is known as the Picard--Fuchs operator, the $C_j(x,\varepsilon)$ are rational functions, and the inhomogeneity is given by a linear combination of master integrals arising from the subsectors of the seed integral. For instance, we have the following Picard--Fuchs operators in $D=2$ for the two-loop equal-mass sunrise~\cite{Laporta:2004rb} and one of its so-called eyeball top sectors~\cite{Giroux:2024yxu},
\begin{align}
\label{eq: PF_sunrise}
    \vcenter{\hbox{\includegraphics[height=1cm]{Sunrise.PNG}}} \ :& \, \qquad \mathcal{L}_2=
    \frac{d^2}{d x^2}+\frac{3 x^2-20 x+9}{x(x-1)(x-9)}\frac{d}{d x}+\frac{x-3}{x(x-1)(x-9)}\,,\\[0.2cm]
    \vcenter{\hbox{\includegraphics[height=0.95cm]{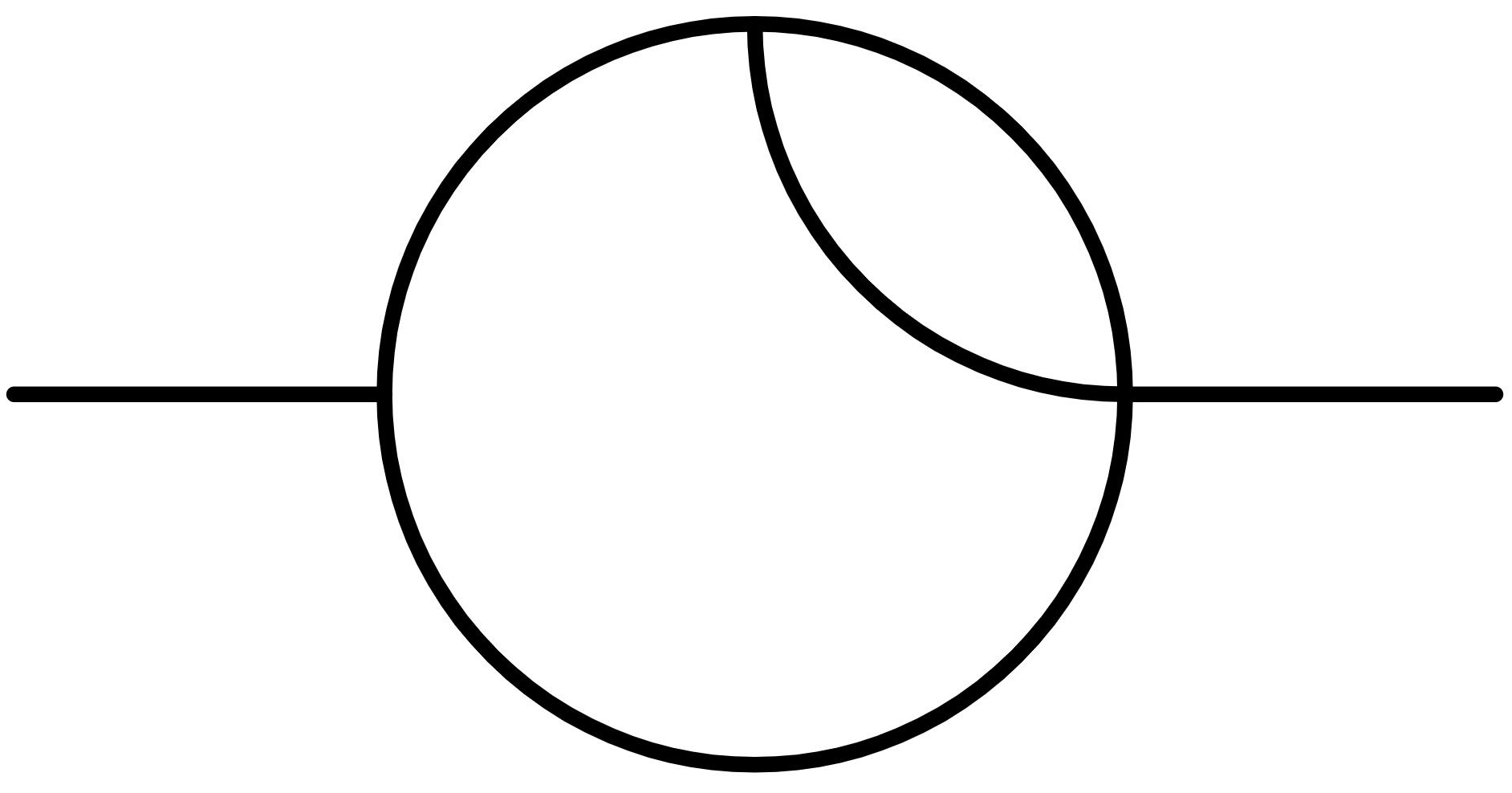}}} \ :& \, \qquad \mathcal{L}_1= \frac{d}{dx} + \frac{x-2}{x(x-4)}\,,
\label{eq: PF_eyeball}
\end{align}
where $x=\frac{p^2}{m^2}$ is the only dimensionless variable, and $m$ is the mass for all internal propagators.

As it turns out, we can detect the presence of non-trivial geometries from the rational factorization\footnote{For Picard--Fuchs operators of low order, and especially for univariate cases, the rational factorization can be easily obtained with the \texttt{DFactor} command in \Maple.} of the Picard--Fuchs operator into a product of lower-order irreducible differential operators $\mathcal{L}_{n_1} \cdots \, \mathcal{L}_{n_m}$, where $n_1 + \dots + n_m = n$. Specifically, if $\mathcal{L}_n$ fully factorizes into a product $\mathcal{L}_1 \, \cdots \, \mathcal{L}_1$ of $n$ (possibly different) operators of order one, then the seed integral can be brought to $d \log$ form. This means that it can be written as an integral with $d \log$ kernels, which typically implies that it can be evaluated in terms of MPLs. See, however, ref.~\cite{Duhr:2020gdd} as well as sec.~\ref{sec:ch2_staggered_elliptic_ladders} for (infinitely many) counterexamples to this claim. Instead, the integral is conclusively polylogarithmic if all subsectors can also be brought to $d \log$ form.

By contrast, the presence of any irreducible operator $\mathcal{L}_{n_i>1}$ indicates a non-trivial geometry. This is because the solutions to the homogeneous part of eq.~\eqref{eq: PF_definition} involve the periods of these geometries (see the next subsection for further details),\footnote{In ref.~\cite{Jockers:2024tpc} it was recently found, however, that the Picard--Fuchs operator for the four-loop banana integral, which depends on a Calabi--Yau threefold, has an alternative solution as a period of a genus-two curve, although with transcendental coefficients.} which as we saw in sec.~\ref{sec:ch1_geometries} lie beyond polylogarithms. In most known cases, the geometry corresponds to a CY $(n_i-1)$-fold, but one can further study the properties of the Picard--Fuchs operator to identify the geometry at hand; see for instance sec.~\ref{sec:ch5_PF_gravity_CY_operator}. 

For the previous example with the sunrise, we find that its Picard--Fuchs operator in eq.~\eqref{eq: PF_sunrise} is irreducible. As it is an irreducible differential operator of order $2$, it correctly indicates that the underlying geometry is a CY onefold, i.e., an elliptic curve; see tab.~\ref{tab:geometries_intro} for reference. By contrast, the operator in eq.~\eqref{eq: PF_eyeball} is of order one. While this factorization implies that the top sector can be written in $d \log$ form, since the elliptic sunrise is one of its subsectors, then it also involves an elliptic curve through the inhomogeneity.

Even though an analysis based on differential equations and Picard--Fuchs operators is very robust and clearly signals the presence of non-trivial geometries, it is also very inefficient. This is particularly true for the first step in the process, which corresponded to solving the system of IBPs. Although algorithmic methods for solving them exist, such as ref.~\cite{Laporta:2000dsw}, with various computer implementations~\cite{Smirnov:2023yhb,vonManteuffel:2012np,Lee:2013mka,Klappert:2020nbg}, solving these enormous coupled systems of equations to find a vector of master integrals can become tremendously time-consuming for high-loop applications. For example, recent calculations in PM gravity at four loops~\cite{Driesse:2024xad,Driesse:2024feo} required over 300,000 core hours in a high-performance cluster to perform the integrand reduction via IBPs. While there also exist alternative methods based on intersection theory to solve the IBPs~\cite{Mastrolia:2018uzb,Frellesvig:2019uqt}, this promising approach is still not competitive with traditional IBPs in state-of-the-art calculations. See also refs.~\cite{Lairez:2022zkj,delaCruz:2024xit} for a method to obtain the Picard--Fuchs operator of a Feynman integral by circumventing standard IBPs. Still, in chapter~\ref{ch:chapter4}, we will use IBPs to compute the Picard--Fuchs operator as a cross-check for verifying the presence of non-trivial geometries due to its reliability.

Given this drawback in the differential equations method, let us therefore introduce in the next subsection a complementary method that also allows us to detect non-trivial geometries -- an analysis of leading singularities. This method will be mostly used in chapter~\ref{ch:chapter4} to systematically classify the Feynman integral geometries appearing in the PM expansion of classical gravity. Besides, it can facilitate finding an $\varepsilon$-factorized form, as it provides a properly normalized master integral basis.

\subsection{Generalized cuts and leading singularities}
\label{sec:ch1_LS}

As we briefly touched upon in sec.~\ref{sec:ch1_geometries} and illustrated with a toy example using Feynman parameters in eq.~\eqref{eq: LS_geometry_intro}, with the discontinuity we can capture the dependence on the non-trivial geometry while dropping the unnecessary logarithmic functions. In general, the geometry is encapsulated by the so-called leading singularity (LS)~\cite{Cachazo:2008vp,Arkani-Hamed:2010pyv}, which is related to the maximally iterated discontinuity. In this section, we will review a related approach to calculate the leading singularity, which is not based on discontinuities but on generalized cuts; see for instance refs.~\cite{Primo:2016ebd,Frellesvig:2017aai,Bosma:2017ens,Bourjaily:2018ycu,Adams:2018kez,Festi:2018qip,Bourjaily:2019hmc,Vergu:2020uur,Bourjaily:2020hjv,Bourjaily:2021vyj} for some applications as well as ref.~\cite{Britto:2024mna} for a recent review.

Discontinuities in Feynman integrals can be calculated when the propagators go on-shell, i.e.~the propagators $\frac{1}{Q_i^2-m_i^2}$ are replaced by delta functions $\delta(Q_i^2-m_i^2)$. They can also be computed using an alternative approach: deforming the integration contour to encircle the pole at $Q_i^2 = m_i^2$ and subsequently taking the residue at the pole. This operation is known as a generalized cut (or simply a cut\footnote{Our notion of cuts should not be confused with unitarity cuts~\cite{Bern:1994zx,Bern:1994cg}, which sew together different vertices in a scattering amplitude and sum over intermediate states. Similarly, generalized cuts do not incorporate Heaviside step functions indicating the energy flow in the propagator~\cite{Cutkosky:1960sp}, and are regarded as a deformation of the contour, without a direct interpretation in terms of a discontinuity; see ref.~\cite{Britto:2024mna} for further distinctions.}), and the maximal cut refers to taking the residues when all propagators go on-shell simultaneously.

For a given integral sector (or diagram), the maximal cut selects those integrals where all propagators are present. By contrast, subsectors -- where some propagators are absent -- vanish under the maximal cut. As a consequence, and up to top sectors, the maximal cut for each diagram isolates the corresponding block in the diagonal of the differential equation~\eqref{eq: matrix_shape_diff_eq}. Similarly, taking the maximal cut in eq.~\eqref{eq: PF_definition} selects the homogeneous part, since the inhomogeneity is related to master integrals of subsectors.

The maximal cut can be extended to the so-called leading singularity, where in addition we deform all further integrals to closed contours. Notably, the leading singularity is a solution to the homogeneous part of the higher-order differential equation~\cite{Primo:2016ebd},
\begin{equation}
    \mathcal{L}_n \Big[ \LS (\mathcal{I}) \Big] = 0\,, \qquad \text{for} \qquad \mathcal{L}_n \, \mathcal{I} = \text{inhomogeneity}\,.
\end{equation}
This can be utilized to create a suitable seed integral for finding an $\varepsilon$-factorized form through the differential equation method; see sec.~\ref{sec:ch5_canonical_form} for details. Concretely, we can divide the master integral by its leading singularity, which in polylogarithmic cases is known to provide a pure basis~\cite{Henn:2014qga, Primo:2016ebd, WasserMSc, Dlapa:2021qsl}.

In addition, the leading singularity also serves as an indication of the Feynman integral geometry~\cite{Primo:2016ebd,Frellesvig:2017aai,Bosma:2017ens,Bourjaily:2018ycu,Adams:2018kez,Festi:2018qip,Bourjaily:2019hmc,Vergu:2020uur,Bourjaily:2020hjv,Bourjaily:2021vyj}. In particular, if the leading singularity of an integral and all of its subsectors is algebraic, which means that the result is a function of the kinematics without further transcendental integrals left, then the integral admits a $d \log$ form. This is related to the previous subsection, where we demanded that $\mathcal{L}_n$ fully factorized into $\mathcal{L}_1 \, \cdots \, \mathcal{L}_1$, since the solutions to the latter are given by algebraic functions. Otherwise, if the leading singularity contains non-trivial integrals, such as eq.~\eqref{eq: LS_geometry_intro}, it normally indicates the presence of a geometry beyond the Riemann sphere, since the result is a period of this geometry. For instance, following sec.~\ref{sec:ch4_Baikov}, and up to irrelevant constant factors, for the eyeball integral we find
\begin{equation}
\label{eq: LS_eyeball}
    \LS \left( \vcenter{\hbox{\includegraphics[height=0.95cm]{Eyeball.PNG}}} \right) \propto \frac{1}{\sqrt{x(x-4)}}\,,
\end{equation}
which can be easily verified to be annihilated by the Picard--Fuchs operator in eq.~\eqref{eq: PF_eyeball}. By contrast, the sunrise integral yields
\begin{equation}
\label{eq: LS_sunrise}
    \LS \left( \vcenter{\hbox{\includegraphics[height=1cm]{Sunrise.PNG}}} \right) \propto \int \frac{dz}{\sqrt{P_4(z)}} \,,
\end{equation}
with $P_4(z) = (z+1)(z-3)(z^2-2xz-4x+x^2)$. This corresponds to the period of an elliptic curve, and is again consistent with this integral involving an integral over a torus.

Finally, let us mention that, if a sector only has leading singularities that vanish, it means that it has no master integrals and is thus reducible to subsectors~\cite{Lee:2013hzt, Bosma:2017ens}. The opposite is more subtle, since depending on the dimension there can be IBP relations that vanish, such as the IBP identity relating the one-loop triangle to the one-loop bubble. While in general dimension the triangle is reducible to its bubble subsector, meaning that there are zero master integrals in the triangle sector, in $D=4$ this IBP relation vanishes, so there exist master integrals for the triangle. From the point of view of the leading singularity, the result is proportional to $0^{D-4}$, resulting in $1/s$ in four dimensions. To avoid complications, we will consider such cases as having zero master integrals within dimensional regularization.

\begin{figure}[tb]
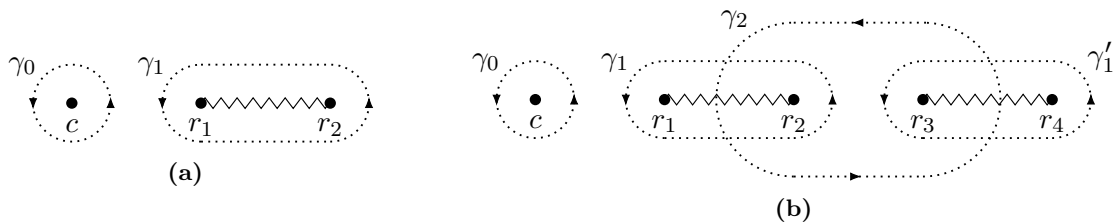

\centering
\subfloat[]{
}
\caption{Examples of closed contours for eq.~\eqref{eq: residues_square_root_example}, in the case of (a) a quadratic polynomial, and (b) a quartic polynomial.}
\label{fig: contours_branch_cuts}
\end{figure}
In the case of simple poles, the leading singularity corresponds to taking a residue, while for branch cuts several possible contours and leading singularities exist. For instance, let us consider
\begin{equation}
\label{eq: residues_square_root_example}
\LS \left( \int \frac{dx}{(z-c) \sqrt{P_n(z)}} \right),
\end{equation}
where $P_n(z)=(z-r_1)\cdots(z-r_n)$ is a polynomial of degree $n$ with different roots $r_i$. In the case of a quadratic polynomial there are two closed contours, see fig.~\ref{fig: contours_branch_cuts}(a). There is $\gamma_0$ around the pole and $\gamma_1$ around the branch cut, which give the same result up to a sign. Instead, for a polynomial of higher degree there exist more contours. For instance, for a quartic polynomial there exist four different contours, see fig.~\ref{fig: contours_branch_cuts}(b). Taking the residue at $z=c$ dictated by the contour $\gamma_0$ leads to an algebraic result $1/\sqrt{P_4(c)}$, without any further integrations left. However, eq.~\eqref{eq: residues_square_root_example} actually yields an elliptic integral in the case of a quartic polynomial. This non-trivial geometry can be explicitly seen by taking the contours $\gamma_1$ or $\gamma_2$, which respectively correspond to the so-called $a$-cycle and $b$-cycle of the elliptic curve. From this example we thus learn that, depending on which contours and residues are taken, the expected Feynman integral geometry can be very different. At this point, an important remark is in order: the distinction between propagators and ISPs becomes crucial. 

Under the leading singularity, residues at the locations where actual propagators go on-shell should always be taken. This is because such poles arise for top sectors, which contain more propagators and allow for more poles and residues than subsectors. While taking the residue omits the dependence on the possibly underlying non-trivial geometry, it is nonetheless taken into account by subsectors where this propagator (and thus the residue) is absent. For example, for the eyeball integral we actually had an intermediate step
\begin{equation}
    \LS \left( \vcenter{\hbox{\includegraphics[height=0.95cm]{Eyeball.PNG}}} \right) \propto \LS \left( \int \frac{dz}{z \sqrt{P_4(z)}} \right) \propto \frac{1}{\sqrt{P_4(0)}}\,,
\end{equation}
where $P_4(z)$ is the same quartic polynomial as in eq.~\eqref{eq: LS_sunrise}, and $z=k_2^2-m^2$ is the extra propagator compared to the sunrise diagram. As such, we take the residue at $z=0$, which results in eq.~\eqref{eq: LS_eyeball}. While taking the residue omits the elliptic curve, it is nonetheless taken into consideration from its sunrise subsector, see eq.~\eqref{eq: LS_sunrise}, where the fourth propagator and thus the residue are not present.

On the other hand, we should not take such residues for ISPs, since there can be master integrals with numerator factors canceling those poles. In the case of eq.~\eqref{eq: residues_square_root_example}, if $z-c$ were to correspond to an ISP and not a propagator, we could always choose a different integral $(z-c)\,\mathcal{I}$ where such a pole is absent. Therefore, taking the residue at $z=c$ in eq.~\eqref{eq: residues_square_root_example} would lead to an incorrect prediction for the geometry, if the degree of the polynomial is $n>2$. To avoid this subtlety with the integration contours, in this thesis we will not take residues on simple poles if there are square roots (of polynomials of ISPs of degree larger than two) involved. Instead, we will first rationalize (if possible) the square roots using changes of variables such as the ones introduced in sec.~\ref{sec:ch1_rationalization}, and only take residues afterwards; see chapter~\ref{ch:chapter4} for examples. 

An analysis of geometries via leading singularities is much simpler than through differential equations, and in sec.~\ref{sec:ch4_Baikov} we will present the loop-by-loop Baikov representation~\cite{Baikov:1996iu,Frellesvig:2017aai} as a convenient way to compute the leading singularity. The downside to this approach is that it relies on identifying suitable changes of variables -- a task that can be extremely challenging, particularly since there is no systematic approach to obtain them.

\newpage
\thispagestyle{plain}

\addtocontents{toc}{%
 \protect\vspace{1em}%
 \protect\noindent
\textcolor{DarkRed}{\textbf{II
\hspace{.4em} Elliptic Feynman integrals in $\boldsymbol{\mathcal{N}=4}$ SYM theory}}
\protect\par
 \protect\vspace{0em}%
}
\part*{Part II \\[1cm] Elliptic Feynman integrals\\ in $\boldsymbol{\mathcal{N}=4}$ SYM theory} 

\chapter{Elliptic ladder integrals}
\label{ch:chapter2}

\begin{info}[\textit{Info:}]
\textit{Part of the content and figures of this chapter have been published together with A.~McLeod, M. von Hippel, M. Wilhelm and C. Zhang in ref.}~\cite{McLeod:2023qdf}\textit{, available at} \href{https://doi.org/10.1007/JHEP05(2023)236}{\textit{JHEP} \textbf{05} (2023) 236} [\href{https://arxiv.org/abs/2301.07965}{2301.07965}].
\end{info}

\section{Motivation}
\label{sec:ch2_intro}

In the first half of the thesis, we focus on massless scalar Feynman integrals in planar 
$\mathcal{N}=4$ SYM theory. As discussed in chapter~\ref{ch:intro}, we will pay particular attention to those involving non-trivial geometries. A natural starting point for this investigation are ladder integrals in $D=4$, which arguably represent one of the simplest and most well-studied Feynman integral families. The basic case occurs at eight points, where ladder integrals are fully understood and can be expressed in terms of classical polylogarithms (defined in eq.~\eqref{eq: classical_polylogs}) to all loop orders~\cite{Usyukina:1993ch}. Schematically, at $L$ loops we have~\cite{Broadhurst:2010ds}
\begin{equation}
\label{eq: 4pt_ladders_result}
    \begin{tikzpicture}[scale=0.25, label distance=-1mm,baseline={([yshift=-0.05cm]current bounding box.center)}]
\node (v20) at (-3.5,2.5) {};
\node (v21) at (-5,3) {};
\node (v19) at (-4,4) {};
\node (v23) at (-3.5,0) {};
\node (v22) at (-5,-0.5) {};
\node (v18) at (-4,-1.5) {};
\node (v26) at (-1.5,2.5) {};
\node (v27) at (0.5,2.5) {};
\node (v42) at (2.5,2.5) {};
\node (v25) at (-1.5,0) {};
\node (v28) at (0.5,0) {};
\node (v41) at (2.5,0) {};
\node (v24) at (-4,-1.5) {};
\node (v45) at (4,-0.5) {};
\node (v16) at (3,-1.5) {};
\node (v44) at (4,3) {};
\node (v17) at (3,4) {};
\draw[thick]  (v20.center) edge (v21.center);
\draw[thick]  (v20.center) edge (v19.center);
\draw[thick]  (v20.center) edge (v23.center);
\draw[thick]  (v23.center) edge (v22.center);
\draw[thick]  (v23.center) edge (v18.center);
\draw[thick]  (v23.center) edge (v25.center);
\draw[thick]  (v20.center) edge (v26.center);
\draw[thick]  (v26.center) edge (v25.center);
\draw[thick]  (v28.center) edge (v27.center);
\draw[thick]  (v28.center) edge (v41.center);
\draw[thick]  (v41.center) edge (v42.center);
\draw[thick]  (v27.center) edge (v42.center);
\draw[thick]  (v42.center) edge (v44.center);
\draw[thick]  (v42.center) edge (v17.center);
\draw[thick]  (v41.center) edge (v45.center);
\draw[thick]  (v41.center) edge (v16.center);
\draw[thick]  (v26.center) edge (v27.center);
\draw[thick]  (v25.center) edge (v28.center);
\draw[fill=black] (-0.5,1.25) circle (2pt);
\draw[fill=black] (-0.1,1.25) circle (2pt);
\draw[fill=black] (-0.9,1.25) circle (2pt);
\end{tikzpicture} = \sum_{j=L}^{2L} c_j(x_1,x_2) \, \Im \text{Li}_j \Bigg( \sqrt{\frac{x_2}{x_1}} \, e^{i \phi(x_1,\,x_2)} \Bigg) \,,
\end{equation}
for some function $\phi(x_1,x_2)$ and coefficients $c_j(x_1,x_2)$, which depend on the kinematic variables $x_1=\frac{p_1^2 p_3^2}{s t}$ and $x_2=\frac{p_2^2 p_4^2}{s t}$, where $s$ and $t$ are Mandelstam variables. Despite their relative simplicity, the 8-pt ladder integrals already offer valuable insight into the all-loop structure of scattering amplitudes. Notably, in ref.~\cite{Broadhurst:2010ds} it was found that, when resummed to all loop orders, the contribution from 8-pt ladders becomes exponentially suppressed at strong coupling. Hence, this motivates extending the study to more general integral families, as already done in refs.~\cite{Caron-Huot:2018dsv,He:2020uxy} for the polylogarithmic 7-pt penta-ladder and 6-pt double penta-ladder families, which respectively correspond to
\begin{equation}
\label{eq: drawing_penta_ladders}
.
\end{equation}

However, this thesis focuses on a different generalization of ladder diagrams: instead of increasing the number of propagators in the boxes at the extremes, we consider the addition of external legs at the vertices of the boxes. This generalization forms the so-called traintrack integral family~\cite{Bourjaily:2018ycu}
\begin{equation}
%
,
\end{equation}
which is significant as each diagram represents the entire leading contribution to a particular component of the N$^{L+1}$MHV amplitude in planar massless $\mathcal{N}=4$ SYM theory at $L$ loops~\cite{Bourjaily:2018ycu}. In addition, they depend on non-trivial geometries, since at $L$ loops these diagrams involve integrals over CY $(L-1)$-folds~\cite{Bourjaily:2018ycu,Bourjaily:2019hmc,Vergu:2020uur,Cao:2023tpx}, making them a rich testing ground for studying Feynman integral geometries. In particular, they allow us to progressively develop tools to tackle the increasingly intricate geometries at higher loop orders, bridging the gap between the simpler polylogarithmic integrals and the elliptic and Calabi--Yau geometries, while also providing examples for multi-scale problems.

In this chapter, we begin our exploration of integrals involving non-trivial geometries by considering the simplest case: traintrack integrals depending on a single elliptic curve. The paradigmatic example is the 2-loop 10-pt double-box diagram
\begin{equation}
\begin{tikzpicture}[scale=0.25, label distance=-1mm,baseline={([yshift=-0.05cm]current bounding box.center)}]
\node (v20) at (-3.5,2.5) {};
\node (v21) at (-5,3) {};
\node (v19) at (-4,4) {};
\node (v23) at (-3.5,0) {};
\node (v22) at (-5,-0.5) {};
\node (v18) at (-4,-1.5) {};
\node (v26) at (-1.5,2.5) {};
\node (v42) at (0.5,2.5) {};
\node (v25) at (-1.5,0) {};
\node (v41) at (0.5,0) {};
\node (v24) at (-4,-1.5) {};
\node (v45) at (2,-0.5) {};
\node (v16) at (1,-1.5) {};
\node (v44) at (2,3) {};
\node (v17) at (1,4) {};
\node (va1) at (-1.5,4) {};
\node (vb1) at (-1.5,-1.5) {};
\draw[thick]  (v20.center) edge (v21.center);
\draw[thick]  (v20.center) edge (v19.center);
\draw[thick]  (v20.center) edge (v23.center);
\draw[thick]  (v23.center) edge (v22.center);
\draw[thick]  (v23.center) edge (v18.center);
\draw[thick]  (v23.center) edge (v25.center);
\draw[thick]  (v20.center) edge (v26.center);
\draw[thick]  (v26.center) edge (v25.center);
\draw[thick]  (v26.center) edge (v42.center);
\draw[thick]  (v25.center) edge (v41.center);
\draw[thick]  (v41.center) edge (v42.center);
\draw[thick]  (v42.center) edge (v44.center);
\draw[thick]  (v42.center) edge (v17.center);
\draw[thick]  (v41.center) edge (v45.center);
\draw[thick]  (v41.center) edge (v16.center);
\draw[thick]  (v26.center) edge (v27.center);
\draw[thick]  (v25.center) edge (v28.center);
\draw[thick]  (va1.center) edge (v26.center);
\draw[thick]  (vb1.center) edge (v25.center);
\end{tikzpicture},
\end{equation}
which has long been known to contain an elliptic curve~\cite{Paulos:2012nu,Caron-Huot:2012awx,Nandan:2013ip,Bourjaily:2017bsb}, and has even been fully computed in terms of elliptic multiple polylogarithms (eMPLs)~\cite{Kristensson:2021ani}. Building upon this, in this chapter we will study elliptic generalizations of the double box. Concretely, we can add loops and rungs to the double box, and modify the relative placement of the external legs in the middle. Through direct integration and linear reducibility, we will identify two families of elliptic Feynman integrals,
\begin{equation}
\label{eq: drawings_elliptic_ladders}
\parbox[c]{0.27\textwidth}{
            \centering
            \raisebox{-0.5\height}{
} \\[0.1cm]
            \small 10-pt elliptic ladder family
        \\[0.2cm]}, \qquad \qquad 
\parbox[c]{0.37\textwidth}{
            \centering
            \raisebox{-0.5\height}{%
} \\[0.1cm]
            \small 10-pt staggered elliptic ladder family
        \\[0.2cm]},
\end{equation}
which depend on the same elliptic curve as the 10-pt double box to all loops. Hence, their result can always be written in terms of the same class of eMPLs, making them the first elliptic Feynman integral families in the literature to exhibit this property. Furthermore, for both families we provide $2L$-fold integral representations that are linearly reducible in all but one variable, enabling the computation in terms of eMPLs to any loop order $L$. As a proof of principle, we explicitly compute a 3-loop result in terms of eMPLs. This makes these integrals exceptional candidates for testing new tools, such as bootstrap techniques for elliptic Feynman integrals, which we will turn to in chapter~\ref{ch:chapter3}. This possibility is further enhanced by the existence of two differential equations relating integrals within these families to lower-loop counterparts. 

Finally, let us highlight that the 10-pt staggered elliptic ladders admit a $d \log$ form on the maximal cut, a trait which commonly indicates that an integral is polylogarithmic. Since these integrals are not expected to be expressible purely in terms of MPLs due to the intrinsic elliptic curve, together with the example of ref.~\cite{Duhr:2020gdd}, they provide the only counterexamples known in the literature to the claim that an integral representation with $d \log$ kernels implies polylogarithmicity.

The remainder of this chapter is structured as follows. First, in sec.~\ref{sec:ch2_one_loop_box} we review the details for the 1-loop box integral, introducing dual-momentum variables. Then, in sec.~\ref{sec:ch2_traintracks}, we introduce traintrack diagrams and their dual momentum and Feynman parametrizations, as well as study in detail the 2-loop 10-pt double box, which will become relevant also for chapter~\ref{ch:chapter3}. In sec.~\ref{sec:ch2_linear_reducibility}, we then present the key technique -- linear reducibility -- that will allow us to identify the two elliptic ladder families. In addition, we introduce elliptic multiple polylogarithms, which can be used to evaluate the result of these elliptic integrals. The 10-pt elliptic ladder family is then studied in sec.~\ref{sec:ch2_elliptic_ladders}, whereas the 10-pt staggered elliptic ladders are analyzed in sec.~\ref{sec:ch2_staggered_elliptic_ladders}. Finally, in sec.~\ref{sec:ch2_diff_eqs}, we review the pair of differential equations satisfied by these elliptic ladder families, and conclude in sec.~\ref{sec:ch2_conclusions} with a discussion.

\section{One-loop box}
\label{sec:ch2_one_loop_box}

In order to illustrate the concepts that will be used throughout this chapter, such as dual coordinates, in this section let us first consider the simple 1-loop 4-mass\footnote{In this chapter, we consider massless scalar Feynman integrals. However, by attaching two (or more) on-shell massless external legs to the same vertex, we can mimic an off-shell (massive) leg. Hence, the 8-pt massless box is commonly referred to as the 4-mass box. Moreover, the 4-mass box becomes finite in strict $D=4$, whereas lower-point box integrals can be regulated while preserving dual-conformal symmetry~\cite{Bourjaily:2013mma}.} box integral in $D=4$. Since this integral will also play an important role in chapter~\ref{ch:chapter3}, we will also calculate its result and its symbol, for which we follow e.g.~ref.~\cite{Bourjaily:2019exo}. In momentum representation, the 1-loop 4-mass box integral is
\begin{equation}
\label{eq: 4mass_box_momentum_repr}
\begin{tikzpicture}[scale=0.55, label distance=-1mm,baseline={([yshift=-0.1cm]current bounding box.center)}]
\node (v1) at (0,0) {};
\node (v2) at (0,-2) {};
\node (v3) at (2,-2) {};
\node (v4) at (2,0) {};
\node[label=above:$1$] (p1) at (2,1) {};
\node[label=right:$2$] (p2) at (3,0) {};
\node[label=right:$3$] (p3) at (3,-2) {};
\node[label=below:$4$] (p4) at (2,-3) {};
\node[label=below:$5$] (p5) at (0,-3) {};
\node[label=left:$6$] (p6) at (-1,-2) {};
\node[label=left:$7$] (p7) at (-1,0) {};
\node[label=above:$8$] (p8) at (0,1) {};
\node[label=above:\textcolor{blue!50}{$x_1$}] (x1) at (1,0.5) {};
\node[label=right:\textcolor{blue!50}{$x_3$}] (x3) at (2.5,-1) {};
\node[label=below:\textcolor{blue!50}{$x_5$}] (x5) at (1,-2.5) {};
\node[label=left:\textcolor{blue!50}{$x_7$}] (x7) at (-0.5,-1) {};
\node[label={[xshift=0.25cm, yshift=-0.1cm]\textcolor{blue!50}{$x_l$}}] (xl) at (1,-1) {};
\draw[very thick]  (p1.center) edge (p4.center);
\draw[very thick]  (p5.center) edge (p8.center);
\draw[very thick]  (p2.center) edge (p7.center);
\draw[very thick]  (p3.center) edge (p6.center);
\draw[thick, blue!50]  (x1.center) edge (x5.center);
\draw[thick, blue!50]  (x3.center) edge (x7.center);
\draw[fill=blue!50] (xl) circle (2pt);
\end{tikzpicture} = \int \frac{d^4l}{l^2 {(l-(p_1+p_2))}^2 {(l-(p_1+\dots+p_4))}^2 {(l-(p_1+\dots+p_6))}^2} \equiv \mathcal{I}_{\text{box}}^{\text{(non-DCI)}}\,,
\end{equation}
where $l$ denotes the loop momentum, and the diagram in momentum space is drawn in black. 

Notably, this integral is not dual-conformal invariant (DCI) yet. To find a DCI representation, we can first express the external momenta as the difference $p_i = x_{i+1} - x_i$ between two dual-momentum $x$-coordinates, where we use a cyclic labeling with $x_9 = x_1$. Then, since $x_j - x_i= p_i + \dots + p_{j-1}$, we have
\begin{equation}
    \mathcal{I}_{\text{box}}^{\text{(non-DCI)}} = \int \frac{d^4l}{l^2 (l+x_1-x_3)^2 (l+x_1-x_5)^2 (l+x_1-x_7)^2} \,.
\end{equation}
Introducing a dual-momentum variable $x_l = l + x_1$ which is associated to the loop momentum, we obtain
\begin{equation}
\label{eq: non-DCI_4mass_box_dual_coords}
    \mathcal{I}_{\text{box}}^{\text{(non-DCI)}} = \int \frac{d^4 x_l}{x_{l,1}^2 x_{l,3}^2 x_{l,5}^2 x_{l,7}^2} \,,
\end{equation}
where we also use the short-hand notation $x_{i,j}^2 \equiv (x_i - x_j)^2$. Note that in dual-momentum representation, the propagators of the integral simply correspond to the edges of the graph dual to the momentum-space Feynman diagram, as represented in blue in eq.~\eqref{eq: 4mass_box_momentum_repr}. 

Now, we can uplift the scalar integral to be DCI by introducing an appropriate numerator,\footnote{Since all remaining integrals will be DCI, we will drop it in the notation in the following.} such as
\begin{equation}
\label{eq: DCI_box_dual_mom_repr}
    \drawbox \, = \int \frac{d^4 x_l \, x_{1,5}^2 x_{3,7}^2}{x_{l,1}^2 x_{l,3}^2 x_{l,5}^2 x_{l,7}^2} \,.
\end{equation}
Then, the result of the integral will only depend on two DCI cross-ratios, which we define as
\begin{equation}
\label{eq: 1-loop_4mass_box_cross-ratios_wrt_xi}
    u \equiv \frac{x_{1,3}^2 x_{5,7}^2}{x_{1,5}^2 x_{3,7}^2}\,, \qquad \qquad v \equiv \frac{x_{1,7}^2 x_{3,5}^2}{x_{1,5}^2 x_{3,7}^2}\,.
\end{equation}
Introducing Feynman parameters (recall sec.~\ref{sec:ch1_Feynman_integrals}), eq.~\eqref{eq: DCI_box_dual_mom_repr} can be rewritten in a manifestly DCI representation which explicitly depends on the DCI cross-ratios, and reads~\cite{Bourjaily:2019jrk}
\begin{equation}
\label{eq: 4mass-box_Feyn_param}
    \drawbox \, = \int_0^\infty \frac{d \alpha_1 \, d \beta_1}{(\alpha_1 + \beta_1 + \alpha_1 \beta_1) (1+\alpha_1 v + \beta_1 u)} \,,
\end{equation}
where the Feynman parameters are $\alpha_1$ and $\beta_1$. The result of this integral is well known~\cite{Hodges1977,Bern:1992em,Bern:1993kr}, see e.g.~ref.~\cite{Bourjaily:2019jrk} for our same notation, and can be easily calculated with modern computer software such as the \Maple~package \HyperInt~\cite{Panzer:2014caa}. To make the result more compact, we can introduce the reparametrization
\begin{equation}
\label{eq: uv_zzbar}
    u=z \bar{z}\,, \qquad \qquad v=(1-z)(1-\bar{z})\,,
\end{equation}
which conversely reads
\begin{equation}
\label{eq: zzbar_uv}
    z, \bar{z} = \frac{1}{2} \Big( 1+u-v \pm \sqrt{\Delta_4} \Big)\,.
\end{equation}
Here, we have a square root $\sqrt{\Delta_4} = z - \bar{z}$, where $\Delta_4$ corresponds to the (properly normalized) Gram determinant of the 4-mass box, given by
\begin{equation}
    \Delta_4 = \frac{\det (x_{i,j}^2)}{{(x_{1,5}^2 x_{3,7}^2)}^2} = (1-u-v)^2-4uv\,.
\end{equation}
The result of eq.~\eqref{eq: 4mass-box_Feyn_param} can be expressed in terms of classical polylogs (recall eq.~\eqref{eq: classical_polylogs}), and is given by
\begin{equation}
\label{eq: 4-mass_box_result}
    \drawbox \, = \frac{1}{\sqrt{\Delta_4}} \Bigg( 2 \text{Li}_2(z) - 2 \text{Li}_2(\bar{z}) + \log(z \bar{z}) \log(\frac{1-z}{1-\bar{z}}) \Bigg) \,.
\end{equation}
In this simple case, we can calculate the leading singularity as a discontinuity, which results in
\begin{equation}
\label{eq: 1-loop_4mass-box_LS}
    \LS \left( \, \drawbox \, \right) \propto \pm \frac{1}{\sqrt{\Delta_4}}\,,
\end{equation}
where the $\pm$ comes from the choice of sign for the square root in eq.~\eqref{eq: zzbar_uv}. Alternatively, we could have obtained the leading singularity from the Feynman parametrization by taking the maximal co-dimension residue, which corresponds to the residue at the point where both denominator factors $f_1 \equiv \alpha_1 + \beta_1 + \alpha_1 \beta_1$ and $g_1\equiv 1+ \alpha_1 v + \beta_1 u$ vanish. This can be easily calculated, e.g.~with the \Mathematica~package \texttt{MultivariateResidues}~\cite{Larsen:2017aqb}, leading to
\begin{equation}
    \mathrm{Res}_{f_1=g_1=0} \left( \frac{1}{(\alpha_1 + \beta_1 + \alpha_1 \beta_1) (1+\alpha_1 v + \beta_1 u)} \right) = \pm \frac{1}{\sqrt{\Delta_4}}\,.
\end{equation}
Since the leading singularity is a square root of the kinematics, without further transcendental integrals, from eq.~\eqref{eq: 4mass-box_Feyn_param} we could have already predicted that the result of the integral had to be polylogarithmic.

From the result in eq.~\eqref{eq: 4-mass_box_result}, we can compute the symbol of the 1-loop 4-mass box, which yields
\begin{equation}
\label{eq: 4-mass_box_symbol}
    \mathcal{S}\left( \, \drawbox \, \right) = \frac{1}{\sqrt{\Delta_4}} \Bigg( \log(u) \otimes \log(\frac{1-z}{1-\bar{z}}) - \log(v) \otimes \log(\frac{z}{\bar{z}}) \Bigg)\,.
\end{equation}
As can be seen, the first entries of the symbol are made of logarithms of the products $\{ z \bar{z}, (1-z)(1-\bar{z}) \}$, which give the two DCI cross-ratios from eq.~\eqref{eq: uv_zzbar}, while the last entries are given by logarithms of the quotients $\{ \frac{1-z}{1-\bar{z}}, \frac{z}{\bar{z}} \}$. This observation will become relevant in sec.~\ref{sec:ch3_Schubert_box}, where we will revisit the symbol of the 1-loop 4-mass box from a bootstrap perspective.

\section{Traintrack diagrams}
\label{sec:ch2_traintracks}

\begin{figure}
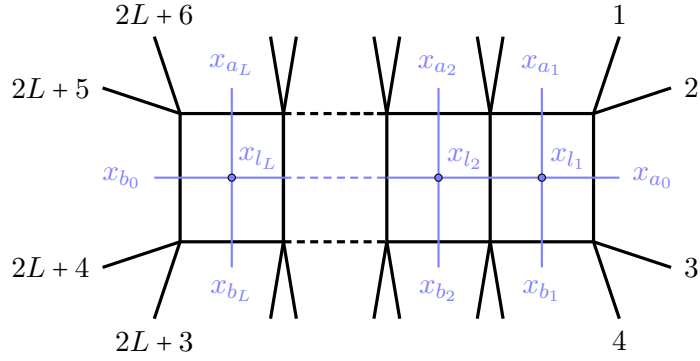

\centering
 
\caption{Momentum-space diagram (in black) and dual graph (in blue) for the $L$-loop $(4L+4)$-pt traintrack diagram, where dashed lines indicate an arbitrary number of loops.}
\label{fig: traintrack_diagram}
\end{figure}
Having exemplified the calculation for the 1-loop box, one can proceed analogously and introduce dual-momentum coordinates for any massless planar integral in $\mathcal{N}=4$ SYM theory. In particular, in the remainder of this chapter we will focus on different limits of the $L$-loop traintrack diagram of fig.~\ref{fig: traintrack_diagram} in $D=4$. Therefore, in this section we provide this integral in terms of dual coordinates, as well as its Feynman parametrization. As a first example, in sec.~\ref{sec:ch2_10pt_double_box} we study the 2-loop 10-pt limit of the traintracks -- the so-called 10-pt double box. This is the simplest integral in the traintracks family that depends on a non-trivial geometry, and it will also play a central role in the analysis at higher loops later on.

In dual coordinates, the integral in fig.~\ref{fig: traintrack_diagram} becomes~\cite{Bourjaily:2018ycu}
\begin{equation}
\label{eq: traintrack_dual_mom_repr}
    \mathcal{I}^{(L)}_{\text{traintrack}} = \int \frac{x_{a_0,b_0}^2 \prod_{i=1}^L d^4 x_{l_i} \, x_{a_i,b_i}^2}{x_{l_1,a_0}^2 \Big[ \prod_{j=1}^{L-1} x_{l_j,l_{j+1}}^2 x_{l_j,a_j}^2 x_{l_j,b_j}^2 \Big] x_{l_L,a_L}^2 x_{l_L,b_L}^2 x_{l_L,b_0}^2} \,.
\end{equation}
As can be seen, it is a simple generalization of eq.~\eqref{eq: DCI_box_dual_mom_repr}, but where we introduce new labels $a_i$ and $b_i$ for the external dual coordinates for compactness of the notation. In addition, we also introduce a numerator, which renders the integral DCI. Hence, the result will only depend on $(L+1)(2L-1)$ DCI cross-ratios\footnote{While for $n$ external dual points there are $4n-15$ algebraically independent cross-ratios, this parametrization uses the $n(n-3)/2$ multiplicatively independent ones~\cite{Bourjaily:2018aeq}, where $n=2L+2$ for the traintracks.} of the type
\begin{equation}
\label{eq: traintracks_cross-ratios_def}
    (ab;cd) \equiv \frac{x_{a,b}^2 x_{c,d}^2}{x_{a,c}^2 x_{b,d}^2}\,.
\end{equation}
As obtained in ref.~\cite{Bourjaily:2018ycu}, we can introduce a loop-by-loop Feynman parametrization which yields a manifestly DCI representation. Specifically, we have
\begin{equation}
\label{eq: traintrack_Feyn_param}
    \mathcal{I}^{(L)}_{\text{traintrack}} = \int_0^\infty \frac{d \alpha_1 \, \cdots \, d \alpha_L \, d \beta_1 \, \cdots \, d \beta_L}{f_1 \, \cdots \, f_L \, g_L} \,,
\end{equation}
with
\begin{subequations}
\begin{align}
    f_k \equiv & \, (a_0 a_{k-1}; a_k b_{k-1}) (a_{k-1} b_k; b_{k-1} a_0) (a_k b_k; b_{k-1} a_{k-1}) f_{k-1} + \alpha_k + \beta_k + \alpha_k \beta_k \nonumber \\
    & \, + \sum_{j=1}^{k-1} \Big[ \alpha_j \alpha_k (b_j a_0; a_j a_k) + \alpha_j \beta_k (b_j a_0; a_j b_k) + \alpha_k \beta_j (a_0 a_j; a_k b_j) + \beta_j \beta_k (a_0 a_j;b_k b_j) \Big]\,,\\
    g_L \equiv & \, 1 + \sum_{j=1}^L \Big[ \alpha_j (b_j a_0; a_j b_0) + \beta_j (a_0 a_j; b_0 b_j) \Big]\,,
\end{align}
\end{subequations}
where $\alpha_i$ and $\beta_i$ are the Feynman parameters and $f_0=0$. One can see that this parametrization is consistent with the 1-loop 4-mass box of eq.~\eqref{eq: 4mass-box_Feyn_param} given the replacements $(a_0 a_1; b_0 b_1) \to u$ and $(b_1 a_0;a_1 b_0) \to v$.

At this point, we proceed similarly to the 1-loop box, and calculate the leading singularity of the $L$-loop traintrack integral by computing the maximal co-dimension residue of its Feynman parametrization, obtaining~\cite{Bourjaily:2018ycu}
\begin{equation}
    \LS \left( \mathcal{I}^{(L)}_{\text{traintrack}} \right) = \int \frac{d \alpha_1 \, \cdots \, d \alpha_{L-1}}{\sqrt{P(\alpha_1, \dots, \alpha_{L-1})}}\,.
\end{equation}
The result is a transcendental integral such as eq.~\eqref{eq: LS_geometry_intro}, which would point towards a non-trivial Feynman integral geometry, in this case a CY $(L-1)$-fold. At $L=2$ and $L=3$ we indeed have an elliptic curve (see the next subsection) and a K3 surface, respectively. However, starting at $L=4$, if we embed the equation $y^2 - P(\alpha_1, \dots, \alpha_{L-1})$ in weighted projective space, see e.g. eq.~\eqref{eq: condition_degree_weight_CY}, the degree of the polynomial is too high and it no longer satisfies the CY condition. This analysis was however refined in ref.~\cite{Cao:2023tpx}, where they used a different parametrization for the traintracks, resulting in a leading singularity that satisfies the CY condition to all loop orders.

\subsection{Two-loop ten-point double box}
\label{sec:ch2_10pt_double_box}

As discussed above, let us study in detail the simplest of the traintrack integrals that depends on a non-trivial geometry: the 10-pt double-box. From eq.~\eqref{eq: traintrack_Feyn_param}, we obtain~\cite{Bourjaily:2017bsb}
\begin{equation}
\label{eq: 10pt_double-box_Feynman_param}
\begin{tikzpicture}[scale=0.55, label distance=-1mm,baseline={([yshift=-0.1cm]current bounding box.center)}]
\node (v1) at (0,0) {};
\node (v2) at (0,-2) {};
\node (v3) at (2,-2) {};
\node (v4) at (2,0) {};
\node (v5) at (4,-2) {};
\node (v6) at (4,-2) {};
\node[label=above:$1$] (p1) at (4,1) {};
\node[label=right:$2$] (p2) at (5,0) {};
\node[label=right:$3$] (p3) at (5,-2) {};
\node[label=below:$4$] (p4) at (4,-3) {};
\node[label=below:$5$] (p5) at (2,-3) {};
\node[label=below:$6$] (p6) at (0,-3) {};
\node[label=left:$7$] (p7) at (-1,-2) {};
\node[label=left:$8$] (p8) at (-1,0) {};
\node[label=above:$9$] (p9) at (0,1) {};
\node[label=above:$10$] (p10) at (2,1) {};
\node[label=above:\textcolor{blue!50}{$x_1$}] (x1) at (3,0.5) {};
\node[label=right:\textcolor{blue!50}{$x_3$}] (x3) at (4.5,-1) {};
\node[label=below:\textcolor{blue!50}{$x_5$}] (x5) at (3,-2.5) {};
\node[label=below:\textcolor{blue!50}{$x_6$}] (x6) at (1,-2.5) {};
\node[label=left:\textcolor{blue!50}{$x_8$}] (x8) at (-0.5,-1) {};
\node[label=above:\textcolor{blue!50}{$x_{10}$}] (x10) at (1,0.5) {};
\node[label={[xshift=0.275cm, yshift=-0.1cm]\textcolor{blue!50}{$x_{l_2}$}}] (xl2) at (1,-1) {};
\node[label={[xshift=0.275cm, yshift=-0.1cm]\textcolor{blue!50}{$x_{l_1}$}}] (xl1) at (3,-1) {};
\draw[very thick]  (p1.center) edge (p4.center);
\draw[very thick]  (p5.center) edge (p10.center);
\draw[very thick]  (p6.center) edge (p9.center);
\draw[very thick]  (p2.center) edge (p8.center);
\draw[very thick]  (p3.center) edge (p7.center);
\draw[thick, blue!50]  (x1.center) edge (x5.center);
\draw[thick, blue!50]  (x6.center) edge (x10.center);
\draw[thick, blue!50]  (x3.center) edge (x7.center);
\draw[fill=blue!50] (xl1) circle (2pt);
\draw[fill=blue!50] (xl2) circle (2pt);
\end{tikzpicture} = \int_0^\infty \frac{d \alpha_1 \, d \alpha_2 \, d \beta_1 \, d \beta_2}{f_1 f_2 \, g_2} \,,
\end{equation}
where
\begin{subequations}
\begin{align}
    f_1 = & \, \alpha_1 + \beta_1 + \alpha_1 \beta_1, \\
    f_2 = & \, u_3 u_4 u_5 f_1 + \alpha_2 + \beta_2 + \alpha_2 \beta_2 + u_4 \alpha_1 \beta_2 + u_3 \alpha_2 \beta_1, \\
    g_2 = & \, 1 + v_1 \alpha_1 + u_1 \beta_1 + u_2 \alpha_2 + v_2 \beta_2.
\end{align}
\end{subequations}
To reach this result, we have used that the legs $p_5$ and $p_{10}$ are massless and on-shell. This is reflected in the dual coordinates by the fact that $x_{1,10}^2=p_{10}^2=0$ and $x_{5,6}^2=p_5^2=0$ are zero. Hence, two (otherwise present) cross-ratios vanish, which reduces the kinematic dependence from 9 to the following 7 cross-ratios:
\begin{subequations}
\label{eq: DCI_cross-ratios_10pt_double-box}
\begin{align}
u_1=& \, \frac{x_{1,3}^2 \, x_{5,8}^2}{x_{1,5}^2 \, x_{3,8}^2} \, , \qquad \, u_2=\frac{x_{3,6}^2 \, x_{8,10}^2}{x_{3,8}^2 \, x_{6,10}^2} \, , \qquad u_3 =  \frac{x_{1,3}^2 \, x_{5,10}^2}{x_{1,5}^2 \, x_{3,10}^2} \, , \qquad u_4=\frac{x_{1,6}^2 \, x_{3,5}^2}{x_{1,5}^2 \, x_{3,6}^2} \, ,\\
u_5=& \, \frac{x_{1,5}^2 \, x_{6,10}^2}{x_{1,6}^2 \, x_{5,10}^2} \, , \qquad  v_1=\frac{x_{1,8}^2 \, x_{3,5}^2}{x_{1,5}^2 \, x_{3,8}^2} \, , \, \; \qquad v_2=\frac{x_{3,10}^2 \, x_{6,8}^2}{x_{3,8}^2 \, x_{6,10}^2} \, .
\end{align}
\end{subequations}

Calculating the maximal co-dimension residue, we obtain a leading singularity
\begin{equation}
\label{eq: LS_10pt_double-box_pre_change_vars}
    \LS \left( \, \drawdbten \, \right) = \int \frac{d \alpha_1}{\sqrt{P_4(\alpha_1)}}\,.
\end{equation}
Here, $P_4(\alpha_1)$ corresponds to a quartic polynomial, which indicates that this integral depends on an elliptic curve. Indeed, if carry out three of the integrations over the Feynman parameters in eq.~\eqref{eq: 10pt_double-box_Feynman_param}, we find~\cite{Bourjaily:2017bsb}
\begin{equation}
\label{eq: 10pt_double-box_old_paper}
    \drawdbten \, = \int_0^\infty \frac{d \alpha_1}{\sqrt{P_4(\alpha_1)}} \, H^{(3)}(\alpha_1) \,,
\end{equation}
where $H^{(3)}(\alpha_1)$ is a linear combination of MPLs of weight 3. In this result, we can explicitly see how the same elliptic curve predicted in eq.~\eqref{eq: LS_10pt_double-box_pre_change_vars} appears in the last integration. 

While this one-fold representation might already be useful, there are, however, two further square roots of quadratic polynomials in the arguments of the MPLs in $H^{(3)}(\alpha_1)$, which prevent us from evaluating the result in terms of elliptic multiple polylogarithms (see the next section). Moreover, trying to rationalize these square roots simultaneously using the technique reviewed in sec.~\ref{sec:ch1_rationalization} does not help, as the required changes of variables increase the degree of the polynomial in the denominator of eq.~\eqref{eq: 10pt_double-box_old_paper}, thus breaking the manifest ellipticity.

To resolve this, we need to take one step back in the calculation~\cite{Kristensson:2021ani}. In particular, after integrating two of the Feynman parameters, we have
\begin{equation}
\label{eq: 10-pt_double-box_intermediate_result}
    \drawdbten \, = \int_0^\infty \frac{d \alpha_1 \, d \alpha_2}{\sqrt{\widetilde{P}(\alpha_1,\alpha_2)}} \, \mathcal{G}^{(2)}(\alpha_1,\alpha_2)\,,
\end{equation}
where $\widetilde{P}(\alpha_1,\alpha_2)$ is of degree 3 in $\alpha_1$ and degree 2 in $\alpha_2$, and $\mathcal{G}^{(2)}(\alpha_1,\alpha_2)$ is a linear combination of weight-2 MPLs. Importantly, their arguments contain polynomials which are linear in $\alpha_1$ and $\alpha_2$, except for
\begin{subequations}
\label{eq: 10-pt_double-box_polynomials_q}
\begin{align}
    q_1 = & \, \alpha_1 (v_1 \alpha_1 + u_2 \alpha_2) + \text{linear}, \\
    q_2 = & \, -u_3 (v_1 \alpha_1 + u_2 \alpha_2) (u_4 u_5 \alpha_1 + \alpha_2) + \text{linear}, \\
    q_3 = & \, u_3 (v_1 \alpha_1 + u_2 \alpha_2) (u_4 u_5 \alpha_1 + \alpha_2) + \text{linear}.
\end{align}
\end{subequations}
These three polynomials are the ones introducing further square roots in the argument of the MPLs in eq.~\eqref{eq: 10pt_double-box_old_paper} after integrating over $\alpha_2$. However, since there is a common factor, it is easy to see that under the change of variables~\cite{Kristensson:2021ani}
\begin{equation}
\label{eq: 10-pt_double-box_change_vars}
    x= v_1 (\alpha_1 + \tilde{\alpha}_2 )\,, \quad \text{for} \quad \tilde{\alpha}_2=\frac{u_2}{v_1}\, \alpha_2\,; \qquad \text{or} \qquad \alpha_1 \to \frac{x}{v_1} - \tilde{\alpha}_2\,, \quad \alpha_2 \to \frac{v_1}{u_2}\, \tilde{\alpha}_2\,,
\end{equation}
these polynomials become linear in $\tilde{\alpha}_2$. Then,
\begin{equation}
\label{eq: 10pt_double-box_intermediate_after_change_vars}
    \drawdbten \, = \frac{1}{u_2} \int_0^\infty dx \int_0^{x/v_1} d \tilde{\alpha}_2 \ \frac{1}{\sqrt{\widetilde{P}\Big( \frac{x}{v_1} - \tilde{\alpha}_2,\frac{v_1}{u_2} \tilde{\alpha}_2 \Big) }} \ \mathcal{G}^{(2)} \Bigg( \frac{x}{v_1} - \tilde{\alpha}_2,\frac{v_1}{u_2} \tilde{\alpha}_2 \Bigg) \,,
\end{equation}
where the prefactor comes from the Jacobian of the transformation. Note also that the integration limits for $\tilde{\alpha}_2$ have changed, which will become a subtlety that we will have to bear in mind in the next sections.

We are therefore in a position where we can safely integrate over $\tilde{\alpha}_2$, as it should not introduce further square roots except for the desired elliptic curve. Indeed, we obtain~\cite{Kristensson:2021ani}
\begin{equation}
\label{eq: 10-pt_double-box_result}
    \drawdbten \, = \int_0^\infty \frac{dx}{y} \, \mathcal{G}^{(3)}(x,y) \,,
\end{equation}
where $\mathcal{G}^{(3)}(x,y)$ is a combination of weight-3 MPLs with arguments being rational functions of $x$ and $y$, and where we have the elliptic curve
\begin{equation}
\label{eq: elliptic_curve_10pt_double-box}
    y^2=\Bigg(\frac{v_1}{u_4}\Big((1-u_4)(1+x-v_2)-u_1+u_3v_2\Big)+h_1+h_2\Bigg)^2-4h_1h_2 \,,
\end{equation}
with
\begin{equation}
h_{1}\equiv \frac{u_2 u_4}{v_1}\Big( x^2+(1 - u_1+v_1)x+v_1\Big) \,, \quad
    h_{2}\equiv \left(x+\frac{v_1}{u_4}\right) \Bigg( (1+x-u_1) \Big( \frac{u_2 u_4}{v_1}-1\Big) +(1-u_3)v_2\Bigg) \,.
\end{equation}
While this one-fold integral representation may seem very similar to eq.~\eqref{eq: 10pt_double-box_old_paper}, by contrast, as it only involves one square root it can be evaluated directly in terms of eMPLs, see the next section for details.

Lastly, let us briefly discuss why does an elliptic curve arise at all for this Feynman integral. For the double-box integral, we can trace its ellipticity to the following first-order differential equation relating it to the 1-loop 10-pt hexagon in $D=6$~\cite{Paulos:2012nu,Nandan:2013ip},
\begin{equation}
\label{eq: diff_eq_10pt_double-box_hex}
    \partial_{u_5} \, \, \drawdbten \, = \frac{1}{\sqrt{-\Delta_6}} \, \, \drawhexten^{(6D)}.
\end{equation}
Here, $\Delta_6=\det (x_{i,j}^2)/{(x_{1,5}^2 x_{3,8}^2 x_{6,10}^2)}^2$ denotes the normalized Gram determinant of the hexagon. As it turns out, $\Delta_6$ is a cubic polynomial in $u_5$. Therefore, if we try to integrate the differential equation we get an elliptic integral, which explains the ellipticity of the 10-pt double box. We will revisit this differential equation, as well as the 1-loop hexagon, in chapter~\ref{ch:chapter3}, where they will play an important role.

\section{Linear reducibility and elliptic polylogarithms}
\label{sec:ch2_linear_reducibility}

In the previous section, we studied the 10-pt double-box integral through Feynman parameters and direct integration. From this example, we can extract two key insights: 
\begin{enumerate}
    \item The non-trivial geometry underlying a Feynman integral might only become explicit at the last integration step, where it manifests as a one-fold integral with a square root in the denominator.
    \item Changes of variables may be required at intermediate stages to reach a final result involving a single square root.
\end{enumerate}
Overall, these observations call for an algorithmic method that can detect such geometric obstructions during direct integration, while accommodating for changes of variables at intermediate steps. This is precisely the aim of the linear reducibility technique~\cite{Brown:2008um,Brown:2009ta,Panzer:2014gra}, which is the focus of sec.~\ref{sec:ch2_polynomial_reduction}. This technique mimics direct integration and can test whether there exist geometric obstructions to it. Furthermore, if satisfied, it provides a sufficient condition for an integral to be expressible in terms of MPLs.

The simplest cases violating linear reducibility are Feynman integrals where an elliptic curve emerges in the last integration step, such as the 10-pt double box that we just studied. Consequently, in secs.~\ref{sec:ch2_eMPLs} and~\ref{sec:ch2_eMPLs_torus}, we respectively introduce elliptic multiple polylogarithms on the elliptic curve and on the torus, which are two complementary formulations for the class of functions that these one-fold elliptic integrals evaluate to.

\subsection{Polynomial reduction}
\label{sec:ch2_polynomial_reduction}

While in general it can be very challenging to determine beforehand whether a given Feynman integral can be evaluated solely in terms of MPLs, there exists a criterion, known as linear reducibility~\cite{Brown:2008um,Brown:2009ta,Panzer:2014gra}, which guarantees that this is possible. In particular, linear reducibility states that all integrals can be carried out sequentially, and that at each step the corresponding integration variable appears at most linearly. Thus, we never introduce any square roots in the integration process.

Linear reducibility can be algorithmically checked with the polynomial reduction method~\cite{Brown:2008um,Brown:2009ta,Panzer:2014gra}. Given a suitable parametric representation for a starting integral, this method generates all possible polynomials that could appear if we were to perform the integrals, and hence it mimics the integration process. 

Concretely, let us assume that we have an integral of the form
\begin{equation}
\int_0^\infty \frac{d^{n} \vec{\alpha}}{F_1(\vec{\alpha}, \vec{x}) \, \cdots \, F_m(\vec{\alpha}, \vec{x})} \,,   
\end{equation}
where the $F_i$ are rational polynomials in the integration variables $\vec{\alpha}$ and in the kinematics $\vec{x}$. In particular, this is relevant for the Feynman parametrization of the traintrack diagrams, see eq.~\eqref{eq: traintrack_Feyn_param}. The first step is to identify one integration variable $\alpha_e$, which appears at most linearly in all polynomials $F_i$. Thus, we can write each polynomial as $F_i = A_i \alpha_e + B_i$. Then, the set of polynomials that could be generated if we were to perform the integral over $\alpha_e$ is given by the set of distinct irreducible factors of $\{ A_k, B_k \}$, $1\leq k \leq m$, and of the resultants $\{ A_k B_l - A_l B_k \}$, $1 \leq k<l \leq m$, where we drop all monomials and purely constant factors. Afterwards, we search for another integration variable which appears at most linearly in these new polynomials, and eliminate it in the same fashion. If there exists a sequence in which we can eliminate all the integration variables, then the integral is linearly reducible and can be expressed in terms of MPLs by performing the integrations in the same order. In addition, the final set of polynomials obtained after the full reduction contains the symbol alphabet for the integral. A computer implementation of the polynomial reduction algorithm can be found in the \Maple~package \HyperInt~\cite{Panzer:2014caa}.

To illustrate this method, let us apply it to the Feynman parametrization of the 1-loop 4-mass box, see eq.~\eqref{eq: 4mass-box_Feyn_param}. In this case, we have
\begin{equation}
    F_1 \equiv \alpha_1 + \beta_1 + \alpha_1 \beta_1, \qquad \qquad F_2 \equiv 1+ \alpha_1 v + \beta_1 u,
\end{equation}
where both polynomials are linear in $\alpha_1$ and $\beta_1$. Performing the polynomial reduction over $\alpha_1$, we obtain the set of irreducible factors
\begin{equation}
    \{ 1 + \beta_1, 1 + u \beta_1, 1 + \beta_1 + u \beta_1 - v \beta_1 + u \beta_1^2 \}.
\end{equation}
As can be seen, the last factor is quadratic in $\beta_1$, which would seem to spoil linear reducibility. Notice, however, that factorizing it would only introduce a square root of the kinematics, and therefore we can still express the result in terms of MPLs. Alternatively, we can explicitly restore linear reducibility by introducing the change of variables to $\{ z, \bar{z} \}$ as in eq.~\eqref{eq: uv_zzbar}, which makes the last polynomial factorize as $(1+z \beta_1)(1+\bar{z} \beta_1)$. Hence, after the transformation, we obtain the set of irreducible factors
\begin{equation}
    \{ 1 + \beta_1, 1 + z \bar{z} \beta_1, 1+z \beta_1, 1+\bar{z} \beta_1 \},
\end{equation}
which are all linear in $\beta_1$. Thus, we can reduce over $\beta_1$, resulting in
\begin{equation}
    \{ z, \bar{z}, 1-z, 1-\bar{z}, z-\bar{z}, 1- z \bar{z} \}.
\end{equation}
Comparing with eqs.~\eqref{eq: 4-mass_box_result} and~\eqref{eq: 4-mass_box_symbol}, we observe that the first four factors make up the symbol of the 1-loop 4-mass box, whereas $z-\bar{z}=\sqrt{\Delta_4}$ appears in the denominator of the result, and the last factor is not present.

In general, although linear reducibility is sensitive to the chosen integration sequence, the algorithm for polynomial reduction is amenable to changes of variables and we can easily adapt the result to the transformation, as shown in the previous example. There is, however, one subtlety regarding changes of variables, and it has to do with the integration limits. Looking for example at eq.~\eqref{eq: 10pt_double-box_intermediate_after_change_vars} for the 10-pt double box, we see that after the change of variables, the integration limits for $\tilde{\alpha}_2$ have been modified to $[0, A]$, with $A=x/v_1$. In order to account for this modification using the polynomial reduction algorithm, we need to restore the original $[0, \infty)$ integration limits. To do so, we can perform a further change of variables $x \to \frac{A x}{1+x}$, which has the effect
\begin{equation}
    \int_0^A dx f(x) = A \int_0^\infty \frac{dx}{(1+x)^2} \, f \left(\frac{Ax}{1+x} \right) \,.
\end{equation}
Hence, we see that it modifies the factors appearing in the reduction. We will use this transformation in secs.~\ref{sec:ch2_elliptic_ladders} and~\ref{sec:ch2_staggered_elliptic_ladders} to amend the polynomial reduction after changes of variables have been performed.

Lastly, let us study what happens in the polynomial reduction of the 10-pt double box. In that case, after the reductions over $\beta_1$ and $\beta_2$ are performed, we obtain 11 irreducible factors, 4 of which are quadratic (or worse) in the remaining two variables. One factor is the polynomial $\widetilde{P}(\alpha_1,\alpha_2)$ and the others are $q_1$, $q_2$ and $q_3$; see eqs.~\eqref{eq: 10-pt_double-box_intermediate_result} and~\eqref{eq: 10-pt_double-box_polynomials_q}. We can then search for a $\mathrm{GL}(2,{\mathbb C})$ rotation
\begin{equation}
\label{eq: GL2_rotation_double_box_general}
\begin{pmatrix}
\tilde{\alpha}_1 \\
\tilde{\alpha}_2
\end{pmatrix} = \begin{pmatrix}
a_{1,1} & a_{1,2}\\
a_{2,1} & a_{2,2}
\end{pmatrix} \begin{pmatrix}
\alpha_1 \\
\alpha_2
\end{pmatrix}\,,
\end{equation}
which eliminates those higher-order terms from the polynomials for one of the variables, say $\tilde{\alpha}_2$. If we demand that all higher-order terms vanish, no change of variables is found. By contrast, demanding that the only quadratic term in $\tilde{\alpha}_2$ that survives is for $\widetilde{P}(\alpha_1,\alpha_2)$, we find the same change of variables as in eq.~\eqref{eq: 10-pt_double-box_change_vars}. 

Integrals like this example, where the penultimate step in the polynomial reduction contains only one polynomial which is quadratic in the integration variable, lead to one-fold integral representations where there is only a single square root. If the roots of the quadratic polynomial contain a square root of a quartic polynomial in the last integration variable, then it results in a one-fold elliptic integral. This happens for instance for the 10-pt double box, see eq.~\eqref{eq: 10-pt_double-box_result}. Since only the elliptic curve obstructs linear reducibility in the last step, this class of integrals are known in the literature as linearly reducible elliptic Feynman integrals~\cite{Hidding:2017jkk}. Their result can be written in terms of elliptic multiple polylogarithms, which is the focus of the next subsection.

\subsection{Elliptic multiple polylogarithms}
\label{sec:ch2_eMPLs}

In sec.~\ref{sec:ch1_MPLs}, we outlined how MPLs are the class of functions describing iterated integrals over rational functions on the Riemann sphere. In this section, we present a generalization: the class of functions describing iterated integrals on an elliptic curve; see refs.~\cite{Broedel:2017kkb,Broedel:2017siw,Broedel:2018iwv,Broedel:2018qkq} for more details.

To begin with, let us proceed analogously as in sec.~\ref{sec:ch1_MPLs}, and first consider the type of functions that can arise in such iterated integrals, following ref.~\cite{Broedel:2017kkb}. In this case, we have iterated integrals over rational functions $R(x,y)$, which are subject to the constraint $y^2=P(x)$ for a cubic or quartic polynomial $P(x)$. Thus, we can write
\begin{equation}
    R(x,y) = \frac{q_1(x) + q_2(x) \, y}{q_3(x) + q_4(x) \, y} \,,
\end{equation}
for $q_i(x)$ being some polynomials. We can now rationalize the denominator by multiplying and dividing by $q_3(x) - q_4(x) y$, which yields
\begin{equation}
    R(x,y) = R_1(x) + \frac{R_2(x)}{y} \,,
\end{equation}
where $R_1(x)$ and $R_2(x)$ are rational functions. As can be seen, iterated integrals over the first term will only generate the rational (over $x$) integration kernels already present in MPLs, defined in sec.~\ref{sec:ch1_MPLs}. By contrast, after partial fraction, the second term leads to the new kernels
\begin{equation}
    \int \frac{x^k dx}{y}\,, \qquad \text{and} \qquad \int \frac{dx}{y (x-c)^k}\,.
\end{equation}
Using integration-by-parts, and focusing on the case of a quartic polynomial $y^2=P_4(x)$, we can reduce the higher powers to a linear combination of rational kernels (in $x$), as well as
\begin{equation}
    \int \frac{dx}{y}\,, \qquad \int \frac{x dx}{y}\,, \qquad \int \frac{x^2 dx}{y}\,, \qquad \int \frac{dx}{y(x-c)}\,.
\end{equation}
The first kernel is a nowhere-vanishing holomorphic differential one-form, thus it corresponds to the so-called differential of the first kind. The second kernel has a simple pole at infinity and is a differential of the third kind. While the third kernel has a double pole at infinity, it also contains a simple pole. Thus, the differential of the second kind is formed by a combination of the second and third kernels. Lastly, the fourth kernel has a simple pole at $x=c$, so it is also a differential of the third kind, and its residue can be normalized to 1 introducing the numerator $y_c \equiv y\big|_{x=c}$. For a cubic polynomial $y^2=P_3(x)$, the second kernel acts as the differential of the second kind, and the third kernel is reducible to the others via integration-by-parts~\cite{Broedel:2017kkb}. 

In general, integrals over these kernels give elliptic integrals, and are analogous to the logarithms we encountered in sec.~\ref{sec:ch1_MPLs} when calculating iterated integrals over rational functions on the Riemann sphere. Therefore, as expected, elliptic multiple polylogarithms (eMPLs) must contain elliptic integrals. Even though we will work with the eMPL construction of ref.~\cite{Broedel:2017kkb}, we will adopt the slightly different conventions of refs.~\cite{Kristensson:2021ani,Wilhelm:2022wow}. Moreover, in this thesis we will only need eMPLs for quartic polynomials, as the elliptic Feynman integrals that we will evaluate naturally depend on quartic polynomials; see, however, ref.~\cite{Broedel:2017kkb} for eMPLs in the cubic case.

Concretely, eMPLs are recursively defined via~\cite{Broedel:2017kkb}
\begin{equation}
\label{eq: definition_eMPLs_on_ell_curve}
    \Ef{n_1 & \ldots & n_k}{c_1 & \ldots& c_k}{x}=
    \int_{0}^{x}d t\,\psi_{n_{1}}(c_{1},t) \, \Ef{n_{2} & \ldots & n_k}{c_{2} & \ldots& c_k}{t} \, , \qquad \text{with} \qquad \mathrm{E}_{4}(;x)=1,
\end{equation}
where the $n_i$ are integer numbers, the $c_i$ are algebraic, and $k$ is the length. While for MPLs the length and the weight are equivalent, it is no longer the case for eMPLs, and the weight is given by $|n_1| + \dots + |n_k|$. As an example, some of the kernels correspond to
\begin{equation}
\label{eq: kernels_psi_eMPLs}
\psi_{1}(c,x) \equiv \frac{1}{x-c}\,, \quad \psi_{0}(0,x) \equiv \frac{1}{y} \,, \quad \psi_{-1}(\infty,x) \equiv \frac{x}{y}\,, \quad \psi_{-1}(c,x) \equiv \frac{y_{c}}{y(x-c)}\,.
\end{equation}
Notice that MPLs are contained in the first kernel. While there is also an infinite tower of further integration kernels $\psi_n$ which deal with $x^2/y$ and its double pole at infinity, the elliptic ladder families considered in secs.~\ref{sec:ch2_elliptic_ladders} and~\ref{sec:ch2_staggered_elliptic_ladders} can be integrated without them. Therefore, we will neglect these kernels in the following; see, however, ref.~\cite{Broedel:2017kkb} for further details.

Now, in order to express any one-fold integral result 
\begin{equation}
\label{eq: example_one-fold_eMPLs}
    \mathcal{I} = \int_0^\infty \frac{dx}{y} \, \mathcal{G}^{(n)}(x,y) \,,
\end{equation}
such as the 10-pt double box of eq.~\eqref{eq: 10-pt_double-box_result}, in terms of eMPLs, we need to rewrite each MPL appearing in the weight-$n$ linear combination $\mathcal{G}^{(n)}$ as an $\mathrm{E}_4$ function. For that, we can first take the differential
\begin{equation}
    \frac{d}{dx} \, \mathcal{G}^{(n)}(x,y)=\sum_{i} \mathcal{G}_i^{(n-1)}(x,y) \, \frac{d}{dx} \, \log R_i(x,y)\,, 
\end{equation}
which is of the form of eq.~\eqref{eq: differential_pre_symbol}, for some lower-weight MPLs $\mathcal{G}_i^{(n-1)}$ and rational functions $R_i$. Integrating back this expression, we can rewrite the MPLs as
\begin{equation}
    \mathcal{G}^{(n)}(x,y)= \mathcal{G}^{(n)}(0,y_{0})+\sum_{i} \int_{0}^{x} dt \, \mathcal{G}^{(n-1)}_i(t,y_{t}) \, \partial_{t} \log R_{i}(t,y_{t}) \, .
\end{equation}
Importantly, since the $\partial_{t} \log R_{i}(t,y_{t})$ only have simple poles and are rational in $t$ and $y_t$, they can be written in terms of the kernels in eq.~\eqref{eq: kernels_psi_eMPLs}. Proceeding recursively for the lower-weight MPLs in $\mathcal{G}_i^{(n-1)}$, we can finally integrate everything back in terms of $\mathrm{E}_4$'s. Lastly, the remaining integral in eq.~\eqref{eq: example_one-fold_eMPLs} is straightforward, since $\int dx/y=\int dx \, \psi_{0}(0,x)$, thus we can also express the final result in terms of $\mathrm{E}_4$ functions. While we will implicitly use the eMPL result for the 10-pt double box in chapter~\ref{ch:chapter3}, we abstain from including the explicit details here as they will not be so relevant for our analysis. See, however, ref.~\cite{Kristensson:2021ani} for details, as well as ref.~\cite{Broedel:2017siw} for an application to the 2-loop sunrise integral.

\subsection{Elliptic multiple polylogarithms on the torus}
\label{sec:ch2_eMPLs_torus}

Although the previous construction of eMPLs is already useful and valid, it hides the fact that the integral can be written as a pure function, that is, as a function where all kinematic dependence appears through the arguments of the eMPLs and not in the prefactor. To achieve this representation, we can formulate eMPLs not as iterated integrals on the elliptic curve, but on the torus~\cite{Broedel:2017kkb,Broedel:2018iwv,Broedel:2018qkq}.

\begin{figure}[t]
\begin{center}
\subfloat[]{
\includegraphics[height=5.5cm]{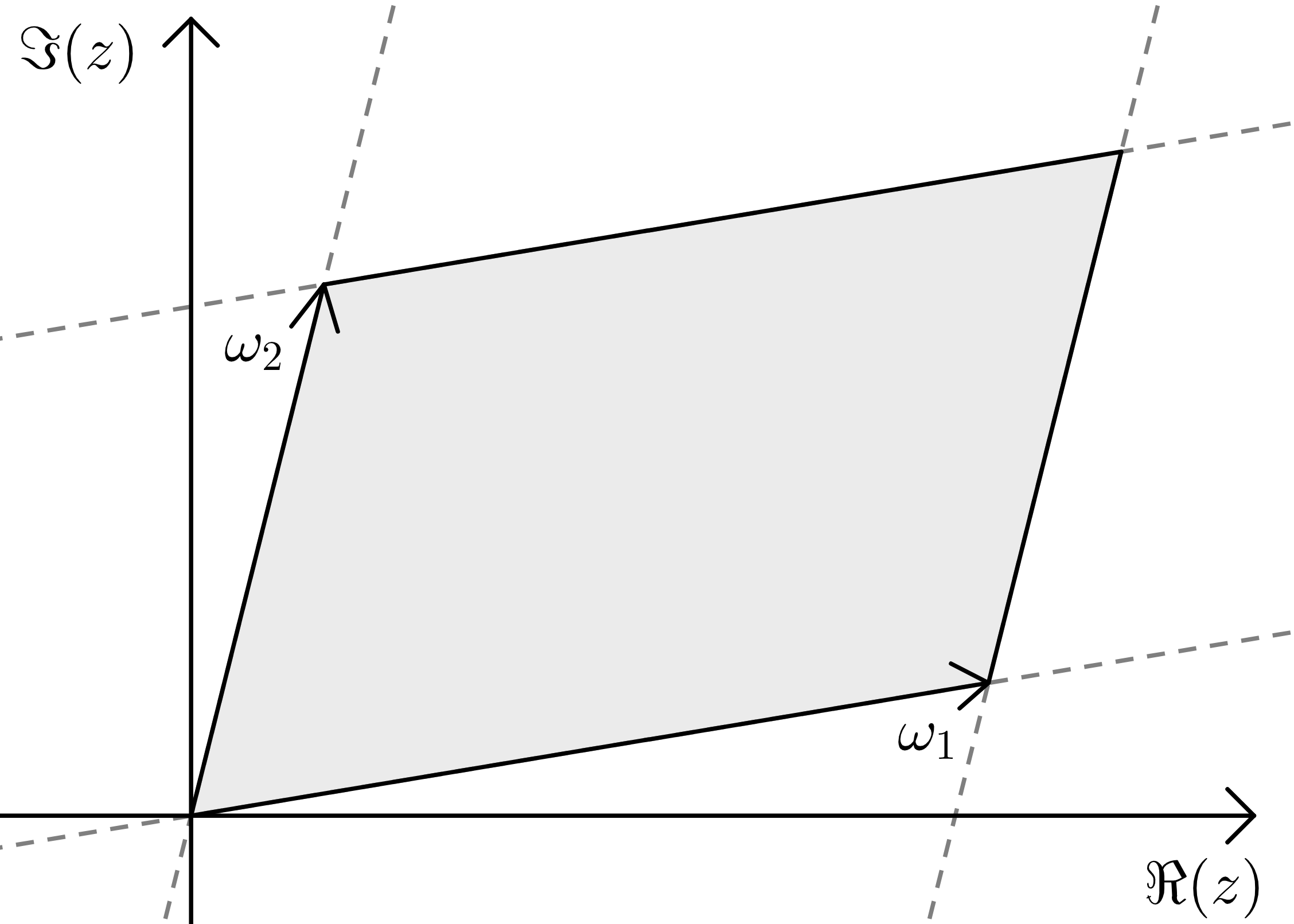}} \qquad \subfloat[]{\includegraphics[height=5.5cm]{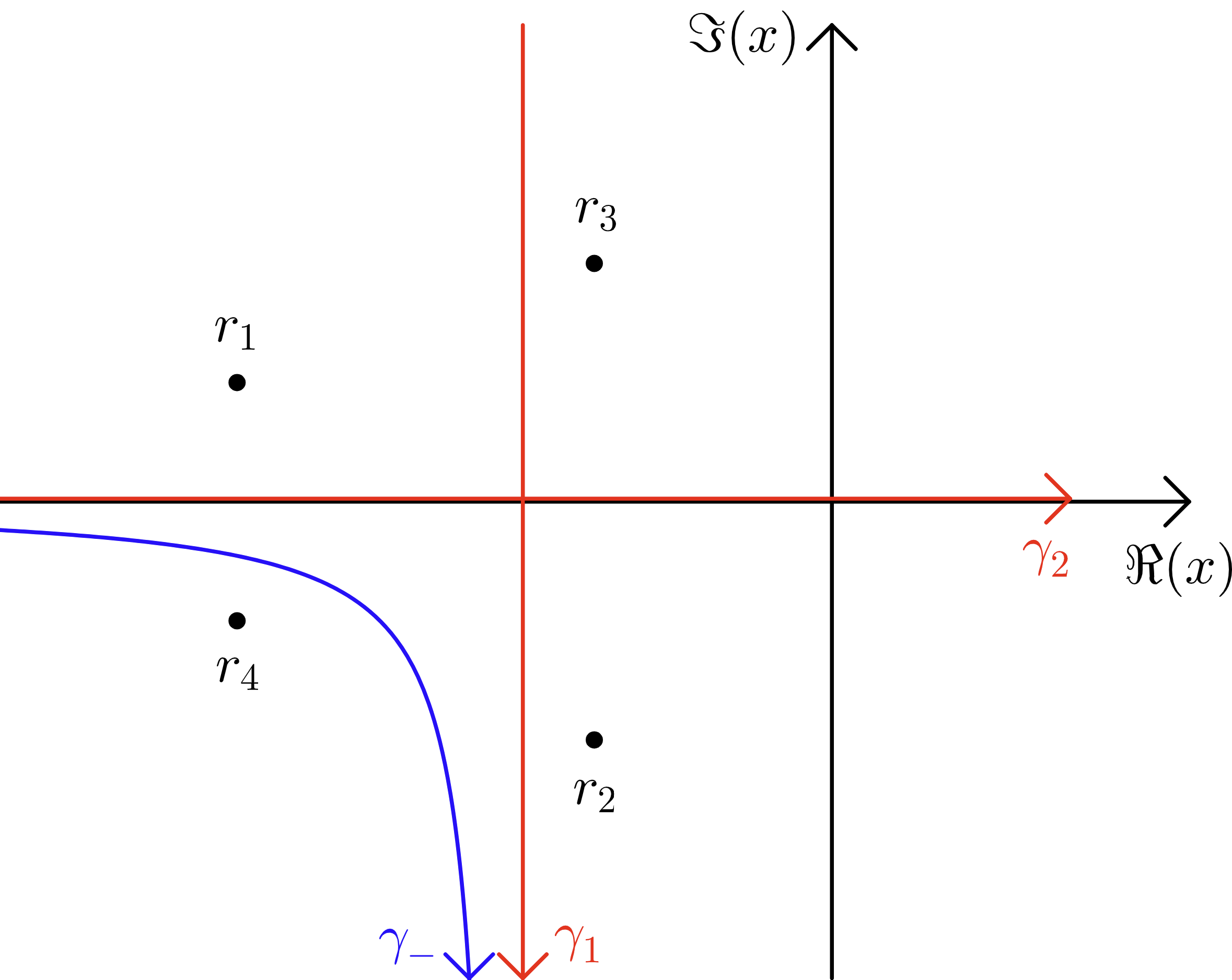}}
\caption{(a) Lattice $\Lambda= \mathbb{Z} \, \omega_1 + \mathbb{Z} \, \omega_2$, spanned by the periods $\omega_1$ and $\omega_2$. The area in gray defines the fundamental parallelogram. (b) Location of the roots $r_i$ of the elliptic curve in kinematic space, and integration contours $\gamma_j$ used in the main text to define Abel's map and the periods.}
\label{fig: lattice_torus}
\end{center}
\end{figure}

The connection between an elliptic curve and the torus can be made explicit if the elliptic curve is first expressed in the so-called Weierstrass normal form
\begin{equation}
\label{eq: Weierstrass_normal_form}
    Y^2 = 4 X^3 - g_2 X - g_3,
\end{equation}
which can always be achieved via appropriate changes of variables; see e.g. chapter 3 of ref.~\cite{Silverman2009}. Now, let us compare the previous expression with the non-linear differential equation 
\begin{equation}
    \wp'(z)^2 = 4 \wp(z)^3 - g_2 \, \wp(z) - g_3.
\end{equation}
This differential equation is satisfied by the so-called Weierstrass $\wp$ function, which is defined as
\begin{equation}
    \wp(z) = \frac{1}{z^2} + \mathlarger{\mathlarger{\sum}}_{(m,n) \neq (0,0)} \left( \frac{1}{(z+m \omega_1 + n \omega_2)^2} - \frac{1}{(m \omega_1 + n \omega_2)^2} \right) \,,
\end{equation}
with $n$ and $m$ being integers. By construction, the Weierstrass $\wp$ function is doubly periodic over $\omega_1$ and $\omega_2$, which are two linearly independent complex numbers called the periods. Thus, they span a lattice $\Lambda = \mathbb{Z} \, \omega_1 + \mathbb{Z} \, \omega_2$, see fig.~\ref{fig: lattice_torus}(a). By gluing together the periodic sides of the fundamental parallelogram, we obtain a torus $\mathbb{C}/\Lambda$. Thus, the Weierstrass $\wp$ function is defined on the torus, as indicated by the variable $z$. Consequently, we see that we can relate an elliptic curve to the torus via $(X,Y) \to (\wp(z),\wp'(z))$. In particular, the explicit connection is given by Abel's map
\begin{equation}
\label{eq: Abel_map}
    z_c^{+} = \int_{-\infty}^c \frac{dx}{y}\,,
\end{equation}
such that we can associate two images $z_c^\pm$ on the torus to each point $x=c$ on the curve (including $c=\infty$), in complete analogy with the two images $\pm y_c$ on the original curve. The second image $z_c^-$ on the torus is obtained via
\begin{equation}
    z_c^+ + z_c^- = z_\infty^- + z_\infty^+ \equiv  z_\infty^- \ \text{mod} \ \Lambda,
\end{equation}
where
\begin{equation}
    z_\infty^- = \int_{\gamma_-} \frac{dx}{y}
\end{equation}
is defined under the conventions and the path $\gamma_-$ specified in fig.~\ref{fig: lattice_torus}(b). Similarly, the periods are given by
\begin{equation}
    \omega_1 = \int_{\gamma_1} \frac{dx}{y}\,, \qquad \text{and} \qquad \omega_2 = \int_{\gamma_2} \frac{dx}{y} = \int_{-\infty}^\infty \frac{dx}{y}\,.
\end{equation}

The inverse operation, which maps the torus to the elliptic curve, is the so-called $\kappa$-map, which can be found e.g.~in refs.~\cite{Broedel:2017kkb,Kristensson:2021ani,Wilhelm:2022wow}. Lastly, it is also customary to normalize the torus such that the periods become $(\omega_1,\omega_2) \to (1,\tau)$, where $\tau \equiv \omega_2/\omega_1$ is the modular parameter. We define the corresponding normalized torus images as 
\begin{equation}
\label{eq: Abel_map_w}
    w_c^{+} \equiv \frac{1}{\omega_1}\, z_c^+ = \frac{1}{\omega_1} \int_{-\infty}^c \frac{dx}{y}\,.
\end{equation}

With these definitions, we can introduce the $\widetilde{\Gamma}$ functions~\cite{Broedel:2017kkb,Broedel:2018iwv}, which are eMPLs defined as iterated integrals on the normalized torus,
\begin{equation}
\label{eq: def_Gamma_tilde}
    \gamt{n_1 & \ldots & n_k}{w_1 & \ldots& w_k}{w}=
    \int_{0}^{w} d w'\,g^{(n_{1})}(w'{-}w_{1}) \, \gamt{n_{2} & \ldots & n_k}{w_{2} & \ldots& w_k}{w'} \, , \qquad \text{with} \qquad \widetilde{\Gamma}(;w)=1.
\end{equation}
The kernels $g^{(n)}(w)$, which have at most simple poles, are generated by the Eisenstein-Kronecker series
\begin{equation}
    \frac{\partial_{w}\theta_{1}(0) \, \theta_{1}(w+\alpha)}{\theta_{1}(w) \, \theta_{1}(\alpha)} = \sum_{n\geq 0}\alpha^{n-1}g^{(n)}(w)\, ,
\end{equation}
where $\theta_{1}(w)$ denotes the odd Jacobi theta function. For instance, $g^{(0)}(w)=1$ and $g^{(1)}(w)=\partial_w \log \theta_1(w)$.

To translate from the $\mathrm{E}_4$ to the $\widetilde{\Gamma}$ representation, we will follow again the conventions in refs.~\cite{Kristensson:2021ani,Wilhelm:2022wow}. In particular, we can rewrite the $\psi_n$ kernels from eq.~\eqref{eq: definition_eMPLs_on_ell_curve} in terms of the $g^{(m)}$ kernels via
\begin{subequations}
\begin{align}
\psi_{1}(c,x) \, d x &=\Bigl(g^{(1)}(w-w_{c}^{+})+g^{(1)}(w-w_{c}^{-}) -g^{(1)}(w-w_{\infty}^{+}) -g^{(1)}(w-w_{\infty}^{-})\Bigr)  \, d w, \label{eq: psi_1_kernel_eMPLs} \\
\psi_{0}(0,x) \, d x&= \omega_{1} \, d w, \\
\psi_{-1}(\infty,x) \, d x &=   \Bigl(g^{(1)}(w-w_{\infty}^{-})-g^{(1)}(w) + g^{(1)}(w_{\infty}^{-})-\omega_{1}a_{3}/4 \Bigr) \, d w, \\
\psi_{-1}(c,x) \, d x &= \Bigl( g^{(1)}(w-w_{c}^{+})-g^{(1)}(w-w_{c}^{-}) + g^{(1)}(w_{c}^{+}) -g^{(1)}(w_{c}^{-}) \Bigr) \, d w,
\end{align}    
\end{subequations}
where $a_3$ is the coefficient of the cubic term in $y^2=P_4(x)$. Hence, with these relations we can also express the result of a one-fold elliptic integral in terms of $\widetilde{\Gamma}$ functions. We will revisit the $\widetilde{\Gamma}$ functions in the next chapter, see sec.~\ref{sec:ch3_elliptic_symbol}, where we will introduce the elliptic symbol.

Lastly, there is a different formulation of eMPLs as iterated integrals on the torus, which allows us to represent the same result much more compactly. These are the $\mathcal{E}_{4}$ functions \cite{Broedel:2018qkq}, defined via the recursion
\begin{equation}
    \cEf{n_1 & \ldots & n_k}{c_1 & \ldots& c_k}{x}=
    \int_{0}^{x}d  t\,\Psi_{n_{1}}(c_{1},t) \, \cEf{n_{2} & \ldots & n_k}{c_{2} & \ldots& c_k}{t} \, , \qquad \text{with} \qquad \mathcal{E}_{4}(;x)=1.
\end{equation}
In particular, we can express a result from $\widetilde{\Gamma}$ functions in terms of $\mathcal{E}_{4}$'s via~\cite{Kristensson:2021ani,Wilhelm:2022wow}
\begin{subequations}
\begin{align}
 \Psi_{\pm (n>0)}(c,x) \, d  x & =\Bigl(g^{(n)}(w-w_{c}^{+})\pm g^{(n)}(w-w_{c}^{-}) - \delta_{\pm n,1}\bigl( g^{(1)}(w-w_{\infty}^{+})+g^{(1)}(w-w_{\infty}^{-} )\bigr) \Bigr) \, d w,\\
 \Psi_0(0,x) \, d x& =d w.
\end{align}
\end{subequations}
As examples, the $\mathcal{E}_{4}$ results for the 2-loop sunrise and 2-loop 10-pt double box can be found in refs.~\cite{Wilhelm:2022wow} and~\cite{Kristensson:2021ani}, respectively.

In summary, in this section we have seen how linear reducibility can be used as a criterion to determine whether or not there exist geometric obstructions to direct integration. The simplest cases that do not satisfy it are the so-called linearly reducible elliptic Feynman integrals, for which a single square root -- the elliptic curve -- appears in the last step in the polynomial reduction. Such cases can be evaluated in terms of eMPLs, which are the class of functions describing iterated integrals on the torus. In the following sections, we will identify two different ladder integral families which can be expressed in terms of eMPLs to all loop orders.

\section{Ten-point elliptic ladder family}
\label{sec:ch2_elliptic_ladders}

Having introduced all the ingredients necessary to evaluate any linearly reducible elliptic Feynman integral, we are finally in a position to investigate generalizations of the 2-loop 10-pt elliptic double box presented in sec.~\ref{sec:ch2_10pt_double_box}. One possibility is adding more external legs in the middle rung of the double box. However, as we will explore in chapter~\ref{ch:chapter3}, we were not able to find a linearly reducible elliptic representation for neither the 11- nor the 12-pt double-box integrals.

Instead, another possible generalization is to introduce more loops. As briefly discussed in sec.~\ref{sec:ch2_traintracks}, if all intermediate rungs of the traintracks have a pair of external legs attached to them, then the resulting geometry is a CY $(L-1)$-fold at $L$-loops. Thus, to stay in the realm of elliptic Feynman integrals, the most we can do is to increase the loop order by adding rungs without further external legs attached. This forms the 10-pt elliptic ladder family, which is characterized by the number $\mathcal{R}$ ($\mathcal{L}$) of loops to the right (left) of the rung with the massless external legs,
\begin{equation}
\label{eq: 10pt_elliptic_ladder_mini_drawing}
    \mathcal{I}_{\mathcal{L},\mathcal{R}} = \, .
\end{equation}

The first question that arises is whether we lose the ellipticity by adding more rungs. However, as we proved in appendix A of ref.~\cite{McLeod:2023qdf}, the leading singularity of the 10-pt elliptic ladder family remains the same as the 2-loop case (recall eq.~\eqref{eq: LS_10pt_double-box_pre_change_vars}) under the addition of rungs, which shows that the same elliptic curve will arise to all loop orders. See also ref.~\cite{Cao:2023tpx} for a similar analysis. The second question, therefore, is whether a linearly reducible elliptic representation can also be achieved for high-loop integrals in this family. To address this question, let us study the $(\mathcal{R}+\mathcal{L})$-loop diagram depicted in fig.~\ref{fig: diagram_elliptic_ladder_family}. Notice that since the diagram only has 10 external legs, the Feynman integral depends on the same external dual coordinates as the 2-loop 10-pt double box, see sec.~\ref{sec:ch2_10pt_double_box}, as well as on the same 7 DCI cross-ratios from eq.~\eqref{eq: DCI_cross-ratios_10pt_double-box}. 

\begin{figure}
\centering
 
\caption{Momentum-space diagram (in black) and dual graph (in blue) for an $(\mathcal{R}+\mathcal{L})$-loop diagram in the 10-pt elliptic ladder family, which has $\mathcal{R}$ ($\mathcal{L}$) loops to the right (left) of the rung with external legs $p_5$ and $p_{10}$. The dashed lines indicate an arbitrary number of loops.}
\label{fig: diagram_elliptic_ladder_family}
\end{figure}
Moreover, there are two reflection symmetries for these diagrams: a reflection $\mathcal{V}$ across the vertical axis, which acts as $\mathcal{V}(\mathcal{I}_{\mathcal{L},\mathcal{R}})=\mathcal{I}_{\mathcal{R},\mathcal{L}}$ on the integrals, and a reflection $\mathcal{H}$ across the horizontal axis, which acts as $\mathcal{H}(\mathcal{I}_{\mathcal{L},\mathcal{R}})=\mathcal{I}_{\mathcal{L},\mathcal{R}}$. On the cross-ratios, these symmetries act as
\begin{subequations}
\label{eq: 10pt_elliptic_ladders_mirror_sym}
\begin{align}
 \mathcal{V}&:\quad u_1 \leftrightarrow u_2,\,v_1 \leftrightarrow v_2,\, u_3 \rightarrow u_2 u_4/v_1,\, u_4 \rightarrow u_3 v_2 / u_1\,,\\[-0.1cm]
 \mathcal{H}&:\quad u_1 \leftrightarrow v_1,\, u_2 \leftrightarrow v_2, \,u_3 \leftrightarrow u_4\, ,
\end{align}
\end{subequations}
while $u_5$ is invariant under both reflections.

Due to the $\mathcal{V}$ symmetry, we can restrict our analysis to $\mathcal{R} \leq \mathcal{L}$, which leads to $\lfloor (\mathcal{R}+\mathcal{L})/2 \rfloor$ distinct diagrams in the family at each loop order. All in all, taking the multi-soft limit of eq.~\eqref{eq: traintrack_Feyn_param}, we obtain the Feynman parametrization
\begin{equation}
\label{eq: 10pt_elliptic_ladder_integrals_Feynman_param}
\mathcal{I}_{\mathcal{L},\mathcal{R}} = \int_0^\infty \frac{d \alpha_1 \, \cdots \, d \alpha_{\mathcal{R}+\mathcal{L}} \, d \beta_1 \, \cdots \, d \beta_{\mathcal{R}+\mathcal{L}}}{f_1 \cdots f_{\mathcal{R}+\mathcal{L}} \, g_{\mathcal{R}+\mathcal{L}}} \, ,
\end{equation}
where
\begin{subequations}
\label{eq: 10pt_elliptic_ladder_integrals_Feynman_param_factors}
\begin{align}
f_{k \leq \mathcal{R}} &= f_{k-1} + \alpha_k + \beta_k + \alpha_k \beta_k + \sum_{j=1}^{k-1} \Big( \alpha_j \beta_k + \alpha_k \beta_j \Big)\,, \\
f_{\mathcal{R}+1} &= u_3 u_4 u_5 f_{\mathcal{R}} + \alpha_{\mathcal{R}+1} + \beta_{\mathcal{R}+1} + \alpha_{\mathcal{R}+1} \beta_{\mathcal{R}+1} + \sum_{j=1}^{\mathcal{R}} \Big( u_4 \alpha_j \beta_{\mathcal{R}+1} + u_3 \alpha_{\mathcal{R}+1} \beta_j \Big)\,, \\
f_{k > \mathcal{R}+1} &= f_{k-1} + \alpha_k + \beta_k + \alpha_k \beta_k + \sum_{j=1}^{\mathcal{R}} \Big( u_4 \alpha_j \beta_k + u_3 \alpha_k \beta_j \Big) + \sum_{j=\mathcal{R}+1}^{k-1} \Big( \alpha_j \beta_k + \alpha_k \beta_j \Big)\,, \\
g_{\mathcal{R}+\mathcal{L}} &= 1 + \sum_{j=1}^{\mathcal{R}} \Big( v_1 \alpha_j + u_1 \beta_j \Big) + \sum_{j=\mathcal{R}+1}^{\mathcal{R}+\mathcal{L}} \Big( u_2 \alpha_j + v_2 \beta_j \Big)\,,
\end{align}
\end{subequations}
with $f_0=0$. For $\mathcal{L}=\mathcal{R}=1$, it results in the 2-loop 10-pt double-box Feynman parametrization of eq.~\eqref{eq: 10pt_double-box_Feynman_param}.

As it is given, this Feynman parametrization is not directly linearly reducible to a one-fold elliptic integral. However, the simplicity of this integral family resides in the fact that none of the terms in eq.~\eqref{eq: 10pt_elliptic_ladder_integrals_Feynman_param_factors} are quadratic in $\beta_i$, and there are no bilinear terms of the type $\beta_i \beta_j$. Therefore, we can perform any $\mathrm{GL}(\mathcal{R}+\mathcal{L},{\mathbb C})$ rotation in the space of $\beta'$s, and it will not generate quadratic or higher powers in $\beta_i$. In particular, we find that the transformation
\begin{subequations}
\label{eq: change_vars_10pt_elliptic_ladders}
\begin{align}
\beta_{1< k \leq \mathcal{R}} &\to \beta_k - \beta_{k-1}\, ,\\
\beta_{\mathcal{R}+1} &\to \beta_{\mathcal{R}+1} - \frac{u_1}{v_2} \beta_{\mathcal{R}} - \beta_{\mathcal{R}+2}\, , \\
\beta_{\mathcal{R}+1< k < \mathcal{R}+\mathcal{L}} &\to \beta_k - \beta_{k+1}\,,
\end{align}
\end{subequations}
grants linear reducibility up to the last step,\footnote{We have verified linear reducibility up to six loops with the {\Maple} package {\HyperInt}~\cite{Panzer:2014caa}, but due to the repetitive Feynman parametrization, we conjecture that the change of variables holds to all loop orders.} where there appears a single square root of a quartic polynomial -- the elliptic curve -- in the last integration variable $x \equiv \beta_{\mathcal{R}+1}$. In addition, we verified that the elliptic curve appears as the leading singularity by computing the maximal co-dimension residue in all diagrams up to 6 loops. Therefore, under the integration sequence $\{ \alpha_1,\dots,\alpha_{\mathcal{R}+\mathcal{L}},\beta_{\mathcal{R}+\mathcal{L}},\dots,\beta_{\mathcal{R}+2},\beta_1,\dots,\beta_{\mathcal{R}+1} \}$ we find a one-fold integral representation
\begin{equation} 
\label{eq: elliptic_ladder_one_fold}
\mathcal{I}_{\mathcal{L},\mathcal{R}} = \int_0^\infty \frac{dx}{y} \, \mathcal{G}_{\mathcal{L},\mathcal{R}}^{(2\mathcal{R}+2\mathcal{L}-1)}(x,y)\, ,
\end{equation}
where $\mathcal{G}_{\mathcal{L},\mathcal{R}}^{(2\mathcal{R}+2\mathcal{L}-1)}(x,y)$ is a linear combination of MPLs of weight $2\mathcal{R}+2\mathcal{L}-1$, with a precise form that depends on the number of loops $\mathcal{L}$ and $\mathcal{R}$ in the diagram. Here, the elliptic curve is given by
\begin{align}
\label{eq: elliptic_curve_10pt_elliptic_ladders}
y^2 &= \bigg(\frac{1}{v_2^2} \Big( (1-v_1+v_2 x) (v_2 u_3 - u_1)+u_1 u_2 (1-u_4) \Big) +h_1+h_2 \bigg)^2 - 4 h_1 h_2\, ,
\end{align}
where
\begin{align}
h_1 & \equiv \frac{1}{v_2^2} \Big( (v_2 x+ u_1) (u_2 u_4 - v_1)+v_1 v_2 (u_3 - 1) \Big)\, ,\\
h_2 & \equiv \frac{-1}{v_2} \Big( v_2 x^2 + (1-u_2+v_2) x + 1 \Big) \,.
\end{align}
Note that, since the change of variables used for the 10-pt elliptic ladder family differs from the transformation used for the 2-loop 10-pt double box, see eq.~\eqref{eq: 10-pt_double-box_change_vars}, the resulting elliptic curve is different from eq.~\eqref{eq: elliptic_curve_10pt_double-box}. However, one can verify that the curves are isomorphic by computing the so-called $j$-invariant, which is given by~\cite{Broedel:2017kkb}
\begin{equation}
j = 256 \,\frac{(1-\lambda (1- \lambda) )^3}{\lambda^2 (1-\lambda)^2} \, ,
\end{equation}
where
\begin{equation}
\lambda \equiv \frac{r_{1,4} \, r_{2,3}}{r_{1,3} \, r_{2,4}} \, , \qquad r_{i,j} \equiv r_i - r_j\,,
\end{equation}
for $r_i$ being the roots of $y^2$.

As a proof of principle for the validity of our change of variables, we explicitly computed the result for the 3-loop integral $\mathcal{I}_{2,1}$. The result can be accessed at ref.~\cite{ancillary_file_elliptic_ladders}, and is given both as a one-fold elliptic integral over MPLs and in terms of $\mathcal{E}_4$ eMPLs. As a cross-check, we found that, while there also appear new arguments at 3 loops, all arguments of the MPL and eMPL functions already present at 2 loops also appear here. This is important, due to the vast swelling of intermediate expressions, especially in terms of $\widetilde{\Gamma}$ functions, which occupied up to 100GB. By contrast, the final expression in terms of $\mathcal{E}_4$ drastically simplified, occupying only a few MB. By the same principle, since we expect the calculation of the corresponding 3-loop elliptic symbol (see sec.~\ref{sec:ch3_elliptic_symbol}) to grow tremendously in intermediate steps, we leave its calculation for future work.

\section{Ten-point staggered elliptic ladder family}
\label{sec:ch2_staggered_elliptic_ladders}

Having identified a first elliptic integral family, a natural question that arises is whether there exist other similar variations. Specifically, whether the relative placement of the external legs in the middle rungs plays any role in the corresponding Feynman integral geometry. This variation leads to the so-called 10-pt staggered elliptic ladder integral family,
\begin{equation}
    \mathcal{I}_{\mathcal{L},\mathcal{M},\mathcal{R}}= \,,
\end{equation}
which is characterized by the number $\mathcal{R}$ ($\mathcal{L}$) of loops to the right (left) of the rungs with the massless external legs, as well as by the number $\mathcal{M}$ of loops between them.

As shown in appendix A of ref.~\cite{McLeod:2023qdf}, the leading singularity for this integral family is again invariant under the addition of rungs in the ladder. Therefore, the leading singularity of any diagram can be related to that of $\mathcal{I}_{1,1,1}$. In this case, surprisingly, the leading singularity is algebraic, meaning that a one-fold representation with $d \log$ kernels is possible. The subsequent question, therefore, is whether this integral family is totally linearly reducible, implying that it can be written solely in terms of MPLs and ceases to be elliptic.

\begin{figure}
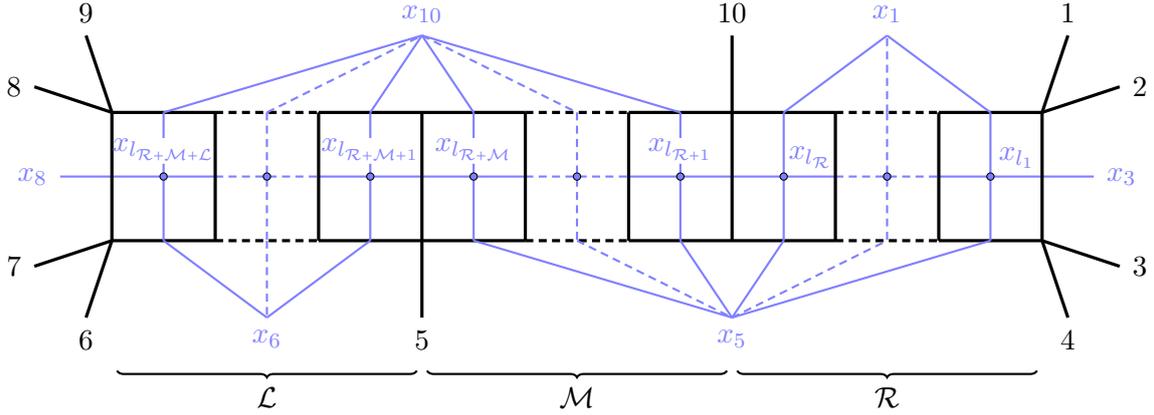

\centering
  
\caption{Momentum-space diagram (in black) and dual graph (in blue) for an $(\mathcal{R}+\mathcal{M}+\mathcal{L})$-loop diagram in the 10-pt staggered elliptic ladder family, which has $\mathcal{R}$ ($\mathcal{L}$) loops to the right (left) of the rung with external legs $p_5$ and $p_{10}$, and $\mathcal{M}$ loops in between. The dashed lines indicate an arbitrary number of loops.}
\label{fig: diagram_staggered_elliptic_ladder_family}
\end{figure}
To address this question, let us study the $(\mathcal{R}+\mathcal{M}+\mathcal{L})$-loop diagram shown in fig.~\ref{fig: diagram_staggered_elliptic_ladder_family}. Once again, the diagram depends on the same 10 external dual coordinates as the 2-loop 10-pt double box, thus it depends on the same 7 DCI cross-ratios given in eq.~\eqref{eq: DCI_cross-ratios_10pt_double-box}. Since $\mathcal{I}_{\mathcal{L},0,\mathcal{R}}=\mathcal{I}_{\mathcal{L},\mathcal{R}}$ is related to the previous family, we can assume $\mathcal{M}>0$ without loss of generality. Moreover, we can restrict our analysis to $\mathcal{R} \leq \mathcal{L}$, since otherwise we can relate it via the mirror symmetries of eq.~\eqref{eq: 10pt_elliptic_ladders_mirror_sym}. Overall, this leaves $\lfloor (\mathcal{R}+\mathcal{M}+\mathcal{L}-1)^2/4 \rfloor$ distinct diagrams in this family at each loop order. Taking the multi-soft limit of eq.~\eqref{eq: traintrack_Feyn_param}, the Feynman parametrization for the 10-pt staggered elliptic ladders is
\begin{equation}
\mathcal{I}_{\mathcal{L},\mathcal{M},\mathcal{R}} = \int_0^\infty \frac{d \alpha_1 \, \cdots \, d \alpha_{\mathcal{R}+\mathcal{M}+\mathcal{L}} \, d \beta_1 \, \cdots \, d \beta_{\mathcal{R}+\mathcal{M}+\mathcal{L}}}{f_1 \cdots f_{\mathcal{R}+\mathcal{M}+\mathcal{L}} \, g_{\mathcal{R}+\mathcal{M}+\mathcal{L}}} \, ,
\end{equation}
where
{\allowdisplaybreaks
\begin{subequations}
\begin{align}
\hspace{-0.2cm} f_{k \leq \mathcal{R}} & = f_{k-1} + \alpha_k + \beta_k + \alpha_k \beta_k + \sum_{j=1}^{k-1} \Big( \alpha_j \beta_k + \alpha_k \beta_j \Big)\,, \\
\hspace{-0.2cm} f_{\mathcal{R}+1} &= u_3 f_\mathcal{R} + \alpha_{\mathcal{R}+1} + \beta_{\mathcal{R}+1} + \alpha_{\mathcal{R}+1} \beta_{\mathcal{R}+1} + \sum_{j=1}^{\mathcal{R}} \Big( \alpha_j \beta_{\mathcal{R}+1} + u_3 \alpha_{\mathcal{R}+1} \beta_j \Big)\,, \\
\hspace{-0.2cm} f_{\mathcal{R}+1<k\leq \mathcal{R}+\mathcal{M}} &= f_{k-1} + \alpha_{k} + \beta_{k} + \alpha_{k} \beta_{k} + \sum_{j=1}^{\mathcal{R}} \Big( \alpha_j \beta_{k} + u_3 \alpha_{k} \beta_j \Big) + \sum_{j=\mathcal{R}+1}^{k-1} \Big( \alpha_j \beta_{k} + \alpha_{k} \beta_j \Big)\,, \\
\hspace{-0.2cm} f_{\mathcal{R}+\mathcal{M}+1} & = u_4 u_5 f_{\mathcal{R}+\mathcal{M}} + \alpha_{\mathcal{R}+\mathcal{M}+1} + \beta_{\mathcal{R}+\mathcal{M}+1} + \alpha_{\mathcal{R}+\mathcal{M}+1} \beta_{\mathcal{R}+\mathcal{M}+1} \nonumber \\
\hspace{-0.2cm} & \ + \sum_{j=1}^{\mathcal{R}} \Big( u_4 \alpha_j \beta_{\mathcal{R}+\mathcal{M}+1} + u_3 \alpha_{\mathcal{R}+\mathcal{M}+1} \beta_j \Big) + \sum_{j=\mathcal{R}+1}^{\mathcal{R}+\mathcal{M}} \Big( u_4 u_5 \alpha_j \beta_{\mathcal{R}+\mathcal{M}+1} + \alpha_{\mathcal{R}+\mathcal{M}+1} \beta_j \Big)\,, \\
\hspace{-0.2cm} f_{k > \mathcal{R}+\mathcal{M}+1} & = f_{k-1} + \alpha_k + \beta_k + \alpha_k \beta_k + \sum_{j=1}^{\mathcal{R}} \Big( u_4 \alpha_j \beta_k + u_3 \alpha_k \beta_j \Big) \nonumber \\
\hspace{-0.2cm} & \ + \sum_{j=\mathcal{R}+1}^{\mathcal{R}+\mathcal{M}} \Big(u_4 u_5 \alpha_j \beta_k + \alpha_k \beta_j \Big) + \sum_{j=\mathcal{R}+\mathcal{M}+1}^{k-1} \Big(\alpha_j \beta_k + \alpha_k \beta_j \Big)\,, \\
\hspace{-0.2cm} g_{\mathcal{R}+\mathcal{M}+\mathcal{L}} &= 1 + \sum_{j=1}^{\mathcal{R}} \Big( v_1 \alpha_j + u_1 \beta_j \Big) + \sum_{j=\mathcal{R}+1}^{\mathcal{R}+\mathcal{M}} \Big( u_2 u_4 u_5 \alpha_j + \frac{u_1}{u_3} \beta_j \Big) + \sum_{j=\mathcal{R}+\mathcal{M}+1}^{\mathcal{R}+\mathcal{M}+\mathcal{L}} \Big( u_2 \alpha_j + v_2 \beta_j \Big)\,.
\end{align}
\end{subequations}
}

Just like for the 10-pt elliptic ladders, since there are no quadratic or bilinear terms in the $\beta'$s, we can perform any $\text{GL}(\mathcal{R}+\mathcal{M}+\mathcal{L},\mathbb{C})$ rotation in the space of $\beta'$s without generating higher-order terms. However, even though the leading singularity is in this case algebraic, we did not find a transformation granting linear reducibility all the way to the end. Instead, with the change of variables
\begin{subequations}
\begin{align}
\beta_{1< k \leq \mathcal{R}} &\to \beta_k - \beta_{k-1}\,, \\
\beta_{\mathcal{R}+1} &\to \beta_{\mathcal{R}+1} - u_3 \beta_{\mathcal{R}}\,,\\
\beta_{\mathcal{R}+1< k \leq \mathcal{R}+\mathcal{M}} &\to \beta_k - \beta_{k-1}\,, \\
\beta_{\mathcal{R}+\mathcal{M}+1} &\to \beta_{\mathcal{R}+\mathcal{M}+1} - \frac{u_1}{u_3 v_2} \beta_{\mathcal{R}+\mathcal{M}} - \beta_{\mathcal{R}+\mathcal{M}+2}\,,\\
\beta_{\mathcal{R}+\mathcal{M}+1< k < \mathcal{R}+\mathcal{M}+\mathcal{L}} &\to \beta_k - \beta_{k+1}\,,
\end{align}
\end{subequations}
we find an elliptic obstruction in the last step, given by a single square root of a quartic polynomial in $x \equiv \beta_{\mathcal{R}+\mathcal{M}+1}$. In fact, we find the same elliptic curve as in eq.~\eqref{eq: elliptic_curve_10pt_elliptic_ladders} for the 10-pt elliptic ladder family. By testing elliptic linear reducibility and computing the maximal co-dimension residue up to 6 loops, we conclude that we obtain the following one-fold representation under the integration sequence $\{ \alpha_1,\dots,\alpha_{\mathcal{R}+\mathcal{M}+\mathcal{L}},\beta_{\mathcal{R}+\mathcal{M}+\mathcal{L}},\dots,\beta_{\mathcal{R}+\mathcal{M}+2},\beta_1,\dots,\beta_{\mathcal{R}+\mathcal{M}+1} \}$:
\begin{equation}
\label{eq: staggered_elliptic_ladder_one_fold}
\mathcal{I}_{\mathcal{L},\mathcal{M},\mathcal{R}} = \int_0^\infty \frac{dx}{(x-z_{+})(x-z_{-})} \, \mathcal{G}^{(2\mathcal{R}+2\mathcal{M}+2\mathcal{L}-1)}_{\mathcal{L},\mathcal{M},\mathcal{R}}(x,y)\, .
\end{equation}
The variables $z_{{\pm}}$ denote the algebraic functions
\begin{equation}
z_{{\pm}}=\frac{u_3 (u_2-u_2 u_4+v_1 -1)-u_1\pm \sqrt{(u_1+u_3(u_2-u_2 u_4+v_1 -1))^2-4u_2 u_3^3 u_4 u_5 v_2}}{2 u_3 v_2}\,,
\end{equation}
and $\mathcal{G}^{(2\mathcal{R}+2\mathcal{M}+2\mathcal{L}-1)}_{\mathcal{L},\mathcal{M},\mathcal{R}}(x,y)$ is a linear combination of weight-$(2\mathcal{R}+2\mathcal{M}+2\mathcal{L}-1)$ MPLs, which depend on the specific diagram and loop order. As a proof of principle, we provide in ref.~\cite{ancillary_file_elliptic_ladders} the MPLs for the one-fold integral representation of the integral $\mathcal{I}_{1,1,1}$.

As the arguments of the MPLs depend on the elliptic curve $y$, in principle the result of the one-fold integral cannot be expressed solely in terms of MPLs. Therefore, together with the example studied in ref.~\cite{Duhr:2020gdd}, this family provides the only other counterexample known in the literature to the expectation that an integral representation with $d \log$ kernels leads to MPLs. To determine if this is truly the case, one could compute the leading singularity for all subsectors, and check whether the ellipticity of this family is inherited through one of them. Even though this would be straightforward, we leave this check for future work.

\section{Second-order differential equation for ladder integrals}
\label{sec:ch2_diff_eqs}

As a final remark, in this section let us study one particular differential equation that ladder integrals satisfy, since it will briefly appear again in the next chapter. In particular, it was shown in ref.~\cite{Drummond:2010cz}, see also ref.~\cite{Drummond:2006rz} for a related work, that ladder diagrams satisfy a second-order differential equation relating them to lower-loop orders. The premise for the existence of this differential equation is that ladder integrals depend on at least one external dual coordinate $x_i$ which only appears in a single propagator in dual space, a condition that is satisfied by all examples studied so far. To obtain the differential equation, we start from the four-dimensional Laplacian operator $\Box_i = \nabla_i \cdot \nabla_i$ in dual space. When acting onto this single propagator (under the integral sign), and removing some conventional prefactors, it yields~\cite{Drummond:2010cz}
\begin{equation}
\Box_i \frac{1}{x_{l,i}^2}=-4 \, \delta^{(4)}(x_l-x_i)\,,
\end{equation}
where recall that $x_{l,i}^2 \equiv (x_l - x_i)^2$, for $x_l$ being the dual coordinate associated with the loop momentum. Therefore, this operation localizes the dual coordinate $x_l$ for the loop momentum attached to $x_i$, which reduces the loop order by one.

For instance, let us apply it to $x_7$ for the non-DCI 1-loop 4-mass box integral from eq.~\eqref{eq: non-DCI_4mass_box_dual_coords},
\begin{equation}
   \Box_7 \, \mathcal{I}_{\text{box}}^{\text{(non-DCI)}} = \Box_7 \int \frac{d^4 x_l}{x_{l,1}^2 x_{l,3}^2 x_{l,5}^2 x_{l,7}^2} = - 4 \int \frac{d^4 x_l \  \delta^{(4)}(x_l-x_7)}{x_{l,1}^2 x_{l,3}^2 x_{l,5}^2} = \frac{-4}{x_{1,7}^2 x_{3,7}^2 x_{5,7}^2}\,.
\end{equation}
On the other hand, we can render the integral DCI as
\begin{equation}
    \mathcal{I}_{\text{box}}^{\text{(non-DCI)}} = \frac{1}{x_{1,5}^2 x_{3,7}^2} \ \drawbox \,.
\end{equation}
Hence,
\begin{align}
    x_{1,7}^2 x_{3,7}^2 x_{5,7}^2 \Box_7 \, \mathcal{I}_{\text{box}}^{\text{(non-DCI)}} =& \, x_{1,7}^2 x_{3,7}^2 x_{5,7}^2 \,  \nabla_7 \cdot \nabla_7 \left(\frac{1}{x_{1,5}^2 x_{3,7}^2} \ \drawbox \, \right) \nonumber \\
    =&\, 4 u v \Big( u \partial_u^2 + v \partial_v^2 + (u+v-1) \partial_u \partial_v +2 \partial_u +2 \partial_v \Big) \ \drawbox \ .
\end{align}
To obtain the second step, we use the chain rule
\begin{equation}
\label{eq: chain_rule_dual_coords_Laplacian}
    \nabla_7^\mu \left( \, \drawbox \, \right) = \frac{\partial}{\partial (x_7)_\mu} \left( \, \drawbox \, \right) = \left( \frac{\partial u}{\partial (x_7)_\mu} \, \partial_u + \frac{\partial v}{\partial (x_7)_\mu} \, \partial_v \right) \ \drawbox \ ,
\end{equation}
with $u = \frac{x_{1,3}^2 x_{5,7}^2}{x_{1,5}^2 x_{3,7}^2}$ and $v = \frac{x_{1,7}^2 x_{3,5}^2}{x_{1,5}^2 x_{3,7}^2}$, as in eq.~\eqref{eq: 1-loop_4mass_box_cross-ratios_wrt_xi}. All in all, we obtain the DCI second-order differential equation~\cite{Drummond:2010cz}
\begin{equation}
    4 u v \Big( u \partial_u^2 + v \partial_v^2 + (u+v-1) \partial_u \partial_v +2 \partial_u +2 \partial_v \Big) \ \drawbox = -4,
\end{equation}
which is satisfied by the result in eq.~\eqref{eq: 4-mass_box_result}.

Proceeding analogously for our elliptic ladder families, and taking the Laplacian with respect to the dual coordinate $x_8$, we obtain the differential equations
\begin{equation}
\label{eq: diff_equation_ladders_DL}
\mathcal{D}_{\mathcal{L}} \,  \mathcal{I}_{\mathcal{L},\mathcal{R}}= \mathcal{I}_{\mathcal{L}-1,\mathcal{R}}\,, \qquad \text{and} \qquad \mathcal{D}_{\mathcal{L}} \,  \mathcal{I}_{\mathcal{L},\mathcal{M},\mathcal{R}}= \mathcal{I}_{\mathcal{L}-1,\mathcal{M},\mathcal{R}}\,,
\end{equation}
where the differential operator is
\begin{align}
\mathcal{D}_{\mathcal{L}} =&\, u_2 v_2\Big[ -2 u_3 u_4 u_5  \left( \partial_{u_1}+\partial_{v_1}\right) - 2 \left(\partial_{u_2} +  \partial_{v_2} \right) - 
  u_1 u_3 u_4 u_5 \partial^2_{u_1} - 
  v_1 u_3 u_4 u_5 \partial^2_{v_1} - 
  u_2  \partial^2_{u_2} - 
  v_2 \partial^2_{v_2} \nonumber\\
 & \, + 
  u_3 u_4 u_5 (1 - u_1 - v_1)  \partial_{v_1}\partial_{u_1} + 
  (1 - u_2 - v_2) \partial_{v_2}\partial_{u_2} - 
  (u_1 - u_3 + u_2 u_3 u_4 u_5) \partial_{u_2}\partial_{u_1} \nonumber\\
 & \, - (v_1-u_4  + v_2 u_3 u_4 u_5 ) \partial_{v_2}\partial_{v_1} - (u_1 + v_2 u_3 u_4 u_5 ) \partial_{v_2}\partial_{u_1}  - 
   (v_1 + u_2 u_3 u_4 u_5 )  \partial_{v_1}\partial_{u_2}     
\Big] \,.
\label{eq: differential operator}
\end{align}
Instead, taking the Laplacian with respect to $x_3$ leads to another set of differential equations,
\begin{equation}
\label{eq: diff_equation_ladders_DR}
\mathcal{D}_{\mathcal{R}} \,  \mathcal{I}_{\mathcal{L},\mathcal{R}}= \mathcal{I}_{\mathcal{L},\mathcal{R}-1}\,, \qquad \text{and} \qquad \mathcal{D}_{\mathcal{R}} \,  \mathcal{I}_{\mathcal{L},\mathcal{M},\mathcal{R}}= \mathcal{I}_{\mathcal{L},\mathcal{M},\mathcal{R}-1}\,,
\end{equation}
where the operator can be obtained from the reflection symmetry $\mathcal{D}_{\mathcal{R}} = \mathcal{V}(\mathcal{D}_{\mathcal{L}})$, which acts on the cross-ratios as given in eq.~\eqref{eq: 10pt_elliptic_ladders_mirror_sym}. Furthermore, as we showed in appendix B of ref.~\cite{McLeod:2023qdf}, the differential equations also hold at the level of the integrand.

As can be seen, these second-order differential equations relate different loop orders within each integral family. For the first family, the boundary values correspond to $\mathcal{I}_{0,\mathcal{R}}$ and $\mathcal{I}_{\mathcal{L},0}$, which are the well-known 8-pt ladders \cite{Usyukina:1993ch} from eq.~\eqref{eq: 4pt_ladders_result}. For the second family, the boundary values are the polylogarithmic 9-pt ladders $\mathcal{I}_{0,\mathcal{M},\mathcal{R}}$ and $\mathcal{I}_{\mathcal{L},\mathcal{M},0}$, which can be obtained as soft limits of the 10-pt elliptic ladders $\mathcal{I}_{\mathcal{L},\mathcal{R}}$. Quite nicely, with the change of variables of eq.~\eqref{eq: change_vars_10pt_elliptic_ladders}, we can obtain a one-fold integral representation analogous to eq.~\eqref{eq: elliptic_ladder_one_fold} but which is totally linearly reducible and can be evaluated in terms of MPLs. For example, when the external leg $p_5$ becomes soft, we have that $x_{5,6}^2=0$, which implies
\begin{equation}
    u_4 \to 1\,, \qquad u_5 \to 1\,, \qquad u_1 \to u_3 v_2\,,
\end{equation}
for the cross-ratios. Under this limit, the elliptic curve from eq.~\eqref{eq: elliptic_curve_10pt_elliptic_ladders} degenerates as
\begin{equation}
    y^2 = \frac{1}{v_2^2} \Big( (1+x)(1-2v_1+v_2 x) + 2u_2^2 (u_3+x) - u_2 x \Big)^2,
\end{equation}
and becomes a perfect square. Thus, ellipticity is lost and, via partial fraction, the result of the one-fold integral can be brought to MPLs.

Lastly, let us point out that the existence of these pairs of second-order differential equations for the elliptic ladder integrals may facilitate bootstrapping their result. First, a similar differential equation satisfied by the double penta-ladders (see eq.~\eqref{eq: drawing_penta_ladders}) has been solved at finite coupling~\cite{Caron-Huot:2018dsv}, thus it would be natural to extend the method to this case. Moreover, as we will explore in the next chapter, advances in bootstrapping techniques for elliptic Feynman integrals may benefit from the input provided by these differential equations, making the elliptic ladder families compelling examples for further development of these tools.

\section{Discussion and open questions}
\label{sec:ch2_conclusions}

In this chapter, we have started the exploration of Feynman integral geometries by considering a simple case: generalizations of ladder integrals in $\mathcal{N}=4$ SYM theory. In particular, through the addition of ladder rungs and reordering of the intermediate massless legs, we have identified two 10-pt ladder integral families which depend on the same elliptic curve to all loop orders. Moreover, using appropriate changes of variables and the polynomial reduction algorithm, we have found that they are linearly reducible into a one-fold integral with a single square root -- the elliptic curve. Thus, they can be written in terms of the same class of eMPLs. In particular, in secs.~\ref{sec:ch2_elliptic_ladders} and~\ref{sec:ch2_staggered_elliptic_ladders}, we found
\begin{equation}
\label{eq: 10pt_elliptic_ladders_conclusion}
 = \int_0^\infty \frac{dx}{(x-z_{+})(x-z_{-})} \, \mathcal{G}^{(2\mathcal{R}+2\mathcal{M}+2\mathcal{L}-1)}_{\mathcal{L},\mathcal{M},\mathcal{R}}(x,y)\,,
\end{equation}
where $\mathcal{G}^{(n)}$ denotes a linear combination of MPLs of weight $n$, and $y$ is the elliptic curve. Interestingly, we have found that the second integral family admits a one-fold integral representation with $d \log$ kernels, which depend on two algebraic functions of the kinematics $z_\pm$. 

We have checked elliptic linear reducibility for all diagrams in the families up to 6 loops and, as a proof of principle, we have calculated a 3-loop representative in terms of eMPLs. While this falls short of an all-loop proof for elliptic linear reducibility, the simplicity of the change of variables and the repetitive loop-by-loop Feynman parametrization make it tempting to expect that it holds in general. We hope that a general proof can be found in the future.

As further future work, it would be interesting to investigate the elliptic symbol alphabet for these integrals and its dependence with the loop order, especially whether it saturates at some point. However, we expect that a direct computation via eMPLs will suffer from expression swelling at intermediate steps. A different approach in this direction could be to use the recently-developed formalism for elliptic symbol integration~\cite{He:2023qld} directly on the previous one-fold integral representations. Alternatively, one could try to apply the elliptic symbol bootstrap method, which we will cover in the next chapter. Overall, the unique properties of these ladder families, such as the pair of second-order differential equations that they satisfy, make them tantalizing test cases for the development and refinement of elliptic technology.

Finally, we also explored integral families with different configurations of external legs, for instance, all intermediate legs on the same side of the ladder, as well as including more legs. However, no change of variables granting linear reducibility up to the last integration variable (and involving a single square root) was found in general. A generic analysis for such cases can be nonetheless found in ref.~\cite{Cao:2023tpx}. Similarly, it would be interesting to apply the methods of this chapter to more exotic generalizations of ladder diagrams, such as the so-called traintrack networks of ref.~\cite{McLeod:2023doa}, which are formed by intersecting different traintrack diagrams.

\chapter{Bootstrapping the twelve-point elliptic double box}
\label{ch:chapter3}

\begin{info}[\textit{Info:}]
\textit{Part of the content and figures of this chapter have been published together with A.~Spiering, M. Wilhelm, Q. Yang and C. Zhang in ref.}~\cite{Morales:2022csr}\textit{, available at} \href{https://doi.org/10.1103/PhysRevLett.131.041601}{\textit{Phys. Rev. Lett.} \textbf{131} (2023) 041601} [\href{https://arxiv.org/abs/2212.09762}{2212.09762}].
\end{info}

\section{Motivation} 
\label{sec:ch3_intro}

In the previous chapter, we explored some examples of elliptic Feynman integrals in planar $\mathcal{N}=4$ SYM theory, focusing on 10-pt ladder integrals in $D=4$. In particular, we studied the 2-loop 10-pt double box in sec.~\ref{sec:ch2_10pt_double_box}, as well as its multiloop generalizations via the addition of ladder rungs, see e.g.~eqs.~\eqref{eq: 10pt_elliptic_ladders_conclusion} and~\eqref{eq: 10pt_staggered_elliptic_ladders_conclusion}. In this chapter, we continue this investigation by studying a different generalization of the double box -- by adding more external legs. 

Specifically, we add more legs to the middle rung of the double box. The most general case is therefore the 2-loop 12-pt double-box integral, which also corresponds to one of the elements in the basis of planar Feynman integrals in $\mathcal{N}=4$ SYM theory at two loops~\cite{Bourjaily:2015jna,Bourjaily:2017wjl}. To understand why this integral requires a new method, let us compare it to the 10-pt double-box integral from the previous chapter and use direct integration with Feynman parameters. Following sec.~\ref{sec:ch2_one_loop_box}, in dual-momentum coordinates this integral is given by
\begin{equation}
\label{eq: 12pt_double-box_dual_coord_param}
 = \int \frac{d^4 x_{l_1} d^4 x_{l_2} \, x_{1,3}^2 x_{2,5}^2 x_{4,6}^2}{x_{l_1,1}^2 x_{l_1,2}^2 x_{l_1,3}^2 x_{l_1,l_2}^2 x_{l_2,4}^2 x_{l_2,5}^2 x_{l_2,6}^2}\,.
\end{equation}
Notice that, compared to chapter~\ref{ch:chapter2}, we have redefined the dual coordinates as $x_i \to x_{(i+1)/2}$, such that the labels run from 1 to 6. This relabeling is done to make the notation more compact and, as we will see, it will simplify the main result of the chapter in sec.~\ref{sec:ch3_symbol_12pt_double_box}. Similarly, we introduce the following notation for the 9 DCI cross-ratios that this integral depends on:
\begin{equation}
\label{eq: cross-ratios_chi_ab}
    \chi_{ab} \equiv (a \, b-1; b \, a-1) = \frac{x_{a,b-1}^2 x_{b,a-1}^2}{x_{a,b}^2 x_{a-1,b-1}^2}\,,
\end{equation}
where we also used the notation $(ab;cd)$ introduced in eq.~\eqref{eq: traintracks_cross-ratios_def}. In this expression, the labels $a$, $b$ are non-adjacent in the cycle $\{1, \dots,6 \}$, with $\chi_{ba} = \chi_{ab}$. In other words, we have the 9 cross-ratios 
\begin{equation}
\label{eq: cross-ratios_chi_ab_all_possibilities}
\{ \chi_{13},\chi_{14},\chi_{15},\chi_{24},\chi_{25},\chi_{26},\chi_{35},\chi_{36},\chi_{46} \}.
\end{equation}

Using the Feynman parametrization from eq.~\eqref{eq: traintrack_Feyn_param}, we obtain
\begin{equation}
\label{eq: 12pt_double-box_Feynman_param}
    \drawdb \, =  \int_0^\infty \frac{d \alpha_1 \, d \alpha_2 \, d \beta_1 \, d \beta_2}{f_1 f_2 \, g_2} \,,
\end{equation}
where
\begin{subequations}
\begin{align}
    \hspace{-0.15cm} f_1 = & \, \alpha_1 + \beta_1 + \alpha_1 \beta_1, \\
    \hspace{-0.15cm} f_2 = & \, \chi_{13} \chi_{14} \chi_{24} f_1 + \alpha_2 + \beta_2 + \alpha_2 \beta_2 + \chi_{24}\chi_{25}\chi_{26} \alpha_1 \alpha_2 + \chi_{24} \alpha_1 \beta_2 + \chi_{13} \alpha_2 \beta_1 + \chi_{13}\chi_{35}\chi_{36} \beta_1 \beta_2, \\
    \hspace{-0.15cm} g_2 = & \, 1 + \chi_{24}\chi_{25} \alpha_1 + \chi_{13}\chi_{14}\chi_{36}\chi_{46} \beta_1 + \chi_{15}\chi_{25} \alpha_2 + \chi_{36}\chi_{46} \beta_2.
\end{align}
\end{subequations}
In the limit where the middle pair of legs $\{p_5,p_6\}$ and $\{p_{11},p_{12} \}$ become massless, the squared differences $x_{1,6}^2=0$ and $x_{3,4}^2=0$ become light-like. As a consequence, $\chi_{26}$ and $\chi_{35}$ also vanish. As expected, in this limit, eq.~\eqref{eq: 12pt_double-box_Feynman_param} reduces to the parametrization of the 10-pt double box of eq.~\eqref{eq: 10pt_double-box_Feynman_param} under the replacements
\begin{align}
    u_1 =& \, \chi_{13}\chi_{14}\chi_{36}\chi_{46}\,, \qquad u_2 = \chi_{15}\chi_{25}\,, \qquad u_3 = \chi_{13}\,, \qquad u_4 = \chi_{24}\,, \nonumber \\
    u_5 =&\, \chi_{14}\,, \qquad \qquad \qquad \hspace{0.175cm}  v_1 = \chi_{24} \chi_{25}\,, \qquad v_2= \chi_{36}\chi_{46}\,.
\end{align}

For the 12-pt double box, we can also calculate the leading singularity as the maximal co-dimension residue, which leads to an integral
\begin{equation}
    \LS \left( \, \drawdb \, \right) = \int \frac{d \alpha_1}{\sqrt{P_4(\alpha_1)}}\,,
\end{equation}
for $P_4(\alpha_1)$ being a quartic polynomial in $\alpha_1$ which depends on the cross-ratios $\chi_{ab}$. Therefore, the 12-pt double-box integral also involves an elliptic curve, as already observed in refs.~\cite{Paulos:2012nu,Caron-Huot:2012awx,Nandan:2013ip,Bourjaily:2017bsb}.

So far, the 12-pt double box seems very similar to the 10-pt case. However, there is a big difference when trying to calculate the result of the Feynman parametrization using direct integration. As explained in secs.~\ref{sec:ch2_10pt_double_box} and~\ref{sec:ch2_polynomial_reduction}, for the 10-pt double box we can perform the integrations over $\beta_1$ and $\beta_2$ without introducing square roots in the remaining variables. Moreover, after the change of variables of eq.~\eqref{eq: 10-pt_double-box_change_vars}, we can integrate over $\tilde{\alpha}_2$ and only introduce a single square root -- the elliptic curve -- in the remaining variable. By contrast, for the 12-pt double box, we can only integrate one variable without introducing square roots, regardless of the integration order. The best outcome for the polynomial reduction algorithm (recall sec.~\ref{sec:ch2_polynomial_reduction}) is obtained after integrating over $\beta_2$. In this case, there are only 4 polynomials in the remaining variables $\{ \alpha_1, \alpha_2, \beta_1 \}$, with only two of the polynomials being quadratic. However, we did not find a change of variables that eliminates all quadratic terms for at least one of the variables. Thus, we cannot perform two integrations without introducing square roots, not to mention trying to integrate three variables to obtain a one-fold elliptic representation. The same conclusion is reached using an alternative Feynman parametrization from ref.~\cite{Loebbert:2019vcj}.

Compared to the Feynman parametrization of the 10-pt double box (see eq.~\eqref{eq: 10pt_double-box_Feynman_param}), the difference in the 12-pt case resides in the fact that in eq.~\eqref{eq: 12pt_double-box_Feynman_param} we also have the bilinear terms $\beta_1 \beta_2$ and $\alpha_1 \alpha_2$. Even though this looks like a small difference, it prevents us from using the changes of variables that granted elliptic linear reducibility for the elliptic ladder families in secs.~\ref{sec:ch2_elliptic_ladders} and~\ref{sec:ch2_staggered_elliptic_ladders}. Consequently, we must search for other methods to calculate this integral.

One approach could be to integrate the first-order differential equation relating this integral to the 1-loop 12-pt hexagon in $D=6$~\cite{Paulos:2012nu,Nandan:2013ip},
\begin{equation}
\label{eq: diff_eq_12pt_double-box_hex}
    \partial_{\chi_{14}} \, \, \drawdb \, = \frac{1}{\sqrt{-\Delta_6}} \, \, \drawhex^{(6D)} \quad \enspace \Longrightarrow \quad \enspace \drawdb \, = \int \frac{d \chi'_{14}}{\sqrt{-\Delta_6(\chi'_{14})}} \, \, \drawhex^{(6D)} \hspace{-0.5cm}(\chi'_{14})\,,
\end{equation}
where $\Delta_6=\det (x_{i,j}^2)/{(x_{1,4}^2 x_{2,5}^2 x_{3,6}^2)}^2$ denotes the normalized Gram determinant of the 12-pt hexagon, which is a cubic polynomial in $\chi_{14}$. Although this approach seems to straightforwardly produce a one-fold elliptic integral representation, the hexagon contains 15 different square roots over Gram determinants of 4-mass box subsectors~\cite{Spradlin:2011wp}. Since we did not find a change of variables that rationalizes simultaneously all of the 15 square roots, we cannot directly evaluate the result in terms of eMPLs, defined in sec.~\ref{sec:ch2_eMPLs}. For the same reason, and also due to the many kinematic scales involved, the differential equations method reviewed in sec.~\ref{sec:ch1_DEs} would be very challenging.

As a consequence, in order to approach this elusive integral we must investigate a new method. In particular, we notice that all of the difficulties are related to the actual calculation of the integral. Perhaps, the final result is not as complicated, and we can try to guess it using physical arguments. Precisely with this objective, we will resort to the so-called symbol bootstrap approach, which has been extensively used in the literature~\cite{Goncharov:2010jf,Dixon:2011pw,Dixon:2011nj,Brandhuber:2012vm,Caron-Huot:2016owq,Almelid:2017qju,Henn:2018cdp,Caron-Huot:2019vjl,Dixon:2020bbt,Guo:2021bym,Dixon:2022rse,Dixon:2022xqh,Hannesdottir:2024hke,Basso:2024hlx}. As briefly introduced in sec.~\ref{sec:ch1_MPLs_symbol}, the bootstrap technique is based on an educated ansatz for the symbol alphabet, together with some physical requirements, and the aim is to deduce the symbol for the final result without performing the full calculation. This way, we may transform the complicated integral to a much simpler linear algebra problem, which amounts to solving for free coefficients in an ansatz that satisfies some constraints. The problem, however, is that the symbol bootstrap approach has only been used for polylogarithmic integrals so far. Consequently, using the construction for the symbol of elliptic Feynman integrals of ref.~\cite{Broedel:2018iwv}, in this chapter we will develop a bootstrap method that also extends to elliptic Feynman integrals, therefore initiating the elliptic symbol bootstrap.

The cornerstone of the bootstrap method is constructing an appropriate ansatz for the symbol alphabet of the integral. In our case, since the result of the 10-point double-box integral in terms of eMPLs and its symbol are already known~\cite{Kristensson:2021ani}, we can leverage this information to guide the ansatz for the 12-pt case. However, we still need to predict some new symbol letters which only appear in the 12-pt symbol and are absent in the 10-pt limit. For polylogarithmic integrals, the task of predicting symbol letters has been traditionally addressed using tools such as cluster algebras and tropical Grassmannians~\cite{Drummond:2018caf,Golden:2013xva,Golden:2014pua,Drummond:2019qjk,Drummond:2019cxm}. More recently, other approaches have been developed based on the Baikov representation and the Gram determinants therein~\cite{Jiang:2024eaj}, intersection theory~\cite{Chen:2023kgw}, and from Landau equations~\cite{Hannesdottir:2021kpd,Dlapa:2023cvx,Fevola:2023kaw,Hannesdottir:2024hke}. For our elliptic integral, however, we found it more useful to rely instead on the recently developed Schubert analysis~\cite{Arkani-Hamed:talk,Yang:2022gko,He:2022tph,He:2023umf}. This approach is based on solving for the leading singularity of the integral in twistor space and, as we will see, extends naturally to accommodate for elliptic integrals.

This chapter is structured as follows. First, in sec.~\ref{sec:ch3_elliptic_symbol}, we introduce the symbol for elliptic Feynman integrals, and review the elliptic symbol of the 10-pt double box. Then, in sec.~\ref{sec:ch3_one_loop_Schubert}, we present the Schubert analysis as a method to solve for the leading singularity of a Feynman integral in twistor space, and use it to predict the symbol letters for the 1-loop 4-mass box. Afterwards, in sec.~\ref{sec:ch3_two_loop_Schubert}, we generalize the Schubert analysis for elliptic integrals, and show how to predict the symbol letters for the double box. Using this prediction for the symbol alphabet, in sec.~\ref{sec:ch3_12pt_double_box}, we then bootstrap the symbol for the 12-pt double box, which results in an elegant one-line formula that satisfies all physical constraints and cross-checks. We end in sec.~\ref{sec:ch3_conclusions} with a discussion and possible future directions.

\section{Elliptic symbology} 
\label{sec:ch3_elliptic_symbol}

In the first chapter, we introduced the symbol for MPLs, which allows us to decompose any polylog into a tensor product of logarithms that encode the singularity structure of the function, see sec.~\ref{sec:ch1_MPLs_symbol}. Here, we introduce the generalization of the symbol to elliptic Feynman integrals and review the type of symbol letters which can appear in this case; see ref.~\cite{Broedel:2018iwv} as well as refs.~\cite{Kristensson:2021ani,Wilhelm:2022wow} for our conventions. Besides, we examine the symbol of the 10-pt double-box integral, which will be the basis for the ansatz of the 12-pt bootstrap.

In sec.~\ref{sec:ch2_eMPLs_torus}, we introduced eMPLs and reviewed how to express an elliptic Feynman integral in terms of $\widetilde{\Gamma}$ functions. As it turns out, as shown in refs.~\cite{Kristensson:2021ani,Wilhelm:2022wow}, the total differential of a $\widetilde{\Gamma}_k^{(n)}$ function of length $k$ and weight $n$ takes the schematic form
\begin{equation}
\label{eq: schematic_dGamma_symbol}
    d \widetilde{\Gamma}_k^{(n)} = \sum_i \, (2 \pi i )^{j_i-1} \, \widetilde{\Gamma}_{k-1}^{(n-j_i)} \, d \Omega^{(j_i)}. 
\end{equation}
The functions $\Omega^{(j)}(w)$ are related to the integration kernels $g^{(n)}(w)$ from eq.~\eqref{eq: def_Gamma_tilde} via
\begin{equation}
\label{eq: def_Omega_g_kernels}
    \partial_w \Omega^{(j)}(w) = (2 \pi i )^{1-j} g^{(j)}(w), \qquad \qquad \partial_\tau \Omega^{(j)}(w) = j (2 \pi i )^{-j} g^{(j+1)}(w),
\end{equation}
recalling that $\tau = \omega_2/\omega_1$ is the modular parameter and $w$ is a point on the torus, defined through Abel's map in eq.~\eqref{eq: Abel_map_w}. In particular, we have
\begin{equation}
\label{eq: values_Omega_eMPLs}
    \Omega^{(-1)}(w) = - 2 \pi i \tau, \qquad \Omega^{(0)}(w) = 2 \pi i w, \qquad \Omega^{(1)}(w) = \log \left( \frac{\theta_1 (w)}{\eta(\tau)} \right), 
\end{equation}
where $\theta_1(w)$ denotes the odd Jacobi theta function, and $\eta(\tau)$ is the Dedekind eta function. For $\Omega^{(j > 1)}(w)$, a closed expression can be found in ref.~\cite{Wilhelm:2022wow}.

Redefining the function in eq.~\eqref{eq: schematic_dGamma_symbol} as $\widetilde{\underline{\Gamma}}_k^{(n)} = (2 \pi i)^{k-n}\widetilde{\Gamma}_k^{(n)}$, the total differential becomes
\begin{equation}
    d \widetilde{\underline{\Gamma}}_k^{(n)} = \sum_i \, \widetilde{\underline{\Gamma}}_{k-1}^{(n-j_i)} \, d \Omega^{(j_i)},
\end{equation}
which is compatible with eq.~\eqref{eq: differential_pre_symbol} by exchanging the weight with the length. Therefore, following eq.~\eqref{eq: symbol}, we can analogously define the symbol for eMPLs recursively as
\begin{equation}
    \mathcal{S} \Big( \widetilde{\underline{\Gamma}}_k^{(n)} \Big) = \sum_i \, \mathcal{S} \Big( \widetilde{\underline{\Gamma}}_{k-1}^{(n-j_i)} \Big)  \otimes  \Omega^{(j_i)}.
\end{equation}
Hence, the symbol of an eMPL of length $k$ generically takes the form
\begin{equation}
    \mathcal{S} \Big( \widetilde{\underline{\Gamma}}_k^{(n)} \Big)= \sum_{i_1, \dots, i_k} c_{i_1 \cdots\, i_k} \, \Omega^{(i_1)} \otimes \cdots \otimes \Omega^{(i_k)},
\end{equation}
where we observe that now the functions $\Omega^{(j)}$ play the role of the symbol letters. Not surprisingly, since $\Omega^{(0)}(w) = 2 \pi i w$ is proportional to an image on the torus through Abel's map, it implies that the symbol letters for elliptic Feynman integrals also involve elliptic integrals.

As explained in sec.~\ref{sec:ch1_MPLs_symbol}, one of the main virtues of the symbol is that it allows us to reduce the complicated functional identities among MPLs to the familiar logarithmic identity $\log(ab)=\log(a)+\log(b)$. In the case of eMPLs this is also true, although the relations obeyed by the functions $\Omega^{(j)}$ are more involved. For instance, integrating eq.~\eqref{eq: psi_1_kernel_eMPLs} and using eqs.~\eqref{eq: kernels_psi_eMPLs} and~\eqref{eq: def_Omega_g_kernels}, we obtain
\begin{align}
    \int_a^b \psi_{1}(c,x) \, d x = &\, \int_a^b \frac{dx}{x-c} = \log \Big( \frac{b-c}{a-c} \Big) \nonumber \\[0.1cm]
    = &\, \int_{w_a^+}^{w_b^+} \Bigl(g^{(1)}(w-w_{c}^{+})+g^{(1)}(w-w_{c}^{-}) -g^{(1)}(w-w_{\infty}^{+}) -g^{(1)}(w-w_{\infty}^{-})\Bigr)  \, d w \nonumber \\[0.1cm]
    = &\, \sum_{\sigma = \pm} \left( \Omega^{(1)}(w_b^+-w_c^\sigma) - \Omega^{(1)}(w_b^+ - w_\infty^\sigma) - \Omega^{(1)}(w_a^+-w_c^\sigma) + \Omega^{(1)}(w_a^+ - w_\infty^\sigma) \right),
\end{align}
which in many cases allows us to reorganize some of the symbol letters of eMPLs to generate ordinary logarithms. In general, such identities can be found via the PSLQ algorithm with the \texttt{FindIntegerNullVector} command in \Mathematica. Otherwise, they can be derived from Abel's addition theorem in combination with the so-called symbol prime, see ref.~\cite{Wilhelm:2022wow}. For example, the symbol of the 2-loop unequal-mass sunrise in $D=2$ takes a very simple form~\cite{Wilhelm:2022wow}, with a symbol alphabet containing logarithms, $\tau$, $\Omega^{(0)}$'s and $\Omega^{(2)}$'s.

\subsection{Symbol of the ten-point elliptic double box} 
\label{sec:ch3_symbol_10pt_double_box}

Having introduced the elliptic symbol, we now focus on the 10-pt double box, which was written as a one-fold elliptic integral in eq.~\eqref{eq: 10-pt_double-box_result}. As discussed in secs.~\ref{sec:ch2_eMPLs} and~\ref{sec:ch2_eMPLs_torus}, this representation can be directly expressed in terms of eMPLs, see ref.~\cite{Kristensson:2021ani} for the result in terms of $\mathrm{E}_4$ and $\mathcal{E}_4$ functions. Re-expressing it in terms of $\widetilde{\Gamma}$ functions, and using the elliptic symbol construction outlined above, the symbol for the 10-pt double-box integral can be calculated. In particular, it obeys the following simple structure~\cite{Kristensson:2021ani,Wilhelm:2022wow}:
\begin{align}
\label{eq: 10pt_double-box_symbol_structure}
    \mathcal{S}\left( \frac{2 \pi i}{\omega_1} \, \drawdbten \, \right) =  \sum_{ikl} c_{ikl}\log({\phi_{k}})\otimes\log({\phi_{l}}) \otimes \! \Biggl[\sum_j \log ({\phi_{ij}}) \otimes \left(2\pi i\,w_{c_j}^+\right) + {\bm\Omega}_{i}\otimes (2 \pi i\,{\tau})   \Biggr] \,.
\end{align}
Let us examine this symbol in detail. First of all, as can be seen, the first two entries of the symbol are ordinary logarithms, where $\phi$ denote rational or algebraic functions of the kinematics, whereas the last entries are always related to elliptic integrals. Depending on whether the last entry corresponds to $2\pi i w_{c_j}^+$ or $2\pi i \tau$, the third entry can either be a logarithm or an elliptic letter ${\bm\Omega}_{i}=\sum_i c_i \, \Omega^{(0,1,2)}$, respectively. In fact, these elliptic letters can be obtained from the rest of the symbol by employing the so-called symbol prime~\cite{Wilhelm:2022wow}, such that
\begin{equation} 
\label{eq:intrep_symbolprime}
 \bm{\Omega}_{i}=\sum_j \, \partial_\tau\int 2\pi i\,w_{c_j}^+ \, d\log({ \phi_{ij}})\,.
\end{equation}
Therefore, we can totally disregard the terms where last entry is $2\pi i \tau$, as they are fixed by the rest of the symbol.

In total, the symbol of the 10-pt double box contains around 10,000 terms~\cite{Kristensson:2021ani}, albeit the symbol alphabet is quite simple. To make this simplicity manifest, let us first note that the 10-pt double-box integral depends on 6 external dual coordinates $x_i$, see eq.~\eqref{eq: 10pt_double-box_Feynman_param}, which are given by $\{ x_1, x_3, x_5, x_6, x_8, x_{10} \}$. Choosing any set of 4 dual coordinates out of the 6, we can form $\binom{6}{4}=15$ 1-loop box integrals.\footnote{For the 10-pt double box, we can form 4-mass as well as 3- and 2-mass box integrals, depending on the 4 dual coordinates that we choose. For the latter cases, since a subset of the dual coordinates are light-like, some of the cross-ratios that they depend on will vanish.} Let us denote these integrals by $\text{Box}_{ij}$, which we label by the 2 dual coordinates $\{i,j \}$ that they do not depend on. Recalling eq.~\eqref{eq: 4-mass_box_symbol}, the corresponding symbol is
\begin{equation}
\label{eq: symbol_Box_ij}
    \mathcal{S}\left( \text{Box}_{ij} \right) = \log(u_{ij}) \otimes \log(\frac{1-z_{ij}}{1-\bar{z}_{ij}}) - \log(v_{ij}) \otimes \log(\frac{z_{ij}}{\bar{z}_{ij}})\,,
\end{equation}
where we normalize $\text{Box}_{ij}$ to have unit leading singularity. The respective DCI cross-ratios are defined analogously to eqs.~\eqref{eq: 1-loop_4mass_box_cross-ratios_wrt_xi} and~\eqref{eq: uv_zzbar}, such that
\begin{equation}
\label{eq: variables_Box_ij}
    u_{ij}=(kl;mn)= z_{ij} \, \bar{z}_{ij}\,, \qquad \text{and} \qquad v_{ij}=(kn;ml)=(1-z_{ij})(1-\bar{z}_{ij}) \,,
\end{equation}
with $\{ k,l,m,n \}=\{1,3,5,6,8,10\} \backslash \{i,j\}$, and $(ab;cd)$ defined in eq.~\eqref{eq: traintracks_cross-ratios_def}.

As it turns out~\cite{Kristensson:2021ani}, the letters in the first two entries of the 10-pt double-box symbol of eq.~\eqref{eq: 10pt_double-box_symbol_structure} can be obtained from the symbols $\mathcal{S}\left( \text{Box}_{ij} \right)$ of the previous 1-loop boxes as well as from the symbols of $\text{Li}_2(1-(ab;cd))$ and $\log((ab;cd)) \log((a'b';c'd'))$, for $(ab;cd)$ being cross-ratios $u_{ij}$ and $v_{ij}$. This has been observed to be a general feature of planar Feynman integrals in $\mathcal{N}=4$ SYM theory~\cite{Gaiotto:2011dt,Caron-Huot:2011zgw,Dennen:2015bet,He:2020lcu}. 

In our case, the first entries of the symbol are given by the rational letters $\{ \log(u_{ij}),\log(v_{ij})\}$. Since the cross-ratios $u_{ij}$, $v_{ij}$ are obtained from the same dual coordinates that the 10-pt double box depends on, they give rise to the 7 DCI cross-ratios $u_i$ and $v_i$ from eq.~\eqref{eq: DCI_cross-ratios_10pt_double-box}. Thus, in the first entry we have the 7 rational letters
\begin{equation}
    \{ \log(u_1), \dots, \log(u_5), \log(v_1), \log(v_2) \}\,,
\end{equation}
which satisfies the so-called first-entry condition~\cite{Gaiotto:2011dt}, i.e.~the letters in the first entries can only be cross-ratios. 

In the second entry, we can again have the same 7 rational letters, in addition to 19 linearly independent algebraic letters formed by the quotients $\Big\{ \log(\frac{z_{ij}}{\bar{z}_{ij}}), \log(\frac{1-z_{ij}}{1-\bar{z}_{ij}}) \Big\}$. Together with the first entries, they satisfy the Steinmann conditions~\cite{Steinmann:1960,Steinmann2:1960}, namely that discontinuities in partially overlapping factorization channels vanish. In practice, this can be verified using eq.~\eqref{eq: disc_symbol} to compute the discontinuities. For instance, in the factorization channels $\{ \textcolor{blue!50}{x_{1,5}^2}, \textcolor{orange!75}{x_{3,6}^2} \}$, we have that
\begin{align}
\label{eq: example_Steinmann}
    0=&\, \mathcal{S}\left( \frac{2 \pi i}{\omega_1} \, \begin{tikzpicture}[scale=0.5,label distance=-1mm,baseline={([yshift=-.5ex]current bounding box.center)}]
\clip (-0.25,13.75) rectangle (3.25,16.25);
		\node (0) at (0.5, 15.5) {};
		\node (1) at (1.5, 16.25) {};
		\node (4) at (2.5, 15.5) {};
		\node (5) at (0.5, 14.5) {};
		\node (7) at (1.5, 13.75) {};
		\node (8) at (2.5, 14.5) {};
        \node (11) at (5, 15) {};
		\node (12) at (6, 15) {};
		\draw[line width=0.225mm] (0.center) to (4.center);
		\draw[line width=0.225mm] (5.center) to (8.center);
		\draw[line width=0.225mm] (0.center) to (5.center);
		\draw[line width=0.225mm] (1.center) to (7.center);
		\draw[line width=0.225mm] (4.center) to (8.center);
		\draw[line width=0.225mm] (0.5,15.5) to (-0.25, 15.8);
		\draw[line width=0.225mm] (0.5,15.5) to (0.2, 16.25);
		\draw[line width=0.225mm] (0.5,14.5) to (-0.25, 14.2);
		\draw[line width=0.225mm] (0.5,14.5) to (0.2, 13.75);
		\draw[line width=0.225mm] (2.5,15.5) to (3.25, 15.8);
		\draw[line width=0.225mm] (2.5,15.5) to (2.95, 16.25);
		\draw[line width=0.225mm] (2.5,14.5) to (3.25, 14.2);
		\draw[line width=0.225mm] (2.5,14.5) to (2.95, 13.75);
        \draw[line width=0.3mm, densely dashed, blue!50] (2,16.25) to (2, 13.75);
        \draw[line width=0.3mm, densely dashed, orange!75] (1.5,15) to (3.25, 15);
        \draw[line width=0.3mm, densely dashed, orange!75] (1,14.5) to (1, 13.75);
        \draw[line width=0.3mm, densely dashed, orange!75] (1.5,15) arc (90:180:0.5);
\end{tikzpicture} \, \right) \nonumber \\[0.2cm]
    = &\, \mathcal{S}\left( \frac{2 \pi i}{\omega_1} \, \text{Disc}_{x_{3,6}^2} \text{Disc}_{x_{1,5}^2} \ \drawdbten \, \right) \nonumber \\[0.2cm]
    =&\,  \sum_{ikl} c_{ikl} \Biggl[\sum_j \log ({\phi_{ij}}) \otimes \left(2\pi i\,w_{c_j}^+\right) + {\bm\Omega}_{i}\otimes (2 \pi i\,{\tau})   \Biggr] \  \text{Disc}_{x_{1,5}^2} \log({\phi_{k}}) \ \text{Disc}_{x_{3,6}^2}\log({\phi_{l}}) \,,
\end{align}
where in the first drawing one can observe that the channels are indeed partially overlapping. To compute the discontinuities in the logarithms, we can recall the definition of the DCI cross-ratios from eq.~\eqref{eq: DCI_cross-ratios_10pt_double-box}. Then, for example, we have
\begin{equation}
    \text{Disc}_{x_{1,5}^2} \log(u_3) = \text{Disc}_{x_{1,5}^2} \log(\frac{x_{1,3}^2 \, x_{5,10}^2}{x_{1,5}^2 \, x_{3,10}^2}) = - \text{Disc}_{x_{1,5}^2} \log(x_{1,5}^2) =  - 2 \pi i\,,
\end{equation}
and similarly for other cross-ratios.

For the third entries, we have 13 linear combinations of elliptic letters ${\bm\Omega}_{i}$, as well as 53 algebraic letters. In this case, however, there are also mixed algebraic letters, which combine $z$ variables from different 1-loop box diagrams, such as $\log(\frac{(z_{ij}-z_{kl})(\bar{z}_{ij}-\bar{z}_{kl})}{(z_{ij}-\bar{z}_{kl})(\bar{z}_{ij}-z_{kl})})$. Lastly, there can be 7 different last entries, all involving elliptic integrals, given by $\{ 2\pi i \tau, 2 \pi i w^+_{u_5}, 2 \pi i w_{c_1}^+, \dots, 2 \pi i w_{c_5}^+ \}$. The explicit symbol letters can be found in ref.~\cite{Kristensson:2021ani}.

Reorganizing the symbol, eq.~\eqref{eq: 10pt_double-box_symbol_structure} can also be written as~\cite{Kristensson:2021ani}
\begin{equation}
    \mathcal{S}\left( \frac{2 \pi i}{\omega_1} \, \drawdbten \, \right) = \mathcal{S} \left( \, \drawhexten \, \right) \otimes ( 2 \pi i \, w_{u_5}^+) + \dots\,,
\end{equation}
where the remaining terms have last entries which do not depend on $u_5$. Recalling from eq.~\eqref{eq: deriv_symbol} that derivatives act on the last entry, and that $w^+_{u_5} = \frac{1}{\omega_1} \int_{-\infty}^{u_5} \frac{dx}{y}$ for $y=\sqrt{-\Delta_6}\,$, we therefore have
\begin{equation}
\label{eq: 10pt_double-box_diff_eq_hexagon_symbol_level}
    \mathcal{S}\left( \frac{2 \pi i}{\omega_1} \, \partial_{u_5} \, \drawdbten \, \right) = \frac{2 \pi i}{\omega_1 \sqrt{-\Delta_6}} \, \mathcal{S} \left( \, \drawhexten \, \right) \,,
\end{equation}
which makes the differential equation in eq.~\eqref{eq: diff_eq_10pt_double-box_hex} manifest at the level of the symbol. Here, the symbol of the hexagon in $D=6$ is given by the so-called Schläfli's formula~\cite{Spradlin:2011wp},
\begin{equation}
	\label{eq: hexagon symbol}
	\mathcal{S}\left( \, \drawhexten \, \right)=\sum_{i<j}\mathcal{S} \left( \text{Box}_{ij} \right) \otimes \log 
	R_{ij}\,, \qquad \text{with} \qquad R_{ij}=\frac{\mathcal{G}_{j}^{i} \, \sqrt{-\mathcal{G}}+ \mathcal{G} \,\sqrt{\mathcal{G}_{ij}}}{\mathcal{G}_{j}^{i} \, \sqrt{-\mathcal{G}}- \mathcal{G} \,\sqrt{\mathcal{G}_{ij}}}\,,
\end{equation}
where $i<j \, \mathlarger{\mathlarger{\in}} \, \{ 1,3,5,6,8,10\}$. Note that, since the last entries contain two different square roots, they correspond to some of the mixed algebraic letters in the third entry of the double box. In this equation, $\mathcal{G}={(x_{1,5}^2 x_{3,8}^2 x_{6,10}^2)}^2 \Delta_6 \,$ denotes the (non-normalized) hexagon Gram determinant. Then, we use
\begin{equation}
    \mathcal{G}_{j}^{i} \equiv (-1)^{i+j}\det x_{a,b}^{2}
\end{equation}
to denote the Gram determinant where we delete row $i$ and column $j$, i.e.~$a \, \mathlarger{\mathlarger{\in}} \, \{1,3,5,6,8,10\}\backslash \{i\}$ and $b\, \mathlarger{\mathlarger{\in}} \, \{1,3,5,6,8,10\}\backslash \{j\}$. Similarly, for $\mathcal{G}_{ij} \equiv \mathcal{G}_{ij}^{ij}$ we remove both rows and columns $\{ i,j \}$, which corresponds to the Gram determinant of the 1-loop box subsector $\text{Box}_{ij}$ of the hexagon.

\section{One-loop Schubert analysis} 
\label{sec:ch3_one_loop_Schubert}

In the previous section, we introduced the symbol for elliptic Feynman integrals, focusing on the case of the 10-pt double box. Our aim is to base the ansatz for the bootstrap of the 12-pt double box on the structure observed in the 10-pt case, recall eq.~\eqref{eq: 10pt_double-box_symbol_structure}. However, as discussed in sec.~\ref{sec:ch3_intro}, we need a method to predict and generate the new letters that will appear in the 12-pt symbol. To this end, we will employ the so-called Schubert analysis~\cite{Arkani-Hamed:talk,Yang:2022gko,He:2022tph,He:2023umf}. This method is based on solving for the leading singularity of the integral from a geometric point of view in momentum-twistor space~\cite{Hodges:2009hk,Mason:2009qx}, and it will generalize naturally to the 2-loop elliptic integral of our interest. 

Before applying the Schubert analysis to the 12-pt double box in sec.~\ref{sec:ch3_12pt_double_box}, in this section let us first review this recently developed method, and apply it to the simplest case: the 1-loop 4-mass box. Specifically, we will use it to predict its symbol letters, given in eq.~\eqref{eq: 4-mass_box_symbol}. With that aim, in sec.~\ref{sec:ch3_one_loop_box_mom_twistors}, we express the 1-loop 4-mass box integral in momentum-twistor space, and explain how to compute the leading singularity in these variables, which will naturally lead to the notion of Schubert problems~\cite{Schubert:1879,Arkani-Hamed:2010pyv}. Subsequently, in sec.~\ref{sec:ch3_Schubert_box}, we review the Schubert analysis for this integral, and show how to predict the corresponding symbol letters.

\subsection{One-loop box in momentum-twistor space} 
\label{sec:ch3_one_loop_box_mom_twistors}

\begin{figure}[t]
\begin{center}
\subfloat[]{
\includegraphics[height=3.5cm]{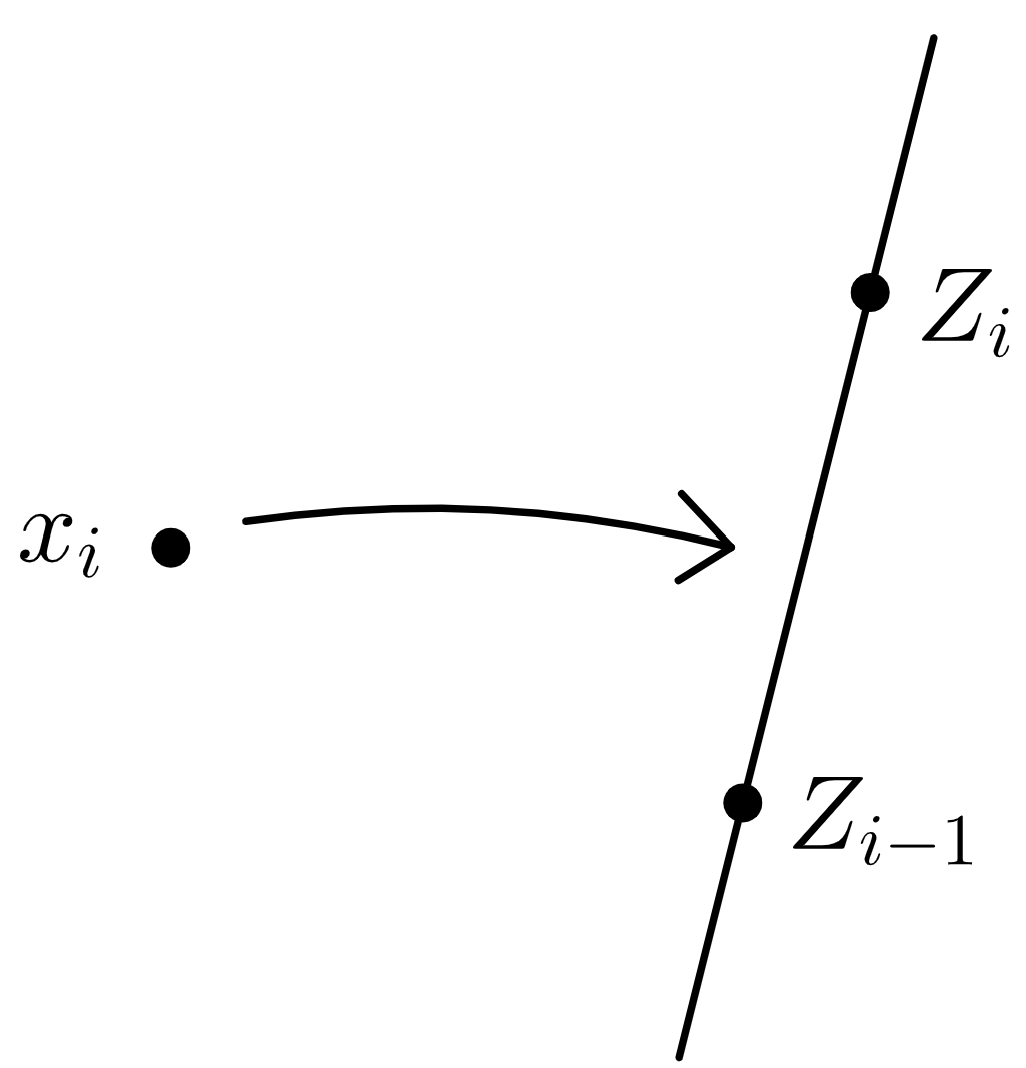}} \qquad \qquad \qquad \qquad \subfloat[]{\includegraphics[height=3.5cm]{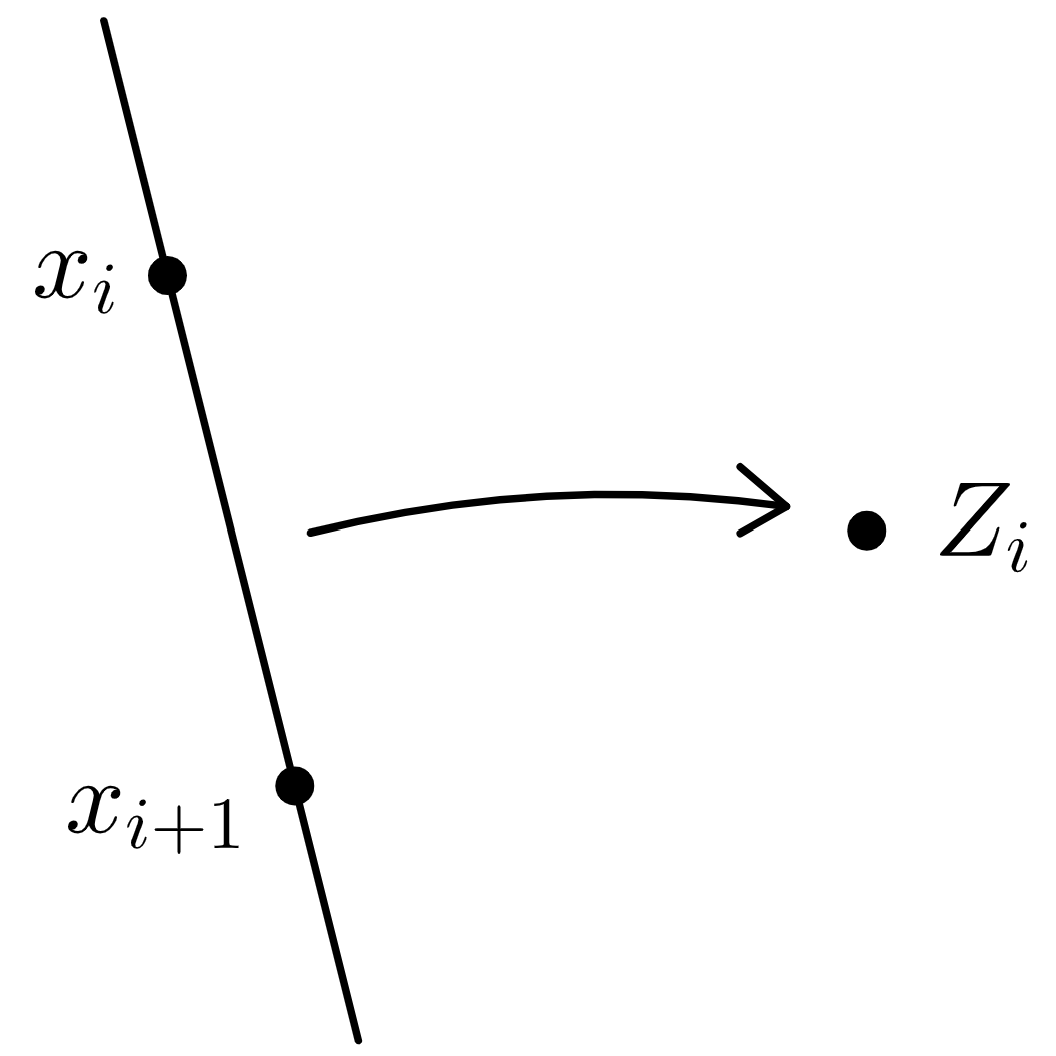}}
\caption{(a) Representation of the map between dual-coordinate space $x_i$ and momentum-twistor space $Z_i$, and (b) vice versa. Figure adapted from ref.~\cite{Elvang:2013cua}.}
\label{fig: momentum_twistors}
\end{center}
\end{figure}

In sec.~\ref{sec:ch2_one_loop_box}, we introduced dual-momentum coordinates, defined via $p_i = x_{i+1}-x_i$ with $x_{n+1}=x_1$ for $n$ external particles, as a way to make momentum conservation manifest, since $p_1 + \dots + p_n = x_{n+1}-x_1=0$. However, dual coordinates do not implement the massless on-shell condition $p_i^2=0$, which has to be enforced as the constraint $(x_{i+1}-x_i)^2=0$ for all external legs $i=1, \dots,n$. To make both conditions manifest simultaneously, we can introduce the momentum-twistor space~\cite{Hodges:2009hk,Mason:2009qx}, which is the twistor space associated with the dual coordinates; see ref.~\cite{Arkani-Hamed:2010pyv} for an introduction. Concretely, in momentum-twistor space each external dual point $x_i$ is associated to a projective line $(Z_{i-1} \ Z_i)$ in $\mathbb{CP}^3$, and vice versa, see fig.~\ref{fig: momentum_twistors}. Similarly, we can associate an unspecified momentum-twistor line $(Z_A \ Z_B)$ to the dual coordinate $x_l$ that is related to the loop momentum. We define the momentum-twistor variables as
\begin{equation}
    Z_i = (\lambda_i^\alpha,x_i^{\alpha \dot{\alpha}} \lambda_{i \alpha})\,,
\end{equation}
where $\lambda_i$, $\tilde{\lambda}_i$ are the spinor-helicity variables, defined via $(p_i)^{\alpha \dot{\alpha}} \equiv p_i^\mu (\sigma_\mu)^{\alpha \dot{\alpha}}=\lambda_i^\alpha \, \tilde{\lambda}_i^{\dot{\alpha}}$, with $\sigma_\mu$ being the Pauli matrices. In terms of momentum-twistor coordinates, the on-shell condition $p_i^2=0$ simply translates to the condition that the lines $(Z_i \ Z_{i+1})$ and $(Z_{i-1} \ Z_i)$ intersect, which is guaranteed since they share a common point $Z_i$.

From the momentum twistors, we can construct the invariant
\begin{equation}
\label{eq: Plucker_coords}
    \langle i\,j\,k\,l \rangle \equiv \det(Z_i, Z_j, Z_k, Z_l) = \epsilon_{ABCD} Z_i^A Z_j^B Z_k^C Z_l^D\,,
\end{equation}
which is also known as a Plücker coordinate, where $\epsilon_{ABCD}$ is the completely antisymmetric tensor. Then, the on-shell condition becomes the trivial statement that $\langle i \, i+1 \, i-1 \, i \rangle = 0$ vanishes. In terms of Plücker coordinates, we can express the squared difference between two dual coordinates as 
\begin{equation}
    x_{i,j}^2 = \frac{\langle i-1 \, i \, j-1 \, j \rangle}{\langle i-1 \, i \rangle \langle j-1 \, j \rangle}\,,
\end{equation}
which also allows us to express the cross-ratios of eq.~\eqref{eq: traintracks_cross-ratios_def} as
\begin{equation}
    (ab;cd) \equiv \frac{x_{a,b}^2 x_{c,d}^2}{x_{a,c}^2 x_{b,d}^2} = \frac{\langle a-1 \, a \, b-1 \, b \rangle \langle c-1 \, c \, d-1 \, d \rangle}{\langle a-1 \, a \, c-1 \, c \rangle \langle b-1 \, b \, d-1 \, d \rangle}\,.
\end{equation}

As an example, let us consider again the 1-loop 4-mass box integral from eq.~\eqref{eq: DCI_box_dual_mom_repr}, and rewrite it in terms of Plücker coordinates,
\begin{equation}
\label{eq: 1loop_4mass-box_Plucker_coords}
    \drawbox \, = \int \frac{d^4 x_l \, x_{1,5}^2 x_{3,7}^2}{x_{l,1}^2 x_{l,3}^2 x_{l,5}^2 x_{l,7}^2} = \int \frac{d^4 Z_A d^4 Z_B}{\text{Vol}\, GL(2)} \frac{\langle 8 \, 1 \, 4 \, 5 \rangle \langle 2 \, 3 \, 6 \, 7 \rangle}{\langle A \, B \, 8 \, 1 \rangle \langle A \, B \, 2 \, 3 \rangle \langle A \, B \, 4 \, 5 \rangle \langle A \, B \, 6 \, 7 \rangle} \,,
\end{equation}
where we integrate over all configurations of momentum twistors $Z_A$ and $Z_B$ up to $GL(2)$ rotations, which leave the projective line invariant. Now, cutting the propagator which lies between the external legs $p_2$ and $p_3$ (see eq.~\eqref{eq: 4mass_box_momentum_repr} for the conventions), and following sec.~\ref{sec:ch1_LS}, we have
\begin{equation}
    x_{l,3}^2 \propto \langle A \, B \, 2 \, 3 \rangle = 0\,.
\end{equation}
Thus, from eq.~\eqref{eq: Plucker_coords}, it implies that the lines $( Z_A \ Z_B )$ and $( Z_2 \ Z_3) $ intersect at some point. Similarly, the maximal-cut condition becomes
\begin{equation}
\label{eq: 1-loop_4mass-box_Plucker_maxcut}
    \langle A \, B \, 8 \, 1 \rangle = \langle A \, B \, 2 \, 3 \rangle = \langle A \, B \, 4 \, 5 \rangle = \langle A \, B \, 6 \, 7 \rangle = 0\,.
\end{equation}

Consequently, we can understand the leading singularity geometrically as the solution for the line $( Z_A \ Z_B )$ which intersects all of the momentum-twistor lines defined by the external dual coordinates; see fig.~\ref{fig: Schubert_problem_4mass-box}. In other words, the leading singularity corresponds to finding the lines in $\mathbb{CP}^3$ which simultaneously intersect four given lines. As it turns out, the number of solutions to this problem was found to be always either two or infinity by the mathematician Hermann Schubert in the 1870's~\cite{Schubert:1879}. Hence, it is commonly referred to as a Schubert problem~\cite{Arkani-Hamed:2010pyv}. In the next subsection, we solve the Schubert problem for the 1-loop 4-mass box integral and, from the solution, introduce a method to predict its symbol letters: the so-called Schubert analysis.

\subsection{Schubert analysis for the one-loop box} 
\label{sec:ch3_Schubert_box}

\begin{figure}[t]
\begin{center}
\includegraphics[height=5.5cm]{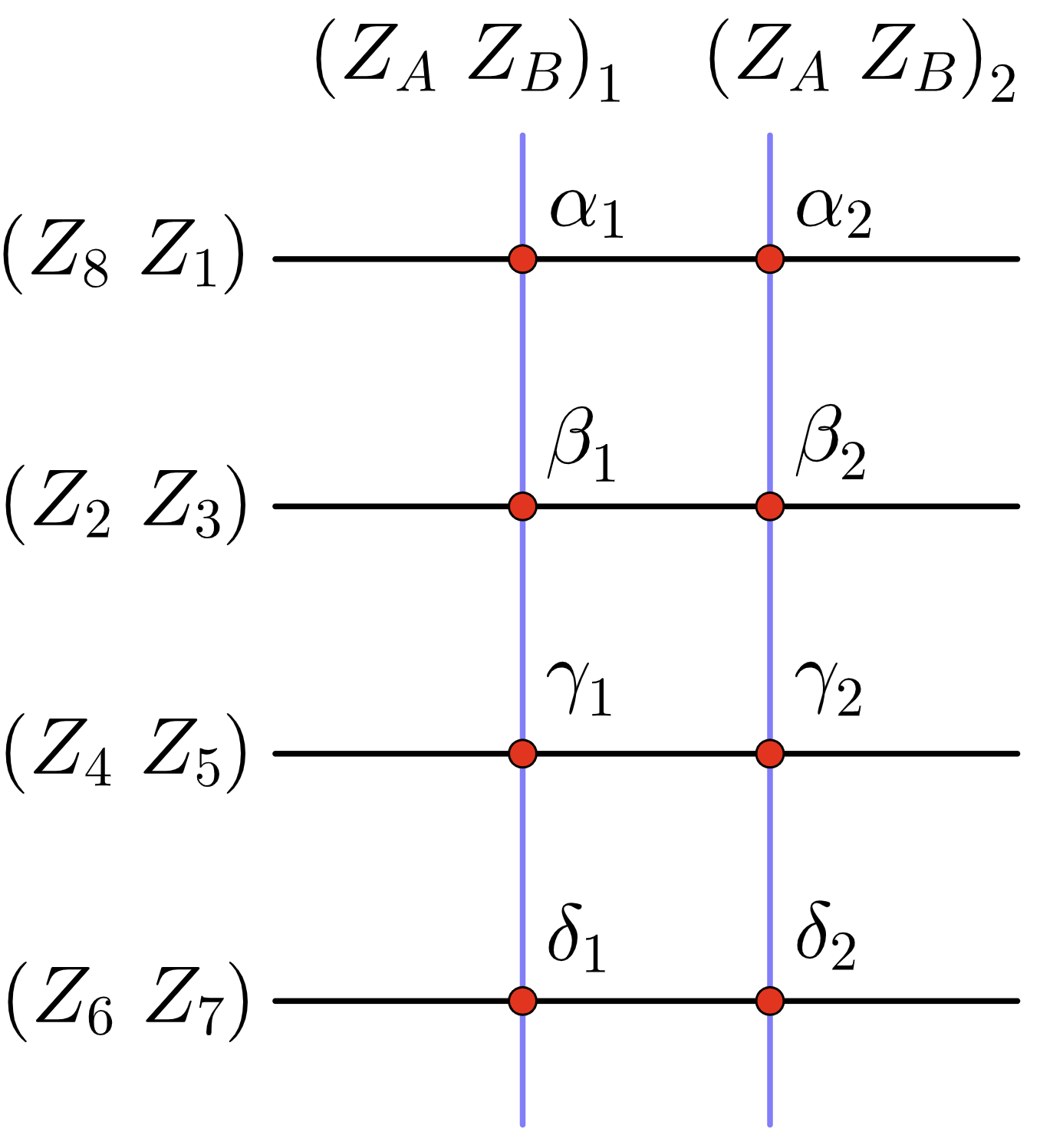}
\caption{Configuration of projective lines in momentum-twistor space as defined by the Schubert problem of the 1-loop 4-mass box integral. Parallel lines indicate that they do not intersect, although in general the lines are skew. Under the maximal cut, there are two solutions ${(Z_A \ Z_B)}_1$ and ${(Z_A \ Z_B)}_2$ (in blue), which simultaneously intersect all of the external lines (in black) at the red dots.}
\label{fig: Schubert_problem_4mass-box}
\end{center}
\end{figure}
In the previous subsection, we introduced the notion of Schubert problems~\cite{Arkani-Hamed:2010pyv} as a method to solve for the leading singularity in momentum-twistor space by finding intersections of projective lines. For the particular case of the 1-loop 4-mass box, we want to solve the maximal-cut condition in eq.~\eqref{eq: 1-loop_4mass-box_Plucker_maxcut}, which is illustrated in fig.~\ref{fig: Schubert_problem_4mass-box}.

In particular, we can parametrize the points $Z_A$ and $Z_B$ with four independent momentum twistors, which can be chosen as~\cite{Arkani-Hamed:2010pyv,Yang:2022gko}
\begin{equation}
    Z_A = \alpha Z_1 + \epsilon Z_2 + Z_8\,, \qquad \text{and} \qquad Z_B = \rho Z_1 + \beta Z_2 + Z_3\,.
\end{equation}
Substituting the expression for $Z_A$ in the condition $\langle A \, B \, 8 \, 1 \rangle = 0$ implies that $\epsilon=0$. Analogously, from $\langle A \, B \, 2 \, 3 \rangle = 0$ we obtain that $\rho = 0$. Then, the remaining conditions $\langle A \, B \, 4 \, 5 \rangle = \langle A \, B \, 6 \, 7 \rangle = 0$ result in quadratic equations for $\alpha$ and $\beta$, such that
\begin{equation}
    \alpha = - \frac{\beta \, \langle 8 \, 2 \, 4 \, 5 \rangle + \langle 8 \, 3 \, 4 \, 5 \rangle}{\beta \, \langle 1 \, 2 \, 4 \, 5 \rangle + \langle 1 \, 3 \, 4 \, 5 \rangle} \,, \qquad \qquad \beta = - \frac{\alpha \, \langle 1 \, 3 \, 6 \, 7 \rangle + \langle 8 \, 3 \, 6 \, 7 \rangle}{\alpha \, \langle 1 \, 2 \, 6 \, 7 \rangle + \langle 8 \, 2 \, 6 \, 7 \rangle}\,.
\end{equation}
From these equations, we obtain two sets of solutions $\{ \alpha_1 , \beta_1\}$ and $\{ \alpha_2 , \beta_2\}$, with explicit expressions that can be found in refs.~\cite{Yang:2022gko,Bourjaily:2013mma}. The $\alpha_i$ parametrize the intersection points between the lines $(Z_A \ Z_B)$ and $(Z_8 \ Z_1)$, whereas the $\beta_i$ yield the intersection points with the line $(Z_2 \ Z_3)$, as illustrated in fig.~\ref{fig: Schubert_problem_4mass-box}. Similarly, we can obtain the intersections $\gamma_i$ with $( Z_4 \ Z_5)$ and $\delta_i$ with $( Z_6 \ Z_7)$, resulting again in two sets of solutions~\cite{Yang:2022gko,Bourjaily:2013mma}. 

Overall, we obtain two different lines ${(Z_A \ Z_B)}_1$ and ${(Z_A \ Z_B)}_2$ which simultaneously intersect the four external lines, as seen in fig.~\ref{fig: Schubert_problem_4mass-box}. These lines represent the two solutions for the leading singularity of the 1-loop 4-mass box, which recall from eq.~\eqref{eq: 1-loop_4mass-box_LS} was given by $\pm \frac{1}{\sqrt{\Delta_4}}$. Indeed, the two solutions correspond to the choice of sign in front of the square root, while the factor of $\frac{1}{\sqrt{\Delta_4}}$ is recovered by taking the residue of eq.~\eqref{eq: 1loop_4mass-box_Plucker_coords} at the solution specified by the lines~\cite{Yang:2022gko}.

The intersection points parametrized by $\{ \alpha_i, \beta_i, \gamma_i, \delta_i \}$ have a definite sign, and their order is always fixed~\cite{Arkani-Hamed:talk,Yang:2022gko}. Hence, as the points lie on a line, we can form 2 multiplicatively independent cross-ratios per line, defined as
\begin{equation}
    \mathcal{U}_i = \frac{(\alpha_i - \beta_i)(\gamma_i - \delta_i)}{(\alpha_i - \gamma_i)(\beta_i - \delta_i)}\,, \qquad \text{and} \qquad \mathcal{V}_i = \frac{(\alpha_i - \delta_i)(\beta_i - \gamma_i)}{(\alpha_i - \gamma_i)(\beta_i - \delta_i)}\,.
\end{equation}
For the 1-loop 4-mass box, these cross-ratios become~\cite{Arkani-Hamed:talk,Yang:2022gko}
\begin{subequations}
\begin{align}
    \mathcal{U}_1 =& \, z\,, \qquad \qquad \mathcal{V}_1=1-z\,, \\
    \mathcal{U}_2 =& \, \bar{z}\,, \qquad \qquad \mathcal{V}_2=1-\bar{z}\,, 
\end{align}
\end{subequations}
where $z$, $\bar{z}$ are defined in eq.~\eqref{eq: zzbar_uv}. Therefore, taking the product $\mathcal{U}_1 \, \mathcal{U}_2=u$, $\mathcal{V}_1 \, \mathcal{V}_2=v$ reveals the arguments of the rational symbol letters which appeared in the first entry of the 1-loop 4-mass box symbol in eq.~\eqref{eq: 4-mass_box_symbol}. Similarly, taking the quotients of the cross-ratios $\mathcal{U}_1 / \mathcal{U}_2=z/\bar{z}$, $\mathcal{V}_1 / \mathcal{V}_2=(1-z)/(1-\bar{z})$ yields the arguments of the algebraic letters which appeared in the second entry of the symbol. Thus, we have been able to predict the symbol letters for this simple integral.

The strategy of constructing cross-ratios from the solutions to the Schubert problem, and subsequently taking their products or quotients in order to generate the arguments of the symbol letters is known in the literature as Schubert analysis~\cite{Arkani-Hamed:talk,Yang:2022gko,He:2022tph,He:2023umf}. Apart from the 1-loop 4-mass box integral, the Schubert analysis has also been successfully used to predict the symbol letters of other polylogarithmic Feynman integrals and amplitudes. For example, in ref.~\cite{Yang:2022gko} it was used to predict the algebraic symbol letters of 8-pt amplitudes up to three loops, as well as the mixed algebraic letters first appearing in the 2-loop 9-pt double box. Similarly, it was used to generate the symbol letters of the 2-loop non-planar double box~\cite{He:2022tph}, and has been extended to massive integrals~\cite{He:2023umf}. In the following section, we show how to use the Schubert analysis for 2-loop Feynman integrals, and introduce its generalization to elliptic cases, with a special emphasis on the 2-loop 10-pt double-box integral.

\section{Two-loop Schubert analysis} 
\label{sec:ch3_two_loop_Schubert}

In the previous section, we introduced Schubert problems as a geometric tool to solve for the leading singularity of a Feynman integral in projective space. Moreover, we introduced a method -- the Schubert analysis -- to predict the corresponding symbol letters from the solution of the Schubert problem, and applied it to the 1-loop 4-mass box. In this section, we describe the generalization of the previous methods to 2-loop integrals, focusing on the case of the 10-pt double box. 

As discussed in sec.~\ref{sec:ch3_intro}, the ansatz for the symbol bootstrap of the 12-pt double box will be greatly inspired by the structure observed for the symbol of the 10-pt case, discussed in sec.~\ref{sec:ch3_symbol_10pt_double_box}. Concretely, in the first two entries of the symbol of the 10-pt double box, we encounter the rational and algebraic symbol letters of the 1-loop 4-mass box. As shown in the previous section, we can generate those with a 1-loop Schubert analysis. By contrast, in the third entries there are mixed algebraic letters, while in the last entries we encounter elliptic letters. Therefore, in sec.~\ref{sec:ch3_Schubert_third_entries} we first describe how to generate the mixed algebraic letters from the third entries using Schubert analysis, and in sec.~\ref{sec:ch3_elliptic_Schubert} we generalize the Schubert analysis for elliptic Feynman integrals, and use it to predict the elliptic last entries.

\subsection{Combining Schubert problems}
\label{sec:ch3_Schubert_third_entries}

The starting point for the Schubert analysis of the 1-loop 4-mass box was to find the projective lines which solved its Schubert problem, and obtain the corresponding intersection points with the external lines, recall sec.~\ref{sec:ch3_Schubert_box} and especially fig.~\ref{fig: Schubert_problem_4mass-box}. Then, by taking products (quotients) of the cross-ratios formed from the intersection points, we generated the arguments of the rational (algebraic) letters of the symbol.

The aim is now to generalize this method to 2-loop integrals. However, as briefly discussed in sec.~\ref{sec:ch3_symbol_10pt_double_box}, the symbol of 2-loop integrals develops mixed algebraic letters in cases with 9 or more external legs~\cite{Kristensson:2021ani,Yang:2022gko}. For instance, for the 10-pt double-box integral there appear letters such as $\log(\frac{(z_{ij}-z_{kl})(\bar{z}_{ij}-\bar{z}_{kl})}{(z_{ij}-\bar{z}_{kl})(\bar{z}_{ij}-z_{kl})})$, which combine $z$ variables from different 1-loop 4-mass box integrals. As shown in ref.~\cite{Yang:2022gko}, these mixed algebraic symbol letters can also be easily predicted from Schubert analysis by combining the Schubert problems of various 1-loop box integrals.

In particular, the 10-pt double box depends on 6 external dual coordinates $x_i$, see eq.~\eqref{eq: 10pt_double-box_Feynman_param} for the conventions. Choosing any set of 4 dual coordinates, we can form $\binom{6}{4}=15$ 1-loop box integrals and, for each of them, solve the respective Schubert problem as in fig.~\ref{fig: Schubert_problem_4mass-box}. Following the previous section, where we described the 1-loop Schubert analysis, the next step would be to construct cross-ratios from the intersection points on the vertical lines. However, as first pointed out in ref.~\cite{Yang:2022gko}, we may also construct cross-ratios from the intersection points on the horizontal lines. This is because the 15 different 1-loop Schubert problems are not completely independent, but they may share some external lines.

\begin{figure}[t]
\begin{center}
\includegraphics[height=6.5cm]{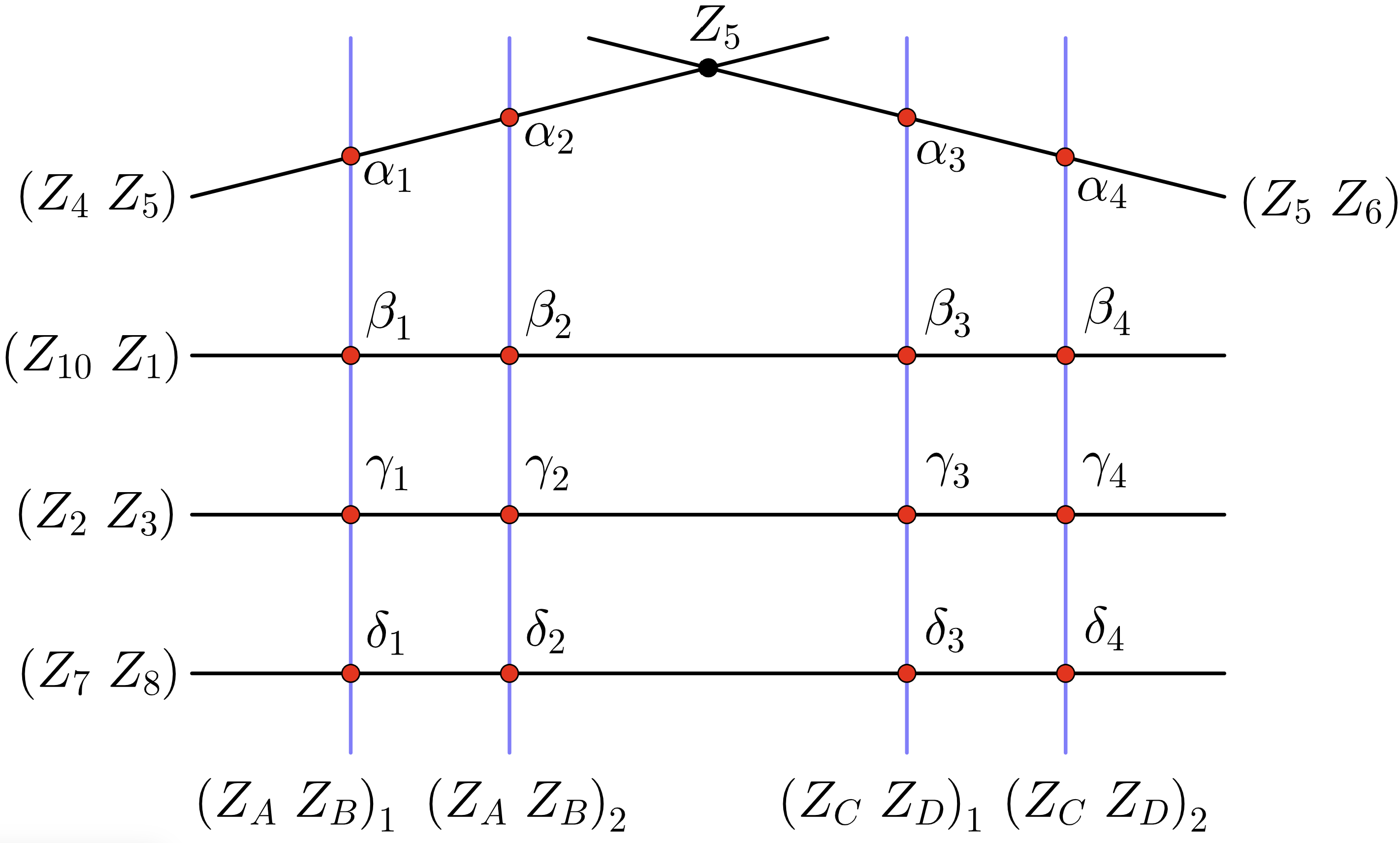}
\caption{Schubert problems for the 1-loop 4-mass box integrals $\{ x_1, x_3, x_5, x_8\}$ and $\{ x_1, x_3, x_6,x_8\}$. For each Schubert problem, there are two lines (in blue), which simultaneously intersect all the respective external lines (in black) at the red dots. Moreover, the integrals have three external lines in common, while the fourth ones intersect at the momentum-twistor coordinate $Z_5$.}
\label{fig: Schubert_problem_mixed_algebraic}
\end{center}
\end{figure}
As an example, let us consider the 1-loop 4-mass box integrals constituted by the dual points $\{ x_1, x_3, x_5, x_8\}$ and $\{ x_1, x_3, x_6,x_8\}$, 
\begin{equation}
\,,
\end{equation}
where the convention for the external legs is the same as in eq.~\eqref{eq: 10pt_double-box_Feynman_param}. The corresponding 1-loop Schubert problems are represented in fig.~\ref{fig: Schubert_problem_mixed_algebraic}, where $(Z_A \ Z_B)_i$ and $(Z_C \ Z_D)_j$ denote the two lines which solve each Schubert problem, respectively. As can be seen, we obtain two copies of the 1-loop Schubert problem of fig.~\ref{fig: Schubert_problem_4mass-box}. However, they also share three of the four external lines, and the remaining lines intersect at the point $Z_5$. Most importantly, the intersection points $\delta_i$, $\gamma_i$ and $\beta_i$ lie on the horizontal lines $(Z_7 \ Z_8)$, $(Z_2 \ Z_3)$ and $(Z_{10} \ Z_1)$, respectively, and have definite sign and are always ordered~\cite{Yang:2022gko}. Consequently, we may also construct two cross-ratios on the horizontal lines, such as
\begin{equation}
\label{eq: cross-ratios_combining_Schubert_problems_box}
    \mathcal{U}_\delta = \frac{(\delta_1 - \delta_2)(\delta_3 - \delta_4)}{(\delta_1 - \delta_3)(\delta_2 - \delta_4)}\,, \qquad \text{and} \qquad \mathcal{V}_\delta = \frac{(\delta_1 - \delta_4)(\delta_2 - \delta_3)}{(\delta_1 - \delta_3)(\delta_2 - \delta_4)}\,,
\end{equation}
and similarly for $\gamma_i$ and $\beta_i$. As they involve the intersection points of different Schubert problems, these cross-ratios naturally mix $z$ variables from the two 1-loop 4-mass box integrals. Concretely, we have that~\cite{Yang:2022gko}
\begin{equation}
    \frac{\mathcal{U}_\delta}{\mathcal{V}_\delta} = \frac{(z_{1358}-z_{1368})(\bar{z}_{1358}-\bar{z}_{1368})}{(z_{1358}-\bar{z}_{1368})(\bar{z}_{1358}-z_{1368})}\,,
\end{equation}
where the subscript indicates the corresponding 1-loop 4-mass box integral, while the product $\mathcal{U}_\delta \, \mathcal{V}_\delta$ becomes rational. The cross-ratios $\{ \mathcal{U}_\gamma, \mathcal{V}_\gamma\}$ and $\{ \mathcal{U}_\beta, \mathcal{V}_\beta\}$ yield the same result. Thus, we are also able to generate mixed algebraic letters by combining Schubert problems. 

\begin{figure}[t]
\begin{center}
\includegraphics[height=6cm]{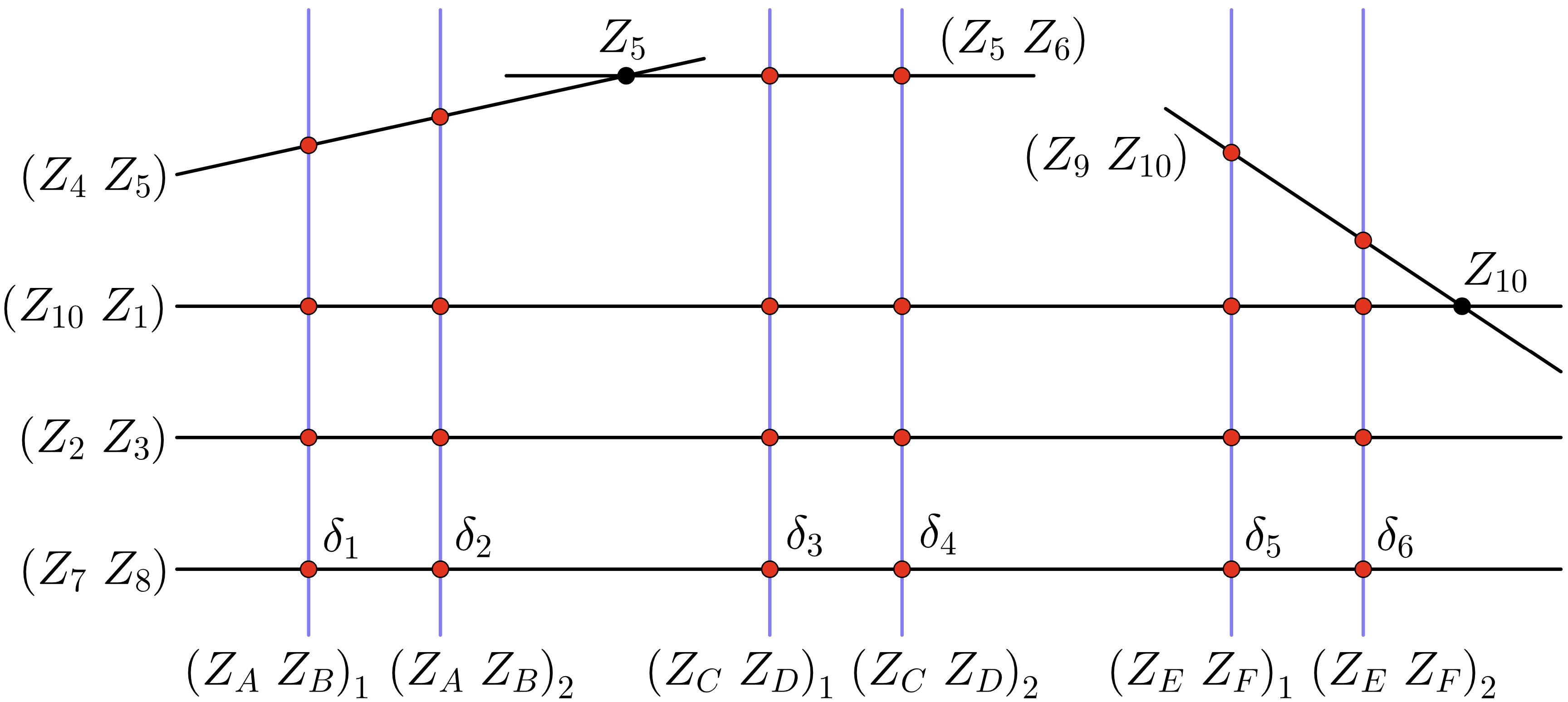}
\caption{Schubert problems for the 1-loop 4-mass box integrals $\{ x_1, x_3, x_5, x_8\}$, $\{ x_1, x_3, x_6,x_8\}$ and $\{ x_1, x_3, x_8,x_{10}\}$, which share three external lines. For each Schubert problem, there are two lines (in blue), which simultaneously intersect all the respective external lines (in black) at the red dots.}
\label{fig: Schubert_problem_hexagon}
\end{center}
\end{figure}

For the particular case of the 10-pt double box, we find that we only need to consider combinations of 1-loop box integrals which share exactly three external lines, as outlined in the previous example. However, in order to generate the mixed algebraic letters of the hexagon symbol (see eq.~\eqref{eq: hexagon symbol}), which appear in the third entries of the double box, we need to consider combinations of three distinct 1-loop Schubert problems which simultaneously share three external lines. For the 10-pt double box, there are $\binom{6}{3}=20$ such configurations, and we illustrate the case for $\{ x_1, x_3, x_5, x_8\}$, $\{ x_1, x_3, x_6,x_8\}$ and $\{ x_1, x_3, x_8,x_{10}\}$ in fig.~\ref{fig: Schubert_problem_hexagon}.

As can be seen, in this case we have 6 intersection points $\delta_i$ which lie on the same projective line, and therefore we can construct more cross-ratios. In general, from $m \geq 4$ points on a line one can define $m(m-3)/2$ multiplicatively independent cross-ratios~\cite{He:2022tph}. In our case, we can define the 9 cross-ratios similarly to eq.~\eqref{eq: cross-ratios_chi_ab} via
\begin{equation}
\mathcal{U}_{ab} = \frac{(\delta_a - \delta_{b-1})(\delta_b - \delta_{a-1})}{(\delta_a - \delta_b)(\delta_{a-1} - \delta_{b-1})}\,,
\end{equation}
where the labels $a$, $b$ are non-adjacent in the cycle $\{ 1, \dots, 6\}$, with $\mathcal{U}_{ba}=\mathcal{U}_{ab}$. Again, the cross-ratios obtained from the intersection points $\gamma_i$ and $\beta_i$ yield the same result. Taking the quotients of these cross-ratios, we finally obtain the mixed algebraic letters that appear in the last entries of the symbol of the hexagon, given in eq.~\eqref{eq: hexagon symbol}. Therefore, together with the 1-loop Schubert analysis of the previous section, we are now able to generate all polylogarithmic symbol letters of the 10-pt double-box integral. In the next subsection, we analyze the remaining elliptic letters, which appear in the last entry of the symbol.

\subsection{Elliptic Schubert analysis} 
\label{sec:ch3_elliptic_Schubert}

In the previous sections, we have described how to predict all polylogarithmic symbol letters of the 10-pt double box using Schubert analysis. Hence, the last step is to generate the elliptic letters, which appear in the last entry of its symbol. According to sec.~\ref{sec:ch3_symbol_10pt_double_box}, these elliptic letters are $\{ 2\pi i \tau, 2 \pi i w^+_{u_5}, 2 \pi i w_{c_1}^+, \dots, 2 \pi i w_{c_5}^+ \}$, where the normalized images on the torus are given by Abel's map, defined in eq.~\eqref{eq: Abel_map_w}.

\begin{figure}[tb]
	\centering
	\includegraphics[height=7cm]{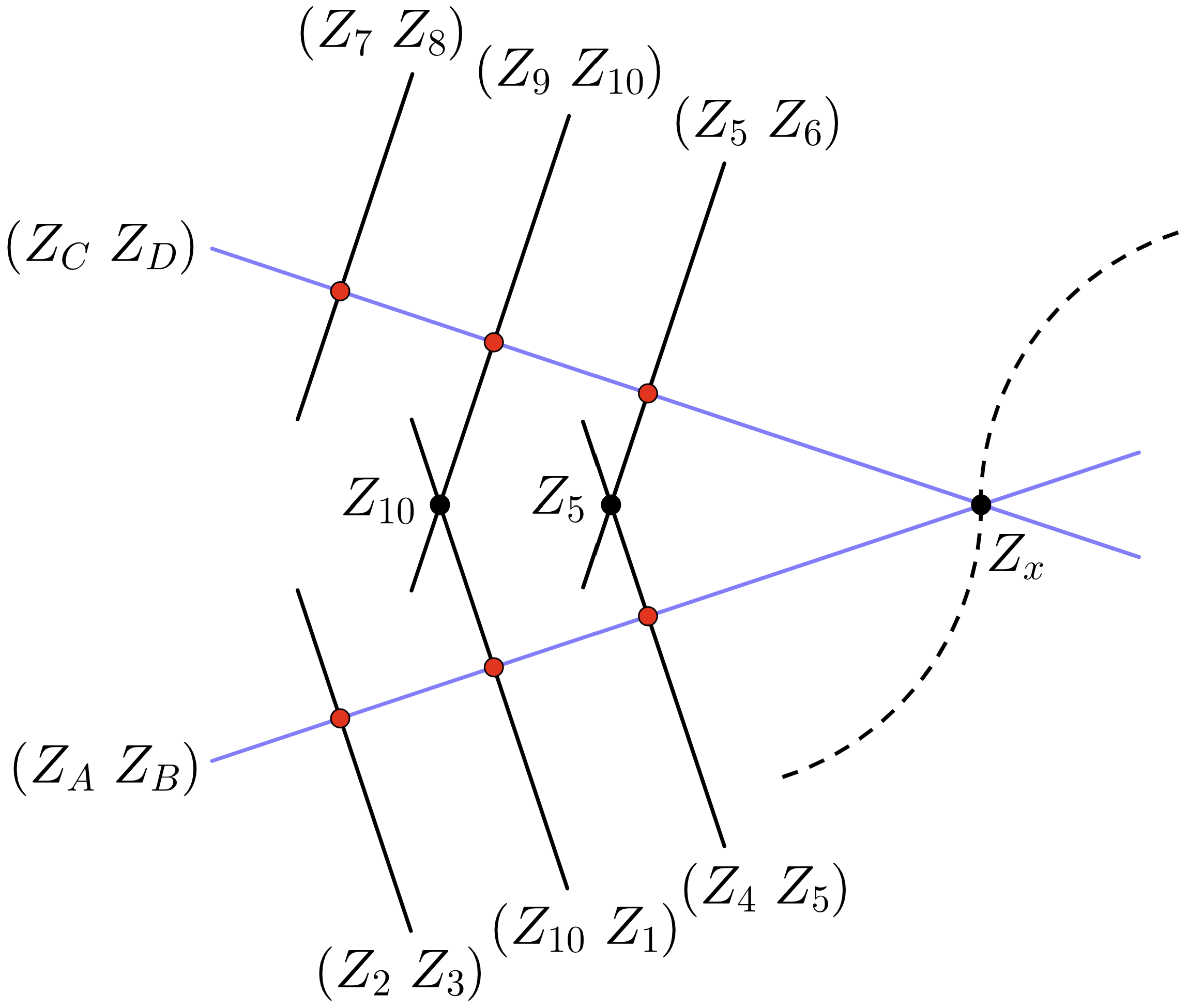}
	\caption{2-loop Schubert problem for the 10-pt double-box integral. There are two lines (in blue) associated with the two loop momenta, which respectively intersect three external lines (in black) at the red dots. The two loop-momentum lines intersect at a point $Z_x$ that lies on an elliptic curve, represented by the dashed curve, which is parametrized by the free parameter $x$. This elliptic curve provides the solution to the 2-loop Schubert problem.}
	\label{fig: Schubert_elliptic}
	\end{figure}
So far, we have considered the Schubert problems of 1-loop 4-mass box integrals, which produce rational and algebraic symbol letters, as well as their combinations, which give rise to mixed algebraic letters. In order to generate the elliptic letters, we must consider instead the Schubert problem of the entire 10-pt double-box integral. As this is a 2-loop integral, we refer to it as a 2-loop Schubert problem. In particular, the maximal cut of the 10-pt double-box (recall the conventions in eq.~\eqref{eq: 10pt_double-box_Feynman_param}) is given in terms of Plücker coordinates by
\begin{equation}
    \langle A \, B \, 10 \, 1 \rangle = \langle A \, B \, 2 \, 3 \rangle = \langle A \, B \, 4 \, 5 \rangle = \langle A \, B \, C \, D \rangle = \langle C \, D \, 5 \, 6 \rangle = \langle C \, D \, 7 \, 8 \rangle = \langle C \, D \, 9 \, 10 \rangle = 0\,,
\end{equation}
where $(Z_A \ Z_B)$ and $(Z_C \ Z_D)$ are the momentum-twistor lines associated to the loop momenta $l_1$ and $l_2$, respectively. As seen from the previous expression, this imposes seven constraints on eight free parameters for the loop momenta. Thus, the solution to the corresponding 2-loop Schubert problem is a curve parametrized by one remaining parameter $x$, see fig.~\ref{fig: Schubert_elliptic}. As it was found in ref.~\cite{Vergu:2020uur}, this curve is given by the intersection of two quadrics, and it defines an elliptic curve given by a quartic polynomial $P_4(x)$. Thus, we can naturally associate to it a normalized one-fold integral
\begin{equation}
\label{eq: Schubert_elliptic_integral}
    \frac{1}{\omega_1} \int_{Z_i}^{Z_f} \frac{dx}{\sqrt{P_4(x)}}\,,
\end{equation}
where the integration limits $Z_i$ and $Z_f$ are so far unspecified. Thus, from the 2-loop Schubert analysis we can recover the ellipticity of the integral.

Now, the task is to predict the elliptic symbol letters. To do so, we can notice that the point $Z_x$ where the two blue lines intersect can move along the elliptic curve, see fig.~\ref{fig: Schubert_elliptic}. In particular, the elliptic curve can intersect some of the external lines (in black), such that $Z_x$ lies in a triple intersection of two blue lines and a black line. In this case, either $(Z_A \ Z_B)$ or $(Z_C \ Z_D)$ solve a 1-loop Schubert problem. For instance, if $Z_x$ lies on the line $(Z_5 \ Z_6)$, then $(Z_A \ Z_B)$ intersects the lines $\{ (Z_{10} \ Z_1),(Z_2 \ Z_3),(Z_4 \ Z_5),(Z_5 \ Z_6)\}$, so it solves the Schubert problem for the 1-loop box $\{x_1,x_3,x_5,x_6 \}$. Then, we claim that this intersection point $Z_x$ gives a possible integration boundary for eq.~\eqref{eq: Schubert_elliptic_integral}. By going through all of the intersections with external lines and combinations of lower and upper bounds in eq.~\eqref{eq: Schubert_elliptic_integral}, we generate a list of candidates $w^+_{c}$ for the last entry of the symbol. Then, the list of candidates can be further reduced via the group law on the elliptic curve, see e.g.~ref.~\cite{Wilhelm:2022wow}, which ultimately yields the set of independent last entries. For instance, integrating from $\alpha_1$ to $\alpha_2$ (recall fig.~\ref{fig: Schubert_problem_4mass-box}) for the previous Schubert problem of the 1-loop box $\{x_1,x_3,x_5,x_6 \}$, we obtain the elliptic symbol letter $w^+_{u_5}$. Notably, for all last entries $w_{c_k}^+$, the $c_k$ turn out to be rational points on the elliptic curve. It would be interesting to explore whether this is a general feature of elliptic Feynman integrals.\footnote{For instance, it is also satisfied by the symbol of the unequal-mass sunrise in $D=2$, see e.g.~ref.~\cite{Wilhelm:2022wow}. The last entries are $\{ 2\pi i \tau, 2\pi i w_0^+ , 2\pi i w^+_{-1} \}$, with $0$ and $-1$ being rational points on the elliptic curve.}

With this extended method, which we may refer to as elliptic Schubert analysis, we are able to successfully predict the last entries of the 10-pt double box.\footnote{Alternatively, we may generate the $c_k$ from eq.~\eqref{eq: 10pt_double-box_diff_eq_hexagon_symbol_level} as the point where the last entries $R_{ij}$ of the hexagon become singular, an approach which is inspired by symbol-level integration~\cite{Caron-Huot:2011dec}.} Together with the 1-loop Schubert analysis from the previous section and the symbol prime~\cite{Wilhelm:2022wow}, which fixes the remaining elliptic symbol letters ${\bm\Omega}_{i}$ in eq.~\eqref{eq: 10pt_double-box_symbol_structure}, we are thus able to generate the entire symbol alphabet of the 10-pt double box. In the next section, we proceed analogously for the 12-pt double box, and generate the corresponding symbol alphabet. Then, assuming that it respects the same structure as eq.~\eqref{eq: 10pt_double-box_symbol_structure}, we bootstrap its symbol.

\section{Twelve-point elliptic double box} 
\label{sec:ch3_12pt_double_box}

In the previous sections, we have introduced the Schubert analysis as a geometric approach to generate the symbol letters of a Feynman integral, and have shown that it has a simple generalization for elliptic cases. In particular, we used it to generate the symbol alphabet of the 10-pt double box, finding agreement with explicit computations~\cite{Kristensson:2021ani}. In this section, we return to the aim of this chapter -- the 12-pt double box -- and use Schubert analysis to predict its symbol alphabet and subsequently bootstrap the result.

The starting point will be to construct the ansatz of the 12-pt symbol following the structure observed in the 10-pt case, see sec.~\ref{sec:ch3_symbol_10pt_double_box}, since we assume that the 12-pt and the 10-pt results are related in the limit where the intermediate legs become massless. Hence, we assume that the symbol of the 12-pt double box respects the same pattern as eq.~\eqref{eq: 10pt_double-box_symbol_structure}, which we repeat here for convenience:
\begin{align}
\label{eq: 12pt_double-box_symbol_structure}
    \mathcal{S}\left( \frac{2 \pi i}{\omega_1} \, \drawdb \, \right) =  \sum_{ikl} c_{ikl}\log({\phi_{k}})\otimes\log({\phi_{l}}) \otimes \! \Biggl[\sum_j \log ({\phi_{ij}}) \otimes \left(2\pi i\,w_{c_j}^+\right) + {\bm\Omega}_{i}\otimes (2 \pi i\,{\tau})   \Biggr] \,.
\end{align}

To predict the corresponding symbol letters, we follow the same prescription for the Schubert analysis as in secs.~\ref{sec:ch3_one_loop_Schubert} and~\ref{sec:ch3_two_loop_Schubert}, but for the particular kinematics of the 12-pt double box. In particular, we denote the external dual coordinates by $\{ x_1, \dots, x_6 \}$, and we obtain the following:
\begin{itemize}
    \item 9 first entries: By construction, we assume that the first-entry condition~\cite{Gaiotto:2011dt} is satisfied; thus, we can have 9 rational letters formed by the DCI cross-ratios of eq.~\eqref{eq: cross-ratios_chi_ab},
    \begin{equation}
        \{ \log \chi_{13},\log \chi_{14},\log \chi_{15},\log \chi_{24},\log \chi_{25},\log \chi_{26},\log \chi_{35},\log \chi_{36},\log \chi_{46} \}\,.
    \end{equation}
    \item 39 second entries: They are obtained from the Schubert analysis of 1-loop 4-mass box integrals, see sec.~\ref{sec:ch3_Schubert_box}. For the 12-pt double box, we have 6 external dual points, so we can form $\binom{6}{4}=15$ of such 4-mass box integrals. By taking products and quotients of the resulting cross-ratios $\mathcal{U}_i$, $\mathcal{V}_i$, we obtain the previous 9 rational letters as well as 30 algebraic letters
    \begin{equation}
        \left\{ \log( \frac{z_{ij}}{\bar{z}_{ij}}), \log( \frac{1-z_{ij}}{1-\bar{z}_{ij}}) \right\},
    \end{equation}
    where $i$, $j$ denote the two dual coordinates from the 12-pt double box which are absent in the respective 4-mass box.
    \item 134 third entries: The polylogarithmic letters are obtained by combining three Schubert problems of 1-loop 4-mass box integrals that share three external lines, see sec.~\ref{sec:ch3_Schubert_third_entries}. For the 12-pt double box we have $\binom{6}{3}=20$ such configurations, each of them giving 9 cross-ratios $\mathcal{U}_{ab}$. From the resulting products and quotients, we obtain 104 multiplicatively independent letters, which appear as the arguments of logarithms. Recalling the notation of sec.~\ref{sec:ch3_symbol_10pt_double_box} for the Gram determinants, we obtain:
    \begin{itemize}
        \item 9 cross-ratios $\chi_{ab}$,
        \item 15 last entries $R_{ij}$ of the 12-pt hexagon,
        \item 5 ratios of pentagon Gram determinants, i.e.~ratios of $\mathcal{G}_{6}/(x_{1,3}^{2}x_{2,4}^{2}x_{3,5}^{2}x_{2,5}^{2}x_{1,4}^{2})$ with its 5 images under the permutations $S_6$ of the external points $x_i$,
        \item 15 letters involving 4-mass box Gram determinants, namely $\mathcal{G}_{56}/(x_{1,3}^{2}x^{2}_{2,4})^{2}$ and its 14 images under $S_6$,
        \item 60 mixed algebraic letters of the form $(\mathcal G_{ij}^{ik}-\sqrt{\mathcal G_{ij}\mathcal G_{ik}})/(\mathcal G_{ij}^{ik}+\sqrt{\mathcal G_{ij}\mathcal G_{ik}})$.
    \end{itemize}
    Lastly, we include the 30 algebraic letters from the second entries. Since we assume the structure of eq.~\eqref{eq: 12pt_double-box_symbol_structure} to hold, the elliptic letters ${\bm\Omega}_{i}$ are fixed by the polylogarithmic parts of the symbol through the symbol prime~\cite{Wilhelm:2022wow}, see also eq.~\eqref{eq:intrep_symbolprime}. Thus, we can neglect them in the following.
    \item 9 last entries: The elliptic letters are obtained from the 2-loop Schubert analysis of the entire 12-pt double-box integral, see sec.~\ref{sec:ch3_elliptic_Schubert}. We obtain $\{ 2 \pi i \tau, 2 \pi i w^+_{\chi_{14}}, 2 \pi i w^+ _0 \}$ and 6 normalized torus images $w^+_{c_k}$, where
    \begin{equation}
    \label{eq: last_entries_ck_12pt_double-box}c_k=\chi_{14}\,\frac{\mathcal{G}_i\,x_{i,k}^4+\mathcal{G}_j\,x_{j,k}^4+2(\mathcal{G}_j^i+\mathcal{G}_{ij}\,x_{i,j}^2)x_{i,k}^2x_{j,k}^2}{2\,\mathcal G_{ij}\,x_{i,j}^2x_{j,k}^2x_{i,k}^2}\,.
    \end{equation}
    Here, $\{i,j,k\}$ are identified with cyclic permutations of either $\{1,2,3\}$ or $\{4,5,6\}$. Moreover, the elliptic curve $y=\sqrt{-\Delta_6}$ corresponds to the Gram determinant of the hexagon, which is a cubic polynomial in $\chi_{14}$, recall eq.~\eqref{eq: diff_eq_12pt_double-box_hex}.
\end{itemize}

With this, we have a complete prediction of the symbol alphabet for the 12-pt double box. The next step is to construct the ansatz following eq.~\eqref{eq: 12pt_double-box_symbol_structure}, and to bootstrap the final result; that is, to uniquely fix more than 400,000 unknown coefficients. In the following subsection, we describe how to constrain these coefficients, showing that it is only necessary to use the integrability condition for the symbol, recall eq.~\eqref{eq: integrability_condition}. Then, in sec.~\ref{sec:ch3_symbol_12pt_double_box}, we present the main result of the chapter: the symbol of the 12-pt double box, along with the consistency checks that it satisfies.

\subsection{Elliptic symbol bootstrap} 
\label{sec:ch3_elliptic_symbol_bootstrap}

Having constructed a complete ansatz for the symbol of the 12-pt double box, the first step in the bootstrap is to lay out the constraints for the symbol, which hopefully restrict all of the free coefficients in the ansatz. As explained in sec.~\ref{sec:ch1_MPLs_symbol}, one of the broadest consistency requirements for a symbol is that it can be associated to a well-defined function. This, in turn, implies that partial derivatives commute at all depths in the symbol, leading to the so-called integrability conditions~\cite{Chen:1977oja}, see eq.~\eqref{eq: integrability_condition}. Since we will mainly base the analysis of this section on these conditions, let us rewrite them here for convenience. For a generic symbol of length $n$
\begin{equation}
    \mathcal{S}(F^{(n)})= \sum_{i_1, \dots, i_n} c_{i_1 \cdots\, i_n} \, S_{i_1} \otimes \cdots \otimes S_{i_n}\,,
\end{equation}
the integrability conditions read
\begin{equation}
\label{eq: integrability_symbol_v2}
0 = \sum_{i_1,\dots,i_n} c_{i_1 \cdots\, i_n} S_{i_1} \otimes \dots \otimes S_{i_{j-1}} \otimes S_{i_{j+2}} \otimes \dots \otimes S_{i_n} \, \Bigg( \frac{\partial S_{i_j}}{\partial x_k} \, \frac{\partial S_{i_{j+1}}}{\partial x_m}-\frac{\partial S_{i_j}}{\partial x_m} \, \frac{\partial S_{i_{j+1}}}{\partial x_k} \Bigg)\,,
\end{equation}
for all $1\leq j \leq n-1$. Thus, these conditions pose $n-1$ constraints for each pair $\{ x_k, x_m \}$ of the independent kinematic variables $\vec{x}$ (not to be confused with the dual points $x_i$).

For our ansatz in eq.~\eqref{eq: 12pt_double-box_symbol_structure}, which is of length $n=4$, some of the integrability conditions can be straightforwardly computed. This is the case for partial derivatives between the first and second entries ($j=1$) and between the second and third entries ($j=2$). As these entries only contain logarithmic letters depending on the 9 DCI cross-ratios $\chi_{ab}$ (recall eq.~\eqref{eq: cross-ratios_chi_ab_all_possibilities}), we can just use $x_k,\,x_m\, \mathlarger{\mathlarger{\in}}\, \{ \chi_{ab} \}$ to compute the derivatives. This is no longer the case for the fourth entries, where elliptic symbol letters appear.

For the last entries, we can have two types of elliptic symbol letters: $2\pi i \tau$ and $2 \pi i w^+_c$, as detailed in the beginning of the section. To simplify the action of the derivatives in eq.~\eqref{eq: integrability_symbol_v2} for $j=3$, we can instead choose to work directly with the images on the torus, $x_k,\,x_m\, \mathlarger{\mathlarger{\in}}\, \{ w^+_{c_1}, \dots, w^+_{c_8},\tau \}$. With this choice, the derivatives act trivially on the last entries, since $\partial w^+_{c_i} /\partial w^+_{c_j} = \delta_{ij}$ and $\partial w^+_{c_i} /\partial \tau = 0$. As shown in ref.~\cite{Wilhelm:2022wow}, the integrability condition for the terms with last entry $2 \pi i \tau$ follows simply from the structure of the ansatz in eq.~\eqref{eq: 12pt_double-box_symbol_structure} together with the symbol prime. Thus, we only need to focus on the last entries being $2 \pi i w^+_c$. For those, the corresponding third entries are logarithms $\log(\phi(\chi_{ab}))$, for some rational or algebraic function $\phi(\chi_{ab})$ depending on the kinematics. Hence, in eq.~\eqref{eq: integrability_symbol_v2} with $j=3$ we need to compute derivatives such as
\begin{equation}
    \frac{\partial \log(\phi(\chi_{ab}))}{\partial w^+_c} = \sum_{\chi} \, \frac{\partial \chi}{\partial w^+_c} \frac{\partial \log(\phi(\chi_{ab}))}{\partial \chi}\,, \qquad \text{and} \qquad \frac{\partial \log(\phi(\chi_{ab}))}{\partial \tau} = \sum_{\chi} \, \frac{\partial \chi}{\partial \tau} \frac{\partial \log(\phi(\chi_{ab}))}{\partial \chi}\,.
\end{equation}
In particular, the difficulty resides in computing the derivatives of the cross-ratios with respect to the torus images $w^+_c$ and the modular parameter $\tau$. In this case, we found it simpler to compute instead the elements of the Jacobian $J=\frac{\partial (w_{c_1}, \dots, w_{c_8}, \tau)}{\partial (\chi_{a_1 b_1}, \dots, \chi_{a_9 b_9})}$, and afterwards invert the matrix to compute the original derivatives. As shown in appendix~\ref{ap:elliptic_integrability}, we have
\begin{align}
\label{eq: integrability_conditions_elliptic_letters_12pt_double-box_w}
\frac{\partial w_c^+}{\partial \chi_{ab}} = & \, \frac{1}{\omega_1 \bar{y}_c} \frac{\partial \bar{c}}{\partial \chi_{ab}} + \frac{1}{\omega_1} \sum_{k=1}^3 \bigg( \frac{\bar{y}_c}{\bar{r}_k - \bar{c}} - \frac{2}{\omega_1} g^{(1)}(w_{\bar{c}}^+) \bigg) \frac{1}{\prod_{j \neq k} (\bar{r}_k - \bar{r}_j)} \frac{\partial \bar{r}_k}{\partial \chi_{ab}}\,, \\[0.2cm]
\frac{\partial \tau}{\partial \chi_{ab}} = & \, \frac{4 \pi i}{\omega_1^2} \sum_{k=1}^3 \frac{1}{\prod_{j \neq k} (\bar{r}_k - \bar{r}_j)} \frac{\partial \bar{r}_k}{\partial \chi_{ab}}\,,
\label{eq: integrability_conditions_elliptic_letters_12pt_double-box_tau}
\end{align}
where the overline is a short-hand notation for $\bar{x} \equiv a_3 x$, with $a_3$ being the coefficient of the cubic term in the elliptic curve. In addition, $y_c$ are the images on the cubic elliptic curve $y=\sqrt{-\Delta_6}\,$, $r_i$ are its roots, and $g^{(1)}(w)$ is one of the kernels for eMPLs on the torus, see eq.~\eqref{eq: def_Gamma_tilde}. In fact, as explained in appendix~\ref{ap:elliptic_integrability}, the terms with $g^{(1)}(w)$ totally drop when inverting to $J^{-1}$. Thus, they also drop in the integrability conditions for elliptic symbol letters, such that the derivatives $\partial \chi_{ab}/\partial w^+_c$ and $\partial \chi_{ab}/\partial \tau$ only depend on the period $\omega_1$, the elliptic curve and its roots.

Introducing these expressions into the integrability conditions of eq.~\eqref{eq: integrability_symbol_v2}, we can finally compute all derivatives analytically, and solve for the unknown coefficients of the ansatz. In practice, however, the integrability conditions are solved numerically with a 300-digit precision at random numerical points for the kinematics using the \Mathematica~commands \texttt{RowReduce} and \texttt{NullSpace}. This is only due to the fact that there are 9 kinematic variables and numerous square roots in this problem, which limit the efficiency of an analytic computation.\footnote{For example, for the unequal-mass sunrise (see ref.~\cite{Wilhelm:2022wow} for our conventions) we can verify the integrability conditions for the last entries analytically, as they only involve 3 kinematic variables and 2 different square roots.}

\begin{table}
\begin{center}
\caption{Number of remaining free parameters in the ansatz of eq.~\eqref{eq: 12pt_double-box_symbol_structure} after imposing the constraints from the integrability conditions, see eq.~\eqref{eq: integrability_symbol_v2}, and the differential equation of eq.~\eqref{eq: diff_eq_12pt_double-box_hex} relating the double box to the hexagon. 
}
\label{tab: bootstrap_12pt}
\begin{tabular}{|l|l|}
\Xhline{0.2ex}
Constraint						& Free parameters\\ 
\hline\vspace{-10pt}
Ansatz					& 9 $\times$ 39 $\times$ 134 $\times$ 8\\[-0.2cm]
								& $\underbrace{\hspace{30pt}}$\\\vspace{-10pt}
Integrability in entries 1 \& 2  	& \hspace{10pt}60\hspace{13pt}$\times$ 134 $\times$ 8 \\[-0.2cm] 
								& \hspace{45pt}$\underbrace{\hspace{34pt}}$\\\vspace{-10pt} 
Integrability in entries 3 \& 4  	& \hspace{10pt}60\hspace{13pt}$\times$\hspace{16pt}19\\[-0.2cm]
								& \hspace{10pt}$\underbrace{\hspace{56pt}}$\\
Integrability in entries 2 \& 3  	& \hspace{35pt}1 \\[0.2cm]
Differential equation hexagon	& \hspace{35pt}0 \\
\Xhline{0.2ex}
\end{tabular}
\end{center}
\end{table}
The results of solving the integrability conditions at each symbol entry are gathered in tab.~\ref{tab: bootstrap_12pt}. The 36 integrability conditions between entries 1 and 2 were simple to solve since the first entries are rational, and the solution is found within 10s with 10 random numerical points. To reduce the number of free coefficients, we can then turn to the 28 integrability conditions between entries 3 and 4, which utilize the expressions from eqs.~\eqref{eq: integrability_conditions_elliptic_letters_12pt_double-box_w}--\eqref{eq: integrability_conditions_elliptic_letters_12pt_double-box_tau}. In this case, we used 40 kinematic points, and it takes $\sim \! 12$min to evaluate the integrability conditions, and another $\sim \! 8$min to find the solution. Finally, the 36 integrability conditions for the middle pair of entries requires just 2 kinematic points, and takes $\sim \! 25$min to evaluate and $\sim \! 35$min to solve.

Overall, demanding integrability for the symbol we are able to reduce the ansatz to a single unknown coefficient, which sits as an overall prefactor and is fixed e.g.~with the differential equation of eq.~\eqref{eq: diff_eq_12pt_double-box_hex}. Thus, we have bootstrapped the symbol of the 12-pt double box! While, at first glance, it might be surprising that integrability alone is capable of fixing the ansatz up to an overall scale, we have a total of 100 conditions to be satisfied. Moreover, most of these conditions depend on different square roots, which analytically cancel separately, thus imposing further constraints. In the next subsection, we present the result for the symbol, along with the numerous consistency checks that it satisfies.

\subsection{Symbol of the twelve-point elliptic double box} 
\label{sec:ch3_symbol_12pt_double_box}

As described in the previous subsection, the integrability conditions are sufficient to uniquely fix all of the 400,000 unknown coefficients in the ansatz of the 12-pt double-box symbol, up to an overall constant. Fixing the constant with the differential equation relating the double-box integral to the 1-loop hexagon in $D=6$, see eq.~\eqref{eq: diff_eq_12pt_double-box_hex}, we thus bootstrap the entire symbol. The result contains $\sim \! 1300$ terms and respects the structure in eq.~\eqref{eq: 12pt_double-box_symbol_structure} by construction. In particular, it can be written in terms of 100 logarithmic letters and 9 different elliptic last entries, and is given by
\begin{align}
\mathcal{S}\bigg(\frac{2\pi i}{\omega_1} \, \drawdb \, \bigg)= & \  \mathcal{S} \bigg(\, \drawhex \, \bigg)\otimes(2\pi i\, w_{\chi_{14}}^+)+\text{F}_\tau\otimes(2\pi i\,\tau)
+\frac{1}{2} \sum_{\substack{i\in\{1,2,3\}\\j\in\{4,5,6\}}} \hspace{-4pt} \text{V}_{ij}\otimes(2\pi i\,w_0^+) \nonumber \\
& \ +\frac{1}{2} \sum_{k\in\{1,...,6\}} \hspace{-4pt} \text{W}_{k}\otimes(2\pi i\,w_{c_{k}}^+)\,,
\label{eq: symbol_12pt_double-box_final}
\end{align}
where the symbol of the hexagon is given in eq.~\eqref{eq: hexagon symbol}, the last entries $w_{c_{k}}^+$ are provided in eq.~\eqref{eq: last_entries_ck_12pt_double-box}, and the length-3 symbol $\text{F}_\tau$ is fixed from the rest of the symbol following eqs.~\eqref{eq:intrep_symbolprime} and~\eqref{eq: 12pt_double-box_symbol_structure}. We also introduce the following definition for the length-3 symbol $\text{V}_{ij}$,
\begin{equation}
\label{eq: symbol_12pt_double-box_Vij}
\text{V}_{ij} \equiv (-1)^{i+j}\bigg( \mathcal{S}(\text{Box}_{ij})\otimes\log \bigg( \frac{z_{ij}^2}{{\bar z}_{ij}^2}\frac{1-\bar{z}_{ij}}{1-z_{ij}} \bigg)-\text{U}_{ij}\otimes \log (v_{ij}) \bigg)\,,
\end{equation}
where the 4-mass box symbol and its variables are defined in eqs.~\eqref{eq: symbol_Box_ij} and~\eqref{eq: variables_Box_ij}, respectively, and the length-2 symbol $\text{U}_{ij}$ is defined as
\begin{align}
\text{U}_{ij} \equiv
- \, \mathcal S\left( \log(u_{ij})\log\left( \frac{v_{ij}}{u_{ij}}\right) \right)\,.
\end{align}
In the last term of eq.~\eqref{eq: symbol_12pt_double-box_final}, we also introduce the length-3 symbol
\begin{align}
\label{eq: symbol_12pt_double-box_Wk}
\text{W}_k = & \, \mathcal{S}(\text{Box}_{ij})\otimes\log\left(\frac{\mathcal G_j\,x_{j,k}^4}{\mathcal G_i\,x_{i,k}^4} \right)+(-1)^{i-j}\hspace{-5pt}\sum_{\substack{l\in\{i,j\}\\ m\not\in\{i,j\}}}\hspace{-4pt}\text{sgn}(m-l) \, \mathcal{S}(\text{Box}_{lm})\otimes\log\left(\frac{\mathcal G_{lm}^{ij}-\sqrt{\mathcal G_{ij}\mathcal G_{lm}}}{\mathcal G_{lm}^{ij}+\sqrt{\mathcal G_{ij}\mathcal G_{lm}}} \right)\nonumber\\
& \, +(-1)^{i+j+k}\hspace{-5pt}\sum_{l\not\in\{1,6,i,j,k\}}\hspace{-5pt}(-1)^l \, \bigr(\text{U}_{il}-\text{U}_{jl}\bigl) \, \otimes \, \log\left(\frac{1-z_{ij}}{1-\bar z_{ij}} \right) \nonumber \\[0.2cm]
&\, -(-1)^{i+j+k}\hspace{-5pt}\sum_{l\not\in\{3,4,i,j,k\}}\hspace{-5pt}(-1)^l\,\bigr(\text{U}_{il}-\text{U}_{jl}\bigl) \, \otimes \, \log\frac{z_{ij}}{\bar z_{ij}}\,,
\end{align}
where the indices $\{i,j,k\}$ are identified with cyclic permutations of either $\{1,2,3\}$ or $\{4,5,6\}$.

To corroborate that the result obtained from the elliptic symbol bootstrap is correct, we verified that several consistency conditions are satisfied. In particular:
\begin{enumerate}
    \item The first-entry conditions~\cite{Gaiotto:2011dt} are satisfied by construction, since the first entries are the DCI cross-ratios $\chi_{ab}$.
    \item The Steinmann conditions~\cite{Steinmann:1960,Steinmann2:1960}, which state that discontinuities in partially overlapping factorization channels vanish (see an example in eq.~\eqref{eq: example_Steinmann}), are satisfied by construction in the first and second entries. This is because the second entries contain algebraic letters of 4-mass box integrals.
    \item The so-called extended Steinmann conditions~\cite{Caron-Huot:2019bsq}, which extend the Steinmann conditions beyond the first pair of entries in the symbol, are also satisfied in all logarithmic entries.
    \item The differential equation relating the 12-pt double box to the 12-pt hexagon (see eq.~\eqref{eq: diff_eq_12pt_double-box_hex}) is manifest at symbol level, analogously to eq.~\eqref{eq: 10pt_double-box_diff_eq_hexagon_symbol_level} for the 10-pt double-box symbol.
    \item In the limit where the pair of intermediate legs $\{p_5,p_6\}$ and $\{p_{11},p_{12}\}$ become massless, it reproduces the symbol of the 10-pt double box~\cite{Kristensson:2021ani}, recall sec.~\ref{sec:ch3_symbol_10pt_double_box}.
    \item The symbol respects the $\mathbb{Z}_2 \times S_3$ symmetry of the double-box integral,\footnote{In eq.~\eqref{eq: 12pt_double-box_dual_coord_param}, this symmetry is broken into $\mathbb{Z}_2 \times \mathbb{Z}_2$ due to the numerator, but is recovered in the symbol with the $\omega_1$ normalization.} which is invariant under a $\mathbb{Z}_2$ vertical mirror $x_i \to x_{7-i}$, and a permutation $S_3$ of the dual coordinates $\{ x_1,x_2,x_3\}$. At the symbol level, the $\mathbb{Z}_2$ symmetry acts on the last entries as $w^+_{\chi_{14}} \to w^+_{\chi_{14}}$, $w^+_0 \to w^+_0$ and $w^+_{c_k} \to -w^+_{c_{7-k}}$, which is reflected in the preceding length-3 entries since $\text{W}_k \to -\text{W}_{7-k}$ while the rest are invariant. Similarly, for example, the permutation $x_5 \leftrightarrow x_6$ acts on the last entries as $w^+_0 \to - w^+_0$, $w^+_{c_4} \to - w^+_{c_4}$ and $w^+_{c_5} \leftrightarrow - w^+_{c_6}$, while the remaining are invariant. The term $\text{F}_\tau\otimes(2\pi i\,\tau)$ also inherits these symmetries via the symbol prime.
    \item The symbol satisfies the pair of second-order differential equations~\cite{Drummond:2006rz,Drummond:2010cz} of sec.~\ref{sec:ch2_diff_eqs}, see eqs.~\eqref{eq: diff_equation_ladders_DL} and~\eqref{eq: diff_equation_ladders_DR}. In particular, acting with the Laplacian operator with respect to the dual coordinates $x_2$ and $x_5$, it reduces to the symbol of 1-loop 4-mass box integrals.
    \item Lastly, the symbol satisfies the conformal Ward identity~\cite{Chicherin:2017bxc}, which states that the 12-pt double-box integral is invariant under conformal boost transformations,
    \begin{equation}
        K_\mu\  \drawdb \, = 0\,,
    \end{equation}
    where the operator is given in dual-momentum space by
    \begin{equation}
        K_\mu = i \, \sum_{i=1}^6 \left( x_i^2 \, \frac{\partial}{\partial x_i^\mu} - 2 x_{i \hspace{1pt} \mu} \, x_i^\nu \, \frac{\partial}{\partial x_i^\nu} - 2 x_{i \hspace{1pt} \mu} \right)\,.
    \end{equation}
    To compute the derivatives, we can use the chain rule as in eq.~\eqref{eq: chain_rule_dual_coords_Laplacian}, and employ the expressions found in eqs.~\eqref{eq: integrability_conditions_elliptic_letters_12pt_double-box_w} and~\eqref{eq: integrability_conditions_elliptic_letters_12pt_double-box_tau} for the derivatives of the elliptic letters.
    \end{enumerate}

Importantly, we can reorganize the symbol in eq.~\eqref{eq: symbol_12pt_double-box_final} much more compactly. For that, we can first realize that certain terms in the symbol are related under the limit $\chi_{14} \to \infty$ with all other cross-ratios $\chi_{ab}$ fixed. For instance, the two terms within $\text{V}_{ij}$, see eq.~\eqref{eq: symbol_12pt_double-box_Vij}, are related to one another, since 
\begin{equation}
\label{eq: x14_to_infty_eq1}
\text{U}_{ij} \equiv \lim_{\chi_{14} \to \infty} \mathcal{S}(\text{Box}_{ij})\,, \qquad \text{and} \qquad \log(v_{ij}) = \lim_{\chi_{14} \to \infty} \log \bigg( \frac{z_{ij}^2}{{\bar z}_{ij}^2}\frac{1-\bar{z}_{ij}}{1-z_{ij}} \bigg)\,.
\end{equation}
As an example, let us consider the case $i=3$ and $j=6$. From eq.~\eqref{eq: variables_Box_ij},
\begin{equation}
    u_{36}=(12;45)=\chi_{13} \chi_{14} \chi_{36}\chi_{46} \equiv u\,, \qquad \text{and} \qquad v_{36} = (15;42) = \chi_{25} \equiv v\,.
\end{equation}
Thus, the limit $\chi_{14} \to \infty$ corresponds to $u \to \infty$ while $v$ stays constant. Hence, we obtain
\begin{subequations}
\begin{align}
    \lim_{\chi_{14} \to \infty} z_{36} = &\ \lim_{u \to \infty} \frac{1}{2} \left( 1+u-v + \sqrt{(1-u-v)^2-4uv} \right) = u-v+ \mathcal{O}(u^{-1})\,, \\
    \lim_{\chi_{14} \to \infty} \bar{z}_{36} = &\ \lim_{u \to \infty} \frac{1}{2} \left( 1+u-v - \sqrt{(1-u-v)^2-4uv} \right) = 1+\frac{v}{u}+ \mathcal{O}(u^{-2})\,.
\end{align}
\end{subequations}
and therefore
\begin{equation}
    \lim_{\chi_{14} \to \infty} \log \bigg( \frac{z_{36}^2}{{\bar z}_{36}^2}\frac{1-\bar{z}_{36}}{1-z_{36}} \bigg) = \log v\,, \qquad
    \lim_{\chi_{14} \to \infty} \log \left( \frac{1-z_{36}}{1-\bar{z}_{36}} \right) = \log \left( \frac{u^2}{v} \right)\,, \qquad
    \lim_{\chi_{14} \to \infty} \log \left( \frac{z_{36}}{\bar{z}_{36}} \right) = \log u\,.
\end{equation}
Thus, the conditions in eq.~\eqref{eq: x14_to_infty_eq1} are satisfied upon rewriting
\begin{equation}
    \text{U}_{36} \equiv - \, \mathcal S\left( \log(u_{36})\log\left( \frac{v_{36}}{u_{36}}\right) \right) = - \log v \, \otimes \, \log u + \log u \, \otimes \, \log \left( \frac{u^2}{v} \right) = \lim_{\chi_{14} \to \infty} \mathcal{S} (\text{Box}_{36}) \,.
\end{equation}
Similarly, the first term in eq.~\eqref{eq: symbol_12pt_double-box_Wk} for $\text{W}_k$ vanishes under $\chi_{14} \to \infty$, as some of the entries become $\log(1)=0$. The same applies to the second term when $m=k$. By contrast, for $m \neq k$ the second term yields the second and third lines of eq.~\eqref{eq: symbol_12pt_double-box_Wk}, up to a sign.

Ultimately, this allows us to rewrite the symbol of the 12-pt double box in eq.~\eqref{eq: symbol_12pt_double-box_final} as
\begin{equation} 
\label{eq: symbol_12pt_double-box_compact}
\vspace{0.1cm}
\mathcal{S} \bigg(\frac{2\pi i}{\omega_1} \, \drawdb \, \bigg) = \sum_{i<j} \ \mathcal{S}(\text{Box}_{ij}) \otimes \mathcal{S} \left( \frac{2\pi i}{\omega_{1}} \mathlarger{\mathlarger{\int}}_{\beta_{ij}}^{\chi_{14}}\frac{d\chi_{14}'\log R_{ij}(\chi_{14}') }{\sqrt{-\Delta_{6}(\chi_{14}')}} \right) 
 - (\chi_{14}{\to} \infty)\,,
 \vspace{0.1cm}
\end{equation}
where $i<j \, \mathlarger{\mathlarger{\in}} \, \{ 1,\dots,6\}$. Here, $\beta_{ij}=\chi_{14}(1+\sqrt{v_{ij}})^{2}/u_{ij}$ for $i \, \mathlarger{\mathlarger{\in}} \, \{1,2,3\}$ and $j \, \mathlarger{\mathlarger{\in}} \, \{4,5,6\}$, which is a root of the 4-mass box Gram determinant $\Delta_4$ with respect to $\chi_{14}$, and $\beta_{ij} = -\infty$ otherwise. As can be seen, the first term manifests the differential equation from eq.~\eqref{eq: diff_eq_12pt_double-box_hex}, while the second term ensures that the symbol is integrable. Moreover, it guarantees that if we take $\chi_{14} \to \infty$ on the first term, the symbol vanishes, as it is the case for the integral itself, see eq.~\eqref{eq: 12pt_double-box_Feynman_param}.

This remarkably compact and elegant formula is the main result of the chapter. Recalling the symbol of the hexagon from eq.~\eqref{eq: hexagon symbol}, we see that it resonates extremely well with the differential equation in eq.~\eqref{eq: diff_eq_12pt_double-box_hex}. In fact, it even looks like we can integrate the differential equation at the level of the symbol to obtain the result in eq.~\eqref{eq: symbol_12pt_double-box_compact}, suggesting that symbol-level integration~\cite{Caron-Huot:2011dec} for this elliptic integral could be possible. This approach was precisely later pursued in ref.~\cite{He:2023qld}, where the elliptic symbol integration method was developed, providing an independent check for our result.

\section{Discussion and open questions}
\label{sec:ch3_conclusions}

In this chapter, we initiated the elliptic symbol bootstrap and, as a proof of principle, used it to determine for the first time the symbol of the 12-pt double-box integral in $D=4$. One crucial ingredient for the bootstrap was the structure of the symbol of the 10-pt double box, which we used to guide the ansatz for the 12-pt case. Moreover, we used Schubert analysis to predict its symbol letters. In particular, we generalized the Schubert analysis to elliptic Feynman integrals, showing how to generate the polylogarithmic as well as the elliptic symbol letters appearing in the last entry of the double-box symbol. Surprisingly, using only the integrability conditions for the symbol and the differential equation relating it to the hexagon, we were able to uniquely bootstrap the result, which satisfies several consistency checks.

One natural follow-up question would be whether we can uplift the symbol to function level, with the aim of expressing the result for the 12-pt double-box integral in terms of eMPLs. However, even though the 12-pt symbol resembles the 10-pt one, it is unknown whether the 12-pt integral can even be expressed in terms of this class of functions. First, as discussed in sec.~\ref{sec:ch3_intro}, linear reducibility fails to provide a one-fold elliptic integral representation. In addition, the presence of 16 distinct square roots makes it unlikely that the integral is expressible in terms of eMPLs, which depend only on a single square root -- the elliptic curve. Therefore, it remains unclear what class of elliptic functions to uplift the symbol to.

Reorganizing the result of the bootstrap, it was possible to write the 12-pt double-box symbol as a simple one-line result, see eq.~\eqref{eq: symbol_12pt_double-box_compact}. In fact, the simplicity of this result has already sparked the development of new elliptic technology, such as elliptic symbol-level integration~\cite{He:2023qld}. Moreover, based on our analysis, in ref.~\cite{Spiering:2024sea} it was possible to essentially write down the symbol of the pentabox and double-pentagon integrals. Together with the 12-pt double box, these three integrals form a basis of all two-loop planar Feynman integrals in $\mathcal{N}=4$ SYM theory in $D=4$~\cite{Bourjaily:2015jna,Bourjaily:2017wjl}. Thus, the symbol of all two-loop planar scattering amplitudes in this theory is now accessible.

More broadly, an exciting future research direction would be to apply the techniques of this chapter to bootstrap other elliptic Feynman integrals and scattering amplitudes, especially in light of the recently discovered sequential discontinuities and genealogical constraints that Feynman integrals obey~\cite{Hannesdottir:2022xki,Hannesdottir:2024cnn}. These conditions, which impose stronger restrictions than the extended Steinmann relations, potentially open the possibility of bootstrapping Feynman integrals that were previously too weakly constrained. Furthermore, it would be interesting to explore the connections between our elliptic bootstrap approach and new developments, such as the novel Landau bootstrap method~\cite{Hannesdottir:2024hke}, particularly since the Landau equations have recently been linked to Schubert analysis~\cite{He:2024fij}. Lastly, it would be exciting to generalize the methods developed in this chapter to Feynman integrals involving more intricate geometries, such as higher-genus curves and Calabi--Yau manifolds.

\newpage\thispagestyle{plain}

\addtocontents{toc}{%
 \protect\vspace{1em}%
 \protect\noindent
\textcolor{DarkRed}{\textbf{III
\hspace{.4em} Special functions in gravitational-wave physics}}
\protect\par
 \protect\vspace{0em}%
}
\part*{Part III \\[1cm] Special functions in gravitational-wave physics}

\chapter{Classification of post-Minkowskian Feynman integral geometries}
\label{ch:chapter4}

\begin{info}[\textit{Info:}]
\textit{Part of the content and figures of this chapter have been published together with H.~Frellesvig and M. Wilhelm in refs.}~\cite{Frellesvig:2023bbf,Frellesvig:2024zph}\textit{, available at} \href{https://doi.org/10.1103/PhysRevLett.132.201602}{\textit{Phys. Rev. Lett.} \textbf{132} (2024)~201602} [\href{https://arxiv.org/abs/2312.11371}{2312.11371}] \textit{and} \href{https://doi.org/10.1007/JHEP08(2024)243}{\textit{JHEP} \textbf{08} (2024) 243} [\href{https://arxiv.org/abs/2405.17255}{2405.17255}]\textit{, respectively; as well as with D. Brammer, H. Frellesvig and M. Wilhelm in ref.}~\cite{Brammer:2025rqo}\textit{, available at} \href{https://arxiv.org/abs/2505.10274}{arXiv:2505.10274}\textit{.}
\end{info}

\section{Motivation}
\label{sec:ch4_intro}

In the second half of the thesis, we study the two-body problem in general relativity, focusing on the gravitational waves emitted during the coalescence of astrophysical compact objects, namely black holes and neutron stars, bound in a binary system. To model the intricate dynamics of this system, we use the so-called post-Minkowskian (PM) expansion~\cite{Buonanno:2022pgc,Damour:2016gwp}, in which Newton's constant $G$ is considered small and the inspiral phase of the merger is treated perturbatively, while accounting for relativistic effects. Thus, within the PM expansion, we can leverage the QFT techniques introduced in chapter~\ref{ch:intro}, specifically scattering amplitudes and Feynman integrals, to compute the corrections to the classical two-body dynamics order by order in $G$. Concretely, at order $G^{L+1}$ we have an $(L+1)$PM correction, which is obtained by computing the $2 \to 2$ scattering of black holes at $L$-loops~\cite{Damour:2016gwp}. Nowadays, this can be most efficiently calculated using a scattering amplitudes-based approach with generalized unitarity and double copy~\cite{Bern:2019nnu,Bern:2019crd,Bern:2024vqs}, as well as from a worldline effective field theory~\cite{Kalin:2019rwq,Kalin:2020mvi,Mogull:2020sak}; see refs.~\cite{Bjerrum-Bohr:2022blt,Buonanno:2022pgc} for a review. In particular, the complete state of the art in PM theory is at three loops~\cite{Bern:2021dqo,Dlapa:2021npj,Bern:2021yeh,Bern:2022jvn,Dlapa:2022lmu,Dlapa:2023hsl,Jakobsen:2023ndj,Damgaard:2023ttc,Jakobsen:2023hig}, i.e.~a 4PM correction, and the four-loop computation is currently under development~\cite{Bern:2023ccb,Klemm:2024wtd,Driesse:2024xad,Bern:2024adl,Driesse:2024feo}.

Notably, while the PM corrections to the two-body problem can be written entirely in terms of polylogarithms up to two loops (3PM order), at three loops (4PM order) there are products of complete elliptic integrals in the result~\cite{Bern:2021dqo,Dlapa:2021npj,Bern:2022jvn,Jakobsen:2023ndj}. In fact, as pointed out in refs.~\cite{Ruf:2021egk,Dlapa:2022wdu}, the source of these elliptic integrals is a non-trivial geometry, as there appears a~K3 surface (recall sec.~\ref{sec:ch1_geometries}) in the respective Feynman integrals.\footnote{While in general an integral over a K3 surface does not give rise to elliptic integrals, this does however happen for single-scale problems, see sec.~\ref{sec:ch4_three_loop_K3} for details.} Therefore, it is natural to suspect that more intricate geometries may appear in the expansion at higher loops, such as higher-genus curves and Calabi--Yau manifolds of higher dimension. With the aim to address this, in this chapter we initiate a systematic exploration of the Feynman integral geometries appearing in the PM expansion that goes beyond the state-of-the-art computations. 

To characterize these geometries, we use the two methods described in sec.~\ref{sec:ch1_DE_LS}: differential equations and leading singularities. Specifically, the aim is to use these tools to detect non-trivial geometries and characterize them, without carrying out the actual calculation of the corresponding Feynman integral. In our case, we found it more efficient to compute the leading singularity (recall sec.~\ref{sec:ch1_LS}) for all topologically-inequivalent integrals at each loop order, since for the PM expansion this calculation is especially streamlined through the so-called Baikov representation~\cite{Baikov:1996iu,Frellesvig:2017aai}, which we will introduce in sec.~\ref{sec:ch4_Baikov}. Moreover, with this representation, we are able to relate the leading singularity and the corresponding Feynman integral geometry across different diagrams and loop orders, which considerably reduces the number of integrals that need to be analyzed. Then, in the cases where a non-trivial geometry is identified with the leading singularity, we also verify it by computing the associated Picard--Fuchs operator via IBP relations and differential equations, recall sec.~\ref{sec:ch1_DEs}.

\begin{figure}[t]
    \centering
\begin{tabular}{c @{\hspace{1cm}} c @{\hspace{1cm}} c @{\hspace{1cm}} c @{\hspace{1cm}} c}
        1SF: & \begin{tikzpicture}[baseline={([yshift=-0.1cm]current bounding box.center)}] 
            \node[] (a) at (-0.5,0) {};
            \node[] (a1) at (0,0) {};
            \node[] (a2) at (1,0) {};
            \node[] (a3) at (1.5,0) {};
            \node[] (b) at (0.5,-0.5) {};
            \node[] (c) at (0,-1) {};
            \node[] (c1) at (1,-1) {};
            \draw[line width=0.15mm] (c.center) --  (a.center);
            \draw[line width=0.15mm] (a1.center) --  (b.center);
            \draw[line width=0.15mm] (c.center) --  (b.center);
            \draw[line width=0.15mm] (b.center) --  (c1.center);
            \draw[line width=0.5mm] (a.center) --  (a3.center);
            \draw[line width=0.15mm] (b.center) --  (a2.center);
            \draw[line width=0.5mm] (c.center) --  (c1.center);
            \draw[line width=0.5mm] (-0.65,0) -- (a.center);
            \draw[line width=0.5mm] (a3.center) -- (1.65,0);
            \draw[line width=0.5mm] (-0.65,-1) -- (c.center);
            \draw[line width=0.15mm] (c1.center) --  (a3.center);
            \draw[line width=0.5mm] (c1.center) -- (1.65,-1);
        \end{tikzpicture} & \begin{tikzpicture}[baseline={([yshift=-0.1cm]current bounding box.center)}] 
	\node[] (a) at (0,0) {};
	\node[] (a1) at (0.5,0) {};
	\node[] (a2) at (1,0) {};
    \node[] (a3) at (1.5,0) {};
	\node[] (b) at (0,-0.5) {};
	\node[] (b1) at (0.75,-0.5) {};
	\node[] (b2) at (0.75,-0.5) {};
    \node[] (b3) at (1.5,-0.5) {};
	\node[] (c) at (0,-1) {};
	\node[] (c1) at (0.5,-1) {};
	\node[] (c2) at (1,-1) {};
    \node[] (c3) at (1.5,-1) {};
	\node[] (p1) at ($(a)+(-0.2,0)$) {};
	\node[] (p2) at ($(c)+(-0.2,0)$) {};
	\node[] (p3) at ($(c3)+(0.2,0)$) {};
	\node[] (p4) at ($(a3)+(0.2,0)$) {};
    \draw[line width=0.5mm] (p1.center) -- (p4.center);
	\draw[line width=0.5mm] (p2.center) -- (p3.center);
    \draw[line width=0.15mm] (c.center) -- (a.center);
	%\draw[line width=0.15mm] (c.center) -- (a1.center);
	\draw[line width=0.15mm] (b.center) -- (b1.center);
	\draw[line width=0.15mm] (b1.center) -- (b2.center);
    \draw[line width=0.15mm] (b1.center) -- (a1.center);
    \draw[line width=0.15mm] (b2.center) -- (c3.center);
	\draw[line width=0.15mm] (b2.center) -- (a2.center);
    \draw[line width=0.15mm] (a3.center) -- (c3.center);
    \draw[line width=0.5mm] (a2.center) --(a3.center);
\end{tikzpicture} & \begin{tikzpicture}[baseline={([yshift=-0.1cm]current bounding box.center)}] 
	\node[] (a) at (0,0) {};
	\node[] (a1) at (0.5,0) {};
	\node[] (a2) at (1,0) {};
    \node[] (a3) at (1.5,0) {};
	\node[] (b) at (0,-0.5) {};
	\node[] (b1) at (0.5,-0.5) {};
	\node[] (b2) at (1,-0.5) {};
    \node[] (b3) at (1.5,-0.5) {};
	\node[] (c) at (0,-1) {};
	\node[] (c1) at (0.5,-1) {};
	\node[] (c2) at (1,-1) {};
    \node[] (c3) at (1.5,-1) {};
	\node[] (p1) at ($(a)+(-0.2,0)$) {};
	\node[] (p2) at ($(c)+(-0.2,0)$) {};
	\node[] (p3) at ($(c3)+(0.2,0)$) {};
	\node[] (p4) at ($(a3)+(0.2,0)$) {};
    \draw[line width=0.5mm] (p1.center) -- (p4.center);
	\draw[line width=0.5mm] (p2.center) -- (p3.center);
    \draw[line width=0.15mm] (c.center) -- (a.center);
	%\draw[line width=0.15mm] (c.center) -- (a1.center);
	\draw[line width=0.15mm] (b.center) -- (b1.center);
	\draw[line width=0.15mm] (b1.center) -- (b2.center);
    \draw[line width=0.15mm] (b.center) -- (a1.center);
    \draw[line width=0.15mm] (b2.center) -- (c3.center);
	\draw[line width=0.15mm] (b2.center) -- (a2.center);
    \draw[line width=0.15mm] (a3.center) -- (c3.center);
     \draw[line width=0.5mm] (a2.center) -- (a3.center);
\end{tikzpicture} & \begin{tikzpicture}[baseline={([yshift=-0.1cm]current bounding box.center)}] 
	\node[] (a) at (0,0) {};
	\node[] (a1) at (0.5,0) {};
	\node[] (a2) at (1,0) {};
    \node[] (a3) at (1.5,0) {};
	\node[] (b) at (0,-0.5) {};
	\node[] (b1) at (0.5,-0.5) {};
	\node[] (b2) at (1,-0.5) {};
    \node[] (b3) at (1.5,-0.5) {};
	\node[] (c) at (0,-1) {};
	\node[] (c1) at (0.5,-1) {};
	\node[] (c2) at (1,-1) {};
    \node[] (c3) at (1.5,-1) {};
	\node[] (p1) at ($(a)+(-0.2,0)$) {};
	\node[] (p2) at ($(c)+(-0.2,0)$) {};
	\node[] (p3) at ($(c3)+(0.2,0)$) {};
	\node[] (p4) at ($(a3)+(0.2,0)$) {};
    \draw[line width=0.5mm] (p1.center) -- (p4.center);
	\draw[line width=0.5mm] (p2.center) -- (p3.center);
    \draw[line width=0.15mm] (c.center) -- (a.center);
	%\draw[line width=0.15mm] (c.center) -- (a1.center);
	\draw[line width=0.15mm] (b.center) -- (b1.center);
	\draw[line width=0.15mm] (b1.center) -- (b2.center);
    \draw[line width=0.15mm] (b.center) -- (a1.center);
    \draw[line width=0.15mm] (b2.center) -- (b3.center);
	\draw[line width=0.15mm] (b3.center) -- (a2.center);
    \draw[line width=0.15mm] (a3.center) -- (c3.center);
\end{tikzpicture}
        \\
\end{tabular}
    \caption{Four-loop PM diagrams depending on non-trivial geometries, organized by their self-force (SF) order, which will be defined in sec.~\ref{sec:ch4_classical_diagrams}. In total, there are two different CY threefolds (labeled by $\text{CY}_3$ and $\text{CY}'_3$) and two K3 surfaces (labeled by K3 and $\text{K3}'$). The geometry labeled by K3 is the same that already appeared at three loops~\cite{Bern:2021dqo,Dlapa:2021npj,Bern:2022jvn,Dlapa:2022wdu,Jakobsen:2023ndj}.}
    \label{fig: diagrams_nontrivial_PM}
\end{figure}
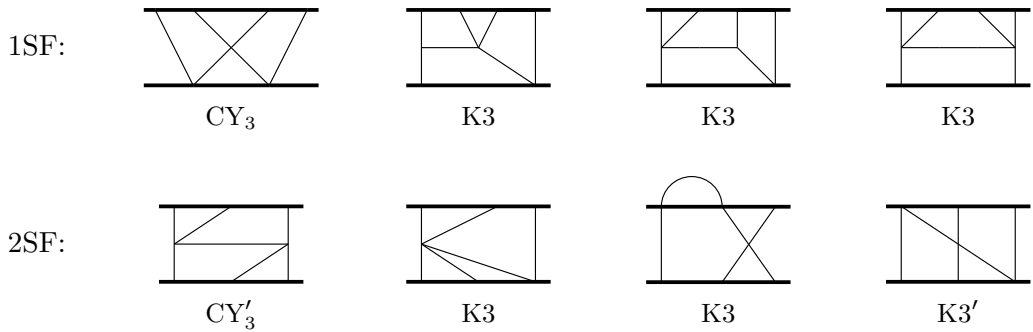
In this chapter, we use this approach to provide the first full classification of Feynman integral geometries relevant to the black-hole scattering process up to four loops, i.e.~5PM order, for both the conservative and the dissipative dynamics. Remarkably, we identify four different non-trivial geometries arising in the expansion at four loops, see fig.~\ref{fig: diagrams_nontrivial_PM}, including the first CY threefold found in gravitational-wave physics (denoted in the figure by $\text{CY}_3'$). In total, we find two different CY threefolds at this order, one in the conservative and one in the dissipative sector, as well as two different K3 surfaces, one of them corresponding to the same K3 surface already present at three-loop order. A subset of these geometries were independently found in a parallel work in ref.~\cite{Klemm:2024wtd}. Classifying the geometries in the PM expansion is crucial, as it determines the class of functions that the 5PM integrals evaluate to, and thus guides the development of analytical and numerical tools required to push the state of the art in the expansion. An example of this can be found in chapter~\ref{ch:chapter5}, where we develop a new technique within differential equations, which is necessary to bring the four-loop $\text{CY}_3'$ integral into $\varepsilon$-factorized form (recall sec.~\ref{sec:ch1_DEs}). 

This chapter is structured as follows. First, in sec.~\ref{sec:ch4_PM_diagrams}, we review the PM expansion, and discuss how to calculate the corrections to the classical dynamics with Feynman diagrams. Then, in sec.~\ref{sec:ch4_classification}, we introduce the Baikov representation for Feynman integrals, and provide a compendium of relations found among leading singularities for PM Feynman integrals. In secs.~\ref{sec:ch4_one_loop} and~\ref{sec:ch4_two_loop}, we then exemplify the calculation of leading singularities at one and two loops, respectively. Thereafter, in secs.~\ref{sec:ch4_three_loop} and~\ref{sec:ch4_four_loop}, we provide a classification of the leading singularities and geometries of the PM integrals at three and four loops, respectively, with a special emphasis on the cases involving non-trivial geometries. Finally, we conclude in sec.~\ref{sec:ch4_conclusions} with a discussion and further directions.

\section{Post-Minkowskian expansion and Feynman diagrams}
\label{sec:ch4_PM_diagrams}

In this section, we review the PM expansion of classical gravity, and discuss how to calculate perturbative corrections to the classical dynamics of a binary system using Feynman diagrams. In particular, we focus on the case of two non-spinning\footnote{Finite-order spin effects appear as numerator factors which, as discussed in sec.~\ref{sec:ch4_modus_operandi}, do not modify our analysis. This may, however, not be true for all-order spin effects, which can resum as exponential factors, see e.g.~ref.~\cite{Chen:2024bpf}. We leave the analysis of such cases for future work.} compact objects, such as black holes and neutron stars, which are inspiraling around each other, bound in a binary system. Although this setting may initially appear different from the unbound scattering process, they are related through analytic continuation~\cite{Damour:2016gwp,Kalin:2019rwq,Kalin:2019inp,Cho:2021arx,Dlapa:2024cje}, which allows us to use the modern scattering amplitude techniques reviewed in chapter~\ref{ch:intro}. Specifically, we will closely follow the formalism and conventions of refs.~\cite{Parra-Martinez:2020dzs,Bern:2021dqo,Herrmann:2021tct,Bern:2021yeh,Bern:2024adl}. Nonetheless, our conclusions will also be valid for PM formulations based on a worldline effective field theory~\cite{Kalin:2020mvi,Mogull:2020sak}, since the leading singularities of the integrals in both formalisms are equivalent.

Concretely, in sec.~\ref{sec:ch4_PM_kinematics}, we introduce the kinematics describing the black-hole scattering process, and in sec.~\ref{sec:ch4_soft_expansion} we review the soft expansion, which is closely tied to the classical dynamics of the system. Lastly, in secs.~\ref{sec:ch4_classical_diagrams} and~\ref{sec:ch4_drawing_classical_diagrams}, we respectively discuss which type of Feynman diagrams contribute in the classical limit, and how to consistently generate all graphs at each loop order.

\subsection{Kinematics}
\label{sec:ch4_PM_kinematics}

For a binary system of black holes in the inspiral phase, we can assume the separation between the bodies to be much larger than their size, such that we have a point-particle approximation and the internal degrees of freedom of the objects can be neglected. Thus, we have that the Schwarzschild radius $r_s \sim G m$ is much smaller than the orbital distance, which is set by the impact parameter $|b|$. In momentum space, the latter is given by $|b| \sim 1/|q|$, where $|q|$ is the momentum transfer for the scattering process. Therefore, in this long-distance regime, we have a natural hierarchy of scales $r_s/|b| \sim G m |q| \ll 1$. Thus, we can set up a perturbative expansion using this small parameter $G m |q| \ll 1$, which in turn is compatible with the PM expansion of classical gravity, where a weak gravitational field is assumed and Newton's constant $G$ is small. In particular, the order $G^n$ in the expansion corresponds to the $n$PM correction.

With these considerations, we can obtain the corrections to the dynamics of the binary system by calculating the $2 \to 2$ scattering of massive scalars (modeling the black holes) minimally coupled to classical gravity~\cite{Cheung:2018wkq,Bern:2019crd}. For example, at tree level, we have the diagram
\begin{equation}
\label{eq: diag_initial_param}
\hspace*{-0.4cm} 
\begin{tikzpicture}[baseline={([yshift=-0.1cm]current bounding box.center)}, scale=1.2] 
	\node[] (a) at (0,0) {};
	\node[] (b) at (0,-1) {};
	\node[label=left:{$p_1$}] (p1) at ($(a)+(-1,0)$) {};
	\node[label=left:{$p_2$}] (p2) at ($(b)+(-1,0)$) {};
	\node[label=right:{$p_3={-}p_2{-}q$}] (p3) at ($(b)+(1,0)$) {};
	\node[label=right:{$p_4={-}p_1{+}q$}] (p4) at ($(a)+(1,0)$) {};
	\draw[line width=0.15mm, postaction={decorate}] (b.center) -- node[sloped, allow upside down, label={[xshift=0.75cm, yshift=0cm]$q$}] {\midarrow} (a.center);
	\draw[line width=0.5mm, postaction={decorate}] (a.center) -- node[sloped, allow upside down] {\midarrow} (p1.center);
	\draw[line width=0.5mm, postaction={decorate}] (a.center) -- node[sloped, allow upside down] {\midarrow} (p4.center);
	\draw[line width=0.5mm, postaction={decorate}] (b.center) -- node[sloped, allow upside down] {\midarrow} (p2.center);
	\draw[line width=0.5mm, postaction={decorate}] (b.center) -- node[sloped, allow upside down] {\midarrow} (p3.center);
\end{tikzpicture}\,,
\end{equation}
where thin and thick lines respectively denote the graviton and massive scalar propagators, while the arrows indicate the direction of the momenta. With this convention of all-outgoing external momenta, the scattering process depends on the five kinematic invariants
\begin{subequations}
\begin{align}
s=& \, (p_1+p_2)^2\,, \qquad \quad t=(p_1+p_4)^2=q^2\,, \qquad \quad u=(p_1+p_3)^2\,, \\[0.2cm]
p_1^2=& \, p_4^2=m_1^2\,, \qquad \quad \,\, p_2^2=p_3^2=m_2^2\,.
\end{align}
\end{subequations}
Here, $m_i$ are the masses of the scalar particles, and $s$, $t$ and $u$ are the usual Mandelstam variables, which in the physical region of the scattering process take values $s>(m_1+m_2)^2$, $t=q^2<0$ and $u<0$. Since the Mandelstam variables satisfy the constraint $s+t+u=2m_1^2+2m_2^2$, in practice we only need four parameters to characterize the scattering process. However, under the classical limit (see the next subsection for details), the masses of the scalars completely factor out as a prefactor, and thus the kinematic dependence of the Feynman integrals reduces to just two parameters. Then, by rescaling, we can render one of them dimensionless, while the dependence on the other dimensionful parameter can be recovered with dimensional analysis. Thus, we can always reduce the kinematic dependence of PM Feynman integrals to a single dimensionless variable. In particular, we can take this variable to be
\begin{equation}
\label{eq: def_sigma_PM}
\sigma = \frac{p_1 \cdot p_2}{m_1 m_2} = \frac{s-m_1^2-m_2^2}{2\, m_1 m_2}\,,
\end{equation}
with $\sigma>1$ in the physical region. Specifically, it corresponds to the relativistic Lorentz factor of particle 1 in the rest frame of particle 2, and vice versa, since $p_1 \cdot p_2 = (E_1,\vec{p}_1) \cdot (m_2,\vec{0})=m_2 \, E_1$.

\subsection{Soft expansion}
\label{sec:ch4_soft_expansion}

In the previous subsection, we have seen that the long-distance regime imposes the hierarchy of scales $r_s \ll |b|$. Here, we will further examine the implications of the classical limit on PM diagrams, where it introduces an additional hierarchy known as the soft expansion. Moreover, we will explore its consequences for the propagators of PM Feynman integrals, revealing a symmetry which we call parity.

In particular, a classical two-body system also satisfies that its angular momentum $J$ is large, $J \gg 1$ in natural units. Thus, the classical limit imposes another hierarchy of scales~\cite{Damour:2016gwp,Damour:2017zjx,Cheung:2018wkq,Bern:2019crd}
\begin{equation}
\label{eq: soft_limit_hierarchy}
m_1^2, \, m_2^2, \, s, \, u \sim J^2 \, |t| \gg |t|=|q|^2\,,
\end{equation}
such that the momentum transfer $|q|$ is small~\cite{Neill:2013wsa}. This is known as the soft expansion and, in practice, it is implemented in perturbation theory with the method of regions~\cite{Beneke:1997zp}, where hard (quantum) loop momenta $k \sim m$ are discarded, while contributions from soft loop momenta $k \sim |q|$ are maintained~\cite{Neill:2013wsa}.

In the soft expansion, it is customary to work with a different parametrization of the external momenta, and decompose it into components that are parallel and perpendicular to the momentum transfer~\cite{Landshoff:1969yyn,Parra-Martinez:2020dzs}. Concretely, we use
\begin{equation}
\label{eq: diag_param_PM}
\,,
\end{equation}
where notice that the arrows for the incoming momenta have been reversed. Here, $\overline{p}_i \cdot q=0$, which can be verified e.g.~from
\begin{equation}
0=p_1^2-p_4^2= p_1^2 - (p_1-q)^2 = 2 \Big( p_1-\frac{q}{2} \Big) \cdot q = - 2 \, \overline{p}_1 \cdot q\,,
\end{equation}
and similarly for $\overline{p}_2 \cdot q =0$. 

With these kinematics, the scaling of graviton propagators in the soft expansion is
\begin{equation}
%
 \ \mathlarger{\in} \ \left\{ \frac{1}{k^2} \,, \frac{1}{(k \pm q)^2} \right\} \sim \frac{1}{|q|^2}\,,
\end{equation}
where $k$ is some loop momentum, while the massive scalar propagators can be expanded as
\begin{equation}
\label{eq: PM_expansion_matter_props}
%
 = \frac{1}{(k+\overline{p}_i \pm \frac{q}{2})^2-m_i^2}=\frac{1}{k^2+2 (\overline{p}_i \pm \frac{q}{2}) \cdot k} = \frac{1}{2\overline{p}_i \cdot k} \left(1-\frac{k^2 \pm q \cdot k}{2\overline{p}_i \cdot k} + \mathcal{O}(q^2) \right).
\end{equation}
Thus, to first order in the soft expansion, the scalar propagators become linearized. Introducing the soft four-velocities 
\begin{equation}
    u_i^\mu=\frac{\overline{p}_i^\mu}{\overline{m}_i}\,, \qquad \text{with} \qquad \overline{m}_i^2=\overline{p}_i^2=m_i^2-\frac{q^2}{4}\,,
\end{equation}
which satisfy that $u_i^2=1$ and $u_i \cdot q=0$, we have
\begin{equation}
\frac{1}{2 \overline{p}_i \cdot k} = \frac{1}{2 u_i \cdot k \sqrt{m_i^2-\frac{q^2}{4}}} = \frac{1}{2u_i \cdot k} \left(\frac{1}{m_i}+ \frac{q^2}{8 m_i^3} + \mathcal{O}(q^4)\right).
\end{equation}
Finally, the matter propagators simply become
\begin{equation}
\begin{tikzpicture}[baseline={([yshift=-0.1cm]current bounding box.center)}] 
	\node[] (a) at (0,0) {};
	\node[] (p4) at ($(a)+(2,0)$) {};
	\draw[line width=0.5mm, postaction={decorate}] (a.center) -- node[sloped, allow upside down] {\midarrow} (p4.center);
\end{tikzpicture} = \frac{1}{m_i}\frac{1}{2u_i\cdot k}+ \mathcal{O}(q^2) \sim \frac{1}{|q|}\,.
\end{equation}
Notably, the mass dependence factors out, which reduces the kinematic dependence of diagrams in the soft expansion from four to two variables, as anticipated in the discussion above. In addition, $q^2$ becomes the only dimensionful parameter, so we could set it to $q^2=-1$ and recover its dependence from dimensional analysis. However, we will keep the full dependence in $q$ as it does not increase the complexity of our analysis. Then, the only dimensionless variable describing the scattering process is
\begin{align}
y &= u_1 \cdot u_2 = \frac{\overline{p}_1 \cdot \overline{p}_2}{\overline{m}_1 \overline{m}_2} = \frac{p_1 \cdot p_2 + \frac{q^2}{4}}{\overline{m}_1 \overline{m}_2}= \sigma + q^2 \, \frac{\sigma (m_1^2+m_2^2)+2m_1m_2}{8 m_1^2 m_2^2} + \mathcal{O}(q^4)\,.
\end{align}
In terms of this variable, the physical region is given by $1 < y < \infty$, where $y \to 1$ corresponds to the static limit, while $y \to \infty$ is the high-energy limit of the scattering. In general, PM Feynman integral results typically involve the square root $\sqrt{y^2-1}$. Thus, it is also common to express them in terms of the variable $x$, defined as $y=\frac{1+x^2}{2x}$, with $0<x<1$ in the physical region. Notably, this change of variables rationalizes the square root, as previously discussed in sec.~\ref{sec:ch1_rationalization}, and particularly shown in eq.~\eqref{eq: rationalization_sqrt_root_PM}.

The fact that the soft expansion results in linearized matter propagators introduces one further symmetry for PM integrals, which is absent for quadratic propagators. If we transform $u_i \to - u_i$ for both $i=1,2$ (and conjugate the corresponding Feynman $i \epsilon$), the entire integral changes at most by a sign, which only depends on the number of linearized propagators and ISPs (recall sec.~\ref{sec:ch1_Feynman_integrals}) involving $u_i$. This transformation, which we refer to as parity, thus separates the PM integrals into contributions that are even or odd under parity. Since linearized propagators are linear in $|q|$, while all other components depend on $|q|^2$, we may also view parity as a distinction between integrals that are even or odd in $|q|$. In fact, as we discussed in detail in appendix~D of ref.~\cite{Frellesvig:2024zph}, the parity transformation splits the PM integrals into two independent sets obeying disconnected IBP relations, which leads to a linear system of differential equations that decouples into two blocks~\cite{Parra-Martinez:2020dzs,DiVecchia:2021bdo,Herrmann:2021tct}, recall also the discussion in sec.~\ref{sec:ch1_DEs}.

As a consequence, the next natural question is whether we need to study both parity contributions for our analysis of non-trivial geometries. To determine this, let us consider the impulse $\Delta p_i^\mu$, which corresponds to the change in momentum for the particle $i$, and is one of the main classical observables that can be obtained from the scattering amplitudes approach~\cite{Kosower:2018adc}. Denoting by $\Delta^{(n)} p_i^\mu$ its $n$PM correction, it can be schematically obtained from the PM Feynman integrals by a Fourier transform~\cite{Dlapa:2023hsl},
\begin{equation}
\label{eq: Fourier_impulse_PM}
\Delta^{(n)} p_i^\mu \sim G^n M^{n+1} \int d^D q \, e^{i q \cdot b} \delta(u_1 \cdot q) \delta(u_2 \cdot q) \left[ q^\mu \, \mathcal{I}^{(n)}_{q} + \sum_{i=1}^2 u_i^\mu \, \mathcal{I}^{(n)}_{u_i} \right],
\end{equation}
where $M$ is a mass, and $\mathcal{I}^{(n)}_{q,u_i}$ denotes an $n$PM integral. Using dimensional analysis, with $[\Delta p_i^\mu]=1$, $[\delta(u_i \cdot q)]=-1$ and $[G]=2-D$ in $D$ dimensions, we find that $[\mathcal{I}^{(n)}_{q}]=n-3$ and $[\mathcal{I}^{(n)}_{u_i}]=n-2$. Thus, we can conclude that $\mathcal{I}^{(n)}_{q}$ are parity odd and $\mathcal{I}^{(n)}_{u_i}$ are parity even for even PM orders, and the opposite for odd PM orders. Now, the result can be written as~\cite{Kalin:2020mvi}
\begin{equation}
 \Delta^{(n)} p_i^\mu=c^{(n)}_{b} \, b^\mu+ \sum_{i=1}^2 c^{(n)}_{u_i} \, u_i^\mu\,,
\end{equation}
such that $\mathcal{I}^{(n)}_{q}$ contribute to $c^{(n)}_{b}$, while $\mathcal{I}^{(n)}_{u_i}$ contribute to $c^{(n)}_{u_i}$. As discussed in ref.~\cite{Dlapa:2023hsl}, for the conservative dynamics, where $\Delta p_1^\mu + \Delta p_2^\mu = 0$, the $c^{(n)}_{u_i}$ are fixed from lower orders in the expansion. Thus, in the conservative regime we would only need $\mathcal{I}^{(n)}_{q}$, which are parity-odd (parity-even) integrals at even (odd) PM orders.\footnote{Note that, compared to the worldline formalism, our definition of parity is slightly different. There, $L$ (parity-odd) matter propagators are replaced by (parity-even) delta functions at $L$ loops, thus modifying the behavior under parity. Then, the $c^{(n)}_{b}$ always depend on parity-even integrals, while $c^{(n)}_{u_i}$ depend on parity-odd.} However, for the dissipative dynamics, where there is a radiation loss, $\Delta p_1^\mu + \Delta p_2^\mu = p_{\text{rad}}^\mu$, the $c^{(n)}_{u_i}$ are unconstrained. Therefore, since we aim to classify the Feynman integral geometries contributing to both regimes, we will need to take into account both parity contributions for each PM diagram, see sec.~\ref{sec:ch4_modus_operandi} for further details.

\subsection{Classical diagrams}
\label{sec:ch4_classical_diagrams}

In the previous subsection, we have studied the soft expansion and its effect on the propagators, which resulted in linearized matter propagators and the associated parity splitting of PM integrals. Now, we turn to the implications of the classical limit on the diagrams, especially when taking into account the scaling of the vertex interactions. This consideration will further divide the diagrams into three categories: quantum, classical, and superclassical.

As discussed above, taking into account powers of $G$ and $|q|$, graviton propagators scale as $|q|^{-2}$, whereas matter propagators become linearized and scale as $|q|^{-1}$. For the graviton self-interaction and the interaction with the scalars, we respectively have
\begin{equation}
\begin{tikzpicture}[baseline={([yshift=-0.1cm]current bounding box.center)}] 
	\node[] (a) at (0,0) {};
	\node[label=left:{$1$}] (a1) at (-1,0) {};
	\node[label={[xshift=0.3cm, yshift=-0.6cm]$2$}] (a2) at (0.71,-0.71) {};
	\node[label={[xshift=0.3cm, yshift=-0.2cm]$n$}] (an) at (0.71,0.71) {};
	\draw[line width=0.15mm] (a.center) -- (a1.center);
	\draw[line width=0.15mm] (a.center) -- (a2.center);
	\draw[line width=0.15mm] (a.center) -- (an.center);
	\node at (0.6,-0.2)[circle,fill,inner sep=0.6pt]{};
	\node at (0.6,0)[circle,fill,inner sep=0.6pt]{};
	\node at (0.6,0.2)[circle,fill,inner sep=0.6pt]{};
\end{tikzpicture} \sim |q|^{2} G^{\frac{n}{2}-1}\,, \qquad \qquad \quad \begin{tikzpicture}[baseline={([yshift=-0.1cm]current bounding box.center)}] 
	\node[] (a) at (0,0) {};
	\node[] (a0) at (-1,0) {};
	\node[] (a1) at (1,0) {};
	\node[label={[xshift=-0.2cm, yshift=-0.6cm]$1$}] (a2) at (-0.71,-0.71) {};
	\node[label={[xshift=0.2cm, yshift=-0.6cm]$n$}] (an) at (0.71,-0.71) {};
	\draw[line width=0.5mm] (a.center) -- (a0.center);
	\draw[line width=0.5mm] (a.center) -- (a1.center);
	\draw[line width=0.15mm] (a.center) -- (a2.center);
	\draw[line width=0.15mm] (a.center) -- (an.center);
	\node at (-0.2,-0.6)[circle,fill,inner sep=0.6pt]{};
	\node at (0,-0.6)[circle,fill,inner sep=0.6pt]{};
	\node at (0.2,-0.6)[circle,fill,inner sep=0.6pt]{};
\end{tikzpicture} \sim G^{\frac{n}{2}}\,,
\end{equation}
where the scalings can be obtained from the Einstein-Hilbert action with a minimally coupled scalar field, see e.g.~ref.~\cite{Bern:2019crd}. Lastly, each integration over the loop momenta scales as $\int d^4k \sim |q|^4$, and external scalar legs contribute with a factor of one.

With this counting, the tree-level diagram in eq.~\eqref{eq: diag_param_PM} scales as $|q|^{-2} G$, which is expected, as it contains the (momentum-space) Newtonian contribution. Then, since we have a perturbative expansion with $G m |q| \ll 1$, recall sec.~\ref{sec:ch4_PM_kinematics}, each loop subsequently adds a factor of $|q|G$. Thus, the expected scaling for an $L$-loop diagram contributing to the classical two-body potential is $|q|^{L-2} G^{L+1}$~\cite{Neill:2013wsa}, which corresponds to an $(L+1)$-PM correction. However, at any given loop order, there also appear diagrams with scalings different than the classical one, see fig.~\ref{fig: diagrams_discarded}.

\begin{figure}[tb]
\centering
\subfloat[]{
}
\caption{Examples of diagrams with a power counting different than the classical one, or which do not contribute to the classical limit. (a) One-loop diagram with quantum power counting; (b) Superclassical one-loop box diagram; (c) and (d) One-loop diagrams which vanish in dimensional regularization.}
\label{fig: diagrams_discarded}
\end{figure}
Firstly, there are PM diagrams with more powers of $|q|$ relative to the classical scaling, which arise due to closed graviton loops, see e.g.~fig.~\ref{fig: diagrams_discarded}(a). These diagrams correspond to quantum corrections, since they are subleading in the soft expansion and become suppressed in the limit of small $|q|$. Thus, we will discard these quantum contributions from the outset. To avoid closed graviton loops, we demand that each loop contains at least one matter line~\cite{Bern:2019crd,Herrmann:2021tct}. In fact, a classical power counting in $|q|$ demands exactly $L$ matter lines at $L$ loops.\footnote{For a connected graph, we can relate the number of loops $L$ with the number of internal edges $I$ and the number of vertices $V$, as $L=I-V+1$. In our case, we can distinguish the contributions from scalars and gravitons, such that $V=V_g+V_s$ and $I=I_g+I_s$. Since there are no scalar self-interactions and the vertices always involve either zero or two scalar particles, we have $V_s=I_s+2$, thus we obtain that $I_g = L+V_g+1$. From the Feynman rules above, classical power counting in $|q|$ demands that $L-2=4L+2V_g-2I_g-I_s$, which finally yields $I_s=L$.} Secondly, there are the so-called superclassical (or iteration) diagrams, which have fewer powers of $|q|$ than the classical power counting, due to the presence of more than $L$ matter propagators at $L$ loops, see e.g.~fig.~\ref{fig: diagrams_discarded}(b). However, these contributions cancel at each loop order when matching the effective field theory to the full theory~\cite{Bern:2019crd}, due to an exponentiation of the action. While in principle they can also be discarded from the beginning, we will maintain superclassical diagrams in our analysis for illustrative purposes and for completeness.

In fact, under the classical limit, we can also discard one further type of diagrams~\cite{Bern:2019crd,Herrmann:2021tct}, see e.g.~fig.~\ref{fig: diagrams_discarded}(c) and fig.~\ref{fig: diagrams_discarded}(d). This applies to cases where at least one of the loop integrals does not depend on any dimensionful scale, and thus vanishes in dimensional regularization. For instance, since the diagram in fig.~\ref{fig: diagrams_discarded}(c) contains a scalar self-interaction, there is no momentum transfer $q$ going through the loop. Similarly, the diagram in fig.~\ref{fig: diagrams_discarded}(d) contains a loop which only involves one matter line, so it does not depend on $q$ either.

Lastly, let us introduce the self-force order, which will be used to further distinguish and classify PM diagrams. The self-force (SF) expansion~\cite{Mino:1996nk,Quinn:1996am,Poisson:2011nh,Barack:2018yvs} corresponds to an expansion in the ratio of the black-hole masses, where we have a heavy black hole and a probe, such that there is a hierarchy $\frac{m_1}{m_2}\ll1$. In particular, the SF expansion can be performed on top of the PM expansion. In the PM diagrams, the self-force classification is then reflected on the number of matter propagators. Explicitly, a diagram with $n_1$ matter propagators on the line $u_1$, and $n_2$ propagators on the line $u_2$, contributes to an SF order given by $\min (n_1,n_2)$. At $L$ loops, the SF order of classical diagrams can therefore take values from 0 to $\lfloor \frac{L}{2} \rfloor$. For instance, at 4 loops we have
\begin{equation}
\parbox[c]{0.15\textwidth}{
            \centering
            \raisebox{-0.5\height}{
} \\[0.1cm]
            \small 2SF order
        \\[0.1cm]}.
\end{equation}
Notice, however, that superclassical diagrams can contribute to different SF orders, since they contain different classical subsectors. For example,
\begin{equation}
\raisebox{-0.25cm}{\parbox[c]{0.15\textwidth}{
            \centering
            %
 \\[0.1cm]
            \small 2SF order
        \\[0.1cm]}} \raisebox{-0.25cm}{.}
\end{equation}

In summary, in these subsections we have studied the implications of the classical limit and the soft expansion on PM diagrams, which enjoy an additional parity symmetry. Moreover, we have seen that power counting can be used at the level of the graphs to separate the classical contributions from the unnecessary quantum corrections. Thus, the focus of the next subsection is to establish which type of diagrams are necessary for our classification of geometries.

\subsection{Generating all classical diagrams}
\label{sec:ch4_drawing_classical_diagrams}

In full-fledged computations of physical observables within the PM expansion, the starting point in the calculation is the gravity integrand. For instance, in approaches based on scattering amplitudes, it is obtained via generalized unitarity and double copy~\cite{Bern:2019nnu,Bern:2019crd,Bern:2024vqs}, while worldline formalisms begin with an effective action and extract the integrand using Feynman rules~\cite{Kalin:2019rwq,Kalin:2020mvi,Mogull:2020sak}. Expanding the gravity integrand then yields the PM integrals which need to be evaluated, with the associated PM diagrams emerging as a by-product of the process. Our aim, however, is to proceed bottom-up, and classify the geometries appearing in the expansion without actually calculating neither the integrals nor the integrand. Thus, we must obtain the relevant diagrams appearing at each loop order by other means.

Unlike in $\mathcal{N}=4$ SYM or $\mathcal{N}=8$ supergravity, where supersymmetry imposes various cancellations among bosonic and fermionic contributions, we can presume that in the PM expansion of general relativity there are no mechanisms forcing such cancellations. Therefore, we will assume that any diagram that contributes in the classical limit also appears in the final result, such that there are no total cancellations of geometries at the end.\footnote{At four loops and conservative 1SF order, it was recently observed in refs.~\cite{Driesse:2024xad,Bern:2024adl} that the complete elliptic integrals appearing in individual diagrams (see sec.~\ref{sec:ch4_four_loop}) drop from the final result of the momentum impulse. Let us note, however, that this is not due to a cancellation of geometries. Rather, it is because the Fourier transform of eq.~\eqref{eq: Fourier_impulse_PM} introduces an $\varepsilon$ prefactor for odd PM orders. Since the particular combination yielding the elliptic integrals is finite in dimensional regularization, these terms simply do not contribute in $D=4$.} Consequently, our task of determining the Feynman integral geometries in the expansion translates to generating all relevant graphs at each loop order and analyzing their geometry individually via leading singularities; see a more detailed discussion in sec.~\ref{sec:ch4_modus_operandi}. Hence, the first step in the process is to find a systematic method to draw all diagrams, which is the focus of this subsection. While our approach may not be the most optimized, it was sufficient for our classification up to four loops. Nonetheless, we could mitigate overcountings at intermediate steps by using considerations e.g.~from generalized unitarity and spanning cuts~\cite{Bern:2024vqs}, but we leave this for future work.

In particular, our strategy is to first draw all relevant top topologies (which only contain cubic vertices), and afterwards generate the list of all subtopologies from them by contracting propagators. To do so, let us first focus on the classical top topologies. Since there are no scalar self-interactions, we can draw the classical top topologies by embedding graviton cubic tree-level graphs inside two matter lines. For instance,
\begin{equation}
 \ ,
\end{equation}
where the graviton 6-pt tree-level graph is shown in blue, and the dots at the external legs attach to the matter lines, shown in black.

Consequently, to generate all classical top topologies, we first have to draw all graviton tree diagrams, and afterwards combine them with the matter lines. It is easy to see that for an $L$-loop classical top topology, we need a graviton tree with $\nex=L+2$ external legs. Up to four loops, we have the following possibilities:
\begin{equation}
\begin{tabular}{|c|c|c|c|c|}
\Xhline{0.2ex}
\enspace $\nex=2$ \enspace & \enspace $\nex=3$ \enspace & \enspace $\nex=4$ \enspace & \enspace $\nex=5$ \enspace & \enspace $\nex=6$ \enspace \\
\hline
& & & & \\
\end{tabular}\ .
\end{equation}

Then, to generate the corresponding $L$-loop classical top topologies, we join the $(L+2)$-pt graviton tree graphs with the matter lines in all possible totally-connected ways. This includes from the case where a single graviton leg is attached to the $u_2$ matter line, which generates a 0SF diagram, to the case where there are $\lfloor \frac{L}{2} \rfloor + 1$ legs attached to the $u_2$ matter line, which provides the maximum SF order.\footnote{To avoid generating graphs that are related by a mirror across the horizontal axis, without loss of generality, we can fix the matter line with the minimal number of propagators to be associated with $u_2$.} Finally, for each case, we draw all permutations of external legs in the graviton graphs. For instance, at two loops, we have
\begin{align}
    & \, \right\}.
\end{align}

Since our aim is to obtain a classification of the geometries appearing in the expansion, already at this stage we can discard from our analysis the top topologies which are topologically equivalent, as they have the same leading singularity. For instance, in the example above, we observe that some of the resulting top topologies can be related through a mirror across the horizontal and vertical axis. In the case of PM diagrams, these symmetries do not modify the leading singularity: the mirror across the vertical axis corresponds to the parity transformation $u_i \to - u_i$, which only affects an overall sign (recall sec.~\ref{sec:ch4_soft_expansion}), while the horizontal reflection swaps $u_1 \leftrightarrow u_2$, but the diagrams depend on $y=u_1 \cdot u_2$, which is invariant. Thus, 
\begingroup
\allowdisplaybreaks
\begin{subequations}
\begin{align}
\LS \left( \right)\,,
\end{align}
\end{subequations}
\endgroup
which reduces the list from 8 to 5 diagrams.\footnote{For graphs with colored edges, one can also discard topologically-equivalent diagrams using the \texttt{IGVF2IsomorphicQ} command in the \texttt{IGraphM} package~\cite{Horvat2023} for {\Mathematica}. However, we found it more efficient to sort through the inequivalent diagrams based on the mirror symmetries.} Moreover, since the matter propagators are linearized, if we invert the sign of either $u_1\to - u_1$ or $u_2\to - u_2$, effectively flipping only one of the matter lines, the integrals change at most by a sign. This way, we are able to relate the leading singularity of the non-planar H diagram to the planar one,
\begin{equation}
\LS \left( \right)\,.
\end{equation}
Lastly, we can also discard the last top topology, as it is not one-particle irreducible (1PI). Thus, it factorizes into the product of lower-loop integrals, 
\begin{equation}
    \LS \left( %
 \right)\,,
\end{equation}
which are included in the lower-loop analysis of geometries. Moreover, all of its 1PI subtopologies are already part of other diagrams; in this particular case, of the H diagram. Hence, from the 8 starting top topologies, only 3 of them are needed to identify the geometries which appear for the first time at two loops in the PM expansion. Proceeding similarly at higher loops, we can obtain the corresponding classical top topologies.

Let us now turn to the superclassical top topologies. To generate them, we essentially follow the same procedure, but allow for several graviton tree-level graphs to form the PM diagram. Specifically, for an $L$-loop top topology, the least we can have is one $(L+2)$-pt graviton graph, which as shown above generates the classical top topology, and at most we can have $L+1$ different $2$-pt graviton graphs. Therefore, we have a decomposition into graviton tree-level graphs given by
\begin{equation}
    L + 1 = \sum_{j=2}^{L+2} (j-1) \,n_j = n_2 + 2n_3+3n_4 + 4 n_5+ \dots + L \, n_{L+1} + (L+1) \, n_{L+2}\,,
\end{equation}
where the $n_j \, \mathlarger{\mathlarger{\in}} \, \mathbb{N}$ denote the number of $j$-pt graviton graphs. In other words, the number of possible decompositions of an $L$-loop top topology into tree-level graphs is given by the integer partitions of $L+1$. For instance, at 3 loops, we have the partitions
\begingroup
\allowdisplaybreaks
\begin{subequations}
    \begin{align}
        4=4 \ (n_5=1): & \ \qquad \ .
    \end{align}
\end{subequations}
\endgroup
Then, just as we did for the classical diagrams, we join the different components, creating one combination per SF order. Afterwards, we draw all permutations of graviton legs, and at the end remove the topologically-equivalent diagrams and those which vanish in dimensional regularization; as detailed in the previous subsection. For example, for $n_4=1$ and $n_2=1$, we obtain
\begin{equation}
    \left\{ \ %
 \ \right\} \ + \ \text{permutations}\,.
\end{equation}

This way, we can create all classical and superclassical top topologies at each loop order. For each diagram in the final list, we then perform all possible edge contractions to generate the subtopologies. The only subtlety is that, unlike for the top topologies, which are already chosen to be classical or superclassical, for the subtopologies we also need to discard quantum contributions, which arise when pinching certain matter propagators. Lastly, we remove those subtopologies with equivalent leading singularities using the selection rules described above, which finally gives us the list of all diagrams that are relevant to the classical potential. Up to four loops, we obtain:
\begin{equation}
\begin{tabular}{|l|cccc|}
\Xhline{0.2ex}
$L$ & 1 & 2 & 3 & 4 \\
\hline
Number of diagrams & 2 & 23 & 531 & 16,596 \\
\Xhline{0.2ex}
\end{tabular}\ .
\label{eq: tab_initial_number_diagrams}
\end{equation}

\section{Classifying geometries with the Baikov representation}
\label{sec:ch4_classification}

In the previous section, we have examined the type of Feynman diagrams which are relevant to the classical dynamics of a two-body system within the PM expansion. Now, we turn to the task of analyzing the corresponding integrals, with the aim of identifying and classifying the non-trivial geometries that arise in the expansion. First, in sec.~\ref{sec:ch4_Baikov}, we review the Baikov representation~\cite{Baikov:1996iu} and its loop-by-loop variation~\cite{Frellesvig:2017aai,Frellesvig:2024ymq}, highlighting their suitability for computing leading singularities of PM Feynman integrals. Then, in sec.~\ref{sec:ch4_modus_operandi}, we explain how to recycle lower-loop results in order to streamline the calculation of leading singularities, and provide more details on our classification method. Finally, in sec.~\ref{sec:ch4_relating_geometries}, we present a collection of relations among leading singularities that connect PM Feynman integral geometries across different loop orders and integral topologies, such as the unraveling of matter propagators and the reduction of superclassical diagrams.

\subsection{The Baikov representation}
\label{sec:ch4_Baikov}

In the Baikov representation, which we briefly mentioned in sec.~\ref{sec:ch1_Feynman_integrals}, the integration variables correspond to the propagators of the diagram; see ref.~\cite{Weinzierl:2022eaz} for an introduction. To derive it, we begin with a Feynman integral in momentum representation such as eq.~\eqref{eq: loop momentum Feynman integral}, and change variables to the Baikov variables $z_i=Q_i^2-m_i^2$ for $i=1 ,\dots, \nint$. Then, the first step is to invert this transformation in order to rewrite the loop momenta $k_i$ in terms of the $z_i$. This is, however, only possible if the number of scalar products $N_\text{V}$ involving loop momenta (see eq.~\eqref{eq: number of total scalar products}) is equal to the number $\nint$ of Baikov variables. In other words, and recalling the discussion in sec.~\ref{sec:ch1_Feynman_integrals}, the change of variables is automatically invertible only if there are no ISPs, as $\nISP = N_{\text{V}} - \nint$. As a consequence, just like we did in sec.~\ref{sec:ch1_Feynman_integrals} to incorporate numerator factors in the integral family framework, to invert the change of variables and obtain the Baikov representation, in general we artificially add extra Baikov variables $z_j$ for $j=\nint + 1 ,\dots, N_{\text{V}}$, which can only appear in the numerator.\footnote{In fact, this addition has the exact same interpretation as ISPs in the integral family framework, recall eq.~\eqref{eq: loop momentum Feynman integral with ISPs}. In essence, we find the Baikov representation for a top sector, in which all scalar products can be identified with a propagator. Then, the induced Baikov representation for the subsectors is obtained by replacing the propagators that are absent by ISPs in the numerator.} After all, we obtain the standard Baikov representation~\cite{Baikov:1996iu}, which is given by
\begin{equation}
\label{eq: standard_Baikov}
\mathcal{I}_{\nu_1\dots\,\nu_\nint}=\mathcal{J}\, \frac{{(\det G(p_1,\dots,p_E))}^{\frac{-D+E+1}{2}}}{\prod_{j=1}^L \Gamma{\Big( \frac{D-E+1-j}{2} \Big)}} \mathlarger{\int}_{\mathfrak{C}} \ \frac{d^{N_{\text{V}}} \! z\  \mathcal{N}(z_{\nint+1},\dots,z_{N_{\text{V}}}) }{z_1^{\nu_1} \, \cdots \, z_{\nint}^{\nu_\nint}} \ {\mathcal{B}(z_1,\dots,z_{N_{\text{V}}})}^{\frac{D-L-E-1}{2}},
\end{equation}
where $\mathcal{J}=\pm 2^{L-\nint}$ is a Jacobian from the change of variables, and where we allow for a generic numerator $\mathcal{N}$ which only depends on the extra Baikov variables. Moreover, $\mathcal{B}(z_1,\dots,z_{N_{\text{V}}}) \equiv \det G(k_1,\dots,k_L,p_1,\dots,p_E)$ denotes the Baikov polynomial, the roots of which determine the integration contour $\mathfrak{C}$, and where $G$ is the Gram matrix, with entries being $G_{ij}(Q_1,\dots,Q_n)=Q_i \cdot Q_j$. Remember as well from sec.~\ref{sec:ch1_Feynman_integrals}, that $E=\text{dim} \langle p_1, \dots, p_{n_{\text{ext}}} \rangle = \nex - 1$ is the number of independent external momenta. For detailed examples using the Baikov representation, we refer to secs.~\ref{sec:ch4_one_loop} and~\ref{sec:ch4_two_loop}.

One of the virtues of the Baikov representation is that computing the maximal cut (recall sec.~\ref{sec:ch1_LS}) is straightforward, as it simply corresponds to calculating the residue at the point where all Baikov variables related to propagators vanish, $z_i=0$ for $i=1 ,\dots, \nint$. For example, in the case where all $\nu_i=1$, we simply have
\begin{align}
\label{eq: Baikov_max_cut_nu_1}
\mathcal{I}_{1\,\cdots\,1} \Big|_{\text{max cut}}= & \ \mathcal{J}\, \frac{{(\det G(p_1,\dots,p_E))}^{\frac{-D+E+1}{2}}}{\prod_{j=1}^L \Gamma{\Big( \frac{D-E+1-j}{2} \Big)}} \nonumber \\
& \, \times \int d^{\nISP}\! z \ \mathcal{N}(z_{\nint+1},\dots,z_{N_{\text{V}}}) \ {\mathcal{B}(\underbrace{0,\dots,0}_{\displaystyle \nint},z_{\nint+1},\dots,z_{N_{\text{V}}})}^{\frac{D-L-E-1}{2}}.
\end{align}
As can be seen, after imposing the maximal cut, there are $\nISP$ integrals remaining. Thus, the complexity of the leading singularity -- and the underlying geometry -- is constrained by the number of ISPs in the integral. For instance, from eq.~\eqref{eq: number of total scalar products} we find that any one-loop integral satisfies $N_{\text{V}}=E+1$, which precisely matches its number of internal propagators, $\nint$. Consequently, at one loop there are no ISPs, and the maximal cut localizes all integrals in eq.~\eqref{eq: standard_Baikov}, resulting in an algebraic leading singularity. Since this property extends to all one-loop subsectors, it follows that one-loop Feynman integrals are polylogarithmic, as we anticipated in sec.~\ref{sec:ch1_MPLs}. 

At higher loops, by contrast, the presence of ISPs introduces the possibility of non-trivial geometries. For example, comparing to tab.~\ref{tab:geometries_intro}, we see that if the leading singularity corresponds to a Calabi--Yau integral, its dimension is bounded to be $\leq \nISP$. Nevertheless, the Baikov representation itself can be redundant, and the underlying geometry can be much simpler than what one would naively infer by the number of ISPs alone. In particular, eq.~\eqref{eq: Baikov_max_cut_nu_1} may reveal further poles, which often happens for top topologies. These extra singularities allow for further residues to be taken, ultimately reducing the complexity of the leading singularity and the geometry. In many cases, however, these poles are only apparent after performing changes of variables, such as those discussed in sec.~\ref{sec:ch1_rationalization}.

To reduce the number of ISPs, and achieve a less redundant representation for the leading singularity, we will employ the loop-by-loop Baikov representation~\cite{Frellesvig:2017aai}, where the Baikov representation from eq.~\eqref{eq: standard_Baikov} is used one loop at a time; see refs.~\cite{Weinzierl:2022eaz,Frellesvig:2024ymq} for an introduction. In this approach, we obtain
\begin{equation}
    N_{\text{V}} = L + \sum_{i=1}^L E_i
\end{equation}
scalar products involving the loop momenta, where $E_i$ denotes the number of independent external momenta with respect to each loop. Compared to eq.~\eqref{eq: number of total scalar products}, in the loop-by-loop approach we therefore reduce the number of ISPs, which simplifies the calculation of the leading singularity. The disadvantage of using a loop-by-loop approach, however, is that it generates $2L-1$ different Baikov polynomials depending on the loop momenta, while in eq.~\eqref{eq: standard_Baikov} we only had one. Therefore, the expression for the leading singularity becomes more convoluted and depends on the parametrization chosen. However, this complexity can be mitigated with an educated loop-by-loop strategy, which we discuss in the next subsection. In practice, a useful implementation of both the standard and loop-by-loop Baikov representation can be found in the {\Mathematica} package \texttt{BaikovPackage}, see ref.~\cite{Frellesvig:2024ymq} for details.

Lastly, let us briefly comment on the interplay between the Baikov representation and the PM expansion. Recalling sec.~\ref{sec:ch4_soft_expansion}, the only novelty of PM integrals compared to usual Feynman integrals is that matter propagators become linearized under the soft expansion. However, this is not an inconvenient for the Baikov representation; as a matter of fact, it even simplifies the expressions. This is because the corresponding Baikov variable is simply $z=2 u_i \cdot k$. Thus, when we invert the change of variables, the scalar product $u_i \cdot k$ maps directly to $z/2$, while a quadratic propagator would involve a linear combination of several Baikov variables.

\subsection{Methodology}
\label{sec:ch4_modus_operandi}

In the previous subsection, we have introduced the loop-by-loop Baikov representation as a convenient approach for computing the leading singularity of PM Feynman integrals, mainly because it inherently reduces the number of ISPs. However, it becomes highly dependent on the loop-by-loop order and parametrization employed, which can significantly impact the complexity of the final result. To address this point, in this subsection we outline the method to obtain the simplest representation, see also ref.~\cite{Frellesvig:2024ymq} for further details. In addition, we provide more details on the methodology used to classify the geometries order by order in the PM expansion.

First of all, let us focus on the strategy for obtaining the optimal loop-by-loop Baikov representation. Overall, the aim is to reduce the number of ISPs, as it simplifies the corresponding leading singularity.\footnote{Let us note, however, that for very few examples, we found that a naively suboptimal parametrization can actually result in a simpler leading singularity; see e.g.~sec.~\ref{sec:ch4_four_loop_CYs_2} for an example at 4 loops, as well as refs.~\cite{Frellesvig:2024ymq,Brammer:2025rqo} for further details.} Therefore, the first consideration is to parametrize the diagrams such that the momentum transfer $q$ is routed through the least amount of loops, preferably only through the last loop of the loop-by-loop representation. This way, we can avoid some scalar products $k_i \cdot q$ with the loop momenta, which only increase the number of ISPs. Similarly, we begin the loop-by-loop ordering with the loops with fewer propagators (bubbles and triangles), and work our way upwards towards box loops. For instance, we would follow the parametrization
\begin{equation}
    \begin{tikzpicture}[baseline={([yshift=-0.1cm]current bounding box.center)}, scale=1.5] 
	\node[] (a) at (0,0) {};
	\node[] (a1) at (0.5,0) {};
	\node[] (a2) at (1,0) {};
    \node[] (a3) at (1.5,0) {};
	\node[] (b) at (0,-0.45) {};
	\node[] (b1) at (0.5,-0.5) {};
	\node[] (b2) at (1,-0.5) {};
    \node[] (b3) at (1.5,-0.5) {};
	\node[] (c) at (0,-1) {};
	\node[] (c1) at (0.5,-1) {};
	\node[] (c2) at (0.6,-1) {};
    \node[] (c3) at (1.5,-1) {};
	\node[] (p1) at ($(a)+(-0.2,0)$) {};
	\node[] (p2) at ($(c)+(-0.2,0)$) {};
	\node[] (p3) at ($(c3)+(0.2,0)$) {};
	\node[] (p4) at ($(a3)+(0.2,0)$) {};
    \draw[line width=0.5mm] (p1.center) -- (p4.center);
	\draw[line width=0.5mm] (p2.center) -- (p3.center);
    \draw[line width=0.15mm] (c.center) -- (a.center);
	\draw[line width=0.15mm] (b.center) -- (a2.center);
    \draw[line width=0.15mm] (b.center) -- (c3.center);
    \draw[line width=0.15mm] (c2.center) -- (b.center);
    \draw[line width=0.15mm, postaction={decorate}] (c3.center) -- node[sloped, allow upside down, label={[xshift=1.3cm, yshift=0cm]$k_4{+}q$}] {\midarrow} (a3.center);
    \node[label={$1$}] (l1) at (0.175,-0.45) {};
    \node[label={$2$}] (l2) at (0.15,-1.08) {};
    \node[label={$3$}] (l3) at (0.65,-1.1) {};
    \node[label={$4$}] (l4) at (0.95,-0.75) {};
\end{tikzpicture}\ ,
\end{equation}
where only the rightmost propagator depends on the momentum transfer $q$, and the numbers specify the loop-by-loop sequence, given by $1 \to 2 \to 3 \to 4$. As can be seen, this order naturally follows the strategy described above, and we minimize the number of ISPs by placing $q$ in the last loop. By contrast, if we were to follow instead the order $1 \to 3 \to 2 \to 4$, since the loop number $3$ sits between $2$ and $4$ in the diagram, we would mix the loop momenta $k_2$ and $k_4$ in its Baikov polynomial, thus creating more ISPs. Similarly, starting with loop number $4$ would mix $k_1$, $k_3$ and $q$, which would otherwise not appear together. In particular, following the parametrization above, in sec.~\ref{sec:ch4_four_loop} we will show that this diagram depends on a non-trivial geometry. For more examples of parametrizations, we refer to secs.~\ref{sec:ch4_one_loop} to~\ref{sec:ch4_four_loop}.

While this parametrization scheme is suitable for most diagrams, starting at four loops, however, we inevitably encounter pentagon loops. For instance, this happens for the non-planar 4-loop 2SF top topology, see diagram \diagramnumberingsign41 in tab.~\ref{tab: 4-loop results 2SF}. The problem with pentagon loops (or higher-sided polygons) is that the number of independent external momenta is $E_i \geq D=4-2\varepsilon$. Hence, the prefactor of the Baikov representation for that loop (see eq.~\eqref{eq: standard_Baikov}) vanishes in strict $D=4$, namely
\begin{equation}
\frac{1}{\Gamma \left(\frac{D-E_i}{2} \right)} = \frac{1}{\Gamma (-\varepsilon)} = -\varepsilon + \mathcal{O}(\varepsilon^2)\,.
\end{equation}
This is a mere artifact of dimensional regularization, as the integration over the Baikov contour $\mathfrak{C}$ compensates it with an $\varepsilon^{-1}$ contribution. For example, ref.~\cite{Frellesvig:2021vdl} showed that, for the elliptic double-box integral, the result for the leading singularity is equivalent to the one derived from a Baikov representation strictly in $D=4$. Nonetheless, in our case we will avoid this nuisance by working in $D=6-2\varepsilon$ for diagrams containing pentagon loops, since the $D=4$ and the $D=6$ integrals -- and thus the geometry -- are related via dimension-shift identities~\cite{Tarasov:1996br,Tarasov:1997kx}.

Before turning to the systematics of the classification of geometries, let us go over a couple of further details of our method. Starting at two loops, see sec.~\ref{sec:ch4_two_loop}, the recursive nature of the loop-by-loop Baikov representation will make its first appearance. Thus, to avoid repeating computations, we will reuse lower-loop results directly at the level of the leading singularity. For instance, we will use the notation
\begin{equation}
\LS \left(
\right).
\end{equation}
In this equation, we use the loop-by-loop order $1 \to 2$, as indicated in the left-hand side. This can be interpreted as doing the Baikov representation for the loop number 1 first, which is a triangle with momentum transfer $k_2$, as indicated by the arrow. From eq.~\eqref{eq: standard_Baikov}, we obtain a Baikov polynomial $\det G(k_1,k_2,u_1)$, which introduces the ISPs $k_2^2$ and $u_1 \cdot k_2$ for the second loop. Thus, when we subsequently do the Baikov representation for the loop number 2, we need to add the extra Baikov variables $z_6=k_2^2$ and $z_7=2u_1 \cdot k_2$ to include the ISPs, as indicated by the dashed propagators. In general, we will use this graphical decomposition to illustrate the integration order and provide a visual representation for the ISPs, see other examples in sec.~\ref{sec:ch4_two_loop}.

Precisely beginning also at two loops, the results for the maximal cut start to contain simple poles. As introduced in sec.~\ref{sec:ch1_LS}, the leading singularity corresponds to deforming the contour $\mathfrak{C}$ to a closed contour around the poles, thus computing the residue; see also sec.~\ref{sec:ch4_Baikov} for a related discussion within Baikov representation. Similarly, there will start appearing simultaneous square roots, which require from the changes of variables reviewed in sec.~\ref{sec:ch1_rationalization} to expose further poles. In practice, these operations can be efficiently implemented together with the {\texttt{\textup{LeadingSingularities}}} command from the {\DlogBasis} package~\cite{Henn:2020lye} in {\Mathematica}. Note, however, that we also encounter examples where such closed contours vanish. As we explained in ref.~\cite{Frellesvig:2024zph}, for those cases, we will replace the closed contour prescription by integrating between the roots of the Baikov polynomials.

Regarding the methodology for classifying the geometries appearing in the PM expansion, we will base our analysis on two crucial assumptions. First, as already mentioned in sec.~\ref{sec:ch4_drawing_classical_diagrams}, we will assume that all individual Feynman diagrams appearing in the expansion ultimately contribute to the observables, such that there are no cancellations of geometries. Second, we will also assume that there is no further decoupling of the differential equations besides the parity splitting, described in sec.~\ref{sec:ch4_soft_expansion}. Under this premise, we can classify the geometries in the PM expansion by analyzing a single candidate integral per parity sector for each of the diagrams in eq.~\eqref{eq: tab_initial_number_diagrams}. This is because, if the corresponding parity block in the differential equation does not decouple further, any master integral in the sector will depend on the same geometry. While it could be the case that these assumptions fail at some high-loop order, so far they are satisfied in the explicit PM computations up to four loops. Importantly, they significantly reduce the number of integrals to be considered: from all master integrals in each parity sector to just one integral, thus streamlining our analysis of geometries.

Consequently, for each diagram in eq.~\eqref{eq: tab_initial_number_diagrams}, we can just consider the integral of each parity that yields the simplest leading singularity. For the natural parity which comes from the drawing of the diagram, i.e.~when all propagator exponents are $\nu_i=1$, we will normally use the scalar integral as a candidate. Still, in some examples, adding an even-parity ISP (such as $k^2$ or $(2 u_i \cdot k)^2$) or dotting a graviton propagator -- neither of which modify the original parity -- lead to a simpler result. Analogously, for the opposite-parity contributions, we will normally add either an odd-parity ISP, such as $2 u_i \cdot k$, or dot a matter propagator in the diagram. Lastly, we will treat cases where the leading singularity contains higher-order poles by adding ISPs that cancel them~\cite{Henn:2020lye}.

\subsection{Relations among PM Feynman integral geometries}
\label{sec:ch4_relating_geometries}

As discussed in the previous subsection, the task of classifying the Feynman integral geometries appearing in the PM expansion can be reduced to computing the leading singularities of one integral of each parity for all diagrams contributing to the classical limit, as obtained in eq.~\eqref{eq: tab_initial_number_diagrams}. While this is already a drastic simplification of the analysis, it is possible to reduce the size of the problem even further. In particular, in this subsection we gather a collection of identities among leading singularities which relate the associated geometries over different topologies and loop orders. Due to these relations, only a much smaller subset of the diagrams needs to be actually studied in detail, which scales down the extent of the analysis by various orders of magnitude at 3 and 4 loops; see tab.~\ref{tab:reduction_diagrams}.

First of all, as we proved in appendix~A of ref.~\cite{Frellesvig:2024zph}, when computing the leading singularity, the vertices along the matter lines can be treated as orderless. Specifically,
\begin{equation}
\label{eq: unraveling_matter_props}
\LS \left( \right),
\end{equation}
where we can have multiple graviton lines (represented by the dots) coming out of the vertices, which should be understood as being part of a bigger diagram (represented by the gray area). We call this operation unraveling, and it allows us to rearrange the vertices and relate the leading singularity -- and thus the geometry -- of numerous non-planar diagrams to planar counterparts. For example, applying the unraveling on the red matter propagators, we have
\begin{align}
\LS \left(
%
 \right)\,.
\end{align}
Since the planar diagrams are also included in the classification, we can discard these non-planar diagrams to begin with. In fact, this property allows us to drop all non-planar diagrams up to 3 loops, and it is only at 4 loops that truly non-planar diagrams arise, see sec.~\ref{sec:ch4_four_loop} for examples.

Secondly, as we showed in appendix~C of ref.~\cite{Frellesvig:2024zph}, we have the following relation:
\begin{equation}
\label{eq: reduction_superclassical}
\LS \left(
%
 \right),
\end{equation}
where the blobs designate any tree-level or loop-level diagram, and can also be empty on one side. This relation applies to superclassical diagrams containing a graviton line that attaches to both matter lines via cubic vertices, and can also be combined with the unraveling of matter propagators of eq.~\eqref{eq: unraveling_matter_props}. For instance,
\begin{equation}
    \LS \left(
%
\right) \propto 
\frac{x^2}{q^2(x^2-1)^2}\,,
\end{equation}
where we used eq.~\eqref{eq: reduction_superclassical} to compute the result. Analogously, we can use this identity to calculate the leading singularity of numerous superclassical diagrams from lower-loop results. Since, by construction, the latter are already included in the classification at lower loops, we can drop these superclassical diagrams from the outset. Notice that this identity is totally analogous to the fact that the leading singularity for the 10-pt elliptic ladder family of chapter~\ref{ch:chapter2} did not change under the addition of rungs, recall the discussion below eq.~\eqref{eq: 10pt_elliptic_ladder_mini_drawing}.

\begin{figure}[t]
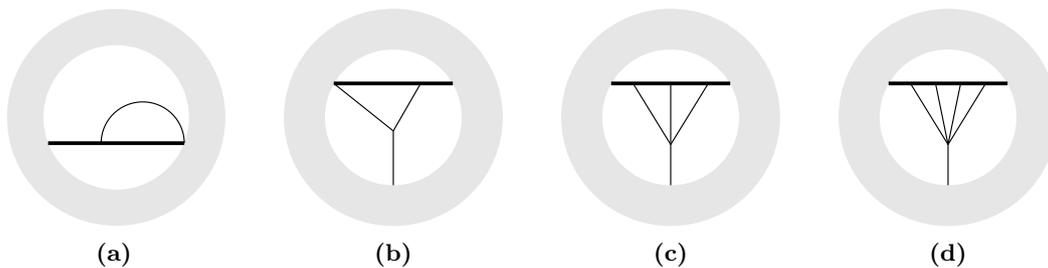

\centering
\subfloat[]{
}
\caption{Subgraphs which have zero master integrals. (a) One-loop bubble with at least one cubic matter vertex; (b) One-loop triangle with a cubic self-interaction and at least one cubic matter vertex at the matter line; (c) Two-loop dangling triangle with a quartic self-interaction and cubic matter vertices; (d) Three-loop dangling triangle with a quintic self-interaction and cubic matter vertices. The subgraphs should be understood as being part of a bigger diagram, as represented by the gray surrounding.}
\label{fig: diagrams_zero_masters}
\end{figure}
Lastly, we can use IBP relations (defined in eq.~\eqref{eq: IBP_definition}) for the individual loops inside of a diagram to determine whether certain configurations have zero master integrals, meaning that they are totally reducible to subsectors (recall the details in sec.~\ref{sec:ch1_LS}). As we showed in appendix~B of ref.~\cite{Frellesvig:2024zph}, this is the case for diagrams containing a one-loop bubble subgraph where at least one of the vertices is cubic, see fig.~\ref{fig: diagrams_zero_masters}(a). Combining it with the unraveling of matter propagators, at two loops it e.g.~applies to
\begin{equation}
\,.
\end{equation}
Similarly, diagrams containing a one-loop triangle subgraph with a cubic graviton self-interaction and at least one cubic vertex at a matter line, see fig.~\ref{fig: diagrams_zero_masters}(b), also have zero master integrals; see appendix~B of ref.~\cite{Frellesvig:2024zph} for a proof. Among others, at three loops it e.g.~applies to
\begin{equation}
%
\,,
\end{equation}
where we highlight the relevant one-loop triangle subgraphs in red, and in the last diagram we combine it with the unraveling of matter propagators. Since these 4 diagrams are completely reducible to subsectors, which are either already included in the classification or do not contribute to the classical limit, we can drop them from the analysis.

Similarly, we noticed that the rule of zero master integrals of fig.~\ref{fig: diagrams_zero_masters}(b) could be extended to two and three loops, see figs.~\ref{fig: diagrams_zero_masters}(c)--(d). In the spirit of ref.~\cite{Frellesvig:2023bbf}, we call these cases multiloop dangling triangles, and they are characterized by having only cubic vertices at the matter line and a single graviton line coming out of the graviton self-interaction vertex. Unlike for the one-loop case, which we could prove using IBPs for the subgraph~\cite{Frellesvig:2024zph}, so far we have not found a general proof for the validity of the reduction for the multiloop dangling triangles. However, we have explicitly verified up to 4 loops that all PM diagrams containing multiloop dangling triangles indeed have zero master integrals in both parity sectors. Hence, it is tempting to suggest that there is a rule extending to all-loop orders, but we leave this investigation for future work.

\begin{table}[t]
\begin{center}
\caption{Number of remaining diagrams after applying each of the identities gathered in the main text, up to 4 loops. The initial number of diagrams follows from eq.~\eqref{eq: tab_initial_number_diagrams}.}
\label{tab:reduction_diagrams}
\begin{tabular}{|l|cccc|}
\Xhline{0.2ex}
& & & & \\[-0.5cm]
$L$ & 1 & 2 & 3 & 4 \\[0.05cm]
\Xhline{0.2ex}
& & & & \\[-0.5cm]
Initial number & 2 & 23 & 531 & 16,596 \\[0.05cm] \hline
& & & & \\[-0.5cm]
Reduction of superclassical diagrams & 1 & 12 & 271 & 8,335 \\[0.05cm] \hline
& & & & \\[-0.5cm]
Reduction of one-loop bubbles & 1 & 7 & 100 & 1,854 \\[0.05cm] \hline
& & & & \\[-0.5cm]
Reduction of one-loop triangles & 1 & 4 & 33 & 476 \\[0.05cm] \hline
& & & & \\[-0.5cm]
Reduction of multiloop dangling triangles & 1 & 4 & 31 & 422 \\[0.05cm] \hline
& & & & \\[-0.5cm]
Unraveling of matter propagators & 1 & 4 & 14 & 70 \\[0.05cm]
\Xhline{0.2ex}
\end{tabular}
\end{center}
\end{table}
All in all, the result of applying the previous simplifying relations is gathered in tab.~\ref{tab:reduction_diagrams}. First, with the reduction of superclassical diagrams of eq.~\eqref{eq: reduction_superclassical}, and in combination with the unraveling of eq.~\eqref{eq: unraveling_matter_props}, we cut the number of diagrams by half. Then, dropping the diagrams which (together with the unraveling) have zero master integrals, see fig.~\ref{fig: diagrams_zero_masters}, further reduces the number of diagrams by one order of magnitude. Lastly, from the remaining diagrams, we can remove those which are topologically equivalent under the unraveling of eq.~\eqref{eq: unraveling_matter_props}, which considerably reduces the number of independent diagrams at 3 and 4 loops. 

With this reduction, we obtain a final list of diagrams to analyze, which is the focus of secs.~\ref{sec:ch4_one_loop} to~\ref{sec:ch4_four_loop}, where we systematically examine them from one to four loops, respectively. Impressively, out of the 16,596 four-loop diagrams contributing to the classical limit, we only need to consider 70 of them. However, these 70 diagrams are not necessarily the only ones containing non-trivial geometries -- they simply form a smaller subset that captures all geometries appearing in the PM expansion at this loop order, but which we can analyze systematically. Any diagram from the original set of 16,596 that is related to any of these 70 by the simplifying rules described above may also depend on non-trivial geometries. Likewise, some leading singularities and geometries within this reduced subset may be related among themselves by more complicated identities, as it seems to be suggested by the four-loop classification; see sec.~\ref{sec:ch4_four_loop}.

Interestingly, all surviving diagrams up to four loops pertain to the following family of PM diagrams (or its superclassical iterations), which following ref.~\cite{Bern:2004kq} we call the Mondrian family:
\begin{equation}
\label{eq: diag_Mondrian}
\begin{tikzpicture}[baseline={([yshift=-0.1cm]current bounding box.center)}, scale=0.7] 
	\node[] (a) at (0,0) {};
	\node[] (a1) at (1,0) {};
	\node[] (a2) at (2,0) {};
	\node[] (a3) at (3,0) {};
	\node[] (a4) at (4,0) {};
	\node[] (a5) at (5,0) {};
	\node[] (a6) at (6,0) {};
	\node[] (b) at (0,-1) {};
	\node[] (b1) at (1,-1) {};
	\node[] (b2) at (2,-1) {};
	\node[] (b3) at (3,-1) {};
	\node[] (b4) at (4,-1) {};
	\node[] (b5) at (5,-1) {};
	\node[] (b6) at (6,-1) {};
	\node[] (c) at (0,-2) {};
	\node[] (c1) at (1,-2) {};
	\node[] (c2) at (2,-2) {};
	\node[] (c3) at (3,-2) {};
	\node[] (c4) at (4,-2) {};
	\node[] (c5) at (5,-2) {};
	\node[] (c6) at (6,-2) {};
	\node[] (p1) at ($(a)+(-0.3,0)$) {};
	\node[] (p2) at ($(c)+(-0.3,0)$) {};
	\node[] (p3) at ($(c6)+(0.3,0)$) {};
	\node[] (p4) at ($(a6)+(0.3,0)$) {};
	\draw[line width=0.15mm] (b.center) -- (a.center);
	\draw[line width=0.15mm] (b1.center) -- (a1.center);
	\draw[line width=0.15mm] (b3.center) -- (a3.center);
	\draw[line width=0.15mm] (b4.center) -- (a4.center);
	\draw[line width=0.15mm] (b6.center) -- (a6.center);
	\draw[line width=0.15mm] (b.center) -- (b6.center);
	\draw[line width=0.15mm] (b.center) -- (c.center);
	\draw[line width=0.15mm] (b2.center) -- (c2.center);
	\draw[line width=0.15mm] (b5.center) -- (c5.center);
	\draw[line width=0.15mm] (b6.center) -- (c6.center);
	\draw[line width=0.5mm] (p1.center) -- (p4.center);
	\draw[line width=0.5mm] (p2.center) -- (p3.center);
	\node at (5,-0.5)[circle,fill,inner sep=0.6pt]{};
	\node at (5.25,-0.5)[circle,fill,inner sep=0.6pt]{};
	\node at (4.75,-0.5)[circle,fill,inner sep=0.6pt]{};
	\node at (5.5,-1.5)[circle,fill,inner sep=0.6pt]{};
	\node at (5.25,-1.5)[circle,fill,inner sep=0.6pt]{};
	\node at (5.75,-1.5)[circle,fill,inner sep=0.6pt]{};
\end{tikzpicture}\,.
\end{equation}
It would be interesting to investigate whether this family comprises all top topologies that are relevant for classifying the geometries in the PM expansion at higher loops. Amusingly, this question seems to be closely tied to the proof that multiloop dangling triangles have zero master integrals. Notice, however, that starting at 4 loops, we also need to include non-planar variations of the Mondrian family to generate all top topologies. One graphical way to obtain these non-planar variations is e.g.~by applying Jacobi-like moves~\cite{Bern:2008qj} to the graviton propagators of the graph, in other words, to exchange the $s$, $t$ and $u$-channels in the corresponding tree-level subgraph. For instance, applying this rule to the propagator in red, we obtain
\begin{equation}
},
\end{equation}
where the last diagram is the 2SF non-planar top topology at 4 loops, see sec.~\ref{sec:ch4_four_loop}.

\section{One-loop diagrams}
\label{sec:ch4_one_loop}

\begin{figure}[t]
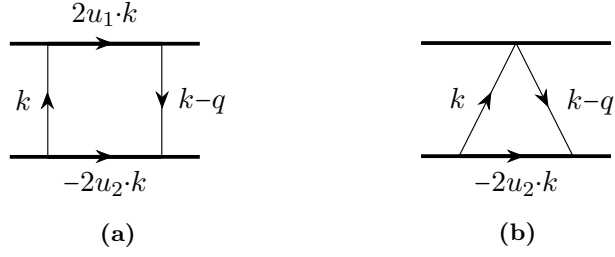

\centering
\subfloat[]{
}
\caption{(a) Parametrization for the one-loop box; (b) Parametrization for the one-loop triangle.}
\label{fig: params_PM_box_and_triangle}
\end{figure}

In this section, we begin the classification of geometries contributing to the PM expansion by studying the one-loop diagrams. As shown in tab.~\ref{tab:reduction_diagrams}, at one loop we actually need just one diagram, which corresponds to the one-loop triangle, see fig.~\ref{fig: params_PM_box_and_triangle}(b). Still, let us go first over the details of the Baikov representation and leading singularity for the superclassical one-loop box in fig.~\ref{fig: params_PM_box_and_triangle}(a), as it serves as a great example to illustrate the method.

\subsection{One-loop box}
\label{sec:ch4_one_loop_box}

Our first example is the one-loop box, which is parametrized in fig.~\ref{fig: params_PM_box_and_triangle}(a). Since we are at one-loop, there are no ISPs (recall the discussion below eq.~\eqref{eq: Baikov_max_cut_nu_1}), and we can use the standard Baikov representation from eq.~\eqref{eq: standard_Baikov}. Using $D=4$ and $E=3$, and introducing the Baikov variables
\begin{equation}
z_1=k^2\,, \qquad z_2=(k-q)^2\,, \qquad z_3=-2u_2 \cdot k\,, \qquad z_4=2u_1 \cdot k\,,
\end{equation}
we have
\begin{equation}
\begin{tikzpicture}[baseline={([yshift=-0.1cm]current bounding box.center)}] 
	\node[] (a) at (0,0) {};
	\node[] (a1) at (0.8,0) {};
	\node[] (b) at (0,-0.8) {};
	\node[] (b1) at (0.8,-0.8) {};
	\node[] (p1) at ($(a)+(-0.2,0)$) {};
	\node[] (p2) at ($(b)+(-0.2,0)$) {};
	\node[] (p3) at ($(b1)+(0.2,0)$) {};
	\node[] (p4) at ($(a1)+(0.2,0)$) {};
	\draw[line width=0.15mm] (b.center) -- (a.center);
	\draw[line width=0.15mm] (b1.center) -- (a1.center);
	\draw[line width=0.5mm] (p1.center) -- (p4.center);
	\draw[line width=0.5mm] (p2.center) -- (p3.center);
\end{tikzpicture} \propto \int \frac{d z_1 d z_2 d z_3 d z_4}{z_1 z_2 z_3 z_4} \frac{1}{\sqrt{\det G(k,u_1,u_2,q)}}\,.
\end{equation}
In this expression, we drop the constant prefactors, and the Baikov polynomial takes the form
\begin{align}
\det G(k,u_1,u_2,q) = & \, (y^2-1) \, \lambda \left( k^2, (k-q)^2,q^2 \right) \nonumber \\
& -4 q^2 \left( (k \cdot u_1)^2 + (k \cdot u_2)^2 - 2 y (k \cdot u_1) (k \cdot u_2) \right) ,
\end{align}
where $\lambda(a,b,c)=a^2+b^2+c^2-2ab-2ac-2bc$ denotes the Källén function. Inverting the change of variables, the Baikov polynomial becomes quadratic in the Baikov variables, $P_2(z_1,\dots,z_4)$, with coefficients depending on $y$ and $q^2$. 

Taking the maximal cut as in eq.~\eqref{eq: Baikov_max_cut_nu_1}, we thus obtain
\begin{equation}
\label{eq: LS_PM_one-loop_box}
\LS \left( \begin{tikzpicture}[baseline={([yshift=-0.1cm]current bounding box.center)}] 
	\node[] (a) at (0,0) {};
	\node[] (a1) at (0.8,0) {};
	\node[] (b) at (0,-0.8) {};
	\node[] (b1) at (0.8,-0.8) {};
	\node[] (p1) at ($(a)+(-0.2,0)$) {};
	\node[] (p2) at ($(b)+(-0.2,0)$) {};
	\node[] (p3) at ($(b1)+(0.2,0)$) {};
	\node[] (p4) at ($(a1)+(0.2,0)$) {};
	\draw[line width=0.15mm] (b.center) -- (a.center);
	\draw[line width=0.15mm] (b1.center) -- (a1.center);
	\draw[line width=0.5mm] (p1.center) -- (p4.center);
	\draw[line width=0.5mm] (p2.center) -- (p3.center);
\end{tikzpicture} \right) \propto \frac{1}{\sqrt{P_2(0,\dots,0)}} \propto \frac{1}{\sqrt{(y^2-1) \, \lambda(0,0,q^2)}} = \frac{1}{q^2 \sqrt{y^2-1}} \propto \frac{x}{q^2 (x^2-1)},
\end{equation}
where in the last step we introduce the change of variables $y=\frac{1+x^2}{2x}$ advocated in sec.~\ref{sec:ch4_soft_expansion} to rationalize the square root. As can be seen, the leading singularity has the $|q|^{-2}$ scaling that would be expected from the power counting of sec.~\ref{sec:ch4_classical_diagrams}, and would moreover have $G^2$ after dressing the interaction vertices. This result with linearized propagators is consistent with eq.~\eqref{eq: reduction_superclassical}, and is in agreement with refs.~\cite{Parra-Martinez:2020dzs,Herrmann:2021tct,DiVecchia:2021bdo}, where a basis of pure integrals (recall sec.~\ref{sec:ch1_LS}) up to two loops is provided. Instead, if we relax the soft expansion and compute the leading singularity with the quadratic matter propagators of eq.~\eqref{eq: PM_expansion_matter_props}, we obtain~\cite{Cachazo:2017jef}
\begin{equation}
\LS^{\text{quadratic}}_{\text{box}} \propto \frac{1}{q^2 m_1 m_2 \sqrt{\sigma^2-1}}\,.
\end{equation}
Parametrizing the result in terms of Mandelstam variables with eq.~\eqref{eq: def_sigma_PM}, in the massless limit it becomes the usual $1/(st)$~\cite{Britto:2004nc}. In addition, the result of eq.~\eqref{eq: LS_PM_one-loop_box} can be recovered with the soft expansion, where recall that the masses have been factored out of the linearized propagators.

As explained in sec.~\ref{sec:ch4_modus_operandi}, for our classification, we must also consider an integral of opposite parity for each diagram. Since at one loop we cannot have ISPs, we only have the possibility of adding a dot to one matter propagator, obtaining
\begin{equation}
\LS \left( \right) = 0\,,
\end{equation}
which implies that the odd-parity integral has zero master integrals in its sector.

\subsection{One-loop triangle}
\label{sec:ch4_one_loop_triangle}

Let us then turn to the one-loop triangle, given in fig.~\ref{fig: params_PM_box_and_triangle}(b). In this case, the external momentum $u_1$ does not enter into the Feynman integral, thus $E=2$. Using the same notation for the Baikov variables as for the one-loop box, we obtain
\begin{align}
\hspace{-0.3cm} %
 \right) & \propto \frac{1}{\sqrt{q^2-(u_2 \cdot q)^2}} = \frac{1}{|q|}\,,
\label{eq: LS_PM_one-loop_triangle}
\end{align}
where remember that $u_i \cdot q=0$. In the case of triangles in $D=4$, the exponent of the Baikov polynomial in eq.~\eqref{eq: standard_Baikov} vanishes, and we only obtain the loop-momentum-independent Gram determinant $\det G(p_1,\dots,p_E)=\det G(u_2,q)$. The result of the leading singularity has the correct scaling in $|q|$, and agrees with refs.~\cite{Parra-Martinez:2020dzs,Herrmann:2021tct,DiVecchia:2021bdo}. Using the quadratic matter propagators of eq.~\eqref{eq: PM_expansion_matter_props}, we obtain~\cite{Cachazo:2017jef}
\begin{equation}
\LS^{\text{quadratic}}_{\text{triangle}} \propto \frac{1}{\sqrt{q^2(4m_2^2-q^2)}}\,,
\end{equation}
which leads to the expected result $1/t$ in the massless limit. Moreover, it is in agreement with eq.~\eqref{eq: LS_PM_one-loop_triangle} in the soft limit, since we had a hierarchy $m_i^2 \gg |q|^2$, see eq.~\eqref{eq: soft_limit_hierarchy}.

For the opposite-parity integral, we obtain
\begin{equation}
\LS \left( \begin{tikzpicture}[baseline={([yshift=-0.1cm]current bounding box.center)}] 
	\node[] (a) at (0.4,0) {};
	\node[] (b) at (0,-0.8) {};
	\node[] (b1) at (0.8,-0.8) {};
	\node[] (p1) at ($(a)+(-0.6,0)$) {};
	\node[] (p2) at ($(b)+(-0.2,0)$) {};
	\node[] (p3) at ($(b1)+(0.2,0)$) {};
	\node[] (p4) at ($(a)+(0.6,0)$) {};
	\draw[line width=0.15mm] (b.center) -- (a.center);
	\draw[line width=0.15mm] (b1.center) -- (a.center);
	\draw[line width=0.5mm] (p1.center) -- (p4.center);
	\draw[line width=0.5mm] (p2.center) -- (p3.center);
	\node at ($(b1)-(0.4,0)$) [circle,fill,inner sep=1.5pt]{};
\end{tikzpicture} \right) \propto \frac{\varepsilon \, (u_2 \cdot q)}{q^2 \sqrt{q^2-(u_2 \cdot q)^2}} = 0\,,
\label{eq: LS_PM_one-loop_triangle_odd_parity}
\end{equation}
where the dependence on $\varepsilon$ is due to having a double propagator, since we need to perform the Laurent expansion of the integrand to calculate the maximal cut. Hence, as expected, at one loop we only obtain algebraic leading singularities.

\section{Two-loop diagrams}
\label{sec:ch4_two_loop}

In this section, we continue the classification of PM geometries to two loops. As shown in tab.~\ref{tab:reduction_diagrams}, out of the 23 initial topologies, at two loops we only need the following 4 diagrams:
\begin{equation}
\label{eq: PM_two_loop_diagrams}
\,.
\end{equation}
In sec.~\ref{sec:ch4_two_loop_0SF}, we will study the first diagram, which is the only one contributing to 0SF order (recall the definition in sec.~\ref{sec:ch4_classical_diagrams}). In addition, we will also use this diagram to illustrate the loop-by-loop Baikov representation. Then, in sec.~\ref{sec:ch4_two_loop_1SF}, we turn to the remaining diagrams, which contribute to 1SF order.

\subsection{Two-loop 0SF diagrams}
\label{sec:ch4_two_loop_0SF}

\begin{figure}[t]
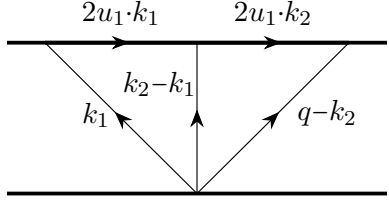

\centering
%
\caption{Parametrization for the two-loop double triangle.}
\label{fig: param_PM_double_triangle}
\end{figure}
Let us begin with the two-loop double-triangle diagram parametrized in fig.~\ref{fig: param_PM_double_triangle}, which is the only diagram from eq.~\eqref{eq: PM_two_loop_diagrams} contributing to 0SF order. As it is the first example where loop-by-loop Baikov makes an appearance, we will go through the details of this computation. First, as can be seen from the parametrization, and as explained in sec.~\ref{sec:ch4_modus_operandi}, we want to follow the loop-by-loop order $\{k_1, k_2\}$, as it avoids mixing $k_1$ with $q$. Thus, starting with $k_1$, we have a one-loop triangle such as the one in sec.~\ref{sec:ch4_one_loop_triangle}, but with momentum transfer $k_2$. In particular, we obtain a Gram determinant $\det G(k_2,u_1)$, which introduces the scalar products $k_2^2$ (an ISP) and $u_1 \cdot k_2$ (a propagator). Therefore, we need to add one extra Baikov variable $z_6=k_2^2$ to perform the integration over the second loop. Using the graphical representation of sec.~\ref{sec:ch4_modus_operandi}, we have
\begin{equation}
\begin{tikzpicture}[baseline={([yshift=-0.1cm]current bounding box.center)}] 
	\node[] (a) at (0,0) {};
	\node[] (a1) at (0.4,0) {};
	\node[] (a2) at (0.8,0) {};
	\node[] (b) at (0.4,-0.8) {};
	\node[] (p1) at ($(a)+(-0.2,0)$) {};
	\node[] (p2) at ($(b)+(-0.6,0)$) {};
	\node[] (p3) at ($(b)+(0.6,0)$) {};
	\node[] (p4) at ($(a2)+(0.2,0)$) {};
	\draw[line width=0.15mm] (b.center) -- (a.center);
	\draw[line width=0.15mm] (b.center) -- (a1.center);
	\draw[line width=0.15mm] (b.center) -- (a2.center);
	\draw[line width=0.5mm] (p1.center) -- (p4.center);
	\draw[line width=0.5mm] (p2.center) -- (p3.center);
\end{tikzpicture} = \begin{tikzpicture}[baseline={([yshift=-0.1cm]current bounding box.center)}] 
	\node[] (a) at (0,0) {};
	\node[] (a1) at (0.8,0) {};
	\node[] (b) at (0.4,-0.8) {};
	\node[] (p1) at ($(a)+(-0.2,0)$) {};
	\node[] (p2) at ($(b)+(-0.6,0)$) {};
	\node[] (p3) at ($(b)+(0.6,0)$) {};
	\node[] (p4) at ($(a1)+(0.2,0)$) {};
	\draw[line width=0.15mm] (b.center) -- (a.center);
	\draw[line width=0.15mm] (a1.center) -- (b.center);
	\draw[line width=0.15mm,-{Latex[length=2.2mm]}](1.1,-0.75) -- (1.1,-0.05);
	\node[label={[xshift=0.35cm, yshift=-0.45cm]$k_2$}] (k) at (1.1,-0.4) {};
	\draw[line width=0.5mm] (p1.center) -- (p4.center);
	\draw[line width=0.5mm] (p2.center) -- (p3.center);
\end{tikzpicture} \times
\begin{tikzpicture}[baseline={([yshift=-0.1cm]current bounding box.center)}] 
	\node[] (a) at (0,0) {};
	\node[] (a1) at (0.8,0) {};
	\node[] (b) at (0.4,-0.8) {};
	\node[] (p1) at ($(a)+(-0.2,0)$) {};
	\node[] (p2) at ($(b)+(-0.6,0)$) {};
	\node[] (p3) at ($(b)+(0.6,0)$) {};
	\node[] (p4) at ($(a1)+(0.2,0)$) {};
	\draw[line width=0.15mm, dashed, postaction={decorate}] (b.center) -- node[sloped, allow upside down, label={[xshift=0.2cm, yshift=0.45cm]$k_2$}] {\midarrow} (a.center);
	\draw[line width=0.15mm] (b.center) -- (a1.center);
	\draw[line width=0.5mm] (p1.center) -- (p4.center);
	\draw[line width=0.5mm] (p2.center) -- (p3.center);
\end{tikzpicture} \propto \int \frac{d z_1 \, \cdots \, d z_5}{z_1 \, \cdots \, z_5} \frac{d z_6}{\sqrt{\det G(k_2,u_1)} \sqrt{\det G(u_1,q)}}\,,
\label{eq: double_triangle_Baikov_intermediate}
\end{equation}
where there are two Gram determinants, $\det G(k_1,k_2,u_1)$ and $\det G(k_2,u_1,q)$, which have vanishing exponent in $D=4$. Inverting the change of variables, the first Gram determinant in eq.~\eqref{eq: double_triangle_Baikov_intermediate} becomes a quadratic polynomial in the Baikov variables,
\begin{equation}
P_2(z_5,z_6) = z_6 - \frac{z_5^2}{4}\,,
\end{equation}
where $z_5=2u_1 \cdot k_2$ is one of the propagators, whereas the second Gram determinant $\det G(u_1,q) = q^2$ is constant. 

Under the maximal cut, we thus obtain an integral over the extra variable $z_6$,
\begin{equation}
\LS \left( \begin{tikzpicture}[baseline={([yshift=-0.1cm]current bounding box.center)}] 
	\node[] (a) at (0,0) {};
	\node[] (a1) at (0.4,0) {};
	\node[] (a2) at (0.8,0) {};
	\node[] (b) at (0.4,-0.8) {};
	\node[] (p1) at ($(a)+(-0.2,0)$) {};
	\node[] (p2) at ($(b)+(-0.6,0)$) {};
	\node[] (p3) at ($(b)+(0.6,0)$) {};
	\node[] (p4) at ($(a2)+(0.2,0)$) {};
	\draw[line width=0.15mm] (b.center) -- (a.center);
	\draw[line width=0.15mm] (b.center) -- (a1.center);
	\draw[line width=0.15mm] (b.center) -- (a2.center);
	\draw[line width=0.5mm] (p1.center) -- (p4.center);
	\draw[line width=0.5mm] (p2.center) -- (p3.center);
\end{tikzpicture} \right) \propto \frac{1}{|q|} \, \LS \left( \int \frac{d z_6}{\sqrt{P_2(0,z_6)}} \right) =  \frac{1}{|q|} \, \LS \left( \int \frac{d z_6}{\sqrt{z_6}} \right)\,.
\label{eq: LS_double_triangle_inf}
\end{equation}
Let us note that, even though it would seem as if the $|q|$-scaling of the leading singularity is not correct, as we expect $|q|^0$, remember that $z_6 \sim q^2$, which amends the power counting. Moreover, we can realize that the integral has a higher-order pole, since under the change of variables $z_6=1/t^2$, we obtain
\begin{equation}
\LS \left( \begin{tikzpicture}[baseline={([yshift=-0.1cm]current bounding box.center)}] 
	\node[] (a) at (0,0) {};
	\node[] (a1) at (0.4,0) {};
	\node[] (a2) at (0.8,0) {};
	\node[] (b) at (0.4,-0.8) {};
	\node[] (p1) at ($(a)+(-0.2,0)$) {};
	\node[] (p2) at ($(b)+(-0.6,0)$) {};
	\node[] (p3) at ($(b)+(0.6,0)$) {};
	\node[] (p4) at ($(a2)+(0.2,0)$) {};
	\draw[line width=0.15mm] (b.center) -- (a.center);
	\draw[line width=0.15mm] (b.center) -- (a1.center);
	\draw[line width=0.15mm] (b.center) -- (a2.center);
	\draw[line width=0.5mm] (p1.center) -- (p4.center);
	\draw[line width=0.5mm] (p2.center) -- (p3.center);
\end{tikzpicture} \right) \propto \frac{1}{|q|} \, \LS \left( \int \frac{d t}{t^2} \right)\,.
\end{equation}
To this end, since the residue at the pole $t=0$ vanishes, we integrate instead the leading singularity between two roots of the Baikov polynomials, as discussed in sec.~\ref{sec:ch4_modus_operandi}. In this case, we have
\begin{subequations}
\begin{align}
\det G(k_1,k_2,u_1)\Big|_{\text{max cut}}& =-\frac{z_6^2}{4}\,, \qquad \qquad \enspace \ \det G(k_2,u_1)\Big|_{\text{max cut}}=z_6\,, \\[0.1cm]
\det G(k_2,u_1,q)\Big|_{\text{max cut}}& =-\frac{(z_6-q^2)^2}{4}\,, \qquad \det G(u_1,q)\Big|_{\text{max cut}}=q^2\,,
\end{align}
\end{subequations}
where we identify the roots $z_6=0$ and $z_6=q^2$. Integrating between them, we obtain~\cite{Parra-Martinez:2020dzs,Herrmann:2021tct,DiVecchia:2021bdo}
\begin{equation}
\label{eq: LS_double_triangle_unit_LS}
\LS \left( \right) \propto \frac{1}{|q|} \, \int_{q^2}^0 \frac{d z_6}{\sqrt{z_6}} \propto 1\,.
\end{equation}

Regarding the odd-parity contribution, we can add a dot to one matter propagator, and use eq.~\eqref{eq: LS_PM_one-loop_triangle_odd_parity} to find that
\begin{align}
\LS \left( %
\right) \nonumber \\
& \propto \LS \left( \int d z_6 \frac{\varepsilon \, (u_1 \cdot k_2)}{z_6 \sqrt{z_6-(u_1 \cdot k_2)^2}} \frac{1}{\sqrt{\det G(u_1,q)}} \right) =  0\,,
\end{align}
which vanishes since $u_1 \cdot k_2=0$ at the maximal cut.

\subsection{Two-loop 1SF diagrams}
\label{sec:ch4_two_loop_1SF}

Having analyzed the 0SF double-triangle diagram, let us now turn to the three 1SF diagrams in eq.~\eqref{eq: PM_two_loop_diagrams}. Following analogous steps as before, and with the loop-by-loop integration order $\{k_1 , k_2\}$, the first diagram leads to
\begin{equation}
\LS \left( \right) \propto \LS \left( \int \frac{x \, d z_8}{|q|^3 z_8 \sqrt{q^2(x^2-1)^2-4x^2 z_8^2}} \right)\,,
\end{equation}
where we add one extra Baikov variable $z_8=2u_1 \cdot k_2$. As can be seen, there is a simple pole at $z_8=0$. Even though we could take the residue at this pole, as explained in sec.~\ref{sec:ch1_LS}, we must be very careful when there are square roots involved. Consequently, we choose to rationalize the square root first. Using the change of variables in eq.~\eqref{eq: change_of_variables_rationalization_r1r2} from $z_8$ to $t_8$, we obtain
\begin{equation}
\LS \left( \begin{tikzpicture}[baseline={([yshift=-0.1cm]current bounding box.center)}] 
	\node[] (a) at (0,0) {};
	\node[] (b) at (0,-0.5) {};
	\node[] (a1) at (0.5,0) {};
	\node[] (b1) at (0.5,-0.5) {};
	\node[] (c) at (0,-1) {};
	\node[] (c1) at (0.5,-1) {};
	\node[] (p1) at ($(a)+(-0.2,0)$) {};
	\node[] (p2) at ($(c)+(-0.2,0)$) {};
	\node[] (p3) at ($(c1)+(0.2,0)$) {};
	\node[] (p4) at ($(a1)+(0.2,0)$) {};
	\draw[line width=0.15mm] (a.center) -- (b.center);
	\draw[line width=0.15mm] (a1.center) -- (b1.center);
	\draw[line width=0.15mm] (b1.center) -- (b.center);
	\draw[line width=0.15mm] (b.center) -- (c.center);
	\draw[line width=0.15mm] (c1.center) -- (b1.center);
	\draw[line width=0.5mm] (a.center) -- (a1.center);
	\draw[line width=0.5mm] (c.center) -- (c1.center);
	\draw[line width=0.5mm] (p1.center) -- (a.center);
	\draw[line width=0.5mm] (a1.center) -- (p4.center);
	\draw[line width=0.5mm] (p2.center) -- (c.center);
	\draw[line width=0.5mm] (c1.center) -- (p3.center);
\end{tikzpicture} \right) \propto \frac{x}{q^4(x^2-1)} \, \LS \left(\int \frac{d t_8}{1+t_8^2} \right) \propto \frac{x}{q^4 (x^2-1)}\,,
\end{equation}
in agreement with refs.~\cite{Parra-Martinez:2020dzs,Herrmann:2021tct,DiVecchia:2021bdo}, and where in the last step we can safely take the residue at either of the poles $t_8=\pm i$. Including the extra factors of $(|q|^2)^2$ that would come from dressing the graviton self-interaction vertices, the result would scale as $|q|^0$, as expected.

For the odd-parity sector, we can add $z_8=2u_1 \cdot k_2$ as an extra ISP in the numerator, which cancels the previous pole. Still, we can rationalize the square root with the same change of variables, leading to
\begin{equation}
\LS \left( {\times} \, 2u_1 {\cdot} k_2 \right) \propto \LS \left( \int \frac{x \, d z_8}{|q|^3 \sqrt{q^2(x^2-1)^2-4x^2 z_8^2}} \right) \propto \frac{1}{|q|^3} \, \LS \left( \int \frac{d t_8}{t_8} \right) \propto \frac{1}{|q|^3}\,.
\end{equation}

Turning to the second 1SF diagram, recall eq.~\eqref{eq: PM_two_loop_diagrams}, we need one extra Baikov variable $z_6=2u_1 \cdot k_2$, which leads to
\begin{equation}
\label{eq: LS_PM_box_bubble}
\LS \left( %
 \right) \propto \LS \left( \int \frac{x \, z_6 \, d z_6}{|q| \sqrt{q^2(x^2-1)^2-4x^2 z_6^2}} \right)\,.
\end{equation}
In this case, the result has one attribute that will also be present in many examples at higher loops: there is a linear factor of $z_6$ in the numerator while the square root only contains even powers of $z_6$. Thus, the change of variables $z_6 = \sqrt{z'_6}$ will simultaneously remove the numerator with the Jacobian and reduce the degree of the polynomial inside of the square root. In this case, it leads to a higher-order pole, as we already saw for the 0SF double-triangle diagram in eq.~\eqref{eq: LS_double_triangle_inf}. Integrating between the roots of the Baikov polynomials, in this case $z'_6=0$ and $z'_6=\frac{q^2 (x^2-1)^2}{4x^2}$, we obtain~\cite{Parra-Martinez:2020dzs,Herrmann:2021tct,DiVecchia:2021bdo}
\begin{equation}
\label{eq: LS_box_with_bubble}
\LS \left( \right) \propto \frac{1}{|q|} \, \int_{\frac{q^2 (x^2-1)^2}{4x^2}}^0 \frac{x \, d z'_6}{|q| \sqrt{q^2(x^2-1)^2-4x^2 z'_6}} \propto \frac{x^2-1}{x}\,.
\end{equation}

For the odd-parity sector, we can add $z_6=2u_1 \cdot k_2$ as an ISP in the numerator, which precludes using the previous trick. However, it allows us instead to use the change of variables in eq.~\eqref{eq: change_of_variables_rationalization_r1r2} from $z_6$ to $t_6$, which exposes a pole at $t_6=0$ with non-vanishing residue:
\begin{align}
\LS \left( %
 {\times} \, 2u_1 {\cdot} k_2 \right) & \propto \LS \left( \int \frac{x \, z_6^2 \, d z_6}{|q| \sqrt{q^2(x^2-1)^2-4x^2 z_6^2}} \right) \nonumber \\[0.1cm]
& \propto \frac{|q| \, (x^2-1)^2}{x^2} \, \LS \left( \int d t_6 \frac{(t_6^2+1)^2}{t_6^3} \right) \nonumber \propto \frac{|q| \, (x^2-1)^2}{x^2}\,.
\end{align}

Finally, we have the last diagram in eq.~\eqref{eq: PM_two_loop_diagrams}. In this case, we need to add two extra Baikov variables $z_6=k_2^2$ and $z_7=2u_1 \cdot k_2$ to match the two ISPs. For the even-parity sector, we obtain
\begin{align}
\LS \left( %
\right) \nonumber \\
& \propto \LS \left( \int \frac{x \, d z_6 d z_7}{\sqrt{4 z_6 - z_7^2} \sqrt{(x^2-1)^2(q^2-z_6)^2-4q^2x^2z_7^2}} \right)\,.
\end{align}
Notably, we encounter a new feature: the presence of two simultaneous square roots. However, as can be seen, the second square root has very simple roots with respect to $z_7$, and they do not introduce further square roots in $z_6$. Thus, we can use again the change of variables in eq.~\eqref{eq: change_of_variables_rationalization_r1r2} from $z_7$ to $t_7$, which gives
\begin{equation}
\LS \left( %
 \right) \propto  \LS \left( \int \frac{x \, d z_6 d t_7}{\sqrt{P_6(z_6,t_7)}} \right).
\end{equation}
The polynomial $P_6$ is of overall degree 6, but is only quadratic in $z_6$. Therefore, we can rationalize it once again with the same change of variables, in this case from $z_6$ to $t_6$, which finally yields two separate simple poles~\cite{Parra-Martinez:2020dzs,Herrmann:2021tct,DiVecchia:2021bdo}
\begin{equation}
\LS \left( %
 \right) \propto \frac{x}{x^2-1} \, \LS \left( \int \frac{d t_6 d t_7}{t_6 (1+t_7^2)} \right) \propto \frac{x}{x^2-1}\,.
\end{equation}

Regarding the odd-parity integral, we can choose to dot one of the matter propagators. In this case, using eq.~\eqref{eq: LS_PM_one-loop_triangle_odd_parity} for the leading singularity of the triangle with a dot, we obtain
\begin{align}
\label{eq: LS_PM_zig_zag_triangle}
\LS \left( %
\right) \nonumber \\
& \, \propto \LS \left( \int \frac{\varepsilon \, x \, z_7 \, d z_6 d z_7}{z_6 \sqrt{4 z_6 - z_7^2} \sqrt{(x^2-1)^2(q^2-z_6)^2-4q^2x^2z_7^2}} \right)\,.
\end{align}
As in eq.~\eqref{eq: LS_PM_box_bubble}, we can notice that the numerator is linear in $z_7$ while both square roots contain even powers. Thus, we can change variables to $z_7 = \sqrt{z_7'}\,$, which removes the numerator and makes both square roots linear in $z_7'$. Subsequently, we can use the transformation from eq.~\eqref{eq: change_of_variables_rationalization_r1r2} to rationalize both square roots simultaneously changing from $z_7'$ to $t_7$, such that~\cite{Parra-Martinez:2020dzs,Herrmann:2021tct,DiVecchia:2021bdo}
\begin{equation}
\LS \left( \begin{tikzpicture}[baseline={([yshift=-0.1cm]current bounding box.center)}] 
	\node[] (a) at (0,0) {};
	\node[] (a1) at (0.8,0) {};
	\node[] (b) at (0.4,-0.8) {};
	\node[] (b1) at (1.2,-0.8) {};
	\node[] (p1) at ($(a)+(-0.2,0)$) {};
	\node[] (p2) at ($(b)+(-0.6,0)$) {};
	\node[] (p3) at ($(b1)+(0.2,0)$) {};
	\node[] (p4) at ($(a1)+(0.6,0)$) {};
	\draw[line width=0.15mm] (b.center) -- (a.center);
	\draw[line width=0.15mm] (b.center) -- (a1.center);
		\draw[line width=0.15mm] (b1.center) -- (a1.center);
	\draw[line width=0.5mm] (p1.center) -- (p4.center);
	\draw[line width=0.5mm] (p2.center) -- (p3.center);
	\node at (0.4,0) [circle,fill,inner sep=1.5pt]{};
\end{tikzpicture} \right) \propto \frac{1}{|q|} \, \LS \left( \int \frac{d z_6 d t_7}{z_6 t_7} \right) \propto \frac{1}{|q|}\,.
\end{equation}

In summary, in this section we have shown that the 4 representative two-loop classical diagrams -- and thus all two-loop classical diagrams -- have algebraic leading singularity in both parity sectors. Therefore, this analysis demonstrates that the PM expansion is polylogarithmic up to two loops, including both conservative and dissipative effects, as already found in explicit computations~\cite{Bern:2019nnu,Bern:2019crd,Kalin:2020mvi,Mogull:2020sak,Parra-Martinez:2020dzs,Herrmann:2021tct,DiVecchia:2021bdo}.

\section{Three-loop diagrams}
\label{sec:ch4_three_loop}

In this section, we extend the classification of Feynman integral geometries appearing in the PM expansion to 3 loops. As shown in the previous sections, up to 2 loops all leading singularities are algebraic, which indicates that the associated function space is spanned by polylogarithms. At 3 loops, however, we will encounter the first non-trivial case: a diagram which depends on a K3 surface, as already identified in refs.~\cite{Dlapa:2022wdu,Ruf:2021egk}.

Following the analysis of tab.~\ref{tab:reduction_diagrams}, at 3 loops we need to compute the leading singularity for an integral of each parity for 14 different topologies. Since we already illustrated the use of the Baikov representation in the previous sections, here we will thus omit the details of the calculations and only provide the full derivation for the diagram involving a K3 surface in sec.~\ref{sec:ch4_three_loop_K3}; see, however, ref.~\cite{Frellesvig:2024zph} for more details. Overall, we obtain the results gathered in tab.~\ref{tab: 3-loop results 0SF and 1SF}, which is organized following the self-force order of the diagrams. Concretely, diagrams \diagramnumberingsign1 and \diagramnumberingsign2 contribute to 0SF order, while diagrams \diagramnumberingsign3 to \diagramnumberingsign14 are of 1SF order.\footnote{As noted in sec.~\ref{sec:ch4_classical_diagrams}, depending on their subsectors, superclassical diagrams can enter at different SF orders. In the table, they are included in the lowest SF order that they contribute to; see rows \diagramnumberingsign1 and \diagramnumberingsign9.} Within each SF order, we order the diagrams based on the number of propagators, from top sectors to subsectors, with the exception of those depending on non-trivial geometries, which we put first for better reference. In particular, this is the case for diagram \diagramnumberingsign3, which depends on a K3 surface given by a degree-6 polynomial $P_6(\vec{z})$; see the next subsection for details.

\newpage

\begin{table}[ht!]
    \centering
    \caption{Results for the leading singularity of one Feynman integral of each parity at 3 loops. The 0SF diagrams are in rows $\diagramnumberingsign1$ and $\diagramnumberingsign2$, and the 1SF diagrams in rows $\diagramnumberingsign3$ to $\diagramnumberingsign14$. To specify which of the loop momenta appear in the ISPs, we include the label for one propagator, where we introduce the short-hand notation $2 u_i \cdot k_j \to k_j$ for the matter propagators.}
    \label{tab: 3-loop results 0SF and 1SF}
\begin{tabular}{|C{0.4cm}|L{3.9cm}|C{2.2cm}|L{3.7cm}|C{2.6cm}|}
    \noalign{\global\arrayrulewidth=0.2ex}\hline
        \noalign{\global\arrayrulewidth=0.4pt} % Reset to default for subsequent lines
    & \multicolumn{2}{c|}{Even parity} & \multicolumn{2}{c|}{Odd parity} \\
    \hline
    \# & \multicolumn{1}{c|}{Integral} & LS & \multicolumn{1}{c|}{Integral} & LS \\
\end{tabular}
\end{table} 

\subsection{A K3 surface at three loops}
\label{sec:ch4_three_loop_K3}

\begin{figure}[t]
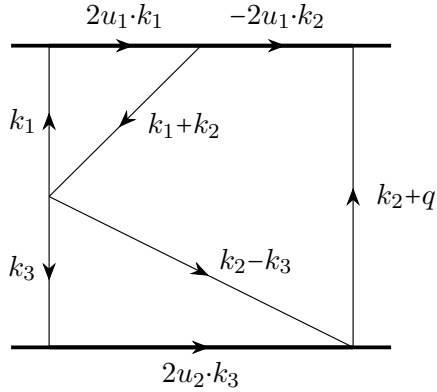

\centering
%
\caption{Parametrization for the diagram \diagramnumberingsign3 of tab.~\ref{tab: 3-loop results 0SF and 1SF}.}
\label{fig: param_3-loop_K3}
\end{figure}
In this subsection, we focus on the diagram in row \diagramnumberingsign3 of tab.~\ref{tab: 3-loop results 0SF and 1SF}, and provide the details for the computation of the leading singularity for one integral of each parity. In particular, with the parametrization in fig.~\ref{fig: param_3-loop_K3} and the loop-by-loop order $\{k_1 , k_3, k_2\}$, for this diagram there are two extra Baikov variables, $z_9=k_2^2$ and $z_{10}=2u_2 \cdot k_2$. 

For the even-parity sector, we can include $z_{10}$ as an extra factor in the numerator, which yields
\begin{equation}
\LS \left( \begin{tikzpicture}[baseline={([yshift=-0.1cm]current bounding box.center)}] 
	\node[] (a) at (0,0) {};
	\node[] (a1) at (0.5,0) {};
	\node[] (a2) at (1,0) {};
	\node[] (b) at (0,-0.5) {};
	\node[] (c) at (0,-1) {};
	\node[] (c1) at (1,-1) {};
	\node[] (p1) at ($(a)+(-0.2,0)$) {};
	\node[] (p2) at ($(c)+(-0.2,0)$) {};
	\node[] (p3) at ($(c1)+(0.2,0)$) {};
	\node[] (p4) at ($(a2)+(0.2,0)$) {};
	\draw[line width=0.15mm] (b.center) -- (a.center);
	\draw[line width=0.15mm] (b.center) -- (a1.center);
	\draw[line width=0.15mm] (c1.center) -- (a2.center);
	\draw[line width=0.15mm] (b.center) -- (c.center);
	\draw[line width=0.15mm] (c1.center) -- (b.center);
	\draw[line width=0.5mm] (p1.center) -- (p4.center);
	\draw[line width=0.5mm] (p2.center) -- (p3.center);
\end{tikzpicture} {\times} \, 2u_2 {\cdot} k_2 \right) \propto \LS \left( \int \frac{x \, z_{10} \, d z_9 d z_{10}}{\sqrt{z_9} \sqrt{4z_9-z_{10}^2} \sqrt{(x^2-1)^2(z_9-1)^2-4x^2z_{10}^2}} \right).
\end{equation}
Here, we have rescaled $\{z_9,z_{10}\} \to \{q^2 z_9, |q| z_{10} \}$ to move the $q$-dependence to the prefactor; in this case, the integral scales as $|q|^0$. As can be seen, the numerator is linear in $z_{10}$ while the square roots depend only quadratically. Thus, we can use the same trick as in eq.~\eqref{eq: LS_PM_zig_zag_triangle}: change variables to $z_{10} = \sqrt{z_{10}'}\,$ and afterwards use the transformation from eq.~\eqref{eq: change_of_variables_rationalization_r1r2} to rationalize both square roots simultaneously changing from $z_{10}'$ to $t_{10}$, leading to
\begin{align}
\LS \left( \begin{tikzpicture}[baseline={([yshift=-0.1cm]current bounding box.center)}] 
	\node[] (a) at (0,0) {};
	\node[] (a1) at (0.5,0) {};
	\node[] (a2) at (1,0) {};
	\node[] (b) at (0,-0.5) {};
	\node[] (c) at (0,-1) {};
	\node[] (c1) at (1,-1) {};
	\node[] (p1) at ($(a)+(-0.2,0)$) {};
	\node[] (p2) at ($(c)+(-0.2,0)$) {};
	\node[] (p3) at ($(c1)+(0.2,0)$) {};
	\node[] (p4) at ($(a2)+(0.2,0)$) {};
	\draw[line width=0.15mm] (b.center) -- (a.center);
	\draw[line width=0.15mm] (b.center) -- (a1.center);
	\draw[line width=0.15mm] (c1.center) -- (a2.center);
	\draw[line width=0.15mm] (b.center) -- (c.center);
	\draw[line width=0.15mm] (c1.center) -- (b.center);
	\draw[line width=0.5mm] (p1.center) -- (p4.center);
	\draw[line width=0.5mm] (p2.center) -- (p3.center);
\end{tikzpicture} {\times} \, 2u_2 {\cdot} k_2 \right) \propto & \ \LS \left( \int \frac{x \, d z_9 d z'_{10}}{\sqrt{z_9} \sqrt{4z_9-z'_{10}} \sqrt{(x^2-1)^2(z_9-1)^2-4x^2z'_{10}}} \right) \nonumber \\[0.1cm]
\propto & \ \LS \left( \int \frac{d z_9 \, d t_{10}}{t_{10} \sqrt{z_9}}  \right) \propto \mathlarger{\int}_{\frac{(x-1)^2}{(x+1)^2}}^{\frac{(x+1)^2}{(x-1)^2}} \frac{dz_9}{\sqrt{z_9}} \propto \frac{x}{x^2-1}\,.
\end{align}
Since we obtain a higher-order pole, in the last step we integrate the leading singularity between two roots of the Baikov polynomials, as discussed in sec.~\ref{sec:ch4_two_loop_0SF}. Hence, for the even-parity sector, the leading singularity is algebraic.

By contrast, for the odd-parity integral we do not have the ISP in the numerator, which prevents using the same trick. Instead, we obtain
\begin{equation}
\label{eq: LS_PM_3loop_K3_pre}
\LS \left( \begin{tikzpicture}[baseline={([yshift=-0.1cm]current bounding box.center)}] 
	\node[] (a) at (0,0) {};
	\node[] (a1) at (0.5,0) {};
	\node[] (a2) at (1,0) {};
	\node[] (b) at (0,-0.5) {};
	\node[] (c) at (0,-1) {};
	\node[] (c1) at (1,-1) {};
	\node[] (p1) at ($(a)+(-0.2,0)$) {};
	\node[] (p2) at ($(c)+(-0.2,0)$) {};
	\node[] (p3) at ($(c1)+(0.2,0)$) {};
	\node[] (p4) at ($(a2)+(0.2,0)$) {};
	\draw[line width=0.15mm] (b.center) -- (a.center);
	\draw[line width=0.15mm] (b.center) -- (a1.center);
	\draw[line width=0.15mm] (c1.center) -- (a2.center);
	\draw[line width=0.15mm] (b.center) -- (c.center);
	\draw[line width=0.15mm] (c1.center) -- (b.center);
	\draw[line width=0.5mm] (p1.center) -- (p4.center);
	\draw[line width=0.5mm] (p2.center) -- (p3.center);
\end{tikzpicture} \right) \propto \frac{x}{|q|} \, \LS \left( \int \frac{d z_9 d z_{10}}{\sqrt{z_9} \sqrt{4z_9-z_{10}^2} \sqrt{(x^2-1)^2(z_9-1)^2-4x^2z_{10}^2}} \right),
\end{equation}
where we transferred the $q$-dependence to the prefactor with the previous rescaling. In this case, we can first rationalize the square root $\sqrt{z_9}$ with the transformation $z_9 = t_9^2$, which leads to
\begin{align}
\LS \left( \begin{tikzpicture}[baseline={([yshift=-0.1cm]current bounding box.center)}] 
	\node[] (a) at (0,0) {};
	\node[] (a1) at (0.5,0) {};
	\node[] (a2) at (1,0) {};
	\node[] (b) at (0,-0.5) {};
	\node[] (c) at (0,-1) {};
	\node[] (c1) at (1,-1) {};
	\node[] (p1) at ($(a)+(-0.2,0)$) {};
	\node[] (p2) at ($(c)+(-0.2,0)$) {};
	\node[] (p3) at ($(c1)+(0.2,0)$) {};
	\node[] (p4) at ($(a2)+(0.2,0)$) {};
	\draw[line width=0.15mm] (b.center) -- (a.center);
	\draw[line width=0.15mm] (b.center) -- (a1.center);
	\draw[line width=0.15mm] (c1.center) -- (a2.center);
	\draw[line width=0.15mm] (b.center) -- (c.center);
	\draw[line width=0.15mm] (c1.center) -- (b.center);
	\draw[line width=0.5mm] (p1.center) -- (p4.center);
	\draw[line width=0.5mm] (p2.center) -- (p3.center);
\end{tikzpicture} \right) & \propto \frac{x}{|q|} \, \LS \left( \int\! \frac{d t_9 d z_{10}}{\sqrt{4 t_9^2 -z_{10}^2} \sqrt{(x^2-1)^2(t_9^2-1)^2-4x^2 z_{10}^2}} \right) \nonumber \\[0.1cm]
& \propto \frac{x}{|q|} \, \int\! \frac{d t_1 d t_2}{\sqrt{t_2^2 (t_1^2-1)^2 (x^2-1)^2 - 4 x^2 t_1^2 (t_2^2+1)^2}} \equiv \frac{x}{|q|} \, \int\! \frac{ dt_1 dt_2}{\sqrt{P_6(t_1,t_2)}}\,,
\label{eq: PM_3loop_K3_poly}
\end{align}
where we used the transformation in eq.~\eqref{eq: change_of_variables_rationalization_r1r2} from $z_{10}$ to $t_{10}$ to rationalize the first square root, and relabeled $t_9 \to t_1$ and $t_{10} \to t_2$ to simplify the notation. As can be seen, we obtain an integral over the square root of a degree-6 polynomial $P_6(t_1,t_2)$ in two variables, and which is quartic in each of them. Comparing with eqs.~\eqref{eq: LS_geometry_intro}--\eqref{eq: condition_degree_weight_CY}, we see that it satisfies the Calabi--Yau condition in $\mathbb{WP}^{1,1,1,3}$. Therefore, the leading singularity defines an integral over a Calabi--Yau twofold, also known as a K3 surface, which is in agreement with refs.~\cite{Ruf:2021egk,Dlapa:2022wdu}. Thus, we have (finally) found the first geometry beyond elliptics in the thesis!

The result above indicates that the corresponding function space for the integral is spanned by integrals over this K3 surface. However, explicit computations at 3 loops have shown that it can be written as a product of complete elliptic integrals~\cite{Bern:2021dqo,Dlapa:2021npj,Bern:2022jvn,Dlapa:2022wdu,Jakobsen:2023ndj}, which seems to contradict the previous result. To clarify the origin of this apparent discrepancy, let us follow refs.~\cite{Ruf:2021egk,Dlapa:2022wdu} and study this geometry using differential equations, recall sec.~\ref{sec:ch1_DEs}. In particular, the associated Picard--Fuchs operator in $D=4$ is given by~\cite{Ruf:2021egk,Dlapa:2022wdu}
\begin{equation}
\label{eq: PF_K3_3loops}
\mathcal{L}_3^{(\text{K3})} =
 \frac{d^3}{d x^3} - \frac{6 x}{1 - x^2} \frac{d^2}{d x^2} + \frac{1 - 4 x^2 + 7 x^4}{x^2 (1 - x^2)^2} \frac{d}{d x} - \frac{1 + x^2}{x^3 (1 - x^2)}\,.
\end{equation}
As it is a differential operator of order 3, it also indicates that the geometry is a K3 surface. 

However, due to the nature of the PM expansion, which only depends on a single dimensionless parameter $x$, the Picard--Fuchs operator is univariate. Importantly, as shown in refs.~\cite{Joyce:1972,Joyce:1973,Verrill:1996,Doran:1998hm,Primo:2017ipr}, the Picard--Fuchs operator $\mathcal{L}_3$ of a univariate K3 surface must be the symmetric square of a second-order operator $\mathcal{L}_2$, which defines an elliptic curve. This means that the three independent solutions to the equation $\mathcal{L}_3 \Psi=0$ are not generic, but take the form $\Psi = \{\psi_1^2,\psi_2^2,\psi_1 \psi_2\}$, where the $\psi_i$ satisfy $\mathcal{L}_2 \psi_i=0$. Hence, despite the initial expectation of an integral over a K3 surface, the symmetric square condition explains why the function space at 3 loops contains products of elliptic integrals, thereby solving the initial discrepancy.

To verify that the Picard--Fuchs operator in eq.~\eqref{eq: PF_K3_3loops} is indeed a symmetric square, let us consider a generic third-order operator
\begin{equation}
    \mathcal{L}_3= \frac{d^3}{dx^3}+ C_2(x) \frac{d^2}{dx^2}+ C_1(x) \frac{d}{dx}+C_0(x)\,.
\end{equation}
As shown in ref.~\cite{Primo:2017ipr}, the operator satisfies the symmetric square condition if the equation
\begin{equation}
\label{eq: symmetric_square_condition}
    C_0(x) = 4\, \alpha_0(x) \alpha_1(x)+2\, \frac{d \alpha_0(x)}{dx}
\end{equation}
is satisfied, where
\begin{equation}
    \alpha_1(x)=\frac{C_2(x)}{3}\,, \qquad \qquad \alpha_0(x)=\frac{C_1(x)}{4}-\frac{\alpha_1(x)^2}{2} - \frac{1}{4} \frac{d \alpha_1(x)}{dx}\,.
\end{equation}
Then, the corresponding second-order differential operator is given by
\begin{equation}
    \mathcal{L}_2= \frac{d^2}{dx^2}+ \alpha_1(x) \frac{d}{dx}+\alpha_0(x)\,.
\end{equation}
In our case, we find that the operator $\mathcal{L}_3^{(\text{K3})}$ in eq.~\eqref{eq: PF_K3_3loops} satisfies the condition in eq.~\eqref{eq: symmetric_square_condition}, and becomes the symmetric square of~\cite{Ruf:2021egk,Dlapa:2022wdu}
\begin{equation}
\label{eq: PF_K3_3loops_L2}
\mathcal{L}_2^{(\text{K3})} =
 \frac{d^2}{d x^2} + \frac{2x}{x^2 -1} \frac{d}{d x} + \frac{1}{4 x^2}\,.
\end{equation}
Since the solutions to $\mathcal{L}_2^{(\text{K3})}$ are complete elliptic functions, the solutions to $\mathcal{L}_3^{(\text{K3})}$ are therefore products of these functions. All in all, this odd-parity integral is the only case in tab.~\ref{tab: 3-loop results 0SF and 1SF} which does not have an algebraic leading singularity. Thus, following the discussion in sec.~\ref{sec:ch4_soft_expansion}, our analysis indicates that the 3-loop integrals contributing to the dissipative dynamics are polylogarithmic, while the conservative regime can also contain products of elliptic integrals, which completely matches explicit calculations~\cite{Bern:2021dqo,Dlapa:2021npj,Bern:2021yeh,Bern:2022jvn,Dlapa:2022lmu,Dlapa:2023hsl,Jakobsen:2023ndj,Damgaard:2023ttc,Jakobsen:2023hig}.

Lastly, let us highlight that the diagram in fig.~\ref{fig: param_3-loop_K3} is not the only one which depends on this K3 surface. Based on sec.~\ref{sec:ch4_relating_geometries}, any non-planar variation obtained using the unraveling of eq.~\eqref{eq: unraveling_matter_props}, as well as its higher-loop superclassical iterations (see eq.~\eqref{eq: reduction_superclassical}), trivially depend on the same geometry. For instance, our analysis directly predicts that the integrals
\begin{equation}
\,,
\end{equation}
depend on the same K3 surface, which is in agreement with explicit computations~\cite{Bern:2022jvn,Klemm:2024wtd}.

\section{Four-loop diagrams}
\label{sec:ch4_four_loop}

In this section, we conclude our classification of Feynman integral geometries in the PM expansion with the analysis at 4 loops. As shown in tab.~\ref{tab:reduction_diagrams}, in this case there are 70 different topologies which need to be individually studied. The results of the analysis are presented in tabs.~\ref{tab: 4-loop results 0SF and 1SF} and~\ref{tab: 4-loop results 2SF}, which are also organized with respect to the SF order and follow the same convention as tab.~\ref{tab: 3-loop results 0SF and 1SF}. Concretely, the first two rows in tab.~\ref{tab: 4-loop results 0SF and 1SF} contain the 0SF diagrams, while the 1SF diagrams are in rows \diagramnumberingsign3 to \diagramnumberingsign36. Tab.~\ref{tab: 4-loop results 2SF} contains all the diagrams contributing to 2SF order. Notably, there are 5 truly non-planar diagrams (in the sense of sec.~\ref{sec:ch4_relating_geometries}), corresponding to rows $\diagramnumberingsign 41$, $\diagramnumberingsign43$, $\diagramnumberingsign47$, $\diagramnumberingsign48$ and $\diagramnumberingsign55$ in tab.~\ref{tab: 4-loop results 2SF}.

As can be seen, at 4 loops there are 8 distinct diagrams involving non-trivial geometries, which we already presented in fig.~\ref{fig: diagrams_nontrivial_PM}. In particular, at 1SF order (tab.~\ref{tab: 4-loop results 0SF and 1SF}), diagram \diagramnumberingsign3 depends on a CY threefold, while diagrams \diagramnumberingsign4 to \diagramnumberingsign6 involve the same K3 surface that appeared at 3 loops (recall sec.~\ref{sec:ch4_three_loop_K3}). At 2SF order (tab.~\ref{tab: 4-loop results 2SF}), diagram \diagramnumberingsign37 involves a different CY threefold. In addition, diagrams \diagramnumberingsign38 and \diagramnumberingsign39 depend on the 3-loop K3 surface, while diagram \diagramnumberingsign40 introduces a second K3 surface. Among these, the geometries of the 1SF diagrams and diagram \diagramnumberingsign40 were first identified in ref.~\cite{Klemm:2024wtd} using differential equations. In the following subsections, we analyze in detail the computation of the leading singularity for these 8 diagrams, especially focusing on the parity sector where the non-trivial geometries arise. 

\newpage

\begin{table}[ht!]
    \centering
    \caption{Results for the leading singularity of one Feynman integral of each parity at 4 loops. The 0SF diagrams are in rows $\diagramnumberingsign1$ and $\diagramnumberingsign2$, and the 1SF diagrams in rows $\diagramnumberingsign3$ to $\diagramnumberingsign36$. To specify which of the loop momenta appear in the ISPs, we include the label for one propagator, where we introduce the short-hand notation $2 u_i \cdot k_j \to k_j$ for the matter propagators.}
    \label{tab: 4-loop results 0SF and 1SF}
    \vspace{0.1cm}
\begin{tabular}{|C{0.4cm}|L{3.9cm}|C{2.2cm}|L{3.7cm}|C{3.4cm}|}
    \noalign{\global\arrayrulewidth=0.2ex}\hline
        \noalign{\global\arrayrulewidth=0.4pt} % Reset to default for subsequent lines
    & \multicolumn{2}{c|}{Even parity} & \multicolumn{2}{c|}{Odd parity} \\
    \hline
    \# & \multicolumn{1}{c|}{Integral} & LS & \multicolumn{1}{c|}{Integral} & LS \\
\end{tabular}
\end{table} 

\newpage

\begin{table}[ht!]
    \centering
    \begin{tabular}{|C{0.4cm}|L{3.9cm}|C{2.5cm}|L{3.7cm}|C{3.1cm}|}
    \noalign{\global\arrayrulewidth=0.2ex}\hline
        \noalign{\global\arrayrulewidth=0.4pt} % Reset to default for subsequent lines
    & \multicolumn{2}{c|}{Even parity} & \multicolumn{2}{c|}{Odd parity} \\
    \hline
    \# & \multicolumn{1}{c|}{Integral} & LS & \multicolumn{1}{c|}{Integral} & LS \\
\end{tabular}
\end{table} 

\newpage

\begin{table}[ht!]
    \centering
    \begin{tabular}{|C{0.4cm}|L{3.9cm}|C{2.5cm}|L{3.7cm}|C{3.1cm}|}
    \noalign{\global\arrayrulewidth=0.2ex}\hline
        \noalign{\global\arrayrulewidth=0.4pt} % Reset to default for subsequent lines
    & \multicolumn{2}{c|}{Even parity} & \multicolumn{2}{c|}{Odd parity} \\
    \hline
    \# & \multicolumn{1}{c|}{Integral} & LS & \multicolumn{1}{c|}{Integral} & LS \\ \noalign{\global\arrayrulewidth=0.2ex}\hline
        \noalign{\global\arrayrulewidth=0.4pt} % Reset to default for subsequent lines

%Diagram 30
& & & & \\
\end{tabular}
\end{table}

\vspace{1cm}

\begin{table}[ht!]
    \centering
    \caption{Results for the leading singularity of one Feynman integral of each parity at 4 loops and 2SF order. To specify which of the loop momenta appear in the ISPs, we include the label for one propagator, where we introduce the short-hand notation $2 u_i \cdot k_j \to k_j$ for the matter propagators. Moreover, the superscript $(6D)$ denotes that the leading singularity has been computed in $D=6$.}
    \label{tab: 4-loop results 2SF}
    \vspace{0.1cm}
    \begin{tabular}{|C{0.4cm}|L{3.85cm}|C{2.95cm}|L{4.1cm}|C{2.4cm}|}
    \noalign{\global\arrayrulewidth=0.2ex}\hline
        \noalign{\global\arrayrulewidth=0.4pt} % Reset to default for subsequent lines
    & \multicolumn{2}{c|}{Even parity} & \multicolumn{2}{c|}{Odd parity} \\
    \hline
    \# & \multicolumn{1}{c|}{Integral} & LS & \multicolumn{1}{c|}{Integral} & LS \\
\end{tabular}
\end{table}

\newpage

\begin{table}[H]
    \centering
    \vspace{0.1cm}
    \begin{tabular}{|C{0.4cm}|L{3.85cm}|C{2.95cm}|L{4.1cm}|C{2.4cm}|}
    \noalign{\global\arrayrulewidth=0.2ex}\hline
        \noalign{\global\arrayrulewidth=0.4pt} % Reset to default for subsequent lines
    & \multicolumn{2}{c|}{Even parity} & \multicolumn{2}{c|}{Odd parity} \\
    \hline
    \# & \multicolumn{1}{c|}{Integral} & LS & \multicolumn{1}{c|}{Integral} & LS \\
\end{tabular}
\end{table}

\newpage

\begin{table}[ht!]
    \centering
    \vspace{0.1cm}
    \begin{tabular}{|C{0.4cm}|L{3.85cm}|C{2.95cm}|L{4.1cm}|C{2.4cm}|}
    \noalign{\global\arrayrulewidth=0.2ex}\hline
        \noalign{\global\arrayrulewidth=0.4pt} % Reset to default for subsequent lines
    & \multicolumn{2}{c|}{Even parity} & \multicolumn{2}{c|}{Odd parity} \\
    \hline
    \# & \multicolumn{1}{c|}{Integral} & LS & \multicolumn{1}{c|}{Integral} & LS \\
\end{tabular}
\end{table}

\subsection{Non-trivial geometries I: Calabi--Yau threefolds}
\label{sec:ch4_four_loop_CYs}

In this subsection, we analyze in detail the 4-loop cases which depend on a Calabi--Yau threefold. In particular, this corresponds to the diagrams in row \diagramnumberingsign3 of tab.~\ref{tab: 4-loop results 0SF and 1SF} and in row \diagramnumberingsign37 of tab.~\ref{tab: 4-loop results 2SF}. Since these are the first three-dimensional Calabi--Yau geometries of the thesis, we will also show how the integrals can be related to such intricate geometries in one parity sector, while having an algebraic leading singularity in the opposite.

In sec.~\ref{sec:ch4_four_loop_CYs_1}, we begin by considering diagram \diagramnumberingsign3 of tab.~\ref{tab: 4-loop results 0SF and 1SF}, which contributes to 1SF order. The corresponding CY integral in the odd-parity sector was first identified using differential equations in ref.~\cite{Klemm:2024wtd}, where it was also brought into $\varepsilon$-factorized form (recall sec.~\ref{sec:ch1_DEs}). Then, in sec.~\ref{sec:ch4_four_loop_CYs_2}, we turn to diagram \diagramnumberingsign37, which contributes to 2SF order. This CY threefold integral, which we reported in ref.~\cite{Frellesvig:2023bbf}, corresponded to the first discovery of such a geometry in gravitational-wave physics. Here, we will analyze the geometry via the leading singularity and, in the next chapter, we will develop a method to bring its differential equation into $\varepsilon$-factorized form.

\subsubsection{A CY threefold at 1SF order}
\label{sec:ch4_four_loop_CYs_1}

\begin{figure}[t]
\centering

\caption{Parametrization for the diagram $\diagramnumberingsign3$ of tab.~\ref{tab: 4-loop results 0SF and 1SF}, which depends on a CY threefold in the odd-parity sector.}
\label{fig: param_CY_1SF}
\end{figure}

Let us then focus first on the diagram in row \diagramnumberingsign3 of tab.~\ref{tab: 4-loop results 0SF and 1SF}, which is parametrized as given in fig.~\ref{fig: param_CY_1SF}. Following the loop-by-loop order $\{k_1 , k_2, k_3, k_4\}$, we need four extra Baikov variables: $z_{11}=k_3^2$, $z_{12}=k_4^2$, $z_{13}=2u_2 \cdot k_3$ and $z_{14}=2u_2 \cdot k_4$. For all forthcoming 4-loop diagrams, we will transfer the $q$-dependence to the prefactor by rescaling the even- and odd-parity variables as $z_i \to q^2\, z_i$ and $z_i \to |q| \, z_i$, respectively, as it will make the final expressions more compact.

For the even parity, we can consider the scalar integral, which leads to
\begin{align}
& \LS \left( \, \begin{tikzpicture}[baseline={([yshift=-0.1cm]current bounding box.center)}] 
    \node[] (a) at (-0.5,0) {};
	\node[] (a1) at (0,0) {};
    \node[] (a2) at (1,0) {};
    \node[] (a3) at (1.5,0) {};
	\node[] (b) at (0.5,-0.5) {};
	\node[] (c) at (0,-1) {};
    \node[] (c1) at (1,-1) {};
    \draw[line width=0.15mm] (c.center) --  (a.center);
	\draw[line width=0.15mm] (a1.center) --  (b.center);
	\draw[line width=0.15mm] (c.center) --  (b.center);
	\draw[line width=0.15mm] (b.center) --  (c1.center);
	\draw[line width=0.5mm] (a.center) --  (a3.center);
	\draw[line width=0.15mm] (b.center) --  (a2.center);
	\draw[line width=0.5mm] (c.center) --  (c1.center);
	\draw[line width=0.5mm] (-0.65,0) -- (a.center);
	\draw[line width=0.5mm] (a3.center) -- (1.65,0);
	\draw[line width=0.5mm] (-0.65,-1) -- (c.center);
	\draw[line width=0.15mm] (c1.center) --  (a3.center);
	\draw[line width=0.5mm] (c1.center) -- (1.65,-1);
    \end{tikzpicture} \, \right) \propto \LS \Bigg( \int \frac{x^2 \, d z_{11} \, \cdots \, d z_{14}}{\sqrt{z_{11}} \sqrt{4z_{11}-z_{13}^2} \sqrt{(x^2-1)^2(z_{12}-1)^2-4x^2z_{14}^2}} \nonumber \\
   & \! \times \frac{1}{\sqrt{(x^2-1)^2 z_{11}^2 + z_{12} \left( (x^2-1)^2 z_{12} - 4 x^2 z_{13} (z_{13} + z_{14}) \right)  - 
 2 z_{11} \left( (x^2-1)^2 z_{12} + 2 x^2 z_{14} (z_{13} + z_{14}) \right)}} \Bigg).
\end{align}
Now, we can use the change of variables in eq.~\eqref{eq: change_of_variables_rationalization_r1r2} to rationalize the first and second square roots simultaneously changing from $z_{11}$ to $t_{11}$, and the third square root changing from $z_{12}$ to $t_{12}$. Then, we obtain a denominator with a single square root, which is quadratic in $z_{14}$. Therefore, we can rationalize it again with the same transformation from $z_{14}$ to $t_{14}$. This reveals a simple pole at $t_{14}=0$, and taking the residue yields a single square root which is quadratic in $z_{13}$. Hence, it can be rationalized again with the same change of variables to $t_{13}$, leading to an algebraic leading singularity,
\begin{equation}
\LS \left( \, \, \right) \propto \frac{x}{x^2-1} \, \LS \left( \int \frac{dt_{11} \, dt_{12} \,  dt_{13}}{\sqrt{t_{11}} (t_{11}-1) t_{12} t_{13}} \right) \propto \frac{x}{x^2-1} \,.
\end{equation}

By contrast, in the odd-parity sector, we can add a dot to one matter propagator, which modifies the scaling to $|q|^{-1}$ and gives
\begin{align}
& \LS \left( \, %
 \, \right) \nonumber \\
    & \propto \frac{1}{|q|} \, \LS \Bigg( \int \frac{\varepsilon \, x^2 \,  \left( z_{13}+2z_{14} \right) \, d z_{11} \, \cdots \, d z_{14}}{\left( z_{11} + z_{14} (z_{13} + z_{14})
 \right)\sqrt{z_{11}} \sqrt{4z_{11}-z_{13}^2} \sqrt{(x^2-1)^2(z_{12}-1)^2-4x^2z_{14}^2}} \nonumber \\
    & \! \times \frac{1}{\sqrt{(x^2-1)^2 z_{11}^2 + z_{12} \left( (x^2-1)^2 z_{12} - 4 x^2 z_{13} (z_{13} + z_{14}) \right)  - 
 2 z_{11} \left( (x^2-1)^2 z_{12} + 2 x^2 z_{14} (z_{13} + z_{14}) \right)}} \Bigg).
\end{align}
As can be seen, the dot generates extra factors both in the numerator and denominator, which come from dotting the triangle subgraph, recall eq.~\eqref{eq: LS_PM_one-loop_triangle_odd_parity}. Notably, this extra factors depend on $z_{11}$, $z_{13}$ and $z_{14}$, which prevents us from using the change of variables employed in the even-parity integral, since here it would produce further square roots at the end. To circumvent this, we can rationalize the first square root with $z_{11} = t_{11}^2$, and use eq.~\eqref{eq: change_of_variables_rationalization_r1r2} to rationalize the second square root by changing from $z_{13}$ to $t_{13}$, and the third square root from $z_{12}$ to $t_{12}$, respectively. The combined effect is a single square root in $z_{14}$ which, after rationalizing with eq.~\eqref{eq: change_of_variables_rationalization_r1r2}, reveals a simple pole in $t_{14}$. Rescaling $t_{12} \to t_{12}/t_{11}$ and relabeling $\{ t_{11}, t_{12}, t_{13} \}  \to \{ t_1, t_2, t_3 \}$, we finally obtain
\begin{equation}
\label{eq: LS_PM_4loop_CY_1}
\LS \left( \, \begin{tikzpicture}[baseline={([yshift=-0.1cm]current bounding box.center)}] 
    \node[] (a) at (-0.5,0) {};
	\node[] (a1) at (0,0) {};
    \node[] (a2) at (1,0) {};
    \node[] (a3) at (1.5,0) {};
	\node[] (b) at (0.5,-0.5) {};
	\node[] (c) at (0,-1) {};
    \node[] (c1) at (1,-1) {};
    \draw[line width=0.15mm] (c.center) --  (a.center);
	\draw[line width=0.15mm] (a1.center) --  (b.center);
	\draw[line width=0.15mm] (c.center) --  (b.center);
	\draw[line width=0.15mm] (b.center) --  (c1.center);
	\draw[line width=0.5mm] (a.center) --  (a3.center);
	\draw[line width=0.15mm] (b.center) --  (a2.center);
	\draw[line width=0.5mm] (c.center) --  (c1.center);
	\draw[line width=0.5mm] (-0.65,0) -- (a.center);
	\draw[line width=0.5mm] (a3.center) -- (1.65,0);
	\draw[line width=0.5mm] (-0.65,-1) -- (c.center);
	\draw[line width=0.15mm] (c1.center) --  (a3.center);
	\draw[line width=0.5mm] (c1.center) -- (1.65,-1);
	\node at (0.5,-1) [circle,fill,inner sep=1.5pt]{};
    \end{tikzpicture} \, \right) \propto \frac{\varepsilon\,x^2}{|q| \sqrt{x^2{-}1}} \int \frac{d t_1 d t_2 d t_3}{\sqrt{P_8(t_1,t_2,t_3)}}\,,
\end{equation}
where
\begin{align}
P_8(t_1,t_2,t_3) \equiv &\,\, -t_2^2 t_3^2 (t_1^2 - 1)^2 (1 - x^6) + 
 2 t_2 t_3^3 (t_1^2 - 1) (t_1^2 + t_2^2) x (1 + x^4) + 
 4 t_2 t_3 (t_1^2 + t_2^2) (t_1^2 + t_3^2) x^3  \nonumber \\[0.1cm]
&\, + \left( 4 t_1^2 t_2^2 - t_2^2 t_3^2 (t_1^2 + 3) (t_1^2 - 1) - t_3^4 (t_1^2 + t_2^2)^2 \right) x^2(1-x^2)\,.
\end{align}
In this case, we obtain an integral over the square root of a degree-8 polynomial $P_8(t_1,t_2,t_3)$ in three variables, which is quartic in each of them. Comparing with eqs.~\eqref{eq: LS_geometry_intro}--\eqref{eq: condition_degree_weight_CY}, as well as with tab.~\ref{tab:geometries_intro}, we see that it satisfies the Calabi--Yau condition in $\mathbb{WP}^{1,1,1,1,4}$. Therefore, the leading singularity indicates that the odd-parity integral depends on a Calabi--Yau threefold.

From the perspective of the Picard--Fuchs operator, in $D=4-2\varepsilon$ dimensions, we have~\cite{Klemm:2024wtd}
\begin{equation}
\label{eq: PF_PM_CY_1_eps}
\mathcal{L}_4^{(\text{CY}_3)} = \frac{d^4}{d x^4} + \sum_{j=0}^{3} \frac{C_j(x,\varepsilon)}{\left( x(1-x^2)(1+x^2) \right)^{4-j}} \frac{d^j}{d x^j}\,,
\end{equation}
where the $C_j(x,\varepsilon)$ are rational polynomials in $x$ and $\varepsilon$. The explicit form of these polynomials will not be relevant to us. Importantly, in the denominator we only have singularities at the physical poles $x=0$ and $x=\pm 1$, which respectively correspond to the high-energy and the static limit, recall sec.~\ref{sec:ch4_soft_expansion}. In the case of $D=4$, the operator becomes~\cite{Klemm:2024wtd}
\begin{align}
\label{eq: PF_PM_CY_1}
&\, \mathcal{L}_4^{(\text{CY}_3)}\Big|_{\varepsilon=0} = 
 \frac{d^4}{d x^4} + \frac{2 - 16 x^2 - 10 x^4}{x(1-x^4)} \frac{d^3}{d x^3} + \frac{1 - 48 x^2 + 70 x^4 + 48 x^6 + 25 x^8}{x^2 (1 - x^4)^2} \frac{d^2}{d x^2} \nonumber \\
& \quad  - \frac{1 + 33 x^2 - 242 x^4 + 78 x^6 -47 x^8 - 15 x^{10}}{x^3 (1 - x^2)^2 (1 + x^2)^3} \frac{d}{d x} + \frac{1 + 2 x^2 + 159 x^4 -452 x^6 + 159 x^8+2x^{10}+x^{12}}{x^4 (1 - x^2)^2 (1+x^2)^4}\,.
\end{align}
As it is an irreducible differential operator of order 4, it is indeed compatible with a three-dimensional CY geometry, as originally identified in ref.~\cite{Klemm:2024wtd}. Importantly, unlike for the 3-loop diagram of sec.~\ref{sec:ch4_three_loop_K3}, here the function space is not spanned by products of elliptic integrals, but actual integrals over the CY threefold. In fact, it was recently found in ref.~\cite{Driesse:2024feo} that the integrals over this CY geometry carry on to the physical observables, and actually contribute to the 5PM 1SF dissipative sector.

\subsubsection{A CY threefold at 2SF order}
\label{sec:ch4_four_loop_CYs_2}

Let us then turn to the diagram in row \diagramnumberingsign37 of tab.~\ref{tab: 4-loop results 2SF}, which we parametrize following fig.~\ref{fig: param_CY_2SF}. For the even-parity sector, we can simply consider the scalar integral. Under the loop-by-loop order $\{k_1 , k_4, k_3, k_2\}$, we need three extra Baikov variables, which we label as $z_{12}=(k_3+q)^2$, $z_{13}=k_2^2$ and $z_{14}=2u_2 \cdot k_2$. Then, the leading singularity is given by
\begin{align}
\LS \left( \begin{tikzpicture}[baseline={([yshift=-0.1cm]current bounding box.center)}] 
	\node[] (a) at (0,0) {};
	\node[] (a1) at (0.75,0) {};
	\node[] (a2) at (1.5,0) {};
	\node[] (b) at (0,-0.5) {};
	\node[] (b1) at (0.5,-0.5) {};
	\node[] (b2) at (1.5,-0.5) {};
	\node[] (c) at (0,-1) {};
	\node[] (c1) at (0.75,-1) {};
	\node[] (c2) at (1.5,-1) {};
	\node[] (p1) at ($(a)+(-0.2,0)$) {};
	\node[] (p2) at ($(c)+(-0.2,0)$) {};
	\node[] (p3) at ($(c2)+(0.2,0)$) {};
	\node[] (p4) at ($(a2)+(0.2,0)$) {};
	\draw[line width=0.15mm] (c.center) -- (a.center);
	\draw[line width=0.15mm] (b.center) -- (b2.center);
	\draw[line width=0.15mm] (c2.center) -- (a2.center);
	\draw[line width=0.15mm] (b2.center) -- (c1.center);
	\draw[line width=0.15mm] (b.center) -- (a1.center);
	\draw[line width=0.5mm] (p1.center) -- (p4.center);
	\draw[line width=0.5mm] (p2.center) -- (p3.center);
\end{tikzpicture} \right) \propto & \, \frac{1}{q^2} \, \LS \left( \int \frac{x \, d z_{12} d z_{13} d z_{14}}{\sqrt{z_{12}} \sqrt{z_{13}} \sqrt{(x^2-1)^2 (z_{13}-1)^2 -4x^2 z_{14}^2}} \right. \nonumber \\
& \, \left. \times \, \frac{1}{\sqrt{z_{14}^2 (z_{12}-1)^2 - 4 z_{12} z_{13}(z_{12} + z_{13} -1)}} \right).
\end{align}
At this point, we can rationalize the third square root changing from $z_{14}$ to $t_{14}$ as in eq.~\eqref{eq: change_of_variables_rationalization_r1r2}. Afterwards, we can rescale $t_{14} \to t_{14}/(\sqrt{z_{12}}\sqrt{z_{13}})$, which removes both the first and the second square roots. Relabeling then $\{ z_{12}, z_{13}, t_{14} \}  \to \{ t_1, t_2, t_3 \}$, we obtain
\begin{equation}
\label{eq: LS_PM_4loop_CY_2}
\LS \left( \begin{tikzpicture}[baseline={([yshift=-0.1cm]current bounding box.center)}] 
	\node[] (a) at (0,0) {};
	\node[] (a1) at (0.75,0) {};
	\node[] (a2) at (1.5,0) {};
	\node[] (b) at (0,-0.5) {};
	\node[] (b1) at (0.5,-0.5) {};
	\node[] (b2) at (1.5,-0.5) {};
	\node[] (c) at (0,-1) {};
	\node[] (c1) at (0.75,-1) {};
	\node[] (c2) at (1.5,-1) {};
	\node[] (p1) at ($(a)+(-0.2,0)$) {};
	\node[] (p2) at ($(c)+(-0.2,0)$) {};
	\node[] (p3) at ($(c2)+(0.2,0)$) {};
	\node[] (p4) at ($(a2)+(0.2,0)$) {};
	\draw[line width=0.15mm] (c.center) -- (a.center);
	\draw[line width=0.15mm] (b.center) -- (b2.center);
	\draw[line width=0.15mm] (c2.center) -- (a2.center);
	\draw[line width=0.15mm] (b2.center) -- (c1.center);
	\draw[line width=0.15mm] (b.center) -- (a1.center);
	\draw[line width=0.5mm] (p1.center) -- (p4.center);
	\draw[line width=0.5mm] (p2.center) -- (p3.center);
\end{tikzpicture} \right) \propto \frac{x}{q^2} \, \int \frac{d t_1 d t_2 d t_3}{\sqrt{Q_8(t_1,t_2,t_3)}}\,,
\end{equation}
where
\begin{align}
 Q_8(t_1,t_2,t_3) \equiv (t_1-1)^2(t_2-1)^2(t_1t_2+t_3^2)^2(1-x^2)^2-64x^2 t_1^2t_2^2t_3^2(t_1+t_2-1)\,.
\end{align}
Again, we have a degree-8 polynomial $Q_8(t_1,t_2,t_3)$ in three variables, which is also quartic in each of them. By eqs.~\eqref{eq: LS_geometry_intro}--\eqref{eq: condition_degree_weight_CY}, it satisfies the Calabi--Yau condition in $\mathbb{WP}^{1,1,1,1,4}$. Consequently, the even-parity integral depends on a different Calabi--Yau threefold compared to eq.~\eqref{eq: LS_PM_4loop_CY_1}, which in this case appears in the 5PM 2SF conservative sector, beyond the current state of the art in PM theory. 

\begin{figure}[t]
\centering

\caption{Parametrization for the diagram $\diagramnumberingsign37$ of tab.~\ref{tab: 4-loop results 2SF}, which depends on a CY threefold in the even-parity sector.}
\label{fig: param_CY_2SF}
\end{figure}
We can also study the Picard--Fuchs operator of this integral, which in $D=4-2\varepsilon$ is given by an order-5 operator,
\begin{equation}
\label{eq: PF_PM_CY_2_eps}
\mathcal{L}_5^{(\text{CY}'_3)} = \frac{d^5}{d x^5} + \sum_{j=0}^{4} \frac{C_j(x,\varepsilon)}{x^5 (1-x^2)^4 P_{\text{apparent}}(x,\varepsilon)} \frac{d^j}{d x^j}\,.
\end{equation}
Here, $C_j(x,\varepsilon)$ are again rational polynomials in $x$ and $\varepsilon$, with the explicit form provided in appendix~A of ref.~\cite{Frellesvig:2024rea}. Besides the physical singularities at $x=0$ and $x=\pm 1$, in this case, we also have a singularity at $P_{\text{apparent}}(x,\varepsilon)=0$, which was absent in the 1SF Calabi--Yau operator of eq.~\eqref{eq: PF_PM_CY_1_eps}. This polynomial takes the form
\begin{equation}
\label{eq: PF_CY_apparent_singularity}
P_{\text{apparent}}(x,\varepsilon) \equiv (3 + 32 \varepsilon + 4 \varepsilon^2) (1 + x^2)^2 - 
 16 x^2 (1 + 5 \varepsilon) (1 + 10 \varepsilon)\,,
\end{equation}
and is quartic in $x$ and quadratic in $\varepsilon$. Since its roots are complicated functions of $\varepsilon$, we do not expect any solutions to be singular at the unphysical values of $x$ for which the polynomial vanishes. Thus, the Picard--Fuchs operator in eq.~\eqref{eq: PF_PM_CY_2_eps} has an $\varepsilon$-dependent apparent singularity. As we will explore in the next chapter, this subtlety impedes bringing the associated differential equation into $\varepsilon$-factorized form with traditional tools, and it requires an extension of the techniques developed in ref.~\cite{Pogel:2022ken,Pogel:2022vat} to accommodate for such cases.

In $D=4$, the Picard--Fuchs operator of eq.~\eqref{eq: PF_PM_CY_2_eps} factorizes into a product of two lower-order operators $\mathcal{L}_5^{(\text{CY}'_3)}\Big|_{\varepsilon=0} = \mathcal{L}^{(\text{CY}'_3)}_1 \cdot \mathcal{L}^{(\text{CY}'_3)}_4$, with
{\allowdisplaybreaks
\begin{align}
\label{eq: PF_PM_CY_2_L1}
\mathcal{L}_1^{(\text{CY}'_3)} = &\, 
 \frac{d}{d x} + \frac{3 + 4 x^2 - 22 x^4 + 84 x^6 - 21 x^8}{x (1 - x^4)(3-x^2)(1-3x^2)}\,, \\[0.2cm]
\mathcal{L}_4^{(\text{CY}'_3)} = &\, 
 \frac{d^4}{d x^4} + \frac{2 - 16 x^2 - 10 x^4}{x(1-x^4)} \frac{d^3}{d x^3} + \frac{1 - 28 x^2 + 46 x^4 + 68 x^6 + 25 x^8}{x^2 (1 - x^4)^2} \frac{d^2}{d x^2} \nonumber \\
& \, - \frac{1 + 11 x^2 - 54 x^4 + 22 x^6 + 37 x^8 + 15 x^{10}}{x^3 (1 - x^2)^3 (1 + x^2)^2} \frac{d}{d x} + \frac{1 + 3 x^2 + 20 x^4 + 3 x^6 + x^8}{x^4 (1 - x^4)^2}\,.
\label{eq: PF_PM_CY_2_L4}
\end{align}
}
Since the rightmost operator in the factorization is an irreducible operator of order 4, it suggests that the underlying geometry is a CY threefold, as we already identified via the leading singularity in eq.~\eqref{eq: LS_PM_4loop_CY_2}. Moreover, we observe that it only contains physical singularities, at $x=0$ and $x=\pm 1$.\footnote{The additional singularity at $x=\pm i$ is due to our choice of variable, $y=\frac{1+x^2}{2x}$. As we show in sec.~\ref{sec:ch5_PF_gravity_CY_operator}, it actually corresponds to a regular point in which none of the solutions become singular, thus an ($\varepsilon$-independent) apparent singularity. In particular, we can remove it by changing variables to $v=y^2$.} To corroborate that we actually have a Calabi--Yau threefold, in the next chapter we will verify that the Picard--Fuchs operator in eq.~\eqref{eq: PF_PM_CY_2_L4} satisfies the necessary conditions~\cite{Bogner:2013kvr}, see especially sec.~\ref{sec:ch5_PF_gravity_CY_operator}. In fact, some of the topological information that is extracted from this consistency check can also be used to streamline the process of bringing the corresponding differential equation into canonical form.

For completeness, let us now show that the odd-parity sector has instead an algebraic leading singularity. For that, we found it simpler to follow instead the loop-by-loop order $\{ k_1, k_2, k_3, k_4\}$, which requires 5 ISPs, chosen as $z_{12}=k_2^2$, $z_{13}=(k_3-q)^2$, $z_{14}=k_4^2$, $z_{15}=2u_1 \cdot k_3$ and $z_{16}=2u_1 \cdot k_4$. Multiplying by $z_{15}=2u_1 \cdot k_3$ in the numerator, we have
\begin{align}
&\LS \left( {\times} \, 2u_1 {\cdot} k_3 \right) \nonumber \\ & \enspace \propto \frac{1}{|q|} \, \LS \left( \int \frac{\varepsilon \, x^2 \, z_{15}\, d z_{12} \, \cdots \, d z_{16}}{\sqrt{z_{12}} \sqrt{(z_{12}-1)^2 z_{15}^2 + 4z_{12}(z_{13}-2)(1+z_{12}-z_{13})} \ P_4(z_{13},\dots,z_{16}) } \right),
\end{align}
where $P_4(z_{13},\dots,z_{16})$ is a polynomial of overall degree 4, but is only quadratic in $z_{16}$. Hence, we can directly compute the residue at either of its poles, which results in
\begin{align}
\LS \left( %
 {\times} \, 2u_1 {\cdot} k_3 \right) \propto & \, \frac{1}{|q|} \, \LS \left( \int \frac{\varepsilon \, x \, z_{15}\, d z_{12} \, \cdots \, d z_{15}}{\sqrt{z_{12}} \sqrt{(z_{12}-1)^2 z_{15}^2 + 4z_{12}(z_{13}-2)(1+z_{12}-z_{13})} } \right. \nonumber \\[0.1cm]
& \, \left. \times \frac{1}{\sqrt{z_{13}-2} \sqrt{z_{14} (1+z_{14} - z_{13})} \sqrt{(x^2-1)^2 (z_{13}-1)^2 -4x^2 z_{15}^2}} \right).
\end{align}
At this point, we can use the transformation in eq.~\eqref{eq: change_of_variables_rationalization_r1r2} to rationalize the fourth square root changing from $z_{14}$ to $t_{14}$. Moreover, we observe that changing variables as $z_{15} = \sqrt{z'_{15}}$ removes the numerator factor, since the remaining square roots contain even powers of $z_{15}$. Thereafter, we can use again eq.~\eqref{eq: change_of_variables_rationalization_r1r2} to change variables from $z'_{15}$ to $t_{15}$ in order to rationalize simultaneously the second and fifth square roots. After all, we obtain an algebraic result,
\begin{equation}
    \LS \left( \begin{tikzpicture}[baseline={([yshift=-0.1cm]current bounding box.center)}] 
	\node[] (a) at (0,0) {};
	\node[] (a1) at (0.75,0) {};
	\node[] (a2) at (1.5,0) {};
	\node[] (b) at (0,-0.5) {};
	\node[] (b1) at (0.5,-0.5) {};
	\node[] (b2) at (1.5,-0.5) {};
	\node[] (c) at (0,-1) {};
	\node[] (c1) at (0.75,-1) {};
	\node[] (c2) at (1.5,-1) {};
	\node[] (p1) at ($(a)+(-0.2,0)$) {};
	\node[] (p2) at ($(c)+(-0.2,0)$) {};
	\node[] (p3) at ($(c2)+(0.2,0)$) {};
	\node[] (p4) at ($(a2)+(0.2,0)$) {};
    \draw[line width=0.15mm] (a.center) -- (b.center);
	\draw[line width=0.15mm] (c2.center) -- (a2.center);
	\draw[line width=0.15mm] (b.center) -- (b2.center);
	\draw[line width=0.15mm] (c2.center) -- (a2.center);
	\draw[line width=0.15mm] (b2.center) -- (c1.center);
    \draw[line width=0.15mm] (b.center) -- (c.center);
	\draw[line width=0.15mm] (b.center) -- (a1.center);
	\draw[line width=0.5mm] (p1.center) -- (p4.center);
	\draw[line width=0.5mm] (p2.center) -- (p3.center);
\end{tikzpicture} {\times} \, 2u_1 {\cdot} k_3 \right) \propto \frac{\varepsilon}{|q|} \, \LS \left( \int \! \frac{dz_{12} dz_{13} dt_{14} dt_{15}}{t_{14} t_{15} (z_{12}-1) \sqrt{z_{12}} \sqrt{z_{13}-2}} \right) \propto \frac{\varepsilon}{|q|} \int_{1}^{2} \! \frac{d z_{13}}{\sqrt{z_{13}-2}} \propto \frac{\varepsilon}{|q|} \,,
\end{equation}
where in the last step we have integrated the leading singularity between two roots of the Baikov polynomials, as discussed in sec.~\ref{sec:ch4_two_loop_0SF}.

\subsection{Non-trivial geometries II: K3 surfaces}
\label{sec:ch4_four_loop_K3s}

Having analyzed the two integrals which depend on three-dimensional Calabi--Yau geometries, in this subsection, we turn to the remaining cases involving non-trivial geometries. In particular, we analyze the 4-loop diagrams in the classification of tabs.~\ref{tab: 4-loop results 0SF and 1SF} and~\ref{tab: 4-loop results 2SF} which depend on K3 surfaces, focusing only on the corresponding parity sector where this geometry arises.

First, in sec.~\ref{sec:ch4_four_loop_K3s_1}, we consider the diagrams contributing to 1SF order, which were first studied in ref.~\cite{Klemm:2024wtd} with the differential equations method. Concretely, we analyze the odd-parity integral for diagrams \diagramnumberingsign4 and \diagramnumberingsign5 in tab.~\ref{tab: 4-loop results 0SF and 1SF}, and the even-parity integral for diagram \diagramnumberingsign6, showing that they depend on the same K3 surface which appeared at 3 loops, recall sec.~\ref{sec:ch4_three_loop_K3}. Moreover, we also show that the 2SF diagrams \diagramnumberingsign38 and \diagramnumberingsign39 in tab.~\ref{tab: 4-loop results 2SF} depend again on the same 3-loop K3 surface for odd parity. Then, in sec.~\ref{sec:ch4_four_loop_K3s_2}, we focus on the 2SF diagram \diagramnumberingsign40, where a different K3 surface appears in the even-parity sector. This last case was first pointed out in ref.~\cite{Ruf:talk} as a non-trivial integral that deserved further study, and in ref.~\cite{Klemm:2024wtd} the underlying geometry was identified using differential equations.

\subsubsection{The three-loop K3 surface at 1SF and 2SF orders}
\label{sec:ch4_four_loop_K3s_1}

\begin{figure}[t]
\centering
\subfloat[]{
}
\caption{Parametrizations for (a) the diagram $\diagramnumberingsign4$ of tab.~\ref{tab: 4-loop results 0SF and 1SF}, and (b) the diagram $\diagramnumberingsign5$ of tab.~\ref{tab: 4-loop results 0SF and 1SF}, which both depend on the 3-loop K3 surface in the odd-parity sector.}
\label{fig: param_K3_1SF_1}
\end{figure}
Let us start with the 1SF diagrams in rows \diagramnumberingsign4 and \diagramnumberingsign5 of tab.~\ref{tab: 4-loop results 0SF and 1SF}, which we parametrize as detailed in fig.~\ref{fig: param_K3_1SF_1}. For both diagrams, we can use the loop-by-loop order $\{ k_1, k_2, k_3, k_4\}$ and choose an identical set of 4 ISPs, given by $z_{12}=k_2^2$, $z_{13}=k_4^2$, $z_{14}=2u_1\cdot k_3$ and $z_{15}=2u_2\cdot k_4$. Then, focusing on the odd-parity sector, they yield the same leading singularity,
\begin{align}
& \LS \left( {\times} \, 2u_1 {\cdot} k_3 \right) \nonumber \\[0.1cm] 
& \, \qquad \qquad \propto \frac{1}{|q|} \, \LS \, \Bigg( \int \frac{ x^2 \, d z_{12} \, \cdots \, d z_{15}}{(z_{13}-z_{12})\sqrt{z_{12}} \sqrt{(x^2-1)^2 (z_{13}-1)^2-4x^2z_{15}^2} }
\nonumber \\[0.1cm]
   &  \, \ \ \, \, \qquad \qquad  \times  
\frac{1}{\sqrt{(x^2-1)^2z_{13}^2+x^2z_{14}^2z_{15}^2-2xz_{13}z_{14}(2xz_{14}+(x^2+1)z_{15})}}  \Bigg)\,,
\end{align}
where the $z_{14}$ that should appear due to the numerator factor has canceled against a similar factor in the denominator. Now, we can directly compute the residue at $z_{12}=z_{13}$, and use eq.~\eqref{eq: change_of_variables_rationalization_r1r2} to rationalize the last square root changing from $z_{14}$ to $t_{14}$. Subsequently, taking the residue at $t_{14}=0$ results in
\begin{align}
\LS \left( {\times} \, 2u_1 {\cdot} k_3 \right) \nonumber \\[0.1cm] 
\, \propto & \ \frac{x}{|q|} \, \LS \left( \int \frac{d z_{13}d z_{15}}{\sqrt{z_{13}} \sqrt{4 z_{13} - z_{15}^2} \sqrt{(x^2-1)^2 (z_{13}-1)^2 - 4 x^2 z_{15}^2}} \right).
\end{align}
Comparing to eq.~\eqref{eq: LS_PM_3loop_K3_pre}, we can directly observe at this stage that the leading singularity depends on the same K3 surface that appeared at 3 loops. Following the same steps, we would explicitly obtain this geometry as defined by the degree-6 polynomial $P_6(t_1,t_2)$ in eq.~\eqref{eq: PM_3loop_K3_poly}.

\begin{figure}[t]
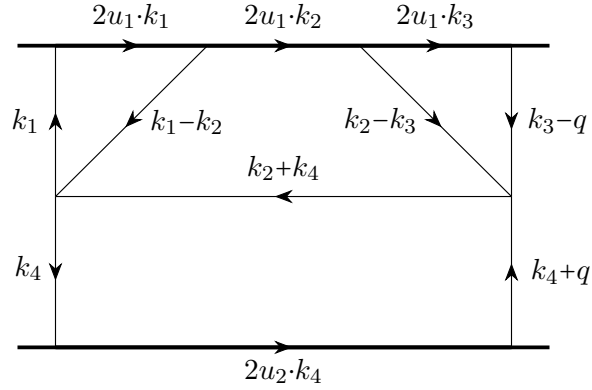

\centering

\caption{Parametrization for the diagram $\diagramnumberingsign6$ of tab.~\ref{tab: 4-loop results 0SF and 1SF}, which depends on the 3-loop K3 surface in the even-parity sector.}
\label{fig: param_K3_1SF_2}
\end{figure}
Let us then turn to the last 1SF diagram, which corresponds to number \diagramnumberingsign6 in tab.~\ref{tab: 4-loop results 0SF and 1SF}. As we will see, proving that this integral is related to the same K3 surface is less trivial. In this case, we can follow the parametrization given in fig.~\ref{fig: param_K3_1SF_2}. With the loop-by-loop order $\{ k_1,k_3,k_2,k_4\}$, we need 3 ISPs, defined as $z_{12}=k_2^2$, $z_{13}=(k_2-q)^2$ and $z_{14}=2u_1 \cdot k_4$. Then, for the even-parity scalar integral, we obtain
\begin{align}
    \LS 
    \left( 
   \begin{tikzpicture}[baseline={([yshift=-0.1cm]current bounding box.center)}] 
	\node[] (a) at (0,0) {};
	\node[] (a1) at (0.5,0) {};
	\node[] (a2) at (1,0) {};
    \node[] (a3) at (1.5,0) {};
	\node[] (b) at (0,-0.5) {};
	\node[] (b1) at (0.5,-0.5) {};
	\node[] (b2) at (1,-0.5) {};
    \node[] (b3) at (1.5,-0.5) {};
	\node[] (c) at (0,-1) {};
	\node[] (c1) at (0.5,-1) {};
	\node[] (c2) at (1,-1) {};
    \node[] (c3) at (1.5,-1) {};
	\node[] (p1) at ($(a)+(-0.2,0)$) {};
	\node[] (p2) at ($(c)+(-0.2,0)$) {};
	\node[] (p3) at ($(c3)+(0.2,0)$) {};
	\node[] (p4) at ($(a3)+(0.2,0)$) {};
    \draw[line width=0.5mm] (p1.center) -- (p4.center);
	\draw[line width=0.5mm] (p2.center) -- (p3.center);
    \draw[line width=0.15mm] (c.center) -- (a.center);
	\draw[line width=0.15mm] (b.center) -- (b1.center);
	\draw[line width=0.15mm] (b1.center) -- (b2.center);
    \draw[line width=0.15mm] (b.center) -- (a1.center);
    \draw[line width=0.15mm] (b2.center) -- (b3.center);
	\draw[line width=0.15mm] (b3.center) -- (a2.center);
    \draw[line width=0.15mm] (a3.center) -- (c3.center);
\end{tikzpicture} 
    \right)
\propto & \, \frac{1}{q^2} \, \LS \, \Bigg( \int  \frac{x\,  d z_{12}d z_{13} d z_{14}}
{\sqrt{z_{12}}\sqrt{z_{13}}\sqrt{(x^2-1)^2-4x^2z_{14}^2}} \nonumber \\
   & \,  \times
 \frac{1}{\sqrt{(z_{12}^2 - 2 z_{12} (z_{13}+1) + (z_{13}-1)^2) z_{14}^2 - 4 z_{12} z_{13}}} \Bigg)\,.
\end{align}
Now, we can shift $z_{12}\to z_{12} +z_{13}$ and apply the change of variables in eq.~\eqref{eq: change_of_variables_rationalization_r1r2} from $z_{13}$ to $t_{13}$ to rationalize the first two square roots. After doing so, the variable $z_{12}$ appears only quadratically in the last square root. Thus, we can rationalize it with the same change of variables, which exposes a residue at $t_{12}=0$. Then, we have
\begin{equation}
    \LS 
    \left( 
 
    \right)
\propto \frac{x}{q^2} \, \LS \left( \int  \frac{d t_{13} d z_{14}}
{\sqrt{(x^2-1)^2-4x^2z_{14}^2} \sqrt{t_{13}^2 (4 z_{14}^2+2)- t_{13}^4-1}} \right).
\end{equation}
Rationalizing the first square root with eq.~\eqref{eq: change_of_variables_rationalization_r1r2} by changing from $z_{14}$ to $t_{14}$, and relabeling $t_{13} \to i\, t_2$ and $t_{14} \to i\, t_1$, we finally obtain
\begin{equation}
    \LS 
    \left( 
   %
 
    \right)
\propto \frac{x}{q^2} \, \int\! \frac{d t_1 d t_2}{\sqrt{t_2^2 (t_1^2-1)^2 (x^2-1)^2 - 4 x^2 t_1^2 (t_2^2+1)^2}}\,.
\end{equation}
As can be seen, we obtain exactly the same leading singularity as in the 3-loop diagram of eq.~\eqref{eq: PM_3loop_K3_poly}. Thus, this even-parity integral also depends on the same K3 surface.

In summary, we have seen that three different 1SF integrals depend on the same K3 surface which already appeared at 3 loops. The corresponding Picard--Fuchs operator is then given in eq.~\eqref{eq: PF_K3_3loops}, which is the symmetric square of the second-order operator of eq.~\eqref{eq: PF_K3_3loops_L2}. Consequently, the solution to these 4-loop integrals is again given by the same products of complete elliptic integrals. However, as observed in refs.~\cite{Driesse:2024xad,Bern:2024adl}, the elliptic functions from the diagram in fig.~\ref{fig: param_K3_1SF_2} drop from the final result in the physical observables in the 1SF conservative sector. As already discussed in sec.~\ref{sec:ch4_drawing_classical_diagrams}, this apparent cancellation is only due to a factor of $\varepsilon$ in the prefactor of these integrals, which makes them not contribute in $D=4$. By contrast, the elliptic integrals from the diagrams in fig.~\ref{fig: param_K3_1SF_1} contribute to the final results in the 1SF dissipative regime.

Let us then continue the classification of non-trivial geometries, and analyze the diagrams contributing at 2SF order. In particular, we focus here on diagrams \diagramnumberingsign38 and \diagramnumberingsign39 of tab.~\ref{tab: 4-loop results 2SF}, and show that they also depend on the same 3-loop K3 surface in the odd-parity sector. The last non-trivial topology, which is in row number \diagramnumberingsign40, will be studied afterwards, as it depends on a different K3 surface at even parity. Thus, we begin with an odd-parity integral for diagram \diagramnumberingsign38, which we parametrize in fig.~\ref{fig: param_K3_2SF_1}(a). With the loop-by-loop order $\{ k_1, k_2, k_3, k_4 \}$ and the 3 ISPs $z_{11}=k_3^2$, $z_{12}=k_4^2$ and $z_{13}=2 u_2 \cdot k_4$, we obtain
\begin{align}
    & \, \LS
    \left( 

    \right) \nonumber \\
& \, \quad \propto \frac{1}{|q|} \, \LS \left(\int\frac{\varepsilon\,x\, (z_{11}+z_{12})\, z_{13} \, dz_{11} dz_{12}dz_{13}}{\sqrt{z_{11}}\sqrt{z_{12}}\sqrt{4z_{12}-z_{13}^2}\sqrt{(x^2-1)^2(z_{12}-1)^2-4x^2z_{13}^2}\left((z_{11}-z_{12})^2+z_{11}z_{13}^2\right)} \right).
\end{align}
Computing the residue at one of the poles for $z_{11}$, the leading singularity becomes
\begin{align}
    & \, \LS
    \left( 
    %
    \right) \nonumber \\
    & \, \propto \frac{1}{|q|} \, \LS \left(\int \frac{\varepsilon\,x\, dz_{12}dz_{13}}
{(4z_{12}-z_{13}^2)\sqrt{z_{12}}\sqrt{(x^2-1)^2(z_{12}-1)^2-4x^2z_{13}^2}} \frac{4z_{12}-z_{13}^2 + z_{13}\sqrt{z_{13}^2-4z_{12}}}{\sqrt{2z_{12}-z_{13}\left(z_{13} - \sqrt{z_{13}^2-4z_{12}}\right)}} \right).
\end{align}
Upon noticing that the numerator and denominator in the second factor can be respectively written as
\begin{equation}
    4z_{12}-z_{13}^2 + z_{13}\sqrt{z_{13}^2-4z_{12}} \, =\sqrt{z_{13}^2 -4 z_{12}} \left(z_{13} - \sqrt{z_{13}^2 -4 z_{12}} \right)\,,
\end{equation}
and
\begin{equation}
    \sqrt{2z_{12}-z_{13}\left(z_{13} - \sqrt{z_{13}^2-4z_{12}}\right)} \,= \frac{z_{13} - \sqrt{z_{13}^2-4z_{12}}}{\sqrt{2}} \,,
\end{equation}
the leading singularity becomes
\begin{equation}
    \LS
    \left( 

    \right)
\propto  \frac{\varepsilon \, x}{|q|} \, \LS \left( \int \frac{dz_{12}dz_{13}}
{\sqrt{z_{12}}\sqrt{4z_{12}-z_{13}^2}\sqrt{(x^2-1)^2(z_{12}-1)^2-4x^2z_{13}^2}} \right).
\end{equation}
As can be seen, this is the same K3 surface that we obtained in eq.~\eqref{eq: LS_PM_3loop_K3_pre} at 3 loops.

\begin{figure}[t]
\centering
\subfloat[]{%
}
\caption{Parametrizations for (a) the diagram $\diagramnumberingsign38$ of tab.~\ref{tab: 4-loop results 2SF}, and (b) the diagram $\diagramnumberingsign39$ of tab.~\ref{tab: 4-loop results 2SF}, which both depend on the 3-loop K3 surface in the odd-parity sector.}
\label{fig: param_K3_2SF_1}
\end{figure}
Lastly, let us consider diagram \diagramnumberingsign39 of tab.~\ref{tab: 4-loop results 2SF}, which we parametrize as in fig.~\ref{fig: param_K3_2SF_1}(b). Taking the loop-by-loop order $\{ k_1, k_2, k_3, k_4\}$ generates two ISPs, $z_{11}=k_4^2$ and $z_{12}=2 u_1 \cdot k_4$. In this case, the leading singularity for the bubble generates a factor of $z_{12}$ in the denominator. However, placing a dot on the matter propagator within the bubble removes this factor, and we directly obtain
\begin{equation}
        \LS \left( \right)
\propto \frac{x}{|q|} \, \LS \left( \int \frac{d z_{11} d z_{12} }{\sqrt{z_{11}} \sqrt{4z_{11}-z_{12}^2} \sqrt{(x^2-1)^2(z_{11}-1)^2-4x^2z_{12}^2}} \right),
\end{equation}
which again is the 3-loop K3 surface from eq.~\eqref{eq: LS_PM_3loop_K3_pre}. 

\subsubsection{A second K3 surface at 2SF order}
\label{sec:ch4_four_loop_K3s_2}

\begin{figure}[t]
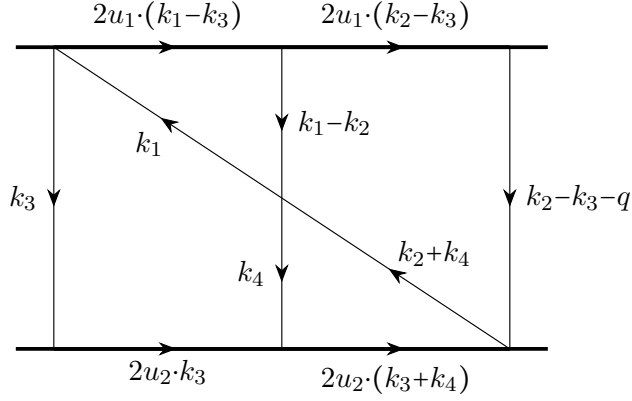

\centering
%
\caption{Parametrization for the diagram $\diagramnumberingsign40$ of tab.~\ref{tab: 4-loop results 2SF}, which depends on a K3 surface (different than the 3-loop one) in the even-parity sector.}
\label{fig: param_K3_2SF_2}
\end{figure}
Let us finally consider diagram \diagramnumberingsign40 of tab.~\ref{tab: 4-loop results 2SF} and show that, in the even-parity sector, there is a K3 surface, which is different from the one appearing at 3 loops. In this case, we can follow the parametrization given in fig.~\ref{fig: param_K3_2SF_2}. With the loop-by-loop order $\{ k_1, k_4, k_3, k_2 \}$, and the 4 ISPs $z_{11}=k_2^2$, $z_{12}=(k_2-q)^2$, $z_{13}=2u_1 \cdot k_2$ and $z_{14}=2u_2 \cdot k_2$, we get
\begin{align}
& \LS \left( \begin{tikzpicture}[baseline={([yshift=-0.1cm]current bounding box.center)}] 
	\node[] (a) at (0,0) {};
	\node[] (a1) at (0.75,0) {};
	\node[] (a2) at (1.5,0) {};
	\node[] (c) at (0,-1) {};
	\node[] (c1) at (0.75,-1) {};
	\node[] (c2) at (1.5,-1) {};
	\node[] (p1) at ($(a)+(-0.2,0)$) {};
	\node[] (p2) at ($(c)+(-0.2,0)$) {};
	\node[] (p3) at ($(c2)+(0.2,0)$) {};
	\node[] (p4) at ($(a2)+(0.2,0)$) {};
	\draw[line width=0.15mm] (a.center) -- (c.center);
	\draw[line width=0.15mm] (a1.center) -- (c1.center);
	\draw[line width=0.15mm] (a2.center) -- (c2.center);
	\draw[line width=0.15mm] (a.center) -- (c2.center);
	\draw[line width=0.5mm] (p1.center) -- (p4.center);
	\draw[line width=0.5mm] (p2.center) -- (p3.center);
\end{tikzpicture} \right) \propto \LS \left( \int \frac{x^2 \, d z_{11} \, \cdots \, d z_{14}}{\sqrt{4z_{11}-z_{13}^2} \sqrt{4z_{11}-z_{14}^2} } \right. \nonumber \\
& \, \enspace \qquad \times \left. \frac{1}{\sqrt{(x^2-1)^2 z_{12}^2 - 2 x (x^2+1) z_{12} z_{13} z_{14} + x^2 z_{13}^2 z_{14}^2} \, \sqrt{P_2(z_{11}, \dots, z_{14})}} \right).
\end{align}
Here, $P_2(z_{11}, \dots, z_{14})$ denotes a polynomial of overall degree 2, which is quadratic in all variables. Using the change of variables in eq.~\eqref{eq: change_of_variables_rationalization_r1r2} to rationalize the first (second) square root by changing from $z_{13}$ ($z_{14}$) to $t_{13}$ ($t_{14}$), we obtain
\begin{align}
& \LS \left( \begin{tikzpicture}[baseline={([yshift=-0.1cm]current bounding box.center)}] 
	\node[] (a) at (0,0) {};
	\node[] (a1) at (0.75,0) {};
	\node[] (a2) at (1.5,0) {};
	\node[] (c) at (0,-1) {};
	\node[] (c1) at (0.75,-1) {};
	\node[] (c2) at (1.5,-1) {};
	\node[] (p1) at ($(a)+(-0.2,0)$) {};
	\node[] (p2) at ($(c)+(-0.2,0)$) {};
	\node[] (p3) at ($(c2)+(0.2,0)$) {};
	\node[] (p4) at ($(a2)+(0.2,0)$) {};
	\draw[line width=0.15mm] (a.center) -- (c.center);
	\draw[line width=0.15mm] (a1.center) -- (c1.center);
	\draw[line width=0.15mm] (a2.center) -- (c2.center);
	\draw[line width=0.15mm] (a.center) -- (c2.center);
	\draw[line width=0.5mm] (p1.center) -- (p4.center);
	\draw[line width=0.5mm] (p2.center) -- (p3.center);
\end{tikzpicture} \right) \propto \LS \left( \int \frac{x^2 \, t_{13} t_{14} \, d z_{11} d z_{12} d t_{13} d t_{14}}{\sqrt{P_7(z_{11}, z_{12}, t_{13}, t_{14})}} \right. \nonumber \\
& \, \ \ \times \left. \frac{1}{\sqrt{\Big( (x-1)^2 z_{12} t_{13} t_{14} -x z_{11}(1 + t_{13}^2) (1 + t_{14}^2) \Big) \Big( (x+1)^2 z_{12} t_{13} t_{14} -x z_{11}(1 + t_{13}^2) (1 + t_{14}^2) \Big)}} \right),
\end{align}
where $P_7(z_{11}, z_{12}, t_{13}, t_{14})$ is a polynomial of overall degree 7 but which is only quadratic in $z_{12}$. Since the roots with respect to $z_{12}$ in the second square root are very simple, we can use eq.~\eqref{eq: change_of_variables_rationalization_r1r2} to change variables to $t_{12}$ and rationalize it. Thus, we get
\begin{equation}
    \LS \left( \begin{tikzpicture}[baseline={([yshift=-0.1cm]current bounding box.center)}] 
	\node[] (a) at (0,0) {};
	\node[] (a1) at (0.75,0) {};
	\node[] (a2) at (1.5,0) {};
	\node[] (c) at (0,-1) {};
	\node[] (c1) at (0.75,-1) {};
	\node[] (c2) at (1.5,-1) {};
	\node[] (p1) at ($(a)+(-0.2,0)$) {};
	\node[] (p2) at ($(c)+(-0.2,0)$) {};
	\node[] (p3) at ($(c2)+(0.2,0)$) {};
	\node[] (p4) at ($(a2)+(0.2,0)$) {};
	\draw[line width=0.15mm] (a.center) -- (c.center);
	\draw[line width=0.15mm] (a1.center) -- (c1.center);
	\draw[line width=0.15mm] (a2.center) -- (c2.center);
	\draw[line width=0.15mm] (a.center) -- (c2.center);
	\draw[line width=0.5mm] (p1.center) -- (p4.center);
	\draw[line width=0.5mm] (p2.center) -- (p3.center);
\end{tikzpicture} \right) \propto \LS \left( \int \frac{x^2 \, dz_{11} dt_{12} dt_{13} dt_{14}}{\sqrt{P_{14}(z_{11},t_{12},t_{13},t_{14})}} \right).
\end{equation}
Surprisingly, even though we obtain a degree-14 polynomial, since the previous changes of variables did not involve higher powers of $z_{11}$, it is only quadratic in $z_{11}$. Thus, we can again use eq.~\eqref{eq: change_of_variables_rationalization_r1r2} to rationalize it by introducing $t_{11}$, which reveals a simple pole at $t_{11}=0$. Taking the residue, we obtain a square root of a degree-6 polynomial that is only quadratic in $t_{12}$, which allows us to repeat the same transformation and take a residue. Afterwards, we obtain a degree-8 polynomial in $t_{13}$ and $t_{14}$. However, rescaling $t_{14} \to t_{14}/t_{13}$, and performing the change of variables $t_{13}\to \sqrt{t_{13}}$, we arrive at the following expression:
\begin{equation}
    \LS \left( \begin{tikzpicture}[baseline={([yshift=-0.1cm]current bounding box.center)}] 
	\node[] (a) at (0,0) {};
	\node[] (a1) at (0.75,0) {};
	\node[] (a2) at (1.5,0) {};
	\node[] (c) at (0,-1) {};
	\node[] (c1) at (0.75,-1) {};
	\node[] (c2) at (1.5,-1) {};
	\node[] (p1) at ($(a)+(-0.2,0)$) {};
	\node[] (p2) at ($(c)+(-0.2,0)$) {};
	\node[] (p3) at ($(c2)+(0.2,0)$) {};
	\node[] (p4) at ($(a2)+(0.2,0)$) {};
	\draw[line width=0.15mm] (a.center) -- (c.center);
	\draw[line width=0.15mm] (a1.center) -- (c1.center);
	\draw[line width=0.15mm] (a2.center) -- (c2.center);
	\draw[line width=0.15mm] (a.center) -- (c2.center);
	\draw[line width=0.5mm] (p1.center) -- (p4.center);
	\draw[line width=0.5mm] (p2.center) -- (p3.center);
\end{tikzpicture} \right) \propto \frac{x^2}{x^2-1} \, \LS \left( \int \frac{dt_{13} dt_{14}}{\sqrt{(x(1+t_{13})(t_{13}+t_{14}^2)-(1+x^2)t_{13}t_{14})^2-4x^2t_{13}^2t_{14}^2}} \right).
\end{equation}
Notably, it reveals a degree-6 polynomial in two variables, which is also quartic in each of them. Based on the discussion in sec.~\ref{sec:ch1_geometries}, and especially in eqs.~\eqref{eq: LS_geometry_intro}--\eqref{eq: condition_degree_weight_CY}, we see that it satisfies the Calabi--Yau condition in $\mathbb{WP}^{1,1,1,3}$. Therefore, the leading singularity indicates that we have an integral over a K3 surface in the even-parity sector. Importantly, this K3 surface is different compared to the one which appeared at 3 loops and in previous 4-loop examples, see eq.~\eqref{eq: PM_3loop_K3_poly}. 

To verify that this is indeed the correct geometry, we can calculate the Picard--Fuchs operator, which in $D=4$ dimensions is given by~\cite{Klemm:2024wtd}
\begin{align}
\label{eq: PF_K3_4loops_2SF}
\mathcal{L}_3^{(\text{K3}')} = & \,
 \frac{d^3}{d x^3} + \frac{3 (1 + x^2 - 69 x^4 + 3 x^6)}{x (x^2-1) (1 - 34 x^2 + x^4)} \frac{d^2}{d x^2} - \frac{7 - 41 x^2 + 245 x^4 - 19 x^6}{x^2 (x^2-1) (1 - 34 x^2 + x^4)} \frac{d}{d x} \nonumber \\[0.1cm]
 & \, + \frac{8 (x^2+1)(1 - 6 x^2 + x^4) }{x^3 (x^2-1) (1 - 34 x^2 + x^4)}\,.
\end{align}
We obtain an irreducible third-order differential operator, which differs from the 3-loop operator in eq.~\eqref{eq: PF_K3_3loops}, and defines a different K3 surface~\cite{Klemm:2024wtd}. Still, as it is a single-scale operator, it necessarily satisfies the symmetric square condition, recall the discussion from sec.~\ref{sec:ch4_three_loop_K3}. In this case, it becomes the symmetric square of the second-order differential operator
\begin{equation}
\label{eq: PF_K3_4loops_2SF_L2}
\mathcal{L}_2^{(\text{K3}')} =
 \frac{d^2}{d x^2} + \frac{1 + x^2 - 69 x^4 + 3 x^6}{x (x^2-1) (1 - 34 x^2 + x^4)} \frac{d}{d x} + \frac{1 - 12 x^2 + 54 x^4 - 12 x^6 + x^8}{x^2 (x^2-1)^2 (1 - 34 x^2 + x^4)}\,.
\end{equation}

In summary, our analysis suggests that in the 5PM 2SF sector there are several geometries, which introduce different special functions in the result. In the conservative regime, we find the CY threefold integral from sec.~\ref{sec:ch4_four_loop_CYs_2}, as well as the K3 surface discussed here, both of which are new compared to the geometries appearing at 5PM 1SF order. In addition, the two 2SF odd-parity integrals of sec.~\ref{sec:ch4_four_loop_K3s_2} introduce the 3-loop K3 surface in the dissipative regime. It will be interesting to see whether the associated special functions contribute to the physical observables at this order, or if they drop in $D=4$, as it was the case for the elliptic integrals in the 5PM 1SF conservative regime~\cite{Driesse:2024xad,Bern:2024adl}.

\section{Discussion and open questions}
\label{sec:ch4_conclusions}

In this chapter, we have initiated a systematic classification of the Feynman integral geometries and special functions which are relevant to black-hole scattering in the PM expansion of classical gravity, performing an analysis up to four loops. Notably, these special functions contribute to the gravitational waves emitted by the two-body system, influencing both the conservative and the dissipative classical dynamics. To detect and characterize the geometries involved, we have analyzed the leading singularities of the integrals, supplementing this approach with the differential equations method in non-trivial cases. As a proof of principle, we have first carried out the analysis up to three loops, finding polylogarithms as well as an integral over a K3 surface, which is in total agreement with explicit computations~\cite{Bern:2021dqo,Dlapa:2021npj,Bern:2021yeh,Bern:2022jvn,Dlapa:2022lmu,Dlapa:2023hsl,Jakobsen:2023ndj,Damgaard:2023ttc,Jakobsen:2023hig}.

Furthermore, we have demonstrated that the loop-by-loop Baikov representation is especially well-suited for computing the leading singularity of PM integrals, as it naturally simplifies the calculation by reducing the number of integration variables. With this framework, we were able to relate the leading singularity -- and thus the underlying geometry -- of numerous Feynman integrals across different topologies and loops orders. These simplifying relations drastically reduced the magnitude of our analysis, see tab.~\ref{tab:reduction_diagrams}, allowing us to do a full classification of geometries one loop beyond the current state of the art.

Specifically, out of the 16,596 initial topologies appearing in the PM expansion at four loops, only 70 needed to be studied in detail in order to characterize the geometries arising at this loop order. Remarkably, among these, we discovered the first Calabi--Yau threefold contributing to gravitational-wave physics in the literature, which introduces new transcendental functions at four loops. Overall, we found integrals depending on two distinct K3 surfaces and two different CY threefolds. Notably, several integrals involve the same K3 surface that already appeared at three loops. However, in some cases, this correspondence was hidden, and it only became apparent that the geometry was the same in the last step of the calculation. In the future, it would be interesting to investigate whether there exist deeper relations among these Feynman integrals, which make this connection between geometries manifest from the outset.

As a result, the 5PM function space becomes especially rich. In the 1SF conservative sector, there appears again the 3-loop K3 surface (see fig.~\ref{fig: param_K3_1SF_2}). The 1SF dissipative sector also depends on this 3-loop geometry (see fig.~\ref{fig: param_K3_1SF_1}), as well as on a CY threefold through the diagram in fig.~\ref{fig: param_CY_1SF}; as confirmed by explicit computations~\cite{Driesse:2024xad,Bern:2024adl,Driesse:2024feo}. Notably, in the dissipative sector, these two geometries mix within the differential equations, since the CY diagram of fig.~\ref{fig: param_CY_1SF} is a subsector of the K3 diagram of fig.~\ref{fig: param_K3_1SF_1}(a). Consequently, the result leads to iterated integrals involving the periods of both geometries~\cite{Driesse:2024feo}. By contrast, the 2SF dissipative sector only depends on the 3-loop K3 surface (see fig.~\ref{fig: param_K3_2SF_1}). Lastly, the 2SF conservative sector depends on a different K3 surface (see fig.~\ref{fig: param_K3_2SF_2}), as well as on a different CY threefold via the diagram in fig.~\ref{fig: param_CY_2SF}. In this case, the geometries do not directly mix, but appear simultaneously through the subtopologies of the diagram in row \diagramnumberingsign42 of tab.~\ref{tab: 4-loop results 2SF}. Still, it would be interesting to study the interplay between the geometries in this sector, and explore whether new features appear via the differential equations method. To analyze this further and facilitate the computation of the 2SF conservative sector, in the next chapter, we bring the differential equation for the 2SF CY integral into $\varepsilon$-factorized form.

A natural future research direction would be to investigate whether the CY geometries found at four loops belong to a CY integral family at higher loops, such as the banana and the traintrack families presented in tab.~\ref{tab:geometries_intro}, which depend on a CY $(L-1)$-fold at $L$ loops. Similarly, it would be exciting to use our analysis to classify the geometries and special functions arising at the next loop order in the PM expansion. One notable feature at five loops is the first appearance of a truly non-planar Abelian diagram,
\begin{equation}
    \begin{tikzpicture}[baseline={([yshift=-0.1cm]current bounding box.center)}] 
    \node[] (a) at (0.75,0) {};
	\node[] (a1) at (1.875,0) {};
    \node[] (a2) at (2.25,0) {};
	\node[] (b) at (0,-1) {};
    \node[] (b1) at (1.5,-1) {};
    \node[] (b2) at (2.25,-1) {};
    \node[] (b3) at (3,-1) {};
    \draw[line width=0.15mm] (a.center) --  (b.center);
    \draw[line width=0.15mm] (a.center) --  (b1.center);
    \draw[line width=0.15mm] (a1.center) --  (b1.center);
    \draw[line width=0.15mm] (a1.center) --  (b2.center);
    \draw[line width=0.15mm] (a2.center) --  (b3.center);
    \draw[line width=0.15mm] (b1.center) --  (1.94,-0.42);
    \draw[line width=0.15mm] (2.06,-0.25) --  (a2.center);
	\draw[line width=0.5mm] (-0.2,0) --  (3.2,0);
	\draw[line width=0.5mm] (-0.2,-1) --  (3.2,-1);
    \end{tikzpicture}\ .
\end{equation}
However, computing its leading singularity requires a better understanding of the cases which have vanishing residues, and thus involve multiple integrations between the roots of the Baikov polynomials. In particular, finding a change of variables that results in a useful integration sequence already hindered some of the calculations at four loops. Thus, we leave the analysis of this five-loop diagram, and the question of whether or not it involves non-trivial geometries, for future work.

Nonetheless, performing a full classification of geometries at five loops could lie beyond the reach of the current method. As seen from tab.~\ref{tab:reduction_diagrams}, the number of relevant topologies grows dramatically with the loop order. At five loops, we would expect around 500,000 topologies in the PM expansion, with approximately 500 diagrams requiring a detailed analysis. Given the complexity of such a classification, it would require an automatization of the process. However, this introduces an additional challenge: automatically finding the necessary changes of variables. In particular, at five loops, the presence of up to 9 simultaneous square roots in the leading singularity makes this task particularly daunting.

More broadly, it would be very interesting to classify the Feynman integral geometries contributing to other observables and theories. In PM gravity, one option could be the waveform, which involves similar diagrams with one emitted graviton, and was only recently computed at one loop~\cite{Brandhuber:2023hhy,Herderschee:2023fxh,Elkhidir:2023dco,Georgoudis:2023lgf,Caron-Huot:2023vxl}. Alternatively, it would be fascinating to apply the tools developed in this chapter to classify the geometries appearing in QCD amplitudes for particle physics phenomenology, which we leave for future work.

\chapter{An \texorpdfstring{\bm{$\varepsilon$}}{eps}-factorized form for the 2SF Calabi--Yau integral}
\label{ch:chapter5}

\begin{info}[\textit{Info:}]
\textit{Part of the content and figures of this chapter have been published together with H. Frellesvig, S. Pögel, S. Weinzierl and M. Wilhelm in ref.}~\cite{Frellesvig:2024rea}\textit{, available at} \href{https://doi.org/10.1007/JHEP02(2025)209}{\textit{JHEP} \textbf{02} (2025) 09} [\href{https://arxiv.org/abs/2412.12057}{2412.12057}].
\end{info}

\section{Motivation}
\label{sec:ch5_intro}

In the previous chapter, we classified the Feynman integral geometries contributing to the two-body classical dynamics within the PM expansion of classical gravity up to four loops. In particular, in sec.~\ref{sec:ch4_four_loop}, we found that the function space at 5PM order becomes especially rich: there appear integrals over two distinct K3 surfaces (labeled K3 and $\text{K3}'$) and two different Calabi--Yau threefolds (labeled $\text{CY}_3$ and $\text{CY}_3'$), see fig.~\ref{fig: diagrams_nontrivial_PM} for the corresponding diagrams.

As we introduced in sec.~\ref{sec:ch1_DEs}, one method to evaluate these Feynman integrals is to solve the corresponding differential equations that the master integrals satisfy. Concretely, we can obtain a linear system of coupled first-order differential equations, given in eq.~\eqref{eq: differential_equation_precanonical_intro}, which we rewrite here for convenience:
\begin{equation}
\label{eq: differential_equation_precanonical}
\frac{d}{dx} \, \vec{\mathcal{I}}=\tilde{A}_x(x,\varepsilon) \, \vec{\mathcal{I}}\,.
\end{equation}
In this equation, $\tilde{A}_x(x,\varepsilon)$ denotes an $n \times n$ matrix whose entries are rational functions of $x$ and $\varepsilon$, and $\vec{\mathcal{I}}$ is the vector of $n$ master integrals. Recalling eq.~\eqref{eq: differential_equation_canonical_intro}, an efficient approach to solving this system is to find a change of basis $\vec{\mathcal{J}}=U(x,\varepsilon)\, \vec{\mathcal{I}}$, such that the $\varepsilon$-dependence of the linear system factorizes as a prefactor,
\begin{equation}
\label{eq: differential_equation_canonical}
\frac{d}{dx} \, \vec{\mathcal{J}}=\varepsilon A_x(x) \, \vec{\mathcal{J}}\,.
\end{equation}
This so-called canonical or $\varepsilon$-factorized form~\cite{Henn:2013pwa} allows us to systematically find the solution for $\vec{\mathcal{J}}$ by expanding order by order in $\varepsilon$. Consequently, once a transformation $U(x,\varepsilon)$ that brings the differential equation into $\varepsilon$-factorized form is found, the corresponding Feynman integrals are essentially solved. Nowadays, several computer implementations have been developed which can systematically find the transformation to $\varepsilon$-factorized form in polylogarithmic cases, see e.g.~refs.~\cite{Lee:2014ioa,Prausa:2017ltv,Gituliar:2017vzm,Meyer:2017joq,Dlapa:2020cwj,Lee:2020zfb}. Furthermore, the concept of an $\varepsilon$-factorized form has also been extended to elliptic~\cite{Adams:2018yfj,Dlapa:2020cwj,Dlapa:2022wdu,Jiang:2023jmk,Gorges:2023zgv}, hyperelliptic~\cite{Duhr:2024uid} and certain univariate CY integrals~\cite{Pogel:2022ken,Pogel:2022vat,Klemm:2024wtd,Driesse:2024feo,Duhr:2025lbz}.

In the case of the PM expansion, the differential equations method also plays a central role in solving the corresponding Feynman integrals in state-of-the-art computations, see e.g. refs.~\cite{Driesse:2024xad,Bern:2024adl,Driesse:2024feo}. In fact, most of the 5PM integrals involving non-trivial geometries that we identified in the previous chapter, recall fig.~\ref{fig: diagrams_nontrivial_PM} and sec.~\ref{sec:ch4_four_loop}, have already been solved with this approach. Specifically, the differential equations for the $\text{CY}_3$ and K3 integrals in the 1SF sector, as well as for the $\text{K3}'$ integral in the 2SF sector, have been brought into $\varepsilon$-factorized form in refs.~\cite{Klemm:2024wtd,Driesse:2024feo,Duhr:2025lbz}. 

However, the 2SF $\text{CY}'_3$ integral in fig.~\ref{fig: diagrams_nontrivial_PM} presents an intrinsic obstacle: its Picard--Fuchs operator contains an $\varepsilon$-dependent apparent singularity, as shown in eqs.~\eqref{eq: PF_PM_CY_2_eps}--\eqref{eq: PF_CY_apparent_singularity}. Importantly, as we will demonstrate in sec.~\ref{sec:ch5_canonical_form}, this type of singularity prevents us from using the method developed in refs.~\cite{Pogel:2022ken,Pogel:2022vat} to obtain the $\varepsilon$-factorized form. In particular, this method employs an ansatz for the transformation matrix that, due to its form, cannot recreate such $\varepsilon$-dependent apparent singularities in the resulting Picard--Fuchs operator. Thus, by construction, it cannot capture such Feynman integrals and bring their differential equation into $\varepsilon$-factorized form. Therefore, the purpose of this chapter is to generalize the method of refs.~\cite{Pogel:2022ken,Pogel:2022vat} to accommodate for this larger class of Feynman integrals. As a proof of principle, we will bring the differential equation for the 2SF $\text{CY}'_3$ integral intro $\varepsilon$-factorized form.\footnote{At the time of writing, ref.~\cite{Duhr:2025lbz} has appeared on the arXiv, presenting an $\varepsilon$-factorized form for the 2SF $\text{CY}'_3$ integral using a different method. However, the result is equivalent, and it does not circumvent the problem of $\varepsilon$-dependent apparent singularities addressed in this chapter.} This way, we facilitate future computations of physical observables at the 5PM 2SF order, which is the next natural landmark in PM theory, recently computed to 5PM 1SF order~\cite{Driesse:2024xad,Bern:2024adl,Driesse:2024feo}.

An apparent singularity of a Picard--Fuchs operator $\mathcal{L}_n$ is a point where the operator becomes singular, while the solutions to $\mathcal{L}_n \Psi=0$ are non-singular. One example are the $\varepsilon$-dependent apparent singularities in eq.~\eqref{eq: PF_CY_apparent_singularity}, which have no physical origin. As a general rule, $\varepsilon$-dependent apparent singularities can occur in the linear system of eq.~\eqref{eq: differential_equation_precanonical} and in the Picard--Fuchs operator. For linear systems, one can change the basis of master integrals and systematically remove these singularities, see e.g.~refs.~\cite{10.1145/220346.220385,10.1145/2755996.2756668}. By contrast, in the case of Picard--Fuchs operators, such transformations are either not algorithmic, or yield an operator of higher order than the original. For instance, see refs.~\cite{TSAI2000747,Abramov:2004,CHEN2016617,Slavyanov:2016,10.1016/j.jsc.2019.02.009} for methods to remove certain $\varepsilon$-dependent apparent singularities in Picard--Fuchs operators, in addition to refs.~\cite{Abramov:2004,Kauers:2013juk} for computer implementations. In many cases, $\varepsilon$-dependent apparent singularities appear in the Picard--Fuchs operator due to a poor choice of master integral basis. However, as we will explore in sec.~\ref{sec:ch5_toy_examples}, it is not guaranteed that there exists a basis where the singularities are absent and, even if it does exist, there is no algorithmic method to find this basis. Therefore, by extending the method of refs.~\cite{Pogel:2022ken,Pogel:2022vat} to allow for cases with $\varepsilon$-dependent apparent singularities, we reveal that it is not necessary to find an optimal basis of master integrals in order to derive an $\varepsilon$-factorized form.

The remainder of this chapter is structured as follows. First, in sec.~\ref{sec:ch5_toy_examples}, we introduce two toy examples which clarify the origin of the $\varepsilon$-dependent apparent singularities. Then, in sec.~\ref{sec:ch5_PF_gravity}, we provide more details on the 2SF CY integral family, and further characterize the underlying geometry by demonstrating that the associated Picard--Fuchs operator is of Calabi--Yau type. In sec.~\ref{sec:ch5_canonical_form}, we then present the method to bring the differential equation into $\varepsilon$-factorized form for cases with $\varepsilon$-dependent apparent singularities in the Picard--Fuchs operator. As a proof of principle, in sec.~\ref{sec:ch5_canonical_form_gravity}, we apply the method to the 2SF CY integral and bring its differential equation into $\varepsilon$-factorized form, including also the subsectors. Lastly, we conclude in sec.~\ref{sec:ch5_conclusions} with a discussion and further directions.

\section{Toy examples with \texorpdfstring{\bm{$\varepsilon$}}{eps}-dependent apparent singularities}
\label{sec:ch5_toy_examples}

Before delving into the specifics of the 2SF CY integral and introducing the general method for obtaining an $\varepsilon$-factorized differential equation for cases with $\varepsilon$-dependent apparent singularities, let us first consider a couple of toy examples in this section. These simple cases will illustrate the origin of the $\varepsilon$-dependence in apparent singularities, and clarify how these singularities can be manipulated through changes of master integral basis.

As the first example, we consider three master integrals $\vec{\mathcal{J}}=(\mathcal{J}_1,\mathcal{J}_2,\mathcal{J}_3)^T$, which satisfy an $\varepsilon$-factorized differential equation of the form of eq.~\eqref{eq: differential_equation_canonical}, given by
\begin{equation}
\label{eq: toy_example_1_linear_system}
 \frac{d}{dx} 
 \left( \begin{array}{c}
  \!\!{\mathcal J}_1 \!\! \\
  \!\!{\mathcal J}_2 \!\! \\
  \!\!{\mathcal J}_3 \!\! \\
 \end{array} \right)
 =
 \varepsilon \left[
  \left( \begin{array}{rrr}
   \!\! -9 & -33 & 56 \!\! \\
   \!\! \frac{9}{2} & 18 & -33 \!\! \\
   \!\! 1 & \frac{9}{2} & -9 \!\! \\ 
  \end{array} \right)
  \frac{1}{x}
  +
  \left( \begin{array}{rrr}
   \!\!12 & 40 & -64\!\! \\
   \!\!-7 & -24 & 40\!\! \\
   \!\!-2 & -7 & 12\!\! \\
  \end{array} \right)
  \frac{1}{x-1} 
  \right] \
 \left( \begin{array}{c}
  \!\!{\mathcal J}_1\!\! \\
  \!\!{\mathcal J}_2\!\! \\
  \!\!{\mathcal J}_3\!\! \\
 \end{array} \right).
\end{equation}
As can be seen, the singularities of this linear system are located at $x=0$ and $x=1$. Now, let us change the master integrals to a derivative basis, such as
\begin{equation}
\label{eq: toy_example_1_deriv_basis}
 \left\{ {\mathcal J}_1, \frac{d}{dx} {\mathcal J}_1, \frac{d^2}{dx^2}{\mathcal J}_1 \right\},
\end{equation}
which results in a third-order Picard--Fuchs operator $\mathcal{L}_3$ for $\mathcal{J}_1$. In particular, we obtain
\begin{equation}
\label{eq: toy_example_1_Picard_Fuchs}
\mathcal{L}_3 =
 C_3(x,\varepsilon) \frac{d^3}{dx^3}
 + C_2(x,\varepsilon) \frac{d^2}{dx^2}
 + C_1(x,\varepsilon) \frac{d}{dx}
 + C_0(x,\varepsilon)\, ,
\end{equation}
where the $C_j(x,\varepsilon)$ are rational polynomials in $x$ and $\varepsilon$, with
\begin{equation}
 C_3(x,\varepsilon)
 =
 x^2 \left(x-1\right)^2 P_{\text{apparent}}(x,\varepsilon)\,.
\end{equation}
Now, we can normalize the Picard--Fuchs operator such that the coefficient of the highest-order derivative is equal to one, which corresponds to rescaling $C_j(x,\varepsilon) \to C_j(x,\varepsilon)/C_3(x,\varepsilon)$. Then, the singularities of the operator become the roots of $C_3(x,\varepsilon)$. Besides the singularities at $x=0$ and $x=1$, which were already present in the linear system of eq.~\eqref{eq: toy_example_1_linear_system}, in the Picard--Fuchs operator there suddenly appear singularities at the three $\varepsilon$-dependent values for which
\begin{equation}
P_{\text{apparent}}(x,\varepsilon) \equiv 128 x \left(1-x\right) + \varepsilon\left(x-9\right) \left(17+46x+x^2\right)
\end{equation}
vanishes. Since these singularities are absent in the linear system, the solutions will be nonetheless regular at these points. Consequently, this implies that they correspond to $\varepsilon$-dependent apparent singularities, which are the focus of this chapter.

Notably, these $\varepsilon$-dependent apparent singularities arise when transforming to the derivative basis of eq.~\eqref{eq: toy_example_1_deriv_basis}. For a general $x$, this change of master integral basis is invertible. However, at the location of these apparent singularities, the determinant of the transformation vanishes, as it is given by
\begin{equation}
    \frac{\varepsilon^2 P_{\text{apparent}}(x,\varepsilon)}{x^3(1-x)^3}\,.
\end{equation}
Thus, at these points, the basis of eq.~\eqref{eq: toy_example_1_deriv_basis} becomes linearly dependent. In fact, following analogous steps, we find that the Picard--Fuchs operators for $\mathcal{J}_2$ and $\mathcal{J}_3$ also contain (different) $\varepsilon$-dependent apparent singularities.

However, in this simple toy example, we are able to find a new basis where neither the linear system nor the Picard--Fuchs operators contain $\varepsilon$-dependent apparent singularities. In particular, we can perform a constant $GL(3,\mathbb{C})$ rotation, and take a basis
\begin{equation}
 \left( \begin{array}{c}
 \!\! {\mathcal J}_1' \!\!\\
 \!\! {\mathcal J}_2' \!\!\\
 \!\! {\mathcal J}_3' \!\!\\
 \end{array} \right)
 =
  \left( \begin{array}{rrr}
   \!\!-\frac{1}{2} & -1 & 1\!\! \\
   \!\!1 & 3 & -4\!\! \\
   \!\!1 & 4 & -8\!\! \\ 
  \end{array} \right) \
 \left( \begin{array}{c}
  \!\!{\mathcal J}_1 \!\!\\
  \!\!{\mathcal J}_2 \!\!\\
  \!\!{\mathcal J}_3 \!\!\\
 \end{array} \right),
\end{equation}
which leads to a linear system
\begin{equation}
 \frac{d}{dx} 
 \left( \begin{array}{c}
  \!\!{\mathcal J}'_1 \!\! \\
  \!\!{\mathcal J}'_2 \!\! \\
  \!\!{\mathcal J}'_3 \!\! \\
 \end{array} \right)
 =
 \varepsilon \left[
  \left( \begin{array}{rrr}
   \!\! 0 & 1 & 0 \!\! \\
   \!\! 1 & 0 & 1 \!\! \\
   \!\! 0 & 1 & 0 \!\! \\ 
  \end{array} \right)
  \frac{1}{x}
  +
  \left( \begin{array}{rrr}
   \!\!0 & -1 & 0\!\! \\
   \!\!0 & 0 & -1\!\! \\
   \!\!0 & 0 & 0\!\! \\
  \end{array} \right)
  \frac{1}{x-1} 
  \right] \
 \left( \begin{array}{c}
  \!\!{\mathcal J}'_1\!\! \\
  \!\!{\mathcal J}'_2\!\! \\
  \!\!{\mathcal J}'_3\!\! \\
 \end{array} \right).
\end{equation}
Then, the leading coefficient for the Picard--Fuchs operator of $\mathcal{J}'_1$ is simply given by $C_3(x,\varepsilon)=x^2(1-x)^2$. Similarly, the Picard--Fuchs operators for $\mathcal{J}'_2$ and $\mathcal{J}'_3$ only contain singularities at $x=0$ and $x=1$. Thus, in this first example, we find a change of basis that completely removes the $\varepsilon$-dependent apparent singularities.

Let us then consider a second toy example, which satisfies a linear differential equation given by
\begin{equation}
\label{eq: toy_example_2_linear_system}
 \frac{d}{dx} 
 \left( \begin{array}{c}
  \!\!{\mathcal J}_1 \!\! \\
  \!\!{\mathcal J}_2 \!\! \\
  \!\!{\mathcal J}_3 \!\! \\
 \end{array} \right)
 =
 \varepsilon
  \left( \begin{array}{ccc}
   \!\! \frac{1}{x-2} & \frac{1}{x-1} & \frac{1}{x} \!\! \\
   \!\! \frac{1}{x-4} & \frac{1}{x-3} & \frac{1}{x-1} \!\! \\
   \!\! \frac{1}{x-5} & \frac{1}{x-4} & \frac{1}{x-2} \!\! \\ 
  \end{array} \right) \
 \left( \begin{array}{c}
  \!\!{\mathcal J}_1\!\! \\
  \!\!{\mathcal J}_2\!\! \\
  \!\!{\mathcal J}_3\!\! \\
 \end{array} \right),
\end{equation}
which has singularities at $x\, \mathlarger{\mathlarger{\in}} \, \{0,\dots,5\}$. Once again, the Picard--Fuchs operators for these master integrals contain $\varepsilon$-dependent apparent singularities. For instance, for $\mathcal{J}_1$ we obtain
\begin{equation}
 C_3(x,\varepsilon)
 =
 - \, x \left(x-1\right)^2 \left(x-2\right)^3 \left(x-3\right) \left(x-4\right)^2 \left(x-5\right)^2 P_{\text{apparent}}(x,\varepsilon)\,,
\end{equation}
with
\begin{equation}
P_{\text{apparent}}(x,\varepsilon) \equiv (x-1)(x-2)(x-3)(x-4) + \varepsilon\left( 6 - 19 x + x^2 + 8 x^3 - 2 x^4 \right) \,.
\end{equation}
In contrast with the first example considered, we now find that the $\varepsilon$-dependent apparent singularities are unavoidable, as there is no constant $GL(3,\mathbb{C})$ rotation which removes them from the Picard--Fuchs operators.

Therefore, from these toy examples, we can infer the following conclusions:
\begin{enumerate}
    \item There can exist a master integral basis with an ordinary $\varepsilon$-factorized linear differential equation, yet with $\varepsilon$-dependent apparent singularities in the Picard--Fuchs operator.
    \item A master integral without $\varepsilon$-dependent apparent singularities in the Picard--Fuchs operator may or may not exist.
\end{enumerate}
In summary, finding an $\varepsilon$-dependent apparent singularity in the Picard--Fuchs operator does not prevent us from obtaining an $\varepsilon$-factorized differential equation in a given master integral basis. Similarly, it does not always imply that this basis is not optimal.

In particular, the 2SF CY integral of sec.~\ref{sec:ch4_four_loop_CYs_2} provides such an example, where the most natural seed integral (the scalar integral) contains an $\varepsilon$-dependent apparent singularity in its Picard--Fuchs operator, recall eqs.~\eqref{eq: PF_PM_CY_2_eps}--\eqref{eq: PF_CY_apparent_singularity}. Furthermore, as we will discuss in the next section, we are not able to find a different seed integral where the $\varepsilon$-dependent apparent singularity is absent, nor can we determine whether such an optimal master integral exists in the first place. Therefore, our aim will not be to find this (possibly non-existent) optimal basis, but to develop a general method to bring its differential equation into $\varepsilon$-factorized form.

\section{Differential equations for the 2SF Calabi--Yau integral}
\label{sec:ch5_PF_gravity}

The focus of this section is the 2SF even-parity CY integral from sec.~\ref{sec:ch4_four_loop_CYs_2}. This is the simplest case that we are aware of, where an $\varepsilon$-dependent apparent singularity arises in the Picard--Fuchs operator, which cannot be avoided by using any other reasonable seed integral. In particular, here we study this Feynman integral in more detail, specifically using the differential equations method. In sec.~\ref{sec:ch5_CY_integral_family}, we provide further details on its integral family and classical subsectors, and discuss the origin of the factorization of its Picard--Fuchs operator in $D=4$. Then, in sec.~\ref{sec:ch5_PF_gravity_CY_operator}, we verify that the resulting fourth-order Picard--Fuchs operator satisfies the necessary conditions for a Calabi--Yau threefold operator~\cite{Bogner:2013kvr}, thereby proving that this is the geometry of the 2SF even-parity integral.

\subsection{Integral family}
\label{sec:ch5_CY_integral_family}

Recalling the discussion in sec.~\ref{sec:ch1_Feynman_integrals}, the integral family for the 2SF CY integral from sec.~\ref{sec:ch4_four_loop_CYs_2} is given by
\begin{equation}
\label{eq: integral_family_CY_2SF}
\mathcal{I}_{\nu_1\dots\,\nu_{22}}= \int \frac{d^D k_1 \, d^D k_2 \, d^D k_3 \, d^D k_4 \ \mathcal{D}_{12}^{-\nu_{12}} \, \cdots \, \mathcal{D}_{22}^{-\nu_{22}}}{\mathcal{D}_1^{\nu_1} \, \cdots \, \mathcal{D}_{11}^{\nu_{11}}}\,,
\end{equation}
where $\nu_1, \dots,\nu_{11} \geq 0$ correspond to the exponents of propagators, while $\nu_{12}, \dots, \nu_{22} \leq 0$ are the exponents of the ISPs. Specifically, we can follow the same parametrization and conventions as in fig.~\ref{fig: param_CY_2SF}, and define the 11 propagators as
\begin{subequations}
\begin{align}
\mathcal{D}_1&=k_1^2\,, & \mathcal{D}_2&=k_3^2\,, & \mathcal{D}_3&=(k_4+q)^2\,, & \mathcal{D}_4&=(k_2-q)^2\,, \\
\mathcal{D}_5&=(k_4-k_3)^2\,, & \mathcal{D}_6&=(k_1-k_2)^2\,, & \mathcal{D}_7&=(k_2+k_3)^2\,, & & \\
\mathcal{D}_8&=2u_1 \cdot k_1\,, & \mathcal{D}_9&=2u_2 \cdot k_3\,, & \mathcal{D}_{10}&=2u_2 \cdot k_4\,, & \mathcal{D}_{11}&=2u_1 \cdot k_2\,.
\end{align}
\end{subequations}
Similarly, the 11 ISPs are given by
\begin{subequations}
\begin{align}
\mathcal{D}_{12}&=k_2^2\,, & \mathcal{D}_{13}&=k_4^2\,, & \mathcal{D}_{14}&=(k_1-q)^2\,, & \mathcal{D}_{15}&=(k_3+q)^2\,, \\
\mathcal{D}_{16}&=(k_1+k_3)^2\,, & \mathcal{D}_{17}&=(k_1+k_4)^2\,, & \mathcal{D}_{18}&=(k_2+k_4)^2\,, & & \\
\mathcal{D}_{19}&=2u_2 \cdot k_1\,, & \mathcal{D}_{20}&=2u_2 \cdot k_2\,, & \mathcal{D}_{21}&=2u_1 \cdot k_3\,, & \mathcal{D}_{22}&=2u_1 \cdot k_4\,.
\end{align}
\end{subequations}

\begin{figure}[tb]
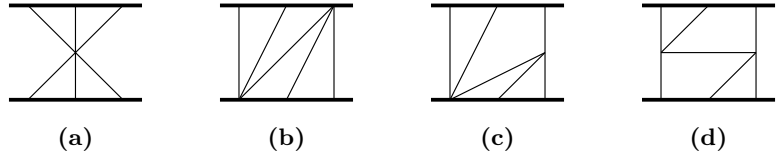

\centering
\subfloat[]{
}
\caption{Feynman integral topologies for the three subsectors (a)--(c) and the top sector (d) of the integral family in eq.~\eqref{eq: integral_family_CY_2SF}.}
\label{fig: diagrams_integral_family_2SF_CY}
\end{figure}
For this integral family, there are three polylogarithmic subsectors and one CY top sector, represented by the topologies in fig.~\ref{fig: diagrams_integral_family_2SF_CY}. For the subsector (a), we pinch the propagator $\mathcal{D}_7$, which corresponds to taking $\nu_7=0$. This topology was not included in the classification of sec.~\ref{sec:ch4_four_loop}, as it factorizes into the product of two 2-loop double triangles, which were already studied in sec.~\ref{sec:ch4_two_loop_0SF}. For the subsectors (b) and (c), we have $\nu_2=\nu_4=0$ and $\nu_2=0$, and they correspond to the diagrams in rows \diagramnumberingsign68 and \diagramnumberingsign57 of tab.~\ref{tab: 4-loop results 2SF}, respectively. Lastly, the top topology in fig.~\ref{fig: diagrams_integral_family_2SF_CY}(d) corresponds to the diagram studied in sec.~\ref{sec:ch4_four_loop_CYs_2}. Using \texttt{FIRE}~\cite{Smirnov:2023yhb} and \texttt{KIRA}~\cite{Klappert:2020nbg}, we can derive IBP relations for the even-parity sector of this integral family, obtaining the following set of pre-canonical master integrals:
{\allowdisplaybreaks
\begin{align*}
		\text{Sector $\sectora{}$ } \left(\!{
}\!\right):&\enspace \begin{cases}  \, \integralprec_6\equiv\integralprec_{111 111 111 11 000 000 000 00}\\*
\, \integralprec_7\equiv\integralprec_{111 211 111 11 000 000 000 00}\\*
\, \integralprec_8\equiv\integralprec_{111 112 111 11 000 000 000 00}\\*
\, \integralprec_9\equiv\integralprec_{111 111 211 11 000 000 000 00}\\*
\, \integralprec_{10}\equiv\integralprec_{111 111 111 11 ({-}1)00 000 000 00}\end{cases}
	\end{align*}}

Our main concern is the top sector for this family (sector $\sectord$ above) which, as discussed in sec.~\ref{sec:ch4_four_loop_CYs_2}, depends on a CY threefold. In particular, let us now elaborate on the choice of seed integral for this sector, which will be the starting point for finding an $\varepsilon$-factorized master integral basis in sec.~\ref{sec:ch5_eps_fact_top_sector}. In eq.~\eqref{eq: LS_PM_4loop_CY_2}, we found that the leading singularity for the scalar integral leads to
\begin{equation}
\LS \left( \mathcal{I}_6 \right) \propto \frac{x}{q^2} \, \int \frac{d t_1 d t_2 d t_3}{\sqrt{Q_8(t_1,t_2,t_3)}}\,,
\end{equation}
which defines an integral over the unique nowhere-vanishing holomorphic top-form of the CY threefold. Thus, even though its Picard--Fuchs operator contains an $\varepsilon$-dependent apparent singularity (recall eq.~\eqref{eq: PF_PM_CY_2_eps}), $\mathcal{I}_6$ is a reasonable candidate for a seed integral in the top sector. 

With the aim of finding one Picard--Fuchs operator free of $\varepsilon$-dependent apparent singularities, we also searched for other seed integrals in this sector. First, we constructed constant linear combinations of master integrals with up to 3 dots or numerator factors, but all 435 different combinations led to Picard--Fuchs operators containing such singularities.\footnote{In principle, one could also consider linear combinations with the coefficients being a function of $x$. This is, however, computationally more demanding, and for the few rational functions that we tried, we reached the same conclusion.} Moreover, we also tried removing the $\varepsilon$-dependent apparent singularity from the Picard--Fuchs operator of eq.~\eqref{eq: PF_PM_CY_2_eps} with the desingularization procedure of refs.~\cite{TSAI2000747,Abramov:2004,CHEN2016617,10.1016/j.jsc.2019.02.009,Kauers:2013juk,Slavyanov:2016}. However, these methods are not algorithmic, and when applied to our operator, they either produce further $\varepsilon$-dependent apparent singularities, or yield a differential operator of higher order than the original Picard--Fuchs operator, such that its solutions are no longer independent. As a consequence, we conclude that either a master integral basis free of $\varepsilon$-dependent apparent singularities may not exist for this top sector, or this basis is non-trivial to find. Hence, we will use the scalar integral $\mathcal{I}_6$ as our seed integral in sec.~\ref{sec:ch5_eps_fact_top_sector} as it has the simplest leading singularity. 

There is, nonetheless, one aspect which we have not addressed so far: the factorization of the Picard--Fuchs operator for $\mathcal{I}_6$ in $D=4$. As discussed in sec.~\ref{sec:ch4_four_loop_CYs_2}, in $D=4-2\varepsilon$, we obtain an irreducible differential operator of order 5, whereas in $D=4$ it factorizes as $\mathcal{L}_5\to \mathcal{L}_1 \cdot \mathcal{L}_4$; see eqs.~\eqref{eq: PF_PM_CY_2_L1}--\eqref{eq: PF_PM_CY_2_L4} for the explicit expressions. In fact, this factorization can be simply understood from the linear system of differential equations. Taking the vector of master integrals $\vec {\mathcal{I}}^{(\text{top})} \equiv (\mathcal{I}_6, \dots, \mathcal{I}_{10})^T$ for the top sector, on the maximal cut, it schematically satisfies a system of differential equations of the form
\begin{equation}
\label{eq: shape_diff_eq_top_sector_CY}
    \frac{d}{dx} \, \vec{\mathcal{I}}^{(\text{top})} = \left( \begin{array}{ccccc}
 \bullet & \bullet & \bullet & \bullet & 0 \\
 \bullet & \bullet & \bullet & \bullet & 0 \\
 \bullet & \bullet & \bullet & \bullet & \mathcal{O}(\varepsilon^2) \\
 \bullet & \bullet & \bullet & \bullet & \mathcal{O}(\varepsilon) \\
 \bullet & \bullet & \bullet & \bullet & \bullet \\
 \end{array} \right)  \ \vec{\mathcal{I}}^{(\text{top})}\,,
\end{equation}
where the dots denote non-zero entries, and $\mathcal{O}(\varepsilon)$ indicates the lowest order in $\varepsilon$. As can be seen, in $D=4-2\varepsilon$ this system is fully coupled: if we try to take further derivatives of $\mathcal{I}_6$, we end up with derivatives of all $\mathcal{I}_6\,$--$\,\mathcal{I}_{10}$. Instead, in $D=4$, all elements of the last column vanish except for the last one. Hence, taking further derivatives of the 4 master integrals $\mathcal{I}_6\,$--$\,\mathcal{I}_9$ does not involve $\mathcal{I}_{10}$. Thus, with regards to the $4 \times 4$ block $\mathcal{I}_6\,$--$\,\mathcal{I}_9$, there is a (quasi-)decoupling, which explains the factorization $\mathcal{L}_5 \to \mathcal{L}_1 \cdot \mathcal{L}_4$ in the Picard--Fuchs operator of eq.~\eqref{eq: PF_PM_CY_2_eps} in $D=4$. Then, 4 of the 5 solutions correspond to the solutions to $\mathcal{L}_4$, which come from the $4 \times 4$ CY block, and the last one is an integral over them.

By contrast, the case of $\mathcal{I}_{10}$ is very different. From the last row of eq.~\eqref{eq: shape_diff_eq_top_sector_CY}, we obtain
\begin{equation}
    \frac{d}{dx} \, \mathcal{I}_{10} = \sum_{i=6}^{9} c_i(x) \, \mathcal{I}_i + c_{10}(x) \, \mathcal{I}_{10}\,,
\end{equation}
where the $c_i(x)$ are rational functions, which leads to
\begin{equation}
    \widetilde{\mathcal{L}}_1 \, \mathcal{I}_{10} \equiv \left( \frac{d}{dx} - c_{10}(x) \right) \, \mathcal{I}_{10} = \sum_{i=6}^{9} c_i(x) \, \mathcal{I}_i\,.
\end{equation}
As discussed above, the right-hand side is annihilated by a fourth-order Picard--Fuchs operator $\widetilde{\mathcal{L}}_4$. Thus, for $\mathcal{I}_{10}$, we obtain instead the factorization $\widetilde{\mathcal{L}}_5 \to \widetilde{\mathcal{L}}_4 \cdot \widetilde{\mathcal{L}}_1$ in $D=4$. Importantly, among the 5 solutions, we find the rational solution to $\widetilde{\mathcal{L}}_1$, given by $x/(x^2-1)$. At first glance, this seems to jeopardize our classification of geometries of chapter~\ref{ch:chapter4}, since an algebraic leading singularity would hide the underlying Calabi--Yau geometry. However, if we carefully calculate the leading singularity for this integral, we obtain
\begin{equation}
\LS \left( \mathcal{I}_{10} \right) \propto \frac{x}{q^2} \, \int \frac{t_2 \, d t_1 d t_2 d t_3}{\sqrt{Q_8(t_1,t_2,t_3)}}\,.
\end{equation}
Under the change of variables $t_2 \to 1/t_2$, the leading singularity reveals a pole at infinity. In fact, this explains the origin of the (quasi-)decoupling for $\mathcal{I}_{10}$, as it is the only master integral with a pole.\footnote{Recall from sec.~\ref{sec:ch1_DEs}, that differential equations have a lower-triangular form because derivatives can remove poles (or propagators), but they cannot create new ones. Thus, this differential equation in $D=4$ has the same shape as if $\mathcal{I}_{10}$ was a top sector, and $\mathcal{I}_6\,$--$\,\mathcal{I}_9$ were its subsectors. In fact, this is the same scenario as for the elliptic integral of the third kind, which contains a pole and (quasi-)decouples from the other two elliptic integrals in the $3 \times 3$ differential equation system that they satisfy.} Precisely, taking the residue at this pole yields the rational solution to $\widetilde{\mathcal{L}}_1$. However, as emphasized in sec.~\ref{sec:ch1_LS}, in our analysis of geometries via leading singularities, we never take such residues in the presence of square roots.

\subsection{Conditions for a Calabi--Yau operator}
\label{sec:ch5_PF_gravity_CY_operator}

Having introduced in detail the 2SF even-parity CY integral and its integral family, in this subsection, let us verify that the fourth-order Picard--Fuchs operator from eq.~\eqref{eq: PF_PM_CY_2_L4} satisfies the conditions of a CY threefold operator. In particular, we follow the conventions given in Definition 3.27 of ref.~\cite{Bogner:2013kvr}; see also ref.~\cite{Pogel:2022vat} for an application in the context of the equal-mass banana integral. According to the previous convention, a CY operator $\mathcal{L}_n$ fulfills the following properties:
\begin{enumerate}
	\item The operator $\mathcal{L}_n$ is self-dual.
	\item $x=0$ is a point of maximal unipotent monodromy (MUM-point).
	\item The holomorphic solution at $x=0$ has a series expansion with $N$-integral coefficients.
	\item The special $q$-coordinate, expressed as a series around $x=0$, is $N$-integral.
	\item The structure series at $x=0$ has an $N$-integral series representation.
\end{enumerate}
In the following, we explain these mathematical conditions and verify that they are satisfied. In general, we will assume that the Picard--Fuchs operator $\mathcal{L}_n$ takes the form
\begin{align}
    \mathcal{L}_n &= \frac{d^n}{d x^n} + \sum_{j=0}^{n-1} a_j(x)\frac{d^j}{d x^j}\,.
\end{align}
Comparing with our definition from eq.~\eqref{eq: PF_definition}, the two conventions are related by setting $a_j(x) \equiv C_j(x,0)/C_n(x,0)$, which normalizes the coefficient of the highest-order derivative to one. As in sec.~\ref{sec:ch5_toy_examples}, the roots of the coefficient $C_n(x,0)$ provide the set of singularities of $\mathcal{L}_n$, which we will denote by $S_{\mathcal{L}_n}$.

\vspace*{0.2cm}
\textbf{\textit{\underline{\smash{Condition 1: Self-duality}}}}

A Picard--Fuchs operator $\mathcal{L}_n$ is self-dual if there exists a rational function $\alpha(x)$ (also known as Yukawa coupling) such that 
\begin{equation}
\label{eq: self-duality_def}
	\mathcal{L}_n\alpha(x)=\alpha(x) \hat{\mathcal{L}}_n\,,
\end{equation}
where $\hat{\mathcal{L}}_n$ is the so-called dual operator, given by
\begin{equation}
	\hat{\mathcal{L}}_n=\frac{d^n}{d x ^n} + \sum_{j=0}^{n-1} (-1)^{n+j}\frac{d^j}{d x ^j} a_j(x)\,.
\end{equation}
Here, note that the derivatives can act both on the coefficients $a_j(x)$ as well as on the function that the operator acts on.

For the case of a fourth-order Picard--Fuchs operator $\mathcal{L}_4$, self-duality is guaranteed if the following condition is satisfied~\cite{Almkvist:2021},
\begin{equation}
\label{eq: check_self_duality_L4}
    0= \frac{a_2(x) a_3(x)}{2} - a_1(x) - \frac{a_3(x)^3}{8} + \frac{d a_2(x)}{dx} - \frac{3 a_3(x)}{4} \frac{d a_3(x)}{dx} - \frac{1}{2} \frac{d^2 a_3(x)}{dx^2}\,.
\end{equation}
Then, the rational function $\alpha(x)$ is the solution to the differential equation
\begin{equation}
\label{eq :alpha_self-adj}
	\frac{d \alpha(x)}{d x}=-\frac{2 }{n} a_{n-1}(x)\alpha(x)\,.
\end{equation}

In our case, we find that the operator $\mathcal{L}_4$ from eq.~\eqref{eq: PF_PM_CY_2_L4} satisfies eq.~\eqref{eq: check_self_duality_L4}, and is self-dual with the Yukawa coupling
\begin{equation}
\label{eq: alpha_L4}
	\alpha(x)=\frac{ x^2+1}{x \left(1-x^2\right)^3}\,.
\end{equation}

\vspace*{0.2cm}
\textbf{\textit{\underline{\smash{Condition 2: MUM-points}}}}

In order to address the second condition, we need to introduce the Riemann-$\mathcal{P}$ symbol for $\mathcal{L}_n$, see e.g.~ref.~\cite{Weinzierl:2022eaz} for details. With that aim, we first define the Euler operator $\theta_x=x\frac{d}{d x}$. Using the identities
\begin{equation}
	\frac{d^k}{d x^k}=\frac{1}{x^k}\sum_{j=1}^{k}S_1(k,j)\theta_x^j\,, \qquad \qquad \theta_x^k=\sum_{j=1}^{k}S_2(k,j) x^j \frac{d^j}{d x^j}\,,
\end{equation}
where $S_1(k,j)$ and $S_2(k,j)$ denote the Stirling numbers of the first and second kind, respectively, we can rewrite $\mathcal{L}_n$ in the $\theta$-form
\begin{equation}
	\mathcal{L}_n = \sum_{j=0}^n P_j(x) \theta^j\,,
\end{equation}
where $P_j(x)$ are polynomials in $x$. Expanding around $x=0$, we can obtain the so-called indicial equation
\begin{equation}
    0= \sum_{j=0}^n P_j(0) \, b^j = C \, \left(b-b_1^{(0)} \right)  \cdots \left( b-b_n^{(0)}\right)\,,
\end{equation}
where $C$ is a constant. This equation defines a polynomial of degree $n$ in $b$, and its $n$ rational roots, denoted as $b_i^{(0)}$, are known as the indicials or local exponents of $\mathcal{L}_n$ at $x=0$. For other points $x=p$, one can proceed analogously by changing variables to $x_p=x-p$ (or $x_\infty=1/x$, for the case of $x=\infty$), and then expand around $x_p=0$ to obtain the indicials $b_i^{(p)}$. 

In general, the local exponents $b_i^{(p)}$ encapsulate the behavior of the solutions of $\mathcal{L}_n$ at $x=p$, as they have local series expansions starting at order $(x-p)^{b^{(p)}_i}$. For example, at regular points $p_{\text{reg}}$ of $\mathcal{L}_n$, the $b_i$ are different and are separated by integer numbers, and we have different convergent holomorphic solutions starting at $(x-p_{\text{reg}})^{b_1},\dots,(x-p_{\text{reg}})^{b_n}$. By contrast, for the singular points of the operator, $p_{\text{sing}} \, \mathlarger{\mathlarger{\in}} \, S_{\mathcal{L}_n}$, some of the indicials can be equal. In the case where all indicials are equal, $b_1=\dots=b_n=b^{(\text{MUM})}$, the corresponding singular point is known as a point of maximal unipotent monodromy (MUM-point), and the solutions at this point are given by a Frobenius basis; as elaborated below in condition 3. For MUM-points, the indicial equation therefore takes the simple form
\begin{equation}
    0= {\left(b - b^{(\text{MUM})} \right)}^n\,.
\end{equation}

For an operator $\mathcal{L}_n$ with $k$ different singular points $p_i \, \mathlarger{\mathlarger{\in}}\, S_{\mathcal{L}_n}$, the collection of indicials can be organized into an $n \times k$ matrix, which is known as the Riemann-$\mathcal{P}$ symbol,
\begin{equation}
	\mathcal{P}(\mathcal{L}_n)=\left\{ \ \, \begin{matrix}
		p_1 & \cdots & p_k \\\hline\hline
        & & \\[-0.4cm]
		b_1^{(p_1)} & \cdots & b_1^{(p_k)}\\
		\!\!\vdots & & \!\!\vdots\\[0.1cm]
		b_n^{(p_1)} & \cdots & b_n^{(p_k)} \\
	\end{matrix} \ \, \right\}\,.
\end{equation}
In particular, the sum of all indicials satisfies the so-called Fuchsian relation $\sum_{i=1}^k \sum_{j=1}^n b_j^{(p_i)} = n(n-1)(k-2)/2$.

For the case of the operator $\mathcal{L}_4$ from eq.~\eqref{eq: PF_PM_CY_2_L4}, the set of singularities is given by $S_{\mathcal{L}_4}=\{\pm 1, \pm i, 0, \infty\}$, and we find a Riemann-$\mathcal{P}$ symbol
\begin{equation}
	\mathcal{P}(\mathcal{L}_4)=\left\{ \ \, \begin{matrix}
		+1 & -1 & +i & -i & 0 & \infty \\\hline\hline
		0 & 0 & 0 & 0 & 1 & 1 \\
        0 & 0 & 1 & 1 & 1 & 1 \\
        0 & 0 & 3 & 3 & 1 & 1 \\
        0 & 0 & 4 & 4 & 1 & 1 \\
	\end{matrix} \ \, \right\}\,.
\end{equation}
Notably, we see that the points $x=\{ \pm1,0,\infty \}$ are MUM-points, since all indicials are equal. Therefore, $\mathcal{L}_4$ also satisfies the second condition for a Calabi--Yau threefold operator. 

However, we can also observe that for the points $x= \pm i$, the indicials are all different and separated by integer numbers. As discussed above, this indicates that the points $x = \pm i$ behave as regular points. Thus, they are not actual singular points of the solutions, but only of the Picard--Fuchs operator. Indeed, the solution space at $x=\pm i$ is spanned by convergent holomorphic series starting at $x^0$, $x^1$, $x^3$ and $x^4$. Consequently, these are ($\varepsilon$-independent) apparent singularities of $\mathcal{L}_4$, and for CY operators they appear at the location of the roots of the Yukawa coupling, see eq.~\eqref{eq: alpha_L4}. Unlike for $\varepsilon$-dependent apparent singularities, these singularities can be easily removed via changes of variables. In the case above, we may introduce $v \equiv y^2=(1+x^2)^2/4x^2$. However, as the $\varepsilon$-independent singularities do not represent an obstacle in the derivation, we will work with the variable $x$, which is the natural choice in PM theory as it rationalizes the square root $\sqrt{y^2-1}$.

\vspace*{0.3cm}
\textbf{\textit{\underline{\smash{Condition 3: Holomorphic solution}}}}

At a MUM-point, the $n$ series solutions $\varpi_i$, $i=1,\dots,n$, are given by the Frobenius method. In particular, we have one holomorphic solution $\varpi_1$, and $(n-1)$ divergent solutions which grow by powers of $\log(x)$,
\begin{subequations}
	\label{eq: Frobenius_MUM}
	\begin{align}
		&\varpi_1=\Sigma_1\,,\\
		&\varpi_2=\frac{1}{2\pi i}\left(\log(x)\Sigma_1+\Sigma_2\right)\,,\\
		&\ \, \vdots \nonumber\\[-0.1cm]
		&\varpi_n=\frac{1}{(2\pi i)^{n-1}}\left(\frac{\log(x)^{n-1}}{(n-1)!}\Sigma_1+\frac{\log(x)^{n-2}}{(n-2)!}\Sigma_2+\dots+\Sigma_n\right)\,.
	\end{align}
\end{subequations}
Here, $\Sigma_i$ denote series expansions in $x$ with rational coefficients. Using these expressions as an ansatz for the solutions and the operator $\mathcal{L}_4$ from eq.~\eqref{eq: PF_PM_CY_2_L4}, we find the series expansions
\begin{subequations}
	\begin{align}
    \label{eq: Frobenius series}
	\Sigma_1&=x+\frac{9 x^3}{2^4}+\frac{1681 x^5}{2^{12}}+\frac{21609 x^7}{2^{16}}+\frac{74805201 x^9}{2^{28}}+\mathcal{O}(x^{11})\,,\\[0.1cm]
	\Sigma_2&=\frac{3 x^3}{2^{3}}+\frac{1517
   x^5}{2^{12}}+\frac{22295 x^7}{2^{16}}+\frac{167072733 x^9}{2^{29}}+\mathcal{O}(x^{11})\,,\\[0.1cm]
   \Sigma_3&=\frac{5 x^3}{2^5}+\frac{7289 x^5}{2^{15}}+\frac{1123325
   x^7}{2^{19}3^2}+\frac{2042158233 x^9}{2^{33}}+\mathcal{O}(x^{11})\,,\\[0.1cm]
   \Sigma_4&=-\frac{5 x^3}{2^5}-\frac{7865 x^5}{2^{16}}-\frac{2448943 x^7}{2^{20}3^3}-\frac{2162890201
   x^9}{2^{35}}+\mathcal{O}(x^{11})\,.
	\end{align}
\end{subequations}

Now, the third condition for a CY operator demands that the holomorphic solution $\varpi_1 = \Sigma_1$ is $N$-integral. In general, a series $\sum_{j=0}^{\infty}c_j \, x^j$ is called integral if the coefficients $c_j \, \mathlarger{\mathlarger{\in}} \, \mathbb{Z}$ are integer numbers. Similarly, a series is called $N$-integral if there exists a rescaling $x'= x/N$ by a natural number $N \mathlarger{\mathlarger{\in}} \, \mathbb{N}$, such that the series $\sum_{j=0}^{\infty}c_j \, (N x')^j$ becomes integral. In our case, the holomorphic solution of eq.~\eqref{eq: Frobenius series} is $N$-integral,\footnote{Here and in the following, $N$-integrality is verified up to $\mathcal{O}(x^{1000})$.} with $N=2^4$.

\newpage
\textbf{\textit{\underline{\smash{Condition 4: Special q-coordinate}}}}

From the previous solutions at the MUM-point $x=0$, we can construct the so-called special $q$-coordinate~\cite{Bogner:2013kvr}, also known in the literature as the mirror map~\cite{Candelas:1990rm}. In particular, defining
\begin{equation}
\label{eq: def_tau_CY_mirror_map}
	\tau\equiv\frac{\varpi_2}{\varpi_1}=\frac{1}{2\pi i}\left(\log(x)+\frac{\Sigma_2}{\Sigma_1}\right),
\end{equation}
the special $q$-coordinate is given by
\begin{equation}
\label{eq: q-coord}
	q \equiv \exp(2\pi i \tau)=x \exp\left(\frac{\Sigma_2}{\Sigma_1}\right).
\end{equation}
In our case, it has a series expansion
\begin{equation}
\label{eq: mirror_map_q_series_expansion}
	q=x+\frac{3 x^3}{2^3}+\frac{941 x^5}{2^{12}}+\frac{5413 x^7}{2^{15}}+\frac{69120717 x^9}{2^{29}}+\frac{452476279 x^{11}}{2^{32}}+\mathcal{O}(x^{12})\,.
\end{equation}
As can be seen, this series is also $N$-integral with $N=2^4$, and therefore the operator $\mathcal{L}_4$ from eq.~\eqref{eq: PF_PM_CY_2_L4} also satisfies the fourth CY condition.

\vspace*{0.2cm}
\textbf{\textit{\underline{\smash{Condition 5: Structure series}}}}

For the last CY condition, we first need to rewrite the Picard--Fuchs operator in a form that manifestly annihilates the solutions $\varpi_i$. With that purpose, we define the operators~\cite{Bogner:2013kvr}
\begin{equation}
  \begin{aligned}
    \mathcal{N}_1 \equiv \theta_x\frac{1}{\varpi_1}\,, \qquad \qquad \mathcal{N}_{i+1} \equiv \theta_x\alpha_{i}\,\mathcal{N}_{i}\,,
  \end{aligned}
\end{equation}
where
\begin{equation}
  \alpha_i \equiv \left( \mathcal{N}_i (\varpi_{i+1})\right)^{-1}
  \,.
\end{equation}
For example, from the definitions above, we immediately see that
\begin{align}
\mathcal{N}_1 (\varpi_1) =& \, \theta_x (1) = 0\,, \\[0.1cm]
\mathcal{N}_2 (\varpi_1) =& \, \theta_x \alpha_1 \mathcal{N}_1 (\varpi_1) = 0\,, \qquad \mathcal{N}_2 (\varpi_2) = \theta_x \underbrace{\alpha_1 \mathcal{N}_1 (\varpi_2)}_{=1} = 0\,, \\
\mathcal{N}_3 (\varpi_1) =& \, \mathcal{N}_3 (\varpi_2) = 0\,, \qquad \qquad \mathcal{N}_3 (\varpi_3) = \theta_x \underbrace{\alpha_2 \mathcal{N}_2 (\varpi_3)}_{=1} = 0\,.
\end{align}
Similarly, the operator $\mathcal{N}_m$ annihilates all of the 
solutions $\varpi_i$ for $1\leq i\leq m$. Therefore, the Picard--Fuchs operator must be related to $\mathcal{N}_n$, which annihilates all $n$ solutions $\varpi_i$. Concretely, introducing the normalization factor $\beta$, we have the so-called local normal form~\cite{Bogner:2013kvr}
\begin{equation}
  \mathcal{L}_n= \beta \,  \mathcal{N}_n=\beta \, \theta_x\alpha_{n-1}\theta_x\alpha_{n-2}\theta_x\ldots\theta_x\alpha_1\theta_x\frac{1}{\varpi_1}\,,
\end{equation}
where the $\alpha_i$ are the so-called structure series of $\mathcal{L}_n$. In order to be a Calabi--Yau operator, in addition to having an $N$-integral series expansion around $x=0$, the structures series also need to satisfy $\alpha_{i}=\alpha_{n-i}$ for $1\leq i\leq n-1$.

For the fourth-order operator of eq.~\eqref{eq: PF_PM_CY_2_L4}, we find that
\begin{equation}
	\mathcal{L}_4=\beta\, \theta_x \alpha_3 \theta_x \alpha_2 \theta_x \alpha_1 \theta_x \frac{1}{\varpi_1}\,,
\end{equation}
where the structure series are given by
\begin{subequations}
\begin{align}
	\alpha_1 &= 1-\frac{3 x^2}{2^2}-\frac{77 x^4}{2^{10}}-\frac{369 x^6}{2^{13}}-\frac{1722877 x^8}{2^{26}}-\frac{1133073 x^{10}}{2^{26}}+\mathcal{O}(x^{11})\,,\\[0.1cm]
	\alpha_2 &= 1-\frac{11 x^2}{2^3}+\frac{2381 x^4}{2^{11}}-\frac{20723 x^6}{2^{14}}+\frac{161802701 x^8}{2^{27}}-\frac{1336327183 x^{10}}{2^{30}}+\mathcal{O}(x^{11})\,, \\[0.1cm]
 	\alpha_3 &= 1-\frac{3 x^2}{2^2}-\frac{77 x^4}{2^{10}}-\frac{369 x^6}{2^{13}}-\frac{1722877 x^8}{2^{26}}-\frac{1133073 x^{10}}{2^{26}}+\mathcal{O}(x^{11})\, .
	\end{align}
\end{subequations}
As can be seen, they fulfill $\alpha_1 = \alpha_3$, and have $N$-integral series representations for $N=2^4$. 

Consequently, the 2SF Picard--Fuchs operator from eq.~\eqref{eq: PF_PM_CY_2_L4} satisfies all conditions of ref.~\cite{Bogner:2013kvr}, and is hence a Calabi--Yau operator. This confirms that the corresponding Feynman integral geometry is a Calabi--Yau threefold, as identified via leading singularities in eq.~\eqref{eq: LS_PM_4loop_CY_2}.

\vspace*{0.2cm}
\textbf{\textit{\underline{\smash{$\boldsymbol{Y}\!$-invariants and instanton numbers}}}}

To further characterize the CY geometry at hand, let us compute a couple of additional quantities. First, we can define the so-called $Y$-invariants~\cite{Bogner:2013kvr}, which following the conventions in ref.~\cite{Pogel:2022vat} are given by
\begin{equation}
    Y_i \equiv \frac{\alpha_1}{\alpha_i}\,, \qquad \text{for} \qquad 1 \leq i \leq n-1\,.
\end{equation}
Importantly, the $Y$-invariants encapsulate the symmetric square condition from sec.~\ref{sec:ch4_three_loop_K3}. By the Proposition 4.6 of ref.~\cite{Bogner:2013kvr}, a CY operator $\mathcal{L}_n$ is the symmetric power of $\mathcal{L}_2$, if and only if, all $Y_i=1$. In such a case, we then have $\mathcal{L}_n = \text{Sym}^{n-1}(\mathcal{L}_2)$. For our operator $\mathcal{L}_4$, we find
\begin{equation}
	\label{eq: Y2_CY}
	Y_2=1+\frac{5 x^2}{2^3}-\frac{775 x^4}{2^{11}}-\frac{445 x^6}{2^{14}}-\frac{5110375 x^8}{2^{27}}-\frac{27054575 x^{10}}{2^{30}}+\mathcal{O}(x^{12})\,.
\end{equation}
Since $Y_2 \neq 1$, we can conclude that $\mathcal{L}_4 \neq \text{Sym}^{3}(\mathcal{L}_2)$. Therefore, the solution space for the 2SF CY integral will not be spanned by products of complete elliptic integrals, as it was the case for the K3 and $\text{K3}'$ integrals in chapter~\ref{ch:chapter4}, but actual integrals over the CY threefold.

Moreover, recalling the Yukawa coupling $\alpha(x)$ from the self-duality condition, see eq.~\eqref{eq: self-duality_def}, we can define the gauge-fixed Yukawa coupling $K(x)$~\cite{Bogner:2013kvr,Candelas:1990rm,Morrison:1991cd},
\begin{equation}
	K(x) \equiv \frac{x^{n-1} \, \alpha_1^{n-1} \, \alpha(x)}{\varpi_1^2}\,.
\end{equation}
For Calabi--Yau threefolds, it holds that $K(x)=Y_2$, which is also satisfied by our operator $\mathcal{L}_4$. In particular, using $K(x)$, one can reformulate the Calabi--Yau conditions from ref.~\cite{Bogner:2013kvr}. Specifically, ref.~\cite{vanStraten:2017} substitutes $N$-integrality by strict integrality in conditions~2--4, and condition~5 is replaced by demanding integer instanton numbers $n_j$. The latter can be obtained from the Lambert series expansion of the gauge-fixed Yukawa coupling in terms of the $q$-coordinate,
\begin{equation}
	K(q)=1+\sum_{j=1}^{\infty} \, j^3 \, n_j \,\frac{q^j}{1-q^j}\,,
\end{equation}
where the expansion $q(x)$ from eq.~\eqref{eq: mirror_map_q_series_expansion} is inverted to obtain the expansion $x(q)$. 

In the case of our operator $\mathcal{L}_4$ from eq.~\eqref{eq: PF_PM_CY_2_L4}, these alternative CY conditions from ref.~\cite{vanStraten:2017} are also satisfied with respect to $x'$, with the rescaling $x'= x/2^4$. In particular, the first ten instanton numbers are given by
\begin{equation}
\begin{tabular}{|l|cccccccccc|}
\Xhline{0.2ex}
$j$ & 1 & 2 & 3 & 4 & 5 & 6 & 7 & 8 & 9 & 10 \\
\hline
$n_j$ & 0 & 20 & 0 & $-$870 & 0 & 67460 & 0 & $-$6821070 & 0 & 800369820 \\
\Xhline{0.2ex}
\end{tabular}\ .
\end{equation}
As first realized in ref.~\cite{Klemm:2024wtd}, these instanton numbers allow us to pinpoint the particular operator $\mathcal{L}_4$ in a table of known CY operators. Specifically, it corresponds to the case 2.33 in ref.~\cite{Almkvist:2021}, which also corresponds to the case $6^{*}$ in ref.~\cite{Almkvist:2005qoo}.

\section{An \texorpdfstring{\bm{$\varepsilon$}}{eps}-factorized form for apparent singularities}
\label{sec:ch5_canonical_form}

In the previous sections, we have studied the origin of $\varepsilon$-dependent apparent singularities in the Picard--Fuchs operator of certain Feynman integrals, focusing on the case of the even-parity 2SF CY integral of sec.~\ref{sec:ch4_four_loop_CYs_2}. In this section, we describe the method to bring the corresponding differential equation into $\varepsilon$-factorized form. First, in sec.~\ref{sec:ch5_degree_eps_apparent_singularities}, we investigate the relation between the shape of an $\varepsilon$-factorized differential equation system and the type of apparent singularity arising in the respective Picard--Fuchs operator. Afterwards, in sec.~\ref{sec:ch5_method_eps_factorized}, we introduce the general method to bring the differential equation into $\varepsilon$-factorized form for cases with $\varepsilon$-dependent apparent singularities in the Picard--Fuchs operator, even beyond CY integrals. Since the apparent singularities occur within a given integral sector, here we will focus on the maximal cut. This will be useful in sec.~\ref{sec:ch5_canonical_form_gravity}, where we will apply this method to obtain an $\varepsilon$-factorized form for the 2SF CY integral, including also its subsectors (recall fig.~\ref{fig: diagrams_integral_family_2SF_CY}).

\subsection{Structure of the \texorpdfstring{\bm{$\varepsilon$}}{eps}-dependent apparent singularities}
\label{sec:ch5_degree_eps_apparent_singularities}

The aim of this subsection is to find a relation between the structure of the $\varepsilon$-factorized differential equation system and the type of apparent singularities arising in the corresponding Picard--Fuchs operator. Our starting point will be a generic $\varepsilon$-factorized differential equation system, recall eq.~\eqref{eq: differential_equation_canonical}, obtained after finding an appropriate change of master integral basis $\vec{\mathcal{J}} = U \vec{\mathcal{I}}$. In practice, we may also change variables in the kinematic space, e.g.~from $x$ to $\tau$. Therefore, in the following we will work with the expression in eq.~\eqref{eq: differential_equation_canonical_tau_intro}, which we rewrite here for convenience:
\begin{equation}
\label{eq: differential_equation_canonical_tau}
\frac{d}{d\tau} \, \vec{\mathcal{J}} = J \frac{d}{dx} \, \vec{\mathcal{J}}=\varepsilon A_\tau(\tau) \, \vec{\mathcal{J}}\,,
\end{equation}
where we introduce the Jacobian $J=\frac{dx}{d\tau}$, as well as the $n \times n$ matrix $A_\tau = J A_x$. 

In general, one can assume that the differential equation is self-dual~\cite{Pogel:2024sdi}. Even though this assumption is not necessary for our method to work, self-duality will reduce the number of unknown coefficients in our ansatz, thus simplifying the process of finding the solution. In particular, it implies that
\begin{equation}
\label{eq: self-duality_def_matrix_diffeq}
 (A_\tau)_{ij} = (A_\tau)_{(n-j+1)(n-i+1)}\,,
\end{equation}
i.e.~the matrix $A_\tau$ is symmetric with respect to the anti-diagonal. Finding an $\varepsilon$-factorized master integral basis with self-duality has been possible in numerous examples, and it has been conjectured to hold for all Feynman integrals~\cite{Pogel:2024sdi}; see ref.~\cite{Duhr:2024xsy} for a first study in this direction.

Now, we can always transform to a derivative basis for one of the master integrals, for instance for $\mathcal{J}_1$. Thereafter, we can derive its Picard--Fuchs operator, which we write with the conventions of eq.~\eqref{eq: PF_definition},
\begin{equation}
\mathcal{L}_n= \sum_{j=0}^{n} C_j(x,\varepsilon) \, \frac{d^j}{d x^j}\,.
\end{equation}
As discussed in sec.~\ref{sec:ch5_toy_examples}, the roots of the polynomial $C_n(x,\varepsilon)$ provide the singularities of the operator. Consequently, this shows that we can transform between the two pictures: a generic $\varepsilon$-factorized differential equation system, and the associated Picard--Fuchs operator for $\mathcal{J}_1$. 

In the following, let us explore the interplay between the shape of the matrix $A_\tau$, and the type of apparent singularities which appear in the corresponding higher-order operator.\footnote{As we show in sec.~\ref{sec:ch5_method_eps_factorized}, the leading coefficient $C_n(x,\varepsilon)$ is the same for the pre-canonical seed integral and the master integral $\mathcal{J}_1$. Thus, we can equivalently study the apparent singularities of the Picard--Fuchs operator in the pre-canonical basis.} Considering a Feynman integral on the maximal cut, the simplest case corresponds to a matrix $A_\tau$ with a lower-triangular form,
\begin{equation}
 A_\tau =
 \left( \begin{array}{ccccc}
  \!\!F_{11} & 0 & 0 & \hdots & 0 \!\!\\
  \!\!F_{21} & F_{22} & 0 & \hdots & 0 \!\!\\
  \!\!\vdots & \vdots & \ddots & \ddots & \vdots \!\!\\
  \!\!F_{(n-1) 1} & F_{(n-1) 1} & \hdots & F_{(n-1)(n-1)} & 0 \!\!\\
  \!\!F_{n 1} & F_{n 2} & \hdots & F_{n (n-1)} & F_{n n} \!\!\\
 \end{array} \right)\,.
\end{equation}
Here, the entries $F_{ij}$ only depend on $x$, since otherwise the system would not be $\varepsilon$-factorized. This case is trivial, since we can follow the same argument as in sec.~\ref{sec:ch5_CY_integral_family} for $\mathcal{I}_{10}$, and conclude that all Picard--Fuchs operators completely factorize into a product of $\mathcal{L}_1$'s.

In order to obtain an irreducible Picard--Fuchs operator of order greater than one, we must at least include non-zero elements in the next-to-diagonal entries:
\begin{equation}
 A_\tau =
 \left( \begin{array}{cccccc}
  \!\!F_{11} & F_{12} & 0 & 0 & \hdots & 0 \!\!\\
  \!\!F_{21} & F_{22} & F_{23} & 0 & \hdots & 0\!\!\\
  \!\!\vdots & \vdots & & \ddots & \ddots & \vdots \!\!\\
  \!\!F_{(n-2) 1} & F_{(n-2) 2} & \hdots & F_{(n-2) (n-2)} & F_{(n-2) (n-1)}  & 0 \!\!\\
  \!\!F_{(n-1) 1} & F_{(n-1) 2} & \hdots & F_{(n-1) (n-2)} & F_{(n-1)(n-1)} & F_{(n-1)n} \!\!\\
  \!\!F_{n 1} & F_{n 2} & \hdots & F_{n (n-2)} & F_{n (n-1)} & F_{n n} \!\!\\
 \end{array} \right)\,.
\end{equation}
This type of differential equation matrix is precisely the basis for the ansatz of refs.~\cite{Pogel:2022ken,Pogel:2022vat}, where they bring the differential equation into $\varepsilon$-factorized form for the equal-mass banana integrals. For such a matrix, we generally obtain an irreducible Picard--Fuchs operator of order $n$, where the coefficient of the highest-order derivative is given by
\begin{equation}
 C_n(x,\varepsilon)
 =
 J^n \prod_{j=1}^{n-1} \left( F_{j(j+1)} \right)^{n-j}\,.
\end{equation}
However, since the $F_{ij}$'s are independent of $\varepsilon$ by construction, a differential equation of this kind will not give rise to $\varepsilon$-dependent apparent singularities.

Consequently, in order to obtain $\varepsilon$-dependent apparent singularities, we must extend the ansatz of refs.~\cite{Pogel:2022ken,Pogel:2022vat}. In particular, we find that these apparent singularities arise when including non-zero elements in the next-to-next-to-diagonal,
\begin{equation}
\label{eq: NND_ansatz}
 A_\tau =
 \left( \begin{array}{cccccc}
  \!\!F_{11} & F_{12} & \hspace{0.5cm} F_{13} & \hspace{0.5cm} 0 & \hdots & 0 \!\!\\
  \!\!F_{21} & F_{22} & \hspace{0.5cm} F_{23} & \hspace{0.5cm} F_{24} & \hdots & 0\!\!\\
  \!\!\vdots & \vdots & & \ddots & \ddots & \vdots \!\!\\
  \!\!F_{(n-3) 1} & F_{(n-3) 2} & \hspace{0.5cm} \hdots & \hspace{0.5cm}F_{(n-3) (n-2)} & F_{(n-3) (n-1)}  & 0 \!\!\\
  \!\!F_{(n-2) 1} & F_{(n-2) 2} & \hspace{0.5cm} \hdots & \hspace{0.5cm} F_{(n-2) (n-2)} & F_{(n-2) (n-1)}  & F_{(n-2) n} \!\!\\
  \!\!F_{(n-1) 1} & F_{(n-1) 2} & \hspace{0.5cm} \hdots & \hspace{0.5cm} F_{(n-1) (n-2)} & F_{(n-1)(n-1)} & F_{(n-1)n} \!\!\\
  \!\!F_{n 1} & F_{n 2} & \hspace{0.5cm} \hdots & \hspace{0.5cm} F_{n (n-2)} & F_{n (n-1)} & F_{n n} \!\!\\
 \end{array} \right)\,.
\end{equation}
Considering the simplest example with $n=3$ and assuming self-duality, recall eq.~\eqref{eq: self-duality_def_matrix_diffeq}, we have a matrix
\begin{equation}
 A_\tau =
 \left( \begin{array}{ccc}
 \!\!F_{11} & F_{12} & F_{13} \!\!\\
 \!\!F_{21} & F_{22} & F_{12} \!\!\\
 \!\!F_{31} & F_{21} & F_{11} \!\!\\
 \end{array} \right)\,.
\end{equation}
In this case, we obtain a leading coefficient
\begin{equation}
 C_3(x,\varepsilon)
 = 
 J^4 \left( F_{12} \frac{d F_{13}}{dx} - F_{13} \frac{d F_{12}}{dx} \right)
 + \varepsilon \, J^3 \left( F_{12}^3 - F_{13}^2 F_{21} + F_{12} F_{13} \left( F_{11} - F_{22} \right)  \right) \,,
\end{equation}
which generates an $\varepsilon$-dependent apparent singularity of the same type as in the toy examples of sec.~\ref{sec:ch5_toy_examples}. In particular, since the matrix above is totally generic for the case $n=3$, we conclude that a $3 \times 3$ 
self-dual differential equation system can lead to an apparent singularity which is, at most, linear in $\varepsilon$.

\begin{figure}[t]
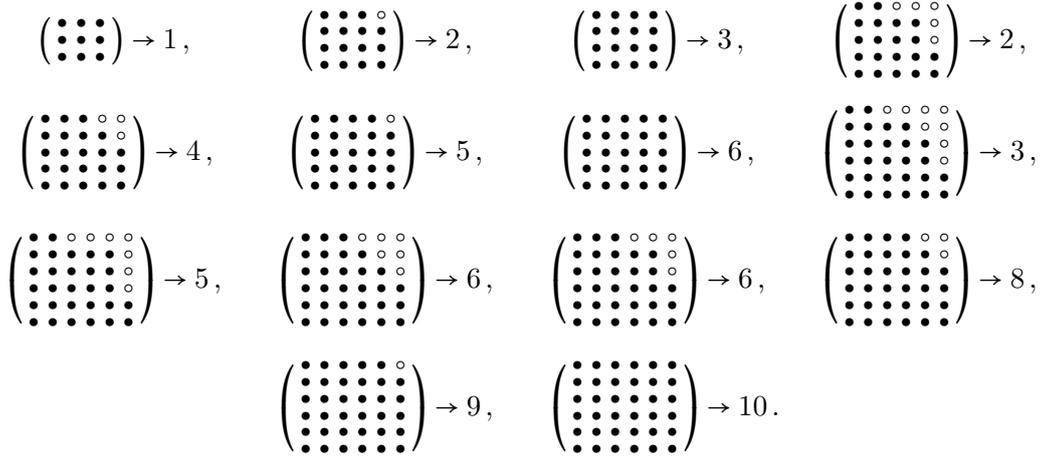

\centering
\begin{tabular}{cccc}
$\left( \begin{smallmatrix} \bullet & \bullet & \bullet\\ \bullet & \bullet & \bullet\\ \bullet & \bullet & \bullet \end{smallmatrix} \right) \to 1\,, \!\quad$ &
$\left( \begin{smallmatrix} \bullet & \bullet & \bullet & \circ \\ \bullet & \bullet & \bullet & \bullet\\ \bullet & \bullet & \bullet & \bullet\\ \bullet & \bullet & \bullet & \bullet \end{smallmatrix} \right) \to 2\,, \!\quad$ &
$\left( \begin{smallmatrix} \bullet & \bullet & \bullet & \bullet\\  \bullet & \bullet & \bullet & \bullet\\ \bullet & \bullet & \bullet & \bullet\\ \bullet & \bullet & \bullet & \bullet \end{smallmatrix} \right) \to 3\,, \!\quad$ & 
$\left( \begin{smallmatrix} \bullet & \bullet & \circ & \circ & \circ \\ \bullet & \bullet & \bullet & \bullet & \circ \\ \bullet & \bullet & \bullet & \bullet & \circ \\ \bullet & \bullet & \bullet & \bullet & \bullet \\ \bullet & \bullet & \bullet & \bullet & \bullet \end{smallmatrix} \right) \to 2\,,$ \\
\end{tabular}
\caption{Different shapes of a self-dual $\varepsilon$-factorized differential equation matrix $A_\tau$, and the resulting degree in $\varepsilon$ for the apparent singularity of the corresponding Picard--Fuchs operator, up to $6 \times 6$ size.}
\label{fig: epsilon_degree_apparent_sing}
\end{figure}

To generalize the previous result, let us explore the degree in $\varepsilon$ of the apparent singularity as a function of the size and the shape of the $\varepsilon$-factorized differential equation matrix. With this purpose, we can use a random ansatz for $A_\tau$, such that each entry can contain an arbitrary degree-5 polynomial in the numerator, and a denominator with possible roots at $x=0$ and $x=\pm1$. Doing so, we obtain the classification of fig.~\ref{fig: epsilon_degree_apparent_sing}, where we represent different shapes of $A_\tau$ and the corresponding degree in $\varepsilon$ for the apparent singularity in the Picard--Fuchs operator, up to a $6 \times 6$ system. As can be seen, the degree in $\varepsilon$ grows both with the system size and with the shape of the matrix, especially when the matrix has fewer zero entries. For instance, a completely generic $6 \times 6$ self-dual $\varepsilon$-factorized differential equation matrix can lead to an apparent singularity of degree 10 in $\varepsilon$.

Finally, let us examine the 2SF CY integral, which led to an apparent singularity that was quadratic in $\varepsilon$, recall eq.~\eqref{eq: PF_CY_apparent_singularity}. 
In this case, using the map from fig.~\ref{fig: epsilon_degree_apparent_sing} in the opposite direction, we can deduce that the corresponding $\varepsilon$-factorized differential equation matrix takes the form
\begin{equation}
\label{eq: ansatz_section3}
 A_\tau =
 \left( \begin{array}{ccccc}
 \!\!F_{11} & F_{12} & 0 & 0 & 0 \!\!\\
 \!\!F_{21} & F_{22} & F_{23} & F_{24} & 0 \!\!\\
 \!\!F_{31} & F_{32} & F_{33} & F_{23} & 0 \!\!\\
 \!\!F_{41} & F_{42} & F_{32} & F_{22} & F_{12}\!\! \\
 \!\!F_{51} & F_{41} & F_{31} & F_{21} & F_{11}\!\! \\
\end{array} \right)\,.
\end{equation}
Thus, we now have an ansatz for the resulting $\varepsilon$-factorized differential equation for our 2SF CY integral on the maximal cut. In order to solve for the entries of the $\varepsilon$-factorized differential equation matrix, the next step is to construct an appropriate change of master integral basis.

\subsection{General method}
\label{sec:ch5_method_eps_factorized}

In the previous subsection, we have found a correspondence between the shape of the final $\varepsilon$-factorized differential equation matrix, and the type of apparent singularity appearing in the Picard--Fuchs operator of $\mathcal{J}_1$. Here, to relate the $\varepsilon$-factorized basis $\vec{\mathcal{J}}$ to the original (pre-canonical) master integral basis $\vec{\mathcal{I}}$, we construct the transformation $\vec{\mathcal{J}}=U \vec{\mathcal{I}}$, for which we follow refs.~\cite{Pogel:2022ken,Pogel:2022vat}. In particular, a natural starting point is to define $\mathcal{J}_1$ as
\begin{equation}
\label{eq: relation_seed_integral}
 {\mathcal J}_1 = \frac{N(\varepsilon)}{\omega(x)}\,  \integralseed \,,
\end{equation}
where $\integralseed$ is the seed master integral, and $N(\varepsilon)$ is a normalization that tunes the leading order in $\varepsilon$ for the series expansion of $\mathcal{J}_1$. The key element is the denominator $\omega(x)$, which corresponds to a solution of the Picard--Fuchs operator in $D=4$; in other words, it is the leading singularity of $\integralseed$. This definition is motivated by the experience from polylogarithmic integrals (see the discussion in sec.~\ref{sec:ch1_LS}), where a pure basis is normally obtained by dividing each master integral by its leading singularity. Importantly, dividing by $\omega(x)$ does not modify the coefficient of the highest-order derivative in the Picard--Fuchs operator, so the type of $\varepsilon$-dependent apparent singularity is the same for $\integralseed$ and for $\mathcal{J}_1$.

With this definition for $\mathcal{J}_1$, we have all the ingredients to solve for the $\varepsilon$-factorized differential equation. On the one hand, we know the pre-canonical basis $\vec{\mathcal{I}}$ and its differential equation matrix $\tilde{A}_x(x,\varepsilon)$ from eq.~\eqref{eq: differential_equation_precanonical}. On the other hand, we have the unknown $\varepsilon$-factorized basis $\vec{\mathcal{J}}$, and the ansatz for its $\varepsilon$-factorized differential equation matrix $A_\tau$ from eq.~\eqref{eq: NND_ansatz}. Using the relation in eq.~\eqref{eq: relation_seed_integral}, we know the type of apparent singularity arising in the Picard--Fuchs operator of $\mathcal{J}_1$ and, as discussed in the previous subsection, the relevant shape for $A_\tau$. Consequently, we can treat the $F_{ij}$'s, $J$ and $\omega(x)$ as unknown functions in an ansatz, and solve for their values to fix the final $\varepsilon$-factorized basis and differential equation. In practice, one can manually input the values for the leading singularity $\omega(x)$ and for $J=\frac{d x}{d \tau}$ from the mirror map in sec.~\ref{sec:ch5_PF_gravity_CY_operator}, which would facilitate finding a solution. However, as this is not necessary, we will use it as a consistency check at the end.

To find the solution for the unknown coefficients, we can employ two equivalent methods:
\begin{enumerate}
    \item Comparison of the linear system of differential equations: 
    
    From the definition for $\mathcal{J}_1$ in eq.~\eqref{eq: relation_seed_integral}, we can construct the transformation matrix $U(x,\varepsilon)$, as well as the rest of the $\varepsilon$-factorized basis. Specifically, we can introduce an intermediate basis $\vec{\mathcal{K}}$, defined as 
    \begin{subequations}
    \label{eq: intermediate_K_basis_def}
    \begin{align}
 {\mathcal K}_1 = & \, {\mathcal J}_1\,, \\[0.1cm]
 {\mathcal K}_j = & \, \frac{J}{\varepsilon} \frac{d {\mathcal K}_{j-1}}{dx}\,, \qquad \text{for} \qquad  2 \leq j \leq n \,,
    \end{align}
    \end{subequations}
    which can be simultaneously related to both bases $\vec{\mathcal{I}}$ and $\vec{\mathcal{J}}$. For instance, we can immediately see that
    \begin{align}
    \mathcal{K}_2 =& \, \frac{J}{\varepsilon} \frac{d \mathcal{J}_1}{dx} \nonumber \\[0.1cm]
    =&\, \frac{J \, N(\varepsilon)}{\varepsilon} \left( \frac{1}{\omega(x)} \frac{d \integralseed}{dx} - \frac{\integralseed}{\omega(x)^2} \frac{d \omega(x)}{dx} \right)\,.
    \end{align}
    Then, we can introduce eqs.~\eqref{eq: NND_ansatz} and~\eqref{eq: differential_equation_precanonical} for $d \mathcal{J}_1/dx$ and $d \integralseed/dx$, respectively, and express $\mathcal{K}_2$ in terms of both $\vec{\mathcal{J}}$ and $\vec{\mathcal{I}}$. Proceeding analogously for the remaining elements in the basis, we find
    \begin{equation}
 \vec{\mathcal{K}} \; = \; U_1 \vec{\mathcal{I}}\,,
 \qquad \qquad
 \vec{\mathcal{K}} \; = \; U_2 \vec{\mathcal{J}}\,,
    \end{equation}
    such that 
    \begin{equation}
 \vec{\mathcal{J}} = U \vec{\mathcal{I}},
 \qquad \text{with} \qquad
 U \; = \; U_2^{-1} U_1.
    \end{equation}
    Hence, we have expressed $U(x,\varepsilon)$ in terms of the unknown functions from eq.~\eqref{eq: NND_ansatz}. Finally, we can use eq.~\eqref{eq: relation_linear_system_intro} together with eq.~\eqref{eq: differential_equation_canonical_tau}, to compare the linear systems and fix the unknown functions from the condition
    \begin{equation}
    \label{eq: condition_1_eps_fact_solution}
    U \tilde{A}_x U^{-1} + \frac{d U}{dx} \, U^{-1} - \frac{\varepsilon}{J} A_\tau 
    = 0\,.
    \end{equation}
    Concretely, we require that the numerator vanishes at each order in $\varepsilon$.
    \item Comparison of the Picard--Fuchs operators: 

    On the one hand, from the linear system in $\vec{\mathcal{J}}$-basis given in eq.~\eqref{eq: NND_ansatz}, we can derive the Picard--Fuchs operator for $\mathcal{J}_1$. Multiplying by the common denominator, we can write it as
    \begin{equation}
    \label{PF_v1}
    \left( \sum_{k=0}^{n} \sum_{j \geq 0} \, p_{k,j}(x) \,  \varepsilon^j \, \frac{d^k}{dx^k} \right)
 {\mathcal J}_1 = 0\,,
    \end{equation}
    where the polynomials $p_{k,j}(x)$ depend on $x$ through the functions $F_{ij}$, and where the upper bound for $j$ depends on $k$, but is finite. On the other hand, from the definition for $\mathcal{J}_1$ in eq.~\eqref{eq: relation_seed_integral}, and using the linear system in $\vec{\mathcal{I}}$-basis from eq.~\eqref{eq: differential_equation_precanonical}, we can derive the alternative form
    \begin{equation}
    \label{PF_v2}
    \left( \sum_{k=0}^{n} \sum_{j \geq 0} \, \tilde{p}_{k,j}(x) \,  \varepsilon^j \, \frac{d^k}{dx^k} \right)
 {\mathcal J}_1 = 0\,.
    \end{equation}
    Hence, matching the operators, we obtain the condition
    \begin{equation}
    \label{eq: condition_2_eps_fact_solution}
    \tilde{p}_{k,j}(x) = c(x) \, p_{k,j} (x)\,,
    \end{equation}
    for all $j$, $k$, and an unknown function $c(x)$.
\end{enumerate}

While the solution obtained from both approaches is equivalent, in the literature the predominant strategy has been to solve for the $\varepsilon$-factorized differential equation using method 1. However, in our case we found it simpler to solve the problem via the comparison of Picard--Fuchs operators, i.e.~with eq.~\eqref{eq: condition_2_eps_fact_solution} from method 2. This is because, in general, the system of equations obtained from eq.~\eqref{eq: condition_1_eps_fact_solution} in method 1 is highly coupled and, at intermediate steps, there can be multiple solutions to the same equation, with some of them leading to incompatible solutions later on. Thus, finding the correct sequence of variables to solve for can become quite burdensome. Instead, the set of equations obtained from eq.~\eqref{eq: condition_2_eps_fact_solution} inherently has a hierarchical structure, allowing us to find the solution by starting with the equations at $k=n$ and decreasing $k$ one step at a time. Therefore, we now have all the tools needed to find the $\varepsilon$-factorized differential equation for the 2SF CY integral, which is the focus of the next section.

\section{Application to the 2SF Calabi--Yau integral}
\label{sec:ch5_canonical_form_gravity}

In this section, we apply the method described in sec.~\ref{sec:ch5_method_eps_factorized} to find an $\varepsilon$-factorized differential equation system for the even-parity sector of the 2SF integral family of eq.~\eqref{eq: integral_family_CY_2SF}. Concretely, we use the 10 master integrals described in sec.~\ref{sec:ch5_CY_integral_family}, which we organize in a vector $\vec{\mathcal{I}}=(\mathcal{I}_1, \dots, \mathcal{I}_{10})^T$. With this basis, we have the pre-canonical differential equation system
\begin{equation}
\label{eq: system_precanonical_CY_family}
	\frac{d}{d x} \, \vec{\mathcal{I}} =\tilde{A}_x(x,\varepsilon) \, \vec{\mathcal{I}} =
	\begin{pNiceMatrix}[columns-width = 0.9em]
\Block[draw=blue,fill=blue!40,rounded-corners]{1-1}{}
0  &  &  &  &   &   & & & &  \\
0 & \Block[draw=blue,fill=blue!40,rounded-corners]{3-3}{} \bullet & \bullet  & \bullet  &   &   &  & & &   \\
0 & \bullet  & \bullet  & \bullet  &   &   &  & & &   \\
0 & \bullet  & \bullet  & \bullet  &  &   &  & & &   \\
0  & \Block[draw=red,fill=red!40,rounded-corners]{1-3}{} \bullet  & \bullet & \bullet  & \Block[draw=blue,fill=blue!40,rounded-corners]{1-1}{}\bullet  &  &  & & &  \\
 \Block[draw=red,fill=red!40,rounded-corners]{5-5}{}\bullet  & \bullet  & \bullet  & \bullet & \bullet &\Block[draw=blue,fill=blue!40,rounded-corners]{5-5}{} \bullet  & \bullet &\bullet &\bullet & 0 \\
\bullet  & \bullet  & \bullet  & \bullet & \bullet & \bullet  & \bullet &\bullet &\bullet & 0 \\
\bullet  & \bullet  & \bullet  & \bullet & \bullet & \bullet  & \bullet &\bullet &\bullet & \bullet \\
\bullet  & \bullet  & \bullet  & \bullet & \bullet & \bullet  & \bullet &\bullet &\bullet & \bullet \\
\bullet  & \bullet  & \bullet  & \bullet & \bullet & \bullet  & \bullet &\bullet &\bullet & \bullet 
\end{pNiceMatrix} \ \vec{\mathcal{I}}\,,
\end{equation}
where the dots indicate non-zero entries. As can be seen, it satisfies the lower-triangular block structure discussed in eq.~\eqref{eq: matrix_shape_diff_eq}, where the blue blocks on the diagonal denote the maximal cut of the 4 different classical sectors $\sectora$--$\sectord$ in fig.~\ref{fig: diagrams_integral_family_2SF_CY}, and the red blocks denote off-diagonal mixings. Recall that sector $\sectora$ only has $\mathcal{I}_1$ as a master integral, sector $\sectorb$ comprises $\mathcal{I}_2\,$--$\, \mathcal{I}_4$, sector $\sectorc$ contains only $\mathcal{I}_5$, and sector $\sectord$ (which depends on the CY) includes $\mathcal{I}_6 \,$--$\, \mathcal{I}_{10}$.

In order to bring this differential equation into $\varepsilon$-factorized form, we will work downwards from the subsectors to the CY top sector, focusing first on the (blue) maximal-cut blocks, and later on the (red) off-diagonal blocks. First, in sec.~\ref{sec:ch5_eps_fact_subsectors}, we consider the polylogarithmic $5 \times 5$ subsector block $\mathcal{I}_1 \,$--$\,\mathcal{I}_5$. Quite conveniently, the $1 \times 1$ maximal-cut block of sector $\sectora$ is trivial, and the $1 \times 1$ block of sector $\sectorc$ is already in $\varepsilon$-factorized form. Thus, the first task consists in bringing the $3 \times 3$ block of sector $\sectorb$ into $\varepsilon$-factorized form. Since its Picard--Fuchs operator does not contain an $\varepsilon$-dependent apparent singularity, one could use standard tools to solve this sector. However, it will serve as a simple example to illustrate how to use the method of sec.~\ref{sec:ch5_method_eps_factorized} and solve the constraints. Afterwards, we bring the off-diagonal mixing between sectors $\sectorb$ and $\sectorc$ into $\varepsilon$-factorized form by using standard polylogarithmic tools.

Subsequently, in sec.~\ref{sec:ch5_eps_fact_top_sector}, we use the method of sec.~\ref{sec:ch5_method_eps_factorized} to bring the $5 \times 5$ CY diagonal block (already introduced in eq.~\eqref{eq: shape_diff_eq_top_sector_CY}) into $\varepsilon$-factorized form, which is the main challenge in the entire calculation. Lastly, we perform a final change of master integral basis to bring the $5 \times 5$ off-diagonal block mixing the top sector with the subsectors into $\varepsilon$-factorized form. In the following, we will use $\vec{\mathcal{J}}$ to denote the basis in which the diagonal blocks become $\varepsilon$-factorized, and $\vec{\mathcal{T}}$ to denote the basis in which the off-diagonal terms also become $\varepsilon$-factorized.

\subsection{An \texorpdfstring{\bm{$\varepsilon$}}{eps}-factorized form for the subsectors}
\label{sec:ch5_eps_fact_subsectors}

In this subsection, we bring the subsectors $\sectora$--$\sectorc$ in eq.~\eqref{eq: system_precanonical_CY_family} into $\varepsilon$-factorized form. Concretely, we focus on the $5 \times 5$ block spanned by the master integrals $\mathcal{I}_1\,$--$\,\mathcal{I}_5$.

As already mentioned, the $1 \times 1$ maximal-cut block for sector $\sectora$ is trivial, while the $1 \times 1$ block for sector $\sectorc$ is already $\varepsilon$-factorized. Thus, we automatically have
\begin{equation}
		\integralFinal_1^{\sectora{}}=\integralprec_1\,, \qquad \qquad
		\integralFinal_1^{\sectorc{}}=\integralprec_5\,,
\end{equation}
where the superscript indicates the respective sector. Notice that these expressions are consistent with eq.~\eqref{eq: relation_seed_integral}, since from the classification of chapter~\ref{ch:chapter4} we found that $\mathcal{I}_1$ and $\mathcal{I}_5$ have unit leading singularity; see eq.~\eqref{eq: LS_double_triangle_unit_LS} and row 57 in tab.~\ref{tab: 4-loop results 2SF}, respectively.

Therefore, the first non-trivial part is the $3 \times 3$ block for sector $\sectorb$, explicitly given by
\begin{equation}
\label{eq: linear_system_sectorb}
\frac{d}{d x} \left( \begin{array}{c}
 \!\! {\mathcal I}_2 \!\!\!\\
 \!\! {\mathcal I}_3 \!\!\!\\
 \!\! {\mathcal I}_4 \!\!\!\\
 \end{array} \right)=
	\left(
\begin{array}{ccc}
 \!\!\frac{x^4 (5 \varepsilon -2)+x^2 (30 \varepsilon -8)+5 \varepsilon -2}{x (x^4-1)} & \frac{4 x (2 \varepsilon +1)}{(x^4-1) (4 \varepsilon -1)} & \frac{(x^2-1) \varepsilon }{x
   (x^2+1) (4 \varepsilon -1)}\!\! \\[0.3cm]
 \!\!\frac{4 (x^2-1) \varepsilon  (4 \varepsilon -1) (5 \varepsilon -1)}{x (x^2+1) (2 \varepsilon +1)} & \frac{4 (x^2-1) \varepsilon }{x(x^2+1)} & -\frac{4 (x^2-1) \varepsilon
   ^2}{x (x^2+1) (2 \varepsilon +1)} \!\!\\[0.3cm]
 \!\!-\frac{(x^2-1) (4 \varepsilon -1) (5 \varepsilon -1)}{x(x^2+1)} & \frac{4 x (2 \varepsilon +1)}{x^4-1} & \frac{x^4 (\varepsilon +1)+x^2 (6 \varepsilon +2)+\varepsilon +1}{x(1-x^4)}\!\!\\
\end{array}
\right) \ \left( \begin{array}{c}
 \!\! {\mathcal I}_2 \!\!\!\\
 \!\! {\mathcal I}_3 \!\!\!\\
 \!\! {\mathcal I}_4 \!\!\!\\
 \end{array} \right)\,.
\end{equation}
In this case, the classification of geometries from chapter~\ref{ch:chapter4} (see specifically row 68 of tab.~\ref{tab: 4-loop results 2SF}) indicates that this sector is polylogarithmic, as it has algebraic leading singularities and no further classical subsectors. Thus, it could be brought into $\varepsilon$-factorized form using standard methods, e.g.~with \texttt{Fuchsia}~\cite{Gituliar:2017vzm}. Nevertheless, we will use the method from sec.~\ref{sec:ch5_method_eps_factorized} to illustrate its application for this simple example, before turning to the more complicated CY top sector in sec.~\ref{sec:ch5_eps_fact_top_sector}.

The first step in the method is to find an appropriate seed integral $\integralseed^{\sectorb{}}$ for sector $\sectorb$. Based on the analysis in tab.~\ref{tab: 4-loop results 2SF}, we expect a good seed integral to be
\begin{equation}
\label{eq: seed_integral_sector_b}
	\integralseed^{\sectorb{}}\equiv \integralprec_4= \integralprec_{1010112111100000000000}\,, \qquad \text{with} \qquad \text{LS}\left( \integralseed^{\sectorb{}} \right) = \frac{x}{x^2-1}\,,
\end{equation}
where we used the notation from the 2SF integral family of eq.~\eqref{eq: integral_family_CY_2SF}. Graphically, this seed integral corresponds to the diagram where we place a dot in the middle graviton propagator. From the linear system in eq.~\eqref{eq: linear_system_sectorb}, we can obtain the Picard--Fuchs operator for the seed integral in $D=4-2\varepsilon$, which corresponds to
\begin{align}
\label{eq:PF_seed_sectorb}
	\mathcal{L}_{\integralseed^{\sectorb{}}}= &\,
	x^3 (x^2-1)^2 \frac{d^3}{d x^3} -2 x^2 (x^2-1) (4 x^2 \varepsilon -4 x^2+4 \varepsilon -1) \frac{d^2}{d x^2} \nonumber\\[0.1cm]
	&\, +2 x \left( 2 - x^2 + 7 x^4 -8 \varepsilon (1 + x^2 + 2 x^4) + 8 \varepsilon^2 (1 - 7 x^2 + x^4) \right) \frac{d}{d x} +4 (x^4-1) (2 \varepsilon -1)^2\,.
\end{align}
In $D=4$, it factorizes into three first-order differential operators, $\mathcal{L}_1 \cdot \mathcal{L}_1 \cdot \mathcal{L}_1$, which confirms the prediction from tab.~\ref{tab: 4-loop results 2SF} that this integral is polylogarithmic in four dimensions. 

As can be seen from eq.~\eqref{eq:PF_seed_sectorb}, the leading coefficient is given by $C_3(x,\varepsilon)=x^3 (x^2-1)^2$. Since it is independent of $\varepsilon$, based on the analysis of sec.~\ref{sec:ch5_degree_eps_apparent_singularities}, we conclude that the ansatz for the $\varepsilon$-factorized differential equation matrix of sector $\sectorb$ is given by
\begin{equation}
\label{eq: linear_system_sectorb_J}
	J\frac{d}{d x} \, \vec{\mathcal{J}}^{\sectorb{}}=\varepsilon  \left( \begin{array}{ccc}
	 F_{11} & F_{12} & 0 \\
	 F_{21} & F_{22} & F_{23} \\
	 F_{31} & F_{32} & F_{33} \\
 \end{array} \right) \, \vec{\mathcal{J}}^{\sectorb{}}\,.
\end{equation}
In this case, we found that self-duality cannot be imposed from the beginning, as it is not manifest with the chosen seed integral. Instead, we will obtain a self-dual differential equation at the end by performing a rotation.

Following the method of sec.~\ref{sec:ch5_method_eps_factorized}, from the linear system in $\vec{\mathcal{J}}^{\sectorb{}}$-basis of eq.~\eqref{eq: linear_system_sectorb_J}, we can derive the corresponding Picard--Fuchs operator for $\mathcal{J}_1^{\sectorb{}}$. In this case, the coefficients $p_{k,j}(x)$ from eq.~\eqref{PF_v1} take the form
{\allowdisplaybreaks
\begin{subequations}
\begin{align}
\label{eq:sec_sectorbp_first}
		p_{3,0}&=J^3 F_{12}^2 F_{23}\,,\\[0.1cm]
		p_{2,1}&=-J^2 F_{12}^2 F_{23} (F_{11}+F_{22}+F_{33})\,,\\[0.1cm]
		p_{2,0}&=-J^2 F_{12} \Big(F_{23} (2 J F_{12}'-3 F_{12} J')+J F_{12} F_{23}'\Big)\,,\\[0.1cm]
		p_{1,2}&=J F_{12}^2 F_{23} \Big(F_{11} (F_{22}+F_{33})-F_{12} F_{21}+F_{22} F_{33}-F_{23} F_{32}\Big)\,,\\[0.1cm]
		p_{1,1}&=J F_{12} \Big[ F_{12} \Big( J \left(F_{11}+F_{22}\right) F_{23}'-F_{23} \left(J' \left(F_{11}+F_{22}+F_{33}\right)+J \left(2 F_{11}'+F_{22}'\right)\right) \Big) \nonumber\\*
		&\hphantom{{}={}}+J F_{23} F_{12}' \left(2F_{11}+F_{22}+F_{33}\right) \Big] \,,\\[0.1cm]
		p_{1,0}&=J \Big[ F_{23} \Big( J^2 \left(2 F_{12}'{}^2-F_{12} F_{12}''\right)+F_{12}^2 J'^2+J F_{12} \left(F_{12} J''-3 J' F_{12}'\right) \Big) \nonumber\\*
		&\hphantom{{}={}} +J F_{12} F_{23}' \left(J F_{12}'-F_{12} J'\right) \Big]\,,\\[0.1cm]
   		p_{0,3}&=F_{12}^2 F_{23} \Big(F_{11} \left(F_{23} F_{32}-F_{22} F_{33}\right)+F_{12} \left(F_{21} F_{33}-F_{23} F_{31}\right)\Big)\,,\\[0.1cm]
   		p_{0,2}&=J F_{12} \Big[ F_{12} \Big(F_{23} \left(F_{11} F_{22}'+F_{33} F_{11}'\right)+F_{22} \left(F_{23} F_{11}'-F_{11} F_{23}'\right)\Big)\nonumber\\*
   		&\hphantom{{}={}}-F_{11} F_{23} F_{12}' \left(F_{22}+F_{33}\right)+F_{12}^2
   \left(F_{21} F_{23}'-F_{23} F_{21}'\right) \Big] \,,\\[0.1cm]
   		p_{0,1}&=J \Big[ F_{12} \Big(F_{23} \left(F_{11} J' F_{12}'+2 J F_{11}' F_{12}'+J F_{11} F_{12}''\right)-J
   F_{11} F_{12}' F_{23}'\Big) -2 J F_{11} F_{23} \left(F_{12}'\right){}^2 \nonumber\\*
   		&\hphantom{{}={}}- F_{12}^2 \Big(F_{23} J' F_{11}'+J \left(F_{23} F_{11}''-F_{11}' F_{23}'\right)\Big) \Big] \,,\\[0.1cm]
   		p_{0,0}&=0\,.   		
\label{eq:sec_sectorbp_last}
\end{align}
\end{subequations}
}

On the other hand, we can express $\mathcal{J}_1^{\sectorb}$ in terms of the seed integral through eq.~\eqref{eq: relation_seed_integral}. In particular, we have that
\begin{equation}
\label{eq: def_J1_sectorb}
 {\mathcal J}_1^{\sectorb} = \frac{1}{\omega(x)}\,  \integralseed^{\sectorb} \,,
\end{equation}
where $\integralseed^{\sectorb}$ is given in eq.~\eqref{eq: seed_integral_sector_b}. Thus, from the Picard--Fuchs operator for $\integralseed^{\sectorb}$, given in eq.~\eqref{eq:PF_seed_sectorb}, we can obtain an alternative expression for the operator annihilating $\mathcal{J}_1^{\sectorb}$. The coefficients $\tilde{p}_{k,j}(x)$ from eq.~\eqref{PF_v2} are then given by
\begin{subequations}
\begin{align}
\label{eq:sec_sectorbptilde_first}
		\tilde{p}_{3,0}&=x^3 (x^2-1)^2 \omega\,,\\[0.1cm]
		\tilde{p}_{2,1}&=-8 x^2 (x^4-1) \omega\,,\\[0.1cm]
		\tilde{p}_{2,0}&=x^2 (x^2-1) \Big(3 x (x^2-1) \omega '+(8 x^2+2) \omega\Big)\,,\\[0.1cm]
		\tilde{p}_{1,2}&=16 x (x^4-7 x^2+1) \omega\,,\\[0.1cm]
		\tilde{p}_{1,1}&=-16 x \Big(x (x^4-1) \omega '+(2 x^4+x^2+1) \omega\Big)\,,\\[0.1cm]
		\tilde{p}_{1,0}&=x \Big[ x (x^2-1) \Big(3 x (x^2-1) \omega ''+4 (4 x^2+1) \omega '\Big)+2 (7 x^4-x^2+2) \omega\Big]\, ,\\[0.1cm]
		\tilde{p}_{0,3}&=0\,,\\[0.1cm]
		\tilde{p}_{0,2}&=16 \Big((x^4-1) \omega+x (x^4-7 x^2+1) \omega '\Big)\,,\\[0.1cm]
		\tilde{p}_{0,1}&=-16 (x^4-1) \omega-8 x \Big(x (x^4-1) \omega ''+2 (2 x^4+x^2+1) \omega '\Big)\,,\\[0.1cm]
		\tilde{p}_{0,0}&=x^3(x^2-1)^2 \omega''' + 2 x^2(x^2-1)(4x^2+1) \omega'' + 2x (7 x^4-x^2+2) \omega '+ 4 (x^4-1) \omega\,.
\label{eq:sec_sectorbptilde_last}
\end{align}
\end{subequations}

Hence, in order to derive the $\varepsilon$-factorized differential equation, we just need to solve the conditions from eq.~\eqref{eq: condition_2_eps_fact_solution}. In this case, we have that
\begin{equation}
	\label{eq:sectorb_matching_condition}
 \tilde{p}_{k,j}(x) = c(x) \, p_{k,j}(x)\,, \qquad \text{for} \qquad 0\leq k\leq 3\,,\ \ 0\leq j\leq 3-k\,,
\end{equation}
where the unknown functions in the ansatz are given by
\begin{equation}
	F_{11}\,,\ F_{12}\,,\ F_{21}\,,\ F_{22}\,,\ F_{23}\,,\ F_{31}\,,\ F_{32}\,,\ F_{33}\,,\ J\,,\ \omega\,,\ c\,.
\end{equation}
In general, solving these constraints can be quite cumbersome, as they are highly coupled. However, we found a hierarchy in the equations, which allows us to disentangle them. As mentioned in sec.~\ref{sec:ch5_method_eps_factorized}, we can start with the equations for $k=3$ and decrease $k$ in steps of one, where for each $k$ we solve the constraints starting also with the largest $j$. Specifically, this is the order given in eqs.~\eqref{eq:sec_sectorbp_first}--\eqref{eq:sec_sectorbp_last} and \eqref{eq:sec_sectorbptilde_first}--\eqref{eq:sec_sectorbptilde_last}. With this sequence, we can solve at each step for the unknown functions specified in tab.~\ref{tab: strategy subsector}. In particular, if we relabel $F_{01} \equiv c$ and $F_{34} \equiv \omega$, the condition from $p_{k,j}$ provides a constraint for the $(3-k-j)$-th derivative of $F_{3-k,4-k-j}$. Notably, these conditions do not fix the value of $J$. Thus, we will assume $J=1$, which means that we do not perform any further changes of variables in the kinematics. However, the freedom in $J$ can also be exploited to normalize one of the entries $F_{ij}$ to be equal to one, as done e.g.~in refs.~\cite{Pogel:2022ken,Pogel:2022vat}. This way, we obtain constraints such as
\begin{subequations}
\begin{align}
	c&= \frac{x^3 (x^2-1)^2 \omega }{F_{12}^2 F_{23}}\,,\\[0.1cm]
	F_{12}'&= \frac{F_{12}}{2} \left(-\frac{F_{23}'}{F_{23}}+\frac{8 x^2+2}{x(1-x^2)}-\frac{3 \omega '}{\omega }\right),\\[0.1cm]
	F_{11}&=\frac{x(1-x^2)(F_{22}+F_{33})+8 (x^2+1)}{x (x^2-1)}\,,\\[0.1cm]
	\omega'''&= - \frac{2 x^2(x^2-1)(4x^2+1) \omega'' + 2x (7 x^4-x^2+2) \omega '+ 4 (x^4-1) \omega}{x^3 (x^2-1)^2}\,,\\[0.1cm]
	F_{23}''&=  F_{23}\left(\frac{ \frac{2 \left(4 x^2+1\right) \omega '}{x (x^2-1)}+3 \omega ''}{\omega }+\frac{2 \left(2 x^4-3 x^2+2\right)}{x^2 (x^2-1)^2}-\frac{3 \omega '^2}{2\omega ^2}+\frac{3 F_{23}'{}^2}{2 F_{23}^2}\right).
\end{align}
\end{subequations}

\begin{table}[t]
\begin{center}
\caption{For given values of $k$ and $j$ in eq.~\eqref{eq:sectorb_matching_condition}, each entry in the table specifies the unknown function to be determined.}
\label{tab: strategy subsector}
$\begin{NiceArray}{|c|cccc|}[columns-width = 2em]
\Xhline{0.2ex}
\diagbox{\,\,k}{j\,\,} & 0 & 1 & 2 & 3 \\ \hline
3 & c &  &  &  \\
2 & F_{12}' & F_{11} &  &  \\
1 & F_{23}'' & F_{22}' & F_{21} &  \\
0 & \omega''' & F_{33}'' & F_{32}' & F_{31}\\
\Xhline{0.2ex}
\end{NiceArray}$
\end{center}
\end{table}
Once again, these constraints can be solved in a specific order. Using the notation from tab.~\ref{tab: strategy subsector}, we can start with $j=0$ and increase it in steps of one, and for each $j$ go from lowest to highest $k$, i.e., we follow
\begin{equation}
	\omega\to F_{23}\to F_{12}\to c\to F_{33}\to F_{22} \to F_{11} \to F_{32} \to F_{21}\to F_{31}\,.
\end{equation}
Choosing appropriate boundary conditions, we finally obtain the solutions
\begin{subequations}
\begin{align}
	\omega(x) =&\, \frac{x}{x^2-1}\,, &
	F_{23}=&\, \frac{4}{x}\,, &
	F_{12}=&\, \frac{4 \sqrt{5}}{x^2-1}\,, &
	c(x) =&\, \frac{x^5 (x^2-1)^3}{320}\,, &
	F_{33}=&\, 0\,,\\[0.1cm]
	F_{22}=&\, \frac{8 (x^2+1)}{x(x^2-1)}\,, &
   F_{11}=&\, 0\,, &
   F_{32}=&\, -\frac{4}{x}\,, &
   F_{21}=&\, \frac{4 \sqrt{5}}{x^2-1}\,, &
   F_{31}=&\, 0\,.
\end{align}
\end{subequations}
As anticipated in sec.~\ref{sec:ch5_method_eps_factorized}, the normalization factor $\omega(x)$ from eq.~\eqref{eq: def_J1_sectorb} corresponds to the leading singularity of the seed integral in $D=4$, recall eq.~\eqref{eq: seed_integral_sector_b}. In fact, comparing eqs.~\eqref{eq:PF_seed_sectorb} and~\eqref{eq:sec_sectorbptilde_last}, we can immediately see that the condition $\tilde{p}_{0,0}(x)=0$ imposes that $\omega(x)$ is a solution to the Picard--Fuchs operator of the seed integral with $\varepsilon=0$.

As a result, we obtain an $\varepsilon$-factorized differential equation,
\begin{equation}
	\frac{d}{d x} \, \vec{\mathcal{J}}^{\sectorb{}}=\varepsilon  \left(
\begin{array}{ccc}
	 \!\!0 & \hspace{0.2cm} \frac{4 \sqrt{5}}{x^2-1} & \hspace{0.2cm} 0\!\! \\[0.1cm]
 \!\!\frac{4 \sqrt{5}}{x^2-1} & \hspace{0.2cm} \frac{8 (x^2+1)}{x(x^2-1)} & \hspace{0.2cm} \frac{4}{x}\!\! \\[0.1cm]
 \!\!0 & \hspace{0.2cm} -\frac{4}{x} & \hspace{0.2cm} 0\!\! \\
 \end{array} \right) \, \vec{\mathcal{J}}^{\sectorb{}}\,.
\end{equation}
The master integral transformation $\vec{\mathcal{J}}^{\sectorb{}} = U^{\sectorb{}} \, \vec{\mathcal{I}}^{\sectorb{}}$, for $\vec{\mathcal{I}}^{\sectorb{}} = (\mathcal{I}_2,\mathcal{I}_3,\mathcal{I}_4)^T$, is given by
\begin{equation}
	U^{\sectorb{}}=\left( \begin{array}{ccc}
	 \!\!0 & 0 & x-\frac{1}{x}\!\! \\[0.2cm]
 \!\!-\frac{(x^2-1)^3 (4 \varepsilon -1) (5 \varepsilon -1)}{4 \sqrt{5} x^2 (x^2+1) \varepsilon } & \frac{(x^2-1) (2 \varepsilon +1)}{\sqrt{5} (x^2+1) \varepsilon } &
 -\frac{(x^2-1) (x^4+6 x^2+1)}{4 \sqrt{5} x^2 (x^2+1)}\!\! \\[0.3cm]
 \!\!\frac{(x^2-1)^2 (4 \varepsilon -1) (5 \varepsilon -1)}{4 \sqrt{5} x^2 \varepsilon } & -\frac{(2 \varepsilon +1) (10 \varepsilon -1)}{4 \sqrt{5} \varepsilon ^2} & \frac{(x^2-1)^2}{4 \sqrt{5}
   x^2}\!\! \\
 \end{array} \right)\,.
\end{equation}
At this point, we already succeeded in bringing the differential equation for sector $\sectorb$ into $\varepsilon$-factorized form. However, we can perform one further rotation of the master integral basis, such that the differential equation is self-dual and obeys Galois symmetry~\cite{Pogel:2024sdi}. Defining a new basis $\vec{\mathcal{J}}^{\sectorb{}}_{\text{SD}} = U^{\sectorb{}}_{\text{SD}} \, \vec{\mathcal{J}}^{\sectorb{}}$, where
\begin{equation}
	U^{\sectorb{}}_{\text{SD}}=\left( \begin{array}{ccc}
	 \!\!1 & i & 0\!\! \\
 \!\!0 & 0 & i \sqrt{2}\!\! \\
 \!\!1 & -i & 0\!\! \\
 \end{array} \right)\,,
\end{equation}
we finally obtain
\begin{equation}
	d \, \vec{\mathcal{J}}^{\sectorb{}}_{\text{SD}}=\varepsilon A^{\sectorb{}} \vec{\mathcal{J}}^{\sectorb{}}_{\text{SD}}=\varepsilon  \left( \begin{array}{ccc}
	 \!\!4 f_2 & 2 \sqrt{2} f_1 & -2(2 f_2- \sqrt{-5} f_3)\!\! \\[0.1cm]
 \!\!-2 \sqrt{2} f_1 & 0 & 2 \sqrt{2} f_1\!\! \\[0.1cm]
 \!\!-2(2 f_2+ \sqrt{-5} f_3) & -2 \sqrt{2} f_1 & 4 f_2\!\! \\
 \end{array} \right) \, \vec{\mathcal{J}}^{\sectorb{}}_{\text{SD}}\,,
\end{equation}
with
\begin{equation}
	f_1=d\log(x)\,,\qquad f_2=d\log\left(\frac{1-x^2}{x}\right)\,, \qquad 
	f_3=d\log\left(\frac{1-x}{1+x}\right)\,.
\end{equation}
As can be seen, the matrix $A^{\sectorb{}}$ is self-dual in the sense of eq.~\eqref{eq: self-duality_def_matrix_diffeq}, i.e.~it is symmetric across the anti-diagonal. Moreover, defining the action of the Galois symmetry $\rho_{-5,2}$ as
\begin{equation}
\label{eq: Galois action}
	\rho_{-5,2}:\ \begin{array}[t]{rl}
		\sqrt{-5}&\mapsto -\sqrt{-5}\,,\\
		\sqrt{2} &\mapsto -\sqrt{2}\,,
	\end{array}
\end{equation}
the differential equation system stays invariant, since
\begin{equation}
	\rho_{-5,2}\left(A^{\sectorb{}}\right)=\left( A^{\sectorb{}} \right) ^T\,, \qquad \qquad
\rho_{-5,2}\left(\mathcal{J}^{\sectorb}_{\text{SD},1}\right)=\mathcal{J}^{\sectorb}_{\text{SD},3}\,.
\end{equation}

The last step for the subsectors is to bring the $1 \times 3$ off-diagonal block mixing sectors $\sectorb$ and $\sectorc$ (recall eq.~\eqref{eq: system_precanonical_CY_family}) into $\varepsilon$-factorized form. Conveniently, after the previous transformations in sector $\sectorb$, this off-diagonal block is already $\varepsilon$-factorized, although with double poles. With the help of \texttt{Fuchsia}~\cite{Gituliar:2017vzm}, we can remove them by performing a final basis transformation. In the end, we find an $\varepsilon$-factorized differential equation for the entire $5 \times 5$ subsector block,
\begin{equation}
	d \, \vec{\mathcal{T}}^{(\text{sub})}=\varepsilon  \left( \begin{array}{ccccc}
	 \!\! 0 & 0 & 0 & 0 & 0 \!\!\\[0.1cm]
    \!\! 0 & 4 f_2   & 2 \sqrt{2} f_1   & -2(2 f_2  - \sqrt{-5} f_3 )  & 0 \!\! \\[0.1cm]
    \!\!0 & -2 \sqrt{2} f_1   & 0 & 2 \sqrt{2} f_1   & 0\!\! \\[0.1cm]
 \!\!0 & -2 \left(2 f_2+ \sqrt{-5} f_3\right)   & -2 \sqrt{2} f_1   & 4 f_2   & 0\!\! \\[0.1cm]
 \!\!0 & 0 & f_2   & 0 & -4 f_2\!\! \\
 \end{array} \right) \, \vec{\mathcal{T}}^{(\text{sub})}\,.
\end{equation}
The final transformation $\vec{\mathcal{T}}^{(\text{sub})} = U^{(\text{sub})} \, \vec{\mathcal{I}}^{(\text{sub})}$, for $\vec{\mathcal{I}}^{(\text{sub})} = (\mathcal{I}_1,\dots,\mathcal{I}_5)^T$, is given by
\begin{align}
	&U^{(\text{sub})}= c_2\left( \begin{array}{ccccc}
    \!\! \frac{c_1}{c_2} & 0 & 0 & 0 & 0 \!\!\\[0.2cm]
    \!\! 0 & \frac{ (x^2-1)^3 (4 \varepsilon -1) (5 \varepsilon -1)}{4 \sqrt{-5} x^2 (x^2+1) \varepsilon } & -\frac{ (x^2-1) (2 \varepsilon +1)}{\sqrt{-5} (x^2+1) \varepsilon } &
   \frac{ (x^2-1) (x^4+6 x^2+1)}{4 \sqrt{-5} (x^2+1) x^2}+x-\frac{1}{x} & 0 \!\!\\[0.2cm]
    \!\! 0 & -\frac{ (x^2-1)^2 (4 \varepsilon -1) (5 \varepsilon -1)}{2 \sqrt{2}\sqrt{-5} x^2 \varepsilon } & \frac{ (2 \varepsilon +1) (10 \varepsilon -1)}{2 \sqrt{2}\sqrt{-5} \varepsilon ^2} & -\frac{ (x^2-1)^2}{2
   \sqrt{2}\sqrt{-5} x^2} & 0 \!\!\\[0.2cm]
    \!\! 0 & -\frac{ (x^2-1)^3 (4 \varepsilon -1) (5 \varepsilon -1)}{4 \sqrt{-5} x^2 (x^2+1) \varepsilon } & \frac{ (x^2-1) (2 \varepsilon +1)}{\sqrt{-5} (x^2+1) \varepsilon } &
   -\frac{ (x^2-1) (x^4+6 x^2+1)}{4 \sqrt{-5} (x^2+1) x^2}+x-\frac{1}{x} & 0 \!\!\\[0.2cm]
    \!\! 0 & -\frac{(x^2-1)^2 (4 \varepsilon -1) (5 \varepsilon -1)}{16 \sqrt{2}\sqrt{-5} x^2 \varepsilon } & \frac{ (2 \varepsilon +1) (26 \varepsilon -3)}{32\sqrt{2} \sqrt{-5} \varepsilon ^2} & -\frac{
   (x^2-1)^2}{16 \sqrt{2}\sqrt{-5} x^2} & -\frac{ (1-6 \varepsilon ) (14 \varepsilon -1)}{16 \sqrt{2}\sqrt{-5} \varepsilon } \!\!\\
 \end{array} \right)\,, \nonumber \\
 &  
\label{eq: def_U_sub_transformation_final}
\end{align}
where we define the normalizations
\begin{equation}
	c_1 \equiv \frac{(1-6 \varepsilon )^2}{2 \varepsilon +1}\,,\qquad c_2 \equiv \frac{\varepsilon ^5}{(4 \varepsilon +1)^2 (6 \varepsilon +1)^2 (14 \varepsilon -1)}\,.
\end{equation}

\subsection{An \texorpdfstring{\bm{$\varepsilon$}}{eps}-factorized form for the top sector}
\label{sec:ch5_eps_fact_top_sector}

Having brought the subsectors into $\varepsilon$-factorized form, in this subsection, we focus on sector $\sectord$, which corresponds to the CY top sector. Since we have already exemplified how to use the method of sec.~\ref{sec:ch5_method_eps_factorized} in the previous subsection, here we will only focus on the new aspects.

In this case, we have 5 master integrals $\vec{\mathcal{I}}^{\sectord} = (\mathcal{I}_6,\dots,\mathcal{I}_{10})^T$, which satisfy a pre-canonical differential equation schematically given in eq.~\eqref{eq: shape_diff_eq_top_sector_CY}. As discussed in sec.~\ref{sec:ch5_CY_integral_family}, for sector $\sectord$, we expect the scalar integral to be a good seed integral candidate. Hence, we define
\begin{equation}
	\integralseed^{\sectord{}} \equiv \integralprec_6=\integralprec_{111 111 111 11 000 000 000 00}\,.
\end{equation}
As we found in eqs.~\eqref{eq: PF_PM_CY_2_eps}--\eqref{eq: PF_CY_apparent_singularity}, in $D=4-2\varepsilon$ dimensions, this integral is annihilated by a fifth-order Picard--Fuchs operator with an $\varepsilon$-dependent apparent singularity that is quadratic in $\varepsilon$. Thus, based on the correspondence of fig.~\ref{fig: epsilon_degree_apparent_sing}, an appropriate ansatz for its $\varepsilon$-factorized differential equation matrix is given by eq.~\eqref{eq: ansatz_section3}. Explicitly, we have a self-dual differential equation
\begin{equation}
\label{eq:sector_d_ansatz}
	J\frac{d}{d x} \, \vec{\mathcal{J}}^{\sectord{}}=\varepsilon \left( \begin{array}{ccccc}
 \!\!F_{11} & F_{12} & 0 & 0 & 0\!\! \\
 \!\!F_{21} & F_{22} & F_{23} & F_{24} & 0\!\! \\
 \!\!F_{31} & F_{32} & F_{33} & F_{23} & 0\!\! \\
 \!\!F_{41} & F_{42} & F_{32} & F_{22} & F_{12}\!\! \\
 \!\!F_{51} & F_{41} & F_{31} & F_{21} & F_{11}\!\! \\
 \end{array} \right) \, \vec{\mathcal{J}}^{\sectord{}}\,,
\end{equation}
where we define $\vec{\mathcal{J}}^{\sectord{}} = (\mathcal{J}^{\sectord{}}_1,\dots,\mathcal{J}^{\sectord{}}_5)^T$. Note that, for this top sector, we define new unknown functions $\omega$, $c$, $J$ and $F_{ij}$.

\begin{table}[t]
\begin{center}
\caption{For given values of $k$ and $j$ in eq.~\eqref{eq:sectord_matching_condition}, each entry in the table specifies the unknown function to be determined. A zero indicates that the corresponding condition is automatically fulfilled.}
\label{tab: strategy topsector}
$\begin{NiceArray}{|c|cccccccc|}[columns-width = 2em]
\Xhline{0.2ex}
\diagbox{\,\,k}{j\,\,} & 0 & 1 & 2 & 3 & 4 & 5 & 6 & 7 \\ \hline
5 & F_{21} & F_{11} & c & &  &  &  & \\
4 & 0 & 0 & J'' & F_{22}'&  &  &  & \\
3 & F_{33} & F_{24}' & F_{23}''' & F_{33}' & F_{31} &  &  & \\
2 & 0 & 0 & 0 & F_{32}'' & 0 & F_{41} &  &  \\
1 & 0 & \omega'''' & 0 & 0 & F_{42}''  & 0 & F_{51} &  \\
0 & 0 & 0 & 0 & 0 & 0 & 0 & 0 & F_{24}\\
\Xhline{0.2ex}
\end{NiceArray}$
\end{center}
\end{table}
Then, we can proceed as described in sec.~\ref{sec:ch5_method_eps_factorized}, and calculate the Picard--Fuchs operator for $\mathcal{J}_1^{\sectord}$ in two different ways, where we take the relation
\begin{equation}
\label{eq:sectorsectorb_matching_integral}
	\integralFinal_1^{\sectord{}}=\frac{\varepsilon}{\omega(x)} \,\integralseed^{\sectord{}}\,.
\end{equation}
From the matching condition in eq.~\eqref{eq: condition_2_eps_fact_solution}, we obtain $33$ constraints,
\begin{equation}
	\label{eq:sectord_matching_condition}
 \tilde{p}_{k,j}(x)= c(x) \, p_{k,j}(x)\,,\qquad 0\leq k\leq 5\,,\ \  0\leq j\leq 7-k\,.
\end{equation}
As in sec.~\ref{sec:ch5_eps_fact_subsectors}, we can solve for the unknowns by starting with $k=5$ and decreasing $k$ in steps of one, proceeding from largest to smallest $j$. Specifically, we follow the order provided in tab.~\ref{tab: strategy topsector}. Noticeably, the table is sparser than in the previous example with sector $\sectorb$, see tab.~\ref{tab: strategy subsector}. This is due to the assumption of self-duality in the differential equation of eq.~\eqref{eq:sector_d_ansatz}, which fixes some $F_{ij}$'s to be equal. Therefore, some of the conditions from eq.~\eqref{eq:sectord_matching_condition} are automatically satisfied, which corresponds to the zero entries in tab.~\ref{tab: strategy topsector}. 

Comparing with tab.~\ref{tab: strategy subsector}, we also observe that the entries in tab.~\ref{tab: strategy topsector} become less systematic. This is because, depending on the unknown functions chosen, the size of the intermediate expressions can grow enormously, while the final constraint is the same. Thus, in some cases, we gave up finding a systematic order in favor of having simpler expressions. For instance, while the condition for $k=1,j=0$ provides the same constraint for $\omega''''$, it is much simpler to obtain via $k=1,j=1$. Similarly, we choose to solve for $J''$ instead of $F_{12}$, as the constraints for $J$ are easier to solve. In fact, $F_{12}$ is left unconstrained and can be set to one, which corresponds to using a different normalization for the differential equation. In the end, some of the simpler constraints are given by
{\allowdisplaybreaks
\begin{subequations}
\begin{align}
    \label{eq: condition_omega_4_top-sector}
	\omega''''&=-\frac{2-16 x^2-10 x^4}{x(1-x^4)}\, \omega'''-\frac{1-28 x^2+46 x^4+68 x^6+25 x^8}{x^2 (1-x^4)^2}\, \omega '' \nonumber \\
   &\hphantom{{}={}}+\frac{1+11 x^2-54 x^4+22 x^6+37 x^8+15 x^{10}}{x^3 (1-x^2)^3
   (1+x^2)^2}\,\omega ' -\frac{1+3 x^2+20 x^4+3 x^6+x^8}{x^4 (1-x^4)^2}\,\omega\,, \\[0.1cm]
   F_{33}&= \frac{F_{23}}{F_{24}}+\frac{40 x J}{x^4-1}\,, \\[0.1cm]
   F_{11}&= \frac{2 F_{24} \Big(x(1-x^4) F_{22}+2 J (x^4-8 x^2+1)\Big)+x(1-x^4)
   F_{23}^2}{2 x (x^4-1) F_{24}}\,, \\[0.1cm]
   F_{24}'&= F_{24} \left(\frac{3 J'}{J}+\frac{5 x^4+8 x^2-1}{x(1-x^4)}-\frac{2 \omega '}{\omega }\right)\,.
   \label{eq: condition_F24_top-sector}
\end{align}
\end{subequations}
}

As in the case of the subsectors, we can solve these highly-coupled constraints with an appropriate order. In particular, we employ
\begin{equation}
	\omega\to J\to F_{23}\to \{F_{24},c,F_{33}\} \to F_{32}\to \{F_{22},F_{42}\}\to \{F_{11},F_{21},F_{31},F_{41},F_{51}\}\,,
\end{equation}
where the braces indicate that any order can be chosen. Importantly, the solution for some of these functions is closely connected to the underlying Calabi--Yau threefold that we studied in sec.~\ref{sec:ch5_PF_gravity_CY_operator}. First of all, comparing eqs.~\eqref{eq: PF_PM_CY_2_L4} and~\eqref{eq: condition_omega_4_top-sector}, we see that the differential equation for $\omega(x)$ is given by the Picard--Fuchs operator $\mathcal{L}_4$ defining the CY geometry in $D=4$. Thus, $\omega(x)$ corresponds to the leading singularity and, more precisely, we can choose it to be the holomorphic Frobenius solution $\varpi_1=\Sigma_1$ from eq.~\eqref{eq: Frobenius series}. Secondly, the constraint from eq.~\eqref{eq: condition_F24_top-sector} imposes that $F_{24}$ is equal to the gauge-fixed Yukawa coupling $Y_2$ of the CY geometry, given in eq.~\eqref{eq: Y2_CY}. Lastly, the differential constraint for $J$ relates it to the mirror map $q=\exp(2\pi i \tau)$ from eqs.~\eqref{eq: def_tau_CY_mirror_map}--\eqref{eq: mirror_map_q_series_expansion}. In summary, we have that
\begin{equation}
	\omega(x)=\varpi_1=\Sigma_1\,,\qquad F_{24}=Y_2\,,\qquad J=\frac{d x}{d \log(q)} = \frac{1}{2 \pi i}\frac{dx}{d \tau}\,.
\end{equation}
In fact, similar relations were found in refs.~\cite{Pogel:2022ken,Pogel:2022vat,Duhr:2022dxb} for the differential equations of the banana and ice-cone integrals.

For the remaining functions, we find a solution as a series expansion. In part, this is due to the lack of a well-defined class of differential forms in which to express the $F_{ij}$'s in the case of a CY geometry. Overall, we find
{\allowdisplaybreaks
\begin{subequations}
\begin{align}
	c(x)&=-\frac{1}{10}\,\sqrt{\frac{-5}{2}}\left(\frac{1}{x}+\frac{13 x}{2^3}+\frac{3441 x^3}{2^{11}}+\frac{25343 x^5}{2^{14}}+\frac{190544849 x^7}{2^{27}}+\mathcal{O}(x^9)\right),\\[0.1cm]
	\omega(x) &=x+\frac{9 x^3}{2^4}+\frac{1681 x^5}{2^{12}}+\frac{21609 x^7}{2^{16}}+\mathcal{O}(x^9)\,,\\[0.1cm]
	J&=x-\frac{3 x^3}{2^2}-\frac{77 x^5}{2^{10}}-\frac{369 x^7}{2^{13}}+\mathcal{O}(x^9)\,,\\[0.1cm]
	F_{11}&=-1+\frac{15 x^2}{2^2}+\frac{1005 x^4}{2^{10}}+\frac{5505 x^6}{2^{13}}+\frac{33070845 x^8}{2^{26}}+\mathcal{O}(x^{9})\,,\\[0.1cm]
	F_{21}&=1+\frac{171 x^2}{2^2}-\frac{18809 x^4}{2^{10}}+\frac{37767 x^6}{2^{13}}+\frac{206196007 x^8}{2^{26}}+\mathcal{O}(x^9)\,,\\[0.1cm]
	F_{22}&=\frac{1}{4}\left(1+\frac{705 x^2}{2^3}-\frac{76795 x^4}{2^{11}}+\frac{4575 x^6}{2^{14}}-\frac{242517275 x^8}{2^{27}}+\mathcal{O}(x^9)\right),\\[0.1cm]
	F_{23}&= \sqrt{\frac{-5}{2}}\left(1+\frac{29 x^2}{2^3}-\frac{3495 x^4}{2^{11}}-\frac{1225 x^6}{2^{14}}-\frac{19494119 x^8}{2^{27}}+\mathcal{O}(x^9)\right),\\[0.1cm]
	F_{24}&= 1+\frac{5 x^2}{2^3}-\frac{775 x^4}{2^{11}}-\frac{445 x^6}{2^{14}}-\frac{5110375 x^8}{2^{27}}+\mathcal{O}(x^9)\,,\\[0.1cm]
	F_{31}&=-\sqrt{\frac{-5}{2}}\left(1+\frac{3 x^2}{2^2}-\frac{264601 x^4}{2^{10}}-\frac{988629 x^6}{2^{13}}-\frac{7257154393 x^8}{2^{26}}+\mathcal{O}(x^9)\right),\\[0.1cm]
	F_{32}&=-\frac{1}{4}\, \sqrt{\frac{-5}{2}}\left(1+\frac{3961 x^2}{2^3}-\frac{321563 x^4}{2^{11}}+\frac{234179 x^6}{2^{14}}+\frac{77183589 x^8}{2^{27}}+\mathcal{O}(x^9)\right),\\[0.1cm]
	F_{33}&=-\frac{5}{2}\left(1+\frac{181 x^2}{2^3}-\frac{12359 x^4}{2^{11}}+\frac{17707 x^6}{2^{14}}+\frac{54628505 x^8}{2^{27}}+\mathcal{O}(x^9)\right),\\[0.1cm]
	F_{41}&=\frac{5}{4}\left(1-\frac{1189 x^2}{4}-\frac{1225145 x^4}{2^{10}}-\frac{4233457 x^6}{2^{13}}-\frac{29885486041 x^8}{2^{26}}+\mathcal{O}(x^9)\right),\\[0.1cm]
		F_{42}&=-\frac{15}{2^4}+\frac{84429 x^2}{2^7}-\frac{8120415 x^4}{2^{15}}+\frac{2919659 x^6}{2^{18}}-\frac{11826290495 x^8}{2^{31}}+\mathcal{O}\left(x^9\right),\\[0.1cm]
	F_{51}&=\frac{881 x^2}{2}+\frac{1734625 x^4}{2^9}+\frac{23147105 x^6}{2^{12}}+\frac{191659062625 x^8}{2^{25}}+\mathcal{O}(x^9)\,.
\end{align}
\end{subequations}
}
Up to the prefactors, these series are $N$-integral (recall sec.~\ref{sec:ch5_PF_gravity_CY_operator}) with $N=2^4$. 

Consequently, we have found an $\varepsilon$-factorized form for the $5 \times 5$ CY block of sector $\sectord$. The corresponding basis transformation $\vec{\mathcal{J}}^{\sectord} = U^{\sectord} \vec{\mathcal{I}}^{\sectord}$ can be found in the ancillary files of ref.~\cite{Frellesvig:2024rea}, but it takes the form
\begin{subequations}
\label{eq: form_canonical_basis_d}
\begin{align}
	\integralFinal_{1}^{\sectord}&=\frac{\varepsilon}{\omega(x)}\integralseed^{\sectord}\,,\\[0.1cm]
	\integralFinal_{2}^{\sectord}&=\frac{J}{\varepsilon}\frac{d \integralFinal_{1}^{\sectord}}{d x}-F_{11}\integralFinal_{1}^{\sectord}\,,\\[0.1cm]
    \integralFinal_3^{\sectord}&=c_1(x,\varepsilon)\,\mathcal{I}_{10}+c_2(x,\varepsilon)\,\integralFinal_{1}^{\sectord}+c_3(x,\varepsilon)\, \integralFinal_{2}^{\sectord}\,,\\[0.1cm]
	\integralFinal_{4}^{\sectord}&=\frac{1}{F_{24}}\left(\frac{J}{\varepsilon}\frac{d \integralFinal_{2}^{\sectord}}{d x}-F_{21}\integralFinal_{1}^{\sectord}-F_{22}\integralFinal_{2}^{\sectord}-F_{23}\integralFinal_{3}^{\sectord}\right),\\[0.1cm]
	\integralFinal_{5}^{\sectord}&=\frac{J}{\varepsilon}\frac{d \integralFinal_{4}^{\sectord}}{d x}-F_{41}\integralFinal_{1}^{\sectord}-F_{42}\integralFinal_{2}^{\sectord}-F_{43}\integralFinal_{3}^{\sectord}-F_{44}\integralFinal_{4}^{\sectord}\,.
\end{align}
\end{subequations}
Here, the coefficients $c_i(x,\varepsilon)$ are given by
\begin{subequations}
\begin{align}
    c_1(x,\varepsilon)=& \, \sqrt{\frac{-5}{2}} \, \frac{1-6\varepsilon}{\varrho}\,, \\[0.1cm]
    c_2(x,\varepsilon)=& \, \frac{1}{\varepsilon}\, \sqrt{\frac{-5}{2}} \left( 1-4 \varepsilon - \frac{x^2 (116 \varepsilon + 3)}{2^4} - \frac{x^4 (14852 \varepsilon + 231)}{2^{12}} - \frac{x^6 (158708 \varepsilon + 1819)}{2^{16}} \right. \nonumber \\[0.1cm]
    & \, \left. - \frac{x^8 (488403524 \varepsilon
   + 4487967)}{2^{28}}+\mathcal{O}(x^9) \right), \\[0.1cm]
    c_3(x,\varepsilon)=& \, \sqrt{\frac{-5}{2}} \left( 3-\frac{3 x^2}{2^3}-\frac{283 x^4}{2^{11}}-\frac{2567 x^6}{2^{15}}-\frac{7084667 x^8}{2^{27}}+\mathcal{O}\left(x^9\right) \right),
\end{align}
\end{subequations}
where we also define $\varrho=x/(x^2-1)$, which corresponds to the rational solution to the operator $\widetilde{\mathcal{L}}_1$ of $\mathcal{I}_{10}$, recall the discussion in the end of sec.~\ref{sec:ch5_CY_integral_family}. In general, the basis from eq.~\eqref{eq: form_canonical_basis_d} is very similar to the derivative basis that would be commonly used for a Calabi--Yau integral~\cite{Pogel:2022ken,Pogel:2022vat}; see also eq.~\eqref{eq: intermediate_K_basis_def}. However, the presence of an $\varepsilon$-dependent apparent singularity introduces a key difference. Specifically, we find an integral insertion -- $\mathcal{J}_3^{\sectord}$ in our case -- that depends on an additional master integral $\mathcal{I}_{10}$ besides the seed integral. In the literature, integral insertions have been observed in elliptic Feynman integrals where a master integral (quasi-)decouples from the rest of the system~\cite{Remiddi:2013joa,Tancredi:2015pta,Duhr:2022dxb,Gorges:2023zgv}. This is completely analogous to the scenario we encountered in sec.~\ref{sec:ch5_CY_integral_family}, where the integral $\mathcal{I}_{10}$ (quasi-)decouples from the rest in $D=4$.

Taking into account the transformations carried out in the previous subsection, we are therefore led to a $10 \times 10$ differential equation matrix where the diagonal blocks are $\varepsilon$-factorized. To leading order in $\varepsilon$, it takes the form
\begin{equation}
	\frac{d}{d x} \, \vec{\mathcal{J}} \sim
	\left(\begin{NiceArray}{ccccc|ccccc}[columns-width = 1.5em]
	\!\!0 & &&&&&&&&\!\!\\
	 \!\!&\varepsilon & \varepsilon &\varepsilon & &&&&\!\!\\	
	\!\! &\varepsilon & 0 &\varepsilon & &&&&\!\!\\	
	\!\! &\varepsilon & \varepsilon &\varepsilon & &&&&\!\!\\
	\!\! & & \varepsilon &  & \varepsilon&&&&\!\!\\\hline
	\!\!\varepsilon &\varepsilon^{-2} &\varepsilon^{-3} & \varepsilon^{-2} &\varepsilon^{-3} & \varepsilon & \varepsilon & & & \!\!\\
	\!\!\varepsilon^0 &\varepsilon^{-3} &\varepsilon^{-4} & \varepsilon^{-3} &\varepsilon^{-4} & \varepsilon & \varepsilon & \varepsilon & \varepsilon & \!\!\\
	\!\!\varepsilon^0 &\varepsilon^{-4} &\varepsilon^{-4} & \varepsilon^{-4} &\varepsilon^{-4} & \varepsilon & \varepsilon & \varepsilon &  \varepsilon& \!\!\\
	\!\!\varepsilon^{-1} &\varepsilon^{-4} &\varepsilon^{-5} & \varepsilon^{-4} &\varepsilon^{-5} & \varepsilon & \varepsilon & \varepsilon & \varepsilon & \varepsilon\!\!\\
	\!\!\varepsilon^{-2} &\varepsilon^{-5} &\varepsilon^{-6} & \varepsilon^{-5} &\varepsilon^{-6} & \varepsilon & \varepsilon & \varepsilon & \varepsilon & \varepsilon \!\!\\
	\end{NiceArray}\right) \ \vec{\mathcal{J}}\,,
\end{equation}
where we define
\begin{equation}
    \vec{\mathcal{J}} \equiv \left( \begin{array}{c}
 \!\! \vec{\mathcal{T}}^{(\text{sub})} \!\!\!\\
 \!\! \vec{\mathcal{J}}^{\sectord} \!\!\!\\
 \end{array} \right) = \left(\begin{array}{cc}
	\!\!U^{(\text{sub})}& 0 \!\!\\
	\!\!0& U^{\sectord} \!\!
	\end{array}\right) \ \vec{\mathcal{I}}\,,
\end{equation}
with $U^{(\text{sub})}$ given in eq.~\eqref{eq: def_U_sub_transformation_final}, and $U^{\sectord}$ in the ancillary files of ref.~\cite{Frellesvig:2024rea}. 

The last step is thus to find an $\varepsilon$-factorized form including the $5 \times 5$ off-diagonal block mixing the subsectors with the top sector. To do so, we can simply shift the top sector integrals $\vec{\mathcal{J}}^{\sectord}$ by adding linear combinations of the master integrals $\vec{\mathcal{T}}^{(\text{sub})}$ from the subsectors. Afterwards, the free coefficients can be fixed by requiring that all off-diagonal terms become $\varepsilon$-factorized. As a result, we finally obtain a transformation
\begin{equation}
    \vec{\mathcal{T}} = U \, \vec{\mathcal{I}}\,,
\end{equation}
such that the differential equation matrix is fully $\varepsilon$-factorized, satisfies Galois symmetry, and the diagonal blocks are self-dual. The explicit expressions for the final change of basis and the differential equation matrix can be found in the ancillary files of ref.~\cite{Frellesvig:2024rea}.

\section{Discussion and open questions}
\label{sec:ch5_conclusions}

In this chapter, we have developed a method to obtain an $\varepsilon$-factorized differential equation for Feynman integrals with $\varepsilon$-dependent apparent singularities in their Picard--Fuchs operator. This extends the range of applicability of $\varepsilon$-factorized differential equations to a broader class of Feynman integrals. In particular, our method becomes especially useful in cases where an optimal basis of master integrals may be difficult to find, as it circumvents the need to construct such a basis explicitly.

A key aspect of our method lies in the interplay between the structure of the differential equation matrix and the nature of the apparent singularity in the associated Picard--Fuchs operator, which guides the appropriate ansatz that needs to be constructed in each case. Specifically, for integrals with $\varepsilon$-dependent apparent singularities, we have shown that the $\varepsilon$-factorized differential equation matrix must include non-zero entries up to the next-to-next-to-diagonal or beyond, thereby enhancing the ansatz of refs.~\cite{Pogel:2022ken,Pogel:2022vat}.

As a proof of principle, we have applied our method to obtain an $\varepsilon$-factorized differential equation for the Calabi--Yau threefold integral from sec.~\ref{sec:ch4_four_loop_CYs_2}, which contributes to black-hole scattering in the PM expansion of classical gravity at four loops and second self-force order. In particular, we have seen how geometric data from the underlying Calabi--Yau -- such as its periods, the $Y$-invariants and the mirror map -- enter through the differential equation. Specifically, this geometric information can be leveraged to facilitate finding the canonical form. The natural next step would be to use this $\varepsilon$-factorized form to evaluate the integral and determine its contribution to gravitational-wave observables, as already done for the other CY integral appearing at first self-force order~\cite{Driesse:2024feo}. However, since this requires computing the boundary values for the master integrals as well as deriving the relevant integrand from classical gravity, we leave this investigation for future work.

For this example in classical gravity, we have obtained a series expansion for the entries of the $\varepsilon$-factorized differential equation matrix. A future research direction would be to explore whether we can find analytic expressions for these entries, which would also determine the corresponding symbol alphabet for this CY integral, as already achieved for the equal-mass banana integrals in refs.~\cite{Pogel:2022ken,Pogel:2022vat}. Likewise, it would be interesting to apply the method developed in this chapter to solve other Feynman integrals using differential equations, especially in cases involving higher-dimensional Calabi--Yau manifolds.

\addtocontents{toc}{%
 \protect\vspace{1em}%
 \protect\noindent
\textcolor{DarkRed}{\textbf{IV
\hspace{.4em} Conclusions}}
\protect\par
 \protect\vspace{0em}%
}

\chapter*{Conclusions and outlook}
\label{ch:conclusions}
\addcontentsline{toc}{chapter}{\protect\numberline{}Conclusions and outlook}
\markboth{Conclusions and outlook}{Conclusions and outlook} % Manually set the left and right marks

In this thesis, we have explored various topics in the fields of scattering amplitudes and Feynman integrals, with a particular focus on the geometries and special functions that they involve. Here, we summarize the main results of the thesis, and outline potential future research directions.

In the first half of the thesis, we have investigated elliptic generalizations of ladder integrals in $\mathcal{N}=4$ SYM theory. First, using a parametric representation and direct integration, we have identified two families of Feynman integrals that depend on the same elliptic curve to all loop orders. Additionally, through the polynomial reduction algorithm, we have shown that these integrals can be expressed in terms of the same class of elliptic multiple polylogarithms. Secondly, we have developed a method for bootstrapping the symbol of elliptic Feynman integrals. In particular, we have extended the Schubert analysis to elliptic cases, and demonstrated how to use it to predict their symbol letters. As a proof of principle, we have applied these methods to calculate for the first time the symbol of the twelve-point double-box integral in $D=4$. 

In the second half of the thesis, we have initiated a systematic classification of the Feynman integral geometries which are relevant for the scattering of black holes in the post-Minkowskian expansion of classical gravity up to four loops. Importantly, the corresponding special functions contribute to the gravitational waves emitted during the inspiral phase of the coalescence, influencing the classical two-body dynamics. Through an analysis of leading singularities, we have identified a total of four distinct non-trivial geometries at four loops -- two K3 surfaces and two Calabi--Yau three-dimensional varieties -- including the first appearance of a Calabi--Yau threefold in the context of gravitational-wave physics in the literature. Subsequently, we have developed a method to evaluate this Calabi--Yau integral by bringing its differential equation into $\varepsilon$-factorized form, a task which required a careful treatment of $\varepsilon$-dependent apparent singularities.

More broadly, we expect the techniques developed in this thesis to have applications in other areas of high-energy theory. One promising direction would be using the methods from the first half of the thesis to bootstrap Feynman integrals and scattering amplitudes relevant to gravitational-wave physics, particularly in the post-Minkowskian expansion. Similarly, these techniques could be employed to study the geometries underlying Feynman integrals in QCD amplitudes for particle physics phenomenology. Another exciting research direction would be to extend these methods to higher-dimensional Calabi--Yau geometries and higher-genus curves.

Despite these advances, a fundamental question remains: why do these intricate geometries emerge in scattering amplitudes? Are they merely artifacts of perturbation theory, introducing an additional complexity that hinders higher-loop calculations, or do they encode deeper physical information that extends beyond the perturbative regime? Notably, all known examples of non-trivial geometries discovered so far involve Calabi--Yau manifolds and higher-genus curves. Is this a universal feature of Feynman integrals, or will more general hypersurfaces appear at higher loops? These intriguing questions suggest a rich interplay between mathematics and theoretical physics, which we hope that future research will continue to unravel.

\addtocontents{toc}{%
 \protect\vspace{1em}%
 \protect\noindent
\textcolor{DarkRed}{\textbf{V
\hspace{.4em} Appendices}}
\protect\par
 \protect\vspace{0em}%
}

\appendix
\numberwithin{equation}{chapter}

\renewcommand{\thechapter}{A}
\chapter{Integrability for elliptic symbol letters}
\label{ap:elliptic_integrability}

In this appendix, we present the details for the computation of the integrability conditions of eq.~\eqref{eq: integrability_symbol_v2} for elliptic symbol letters, which are a crucial ingredient in the analysis of chapter~\ref{ch:chapter3}. In particular, the goal is to obtain expressions for the derivatives of the cross-ratios $\chi_{ab}$ with respect to the last entries of the symbol, $\partial \chi_{ab}/\partial w^+_c$ and $\partial \chi_{ab}/\partial \tau$, see sec.~\ref{sec:ch3_elliptic_symbol_bootstrap} for further details. With this purpose, we found it simpler to compute instead the elements of the $9\times9$ Jacobian $J=\frac{\partial (w_{c_1}, \dots, w_{c_8}, \tau)}{\partial (\chi_{a_1 b_1}, \dots, \chi_{a_9 b_9})}$, and later invert the matrix, $J^{-1}=\frac{1}{\det(J)} \text{adj}(J)^T$. Since the integrability conditions are bilinear in the derivatives and involve a difference, see eq.~\eqref{eq: integrability_symbol_v2}, at this stage we already observe that the determinant $\frac{1}{\det(J)}$ drops from the conditions, and thus we only need the adjoint matrix $\text{adj}(J)$.

The first step is to find an expression for the elements $\partial w^+_c/\partial \chi_{ab}$ and $\partial \tau/\partial \chi_{ab}$ of the Jacobian. For that, we can start from Abel's map, see eq.~\eqref{eq: Abel_map_w}. The normalized torus images are therefore given by
\begin{equation}
w_c^+ = \frac{1}{\omega_1}\, z_c^+= \frac{1}{\omega_1} \int_{-\infty}^c \frac{dx}{y(x)}\,,
\end{equation}
where $y(x)$ is the elliptic curve. Therefore, taking a derivative we have
\begin{equation}
\label{eq_derivatives_Abels_map}
\frac{\partial w_c^+}{\partial \chi_{ab}} = \frac{1}{\omega_1 y_c} \frac{\partial c}{\partial \chi_{ab}} - \frac{z_c^+}{\omega_1^2} \frac{\partial \omega_1}{\partial \chi_{ab}} + \frac{1}{\omega_1} \int_{-\infty}^c \frac{\partial}{\partial \chi_{ab}} \left( \frac{dx}{y}  \right).
\end{equation}
To calculate the second and third terms, we need to introduce the explicit expressions for the elliptic curve and the periods, which will differ depending on the degree of the polynomial defining the elliptic curve, $y(x)^2=P(x)$. Therefore, for completeness, we will consider the two cases separately. In sec.~\ref{sec: app_integrability_cubic_elliptic} we treat the case of an elliptic curve given by a cubic polynomial, which is relevant for the symbol of the 12-pt double box, whereas in sec.~\ref{sec: app_integrability_quartic_elliptic} we focus on a quartic elliptic curve, which is relevant for the symbol of the 10-pt double box.

\section{Elliptic curve given by a cubic polynomial}
\label{sec: app_integrability_cubic_elliptic}

Let us consider an elliptic curve defined by a monic\footnote{For the case of the 12-pt double box, where we obtain the elliptic curve as $y(x)=\sqrt{\Delta_6(x)}\,$, the corresponding polynomial is not monic. Thus, we will need to transform it via $x \to x/a_3$ and $y \to y/a_3$, which normalizes the results at the end by the coefficient $a_3$ of the cubic term, see eqs.~\eqref{eq: integrability_conditions_elliptic_letters_12pt_double-box_w} and~\eqref{eq: integrability_conditions_elliptic_letters_12pt_double-box_tau}.} cubic polynomial
\begin{equation}
y(x)^2= x^3 + a_2 x^2 + a_1 x + a_0 = \prod_{i=1}^3 (x-r_i),
\end{equation}
with $r_1<r_2<r_3$ being the roots of the elliptic curve. In this case, we can express the periods $\omega_i$ and the quasi-periods $\eta_i$ as~\cite{Broedel:2017kkb}
\begin{align}
\label{eq: periods_quasiperiods_cubic}
\omega_1 =& \, \frac{2}{c_3} K(\lambda)\,, \qquad \qquad \quad \enspace \eta_1= 2 c_3 \Big( E(\lambda) - \frac{2-\lambda}{3} K(\lambda) \Big)\,, \\
\omega_2 =& \, \frac{2 i}{c_3} K(1-\lambda)\,, \qquad \qquad \eta_2 = 2 c_3 \Big( -i E(1-\lambda) + i \frac{1+\lambda}{3} K(1-\lambda) \Big)\,,
\end{align}
where $K(\lambda)$ and $E(\lambda)$ denote the complete elliptic integrals of the first and of the second kind, respectively, and where we introduce the notation
\begin{equation}
\lambda \equiv \frac{r_2 - r_1}{r_3 - r_1}\,, \qquad c_3 \equiv\frac{1}{2} \sqrt{r_3 - r_1}\,.
\end{equation}
In these expressions, we take into account some difference in conventions from refs.~\cite{Kristensson:2021ani,Wilhelm:2022wow} compared to ref.~\cite{Broedel:2017kkb}. In addition, the periods and quasi-periods obey the Legendre relation $\eta_1 \omega_2 - \eta_2 \omega_1 = 2 \pi i$.

With the following differential equations for the complete elliptic integrals,
\begin{equation}
\label{eq: diff_eq_complete_elliptic_integrals}
    \frac{d K(\lambda)}{d \lambda} = \frac{E(\lambda)}{\lambda(1-\lambda^2)}- \frac{K(\lambda)}{\lambda}\,, \qquad \text{and} \qquad \frac{d E(\lambda)}{d \lambda} = \frac{E(\lambda)-K(\lambda)}{\lambda}\,,
\end{equation}
and using eq.~\eqref{eq: periods_quasiperiods_cubic} for the periods and quasi-periods, it can be easily calculated that
\begin{equation}
\label{eq_deriv_omega1_cubic}
\frac{\partial \omega_1}{\partial \chi_{ab}} = \sum_{k=1}^3 \bigg( -2 \eta_1 - \frac{\omega_1}{6} \sum_{j \neq k} (r_k - r_j) \bigg) \frac{1}{\prod_{j \neq k}(r_k - r_j)} \frac{\partial r_k}{\partial \chi_{ab}} \,,
\end{equation}
which gives an expression for the second term in eq.~\eqref{eq_derivatives_Abels_map}. The equation for $\omega_2$ follows under the replacements $\omega_1 \to \omega_2$ and $\eta_1 \to \eta_2$. At this stage, we can already calculate the element $\partial \tau/\partial \chi_{ab}$ of the Jacobian,
\begin{align}
    \frac{\partial \tau}{\partial \chi_{ab}} = &\, \frac{\partial}{\partial \chi_{ab}} \frac{\omega_2}{\omega_1} = \frac{1}{\omega_1} \frac{\partial \omega_2}{\partial \chi_{ab}} - \frac{\omega_2}{\omega_1^2} \frac{\partial \omega_1}{\partial \chi_{ab}} \nonumber \\[0.2cm]
    = &\, \sum_{k=1}^3 \bigg( \frac{2}{\omega_1^2} \underbrace{(-\eta_2 \omega_1 + \eta_1 \omega_2)}_{\displaystyle 2 \pi i} - \underbrace{\frac{\frac{\omega_2}{\omega_1}-\frac{\omega_2}{\omega_1}}{6}}_{\displaystyle 0} \sum_{j \neq k} (r_k - r_j) \bigg) \frac{1}{\prod_{j \neq k}(r_k - r_j)} \frac{\partial r_k}{\partial \chi_{ab}} \nonumber \\
    = &\, \frac{4 \pi i}{\omega_1^2} \sum_{k=1}^3 \frac{1}{\prod_{j \neq k} (r_k - r_j)} \frac{\partial r_k}{\partial \chi_{ab}}\,,
    \label{eq: integrability_conditions_elliptic_letters_12pt_double-box_tau_proof}
\end{align}
which yields the result from eq.~\eqref{eq: integrability_conditions_elliptic_letters_12pt_double-box_tau}.

For the last term in eq.~\eqref{eq_derivatives_Abels_map}, we can first introduce the roots of the elliptic curve,
\begin{equation}
\label{eq_deriv_integrand_cubic}
\int_{-\infty}^c \frac{\partial}{\partial \chi_{ab}} \left( \frac{dx}{y}  \right) = \frac{1}{2} \sum_{k=1}^3 \left( \frac{\partial r_k}{\partial \chi_{ab}} \right) \int_{-\infty}^c \frac{dx}{y \, (x-r_k)}\,.
\end{equation}
However, the resulting integral is not suitable for an integral basis, as it explicitly depends on the roots of the particular elliptic curve at hand. To this end, we can use the integration-by-parts relation given in eq.~(6.6) of ref.~\cite{Broedel:2017kkb}, which for $k=1$, $\chi=1$ and $a_1=r_k$ becomes
\begin{equation}
\label{eq: intermediate_step_integrability_cubic}
\int_{-\infty}^c \frac{dx}{y \, (x-r_k)} = \frac{1}{\prod_{j\neq k}(r_k-r_j)} \left( - 2 \, \frac{y}{x-r_k} \bigg|_{-\infty}^c + \int_{-\infty}^c \frac{x \, dx}{y} - r_k \, z_c^+ \right).
\end{equation}
We can observe that the first two terms are divergent in the $x \to -\infty$ limit, although the divergence exactly cancels between them. Thus, we need to rewrite these terms in a basis of finite integrals. To do so, we can go to the torus, for which we need to transform the elliptic curve to the Weierstrass normal form $Y^2=4 X^3 - g_2 X - g_3$, recall eq.~\eqref{eq: Weierstrass_normal_form}. In our case, this can be achieved via the map $x = 4X - a_2/3,$ $y=4Y$. Moreover, on the torus we can use the identification $(X,Y) \to (\wp(z),\wp'(z))$ with the Weierstrass $\wp$ function, see sec.~\ref{sec:ch2_eMPLs_torus} for further details.

For the first term in eq.~\eqref{eq: intermediate_step_integrability_cubic}, we have
\begin{align}
- 2 \frac{y}{x-r_k} \bigg|_{-\infty}^c = & \, - 2 \frac{y_c}{c-r_k} + 2 \frac{y}{x-r_k} \bigg|_{x=-\infty} \nonumber \\[0.1cm]
= & \, - 2 \frac{y_c}{c-r_k} + 8 \frac{Y}{4X - \frac{a_2}{3} -r_k} \bigg|_{X=-\infty} \nonumber \\[0.1cm]
= & \, - 2 \frac{y_c}{c-r_k} + 8 \frac{\wp'(z)}{4\wp(z) - \frac{a_2}{3} -r_k} \bigg|_{z=0} \nonumber \\[0.1cm]
= & \, - 2 \frac{y_c}{c-r_k} + \bigg[ - \frac{4}{z} + \mathcal{O}(z) \bigg]_{z=0} \nonumber \\[0.1cm]
= & \, - 2 \frac{y_c}{c-r_k} - \frac{4}{z} \bigg|_{z=0}\,,
\end{align}
where we first transform to Weierstrass normal form, then to the torus with the Weierstrass $\wp$ function, and finally expand it around $z=0$.

For the second term in eq.~\eqref{eq: intermediate_step_integrability_cubic} we proceed analogously, regulating the $z=0$ divergence of the Weierstrass $\wp$ function,
\begin{align}
\int_{-\infty}^c \frac{x \, dx}{y} = & \, 4 \int_{-\infty}^{\frac{1}{4} (c+ a_2/3)} \frac{X \, dX}{Y} - \frac{a_2}{3} \int_{-\infty}^{\frac{1}{4} (c+ a_2/3)} \frac{dX}{Y} \nonumber \\[0.1cm]
= & \, 4 \int_0^{z_c^+} \wp(z) dz - \frac{a_2}{3} \int_{-\infty}^c \frac{dx}{y} \nonumber \\[0.1cm]
= & \, 4 \int_0^{z_c^+} \left( \wp(z) - \frac{1}{z^2} \right) dz  - \frac{4}{z}\bigg|^{z_c^+}_{0} - \frac{a_2}{3} \int_{-\infty}^c \frac{dx}{y} \nonumber \\[0.1cm]
= & \, -4 \zeta(z^+_c) + \frac{4}{z}\bigg|_{z=0} - \frac{a_2}{3} \int_{-\infty}^c \frac{dx}{y}\,.
\end{align}
In the last step, we have introduced the definition of the Weierstrass zeta function $\zeta(z)$, see e.g.~ref.~\cite{Broedel:2017kkb},
\begin{equation}
    \zeta(z) = \frac{1}{z} - \int_0^{z} \left( \wp(z') - \frac{1}{z'^2} \right) dz'\,.
\end{equation}

Replacing these expressions in eq.~\eqref{eq: intermediate_step_integrability_cubic}, and using that $a_2=-r_1-r_2-r_3$, we obtain
\begin{equation}
\int_{-\infty}^c \frac{dx}{y \, (x-r_k)} = \frac{1}{\prod_{j\neq k}(r_k-r_j)} \left( - 2 \frac{y_c}{c-r_k} -4 \zeta(z_c^+) - \frac{1}{3} \sum_{j \neq k} (r_k - r_j) \, z_c^+ \right) \,.
\end{equation}
Noticeably, we observe that all divergences drop out.

Finally, substituting this result in eq.~\eqref{eq_derivatives_Abels_map}, we obtain that
\begin{equation}
\frac{\partial w_c^+}{\partial \chi_{ab}} =  \frac{1}{\omega_1 y_c} \frac{\partial c}{\partial \chi_{ab}} + \frac{1}{\omega_1} \sum_{k=1}^3 \bigg( \frac{y_c}{r_k-c} - 2 \underbrace{\Big( \zeta(z_c^+) - \frac{\eta_1}{\omega_1} \, z_c^+ \Big)}_{\displaystyle= \frac{1}{\omega_1} g^{(1)}(w_c^+)} \bigg) \frac{1}{\prod_{j \neq k} (r_k - r_j)} \frac{\partial r_k}{\partial \chi_{ab}}\,,
 \label{eq: integrability_conditions_elliptic_letters_12pt_double-box_w_proof}
\end{equation}
which yields the result used in eq.~\eqref{eq: integrability_conditions_elliptic_letters_12pt_double-box_w}. In the last step, we introduced an alternative definition for the eMPL kernel $g^{(1)}(w_c^+)$~\cite{Broedel:2017kkb}. Notably, since the terms with $g^{(1)}(w_c^+)$ contain the same prefactor as eq.~\eqref{eq: integrability_conditions_elliptic_letters_12pt_double-box_tau_proof}, they drop in the adjoint $\text{adj}(J)$. As a cross-check, we have verified the validity of eqs.~\eqref{eq: integrability_conditions_elliptic_letters_12pt_double-box_tau_proof} and~\eqref{eq: integrability_conditions_elliptic_letters_12pt_double-box_w_proof} by comparing with numerical differentiation, finding agreement up to 300 digits.

\section{Elliptic curve given by a quartic polynomial}
\label{sec: app_integrability_quartic_elliptic}

Let us now turn to the case of an elliptic curve given by a monic quartic polynomial, and consider
\begin{equation}
y(x)^2= x^4 + a_3x^3 + a_2 x^2 + a_1 x + a_0 = \prod_{i=1}^4 (x-r_i),
\end{equation}
with $r_i$ being the roots of the elliptic curve, given by two complex conjugate pairs. Analogously to the cubic case, we can express the periods $\omega_i$ and the quasi-periods $\eta_i$ as~\cite{Broedel:2017kkb}
\begin{align}
\label{eq: periods_quasiperiods_quartic}
\omega_1 =& \, \frac{2}{c_4} K(\lambda)\,, \qquad \qquad \quad \enspace \eta_1= 2 c_4 \Big( E(\lambda) - \frac{2-\lambda}{3} K(\lambda) \Big)\,, \\
\omega_2 =& \, \frac{2 i}{c_4} K(1-\lambda)\,, \qquad \qquad \eta_2 = 2 c_4 \Big( -i E(1-\lambda) + i \frac{1+\lambda}{3} K(1-\lambda) \Big)\,,
\end{align}
where we now use the notation
\begin{equation}
\lambda \equiv \frac{(r_1 - r_4) (r_2 - r_3)}{(r_1 - r_3) (r_2 - r_4)}\,, \qquad c_4 \equiv \frac{1}{2} \sqrt{(r_1 - r_3) (r_2 - r_4)}\,.
\end{equation}
The periods and quasi-periods still obey the same Legendre relation $\eta_1 \omega_2 - \eta_2 \omega_1 = 2 \pi i$.

With the differential equations for the complete elliptic integrals from eq.~\eqref{eq: diff_eq_complete_elliptic_integrals}, we find that
\begin{equation}
\label{eq_deriv_omega1_quartic}
\frac{\partial \omega_1}{\partial \chi_{ab}} = \sum_{k=1}^4 \bigg( -2 \eta_1 - \frac{\omega_1}{6} \Big( S_2(\vec{r}) - 3 r_k S_1(\vec{r}) + 6 r_k^2 \Big) \bigg) \frac{1}{\prod_{j \neq k}(r_k - r_j)} \frac{\partial r_k}{\partial \chi_{ab}} \,,
\end{equation}
and similarly for $\omega_2$, where $S_i(\vec{r})$ denotes the elementary symmetric polynomial of degree $i$ in the roots. We can then calculate the element $\partial \tau/\partial \chi_{ab}$ of the Jacobian,
\begin{align}
    \frac{\partial \tau}{\partial \chi_{ab}} = &\, \frac{\partial}{\partial \chi_{ab}} \frac{\omega_2}{\omega_1} = \frac{1}{\omega_1} \frac{\partial \omega_2}{\partial \chi_{ab}} - \frac{\omega_2}{\omega_1^2} \frac{\partial \omega_1}{\partial \chi_{ab}} \nonumber \\[0.2cm]
    = &\, \sum_{k=1}^4 \bigg( \frac{2}{\omega_1^2} \underbrace{(-\eta_2 \omega_1 + \eta_1 \omega_2)}_{\displaystyle 2 \pi i} - \underbrace{\frac{\frac{\omega_2}{\omega_1}-\frac{\omega_2}{\omega_1}}{6}}_{\displaystyle 0} \Big( S_2(\vec{r}) - 3 r_k S_1(\vec{r}) + 6 r_k^2 \Big) \bigg) \frac{1}{\prod_{j \neq k}(r_k - r_j)} \frac{\partial r_k}{\partial \chi_{ab}} \nonumber \\
    = &\, \frac{4 \pi i}{\omega_1^2} \sum_{k=1}^4 \frac{1}{\prod_{j \neq k} (r_k - r_j)} \frac{\partial r_k}{\partial \chi_{ab}}\,,
    \label{eq: integrability_conditions_elliptic_letters_10pt_double-box_tau_proof}
\end{align}
which yields an almost identical result to eq.~\eqref{eq: integrability_conditions_elliptic_letters_12pt_double-box_tau_proof}.

With this, we again have the last term in eq. \eqref{eq_derivatives_Abels_map} left to calculate. Introducing the roots of the elliptic curve,
\begin{equation}
\label{eq_deriv_integrand_quartic}
\int_{-\infty}^c \frac{\partial}{\partial \chi_{ab}} \left( \frac{dx}{y}  \right) = \frac{1}{2} \sum_{k=1}^4 \left( \frac{\partial r_k}{\partial \chi_{ab}} \right) \int_{-\infty}^c \frac{dx}{y \, (x-r_k)}\,.
\end{equation}
Now, we can use a generalization to the quartic case of the integration-by-parts relations of ref.~\cite{Broedel:2017kkb}, such that
\begin{equation}
\label{eq: intermediate_step_integrability_quartic}
\int_{-\infty}^c \frac{dx}{y \, (x-r_k)} = \frac{1}{\prod_{j\neq k}(r_k-r_j)} \bigg[ - 2 \frac{y}{x-r_k} \bigg|_{-\infty}^c + \sum_{j \neq k} (r_k - r_j) \int_{-\infty}^c \frac{dx}{y} (x - r_k) + 2 \int_{-\infty}^c \frac{dx}{y} (x-r_k)^2 \bigg].
\end{equation}
Once again, we can observe that the first and last terms are divergent in the $x \to -\infty$ limit, and that the divergence cancels between them. Thus, we need to rewrite them in a basis of finite integrals. To do so, we can replace them by the finite expressions
\begin{equation}
-2 \frac{y}{x-r_1} \bigg|_{-\infty}^c + 2 \int_{-\infty}^c \frac{x^2 \, dx}{y} = - 2 \Big( \frac{y}{x-r_1} - x \Big) \bigg|_{-\infty}^c + 2 \int_{-\infty}^c \frac{x^2 - y}{y} dx\,.
\end{equation}

Introducing the following integral representation of the Weierstrass zeta function,
\begin{equation}
\zeta (z_c^+) = \int_{-\infty}^c \frac{x^2 + \frac{a_3}{2} x + \frac{a_2}{6} - y}{-2y} dx - c - \frac{a_3}{4}\,,
\end{equation}
we can write
\begin{equation}
\int_{-\infty}^c \frac{x^2 - y}{y} dx = -2 \Big( \zeta(z_c^+) + c + \frac{a_3}{4} \Big) - \frac{a_3}{2} \int_{-\infty}^c \frac{x \, dx}{y} - \frac{a_2}{6} \int_{-\infty}^c \frac{dx}{y}\,.
\end{equation}
Taking the previous expressions into account, and using that $a_2=S_2(\vec{r})$ and $a_3=-S_1(\vec{r})$, we have that
\begin{align}
\int_{-\infty}^c \frac{dx}{y \, (x-r_k)} = \frac{1}{\prod_{j\neq k}(r_k-r_j)} \bigg[ 2 \frac{(r_k - c)^2 + y_c}{r_k-c} -4 \zeta(z_c^+) - \frac{1}{3} \Big( S_2(\vec{r}) - 3 r_k S_1(\vec{r}) + 6 r_k^2 \Big) z_c^+ \bigg]\,,
\end{align}
where the divergences and the integrals $\int \frac{x \, dx}{y}$ have dropped.

Finally, substituting the expressions in eq.~\eqref{eq_derivatives_Abels_map}, we obtain that
\begin{equation}
\frac{\partial w_c^+}{\partial \chi_{ab}} =  \frac{1}{\omega_1 y_c} \frac{\partial c}{\partial \chi_{ab}} + \frac{1}{\omega_1} \sum_{k=1}^4 \bigg( \frac{y_c+(r_k-c)^2}{r_k-c} - 2 \underbrace{\Big( \zeta(z_c^+) - \frac{\eta_1}{\omega_1} \, z_c^+ \Big)}_{\displaystyle= \frac{1}{\omega_1} g^{(1)}(w_c^+)} \bigg) \frac{1}{\prod_{j \neq k} (r_k - r_j)} \frac{\partial r_k}{\partial \chi_{ab}}\,,
 \label{eq: integrability_conditions_elliptic_letters_10pt_double-box_w_proof}
\end{equation}
which yields a result very similar to eq.~\eqref{eq: integrability_conditions_elliptic_letters_12pt_double-box_w_proof}. Since the terms with $g^{(1)}(w_c^+)$ contain the same prefactor as eq.~\eqref{eq: integrability_conditions_elliptic_letters_10pt_double-box_tau_proof}, they also drop in the adjoint $\text{adj}(J)$. As in the cubic case, we have also verified the validity of eqs.~\eqref{eq: integrability_conditions_elliptic_letters_10pt_double-box_tau_proof} and~\eqref{eq: integrability_conditions_elliptic_letters_10pt_double-box_w_proof} by comparing with numerical differentiation.

In particular, these expressions become relevant for the integrability conditions of the 10-pt double-box symbol, given in sec.~\ref{sec:ch3_symbol_10pt_double_box}. This is because the symbol in eq.~\eqref{eq: 10pt_double-box_symbol_structure} was obtained using direct integration, recall eq.~\eqref{eq: 10-pt_double-box_result}, where the resulting elliptic curve was defined by a quartic polynomial. The only difference compared to the 12-pt case is that the last entries also contain a negative image on the torus, $w^-_\infty$. To take care of this elliptic letter using the previous formalism, we can use the relation provided in eq.~(3.22) of ref.~\cite{Wilhelm:2022wow},
\begin{equation}
2 w_\infty^- = 1 + w_\xi^+\,, \qquad \text{with} \qquad \xi=\frac{a_3^4 - 8a_2 a_3^2 + 16 a_2^2-64 a_0}{8 (a_3^3 - 4 a_2 a_3 + 8 a_1)}\,,
\end{equation}
so that 
\begin{equation}
    \frac{\partial w_\infty^-}{\partial \chi_{ab}} = \frac{1}{2} \frac{\partial w_\xi^+}{\partial \chi_{ab}}
\end{equation}
can be calculated with the expressions above.

\newpage
%% ------------------------------------------------------------------------------------------------
\label{Bibliography}
%\lhead{\emph{Bibliography}}  % Change the left side page header to "Bibliography"
%\bibliographystyle{unsrtnat}  % Use the "unsrtnat" BibTeX style for formatting the Bibliography
%\bibliographystyle{apsrev4-1}
\bibliography{references}  % The references (bibliography) information are stored in the file named "Bibliography.bib"

\providecommand{\noopsort}[1]{}\providecommand{\singleletter}[1] {#1}%

\providecommand{\href}[2]{#2}\begingroup\raggedright\begin{thebibliography}{100}

\bibitem{Morales:2022csr}
R.~Morales, A.~Spiering, M.~Wilhelm, Q.~Yang and C.~Zhang, \emph{{Bootstrapping
  Elliptic Feynman Integrals Using Schubert Analysis}},
  \href{https://doi.org/10.1103/PhysRevLett.131.041601}{\emph{Phys. Rev. Lett.}
  {\bfseries 131} (2023) 041601}
  [\href{https://arxiv.org/abs/2212.09762}{{2212.09762}}].

\bibitem{McLeod:2023qdf}
A.~McLeod, R.~Morales, M.~von Hippel, M.~Wilhelm and C.~Zhang, \emph{{An
  infinite family of elliptic ladder integrals}},
  \href{https://doi.org/10.1007/JHEP05(2023)236}{\emph{JHEP} {\bfseries 05}
  (2023) 236} [\href{https://arxiv.org/abs/2301.07965}{{2301.07965}}].

\bibitem{Frellesvig:2023bbf}
H.~Frellesvig, R.~Morales and M.~Wilhelm, \emph{{Calabi-Yau Meets Gravity: A
  Calabi-Yau Threefold at Fifth Post-Minkowskian Order}},
  \href{https://doi.org/10.1103/PhysRevLett.132.201602}{\emph{Phys. Rev. Lett.}
  {\bfseries 132} (2024) 201602}
  [\href{https://arxiv.org/abs/2312.11371}{{2312.11371}}].

\bibitem{Frellesvig:2024zph}
H.~Frellesvig, R.~Morales and M.~Wilhelm, \emph{{Classifying post-Minkowskian
  geometries for gravitational waves via loop-by-loop Baikov}},
  \href{https://doi.org/10.1007/JHEP08(2024)243}{\emph{JHEP} {\bfseries 08}
  (2024) 243} [\href{https://arxiv.org/abs/2405.17255}{{2405.17255}}].

\bibitem{Frellesvig:2024rea}
H.~Frellesvig, R.~Morales, S.~P\"ogel, S.~Weinzierl and M.~Wilhelm,
  \emph{{Calabi-Yau Feynman integrals in gravity: $\varepsilon$-factorized form
  for apparent singularities}},
  \href{https://doi.org/10.1007/JHEP02(2025)209}{\emph{JHEP} {\bfseries 02}
  (2025) 209} [\href{https://arxiv.org/abs/2412.12057}{{2412.12057}}].

\bibitem{Brammer:2025rqo}
D.~Brammer, H.~Frellesvig, R.~Morales and M.~Wilhelm, \emph{{Classification of
  Feynman integral geometries for black-hole scattering at 5PM order}},
  \href{https://arxiv.org/abs/2505.10274}{{2505.10274}}.

\bibitem{Bern:2022jnl}
Z.~Bern and J.~Trnka, \emph{{Snowmass TF04 Report: Scattering Amplitudes and
  their Applications}},  \href{https://arxiv.org/abs/2210.03146}{{2210.03146}}.

\bibitem{ATLAS:2012yve}
{\scshape ATLAS} collaboration, \emph{{Observation of a new particle in the
  search for the Standard Model Higgs boson with the ATLAS detector at the
  LHC}}, \href{https://doi.org/10.1016/j.physletb.2012.08.020}{\emph{Phys.
  Lett. B} {\bfseries 716} (2012) 1}
  [\href{https://arxiv.org/abs/1207.7214}{{1207.7214}}].

\bibitem{CMS:2012qbp}
{\scshape CMS} collaboration, \emph{{Observation of a New Boson at a Mass of
  125 GeV with the CMS Experiment at the LHC}},
  \href{https://doi.org/10.1016/j.physletb.2012.08.021}{\emph{Phys. Lett. B}
  {\bfseries 716} (2012) 30}
  [\href{https://arxiv.org/abs/1207.7235}{{1207.7235}}].

\bibitem{LIGOScientific:2016aoc}
{\scshape LIGO Scientific, Virgo} collaboration, \emph{{Observation of
  Gravitational Waves from a Binary Black Hole Merger}},
  \href{https://doi.org/10.1103/PhysRevLett.116.061102}{\emph{Phys. Rev. Lett.}
  {\bfseries 116} (2016) 061102}
  [\href{https://arxiv.org/abs/1602.03837}{{1602.03837}}].

\bibitem{Bruning:2019}
O.~Brüning and L.~Rossi, \emph{{The High-Luminosity Large Hadron Collider}},
  \href{https://doi.org/10.1038/s42254-019-0050-6}{\emph{Nat. Rev. Phys.}
  {\bfseries 1} (2019) 241}.

\bibitem{Ballmer:2022uxx}
{S.W. Ballmer, R. Adhikari, L. Badurina, D.A. Brown, S. Chattopadhyay, M.
  Evans, P.~Fritschel, E. Hall, J.M. Hogan, K. Jani, T. Kovachy, K. Kuns, A.
  Schwartzman, D.~Sigg, B. Slagmolen, S. Vitale and C. Wipf},
  \emph{{Snowmass2021 Cosmic Frontier White Paper: Future Gravitational-Wave
  Detector Facilities}},  in \emph{{Snowmass 2021}}, 3, 2022
  [\href{https://arxiv.org/abs/2203.08228}{{2203.08228}}].

\bibitem{Purrer:2019jcp}
M.~P\"urrer and C.-J.~Haster, \emph{{Gravitational waveform accuracy
  requirements for future ground-based detectors}},
  \href{https://doi.org/10.1103/PhysRevResearch.2.023151}{\emph{Phys. Rev.
  Res.} {\bfseries 2} (2020) 023151}
  [\href{https://arxiv.org/abs/1912.10055}{{1912.10055}}].

\bibitem{FebresCordero:2022psq}
F.~Febres~Cordero, A.~von Manteuffel and T.~Neumann, \emph{{Computational
  Challenges for Multi-loop Collider Phenomenology: A Snowmass 2021 White
  Paper}}, \href{https://doi.org/10.1007/s41781-022-00088-0}{\emph{Comput.
  Softw. Big Sci.} {\bfseries 6} (2022) 14}
  [\href{https://arxiv.org/abs/2204.04200}{{2204.04200}}].

\bibitem{Buonanno:2022pgc}
A.~Buonanno, M.~Khalil, D.~O'Connell, R.~Roiban, M.P.~Solon and M.~Zeng,
  \emph{{Snowmass White Paper: Gravitational Waves and Scattering Amplitudes}},
   in \emph{{Snowmass 2021}}, 4, 2022
  [\href{https://arxiv.org/abs/2204.05194}{{2204.05194}}].

\bibitem{Dixon:1996wi}
L.J.~Dixon, \emph{{Calculating scattering amplitudes efficiently}},  in
  \emph{{Theoretical Advanced Study Institute in Elementary Particle Physics
  (TASI 95): QCD and Beyond}}, pp.~539--584, 1, 1996
  [\href{https://arxiv.org/abs/hep-ph/9601359}{{hep-ph/9601359}}].

\bibitem{Weinzierl:2016bus}
S.~Weinzierl, \emph{{Tales of 1001 Gluons}},
  \href{https://doi.org/10.1016/j.physrep.2017.01.004}{\emph{Phys. Rept.}
  {\bfseries 676} (2017) 1}
  [\href{https://arxiv.org/abs/1610.05318}{{1610.05318}}].

\bibitem{Dixon:2011xs}
L.J.~Dixon, \emph{{Scattering amplitudes: the most perfect microscopic
  structures in the universe}},
  \href{https://doi.org/10.1088/1751-8113/44/45/454001}{\emph{J. Phys. A}
  {\bfseries 44} (2011) 454001}
  [\href{https://arxiv.org/abs/1105.0771}{{1105.0771}}].

\bibitem{Arkani-Hamed:2012zlh}
N.~Arkani-Hamed, J.L.~Bourjaily, F.~Cachazo, A.B.~Goncharov, A.~Postnikov and
  J.~Trnka, \emph{{Grassmannian Geometry of Scattering Amplitudes}}, Cambridge
  University Press (2016),
  \href{https://doi.org/10.1017/CBO9781316091548}{10.1017/CBO9781316091548},
  [\href{https://arxiv.org/abs/1212.5605}{{1212.5605}}].

\bibitem{Parke:1986gb}
S.J.~Parke and T.R.~Taylor, \emph{{An Amplitude for $n$ Gluon Scattering}},
  \href{https://doi.org/10.1103/PhysRevLett.56.2459}{\emph{Phys. Rev. Lett.}
  {\bfseries 56} (1986) 2459}.

\bibitem{Laporta:2017okg}
S.~Laporta, \emph{{High-precision calculation of the 4-loop contribution to the
  electron g-2 in QED}},
  \href{https://doi.org/10.1016/j.physletb.2017.06.056}{\emph{Phys. Lett. B}
  {\bfseries 772} (2017) 232}
  [\href{https://arxiv.org/abs/1704.06996}{{1704.06996}}].

\bibitem{Smirnov:2021rhf}
A.V.~Smirnov, N.D.~Shapurov and L.I.~Vysotsky, \emph{{FIESTA5: Numerical
  high-performance Feynman integral evaluation}},
  \href{https://doi.org/10.1016/j.cpc.2022.108386}{\emph{Comput. Phys. Commun.}
  {\bfseries 277} (2022) 108386}
  [\href{https://arxiv.org/abs/2110.11660}{{2110.11660}}].

\bibitem{Heinrich:2023til}
G.~Heinrich, S.P.~Jones, M.~Kerner, V.~Magerya, A.~Olsson and J.~Schlenk,
  \emph{{Numerical scattering amplitudes with pySecDec}},
  \href{https://doi.org/10.1016/j.cpc.2023.108956}{\emph{Comput. Phys. Commun.}
  {\bfseries 295} (2024) 108956}
  [\href{https://arxiv.org/abs/2305.19768}{{2305.19768}}].

\bibitem{Vollinga:2004sn}
J.~Vollinga and S.~Weinzierl, \emph{{Numerical evaluation of multiple
  polylogarithms}},
  \href{https://doi.org/10.1016/j.cpc.2004.12.009}{\emph{Comput. Phys. Commun.}
  {\bfseries 167} (2005) 177}
  [\href{https://arxiv.org/abs/hep-ph/0410259}{{hep-ph/0410259}}].

\bibitem{Walden:2020odh}
M.~Walden and S.~Weinzierl, \emph{{Numerical evaluation of iterated integrals
  related to elliptic Feynman integrals}},
  \href{https://doi.org/10.1016/j.cpc.2021.108020}{\emph{Comput. Phys. Commun.}
  {\bfseries 265} (2021) 108020}
  [\href{https://arxiv.org/abs/2010.05271}{{2010.05271}}].

\bibitem{Boyle:2019kee}
M.~Boyle et~al., \emph{{The SXS Collaboration catalog of binary black hole
  simulations}}, \href{https://doi.org/10.1088/1361-6382/ab34e2}{\emph{Class.
  Quant. Grav.} {\bfseries 36} (2019) 195006}
  [\href{https://arxiv.org/abs/1904.04831}{{1904.04831}}].

\bibitem{Chen:1977oja}
K.-T.~Chen, \emph{{Iterated path integrals}},
  \href{https://doi.org/10.1090/S0002-9904-1977-14320-6}{\emph{Bull. Am. Math.
  Soc.} {\bfseries 83} (1977) 831}.

\bibitem{Goncharov:1995ifj}
A.B.~Goncharov, \emph{{Geometry of Configurations, Polylogarithms, and Motivic
  Cohomology}}, \href{https://doi.org/10.1006/aima.1995.1045}{\emph{Adv. Math.}
  {\bfseries 114} (1995) 197}.

\bibitem{Bourjaily:2022bwx}
{J.L. Bourjaily, J. Broedel, E. Chaubey, C. Duhr, H. Frellesvig, M. Hidding, R.
  Marzucca, A.J. McLeod, M. Spradlin, L. Tancredi, C. Vergu, M. Volk, A.
  Volovich, M. von Hippel, S. Weinzierl, M. Wilhelm and C. Zhang},
  \emph{{Functions Beyond Multiple Polylogarithms for Precision Collider
  Physics}},  in \emph{{Snowmass 2021}}, 3, 2022
  [\href{https://arxiv.org/abs/2203.07088}{{2203.07088}}].

\bibitem{Sabry:1962rge}
A.~Sabry, \emph{{Fourth order spectral functions for the electron propagator}},
  \href{https://doi.org/10.1016/0029-5582(62)90535-7}{\emph{Nucl. Phys.}
  {\bfseries 33} (1962) 401}.

\bibitem{Broadhurst:1993mw}
D.J.~Broadhurst, J.~Fleischer and O.V.~Tarasov, \emph{{Two loop two point
  functions with masses: Asymptotic expansions and Taylor series, in any
  dimension}}, \href{https://doi.org/10.1007/BF01474625}{\emph{Z. Phys. C}
  {\bfseries 60} (1993) 287}
  [\href{https://arxiv.org/abs/hep-ph/9304303}{{hep-ph/9304303}}].

\bibitem{Laporta:2004rb}
S.~Laporta and E.~Remiddi, \emph{{Analytic treatment of the two loop equal mass
  sunrise graph}},
  \href{https://doi.org/10.1016/j.nuclphysb.2004.10.044}{\emph{Nucl. Phys. B}
  {\bfseries 704} (2005) 349}
  [\href{https://arxiv.org/abs/hep-ph/0406160}{{hep-ph/0406160}}].

\bibitem{Caron-Huot:2012awx}
S.~Caron-Huot and K.J.~Larsen, \emph{{Uniqueness of two-loop master contours}},
  \href{https://doi.org/10.1007/JHEP10(2012)026}{\emph{JHEP} {\bfseries 10}
  (2012) 026} [\href{https://arxiv.org/abs/1205.0801}{{1205.0801}}].

\bibitem{Adams:2013nia}
L.~Adams, C.~Bogner and S.~Weinzierl, \emph{{The two-loop sunrise graph with
  arbitrary masses}}, \href{https://doi.org/10.1063/1.4804996}{\emph{J. Math.
  Phys.} {\bfseries 54} (2013) 052303}
  [\href{https://arxiv.org/abs/1302.7004}{{1302.7004}}].

\bibitem{Bloch:2013tra}
S.~Bloch and P.~Vanhove, \emph{{The elliptic dilogarithm for the sunset
  graph}}, \href{https://doi.org/10.1016/j.jnt.2014.09.032}{\emph{J. Number
  Theor.} {\bfseries 148} (2015) 328}
  [\href{https://arxiv.org/abs/1309.5865}{{1309.5865}}].

\bibitem{Adams:2014vja}
L.~Adams, C.~Bogner and S.~Weinzierl, \emph{{The two-loop sunrise graph in two
  space-time dimensions with arbitrary masses in terms of elliptic
  dilogarithms}}, \href{https://doi.org/10.1063/1.4896563}{\emph{J. Math.
  Phys.} {\bfseries 55} (2014) 102301}
  [\href{https://arxiv.org/abs/1405.5640}{{1405.5640}}].

\bibitem{Remiddi:2016gno}
E.~Remiddi and L.~Tancredi, \emph{{Differential equations and dispersion
  relations for Feynman amplitudes. The two-loop massive sunrise and the kite
  integral}},
  \href{https://doi.org/10.1016/j.nuclphysb.2016.04.013}{\emph{Nucl. Phys. B}
  {\bfseries 907} (2016) 400}
  [\href{https://arxiv.org/abs/1602.01481}{{1602.01481}}].

\bibitem{Adams:2016xah}
L.~Adams, C.~Bogner, A.~Schweitzer and S.~Weinzierl, \emph{{The kite integral
  to all orders in terms of elliptic polylogarithms}},
  \href{https://doi.org/10.1063/1.4969060}{\emph{J. Math. Phys.} {\bfseries 57}
  (2016) 122302} [\href{https://arxiv.org/abs/1607.01571}{{1607.01571}}].

\bibitem{Broedel:2017siw}
J.~Broedel, C.~Duhr, F.~Dulat and L.~Tancredi, \emph{{Elliptic polylogarithms
  and iterated integrals on elliptic curves II: an application to the sunrise
  integral}}, \href{https://doi.org/10.1103/PhysRevD.97.116009}{\emph{Phys.
  Rev. D} {\bfseries 97} (2018) 116009}
  [\href{https://arxiv.org/abs/1712.07095}{{1712.07095}}].

\bibitem{Kristensson:2021ani}
A.~Kristensson, M.~Wilhelm and C.~Zhang, \emph{{Elliptic Double Box and
  Symbology Beyond Polylogarithms}},
  \href{https://doi.org/10.1103/PhysRevLett.127.251603}{\emph{Phys. Rev. Lett.}
  {\bfseries 127} (2021) 251603}
  [\href{https://arxiv.org/abs/2106.14902}{{2106.14902}}].

\bibitem{Giroux:2022wav}
M.~Giroux and A.~Pokraka, \emph{{Loop-by-loop differential equations for dual
  (elliptic) Feynman integrals}},
  \href{https://doi.org/10.1007/JHEP03(2023)155}{\emph{JHEP} {\bfseries 03}
  (2023) 155} [\href{https://arxiv.org/abs/2210.09898}{{2210.09898}}].

\bibitem{Stawinski:2023qtw}
S.F.~Stawinski, \emph{{An elliptic one-loop amplitude in anti-de-Sitter
  space}}, \href{https://doi.org/10.1007/JHEP02(2024)208}{\emph{JHEP}
  {\bfseries 02} (2024) 208}
  [\href{https://arxiv.org/abs/2309.15059}{{2309.15059}}].

\bibitem{Giroux:2024yxu}
M.~Giroux, A.~Pokraka, F.~Porkert and Y.~Sohnle, \emph{{The soaring kite: a
  tale of two punctured tori}},
  \href{https://doi.org/10.1007/JHEP05(2024)239}{\emph{JHEP} {\bfseries 05}
  (2024) 239} [\href{https://arxiv.org/abs/2401.14307}{{2401.14307}}].

\bibitem{Huang:2013kh}
R.~Huang and Y.~Zhang, \emph{{On Genera of Curves from High-loop Generalized
  Unitarity Cuts}}, \href{https://doi.org/10.1007/JHEP04(2013)080}{\emph{JHEP}
  {\bfseries 04} (2013) 080}
  [\href{https://arxiv.org/abs/1302.1023}{{1302.1023}}].

\bibitem{Marzucca:2023gto}
R.~Marzucca, A.J.~McLeod, B.~Page, S.~P\"ogel and S.~Weinzierl, \emph{{Genus
  drop in hyperelliptic Feynman integrals}},
  \href{https://doi.org/10.1103/PhysRevD.109.L031901}{\emph{Phys. Rev. D}
  {\bfseries 109} (2024) L031901}
  [\href{https://arxiv.org/abs/2307.11497}{{2307.11497}}].

\bibitem{Duhr:2024uid}
C.~Duhr, F.~Porkert and S.F.~Stawinski, \emph{{Canonical differential equations
  beyond genus one}},
  \href{https://doi.org/10.1007/JHEP02(2025)014}{\emph{JHEP} {\bfseries 02}
  (2025) 014} [\href{https://arxiv.org/abs/2412.02300}{{2412.02300}}].

\bibitem{Primo:2017ipr}
A.~Primo and L.~Tancredi, \emph{{Maximal cuts and differential equations for
  Feynman integrals. An application to the three-loop massive banana graph}},
  \href{https://doi.org/10.1016/j.nuclphysb.2017.05.018}{\emph{Nucl. Phys. B}
  {\bfseries 921} (2017) 316}
  [\href{https://arxiv.org/abs/1704.05465}{{1704.05465}}].

\bibitem{Bourjaily:2018ycu}
J.L.~Bourjaily, Y.-H.~He, A.J.~Mcleod, M.~Von~Hippel and M.~Wilhelm,
  \emph{{Traintracks through Calabi-Yau Manifolds: Scattering Amplitudes beyond
  Elliptic Polylogarithms}},
  \href{https://doi.org/10.1103/PhysRevLett.121.071603}{\emph{Phys. Rev. Lett.}
  {\bfseries 121} (2018) 071603}
  [\href{https://arxiv.org/abs/1805.09326}{{1805.09326}}].

\bibitem{Bourjaily:2018yfy}
J.L.~Bourjaily, A.J.~McLeod, M.~von Hippel and M.~Wilhelm, \emph{{Bounded
  Collection of Feynman Integral Calabi-Yau Geometries}},
  \href{https://doi.org/10.1103/PhysRevLett.122.031601}{\emph{Phys. Rev. Lett.}
  {\bfseries 122} (2019) 031601}
  [\href{https://arxiv.org/abs/1810.07689}{{1810.07689}}].

\bibitem{Bonisch:2021yfw}
K.~B\"onisch, C.~Duhr, F.~Fischbach, A.~Klemm and C.~Nega, \emph{{Feynman
  integrals in dimensional regularization and extensions of Calabi-Yau
  motives}}, \href{https://doi.org/10.1007/JHEP09(2022)156}{\emph{JHEP}
  {\bfseries 09} (2022) 156}
  [\href{https://arxiv.org/abs/2108.05310}{{2108.05310}}].

\bibitem{Broedel:2021zij}
J.~Broedel, C.~Duhr and N.~Matthes, \emph{{Meromorphic modular forms and the
  three-loop equal-mass banana integral}},
  \href{https://doi.org/10.1007/JHEP02(2022)184}{\emph{JHEP} {\bfseries 02}
  (2022) 184} [\href{https://arxiv.org/abs/2109.15251}{{2109.15251}}].

\bibitem{Duhr:2022pch}
C.~Duhr, A.~Klemm, F.~Loebbert, C.~Nega and F.~Porkert,
  \emph{{Yangian-Invariant Fishnet Integrals in Two Dimensions as Volumes of
  Calabi-Yau Varieties}},
  \href{https://doi.org/10.1103/PhysRevLett.130.041602}{\emph{Phys. Rev. Lett.}
  {\bfseries 130} (2023) 041602}
  [\href{https://arxiv.org/abs/2209.05291}{{2209.05291}}].

\bibitem{Lairez:2022zkj}
P.~Lairez and P.~Vanhove, \emph{{Algorithms for minimal
  Picard\textendash{}Fuchs operators of Feynman integrals}},
  \href{https://doi.org/10.1007/s11005-023-01661-3}{\emph{Lett. Math. Phys.}
  {\bfseries 113} (2023) 37}
  [\href{https://arxiv.org/abs/2209.10962}{{2209.10962}}].

\bibitem{Pogel:2022vat}
S.~P\"ogel, X.~Wang and S.~Weinzierl, \emph{{Bananas of equal mass: any loop,
  any order in the dimensional regularisation parameter}},
  \href{https://doi.org/10.1007/JHEP04(2023)117}{\emph{JHEP} {\bfseries 04}
  (2023) 117} [\href{https://arxiv.org/abs/2212.08908}{{2212.08908}}].

\bibitem{Duhr:2022dxb}
C.~Duhr, A.~Klemm, C.~Nega and L.~Tancredi, \emph{{The ice cone family and
  iterated integrals for Calabi-Yau varieties}},
  \href{https://doi.org/10.1007/JHEP02(2023)228}{\emph{JHEP} {\bfseries 02}
  (2023) 228} [\href{https://arxiv.org/abs/2212.09550}{{2212.09550}}].

\bibitem{Cao:2023tpx}
Q.~Cao, S.~He and Y.~Tang, \emph{{Cutting the traintracks: Cauchy, Schubert and
  Calabi-Yau}}, \href{https://doi.org/10.1007/JHEP04(2023)072}{\emph{JHEP}
  {\bfseries 04} (2023) 072}
  [\href{https://arxiv.org/abs/2301.07834}{{2301.07834}}].

\bibitem{McLeod:2023doa}
A.J.~McLeod and M.~von Hippel, \emph{{Traintracks All the Way Down}},
  \href{https://arxiv.org/abs/2306.11780}{{2306.11780}}.

\bibitem{Duhr:2023eld}
C.~Duhr, A.~Klemm, F.~Loebbert, C.~Nega and F.~Porkert, \emph{{The Basso-Dixon
  formula and Calabi-Yau geometry}},
  \href{https://doi.org/10.1007/JHEP03(2024)177}{\emph{JHEP} {\bfseries 03}
  (2024) 177} [\href{https://arxiv.org/abs/2310.08625}{{2310.08625}}].

\bibitem{Duhr:2024hjf}
C.~Duhr, A.~Klemm, F.~Loebbert, C.~Nega and F.~Porkert, \emph{{Geometry from
  integrability: multi-leg fishnet integrals in two dimensions}},
  \href{https://doi.org/10.1007/JHEP07(2024)008}{\emph{JHEP} {\bfseries 07}
  (2024) 008} [\href{https://arxiv.org/abs/2402.19034}{{2402.19034}}].

\bibitem{Bern:2021dqo}
{Z. Bern, J. Parra-Martinez, R. Roiban, M. S. Ruf, C.-H. Shen, M. P. Solon and
  M. Zeng}, \emph{{Scattering Amplitudes and Conservative Binary Dynamics at
  ${\cal O}(G^4)$}},
  \href{https://doi.org/10.1103/PhysRevLett.126.171601}{\emph{Phys. Rev. Lett.}
  {\bfseries 126} (2021) 171601}
  [\href{https://arxiv.org/abs/2101.07254}{{2101.07254}}].

\bibitem{Dlapa:2021npj}
C.~Dlapa, G.~K\"alin, Z.~Liu and R.A.~Porto, \emph{{Dynamics of binary systems
  to fourth Post-Minkowskian order from the effective field theory approach}},
  \href{https://doi.org/10.1016/j.physletb.2022.137203}{\emph{Phys. Lett. B}
  {\bfseries 831} (2022) 137203}
  [\href{https://arxiv.org/abs/2106.08276}{{2106.08276}}].

\bibitem{Bern:2022jvn}
{Z. Bern, J. Parra-Martinez, R. Roiban, M.S. Ruf, C.-H. Shen, M.P. Solon and M.
  Zeng}, \emph{{Scattering amplitudes and conservative dynamics at the fourth
  post-Minkowskian order}},
  \href{https://doi.org/10.22323/1.416.0051}{\emph{PoS} {\bfseries LL2022}
  (2022) 051}.

\bibitem{Dlapa:2022wdu}
C.~Dlapa, J.M.~Henn and F.J.~Wagner, \emph{{An algorithmic approach to finding
  canonical differential equations for elliptic Feynman integrals}},
  \href{https://doi.org/10.1007/JHEP08(2023)120}{\emph{JHEP} {\bfseries 08}
  (2023) 120} [\href{https://arxiv.org/abs/2211.16357}{{2211.16357}}].

\bibitem{Jakobsen:2023ndj}
G.U.~Jakobsen, G.~Mogull, J.~Plefka, B.~Sauer and Y.~Xu, \emph{{Conservative
  Scattering of Spinning Black Holes at Fourth Post-Minkowskian Order}},
  \href{https://doi.org/10.1103/PhysRevLett.131.151401}{\emph{Phys. Rev. Lett.}
  {\bfseries 131} (2023) 151401}
  [\href{https://arxiv.org/abs/2306.01714}{{2306.01714}}].

\bibitem{Klemm:2024wtd}
A.~Klemm, C.~Nega, B.~Sauer and J.~Plefka, \emph{{Calabi-Yau periods for black
  hole scattering in classical general relativity}},
  \href{https://doi.org/10.1103/PhysRevD.109.124046}{\emph{Phys. Rev. D}
  {\bfseries 109} (2024) 124046}
  [\href{https://arxiv.org/abs/2401.07899}{{2401.07899}}].

\bibitem{Driesse:2024xad}
M.~Driesse, G.U.~Jakobsen, G.~Mogull, J.~Plefka, B.~Sauer and J.~Usovitsch,
  \emph{{Conservative Black Hole Scattering at Fifth Post-Minkowskian and First
  Self-Force Order}},
  \href{https://doi.org/10.1103/PhysRevLett.132.241402}{\emph{Phys. Rev. Lett.}
  {\bfseries 132} (2024) 241402}
  [\href{https://arxiv.org/abs/2403.07781}{{2403.07781}}].

\bibitem{Bern:2024adl}
{Z. Bern, E. Herrmann, R. Roiban, M.S. Ruf, A.V. Smirnov, V.A. Smirnov and M.
  Zeng}, \emph{{Amplitudes, supersymmetric black hole scattering at $
  \mathcal{O}({G}^5) $, and loop integration}},
  \href{https://doi.org/10.1007/JHEP10(2024)023}{\emph{JHEP} {\bfseries 10}
  (2024) 023} [\href{https://arxiv.org/abs/2406.01554}{{2406.01554}}].

\bibitem{Driesse:2024feo}
{M. Driesse, G.U. Jakobsen, A. Klemm, G. Mogull, C. Nega, J. Plefka, B. Sauer
  and J.~Usovitsch}, \emph{{Emergence of Calabi\textendash{}Yau manifolds in
  high-precision black-hole scattering}},
  \href{https://doi.org/10.1038/s41586-025-08984-2}{\emph{Nature} {\bfseries
  641} (2025) 603} [\href{https://arxiv.org/abs/2411.11846}{{2411.11846}}].

\bibitem{Brown:2010bw}
F.~Brown and O.~Schnetz, \emph{{A K3 in $\phi^4$}},
  \href{https://doi.org/10.1215/00127094-1644201}{\emph{Duke Math. J.}
  {\bfseries 161} (2012) 1817}
  [\href{https://arxiv.org/abs/1006.4064}{{1006.4064}}].

\bibitem{Schnetz:2019cab}
O.~Schnetz, \emph{{Geometries in perturbative quantum field theory}},
  \href{https://doi.org/10.4310/CNTP.2021.v15.n4.a2}{\emph{Commun. Num. Theor.
  Phys.} {\bfseries 15} (2021) 743}
  [\href{https://arxiv.org/abs/1905.08083}{{1905.08083}}].

\bibitem{LevinRacinet2007}
A.~{Levin} and G.~{Racinet}, \emph{{Towards multiple elliptic polylogarithms}},
   \href{https://arxiv.org/abs/math/0703237}{{math/0703237}}.

\bibitem{brown2011multiple}
F.~Brown and A.~Levin, \emph{{Multiple Elliptic Polylogarithms}},
  \href{https://arxiv.org/abs/1110.6917}{{1110.6917}}.

\bibitem{Broedel:2014vla}
J.~Broedel, C.R.~Mafra, N.~Matthes and O.~Schlotterer, \emph{{Elliptic multiple
  zeta values and one-loop superstring amplitudes}},
  \href{https://doi.org/10.1007/JHEP07(2015)112}{\emph{JHEP} {\bfseries 07}
  (2015) 112} [\href{https://arxiv.org/abs/1412.5535}{{1412.5535}}].

\bibitem{Broedel:2017kkb}
J.~Broedel, C.~Duhr, F.~Dulat and L.~Tancredi, \emph{{Elliptic polylogarithms
  and iterated integrals on elliptic curves. Part I: general formalism}},
  \href{https://doi.org/10.1007/JHEP05(2018)093}{\emph{JHEP} {\bfseries 05}
  (2018) 093} [\href{https://arxiv.org/abs/1712.07089}{{1712.07089}}].

\bibitem{Broedel:2018iwv}
J.~Broedel, C.~Duhr, F.~Dulat, B.~Penante and L.~Tancredi, \emph{{Elliptic
  symbol calculus: from elliptic polylogarithms to iterated integrals of
  Eisenstein series}},
  \href{https://doi.org/10.1007/JHEP08(2018)014}{\emph{JHEP} {\bfseries 08}
  (2018) 014} [\href{https://arxiv.org/abs/1803.10256}{{1803.10256}}].

\bibitem{Broedel:2018qkq}
J.~Broedel, C.~Duhr, F.~Dulat, B.~Penante and L.~Tancredi, \emph{{Elliptic
  Feynman integrals and pure functions}},
  \href{https://doi.org/10.1007/JHEP01(2019)023}{\emph{JHEP} {\bfseries 01}
  (2019) 023} [\href{https://arxiv.org/abs/1809.10698}{{1809.10698}}].

\bibitem{DHoker:2023vax}
E.~D'Hoker, M.~Hidding and O.~Schlotterer, \emph{{Constructing polylogarithms
  on higher-genus Riemann surfaces}},
  \href{https://arxiv.org/abs/2306.08644}{{2306.08644}}.

\bibitem{Baune:2024biq}
K.~Baune, J.~Broedel, E.~Im, A.~Lisitsyn and F.~Zerbini,
  \emph{{Schottky\textendash{}Kronecker forms and hyperelliptic
  polylogarithms}}, \href{https://doi.org/10.1088/1751-8121/ad8197}{\emph{J.
  Phys. A} {\bfseries 57} (2024) 445202}
  [\href{https://arxiv.org/abs/2406.10051}{{2406.10051}}].

\bibitem{DHoker:2024ozn}
E.~D'Hoker and O.~Schlotterer, \emph{{Fay identities for polylogarithms on
  higher-genus Riemann surfaces}},
  \href{https://arxiv.org/abs/2407.11476}{{2407.11476}}.

\bibitem{Baune:2024ber}
K.~Baune, J.~Broedel, E.~Im, A.~Lisitsyn and Y.~Moeckli, \emph{{Higher-genus
  Fay-like identities from meromorphic generating functions}},
  \href{https://doi.org/10.21468/SciPostPhys.18.3.093}{\emph{SciPost Phys.}
  {\bfseries 18} (2025) 093}
  [\href{https://arxiv.org/abs/2409.08208}{{2409.08208}}].

\bibitem{Adams:2018bsn}
L.~Adams, E.~Chaubey and S.~Weinzierl, \emph{{Planar Double Box Integral for
  Top Pair Production with a Closed Top Loop to all orders in the Dimensional
  Regularization Parameter}},
  \href{https://doi.org/10.1103/PhysRevLett.121.142001}{\emph{Phys. Rev. Lett.}
  {\bfseries 121} (2018) 142001}
  [\href{https://arxiv.org/abs/1804.11144}{{1804.11144}}].

\bibitem{Adams:2018kez}
L.~Adams, E.~Chaubey and S.~Weinzierl, \emph{{Analytic results for the planar
  double box integral relevant to top-pair production with a closed top loop}},
  \href{https://doi.org/10.1007/JHEP10(2018)206}{\emph{JHEP} {\bfseries 10}
  (2018) 206} [\href{https://arxiv.org/abs/1806.04981}{{1806.04981}}].

\bibitem{Broedel:2019hyg}
J.~Broedel, C.~Duhr, F.~Dulat, B.~Penante and L.~Tancredi, \emph{{Elliptic
  polylogarithms and Feynman parameter integrals}},
  \href{https://doi.org/10.1007/JHEP05(2019)120}{\emph{JHEP} {\bfseries 05}
  (2019) 120} [\href{https://arxiv.org/abs/1902.09971}{{1902.09971}}].

\bibitem{Abreu:2019fgk}
S.~Abreu, M.~Becchetti, C.~Duhr and R.~Marzucca, \emph{{Three-loop
  contributions to the $\rho$ parameter and iterated integrals of modular
  forms}}, \href{https://doi.org/10.1007/JHEP02(2020)050}{\emph{JHEP}
  {\bfseries 02} (2020) 050}
  [\href{https://arxiv.org/abs/1912.02747}{{1912.02747}}].

\bibitem{Duhr:2024bzt}
C.~Duhr, F.~Gasparotto, C.~Nega, L.~Tancredi and S.~Weinzierl, \emph{{On the
  electron self-energy to three loops in QED}},
  \href{https://doi.org/10.1007/JHEP11(2024)020}{\emph{JHEP} {\bfseries 11}
  (2024) 020} [\href{https://arxiv.org/abs/2408.05154}{{2408.05154}}].

\bibitem{Forner:2024ojj}
F.~Forner, C.~Nega and L.~Tancredi, \emph{{On the photon self-energy to three
  loops in QED}}, \href{https://doi.org/10.1007/JHEP03(2025)148}{\emph{JHEP}
  {\bfseries 03} (2025) 148}
  [\href{https://arxiv.org/abs/2411.19042}{{2411.19042}}].

\bibitem{Weinzierl:2022eaz}
S.~Weinzierl, \emph{{Feynman Integrals. A Comprehensive Treatment for Students
  and Researchers}}, UNITEXT for Physics, Springer (2022),
  \href{https://doi.org/10.1007/978-3-030-99558-4}{10.1007/978-3-030-99558-4},
  [\href{https://arxiv.org/abs/2201.03593}{{2201.03593}}].

\bibitem{Brink:1976bc}
L.~Brink, J.H.~Schwarz and J.~Scherk, \emph{{Supersymmetric Yang-Mills
  Theories}}, \href{https://doi.org/10.1016/0550-3213(77)90328-5}{\emph{Nucl.
  Phys. B} {\bfseries 121} (1977) 77}.

\bibitem{Drummond:2007au}
J.M.~Drummond, J.~Henn, G.P.~Korchemsky and E.~Sokatchev, \emph{{Conformal Ward
  identities for Wilson loops and a test of the duality with gluon
  amplitudes}},
  \href{https://doi.org/10.1016/j.nuclphysb.2009.10.013}{\emph{Nucl. Phys. B}
  {\bfseries 826} (2010) 337}
  [\href{https://arxiv.org/abs/0712.1223}{{0712.1223}}].

\bibitem{Drummond:2008vq}
J.M.~Drummond, J.~Henn, G.P.~Korchemsky and E.~Sokatchev, \emph{{Dual
  superconformal symmetry of scattering amplitudes in N=4 super-Yang-Mills
  theory}}, \href{https://doi.org/10.1016/j.nuclphysb.2009.11.022}{\emph{Nucl.
  Phys. B} {\bfseries 828} (2010) 317}
  [\href{https://arxiv.org/abs/0807.1095}{{0807.1095}}].

\bibitem{Usyukina:1993ch}
N.I.~Usyukina and A.I.~Davydychev, \emph{{Exact results for three and four
  point ladder diagrams with an arbitrary number of rungs}},
  \href{https://doi.org/10.1016/0370-2693(93)91118-7}{\emph{Phys. Lett. B}
  {\bfseries 305} (1993) 136}.

\bibitem{Broadhurst:2010ds}
D.J.~Broadhurst and A.I.~Davydychev, \emph{{Exponential suppression with four
  legs and an infinity of loops}},
  \href{https://doi.org/10.1016/j.nuclphysbps.2010.09.014}{\emph{Nucl. Phys. B
  Proc. Suppl.} {\bfseries 205-206} (2010) 326}
  [\href{https://arxiv.org/abs/1007.0237}{{1007.0237}}].

\bibitem{Caron-Huot:2018dsv}
S.~Caron-Huot, L.J.~Dixon, M.~von Hippel, A.J.~McLeod and G.~Papathanasiou,
  \emph{{The Double Pentaladder Integral to All Orders}},
  \href{https://doi.org/10.1007/JHEP07(2018)170}{\emph{JHEP} {\bfseries 07}
  (2018) 170} [\href{https://arxiv.org/abs/1806.01361}{{1806.01361}}].

\bibitem{He:2020uxy}
S.~He, Z.~Li, Y.~Tang and Q.~Yang, \emph{{The Wilson-loop $d$ log
  representation for Feynman integrals}},
  \href{https://doi.org/10.1007/JHEP05(2021)052}{\emph{JHEP} {\bfseries 05}
  (2021) 052} [\href{https://arxiv.org/abs/2012.13094}{{2012.13094}}].

\bibitem{Goncharov:2010jf}
A.B.~Goncharov, M.~Spradlin, C.~Vergu and A.~Volovich, \emph{{Classical
  Polylogarithms for Amplitudes and Wilson Loops}},
  \href{https://doi.org/10.1103/PhysRevLett.105.151605}{\emph{Phys. Rev. Lett.}
  {\bfseries 105} (2010) 151605}
  [\href{https://arxiv.org/abs/1006.5703}{{1006.5703}}].

\bibitem{Duhr:2011zq}
C.~Duhr, H.~Gangl and J.R.~Rhodes, \emph{{From polygons and symbols to
  polylogarithmic functions}},
  \href{https://doi.org/10.1007/JHEP10(2012)075}{\emph{JHEP} {\bfseries 10}
  (2012) 075} [\href{https://arxiv.org/abs/1110.0458}{{1110.0458}}].

\bibitem{Dixon:2011pw}
L.J.~Dixon, J.M.~Drummond and J.M.~Henn, \emph{{Bootstrapping the three-loop
  hexagon}}, \href{https://doi.org/10.1007/JHEP11(2011)023}{\emph{JHEP}
  {\bfseries 11} (2011) 023}
  [\href{https://arxiv.org/abs/1108.4461}{{1108.4461}}].

\bibitem{Dixon:2011nj}
L.J.~Dixon, J.M.~Drummond and J.M.~Henn, \emph{{Analytic result for the
  two-loop six-point NMHV amplitude in N=4 super Yang-Mills theory}},
  \href{https://doi.org/10.1007/JHEP01(2012)024}{\emph{JHEP} {\bfseries 01}
  (2012) 024} [\href{https://arxiv.org/abs/1111.1704}{{1111.1704}}].

\bibitem{Brandhuber:2012vm}
A.~Brandhuber, G.~Travaglini and G.~Yang, \emph{{Analytic two-loop form factors
  in N=4 SYM}}, \href{https://doi.org/10.1007/JHEP05(2012)082}{\emph{JHEP}
  {\bfseries 05} (2012) 082}
  [\href{https://arxiv.org/abs/1201.4170}{{1201.4170}}].

\bibitem{Caron-Huot:2016owq}
S.~Caron-Huot, L.J.~Dixon, A.~McLeod and M.~von Hippel, \emph{{Bootstrapping a
  Five-Loop Amplitude Using Steinmann Relations}},
  \href{https://doi.org/10.1103/PhysRevLett.117.241601}{\emph{Phys. Rev. Lett.}
  {\bfseries 117} (2016) 241601}
  [\href{https://arxiv.org/abs/1609.00669}{{1609.00669}}].

\bibitem{Almelid:2017qju}
O.~Almelid, C.~Duhr, E.~Gardi, A.~McLeod and C.D.~White, \emph{{Bootstrapping
  the QCD soft anomalous dimension}},
  \href{https://doi.org/10.1007/JHEP09(2017)073}{\emph{JHEP} {\bfseries 09}
  (2017) 073} [\href{https://arxiv.org/abs/1706.10162}{{1706.10162}}].

\bibitem{Henn:2018cdp}
J.~Henn, E.~Herrmann and J.~Parra-Martinez, \emph{{Bootstrapping two-loop
  Feynman integrals for planar $ \mathcal{N}=4 $ sYM}},
  \href{https://doi.org/10.1007/JHEP10(2018)059}{\emph{JHEP} {\bfseries 10}
  (2018) 059} [\href{https://arxiv.org/abs/1806.06072}{{1806.06072}}].

\bibitem{Caron-Huot:2019vjl}
S.~Caron-Huot, L.J.~Dixon, F.~Dulat, M.~von Hippel, A.J.~McLeod and
  G.~Papathanasiou, \emph{{Six-Gluon amplitudes in planar $ \mathcal{N} $ = 4
  super-Yang-Mills theory at six and seven loops}},
  \href{https://doi.org/10.1007/JHEP08(2019)016}{\emph{JHEP} {\bfseries 08}
  (2019) 016} [\href{https://arxiv.org/abs/1903.10890}{{1903.10890}}].

\bibitem{Dixon:2020bbt}
L.J.~Dixon, A.J.~McLeod and M.~Wilhelm, \emph{{A Three-Point Form Factor
  Through Five Loops}},
  \href{https://doi.org/10.1007/JHEP04(2021)147}{\emph{JHEP} {\bfseries 04}
  (2021) 147} [\href{https://arxiv.org/abs/2012.12286}{{2012.12286}}].

\bibitem{Guo:2021bym}
Y.~Guo, L.~Wang and G.~Yang, \emph{{Bootstrapping a Two-Loop Four-Point Form
  Factor}}, \href{https://doi.org/10.1103/PhysRevLett.127.151602}{\emph{Phys.
  Rev. Lett.} {\bfseries 127} (2021) 151602}
  [\href{https://arxiv.org/abs/2106.01374}{{2106.01374}}].

\bibitem{Dixon:2022rse}
L.J.~Dixon, O.~G\"urdo\u{g}an, A.J.~McLeod and M.~Wilhelm, \emph{{Bootstrapping
  a stress-tensor form factor through eight loops}},
  \href{https://doi.org/10.1007/JHEP07(2022)153}{\emph{JHEP} {\bfseries 07}
  (2022) 153} [\href{https://arxiv.org/abs/2204.11901}{{2204.11901}}].

\bibitem{Dixon:2022xqh}
L.J.~Dixon, O.~G\"urdo\u{g}an, Y.-T.~Liu, A.J.~McLeod and M.~Wilhelm,
  \emph{{Antipodal Self-Duality for a Four-Particle Form Factor}},
  \href{https://doi.org/10.1103/PhysRevLett.130.111601}{\emph{Phys. Rev. Lett.}
  {\bfseries 130} (2023) 111601}
  [\href{https://arxiv.org/abs/2212.02410}{{2212.02410}}].

\bibitem{Hannesdottir:2024hke}
H.S.~Hannesdottir, A.J.~McLeod, M.D.~Schwartz and C.~Vergu, \emph{{Applications
  of the Landau bootstrap}},
  \href{https://doi.org/10.1103/PhysRevD.111.085003}{\emph{Phys. Rev. D}
  {\bfseries 111} (2025) 085003}
  [\href{https://arxiv.org/abs/2410.02424}{{2410.02424}}].

\bibitem{Basso:2024hlx}
B.~Basso, L.J.~Dixon and A.G.~Tumanov, \emph{{The three-point form factor of Tr
  \ensuremath{\phi}$^{3}$ to six loops}},
  \href{https://doi.org/10.1007/JHEP02(2025)034}{\emph{JHEP} {\bfseries 02}
  (2025) 034} [\href{https://arxiv.org/abs/2410.22402}{{2410.22402}}].

\bibitem{Caron-Huot:2020bkp}
{S. Caron-Huot, L.J. Dixon, J.M. Drummond, F. Dulat, J. Foster, \"O.
  G\"urdo\u{g}an, M. von Hippel, A.J. McLeod and G. Papathanasiou}, \emph{{The
  Steinmann Cluster Bootstrap for $N$ = 4 Super Yang-Mills Amplitudes}},
  \href{https://doi.org/10.22323/1.376.0003}{\emph{PoS} {\bfseries CORFU2019}
  (2020) 003} [\href{https://arxiv.org/abs/2005.06735}{{2005.06735}}].

\bibitem{Arkani-Hamed:2022rwr}
N.~Arkani-Hamed, L.J.~Dixon, A.J.~McLeod, M.~Spradlin, J.~Trnka and
  A.~Volovich, \emph{{Solving Scattering in $N$ = 4 Super-Yang-Mills Theory}},
  \href{https://arxiv.org/abs/2207.10636}{{2207.10636}}.

\bibitem{Pretorius:2005gq}
F.~Pretorius, \emph{{Evolution of binary black hole spacetimes}},
  \href{https://doi.org/10.1103/PhysRevLett.95.121101}{\emph{Phys. Rev. Lett.}
  {\bfseries 95} (2005) 121101}
  [\href{https://arxiv.org/abs/gr-qc/0507014}{{gr-qc/0507014}}].

\bibitem{Campanelli:2005dd}
M.~Campanelli, C.O.~Lousto, P.~Marronetti and Y.~Zlochower, \emph{{Accurate
  evolutions of orbiting black-hole binaries without excision}},
  \href{https://doi.org/10.1103/PhysRevLett.96.111101}{\emph{Phys. Rev. Lett.}
  {\bfseries 96} (2006) 111101}
  [\href{https://arxiv.org/abs/gr-qc/0511048}{{gr-qc/0511048}}].

\bibitem{Baker:2005vv}
J.G.~Baker, J.~Centrella, D.-I.~Choi, M.~Koppitz and J.~van Meter,
  \emph{{Gravitational wave extraction from an inspiraling configuration of
  merging black holes}},
  \href{https://doi.org/10.1103/PhysRevLett.96.111102}{\emph{Phys. Rev. Lett.}
  {\bfseries 96} (2006) 111102}
  [\href{https://arxiv.org/abs/gr-qc/0511103}{{gr-qc/0511103}}].

\bibitem{Goldberger:2004jt}
W.D.~Goldberger and I.Z.~Rothstein, \emph{{An Effective field theory of gravity
  for extended objects}},
  \href{https://doi.org/10.1103/PhysRevD.73.104029}{\emph{Phys. Rev. D}
  {\bfseries 73} (2006) 104029}
  [\href{https://arxiv.org/abs/hep-th/0409156}{{hep-th/0409156}}].

\bibitem{Blanchet:2013haa}
L.~Blanchet, \emph{{Gravitational Radiation from Post-Newtonian Sources and
  Inspiralling Compact Binaries}},
  \href{https://doi.org/10.12942/lrr-2014-2}{\emph{Living Rev. Rel.} {\bfseries
  17} (2014) 2} [\href{https://arxiv.org/abs/1310.1528}{{1310.1528}}].

\bibitem{Levi:2018nxp}
M.~Levi, \emph{{Effective Field Theories of Post-Newtonian Gravity: A
  comprehensive review}},
  \href{https://doi.org/10.1088/1361-6633/ab12bc}{\emph{Rept. Prog. Phys.}
  {\bfseries 83} (2020) 075901}
  [\href{https://arxiv.org/abs/1807.01699}{{1807.01699}}].

\bibitem{Damour:2016gwp}
T.~Damour, \emph{{Gravitational scattering, post-Minkowskian approximation and
  Effective One-Body theory}},
  \href{https://doi.org/10.1103/PhysRevD.94.104015}{\emph{Phys. Rev. D}
  {\bfseries 94} (2016) 104015}
  [\href{https://arxiv.org/abs/1609.00354}{{1609.00354}}].

\bibitem{Mino:1996nk}
Y.~Mino, M.~Sasaki and T.~Tanaka, \emph{{Gravitational radiation reaction to a
  particle motion}},
  \href{https://doi.org/10.1103/PhysRevD.55.3457}{\emph{Phys. Rev. D}
  {\bfseries 55} (1997) 3457}
  [\href{https://arxiv.org/abs/gr-qc/9606018}{{gr-qc/9606018}}].

\bibitem{Quinn:1996am}
T.C.~Quinn and R.M.~Wald, \emph{{An Axiomatic approach to electromagnetic and
  gravitational radiation reaction of particles in curved space-time}},
  \href{https://doi.org/10.1103/PhysRevD.56.3381}{\emph{Phys. Rev. D}
  {\bfseries 56} (1997) 3381}
  [\href{https://arxiv.org/abs/gr-qc/9610053}{{gr-qc/9610053}}].

\bibitem{Poisson:2011nh}
E.~Poisson, A.~Pound and I.~Vega, \emph{{The Motion of point particles in
  curved spacetime}}, \href{https://doi.org/10.12942/lrr-2011-7}{\emph{Living
  Rev. Rel.} {\bfseries 14} (2011) 7}
  [\href{https://arxiv.org/abs/1102.0529}{{1102.0529}}].

\bibitem{Barack:2018yvs}
L.~Barack and A.~Pound, \emph{{Self-force and radiation reaction in general
  relativity}}, \href{https://doi.org/10.1088/1361-6633/aae552}{\emph{Rept.
  Prog. Phys.} {\bfseries 82} (2019) 016904}
  [\href{https://arxiv.org/abs/1805.10385}{{1805.10385}}].

\bibitem{Kalin:2019rwq}
G.~K\"alin and R.A.~Porto, \emph{{From Boundary Data to Bound States}},
  \href{https://doi.org/10.1007/JHEP01(2020)072}{\emph{JHEP} {\bfseries 01}
  (2020) 072} [\href{https://arxiv.org/abs/1910.03008}{{1910.03008}}].

\bibitem{Kalin:2019inp}
G.~K\"alin and R.A.~Porto, \emph{{From boundary data to bound states. Part II.
  Scattering angle to dynamical invariants (with twist)}},
  \href{https://doi.org/10.1007/JHEP02(2020)120}{\emph{JHEP} {\bfseries 02}
  (2020) 120} [\href{https://arxiv.org/abs/1911.09130}{{1911.09130}}].

\bibitem{Cho:2021arx}
G.~Cho, G.~K\"alin and R.A.~Porto, \emph{{From boundary data to bound states.
  Part III. Radiative effects}},
  \href{https://doi.org/10.1007/JHEP04(2022)154}{\emph{JHEP} {\bfseries 04}
  (2022) 154} [\href{https://arxiv.org/abs/2112.03976}{{2112.03976}}].

\bibitem{Dlapa:2024cje}
C.~Dlapa, G.~K\"alin, Z.~Liu and R.A.~Porto, \emph{{Local in Time Conservative
  Binary Dynamics at Fourth Post-Minkowskian Order}},
  \href{https://doi.org/10.1103/PhysRevLett.132.221401}{\emph{Phys. Rev. Lett.}
  {\bfseries 132} (2024) 221401}
  [\href{https://arxiv.org/abs/2403.04853}{{2403.04853}}].

\bibitem{Bern:2019nnu}
Z.~Bern, C.~Cheung, R.~Roiban, C.-H.~Shen, M.P.~Solon and M.~Zeng,
  \emph{{Scattering Amplitudes and the Conservative Hamiltonian for Binary
  Systems at Third Post-Minkowskian Order}},
  \href{https://doi.org/10.1103/PhysRevLett.122.201603}{\emph{Phys. Rev. Lett.}
  {\bfseries 122} (2019) 201603}
  [\href{https://arxiv.org/abs/1901.04424}{{1901.04424}}].

\bibitem{Bern:2019crd}
Z.~Bern, C.~Cheung, R.~Roiban, C.-H.~Shen, M.P.~Solon and M.~Zeng, \emph{{Black
  Hole Binary Dynamics from the Double Copy and Effective Theory}},
  \href{https://doi.org/10.1007/JHEP10(2019)206}{\emph{JHEP} {\bfseries 10}
  (2019) 206} [\href{https://arxiv.org/abs/1908.01493}{{1908.01493}}].

\bibitem{Kalin:2020mvi}
G.~K\"alin and R.A.~Porto, \emph{{Post-Minkowskian Effective Field Theory for
  Conservative Binary Dynamics}},
  \href{https://doi.org/10.1007/JHEP11(2020)106}{\emph{JHEP} {\bfseries 11}
  (2020) 106} [\href{https://arxiv.org/abs/2006.01184}{{2006.01184}}].

\bibitem{Mogull:2020sak}
G.~Mogull, J.~Plefka and J.~Steinhoff, \emph{{Classical black hole scattering
  from a worldline quantum field theory}},
  \href{https://doi.org/10.1007/JHEP02(2021)048}{\emph{JHEP} {\bfseries 02}
  (2021) 048} [\href{https://arxiv.org/abs/2010.02865}{{2010.02865}}].

\bibitem{Bjerrum-Bohr:2022blt}
N.E.J.~Bjerrum-Bohr, P.H.~Damgaard, L.~Plante and P.~Vanhove, \emph{{The SAGEX
  review on scattering amplitudes Chapter 13: Post-Minkowskian expansion from
  scattering amplitudes}},
  \href{https://doi.org/10.1088/1751-8121/ac7a78}{\emph{J. Phys. A} {\bfseries
  55} (2022) 443014} [\href{https://arxiv.org/abs/2203.13024}{{2203.13024}}].

\bibitem{Kotikov:1990kg}
A.V.~Kotikov, \emph{{Differential equations method: New technique for massive
  Feynman diagrams calculation}},
  \href{https://doi.org/10.1016/0370-2693(91)90413-K}{\emph{Phys. Lett. B}
  {\bfseries 254} (1991) 158}.

\bibitem{Cachazo:2008vp}
F.~Cachazo, \emph{{Sharpening The Leading Singularity}},
  \href{https://arxiv.org/abs/0803.1988}{{0803.1988}}.

\bibitem{Arkani-Hamed:2010pyv}
N.~Arkani-Hamed, J.L.~Bourjaily, F.~Cachazo and J.~Trnka, \emph{{Local
  Integrals for Planar Scattering Amplitudes}},
  \href{https://doi.org/10.1007/JHEP06(2012)125}{\emph{JHEP} {\bfseries 06}
  (2012) 125} [\href{https://arxiv.org/abs/1012.6032}{{1012.6032}}].

\bibitem{Henn:2013pwa}
J.M.~Henn, \emph{{Multiloop integrals in dimensional regularization made
  simple}}, \href{https://doi.org/10.1103/PhysRevLett.110.251601}{\emph{Phys.
  Rev. Lett.} {\bfseries 110} (2013) 251601}
  [\href{https://arxiv.org/abs/1304.1806}{{1304.1806}}].

\bibitem{Tan12}
T.~Tantau, \emph{Graph drawing in {TikZ}},  in \emph{Proceedings of the 20th
  International Conference on Graph Drawing}, GD'12, pp.~517--528,
  Springer-Verlag, 2013,
  \href{https://doi.org/10.1007/978-3-642-36763-2_46}{DOI}.

\bibitem{Tarasov:1996br}
O.V.~Tarasov, \emph{{Connection between Feynman integrals having different
  values of the space-time dimension}},
  \href{https://doi.org/10.1103/PhysRevD.54.6479}{\emph{Phys. Rev. D}
  {\bfseries 54} (1996) 6479}
  [\href{https://arxiv.org/abs/hep-th/9606018}{{hep-th/9606018}}].

\bibitem{Tarasov:1997kx}
O.V.~Tarasov, \emph{{Generalized recurrence relations for two loop propagator
  integrals with arbitrary masses}},
  \href{https://doi.org/10.1016/S0550-3213(97)00376-3}{\emph{Nucl. Phys. B}
  {\bfseries 502} (1997) 455}
  [\href{https://arxiv.org/abs/hep-ph/9703319}{{hep-ph/9703319}}].

\bibitem{Passarino:1978jh}
G.~Passarino and M.J.G.~Veltman, \emph{{One Loop Corrections for e+ e-
  Annihilation Into mu+ mu- in the Weinberg Model}},
  \href{https://doi.org/10.1016/0550-3213(79)90234-7}{\emph{Nucl. Phys. B}
  {\bfseries 160} (1979) 151}.

\bibitem{Baikov:1996iu}
P.A.~Baikov, \emph{{Explicit solutions of the multiloop integral recurrence
  relations and its application}},
  \href{https://doi.org/10.1016/S0168-9002(97)00126-5}{\emph{Nucl. Instrum.
  Meth. A} {\bfseries 389} (1997) 347}
  [\href{https://arxiv.org/abs/hep-ph/9611449}{{hep-ph/9611449}}].

\bibitem{Frellesvig:2017aai}
H.~Frellesvig and C.G.~Papadopoulos, \emph{{Cuts of Feynman Integrals in Baikov
  representation}}, \href{https://doi.org/10.1007/JHEP04(2017)083}{\emph{JHEP}
  {\bfseries 04} (2017) 083}
  [\href{https://arxiv.org/abs/1701.07356}{{1701.07356}}].

\bibitem{Chetyrkin:1981qh}
K.G.~Chetyrkin and F.V.~Tkachov, \emph{{Integration by parts: The algorithm to
  calculate $\beta$-functions in 4 loops}},
  \href{https://doi.org/10.1016/0550-3213(81)90199-1}{\emph{Nucl. Phys. B}
  {\bfseries 192} (1981) 159}.

\bibitem{Smirnov:2010hn}
A.V.~Smirnov and A.V.~Petukhov, \emph{{The Number of Master Integrals is
  Finite}}, \href{https://doi.org/10.1007/s11005-010-0450-0}{\emph{Lett. Math.
  Phys.} {\bfseries 97} (2011) 37}
  [\href{https://arxiv.org/abs/1004.4199}{{1004.4199}}].

\bibitem{Bitoun:2017nre}
T.~Bitoun, C.~Bogner, R.P.~Klausen and E.~Panzer, \emph{{Feynman integral
  relations from parametric annihilators}},
  \href{https://doi.org/10.1007/s11005-018-1114-8}{\emph{Lett. Math. Phys.}
  {\bfseries 109} (2019) 497}
  [\href{https://arxiv.org/abs/1712.09215}{{1712.09215}}].

\bibitem{Bourjaily:2018aeq}
J.L.~Bourjaily, A.J.~McLeod, M.~von Hippel and M.~Wilhelm, \emph{{Rationalizing
  Loop Integration}},
  \href{https://doi.org/10.1007/JHEP08(2018)184}{\emph{JHEP} {\bfseries 08}
  (2018) 184} [\href{https://arxiv.org/abs/1805.10281}{{1805.10281}}].

\bibitem{Britto:2024mna}
R.~Britto, C.~Duhr, H.S.~Hannesdottir and S.~Mizera, \emph{{Cutting-Edge Tools
  for Cutting Edges}},  in \emph{Encyclopedia of Mathematical Physics (Second
  Edition)}, pp.~595--620, Academic Press, 2025,
  \href{https://doi.org/10.1016/B978-0-323-95703-8.00097-5}{DOI}
  [\href{https://arxiv.org/abs/2402.19415}{{2402.19415}}].

\bibitem{Hubsch:1992nu}
T.~Hubsch, \emph{{Calabi-Yau manifolds: A Bestiary for physicists}}, World
  Scientific (1994), \href{https://doi.org/10.1142/1410}{10.1142/1410}.

\bibitem{Bourjaily:2019hmc}
J.L.~Bourjaily, A.J.~McLeod, C.~Vergu, M.~Volk, M.~Von~Hippel and M.~Wilhelm,
  \emph{{Embedding Feynman Integral (Calabi-Yau) Geometries in Weighted
  Projective Space}},
  \href{https://doi.org/10.1007/JHEP01(2020)078}{\emph{JHEP} {\bfseries 01}
  (2020) 078} [\href{https://arxiv.org/abs/1910.01534}{{1910.01534}}].

\bibitem{Asymptote}
A.~Hammerlindl, J.~Bowman and T.~Prince, ``{Asymptote: The Vector Graphics
  Language}.'' \url{https://asymptote.sourceforge.io/}.

\bibitem{Schimmrigk:2024xid}
R.~Schimmrigk, \emph{{Special Fano geometry from Feynman integrals}},
  \href{https://doi.org/10.1016/j.physletb.2025.139420}{\emph{Phys. Lett. B}
  {\bfseries 864} (2025) 139420}
  [\href{https://arxiv.org/abs/2412.20236}{{2412.20236}}].

\bibitem{Papathanasiou:2025stn}
G.~Papathanasiou, S.~Weinzierl, K.~Wu and Y.~Zhang, \emph{{Rationalisation of
  multiple square roots in Feynman integrals}},
  \href{https://doi.org/10.1007/JHEP05(2025)078}{\emph{JHEP} {\bfseries 05}
  (2025) 078} [\href{https://arxiv.org/abs/2501.07490}{{2501.07490}}].

\bibitem{Parra-Martinez:2020dzs}
J.~Parra-Martinez, M.S.~Ruf and M.~Zeng, \emph{{Extremal black hole scattering
  at $\mathcal{O}(G^3)$: graviton dominance, eikonal exponentiation, and
  differential equations}},
  \href{https://doi.org/10.1007/JHEP11(2020)023}{\emph{JHEP} {\bfseries 11}
  (2020) 023} [\href{https://arxiv.org/abs/2005.04236}{{2005.04236}}].

\bibitem{Bonciani:2010ms}
R.~Bonciani, G.~Degrassi and A.~Vicini, \emph{{On the Generalized Harmonic
  Polylogarithms of One Complex Variable}},
  \href{https://doi.org/10.1016/j.cpc.2011.02.011}{\emph{Comput. Phys. Commun.}
  {\bfseries 182} (2011) 1253}
  [\href{https://arxiv.org/abs/1007.1891}{{1007.1891}}].

\bibitem{Adams:2018yfj}
L.~Adams and S.~Weinzierl, \emph{{The $\varepsilon$-form of the differential
  equations for Feynman integrals in the elliptic case}},
  \href{https://doi.org/10.1016/j.physletb.2018.04.002}{\emph{Phys. Lett. B}
  {\bfseries 781} (2018) 270}
  [\href{https://arxiv.org/abs/1802.05020}{{1802.05020}}].

\bibitem{Besier:2018jen}
M.~Besier, D.~Van~Straten and S.~Weinzierl, \emph{{Rationalizing roots: an
  algorithmic approach}},
  \href{https://doi.org/10.4310/CNTP.2019.v13.n2.a1}{\emph{Commun. Num. Theor.
  Phys.} {\bfseries 13} (2019) 253}
  [\href{https://arxiv.org/abs/1809.10983}{{1809.10983}}].

\bibitem{Besier:2019kco}
M.~Besier, P.~Wasser and S.~Weinzierl, \emph{{RationalizeRoots: Software
  Package for the Rationalization of Square Roots}},
  \href{https://doi.org/10.1016/j.cpc.2020.107197}{\emph{Comput. Phys. Commun.}
  {\bfseries 253} (2020) 107197}
  [\href{https://arxiv.org/abs/1910.13251}{{1910.13251}}].

\bibitem{Duhr:2014woa}
C.~Duhr, \emph{{Mathematical aspects of scattering amplitudes}},  in
  \emph{{Theoretical Advanced Study Institute in Elementary Particle Physics}:
  {Journeys Through the Precision Frontier: Amplitudes for Colliders}},
  pp.~419--476, 2015, \href{https://doi.org/10.1142/9789814678766_0010}{DOI}
  [\href{https://arxiv.org/abs/1411.7538}{{1411.7538}}].

\bibitem{Duhr:2019tlz}
C.~Duhr and F.~Dulat, \emph{{PolyLogTools \textemdash{} polylogs for the
  masses}}, \href{https://doi.org/10.1007/JHEP08(2019)135}{\emph{JHEP}
  {\bfseries 08} (2019) 135}
  [\href{https://arxiv.org/abs/1904.07279}{{1904.07279}}].

\bibitem{Goncharov:2005sla}
A.B.~Goncharov, \emph{{Galois symmetries of fundamental groupoids and
  noncommutative geometry}},
  \href{https://doi.org/10.1215/S0012-7094-04-12822-2}{\emph{Duke Math. J.}
  {\bfseries 128} (2005) 209}
  [\href{https://arxiv.org/abs/math/0208144}{{math/0208144}}].

\bibitem{DelDuca:2010zg}
V.~Del~Duca, C.~Duhr and V.A.~Smirnov, \emph{{The Two-Loop Hexagon Wilson Loop
  in N = 4 SYM}}, \href{https://doi.org/10.1007/JHEP05(2010)084}{\emph{JHEP}
  {\bfseries 05} (2010) 084}
  [\href{https://arxiv.org/abs/1003.1702}{{1003.1702}}].

\bibitem{Brown:2008um}
F.~Brown, \emph{{The Massless higher-loop two-point function}},
  \href{https://doi.org/10.1007/s00220-009-0740-5}{\emph{Commun. Math. Phys.}
  {\bfseries 287} (2009) 925}
  [\href{https://arxiv.org/abs/0804.1660}{{0804.1660}}].

\bibitem{Brown:2009ta}
F.C.S.~Brown, \emph{{On the periods of some Feynman integrals}},
  \href{https://arxiv.org/abs/0910.0114}{{0910.0114}}.

\bibitem{Panzer:2014gra}
E.~Panzer, \emph{{On hyperlogarithms and Feynman integrals with divergences and
  many scales}}, \href{https://doi.org/10.1007/JHEP03(2014)071}{\emph{JHEP}
  {\bfseries 03} (2014) 071}
  [\href{https://arxiv.org/abs/1401.4361}{{1401.4361}}].

\bibitem{DiVecchia:2021bdo}
P.~Di~Vecchia, C.~Heissenberg, R.~Russo and G.~Veneziano, \emph{{The eikonal
  approach to gravitational scattering and radiation at $ \mathcal{O}
  $(G$^{3}$)}}, \href{https://doi.org/10.1007/JHEP07(2021)169}{\emph{JHEP}
  {\bfseries 07} (2021) 169}
  [\href{https://arxiv.org/abs/2104.03256}{{2104.03256}}].

\bibitem{Herrmann:2021tct}
E.~Herrmann, J.~Parra-Martinez, M.S.~Ruf and M.~Zeng, \emph{{Radiative
  classical gravitational observables at $ \mathcal{O} (G^{3})$ from scattering
  amplitudes}}, \href{https://doi.org/10.1007/JHEP10(2021)148}{\emph{JHEP}
  {\bfseries 10} (2021) 148}
  [\href{https://arxiv.org/abs/2104.03957}{{2104.03957}}].

\bibitem{Lee:2014ioa}
R.N.~Lee, \emph{{Reducing differential equations for multiloop master
  integrals}}, \href{https://doi.org/10.1007/JHEP04(2015)108}{\emph{JHEP}
  {\bfseries 04} (2015) 108}
  [\href{https://arxiv.org/abs/1411.0911}{{1411.0911}}].

\bibitem{Prausa:2017ltv}
M.~Prausa, \emph{{epsilon: A tool to find a canonical basis of master
  integrals}}, \href{https://doi.org/10.1016/j.cpc.2017.05.026}{\emph{Comput.
  Phys. Commun.} {\bfseries 219} (2017) 361}
  [\href{https://arxiv.org/abs/1701.00725}{{1701.00725}}].

\bibitem{Gituliar:2017vzm}
O.~Gituliar and V.~Magerya, \emph{{Fuchsia: a tool for reducing differential
  equations for Feynman master integrals to epsilon form}},
  \href{https://doi.org/10.1016/j.cpc.2017.05.004}{\emph{Comput. Phys. Commun.}
  {\bfseries 219} (2017) 329}
  [\href{https://arxiv.org/abs/1701.04269}{{1701.04269}}].

\bibitem{Meyer:2017joq}
C.~Meyer, \emph{{Algorithmic transformation of multi-loop master integrals to a
  canonical basis with CANONICA}},
  \href{https://doi.org/10.1016/j.cpc.2017.09.014}{\emph{Comput. Phys. Commun.}
  {\bfseries 222} (2018) 295}
  [\href{https://arxiv.org/abs/1705.06252}{{1705.06252}}].

\bibitem{Dlapa:2020cwj}
C.~Dlapa, J.~Henn and K.~Yan, \emph{{Deriving canonical differential equations
  for Feynman integrals from a single uniform weight integral}},
  \href{https://doi.org/10.1007/JHEP05(2020)025}{\emph{JHEP} {\bfseries 05}
  (2020) 025} [\href{https://arxiv.org/abs/2002.02340}{{2002.02340}}].

\bibitem{Lee:2020zfb}
R.N.~Lee, \emph{{Libra: A package for transformation of differential systems
  for multiloop integrals}},
  \href{https://doi.org/10.1016/j.cpc.2021.108058}{\emph{Comput. Phys. Commun.}
  {\bfseries 267} (2021) 108058}
  [\href{https://arxiv.org/abs/2012.00279}{{2012.00279}}].

\bibitem{Jiang:2023jmk}
X.~Jiang, X.~Wang, L.L.~Yang and J.~Zhao,
  \emph{{\ensuremath{\varepsilon}-factorized differential equations for
  two-loop non-planar triangle Feynman integrals with elliptic curves}},
  \href{https://doi.org/10.1007/JHEP09(2023)187}{\emph{JHEP} {\bfseries 09}
  (2023) 187} [\href{https://arxiv.org/abs/2305.13951}{{2305.13951}}].

\bibitem{Gorges:2023zgv}
L.~G\"orges, C.~Nega, L.~Tancredi and F.J.~Wagner, \emph{{On a procedure to
  derive \ensuremath{\epsilon}-factorised differential equations beyond
  polylogarithms}}, \href{https://doi.org/10.1007/JHEP07(2023)206}{\emph{JHEP}
  {\bfseries 07} (2023) 206}
  [\href{https://arxiv.org/abs/2305.14090}{{2305.14090}}].

\bibitem{Pogel:2022ken}
S.~P\"ogel, X.~Wang and S.~Weinzierl, \emph{{Taming Calabi-Yau Feynman
  Integrals: The Four-Loop Equal-Mass Banana Integral}},
  \href{https://doi.org/10.1103/PhysRevLett.130.101601}{\emph{Phys. Rev. Lett.}
  {\bfseries 130} (2023) 101601}
  [\href{https://arxiv.org/abs/2211.04292}{{2211.04292}}].

\bibitem{Duhr:2025lbz}
C.~Duhr, S.~Maggio, C.~Nega, B.~Sauer, L.~Tancredi and F.J.~Wagner,
  \emph{{Aspects of canonical differential equations for Calabi-Yau geometries
  and beyond}},  \href{https://arxiv.org/abs/2503.20655}{{2503.20655}}.

\bibitem{Muller-Stach:2012tgj}
S.~M\"uller-Stach, S.~Weinzierl and R.~Zayadeh, \emph{{Picard-Fuchs equations
  for Feynman integrals}},
  \href{https://doi.org/10.1007/s00220-013-1838-3}{\emph{Commun. Math. Phys.}
  {\bfseries 326} (2014) 237}
  [\href{https://arxiv.org/abs/1212.4389}{{1212.4389}}].

\bibitem{Adams:2017tga}
L.~Adams, E.~Chaubey and S.~Weinzierl, \emph{{Simplifying Differential
  Equations for Multiscale Feynman Integrals beyond Multiple Polylogarithms}},
  \href{https://doi.org/10.1103/PhysRevLett.118.141602}{\emph{Phys. Rev. Lett.}
  {\bfseries 118} (2017) 141602}
  [\href{https://arxiv.org/abs/1702.04279}{{1702.04279}}].

\bibitem{Duhr:2020gdd}
C.~Duhr and F.~Brown, \emph{{A double integral of dlog forms which is not
  polylogarithmic}}, \href{https://doi.org/10.22323/1.383.0005}{\emph{PoS}
  {\bfseries MA2019} (2022) 005}
  [\href{https://arxiv.org/abs/2006.09413}{{2006.09413}}].

\bibitem{Jockers:2024tpc}
{H. Jockers, S. Kotlewski, P. Kuusela, A.J. McLeod, S. P\"ogel, M. Sarve, X.
  Wang and S.~Weinzierl}, \emph{{A Calabi-Yau-to-curve correspondence for
  Feynman integrals}},
  \href{https://doi.org/10.1007/JHEP01(2025)030}{\emph{JHEP} {\bfseries 25}
  (2020) 030} [\href{https://arxiv.org/abs/2404.05785}{{2404.05785}}].

\bibitem{Laporta:2000dsw}
S.~Laporta, \emph{{High precision calculation of multiloop Feynman integrals by
  difference equations}},
  \href{https://doi.org/10.1142/S0217751X00002159}{\emph{Int. J. Mod. Phys. A}
  {\bfseries 15} (2000) 5087}
  [\href{https://arxiv.org/abs/hep-ph/0102033}{{hep-ph/0102033}}].

\bibitem{Smirnov:2023yhb}
A.V.~Smirnov and M.~Zeng, \emph{{FIRE 6.5: Feynman integral reduction with new
  simplification library}},
  \href{https://doi.org/10.1016/j.cpc.2024.109261}{\emph{Comput. Phys. Commun.}
  {\bfseries 302} (2024) 109261}
  [\href{https://arxiv.org/abs/2311.02370}{{2311.02370}}].

\bibitem{vonManteuffel:2012np}
A.~von Manteuffel and C.~Studerus, \emph{{Reduze 2 - Distributed Feynman
  Integral Reduction}},  \href{https://arxiv.org/abs/1201.4330}{{1201.4330}}.

\bibitem{Lee:2013mka}
R.N.~Lee, \emph{{LiteRed 1.4: a powerful tool for reduction of multiloop
  integrals}}, \href{https://doi.org/10.1088/1742-6596/523/1/012059}{\emph{J.
  Phys. Conf. Ser.} {\bfseries 523} (2014) 012059}
  [\href{https://arxiv.org/abs/1310.1145}{{1310.1145}}].

\bibitem{Klappert:2020nbg}
J.~Klappert, F.~Lange, P.~Maierh\"ofer and J.~Usovitsch, \emph{{Integral
  reduction with Kira 2.0 and finite field methods}},
  \href{https://doi.org/10.1016/j.cpc.2021.108024}{\emph{Comput. Phys. Commun.}
  {\bfseries 266} (2021) 108024}
  [\href{https://arxiv.org/abs/2008.06494}{{2008.06494}}].

\bibitem{Mastrolia:2018uzb}
P.~Mastrolia and S.~Mizera, \emph{{Feynman Integrals and Intersection Theory}},
  \href{https://doi.org/10.1007/JHEP02(2019)139}{\emph{JHEP} {\bfseries 02}
  (2019) 139} [\href{https://arxiv.org/abs/1810.03818}{{1810.03818}}].

\bibitem{Frellesvig:2019uqt}
H.~Frellesvig, F.~Gasparotto, M.K.~Mandal, P.~Mastrolia, L.~Mattiazzi and
  S.~Mizera, \emph{{Vector Space of Feynman Integrals and Multivariate
  Intersection Numbers}},
  \href{https://doi.org/10.1103/PhysRevLett.123.201602}{\emph{Phys. Rev. Lett.}
  {\bfseries 123} (2019) 201602}
  [\href{https://arxiv.org/abs/1907.02000}{{1907.02000}}].

\bibitem{delaCruz:2024xit}
L.~de~la Cruz and P.~Vanhove, \emph{{Algorithm for differential equations for
  Feynman integrals in general dimensions}},
  \href{https://doi.org/10.1007/s11005-024-01832-w}{\emph{Lett. Math. Phys.}
  {\bfseries 114} (2024) 89}
  [\href{https://arxiv.org/abs/2401.09908}{{2401.09908}}].

\bibitem{Primo:2016ebd}
A.~Primo and L.~Tancredi, \emph{{On the maximal cut of Feynman integrals and
  the solution of their differential equations}},
  \href{https://doi.org/10.1016/j.nuclphysb.2016.12.021}{\emph{Nucl. Phys. B}
  {\bfseries 916} (2017) 94}
  [\href{https://arxiv.org/abs/1610.08397}{{1610.08397}}].

\bibitem{Bosma:2017ens}
J.~Bosma, M.~Sogaard and Y.~Zhang, \emph{{Maximal Cuts in Arbitrary
  Dimension}}, \href{https://doi.org/10.1007/JHEP08(2017)051}{\emph{JHEP}
  {\bfseries 08} (2017) 051}
  [\href{https://arxiv.org/abs/1704.04255}{{1704.04255}}].

\bibitem{Festi:2018qip}
D.~Festi and D.~van Straten, \emph{{Bhabha Scattering and a special pencil of
  K3 surfaces}},
  \href{https://doi.org/10.4310/CNTP.2019.v13.n2.a4}{\emph{Commun. Num. Theor.
  Phys.} {\bfseries 13} (2019) 463}
  [\href{https://arxiv.org/abs/1809.04970}{{1809.04970}}].

\bibitem{Vergu:2020uur}
C.~Vergu and M.~Volk, \emph{{Traintrack Calabi-Yaus from Twistor Geometry}},
  \href{https://doi.org/10.1007/JHEP07(2020)160}{\emph{JHEP} {\bfseries 07}
  (2020) 160} [\href{https://arxiv.org/abs/2005.08771}{{2005.08771}}].

\bibitem{Bourjaily:2020hjv}
J.L.~Bourjaily, N.~Kalyanapuram, C.~Langer, K.~Patatoukos and M.~Spradlin,
  \emph{{Elliptic, Yangian-Invariant \textquotedblleft{}Leading
  Singularity\textquotedblright{}}},
  \href{https://doi.org/10.1103/PhysRevLett.126.201601}{\emph{Phys. Rev. Lett.}
  {\bfseries 126} (2021) 201601}
  [\href{https://arxiv.org/abs/2012.14438}{{2012.14438}}].

\bibitem{Bourjaily:2021vyj}
J.L.~Bourjaily, N.~Kalyanapuram, C.~Langer and K.~Patatoukos,
  \emph{{Prescriptive unitarity with elliptic leading singularities}},
  \href{https://doi.org/10.1103/PhysRevD.104.125009}{\emph{Phys. Rev. D}
  {\bfseries 104} (2021) 125009}
  [\href{https://arxiv.org/abs/2102.02210}{{2102.02210}}].

\bibitem{Bern:1994zx}
Z.~Bern, L.J.~Dixon, D.C.~Dunbar and D.A.~Kosower, \emph{{One loop n point
  gauge theory amplitudes, unitarity and collinear limits}},
  \href{https://doi.org/10.1016/0550-3213(94)90179-1}{\emph{Nucl. Phys. B}
  {\bfseries 425} (1994) 217}
  [\href{https://arxiv.org/abs/hep-ph/9403226}{{hep-ph/9403226}}].

\bibitem{Bern:1994cg}
Z.~Bern, L.J.~Dixon, D.C.~Dunbar and D.A.~Kosower, \emph{{Fusing gauge theory
  tree amplitudes into loop amplitudes}},
  \href{https://doi.org/10.1016/0550-3213(94)00488-Z}{\emph{Nucl. Phys. B}
  {\bfseries 435} (1995) 59}
  [\href{https://arxiv.org/abs/hep-ph/9409265}{{hep-ph/9409265}}].

\bibitem{Cutkosky:1960sp}
R.E.~Cutkosky, \emph{{Singularities and discontinuities of Feynman
  amplitudes}}, \href{https://doi.org/10.1063/1.1703676}{\emph{J. Math. Phys.}
  {\bfseries 1} (1960) 429}.

\bibitem{Henn:2014qga}
J.M.~Henn, \emph{{Lectures on differential equations for Feynman integrals}},
  \href{https://doi.org/10.1088/1751-8113/48/15/153001}{\emph{J. Phys. A}
  {\bfseries 48} (2015) 153001}
  [\href{https://arxiv.org/abs/1412.2296}{{1412.2296}}].

\bibitem{WasserMSc}
P.~Wasser, \emph{Analytic properties of feynman integrals for scattering
  amplitudes}, \href{https://doi.org/10.25358/openscience-2655}{\emph{M.Sc.
  thesis, Johannes Gutenberg-Universität Mainz, Germany} (2016) }.

\bibitem{Dlapa:2021qsl}
C.~Dlapa, X.~Li and Y.~Zhang, \emph{{Leading singularities in Baikov
  representation and Feynman integrals with uniform transcendental weight}},
  \href{https://doi.org/10.1007/JHEP07(2021)227}{\emph{JHEP} {\bfseries 07}
  (2021) 227} [\href{https://arxiv.org/abs/2103.04638}{{2103.04638}}].

\bibitem{Lee:2013hzt}
R.N.~Lee and A.A.~Pomeransky, \emph{{Critical points and number of master
  integrals}}, \href{https://doi.org/10.1007/JHEP11(2013)165}{\emph{JHEP}
  {\bfseries 11} (2013) 165}
  [\href{https://arxiv.org/abs/1308.6676}{{1308.6676}}].

\bibitem{Paulos:2012nu}
M.F.~Paulos, M.~Spradlin and A.~Volovich, \emph{{Mellin Amplitudes for Dual
  Conformal Integrals}},
  \href{https://doi.org/10.1007/JHEP08(2012)072}{\emph{JHEP} {\bfseries 08}
  (2012) 072} [\href{https://arxiv.org/abs/1203.6362}{{1203.6362}}].

\bibitem{Nandan:2013ip}
D.~Nandan, M.F.~Paulos, M.~Spradlin and A.~Volovich, \emph{{Star Integrals,
  Convolutions and Simplices}},
  \href{https://doi.org/10.1007/JHEP05(2013)105}{\emph{JHEP} {\bfseries 05}
  (2013) 105} [\href{https://arxiv.org/abs/1301.2500}{{1301.2500}}].

\bibitem{Bourjaily:2017bsb}
J.L.~Bourjaily, A.J.~McLeod, M.~Spradlin, M.~von Hippel and M.~Wilhelm,
  \emph{{Elliptic Double-Box Integrals: Massless Scattering Amplitudes beyond
  Polylogarithms}},
  \href{https://doi.org/10.1103/PhysRevLett.120.121603}{\emph{Phys. Rev. Lett.}
  {\bfseries 120} (2018) 121603}
  [\href{https://arxiv.org/abs/1712.02785}{{1712.02785}}].

\bibitem{Bourjaily:2013mma}
J.L.~Bourjaily, S.~Caron-Huot and J.~Trnka, \emph{{Dual-Conformal
  Regularization of Infrared Loop Divergences and the Chiral Box Expansion}},
  \href{https://doi.org/10.1007/JHEP01(2015)001}{\emph{JHEP} {\bfseries 01}
  (2015) 001} [\href{https://arxiv.org/abs/1303.4734}{{1303.4734}}].

\bibitem{Bourjaily:2019exo}
J.L.~Bourjaily, E.~Gardi, A.J.~McLeod and C.~Vergu, \emph{{All-mass $n$-gon
  integrals in $n$ dimensions}},
  \href{https://doi.org/10.1007/JHEP08(2020)029}{\emph{JHEP} {\bfseries 08}
  (2020) 029} [\href{https://arxiv.org/abs/1912.11067}{{1912.11067}}].

\bibitem{Bourjaily:2019jrk}
J.L.~Bourjaily, F.~Dulat and E.~Panzer, \emph{{Manifestly Dual-Conformal Loop
  Integration}},
  \href{https://doi.org/10.1016/j.nuclphysb.2019.03.022}{\emph{Nucl. Phys. B}
  {\bfseries 942} (2019) 251}
  [\href{https://arxiv.org/abs/1901.02887}{{1901.02887}}].

\bibitem{Hodges1977}
A.~Hodges, \emph{{Crossing and Twistor Diagrams}},  in \emph{{Twistor
  Newsletter}}, vol.~5, p.~4, 1977.

\bibitem{Bern:1992em}
Z.~Bern, L.J.~Dixon and D.A.~Kosower, \emph{{Dimensionally regulated one loop
  integrals}}, \href{https://doi.org/10.1016/0370-2693(93)90400-C}{\emph{Phys.
  Lett. B} {\bfseries 302} (1993) 299}
  [\href{https://arxiv.org/abs/hep-ph/9212308}{{hep-ph/9212308}}].

\bibitem{Bern:1993kr}
Z.~Bern, L.J.~Dixon and D.A.~Kosower, \emph{{Dimensionally regulated pentagon
  integrals}}, \href{https://doi.org/10.1016/0550-3213(94)90398-0}{\emph{Nucl.
  Phys. B} {\bfseries 412} (1994) 751}
  [\href{https://arxiv.org/abs/hep-ph/9306240}{{hep-ph/9306240}}].

\bibitem{Panzer:2014caa}
E.~Panzer, \emph{{Algorithms for the symbolic integration of hyperlogarithms
  with applications to Feynman integrals}},
  \href{https://doi.org/10.1016/j.cpc.2014.10.019}{\emph{Comput. Phys. Commun.}
  {\bfseries 188} (2015) 148}
  [\href{https://arxiv.org/abs/1403.3385}{{1403.3385}}].

\bibitem{Larsen:2017aqb}
K.J.~Larsen and R.~Rietkerk, \emph{{MultivariateResidues: a Mathematica package
  for computing multivariate residues}},
  \href{https://doi.org/10.1016/j.cpc.2017.08.025}{\emph{Comput. Phys. Commun.}
  {\bfseries 222} (2018) 250}
  [\href{https://arxiv.org/abs/1701.01040}{{1701.01040}}].

\bibitem{Hidding:2017jkk}
M.~Hidding and F.~Moriello, \emph{{All orders structure and efficient
  computation of linearly reducible elliptic Feynman integrals}},
  \href{https://doi.org/10.1007/JHEP01(2019)169}{\emph{JHEP} {\bfseries 01}
  (2019) 169} [\href{https://arxiv.org/abs/1712.04441}{{1712.04441}}].

\bibitem{Wilhelm:2022wow}
M.~Wilhelm and C.~Zhang, \emph{{Symbology for elliptic multiple polylogarithms
  and the symbol prime}},
  \href{https://doi.org/10.1007/JHEP01(2023)089}{\emph{JHEP} {\bfseries 01}
  (2023) 089} [\href{https://arxiv.org/abs/2206.08378}{{2206.08378}}].

\bibitem{Silverman2009}
J.H.~Silverman, \emph{{The Arithmetic of Elliptic Curves, 2nd Edition}},
  Graduate Texts in Mathematics, Springer (2009),
  \href{https://doi.org/10.1007/978-0-387-09494-6}{10.1007/978-0-387-09494-6}.

\bibitem{ancillary_file_elliptic_ladders}
\url{https://erda.ku.dk/archives/79261af03bcac014086f909f1f527379/published-archive.html}.

\bibitem{Drummond:2010cz}
J.M.~Drummond, J.M.~Henn and J.~Trnka, \emph{{New differential equations for
  on-shell loop integrals}},
  \href{https://doi.org/10.1007/JHEP04(2011)083}{\emph{JHEP} {\bfseries 04}
  (2011) 083} [\href{https://arxiv.org/abs/1010.3679}{{1010.3679}}].

\bibitem{Drummond:2006rz}
J.M.~Drummond, J.~Henn, V.A.~Smirnov and E.~Sokatchev, \emph{{Magic identities
  for conformal four-point integrals}},
  \href{https://doi.org/10.1088/1126-6708/2007/01/064}{\emph{JHEP} {\bfseries
  01} (2007) 064}
  [\href{https://arxiv.org/abs/hep-th/0607160}{{hep-th/0607160}}].

\bibitem{He:2023qld}
S.~He and Y.~Tang, \emph{{Algorithm for symbol integrations for loop
  integrals}}, \href{https://doi.org/10.1103/PhysRevD.108.L041702}{\emph{Phys.
  Rev. D} {\bfseries 108} (2023) L041702}
  [\href{https://arxiv.org/abs/2304.01776}{{2304.01776}}].

\bibitem{Bourjaily:2015jna}
J.L.~Bourjaily and J.~Trnka, \emph{{Local Integrand Representations of All
  Two-Loop Amplitudes in Planar SYM}},
  \href{https://doi.org/10.1007/JHEP08(2015)119}{\emph{JHEP} {\bfseries 08}
  (2015) 119} [\href{https://arxiv.org/abs/1505.05886}{{1505.05886}}].

\bibitem{Bourjaily:2017wjl}
J.L.~Bourjaily, E.~Herrmann and J.~Trnka, \emph{{Prescriptive Unitarity}},
  \href{https://doi.org/10.1007/JHEP06(2017)059}{\emph{JHEP} {\bfseries 06}
  (2017) 059} [\href{https://arxiv.org/abs/1704.05460}{{1704.05460}}].

\bibitem{Loebbert:2019vcj}
F.~Loebbert, D.~M\"uller and H.~M\"unkler, \emph{{Yangian Bootstrap for
  Conformal Feynman Integrals}},
  \href{https://doi.org/10.1103/PhysRevD.101.066006}{\emph{Phys. Rev. D}
  {\bfseries 101} (2020) 066006}
  [\href{https://arxiv.org/abs/1912.05561}{{1912.05561}}].

\bibitem{Spradlin:2011wp}
M.~Spradlin and A.~Volovich, \emph{{Symbols of One-Loop Integrals From Mixed
  Tate Motives}}, \href{https://doi.org/10.1007/JHEP11(2011)084}{\emph{JHEP}
  {\bfseries 11} (2011) 084}
  [\href{https://arxiv.org/abs/1105.2024}{{1105.2024}}].

\bibitem{Drummond:2018caf}
J.~Drummond, J.~Foster, O.~G\"urdo\u{g}an and G.~Papathanasiou, \emph{{Cluster
  adjacency and the four-loop NMHV heptagon}},
  \href{https://doi.org/10.1007/JHEP03(2019)087}{\emph{JHEP} {\bfseries 03}
  (2019) 087} [\href{https://arxiv.org/abs/1812.04640}{{1812.04640}}].

\bibitem{Golden:2013xva}
J.~Golden, A.B.~Goncharov, M.~Spradlin, C.~Vergu and A.~Volovich,
  \emph{{Motivic Amplitudes and Cluster Coordinates}},
  \href{https://doi.org/10.1007/JHEP01(2014)091}{\emph{JHEP} {\bfseries 01}
  (2014) 091} [\href{https://arxiv.org/abs/1305.1617}{{1305.1617}}].

\bibitem{Golden:2014pua}
J.~Golden and M.~Spradlin, \emph{{A Cluster Bootstrap for Two-Loop MHV
  Amplitudes}}, \href{https://doi.org/10.1007/JHEP02(2015)002}{\emph{JHEP}
  {\bfseries 02} (2015) 002}
  [\href{https://arxiv.org/abs/1411.3289}{{1411.3289}}].

\bibitem{Drummond:2019qjk}
J.~Drummond, J.~Foster, O.~G\"urdo\u{g}an and C.~Kalousios, \emph{{Tropical
  Grassmannians, cluster algebras and scattering amplitudes}},
  \href{https://doi.org/10.1007/JHEP04(2020)146}{\emph{JHEP} {\bfseries 04}
  (2020) 146} [\href{https://arxiv.org/abs/1907.01053}{{1907.01053}}].

\bibitem{Drummond:2019cxm}
J.~Drummond, J.~Foster, O.~G\"urdo\u{g}an and C.~Kalousios, \emph{{Algebraic
  singularities of scattering amplitudes from tropical geometry}},
  \href{https://doi.org/10.1007/JHEP04(2021)002}{\emph{JHEP} {\bfseries 04}
  (2021) 002} [\href{https://arxiv.org/abs/1912.08217}{{1912.08217}}].

\bibitem{Jiang:2024eaj}
X.~Jiang, J.~Liu, X.~Xu and L.L.~Yang, \emph{{Symbol letters of Feynman
  integrals from Gram determinants}},
  \href{https://doi.org/10.1016/j.physletb.2025.139443}{\emph{Phys. Lett. B}
  {\bfseries 864} (2025) 139443}
  [\href{https://arxiv.org/abs/2401.07632}{{2401.07632}}].

\bibitem{Chen:2023kgw}
J.~Chen, B.~Feng and L.L.~Yang, \emph{{Intersection theory rules symbology}},
  \href{https://doi.org/10.1007/s11433-023-2239-8}{\emph{Sci. China Phys. Mech.
  Astron.} {\bfseries 67} (2024) 221011}
  [\href{https://arxiv.org/abs/2305.01283}{{2305.01283}}].

\bibitem{Hannesdottir:2021kpd}
H.S.~Hannesdottir, A.J.~McLeod, M.D.~Schwartz and C.~Vergu, \emph{{Implications
  of the Landau equations for iterated integrals}},
  \href{https://doi.org/10.1103/PhysRevD.105.L061701}{\emph{Phys. Rev. D}
  {\bfseries 105} (2022) L061701}
  [\href{https://arxiv.org/abs/2109.09744}{{2109.09744}}].

\bibitem{Dlapa:2023cvx}
C.~Dlapa, M.~Helmer, G.~Papathanasiou and F.~Tellander, \emph{{Symbol alphabets
  from the Landau singular locus}},
  \href{https://doi.org/10.1007/JHEP10(2023)161}{\emph{JHEP} {\bfseries 10}
  (2023) 161} [\href{https://arxiv.org/abs/2304.02629}{{2304.02629}}].

\bibitem{Fevola:2023kaw}
C.~Fevola, S.~Mizera and S.~Telen, \emph{{Landau Singularities Revisited:
  Computational Algebraic Geometry for Feynman Integrals}},
  \href{https://doi.org/10.1103/PhysRevLett.132.101601}{\emph{Phys. Rev. Lett.}
  {\bfseries 132} (2024) 101601}
  [\href{https://arxiv.org/abs/2311.14669}{{2311.14669}}].

\bibitem{Arkani-Hamed:talk}
N.~Arkani-Hamed, ``{talk in the conference Positive Geometries in Scattering
  Amplitudes and Beyond, Mainz Institute for Theoretical Physics, 2021}.''
  \url{https://indico.mitp.uni-mainz.de/event/236/overview}.

\bibitem{Yang:2022gko}
Q.~Yang, \emph{{Schubert problems, positivity and symbol letters}},
  \href{https://doi.org/10.1007/JHEP08(2022)168}{\emph{JHEP} {\bfseries 08}
  (2022) 168} [\href{https://arxiv.org/abs/2203.16112}{{2203.16112}}].

\bibitem{He:2022tph}
S.~He, J.~Liu, Y.~Tang and Q.~Yang, \emph{{Symbology of Feynman integrals from
  twistor geometries}},
  \href{https://doi.org/10.1007/s11433-023-2264-8}{\emph{Sci. China Phys. Mech.
  Astron.} {\bfseries 67} (2024) 231011}
  [\href{https://arxiv.org/abs/2207.13482}{{2207.13482}}].

\bibitem{He:2023umf}
S.~He, X.~Jiang, J.~Liu and Q.~Yang, \emph{{On symbology and differential
  equations of Feynman integrals from Schubert analysis}},
  \href{https://doi.org/10.1007/JHEP12(2023)140}{\emph{JHEP} {\bfseries 12}
  (2023) 140} [\href{https://arxiv.org/abs/2309.16441}{{2309.16441}}].

\bibitem{Gaiotto:2011dt}
D.~Gaiotto, J.~Maldacena, A.~Sever and P.~Vieira, \emph{{Pulling the straps of
  polygons}}, \href{https://doi.org/10.1007/JHEP12(2011)011}{\emph{JHEP}
  {\bfseries 12} (2011) 011}
  [\href{https://arxiv.org/abs/1102.0062}{{1102.0062}}].

\bibitem{Caron-Huot:2011zgw}
S.~Caron-Huot, \emph{{Superconformal symmetry and two-loop amplitudes in planar
  N=4 super Yang-Mills}},
  \href{https://doi.org/10.1007/JHEP12(2011)066}{\emph{JHEP} {\bfseries 12}
  (2011) 066} [\href{https://arxiv.org/abs/1105.5606}{{1105.5606}}].

\bibitem{Dennen:2015bet}
T.~Dennen, M.~Spradlin and A.~Volovich, \emph{{Landau Singularities and
  Symbology: One- and Two-loop MHV Amplitudes in SYM Theory}},
  \href{https://doi.org/10.1007/JHEP03(2016)069}{\emph{JHEP} {\bfseries 03}
  (2016) 069} [\href{https://arxiv.org/abs/1512.07909}{{1512.07909}}].

\bibitem{He:2020lcu}
S.~He, Z.~Li, Q.~Yang and C.~Zhang, \emph{{Feynman Integrals and Scattering
  Amplitudes from Wilson Loops}},
  \href{https://doi.org/10.1103/PhysRevLett.126.231601}{\emph{Phys. Rev. Lett.}
  {\bfseries 126} (2021) 231601}
  [\href{https://arxiv.org/abs/2012.15042}{{2012.15042}}].

\bibitem{Steinmann:1960}
O.~Steinmann, \emph{{\"Uber den Zusammenhang zwischen den Wightmanfunktionen
  und der retardierten Kommutatoren}}, {\emph{Helv. Physica Acta} {\bfseries
  33} (1960) 257}.

\bibitem{Steinmann2:1960}
O.~Steinmann, \emph{{Wightman-Funktionen und retardierten Kommutatoren. II}},
  {\emph{Helv. Physica Acta} {\bfseries 33} (1960) 347}.

\bibitem{Hodges:2009hk}
A.~Hodges, \emph{{Eliminating spurious poles from gauge-theoretic amplitudes}},
  \href{https://doi.org/10.1007/JHEP05(2013)135}{\emph{JHEP} {\bfseries 05}
  (2013) 135} [\href{https://arxiv.org/abs/0905.1473}{{0905.1473}}].

\bibitem{Mason:2009qx}
L.J.~Mason and D.~Skinner, \emph{{Dual Superconformal Invariance, Momentum
  Twistors and Grassmannians}},
  \href{https://doi.org/10.1088/1126-6708/2009/11/045}{\emph{JHEP} {\bfseries
  11} (2009) 045} [\href{https://arxiv.org/abs/0909.0250}{{0909.0250}}].

\bibitem{Schubert:1879}
H.~Schubert, \emph{{Kalkül der Abzählenden Geometrie}}, Verlag von B. G.
  Teubner (1879).

\bibitem{Elvang:2013cua}
H.~Elvang and Y.-t.~Huang, \emph{{Scattering Amplitudes in Gauge Theory and
  Gravity}}, Cambridge University Press (2015),
  \href{https://doi.org/10.1017/CBO9781107706620}{10.1017/CBO9781107706620},
  [\href{https://arxiv.org/abs/1308.1697}{{1308.1697}}].

\bibitem{Caron-Huot:2011dec}
S.~Caron-Huot and S.~He, \emph{{Jumpstarting the All-Loop S-Matrix of Planar
  N=4 Super Yang-Mills}},
  \href{https://doi.org/10.1007/JHEP07(2012)174}{\emph{JHEP} {\bfseries 07}
  (2012) 174} [\href{https://arxiv.org/abs/1112.1060}{{1112.1060}}].

\bibitem{Caron-Huot:2019bsq}
S.~Caron-Huot, L.J.~Dixon, F.~Dulat, M.~Von~Hippel, A.J.~McLeod and
  G.~Papathanasiou, \emph{{The Cosmic Galois Group and Extended Steinmann
  Relations for Planar $\mathcal{N} = 4$ SYM Amplitudes}},
  \href{https://doi.org/10.1007/JHEP09(2019)061}{\emph{JHEP} {\bfseries 09}
  (2019) 061} [\href{https://arxiv.org/abs/1906.07116}{{1906.07116}}].

\bibitem{Chicherin:2017bxc}
D.~Chicherin and E.~Sokatchev, \emph{{Conformal anomaly of generalized form
  factors and finite loop integrals}},
  \href{https://doi.org/10.1007/JHEP04(2018)082}{\emph{JHEP} {\bfseries 04}
  (2018) 082} [\href{https://arxiv.org/abs/1709.03511}{{1709.03511}}].

\bibitem{Spiering:2024sea}
A.~Spiering, M.~Wilhelm and C.~Zhang, \emph{{All Planar Two-Loop Amplitudes in
  Maximally Supersymmetric Yang-Mills Theory}},
  \href{https://doi.org/10.1103/PhysRevLett.134.071602}{\emph{Phys. Rev. Lett.}
  {\bfseries 134} (2025) 071602}
  [\href{https://arxiv.org/abs/2406.15549}{{2406.15549}}].

\bibitem{Hannesdottir:2022xki}
H.S.~Hannesdottir, A.J.~McLeod, M.D.~Schwartz and C.~Vergu, \emph{{Constraints
  on sequential discontinuities from the geometry of on-shell spaces}},
  \href{https://doi.org/10.1007/JHEP07(2023)236}{\emph{JHEP} {\bfseries 07}
  (2023) 236} [\href{https://arxiv.org/abs/2211.07633}{{2211.07633}}].

\bibitem{Hannesdottir:2024cnn}
H.S.~Hannesdottir, L.~Lippstreu, A.J.~McLeod and M.~Polackova, \emph{{Minimal
  Cuts and Genealogical Constraints on Feynman Integrals}},
  \href{https://arxiv.org/abs/2406.05943}{{2406.05943}}.

\bibitem{He:2024fij}
S.~He, X.~Jiang, J.~Liu and Q.~Yang, \emph{{Landau-based Schubert analysis}},
  \href{https://arxiv.org/abs/2410.11423}{{2410.11423}}.

\bibitem{Bern:2024vqs}
Z.~Bern, E.~Herrmann, R.~Roiban, M.S.~Ruf and M.~Zeng, \emph{{Global Bases for
  Nonplanar Loop Integrands, Generalized Unitarity, and the Double Copy to All
  Loop Orders}},  \href{https://arxiv.org/abs/2408.06686}{{2408.06686}}.

\bibitem{Bern:2021yeh}
{Z. Bern, J. Parra-Martinez, R. Roiban, M. S. Ruf, C.-H. Shen, M. P. Solon and
  M. Zeng}, \emph{{Scattering Amplitudes, the Tail Effect, and Conservative
  Binary Dynamics at ${\cal O}(G^4)$}},
  \href{https://doi.org/10.1103/PhysRevLett.128.161103}{\emph{Phys. Rev. Lett.}
  {\bfseries 128} (2022) 161103}
  [\href{https://arxiv.org/abs/2112.10750}{{2112.10750}}].

\bibitem{Dlapa:2022lmu}
C.~Dlapa, G.~K\"alin, Z.~Liu, J.~Neef and R.A.~Porto, \emph{{Radiation Reaction
  and Gravitational Waves at Fourth Post-Minkowskian Order}},
  \href{https://doi.org/10.1103/PhysRevLett.130.101401}{\emph{Phys. Rev. Lett.}
  {\bfseries 130} (2023) 101401}
  [\href{https://arxiv.org/abs/2210.05541}{{2210.05541}}].

\bibitem{Dlapa:2023hsl}
C.~Dlapa, G.~K\"alin, Z.~Liu and R.A.~Porto, \emph{{Bootstrapping the
  relativistic two-body problem}},
  \href{https://doi.org/10.1007/JHEP08(2023)109}{\emph{JHEP} {\bfseries 08}
  (2023) 109} [\href{https://arxiv.org/abs/2304.01275}{{2304.01275}}].

\bibitem{Damgaard:2023ttc}
P.H.~Damgaard, E.R.~Hansen, L.~Plant\'e and P.~Vanhove, \emph{{Classical
  observables from the exponential representation of the gravitational
  S-matrix}}, \href{https://doi.org/10.1007/JHEP09(2023)183}{\emph{JHEP}
  {\bfseries 09} (2023) 183}
  [\href{https://arxiv.org/abs/2307.04746}{{2307.04746}}].

\bibitem{Jakobsen:2023hig}
G.U.~Jakobsen, G.~Mogull, J.~Plefka and B.~Sauer, \emph{{Dissipative Scattering
  of Spinning Black Holes at Fourth Post-Minkowskian Order}},
  \href{https://doi.org/10.1103/PhysRevLett.131.241402}{\emph{Phys. Rev. Lett.}
  {\bfseries 131} (2023) 241402}
  [\href{https://arxiv.org/abs/2308.11514}{{2308.11514}}].

\bibitem{Bern:2023ccb}
{Z. Bern, E. Herrmann, R. Roiban, M.S. Ruf, A.V. Smirnov, V.A. Smirnov and M.
  Zeng}, \emph{{Conservative binary dynamics at order ${\cal O}(\alpha^5)$ in
  electrodynamics}},
  \href{https://doi.org/10.1103/PhysRevLett.132.251601}{\emph{Phys. Rev. Lett.}
  {\bfseries 132} (2024) 251601}
  [\href{https://arxiv.org/abs/2305.08981}{{2305.08981}}].

\bibitem{Ruf:2021egk}
M.~Ruf, \emph{{Precise Predictions for Gravitational Binary Systems from
  Scattering Amplitudes}},
  \href{https://doi.org/10.6094/UNIFR/223366}{\emph{Ph.D. thesis, Freiburg U.,
  Germany} (2021) }.

\bibitem{Chen:2024bpf}
G.~Chen, J.-W.~Kim and T.~Wang, \emph{{Systematic integral evaluation for
  spin-resummed binary dynamics}},
  \href{https://doi.org/10.1103/PhysRevD.111.L021701}{\emph{Phys. Rev. D}
  {\bfseries 111} (2025) L021701}
  [\href{https://arxiv.org/abs/2406.17658}{{2406.17658}}].

\bibitem{Cheung:2018wkq}
C.~Cheung, I.Z.~Rothstein and M.P.~Solon, \emph{{From Scattering Amplitudes to
  Classical Potentials in the Post-Minkowskian Expansion}},
  \href{https://doi.org/10.1103/PhysRevLett.121.251101}{\emph{Phys. Rev. Lett.}
  {\bfseries 121} (2018) 251101}
  [\href{https://arxiv.org/abs/1808.02489}{{1808.02489}}].

\bibitem{Damour:2017zjx}
T.~Damour, \emph{{High-energy gravitational scattering and the general
  relativistic two-body problem}},
  \href{https://doi.org/10.1103/PhysRevD.97.044038}{\emph{Phys. Rev. D}
  {\bfseries 97} (2018) 044038}
  [\href{https://arxiv.org/abs/1710.10599}{{1710.10599}}].

\bibitem{Neill:2013wsa}
D.~Neill and I.Z.~Rothstein, \emph{{Classical Space-Times from the S Matrix}},
  \href{https://doi.org/10.1016/j.nuclphysb.2013.09.007}{\emph{Nucl. Phys. B}
  {\bfseries 877} (2013) 177}
  [\href{https://arxiv.org/abs/1304.7263}{{1304.7263}}].

\bibitem{Beneke:1997zp}
M.~Beneke and V.A.~Smirnov, \emph{{Asymptotic expansion of Feynman integrals
  near threshold}},
  \href{https://doi.org/10.1016/S0550-3213(98)00138-2}{\emph{Nucl. Phys. B}
  {\bfseries 522} (1998) 321}
  [\href{https://arxiv.org/abs/hep-ph/9711391}{{hep-ph/9711391}}].

\bibitem{Landshoff:1969yyn}
P.V.~Landshoff and J.C.~Polkinghorne, \emph{{Iterations of regge cuts}},
  \href{https://doi.org/10.1103/PhysRev.181.1989}{\emph{Phys. Rev.} {\bfseries
  181} (1969) 1989}.

\bibitem{Kosower:2018adc}
D.A.~Kosower, B.~Maybee and D.~O'Connell, \emph{{Amplitudes, Observables, and
  Classical Scattering}},
  \href{https://doi.org/10.1007/JHEP02(2019)137}{\emph{JHEP} {\bfseries 02}
  (2019) 137} [\href{https://arxiv.org/abs/1811.10950}{{1811.10950}}].

\bibitem{Horvat2023}
{S. Horvát, J. Podkalicki, G. Csárdi, T. Nepusz, V. Traag, F. Zanini and D.
  Noom}, \emph{{IGraph/M: graph theory and network analysis for Mathematica}},
  \href{https://doi.org/10.21105/joss.04899}{\emph{Journal of Open Source
  Software} {\bfseries 8} (2023) 4899}
  [\href{https://arxiv.org/abs/2209.09145}{{2209.09145}}].

\bibitem{Frellesvig:2024ymq}
H.~Frellesvig, \emph{{The loop-by-loop Baikov representation \textemdash{}
  Strategies and implementation}},
  \href{https://doi.org/10.1007/JHEP04(2025)111}{\emph{JHEP} {\bfseries 04}
  (2025) 111} [\href{https://arxiv.org/abs/2412.01804}{{2412.01804}}].

\bibitem{Frellesvig:2021vdl}
H.~Frellesvig, C.~Vergu, M.~Volk and M.~von Hippel, \emph{{Cuts and
  Isogenies}}, \href{https://doi.org/10.1007/JHEP05(2021)064}{\emph{JHEP}
  {\bfseries 05} (2021) 064}
  [\href{https://arxiv.org/abs/2102.02769}{{2102.02769}}].

\bibitem{Henn:2020lye}
J.~Henn, B.~Mistlberger, V.A.~Smirnov and P.~Wasser, \emph{{Constructing d-log
  integrands and computing master integrals for three-loop four-particle
  scattering}}, \href{https://doi.org/10.1007/JHEP04(2020)167}{\emph{JHEP}
  {\bfseries 04} (2020) 167}
  [\href{https://arxiv.org/abs/2002.09492}{{2002.09492}}].

\bibitem{Bern:2004kq}
Z.~Bern, L.J.~Dixon and D.A.~Kosower, \emph{{N=4 super-Yang-Mills theory, QCD
  and collider physics}},
  \href{https://doi.org/10.1016/j.crhy.2004.09.007}{\emph{Comptes Rendus
  Physique} {\bfseries 5} (2004) 955}
  [\href{https://arxiv.org/abs/hep-th/0410021}{{hep-th/0410021}}].

\bibitem{Bern:2008qj}
Z.~Bern, J.J.M.~Carrasco and H.~Johansson, \emph{{New Relations for
  Gauge-Theory Amplitudes}},
  \href{https://doi.org/10.1103/PhysRevD.78.085011}{\emph{Phys. Rev. D}
  {\bfseries 78} (2008) 085011}
  [\href{https://arxiv.org/abs/0805.3993}{{0805.3993}}].

\bibitem{Cachazo:2017jef}
F.~Cachazo and A.~Guevara, \emph{{Leading Singularities and Classical
  Gravitational Scattering}},
  \href{https://doi.org/10.1007/JHEP02(2020)181}{\emph{JHEP} {\bfseries 02}
  (2020) 181} [\href{https://arxiv.org/abs/1705.10262}{{1705.10262}}].

\bibitem{Britto:2004nc}
R.~Britto, F.~Cachazo and B.~Feng, \emph{{Generalized unitarity and one-loop
  amplitudes in N=4 super-Yang-Mills}},
  \href{https://doi.org/10.1016/j.nuclphysb.2005.07.014}{\emph{Nucl. Phys. B}
  {\bfseries 725} (2005) 275}
  [\href{https://arxiv.org/abs/hep-th/0412103}{{hep-th/0412103}}].

\bibitem{Joyce:1972}
G.S.~Joyce, \emph{{Lattice Green function for the simple cubic lattice}},
  \href{https://doi.org/10.1088/0305-4470/5/8/001}{\emph{J. Phys. A: Gen.
  Phys.} {\bfseries 5} (1972) L65}.

\bibitem{Joyce:1973}
G.S.~Joyce, \emph{{On the simple cubic lattice Green function}},
  \href{https://doi.org/10.1098/rsta.1973.0018}{\emph{Philosophical
  Transactions of the Royal Society of London. Series A, Mathematical and
  Physical Sciences} {\bfseries 273} (1973) 583–610}.

\bibitem{Verrill:1996}
H.A.~Verrill, \emph{{Root lattices and pencils of varieties}},
  \href{https://doi.org/10.1215/kjm/1250518557}{\emph{Journal of Mathematics of
  Kyoto University} {\bfseries 36} (1996) 423 }.

\bibitem{Doran:1998hm}
C.F.~Doran, \emph{{Picard-Fuchs uniformization: Modularity of the mirror map
  and mirror moonshine}},
  \href{https://arxiv.org/abs/math/9812162}{{math/9812162}}.

\bibitem{Bogner:2013kvr}
M.~Bogner, \emph{{Algebraic characterization of differential operators of
  Calabi-Yau type}},  \href{https://arxiv.org/abs/1304.5434}{{1304.5434}}.

\bibitem{Ruf:talk}
M.S.~Ruf, ``{talk in the conference Amplitudes, CERN, 2023}.''
  \url{https://indico.cern.ch/event/1228963/overview}.

\bibitem{Brandhuber:2023hhy}
A.~Brandhuber, G.R.~Brown, G.~Chen, S.~De~Angelis, J.~Gowdy and G.~Travaglini,
  \emph{{One-loop gravitational bremsstrahlung and waveforms from a heavy-mass
  effective field theory}},
  \href{https://doi.org/10.1007/JHEP06(2023)048}{\emph{JHEP} {\bfseries 06}
  (2023) 048} [\href{https://arxiv.org/abs/2303.06111}{{2303.06111}}].

\bibitem{Herderschee:2023fxh}
A.~Herderschee, R.~Roiban and F.~Teng, \emph{{The sub-leading scattering
  waveform from amplitudes}},
  \href{https://doi.org/10.1007/JHEP06(2023)004}{\emph{JHEP} {\bfseries 06}
  (2023) 004} [\href{https://arxiv.org/abs/2303.06112}{{2303.06112}}].

\bibitem{Elkhidir:2023dco}
A.~Elkhidir, D.~O'Connell, M.~Sergola and I.A.~Vazquez-Holm, \emph{{Radiation
  and reaction at one loop}},
  \href{https://doi.org/10.1007/JHEP07(2024)272}{\emph{JHEP} {\bfseries 07}
  (2024) 272} [\href{https://arxiv.org/abs/2303.06211}{{2303.06211}}].

\bibitem{Georgoudis:2023lgf}
A.~Georgoudis, C.~Heissenberg and I.~Vazquez-Holm, \emph{{Inelastic
  exponentiation and classical gravitational scattering at one loop}},
  \href{https://doi.org/10.1007/JHEP06(2023)126}{\emph{JHEP} {\bfseries 06}
  (2023) 126} [\href{https://arxiv.org/abs/2303.07006}{{2303.07006}}].

\bibitem{Caron-Huot:2023vxl}
S.~Caron-Huot, M.~Giroux, H.S.~Hannesdottir and S.~Mizera, \emph{{What can be
  measured asymptotically?}},
  \href{https://doi.org/10.1007/JHEP01(2024)139}{\emph{JHEP} {\bfseries 01}
  (2024) 139} [\href{https://arxiv.org/abs/2308.02125}{{2308.02125}}].

\bibitem{10.1145/220346.220385}
A.~Barkatou, \emph{A rational version of moser's algorithm},  in
  \emph{Proceedings of the 1995 International Symposium on Symbolic and
  Algebraic Computation}, p.~297–302, Association for Computing Machinery,
  1995, \href{https://doi.org/10.1145/220346.220385}{DOI}.

\bibitem{10.1145/2755996.2756668}
M.A.~Barkatou and S.S.~Maddah, \emph{Removing apparent singularities of systems
  of linear differential equations with rational function coefficients},  in
  \emph{Proceedings of the 2015 ACM International Symposium on Symbolic and
  Algebraic Computation}, p.~53–60, Association for Computing Machinery,
  2015, \href{https://doi.org/10.1145/2755996.2756668}{DOI}.

\bibitem{TSAI2000747}
H.~Tsai, \emph{Weyl closure of a linear differential operator},
  \href{https://doi.org/10.1006/jsco.1999.0400}{\emph{Journal of Symbolic
  Computation} {\bfseries 29} (2000) 747}.

\bibitem{Abramov:2004}
S.A.~Abramov, M.A.~Barkatou and M.~Van~Hoeij, \emph{Apparent singularities of
  linear difference equations with polynomial coefficients},
  \href{https://doi.org/10.1007/s00200-005-0193-9}{\emph{Appl. Algebra Eng.,
  Commun. Comput.} {\bfseries 17} (2006) 117–133}
  [\href{https://arxiv.org/abs/math/0409508}{{math/0409508}}].

\bibitem{CHEN2016617}
S.~Chen, M.~Kauers and M.F.~Singer, \emph{Desingularization of ore operators},
  \href{https://doi.org/10.1016/j.jsc.2015.11.001}{\emph{Journal of Symbolic
  Computation} {\bfseries 74} (2016) 617}
  [\href{https://arxiv.org/abs/1408.5512}{{1408.5512}}].

\bibitem{Slavyanov:2016}
S.Y.~Slavyanov, D.A.~Satco, A.M.~Ishkhanyan and T.A.~Rotinyan,
  \emph{{Generation and removal of apparent singularities in linear ordinary
  differential equations with polynomial coefficients}},
  \href{https://doi.org/10.1134/S0040577916120059}{\emph{Theor. Math. Phys.}
  {\bfseries 189} (2016) 1726–1733}
  [\href{https://arxiv.org/abs/1606.01476}{{1606.01476}}].

\bibitem{10.1016/j.jsc.2019.02.009}
S.~Chen, M.~Kauers, Z.~Li and Y.~Zhang, \emph{Apparent singularities of
  d-finite systems}, \href{https://doi.org/10.1016/j.jsc.2019.02.009}{\emph{J.
  Symb. Comput.} {\bfseries 95} (2019) 217–237}
  [\href{https://arxiv.org/abs/1705.00838}{{1705.00838}}].

\bibitem{Kauers:2013juk}
M.~Kauers, M.~Jaroschek and F.~Johansson, \emph{Ore polynomials in sage},  in
  \emph{Computer Algebra and Polynomials: Applications of Algebra and Number
  Theory}, pp.~105--125, Springer International Publishing, 2015,
  \href{https://doi.org/10.1007/978-3-319-15081-9_6}{DOI}
  [\href{https://arxiv.org/abs/1306.4263}{{1306.4263}}].

\bibitem{Almkvist:2021}
G.~Almkvist and D.~van Straten, \emph{{Calabi-Yau operators of degree two}},
  \href{https://doi.org/10.1007/s10801-023-01272-0}{\emph{J. Algebr. Comb.}
  {\bfseries 58} (2023) 1203}
  [\href{https://arxiv.org/abs/2103.08651}{{2103.08651}}].

\bibitem{Candelas:1990rm}
P.~Candelas, X.C.~De~La~Ossa, P.S.~Green and L.~Parkes, \emph{{A Pair of
  Calabi-Yau manifolds as an exactly soluble superconformal theory}},
  \href{https://doi.org/10.1016/0550-3213(91)90292-6}{\emph{Nucl. Phys. B}
  {\bfseries 359} (1991) 21}.

\bibitem{Morrison:1991cd}
D.R.~Morrison, \emph{{Picard-Fuchs equations and mirror maps for
  hypersurfaces}}, {\emph{AMS/IP Stud. Adv. Math.} {\bfseries 9} (1998) 185}
  [\href{https://arxiv.org/abs/hep-th/9111025}{{hep-th/9111025}}].

\bibitem{vanStraten:2017}
D.~van Straten, \emph{{Calabi--Yau Operators}},  in \emph{{Uniformization,
  Riemann-Hilbert Correspondence, Calabi-Yau Manifolds, and Picard-Fuchs
  Equations}}, pp.~401--451, {Int. Press}, 2017
  [\href{https://arxiv.org/abs/1704.00164}{{1704.00164}}].

\bibitem{Almkvist:2005qoo}
G.~Almkvist, C.~van Enckevort, D.~van Straten and W.~Zudilin, \emph{{Tables of
  Calabi--Yau equations}},
  \href{https://arxiv.org/abs/math/0507430}{{math/0507430}}.

\bibitem{Pogel:2024sdi}
S.~P\"ogel, X.~Wang, S.~Weinzierl, K.~Wu and X.~Xu, \emph{{Self-dualities and
  Galois symmetries in Feynman integrals}},
  \href{https://doi.org/10.1007/JHEP09(2024)084}{\emph{JHEP} {\bfseries 09}
  (2024) 084} [\href{https://arxiv.org/abs/2407.08799}{{2407.08799}}].

\bibitem{Duhr:2024xsy}
C.~Duhr, F.~Porkert, C.~Semper and S.F.~Stawinski, \emph{{Self-duality from
  twisted cohomology}},
  \href{https://doi.org/10.1007/JHEP03(2025)053}{\emph{JHEP} {\bfseries 03}
  (2025) 053} [\href{https://arxiv.org/abs/2408.04904}{{2408.04904}}].

\bibitem{Remiddi:2013joa}
E.~Remiddi and L.~Tancredi, \emph{{Schouten identities for Feynman graph
  amplitudes; The Master Integrals for the two-loop massive sunrise graph}},
  \href{https://doi.org/10.1016/j.nuclphysb.2014.01.009}{\emph{Nucl. Phys. B}
  {\bfseries 880} (2014) 343}
  [\href{https://arxiv.org/abs/1311.3342}{{1311.3342}}].

\bibitem{Tancredi:2015pta}
L.~Tancredi, \emph{{Integration by parts identities in integer numbers of
  dimensions. A criterion for decoupling systems of differential equations}},
  \href{https://doi.org/10.1016/j.nuclphysb.2015.10.015}{\emph{Nucl. Phys. B}
  {\bfseries 901} (2015) 282}
  [\href{https://arxiv.org/abs/1509.03330}{{1509.03330}}].

\end{thebibliography}\endgroup
%\bibliography{Ref}
\bibliographystyle{JHEP}
%% ------------------------------------------------------------------------------------------------

\end{document}